\documentclass[final,letterpaper,12pt,oneside]{class_diss}

\usepackage{graphicx} 
\usepackage{amsmath} 
\usepackage{amsxtra} 
\usepackage{amssymb} 
\usepackage{amsthm} 
\usepackage{latexsym} 
\usepackage{setspace} 
\usepackage[margin=1in]{geometry} 
\usepackage[titles]{tocloft} 
\usepackage[cal=boondox]{mathalfa}
\usepackage{threeparttable}
\usepackage{booktabs}
\usepackage[figuresright]{rotating}
\usepackage{multirow}
\usepackage{subfig}
\usepackage{comment}
\usepackage{lineno}
\usepackage{lscape}
\usepackage{placeins}
\usepackage{longtable}

\newcommand{\be}{\begin{equation}}
\newcommand{\ee}{\end{equation}}
\newcommand{\bea}{\begin{eqnarray}}
\newcommand{\eea}{\end{eqnarray}}
\newcommand{\hunit}{$\rm{km \ s^{-1} \ Mpc^{-1}}$}
\newcommand{\lcdm}{$\Lambda$CDM}
\newcommand{\pcdm}{$\phi$CDM}
\usepackage{array}

\newcommand{\hiig}{H\,\textsc{ii}G}
\newcommand{\hii}{H\,\textsc{ii}}
\newcommand{\Om}{\Omega_{m0}}
\newcommand{\Ok}{\Omega_{k0}}
\newcommand{\om}{$\Omega_{m0}$}
\newcommand{\ok}{$\Omega_{k0}$}
\newcommand{\wx}{$w_{\rm X}$}
\newcommand{\wX}{w_{\rm X}}
\newcommand{\mq}{Mg\,\textsc{ii} QSO}
\newcommand{\mii}{Mg\,\textsc{ii}}
\newcommand{\cq}{C\,\textsc{iv} QSO}
\newcommand{\civ}{C\,\textsc{iv}}
\newcommand{\rfe}{${\cal R}_{\rm{Fe\,\textsc{ii}}}$}
\newcommand{\Feii}{Fe\,\textsc{ii}}
\newcommand{\obh}{\Omega_{b}h^2}
\newcommand{\och}{\Omega_{c}h^2}
\newcommand{\onh}{\Omega_{\nu}h^2}
\newcommand{\onhs}{$\Omega_{\nu}h^2$}
\newcommand{\obhs}{$\Omega_{b}h^2$}
\newcommand{\ochs}{$\Omega_{c}h^2$}

\usepackage[T1]{fontenc}
\usepackage{scalerel}
\usepackage{tikz}
\usetikzlibrary{svg.path}

\definecolor{orcidlogocol}{HTML}{A6CE39}
\tikzset{
  orcidlogo/.pic={
    \fill[orcidlogocol] svg{M256,128c0,70.7-57.3,128-128,128C57.3,256,0,198.7,0,128C0,57.3,57.3,0,128,0C198.7,0,256,57.3,256,128z};
    \fill[white] svg{M86.3,186.2H70.9V79.1h15.4v48.4V186.2z}
                 svg{M108.9,79.1h41.6c39.6,0,57,28.3,57,53.6c0,27.5-21.5,53.6-56.8,53.6h-41.8V79.1z M124.3,172.4h24.5c34.9,0,42.9-26.5,42.9-39.7c0-21.5-13.7-39.7-43.7-39.7h-23.7V172.4z}
                 svg{M88.7,56.8c0,5.5-4.5,10.1-10.1,10.1c-5.6,0-10.1-4.6-10.1-10.1c0-5.6,4.5-10.1,10.1-10.1C84.2,46.7,88.7,51.3,88.7,56.8z};
  }
}

\newcommand\orcidicon[1]{\href{https://orcid.org/#1}{\mbox{\scalerel*{
\begin{tikzpicture}[yscale=-1,transform shape]
\pic{orcidlogo};
\end{tikzpicture}
}{|}}}}

\usepackage[authoryear,sort&compress]{natbib}
\usepackage{mathtools}
\setcitestyle{authoryear}

\usepackage[pdftex, plainpages=false, pdfpagelabels]{hyperref}

\hypersetup{
    linktocpage=true,
    colorlinks=true,
    citecolor=blue,
    urlcolor=blue,
    linkcolor=blue,
    citebordercolor={1 0 0},
    urlbordercolor={1 0 0},
    linkbordercolor={.7 .8 .8},
    }

\begin{document}




\thispagestyle{empty}


\pdfbookmark[0]{Title Page}{PDFTitlePage}

\begin{center}

   \vspace{1cm}


   \large Cosmological constraints from standardized non-CMB observations\\

   \vspace{0.5cm}

   by\\

   \vspace{0.5cm}


   \large Shulei Cao\\

   \vspace{0.5cm}


   M.S., Beijing Normal University, 2016\\

   \vspace{0.55cm}
   \rule{2in}{0.5pt}\\
   \vspace{0.75cm}

   {\large AN ABSTRACT OF A DISSERTATION}\\

   \vspace{0.5cm}
   \begin{singlespace}
   submitted in partial fulfillment of the\\
   requirements for the degree\\
   \end{singlespace}

   \vspace{0.5cm}


   {\large DOCTOR OF PHILOSOPHY}\\
   \vspace{0.5cm}


   \begin{singlespace}
   Department of Physics\\
   College of Arts and Sciences\\
   \end{singlespace}

   \vspace{0.5cm}

   \begin{singlespace}
   {\Large KANSAS STATE UNIVERSITY}\\
   Manhattan, Kansas\\
   \end{singlespace}


   2023\\
   \vspace{1cm}

\end{center}

\begin{abstract} 
   \setcounter{page}{-1} 
   \pdfbookmark[0]{Abstract}{PDFAbstractPage} 

\pagestyle{empty}
\setlength{\baselineskip}{0.8cm}



The current expansion of the Universe has been observed to be accelerating, and the widely accepted spatially-flat concordance model of general relativistic cosmology attributes this phenomenon to a constant dark energy, a cosmological constant, which is measured to comprise about 70\% of the total energy budget of the current Universe. However, observational discrepancies and theoretical puzzles have raised questions about this model, suggesting that alternative cosmological models with non-zero spatial curvature and/or dark energy dynamics might provide better explanations.

To explore these possibilities, we have conducted a series of studies using standardized, lower-redshift observations to constrain six different cosmological models with varying degrees of flatness and dark energy dynamics. Through comparing these observations with theoretical predictions, we aim to deepen our understanding of the evolution of the Universe and shed new light on its mysteries. Our data provide consistent cosmological constraints across all six models, with some suggesting the possibility of mild dark energy dynamics and slight spatial curvature. However, these joint constraints do not rule out the possibility of dark energy being a cosmological constant and the spatial hypersurfaces being flat. Overall, our findings contribute to the ongoing efforts to refine our understanding of the Universe and its properties, and suggest that multiple cosmological models remain viable.

In addition, our research demonstrates that certain types of astronomical observations, such as gamma ray bursts and reverberation-mapped quasar measurements, can be standardized and utilized in cosmology. This allows us to gain valuable insights into the properties of the Universe and its evolution by using these observations as distance indicators. By using these new standardized observations to constrain our cosmological models, we can improve our understanding of the nature of dark energy and other fundamental components of the Universe.
   \vfill 
\end{abstract} 



\newpage


\thispagestyle{empty}


\begin{center}

   \vspace{1cm}


\large Cosmological constraints from standardized non-CMB observations\\

\vspace{0.5cm}

by\\

\vspace{0.5cm}


\large Shulei Cao\\

 \vspace{0.3cm}


   M.S., Beijing Normal University, 2016\\

   \vspace{0.35cm}
   \rule{2in}{0.5pt}\\
   \vspace{0.65cm}

   {\large A DISSERTATION}\\

   \vspace{0.3cm}
   \begin{singlespace}
   submitted in partial fulfillment of the\\
   requirements for the degree\\
   \end{singlespace}

   \vspace{0.3cm}


   {\large DOCTOR OF PHILOSOPHY}\\
   \vspace{0.3cm}


   \begin{singlespace}
   Department of Physics\\
   College of Arts and Sciences\\
   \end{singlespace}

   \vspace{0.3cm}

   \begin{singlespace}
   {\large KANSAS STATE UNIVERSITY}\\
   Manhattan, Kansas\\
   \end{singlespace}


   2023\\
   \vspace{0.3cm}

    \end{center}

    \begin{flushright}
    Approved by:\\
    \vspace{0.3cm}
    \begin{singlespace}
    Major Professor


    Bharat Ratra\\
    \end{singlespace}
    \end{flushright}





\newpage

\thispagestyle{empty}

\vspace*{0.9cm}

\begin{center}

{\bf \Huge Copyright}

\vspace{1cm}


\Large\copyright\ Shulei Cao 2023.\\

\vspace{0.5cm}

\end{center}


\begin{abstract}



\vfill
\end{abstract}


\newpage
\pagenumbering{roman}


\setcounter{page}{6}


\pdfbookmark[0]{\contentsname}{contents}


\renewcommand{\cftchapleader}{\cftdotfill{\cftdotsep}}


\renewcommand{\cftchapfont}{\mdseries}
\renewcommand{\cftchappagefont}{\mdseries}



\tableofcontents
\listoffigures
\listoftables




\newpage
\phantomsection
\addcontentsline{toc}{chapter}{Acknowledgements}

\newpage
\vspace*{0.9cm}
\begin{center}
{\bf \Huge Acknowledgments}
\end{center}

\setlength{\baselineskip}{0.8cm}



I am deeply grateful to my major professor, Dr. Bharat Ratra, for his unwavering support, exceptional guidance, and invaluable mentorship throughout my doctoral journey. Dr. Ratra's expertise, patience, and constant encouragement have been instrumental in shaping me as a researcher and scholar. I owe a great deal of my academic success to his insightful feedback and constructive criticism. 

I would also like to express my gratitude to my committee members, Dr. Glenn Horton-Smith, Dr. Lado Samushia, and Dr. David Auckly, for their unwavering support, including their attendance at my oral exam and defense. Their presence and thoughtful questions helped me to better understand and articulate my research, and their feedback and guidance were invaluable in ensuring its success.

In addition, I would like to thank my co-authors, Joseph Ryan, Maria Dainotti, Michal Zaja\v{c}ek, Swayamtrupta Panda, Narayan Khadka, Mary Loli Mart\'inez-Aldama, and Bo\.zena Czerny, for their valuable contributions to this project. Our collaboration has been an enriching and rewarding experience, and I look forward to continuing to work with them in the future.

Furthermore, I extend my thanks to Javier de Cruz P\'{e}rez, Chan-Gyung Park, Ana Luisa Gonz\'{a}lez-Mor\'{a}n, Ricardo Ch\'{a}vez, F. Y. Wang, and J. P. Hu for their insightful discussions on data analysis. I am also grateful for the financial support received from the US Department of Energy grant DE-SC0011840 and to the Beocat Research Cluster at Kansas State University, funded in part by NSF grants CNS-1006860, EPS-1006860, EPS-0919443, ACI-1440548, CHE-1726332, and NIH P20GM113109, for providing the computational resources necessary for these projects.

My sincere gratitude also goes to the Department of Physics at Kansas State University, where I conducted the research for this thesis. Throughout my PhD journey, I had the privilege of learning from and working alongside exceptional professors who provided invaluable guidance and support. My fellow graduate students also deserve special thanks for their camaraderie and encouragement, which made this experience all the more enriching. I am particularly grateful to Dr. Lado Samushia, who supported me and wrote me a recommendation letter to help me find a job, and to Dr. Mick O’Shea and Kim Coy, whose unwavering commitment to the well-being and success of graduate students is truly commendable. They graciously lent an ear to our concerns and provided invaluable assistance whenever needed. Their dedication to fostering a supportive and inclusive academic community is deeply appreciated.

Finally, I wish to express my deep appreciation to my parents, Xing-Hong Cao and Bai-Lian Zhang, and my sister, Rong-Ju Gao. Their unconditional love, encouragement, and support have been my constant source of inspiration throughout this journey. This work is dedicated to them.






\newpage
\pagenumbering{arabic}
\setcounter{page}{1}



\cleardoublepage


\chapter{Introduction}
\label{introduction}

Physical cosmology is the branch of physics, astrophysics, and astronomy that studies the structure, origin, evolution, and overall properties of the Universe on large scales. It seeks to answer fundamental questions about the Universe, such as its age, size, shape, composition, and how it has changed over time.

Physical cosmology is based on the principles of general relativity and the other laws of physics, and it uses observations and data from various sources, including telescopes and spacecraft, to test and refine models of the Universe. It also uses theoretical models and computer simulations to understand the behavior of the Universe on a large scale.

Some of the key areas of study in physical cosmology include the Big Bang model, dark matter and dark energy, the cosmic microwave background (CMB) radiation, the formation and evolution of galaxies and galaxy clusters, and the large-scale structure of the Universe. Physical cosmology is a rapidly evolving field, with new discoveries and insights emerging all the time.

Theoretical aspects of physical cosmology are discussed in this chapter, with a focus on the significance of distance measures in establishing links between theory and observation.

\section{General relativity}
\label{makereference1.1}

\subsection{Tensors in differential geometry}
\label{makereference1.1.1}

Differential geometry is a branch of mathematics that studies the geometry of curved spaces and the properties of objects such as curves and surfaces in these spaces. It uses the tools of calculus, linear algebra, and topology to study the properties of curves, surfaces, and other geometric objects in a rigorous and precise way.

At the heart of differential geometry is the study of manifolds, which are spaces that locally look like Euclidean space but may have global properties that are significantly different. Differential geometry is concerned with the geometric properties of manifolds and their relationships with other mathematical objects such as vector fields, differential forms, and connections.

One of the central concepts in differential geometry is the notion of a tangent space, which is a vector space that is tangent to a manifold at a particular point. Tangent spaces allow us to define notions such as tangent vectors and tangent bundles, which are crucial for describing the geometry of curves and surfaces on manifolds.

In differential geometry, a vector is defined as an element of a tangent space at a point on a manifold. A tangent space is a vector space that is tangent to the manifold at that point. A vector in differential geometry is denoted as a directed line segment with a magnitude and a direction. However, unlike in Euclidean geometry where vectors are represented as ordered pairs or triplets of numbers, in differential geometry, a vector is represented as a differential operator acting on smooth functions defined on the manifold.

A manifold is a topological space that is locally homeomorphic to Euclidean space. This means that every point on the manifold has a neighborhood that can be mapped onto an open set in Euclidean space by a continuous function called a chart (coordinate system). Manifolds can be of any dimension, and they can be either finite or infinite in size. For example, a curve in the plane or a surface in three-dimensional space is a two-dimensional manifold, while a sphere is a two-dimensional manifold that is closed (i.e., compact without boundary).

Given a point on a manifold, a vector at that point can be thought of as a directional derivative of a function defined on the manifold. In other words, a vector at a point specifies the rate and direction of change of a function as one moves in that direction from the point.

In a coordinate basis $\{X_{\mu}\coloneqq \partial/\partial x^{\mu}\}$ with dimension of $n$, an arbitrary tangent vector $v$ can be expressed as
\begin{equation}
    v=\sum^n_{\mu=1} v^{\mu}X_{\mu}\equiv v^{\mu}X_{\mu}=v^{\mu}\frac{\partial}{\partial x^{\mu}},
\end{equation}
where $v^{\mu}$ is the $\mu$th component of $v$ in $\{X_{\mu}\}$, $\mu=1,2,\dots,n$, and the equivalent sign implies the Einstein summation convention of implicitly summing over pairs of repeated upper and lower indices that is used throughout the thesis. By choosing a new coordinate basis $\{X^{\prime}_{\nu}\}$, $X_{\mu}$ can be expressed as
\begin{equation}
\label{Xu}
    X_{\mu}=\frac{\partial x^{\prime}_{\nu}}{\partial x^{\mu}}X^{\prime}_{\nu},
\end{equation}
so that the $\nu$th component of $v$ in $\{X^{\prime}_{\nu}\}$
\begin{equation}
\label{vl}
    v^{\prime\nu}=v^{\mu}\frac{\partial x^{\prime}_{\nu}}{\partial x^{\mu}}.
\end{equation}
This transformation law allows us to compare vectors at different points on a manifold and to define vector fields, which are collections of vectors defined on a manifold.

A dual vector space $V^{*}$ (with a dual basis of $v^{1*},v^{2*},\dots,v^{n*}\in V^{*}$) to a tangent vector space $V$ (with a basis of $v_{1},v_{2},\dots,v_{n}\in V$) is defined as
\begin{equation}
    v^{\mu*}(v_{\nu})=\delta^{\mu}_{\;\,\nu}=
    \begin{cases}
    1, & \text{if}\ \mu=\nu , \\
    0, & \text{else}.
    \end{cases}
\end{equation}

A tensor, $T$, of type $(k,l)$ over $V$ is a multilinear map~\citep{Wald:1984rg}
\begin{equation}
    T:\underbrace{V^{*}\times \dots\times V^{*}}_{k}\times\underbrace{V\times \dots\times V}_{l}\rightarrow\mathbb{R},
\end{equation}
where $\mathbb{R}$ is the set of real numbers. Thus, a tensor of type $(0,1)$ or $(1,0)$ is a dual (cotangent) vector or an ordinary (tangent) vector. The contraction operation with respect to $i$th dual vector and $j$th vector, defined below, is a map $C:\mathcal{T}(k,l)\rightarrow\mathcal{T}(k-1,l-1)$, where $\mathcal{T}(k,l)$ is a vector space with a collection of all tensors of type $(k,l)$, and if $T\in\mathcal{T}(k,l)$, then
\begin{equation}
\label{CT}
CT=\sum^n_{\alpha}T(\dots,\overbrace{v^{\alpha*}}^{i\text{th}},\dots;\dots,\underbrace{v_{\alpha}}_{j\text{th}},\dots),
\end{equation}
where $\{v_\mu\}$ is a basis of $V$ and $\{v^{\nu*}\}$ is its dual basis. The outer product of $T$ and $T^{\prime}$ is denoted as $S=T\otimes T^{\prime}$. The $n^{k+l}$ simple tensors $\{v_{\mu_1}\otimes\cdots\otimes v_{\mu_k}\otimes v^{\nu_1*}\otimes\cdots\otimes v^{\nu_l*}\}$ is a basis of $\mathcal{T}(k,l)$, then for $T\in\mathcal{T}(k,l)$
\begin{equation}
T=T^{\mu_1\cdots\mu_k}_{\qquad\ \nu_1\cdots\nu_l}v_{\mu_1}\otimes\cdots\otimes v_{\mu_k}\otimes v^{\nu_1*}\otimes\cdots\otimes v^{\nu_l*},
\end{equation}
where $T^{\mu_1\cdots\mu_k}_{\qquad\ \nu_1\cdots\nu_l}$ are called the components of the tensor $T$ with respect to the basis ${v_\mu}$. Consequently, the components of $CT$ in equation \eqref{CT}
\begin{equation}
    (CT)^{\mu_1\cdots\mu_{k-1}}_{\qquad\quad\nu_1\cdots\nu_{l-1}}=T^{\mu_1\cdots\alpha\cdots\mu_{k-1}}_{\qquad\qquad\ \nu_1\cdots\alpha\cdots\nu_{l-1}},
\end{equation}
and the outer product $S=T\otimes T^{\prime}$ has components
\begin{equation}
    S^{\mu_1\cdots\mu_{k+k^{\prime}}}_{\qquad\quad\ \nu_1\cdots\nu_{l+l^{\prime}}}=T^{\mu_1\cdots\mu_k}_{\qquad\ \nu_1\cdots\nu_l}T^{\prime}{}^{\mu_{k+1}\cdots\mu_{k+k^{\prime}}}_{\qquad\qquad\nu_{l+1}\cdots\nu_{l+l^{\prime}}}.
\end{equation}

The dual basis of $\{X_{\mu}\}$ in equation \eqref{Xu} is $\{dx^\mu\}$ since $dx^\mu(\partial/\partial x^\nu)=\delta^{\mu}_{\;\,\nu}$. Therefore, for a dual vector $\omega$ with components of $\omega_{\mu}$ in $\{dx^\mu\}$, the components in a new basis of $\{dx^\prime{}^{\mu^\prime}\}$ become
\begin{equation}
    \omega^\prime_{\mu^\prime}=\omega_{\mu}\frac{\partial x^\mu}{\partial x^\prime{}^{\mu^\prime}}.
\end{equation}
Consequently, in general, the components of a type $(k,l)$ tensor $T$ follow the tensor transformation law
\begin{equation}
\label{ttl}
    T^\prime{}^{\mu^\prime_1\cdots\mu^\prime_k}_{\qquad\,\nu^\prime_1\cdots\nu^\prime_l}=T^{\mu_1\cdots\mu_k}_{\qquad\ \nu_1\cdots\nu_l}\frac{\partial x^\prime{}^{\mu_1^\prime}}{\partial x^{\mu_1}}\cdots\frac{\partial x^{\nu_l}}{\partial x^\prime{}^{\nu_l^\prime}}.
\end{equation}
A tensor field is defined as an assignment of a tensor over $V_p$ for every point $p$ in the manifold $M$.

Following~\cite{Wald:1984rg}, we use the following abstract index notation rules. A type $(k,l)$ tensor is denoted by a letter followed by $k$ contravariant (tangent) and $l$ covariant (cotangent) indices. These indices are represented using lower case Latin letters, and the tensor itself is denoted as $T^{a_1\cdots a_k}_{\qquad\,b_1\cdots b_l}$. The contraction of a tensor is indicated by repeating the letters of the contracted slots. For instance, the type $(2,2)$ tensor $T^{abc}_{\quad\, bef}$ is obtained by contracting the second contravariant slot and the first covariant slot of $T^{abc}_{\quad\, def}$. The outer product of the tensors $T^{abc}_{\quad\, def}$ and $S^{ab}_{\ \ c}$ is simply $T^{abc}_{\quad def}S^{gh}_{\ \ i}$. The labels of components are indicated by using Greek letters. For example, $T^{\alpha\beta\gamma}_{\quad\ \sigma\mu\nu}$ is a basis component of the tensor $T^{abc}_{\quad\, def}$.


\subsection{Metric tensor and curvature}
\label{makereference1.1.2}

A metric $g$ on a manifold $M$ is a type $(0,2)$ tensor field that is nondegenerate and symmetric. Nondegeneracy means that $g(v,v_2)=0$ for all $v\in V_p$ if and only if $v_2=0$. Symmetry means that $g(v_1,v_2)=g(v_2,v_1)$ for any $v_1,v_2\in V_p$. Geometrically, the metric $g$ can be interpreted as an ``infinitesimal squared distance'' or a (not necessarily positive definite) inner product on the tangent space at each point.

In a coordinate basis of $\{dx^\mu\}$, the metric tensor $g_{ab}$ (or equivalently the line element $ds^2$) is
\begin{equation}
    g_{ab}\equiv ds^2=g_{\mu\nu}dx^{\mu}dx^{\nu}.
\end{equation}
One can always find an orthonormal basis $\{v_{\mu}\}$ so that $g(v_\alpha,v_\beta)=\pm\delta_{\alpha\beta}$. The signature of a metric is determined by counting the number of ``$+$'' and ``$-$'' signs that appear in this expression. Metrics with all ``$+$'' signs are called Euclidean (or Riemannian in the context of general relativity, \citealp{Wald:1984rg}), while those with signatures like the spacetime metric ($-,+,+,+$) are called Lorentzian. Lorentzian metrics are commonly used in general relativity to describe the geometry of spacetime.

A derivative operator $\nabla$ (or covariant derivative) on a manifold $M$ is a linear map that takes each smooth (or differentiable) type $(k,l)$ tensor field to a smooth type $(k,l+1)$ tensor field. It satisfies the Leibniz rule, commutativity with contraction, consistency with the notation of tangent vectors as directional derivatives on scalar fields, and sometimes the torsion-free condition. Geometrically, the covariant derivative provides a way of differentiating tensor fields on curved manifolds without reference to a specific coordinate system.

The commutator of two vector fields $v^a$ and $\omega^b$ in terms of any derivative operator $\nabla_a$ is
\begin{equation}
    [v,\omega]^b=v^a\nabla_a\omega^b-\omega^a\nabla_a v^b.
\end{equation}
For any two derivative operators $\tilde{\nabla}$ and $\nabla$ defined on a smooth manifold, there exists a tensor field $C^{c}_{\ ab}\,(=C^{c}_{\ ba})$ of type $(1,2)$ such that, for $T\in\mathcal{T}(k,l)$ the following identity holds
\begin{equation}
    \nabla_aT^{b_1\cdots b_k}_{\qquad\,c_1\cdots c_l}=\tilde\nabla_aT^{b_1\cdots b_k}_{\qquad\,c_1\cdots c_l}+\sum_iC^{b_i}_{\;\ ad}T^{b_1\cdots d\cdots b_k}_{\qquad\quad c_1\cdots c_l}-\sum_iC^{d}_{\ ac_j}T^{b_1\cdots b_k}_{\qquad c_1\cdots d\cdots c_l}.
\end{equation}
One significant use of the equation mentioned above arises when the operator $\tilde\nabla_a$ is substituted by the conventional derivative operator $\partial_a$. In this case, the tensor field $C^{c}_{\ ab}$ is denoted as the Christoffel symbol $\Gamma^{c}_{\ ab}$. For example, the equation
\begin{equation}
    \nabla_a v^b=\partial_a v^b+\Gamma^{b}_{\ ac}v^c.
\end{equation}
expresses the covariant derivative of the vector field $v^b$ in terms of its partial derivatives and the Christoffel symbol. This equation plays a crucial role in the study of curved spacetimes, as it allows for the generalization of the concept of differentiation to curved manifolds. A Christoffel symbol $\Gamma^{c}_{\ ab}$ is a tensor field associated with the derivative operator $\nabla_a$ and the coordinate system used to define $\partial_a$. It is important to note that a different coordinate system with a different ordinary derivative operator $\partial^\prime_a$ would have a different Christoffel symbol $\Gamma^\prime{}^{c}_{\ ab}$.

Parallel transport of a vector $v^a$ along a curve $C$ with a tangent $t^a$ can be defined by
\begin{equation}
\label{paratrans}
    t^a\nabla_av^b=t^a\partial_av^b+t^a\Gamma^{b}_{\ ac}v^c=0,
\end{equation}
and in terms of components in the coordinate basis and parameter $t$ along the curve, it takes the form
\begin{equation}
\label{paratranscomp}
    \frac{dv^\nu}{dt}+t^\mu\Gamma^{\nu}_{\ \mu\lambda}v^\lambda=0,
\end{equation}
where $v^\nu$ are the components of the vector $v^a$ in the coordinate basis. More generally, the parallel transport of an arbitrary type $(k,l)$ tensor $T\in\mathcal{T}(k,l)$ can be defined as
\begin{equation}
    t^a\nabla_aT^{b_1\cdots b_k}_{\qquad\,c_1\cdots c_l}=0.
\end{equation}

Let $g_{ab}$ be a metric on a manifold and let $v^a$ and $\omega^a$ be two vectors on the manifold. The inner product of $g_{ab}v^a\omega^b$ is a scalar function on the manifold. If we parallel-transport $v^a$ and $\omega^a$ along any curve, the inner product remains unchanged, which means that its covariant derivative with respect to any tangent vector $t^a$ must be zero: $t^a\nabla_a(g_{bc}v^b\omega^c)=0$. Now, suppose that $v^a$ and $\omega^a$ follow the parallel transport equation \eqref{paratrans}. Since $t^a$ is an arbitrary tangent vector and $v^a$ and $\omega^a$ are non-zero, we must have $\nabla_ag_{bc}=0$. Consequently
\begin{equation}
    \nabla_ag_{bc}=0\Rightarrow\tilde\nabla_ag_{bc}-C^{d}_{\ ab}g_{dc}-C^{d}_{\ ac}g_{bd}\Rightarrow C^{d}_{\ ab}=\frac{1}{2}g^{cd}\{\tilde\nabla_a g_{bd}+\tilde\nabla_b g_{ad}-\tilde\nabla_d g_{ab}\}.
\end{equation}
Therefore, we have shown that if $\nabla_ag_{bc}=0$, then there exists a unique derivative operator $\nabla_a$ satisfying this condition with the above choice of $C^{c}_{\ ab}$. If we adopt $\tilde\nabla_a$ as the coordinate derivative $\partial_a$, then the corresponding Christoffel symbols can be computed as
\begin{equation}
    \Gamma^{d}_{\ ab}=\frac{1}{2}g^{cd}\{\partial_a g_{bd}+\partial_b g_{ad}-\partial_d g_{ab}\},
\end{equation}
which implies that the coordinate components of $\Gamma^{\alpha}_{\ \mu\nu}$ are given by
\begin{equation}
\label{christoffel}
    \Gamma^{\alpha}_{\ \mu\nu}=\frac{1}{2}g^{\alpha\lambda}\{\partial_\mu g_{\nu\lambda}+\partial_\nu g_{\mu\lambda}-\partial_\lambda g_{\mu\nu}\}=\frac{1}{2}g^{\alpha\lambda}\left\{\frac{\partial g_{\nu\lambda}}{\partial x^\mu}+\frac{\partial g_{\mu\lambda}}{\partial x^\nu}-\frac{\partial g_{\mu\nu}}{\partial x^\lambda}\right\}.
\end{equation}
Here, $x^\mu$ represents the coordinates of a point on the manifold, and the indices $\alpha, \mu,$ and $\nu$ range over all possible values.

Let $\nabla_a$ denote a derivative operator, $\omega_a$ a dual vector field, and $f$ a smooth function. The action of two derivative operators onto $f\omega_c$ can be expressed as
\begin{equation}
    \nabla_a\nabla_b(f\omega_c)=\nabla_a(\omega_c\nabla_bf+f\nabla_b\omega_c)=(\nabla_a\nabla_bf)\omega_c+\nabla_bf\nabla_a\omega_c+\nabla_af\nabla_b\omega_c+f\nabla_a\nabla_b\omega_c.
\end{equation}
By subtracting the tensor $\nabla_b\nabla_a(f\omega_c)$ from both sides, we get
\begin{equation}
    (\nabla_a\nabla_b-\nabla_b\nabla_a)(f\omega_c)=f(\nabla_a\nabla_b-\nabla_b\nabla_a)\omega_c.
\end{equation}
This implies that the difference of two derivative operators acting on $f\omega_c$ is proportional to $f(\nabla_a\nabla_b-\nabla_b\nabla_a)\omega_c$, which suggests the existence of a tensor field $R_{abc}^{\quad d}$, known as the Riemann curvature tensor, such that
\begin{equation}
\label{riemann}
    \nabla_a\nabla_b\omega_c-\nabla_b\nabla_a\omega_c=R_{abc}^{\quad d}\omega_d,
\end{equation}
for all dual vector fields $\omega_c$.

The Riemann curvature tensor $R_{abc}^{\quad d}$ has four key properties:
\begin{enumerate}
\item \emph{Antisymmetry in the first two indices}: $R_{abc}^{\quad d}=-R_{bac}^{\quad d}$.
\item \emph{Total antisymmetry}: The totally antisymmetric part of $R_{abc}^{\quad d}$ is zero, i.e., $R_{[abc]}^{\quad\ d}=0$. Here, the squared brackets denote total antisymmetrization, as defined in equation (2.4.4) of \cite{Wald:1984rg}.
\item \emph{Skew symmetry}: For the derivative operator $\nabla_a$ associated with the metric $\nabla_ag_{bc}=0$, we have $R_{abcd}=-R_{abdc}$.
\item \emph{First Bianchi identity}: The Bianchi identity holds, which states that the cyclic sum of covariant derivatives of $R_{abc}^{\quad d}$ vanishes, i.e.,
\begin{equation}
\label{1bianchi}
\nabla_{[a}R_{bc]d}^{\quad\ e}=0.
\end{equation}
\end{enumerate}
The Ricci tensor $R_{ab}=R_{acb}^{\quad c}$ is a rank-2 tensor that encodes important geometric information about a Riemannian manifold. The Ricci tensor is symmetric, i.e., $R_{ab}=R_{ba}$, which follows from the symmetry of the Riemann tensor. The scalar curvature $R$ is obtained by taking the trace of the Ricci tensor, i.e., $R=R^{\ a}_a$. It is a scalar function that describes the intrinsic curvature of the manifold at a given point. Contraction of the first Bianchi identity \eqref{1bianchi} leads to
\begin{equation}
    \nabla_aR_{bcd}^{\quad a}+\nabla_bR_{cd}-\nabla_cR_{bd}=\nabla_aR_{c}^{\ a}+\nabla_bR_{c}^{\ b}-\nabla_cR=0,
\end{equation}
or
\begin{equation}
    \nabla^a(R_{ab}-\frac{1}{2}Rg_{ab})\equiv\nabla^aG_{ab}=0,
\end{equation}
where $G_{ab}$ is called the Einstein tensor.

By choosing a coordinate system, we can express the derivative operator in terms of the ordinary derivative $\partial_a$ and the Christoffel symbol $\Gamma^{c}_{\ ab}$ as discussed above. For a dual vector field $\omega_a$, we have
\begin{equation}
    \nabla_b\omega_c=\partial_b\omega_c-\Gamma^{d}_{\ bc}\omega_d,
\end{equation}
so the equation \eqref{riemann} can be expressed as
\begin{equation}
    R_{abc}^{\quad d}\omega_d=\{-2\partial_{[a}\Gamma^{d}_{\ b]c}+2\Gamma^{e}_{\ c[a}\Gamma^{d}_{\ b]e}\}\omega_d\Rightarrow R_{abc}^{\quad d}=-2\partial_{[a}\Gamma^{d}_{\ b]c}+2\Gamma^{e}_{\ c[a}\Gamma^{d}_{\ b]e}.
\end{equation}
The coordinate basis components of Riemann tensor is then
\begin{equation}
    R_{\alpha\beta\gamma}^{\quad\ \lambda}=-2\partial_{[\alpha}\Gamma^{\lambda}_{\ \beta]\gamma}+2\Gamma^{\sigma}_{\ \gamma[\alpha}\Gamma^{\lambda}_{\ \beta]\sigma},
\end{equation}
or explicitly
\begin{equation}
    R_{\alpha\beta\gamma}^{\quad\ \lambda}=\frac{\partial}{\partial x^{\beta}}\Gamma^{\lambda}_{\ \alpha\gamma}-\frac{\partial}{\partial x^{\alpha}}\Gamma^{\lambda}_{\ \beta\gamma}+\Gamma^{\sigma}_{\ \gamma\alpha}\Gamma^{\lambda}_{\ \beta\sigma}-\Gamma^{\sigma}_{\ \gamma\beta}\Gamma^{\lambda}_{\ \alpha\sigma}.
\end{equation}
If we define $\nabla_a\omega_c\equiv\omega_{c;a}$ and $\partial_a\omega_c\equiv\omega_{c,a}$, and for coordinate basis components $\nabla_\alpha\omega_\beta\equiv\omega_{\beta;\alpha}$ and $\partial_\alpha\omega_\beta=\partial\omega_\beta/\partial x^{\alpha}\equiv\omega_{\beta,\alpha}$, the above equation becomes
\begin{equation}
\label{RiemannT}
    R_{\alpha\beta\gamma}^{\quad\ \lambda}=\Gamma^{\lambda}_{\ \alpha\gamma,\beta}-\Gamma^{\lambda}_{\ \beta\gamma,\alpha}+\Gamma^{\sigma}_{\ \alpha\gamma}\Gamma^{\lambda}_{\ \beta\sigma}-\Gamma^{\sigma}_{\ \beta\gamma}\Gamma^{\lambda}_{\ \alpha\sigma}.
\end{equation}

A geodesic is a curve on a manifold, associated with a derivative operator $\nabla_a$, whose tangent vector $T^a$ is parallel transported along itself, and satisfies the equation $T^a\nabla_aT^b=0$. When a coordinate system is introduced, the geodesic is mapped to a curve $x^\mu(t)$ in $\mathbb{R}^n$. Using equation \eqref{paratranscomp}, the components of $T^a$, denoted $T^\mu=dx^\mu/dt$, satisfy the geodesic equation
\begin{equation}
    \frac{dT^\nu}{dt}+\Gamma^{\nu}_{\ \mu\lambda}T^\mu T^\lambda=\frac{d^2x^\nu}{dt^2}+\Gamma^{\nu}_{\ \mu\lambda}\frac{dx^\mu}{dt}\frac{dx^\lambda}{dt}=0.
\end{equation}

\subsection{Einstein's field equation}
\label{makereference1.1.3}

The theory of general relativity is built on two fundamental principles: the equivalence principle and the principle of general covariance. The equivalence principle states that the effects of gravity and acceleration are indistinguishable. Specifically, a particle in a gravitational field experiences a force equivalent to the force it would experience if it were uniformly accelerated in the opposite direction. This principle is the foundation of general relativity, which describes gravity as a curvature of spacetime caused by the presence of matter and energy.

The principle of general covariance requires that the laws of physics be formulated independently of the choice of coordinates used to describe physical phenomena. In other words, the laws of physics should be the same in all coordinate systems related by a smooth transformation. This principle allows us to describe the laws of physics in a coordinate-independent manner and is essential for a theory of gravity consistent with general relativity. By imposing the principle of general covariance, we ensure that the laws of physics are consistent with the equivalence principle and that they hold true in any coordinate system, regardless of how the coordinates are chosen.

In general relativity, spacetime is a manifold $M$ with a Lorentz metric $g_{ab}$ of signature $- + \cdots +$. A curve is considered timelike if the norm of its tangent is everywhere negative, $g_{ab}T^aT^b<0$. Particle motion is represented by a timelike curve, and perfect fluids are described using a 4-velocity $u^a$, density $\rho$, and pressure $p$. The electromagnetic field is represented by an antisymmetric tensor $F_{ab}$. A particle with rest mass $m$ and charge $q$ placed in an electromagnetic field $F_{ab}$ satisfies the Lorentz force equation
\begin{equation}
mu^a\nabla_au^b=qF^b_{\ c}u^c,
\end{equation}
where $F^b_{\ c}=g^{bd}F_{dc}$. The 4-velocity $u^a$ of a free particle satisfies the geodesic equation of motion
\begin{equation}
u^a\nabla_au^b=0,
\end{equation}
where $\nabla_a$ is the derivative operator associated with $g_{ab}$. If the acceleration $a^b=u^a\nabla_au^b$ is non-zero, a force $f^b=ma^b$ acts on the particle, where $m$ is its rest mass.
The 4-momentum of the particle can be defined as
\begin{equation}
    P^a=mu^a.
\end{equation}
The energy of the particle, as measured by an observer with 4-velocity $v^a$ at the event on the particle's world line is given by
\begin{equation}
    E=-P_av^a.
\end{equation}
A stress-energy tensor $T_{ab}$ describes continuous matter distributions and fields. For a perfect fluid the stress-energy tensor is given by
\begin{equation}
\label{set}
    T_{ab}=\rho u_au_b+p(g_{ab}+u_au_b),
\end{equation}
and it satisfies the equation of motion
\begin{equation}
\label{energycons}
    \nabla^aT_{ab}=0,
\end{equation}
which yields the following two equations:
\begin{equation}
    u^a\nabla_a\rho+(\rho+p)\nabla^au_a=0,
\end{equation}
and
\begin{equation}
    (p+\rho)u^a\nabla_au_b+(g_{ab}+u_au_b)\nabla^ap=mu^a,
\end{equation}
where $u_b=g_{ab}u^a$.
Note that in the special case where the fluid is at rest, i.e., $u^\mu=(1,0,0,0)$, the stress-energy tensor reduces to the familiar form $T_{00}=\rho$, $T_{ij}=p\delta_{ij}$, where $\delta_{ij}$ is the Kronecker delta. In the context of general relativity, the spacetime manifold $M$ is typically represented using a coordinate system that maps it onto $\mathbb{R}^4$. In this coordinate system, Latin letters are used to represent the spatial coordinates, with $i,j,k,\cdots=1,2,3$, while Greek letters are used to represent the time coordinate plus the spatial coordinates, with $\alpha,\beta,\gamma,\cdots=0,1,2,3$. Here, the number 0 represents the time coordinate, while 1, 2, and 3 represent the spatial coordinates. This convention is useful for representing the components of tensors in a coordinate basis and for performing calculations involving them.

The Klein-Gordon scalar field $\phi$ in curved spacetime satisfies the equation
\begin{equation}
    \nabla^a\nabla_a\phi-m^2\phi=0.
\end{equation}
The stress-energy tensor of this field is
\begin{equation}
    T_{ab}=\nabla_a\phi\nabla_b\phi-\frac{1}{2}g_{ab}(\nabla_c\phi\nabla^c\phi+m^2\phi^2),
\end{equation}
and satisfies the conservation equation \eqref{energycons}.

Maxwell's equations in curved spacetime govern the behavior of the electromagnetic field. They consist of two equations:
\begin{equation}
\label{maxwelleq1}
    \nabla^aF_{ab}=-4\pi j_b,
\end{equation}
which relates the current density 4-vector $j^a$ to the field strength tensor $F_{ab}$, and
\begin{equation}
    \nabla_{[a}F_{bc]}=0,
\end{equation}
which expresses the fact that the electromagnetic field is divergence-free. Here $j_b=g_{ab}j^a$ and $j^a$ is the current density 4-vector of electric charge.
The electromagnetic stress-energy tensor is a measure of the energy and momentum density of the electromagnetic field. It is given by
\begin{equation}
    T_{ab}=\frac{1}{4\pi}\left\{F_{ac}F_b^{\ c}-\frac{1}{4}g_{ab}F_{de}F^{de}\right\}.
\end{equation}
The vector potential $A_a$ satisfies the wave equation in curved spacetime,
\begin{equation}
    \nabla^a\nabla_aA_b-R^{d}_{\ b}A_d=-4\pi j_b,
\end{equation}
in Lorenz gauge where $\nabla_aA^a=0$. $R^{d}_{\ b}=R_{ab}g^{ad}$ is the Ricci curvature tensor. This equation describes the propagation of electromagnetic waves in a curved spacetime.

Einstein's field equation is the cornerstone of general relativity, relating the geometry of spacetime to the distribution of matter and energy within it. It takes the form
\begin{equation}
\label{EFE}
G_{ab}\equiv R_{ab}-\frac{1}{2}Rg_{ab}=8\pi T_{ab},
\end{equation}
where $G_{ab}$ is the Einstein tensor, $R$ is the Ricci scalar, and we use units for which the gravitational constant $G$ and the speed of light $c$ are 1. The left-hand side represents the curvature of spacetime, while the right-hand side represents the distribution of matter and energy in spacetime. This equation encodes the fundamental principle that matter tells spacetime how to curve, and through the geodesic equation or other matter equation og motion, spacetime tells matter how to move. Einstein first proposed this equation in 1915, and it remains a fundamental equation of modern physics.

Einstein's field equation can be derived using the Lagrangian formalism in a coordinate basis $\{dx^\mu\}$, where the Einstein-Hilbert action is given by
\begin{equation}
S_{\rm EH}=\int\frac{1}{16\pi}R\sqrt{-g}d^4x,
\end{equation}
with $g$ being the determinant of the metric tensor's coordinate basis components $g_{\mu\nu}$ and $G=c=1$. The complete action is defined as
\begin{equation}
S=S_{\rm EH}+S_{\rm M}=\int\left(\frac{1}{16\pi}R+\mathcal{L}_{\rm M}\right)\sqrt{-g}d^4x,
\end{equation}
where $S_{\rm M}$ and $\mathcal{L}_{\rm M}$ are the action and Lagrangian of matter. 

Using the principle of least action, we vary the action with respect to the inverse metric tensor $g^{\mu\nu}$, which yields
\begin{align}
\label{lap}
\nonumber
    0 &=\delta S \\
    \nonumber
      &= \int\left[\frac{1}{16\pi}\frac{\delta (R\sqrt{-g})}{\delta g^{\mu\nu}}+\frac{\delta(\mathcal{L}_{\rm M}\sqrt{-g})}{\delta g^{\mu\nu}}\right]\delta g^{\mu\nu}d^4x \\
      &=\int\left[\frac{1}{16\pi}\left(\frac{\delta R}{\delta g^{\mu\nu}}+\frac{R}{\sqrt{-g}}\frac{\delta \sqrt{-g}}{\delta g^{\mu\nu}}\right)+\frac{1}{\sqrt{-g}}\frac{\delta(\mathcal{L}_{\rm M}\sqrt{-g})}{\delta g^{\mu\nu}}\right]\sqrt{-g}\delta g^{\mu\nu}d^4x,
\end{align}
where $\delta$ is the variation operator. This equation should be hold for any variation $\delta g_{\mu\nu}$, thus it becomes
\begin{equation}
\label{variation}
    \frac{\delta R}{\delta g^{\mu\nu}}+\frac{R}{\sqrt{-g}}\frac{\delta \sqrt{-g}}{\delta g^{\mu\nu}}=-\frac{16\pi}{\sqrt{-g}}\frac{\delta(\mathcal{L}_{\rm M}\sqrt{-g})}{\delta g^{\mu\nu}}.
\end{equation}
The first term on the left-hand-side of equation \eqref{variation} requires the variation of the Ricci scalar
\begin{equation}
\label{variationR}
    \delta R=R_{\alpha\beta}\delta g^{\alpha\beta}+g^{\alpha\beta}\delta R_{\alpha\beta},
\end{equation}
which requires the variation of the Ricci tensor $\delta R_{\alpha\beta}$.
From equation \eqref{RiemannT}, we get the variation of the Riemann tensor
\begin{align}
\nonumber
    \delta R_{\alpha\beta\gamma}^{\quad\ \lambda}=&\delta\Gamma^{\lambda}_{\ \alpha\gamma,\beta}-\delta\Gamma^{\lambda}_{\ \beta\gamma,\alpha}+\Gamma^{\lambda}_{\ \beta\sigma}\delta\Gamma^{\sigma}_{\ \alpha\gamma}+\Gamma^{\sigma}_{\ \alpha\gamma}\delta\Gamma^{\lambda}_{\ \beta\sigma}-\Gamma^{\lambda}_{\ \alpha\sigma}\delta\Gamma^{\sigma}_{\ \beta\gamma}-\Gamma^{\sigma}_{\ \beta\gamma}\delta\Gamma^{\lambda}_{\ \alpha\sigma} \\
    \nonumber
    =&\delta\Gamma^{\lambda}_{\ \alpha\gamma;\beta}-\Gamma^{\lambda}_{\ \beta\sigma}\delta\Gamma^{\sigma}_{\ \alpha\gamma}+\Gamma^{\sigma}_{\ \beta\gamma}\delta\Gamma^{\lambda}_{\ \alpha\sigma}+\Gamma^{\sigma}_{\ \beta\alpha}\delta\Gamma^{\lambda}_{\ \sigma\gamma}-\delta\Gamma^{\lambda}_{\ \beta\gamma;\alpha}+\Gamma^{\lambda}_{\ \alpha\sigma}\delta\Gamma^{\sigma}_{\ \beta\gamma}-\Gamma^{\sigma}_{\ \alpha\gamma}\delta\Gamma^{\lambda}_{\ \beta\sigma} \\
    \nonumber
    &-\Gamma^{\sigma}_{\ \beta\alpha}\delta\Gamma^{\lambda}_{\ \sigma\gamma}+\Gamma^{\lambda}_{\ \beta\sigma}\delta\Gamma^{\sigma}_{\ \alpha\gamma}+\Gamma^{\sigma}_{\ \alpha\gamma}\delta\Gamma^{\lambda}_{\ \beta\sigma}-\Gamma^{\lambda}_{\ \alpha\sigma}\delta\Gamma^{\sigma}_{\ \beta\gamma}-\Gamma^{\sigma}_{\ \beta\gamma}\delta\Gamma^{\lambda}_{\ \alpha\sigma} \\
    =&\delta\Gamma^{\lambda}_{\ \alpha\gamma;\beta}-\delta\Gamma^{\lambda}_{\ \beta\gamma;\alpha},
\end{align}
so the variation of the Ricci tensor is
\begin{equation}
    \delta R_{\alpha\gamma}=\delta R_{\alpha\lambda\gamma}^{\quad\ \lambda}=\delta\Gamma^{\lambda}_{\ \alpha\gamma;\lambda}-\delta\Gamma^{\lambda}_{\ \lambda\gamma;\alpha}.
\end{equation}
Therefore, equation \eqref{variationR} becomes
\begin{equation}
\label{variationR2}
    \delta R=R_{\alpha\beta}\delta g^{\alpha\beta}+(g^{\alpha\beta}\delta\Gamma^{\lambda}_{\ \alpha\beta}-g^{\lambda\beta}\delta\Gamma^{\gamma}_{\ \gamma\beta})_{;\lambda}\equiv R_{\alpha\beta}\delta g^{\alpha\beta}+\nabla_\lambda v^\lambda,
\end{equation}
where the second term can be substituted back into equation \eqref{lap} to get a boundary term
\begin{equation}
\label{bund}
    \frac{1}{16\pi}\int \sqrt{-g}\nabla_\lambda v^\lambda dx^4.
\end{equation}
This boundary term does not contribute to the variation of the action when $\delta g^{\mu\nu}$ vanishes in a neighborhood of the boundary or when there is no boundary. In this case, the first term on the left-hand-side of equation \eqref{variation} becomes
\begin{equation}
    \frac{\delta R}{\delta g^{\mu\nu}}=R_{\mu\nu}.
\end{equation}
In the second term on the left-hand-side of equation \eqref{variation}
\begin{equation}
    \delta \sqrt{-g}=-\frac{1}{2\sqrt{-g}}\delta g=-\frac{1}{2}\sqrt{-g}g_{\mu\nu}\delta g^{\mu\nu},
\end{equation}
where $\delta g=gg^{\mu\nu}\delta g_{\mu\nu}$ and $g^{\mu\nu}\delta g_{\mu\nu}=-g_{\mu\nu}\delta g^{\mu\nu}$ are used. Thus the second term of the left-hand-side of equation \eqref{variation} becomes
\begin{equation}
    \frac{R}{\sqrt{-g}}\frac{\delta \sqrt{-g}}{\delta g^{\mu\nu}}=-\frac{1}{2}Rg_{\mu\nu},
\end{equation}
and then equation \eqref{variation} becomes
\begin{equation}
\label{EFE1}
    R_{\mu\nu}-\frac{1}{2}Rg_{\mu\nu}=-\frac{16\pi}{\sqrt{-g}}\frac{\delta(\mathcal{L}_{\rm M}\sqrt{-g})}{\delta g^{\mu\nu}}.
\end{equation}
In the coordinate basis, the stress-energy tensor's components can be defined using the Lagrangian density $\mathcal{L}_{\rm M}$ for matter as follows:
\begin{equation}
\label{SET1}
    T_{\mu\nu}\coloneqq -\frac{2}{\sqrt{-g}}\frac{\delta(\mathcal{L}_{\rm M}\sqrt{-g})}{\delta g^{\mu\nu}}=-2\frac{\delta\mathcal{L}_{\rm M}}{\delta g^{\mu\nu}}+g_{\mu\nu}\mathcal{L}_{\rm M}.
\end{equation}
With this definition, equation \eqref{EFE1} becomes Einstein's field equation \eqref{EFE} in the coordinate basis
\begin{equation}
\label{EFE2}
    R_{\mu\nu}-\frac{1}{2}Rg_{\mu\nu}=8\pi T_{\mu\nu}.
\end{equation}
We can modify the Einstein-Hilbert action by adding a cosmological constant term:
\begin{equation}
    \tilde S_{\rm HE}=\int\frac{1}{16\pi}(R-2\Lambda)\sqrt{-g}d^4x,
\end{equation}
where $\Lambda$ is commonly know as a cosmological constant, which is a form of dark energy in the standard $\Lambda$CDM model \citep{peeb84}. When we vary this modified action, we obtain a modified version of Einstein's field equation \eqref{EFE2}:
\begin{equation}
\label{EFE3}
    R_{\mu\nu}-\frac{1}{2}Rg_{\mu\nu}+\Lambda g_{\mu\nu}=G_{\mu\nu}+\Lambda g_{\mu\nu}=8\pi T_{\mu\nu}.
\end{equation}
Note that the property of the Einstein tensor, $\nabla^\mu G_{\mu\nu}=0$, allows the addition of the cosmological constant term $\Lambda g_{\mu\nu}$ into Einstein's field equation. We can also interpret Einstein's field equation \eqref{EFE} with the stress-energy tensor defined as:
\begin{equation}
    T_{\mu\nu}= T^{\rm M}_{\mu\nu}+T^{\Lambda}_{\mu\nu},
\end{equation}
where $T^{\rm M}_{\mu\nu}$ is the stress-energy tensor defined in equation \eqref{SET1}, and 
\begin{equation}
    T^{\Lambda}_{\mu\nu}\coloneqq -\frac{\Lambda}{8\pi}g_{\mu\nu}.
\end{equation}

\section{Friedmann equations}
\label{makereference1.2}

The cosmological principle asserts that the Universe is spatially homogeneous and isotropic on large spatial scales. This means that the Friedmann-Lema\^itre-Robertson-Walker (FLRW) metric can be used to describe the Universe, which takes the form:
\begin{equation}
\label{FLRW}
    ds^2=g_{\mu\nu}dx^\mu dx^\nu=-dt^2+a^2(t)\left[\frac{dr^2}{1-kr^2}+r^2(d\theta^2+\sin^2\theta d\phi^2)\right],
\end{equation}
where $g_{\mu\nu}$ are the metric tensor components, $c=1$, $t$ is cosmic time, $(r,\theta,\phi)$ are spherical coordinates, and $a(t)$ is the scale factor. The scale factor is defined as:
\begin{equation}
    \mathbf{r}(t)=a(t)\mathbf{x},
\end{equation}
where $\mathbf{r}(t)$ is the physical distance between two points at a given time, and $\mathbf{x}$ is the comoving distance between those points.
The curvature of the spatial hypersurfaces is described by the value of $k$, which can be positive ($+1$), zero, or negative ($-1$). If $k$ is positive, space is positively curved like a sphere. If $k$ is negative, space is negatively curved like a saddle. If $k$ is zero, space is flat like a sheet of paper.

From equation \eqref{FLRW}, we can see that the non-zero components of the metric tensor in coordinates $(x^0,x^1,x^2,x^3)=(t,r,\theta,\phi)$ are
\begin{equation}
    g_{00}=-1,\ g_{11}=\frac{a^2}{1-kr^2},\ g_{22}=a^2r^2,\ g_{33}=a^2r^2\sin^2\theta.
\end{equation}
Therefore by using equation \eqref{christoffel}, we obtain the non-zero components of the Christoffel symbol $\Gamma^{\alpha}_{\ \mu\nu}$
\begin{align}
\nonumber
    &\Gamma^{0}_{\ 11}=\frac{a\dot{a}}{1-kr^2}=\frac{\dot{a}}{a}g_{11},\ \Gamma^{0}_{\ 22}=a\dot{a}r^2=\frac{\dot{a}}{a}g_{22},\ \Gamma^{0}_{\ 33}=a\dot{a}r^2\sin^2\theta=\frac{\dot{a}}{a}g_{33}, \iff \Gamma^{0}_{\ ij}=\frac{\dot{a}}{a}g_{ij},\\
    \nonumber
    &\Gamma^{i}_{\ 0j}=\Gamma^{i}_{\ j0}=\frac{\dot{a}}{a}\delta^{i}_{\ j},\ \Gamma^{1}_{\ 11}=\frac{kr}{1-kr^2},\ \Gamma^{1}_{\ 22}=-r(1-kr^2),\ \Gamma^{1}_{\ 33}=-r(1-kr^2)\sin^2\theta, \\
    &\Gamma^{2}_{\ 12}=\Gamma^{2}_{\ 21}=\Gamma^{3}_{\ 13}=\Gamma^{3}_{\ 31}=\frac{1}{r},\ \Gamma^{2}_{\ 33}=-\sin\theta\cos\theta,\ \Gamma^{3}_{\ 23}=\Gamma^{3}_{\ 32}=\cot\theta. \label{chrisym}
\end{align}

The stress-energy tensor of a perfect fluid takes the form of equation \eqref{set}, whose components in the coordinate basis is
\begin{equation}
\label{setc}
    T_{\mu\nu}=\rho u_\mu u_\nu+p(g_{\mu\nu}+u_\mu u_\nu),
\end{equation}
where the symbols are introduced in Section \ref{makereference1.1.3}. Since $g_{0i}=g_{i0}=u_i=u^i=0$, the non-zero parts of $T_{\mu\nu}$ are
\begin{equation}
\label{Tflrw}
    T_{00}=\rho,\ T_{ij}=pg_{ij}.
\end{equation}

Thus Einstein's field equations \eqref{EFE3} become the following two equations
\begin{equation}
\label{feqt}
    R_{00}+\frac{1}{2}R=8\pi\rho+\Lambda,
\end{equation}
and
\begin{equation}
\label{feqs}
    R_{ij}-\frac{1}{2}Rg_{ij}=8\pi pg_{ij}-\Lambda g_{ij}.
\end{equation}
Note that the gravitational constant $G=1$ is applied here.

From equation \eqref{chrisym} we can derive the desired components of the Ricci tensor $R_{\mu\nu}$ and the Ricci scalar $R$ as follows
\begin{equation}
\label{ricci}
    R_{00}=-3\frac{\ddot{a}}{a},\ R_{ij}=\frac{1}{a^2}(2\dot{a}^2+a\ddot{a}+2k)g_{ij},\ R=6\left(\frac{\ddot{a}}{a}+\frac{\dot{a}^2}{a^2}+\frac{k}{a^2}\right).
\end{equation}

Substituting equation \eqref{ricci} into equations \eqref{feqt} and \eqref{feqs}, we get
\begin{equation}
\label{feq1}
    \left(\frac{\dot{a}}{a}\right)^2=\frac{8\pi}{3}\rho+\frac{\Lambda}{3}-\frac{k}{a^2},
\end{equation}
and
\begin{equation}
\label{feq0}
    2\frac{\ddot{a}}{a}+\frac{\dot{a}^2}{a^2}+\frac{k}{a^2}=-8\pi p+\Lambda.
\end{equation}
Equation \eqref{feq1} is known as the first Friedmann equation. By combining equations \eqref{feq1} and \eqref{feq0}, we can get the second Friedmann equation
\begin{equation}
\label{feq2}
    \frac{\ddot{a}}{a}=-\frac{4\pi}{3}(\rho+3p)+\frac{\Lambda}{3}.
\end{equation}
Friedmann equations \eqref{feq1} and \eqref{feq2} can yield
\begin{equation}
\label{cons}
    \dot{\rho}+3\frac{\dot{a}}{a}(\rho+p)=0,
\end{equation}
which expresses the conservation of the stress-energy $T^{\mu\nu}_{\ \ \,;\nu}=0$.
If we define an equation of state parameter
\begin{equation}
    w\coloneqq \frac{p}{\rho},
\end{equation}
then equation \eqref{cons} has a solution
\begin{equation}
    \rho=\rho_0\left(\frac{a_0}{a}\right)^{3(1+w)},
\end{equation}
where at present time $t=t_0$, $\rho(t_0)=\rho_0$, and $a(t_0)=a_0$.
For dust ($p=0$), we find $\rho\propto a^{-3}$; for radiation ($p=\rho/3$), we find $\rho\propto a^{-4}$; and for the cosmological constant $\rho_\Lambda=\Lambda/8\pi$, $w_\Lambda=-1$. 

The Hubble parameter is a measure of the expansion rate of the Universe and is defined as the time derivative of the scale factor $a$ normalized by the scale factor itself, i.e.,
\begin{equation}
H\coloneqq\frac{\dot a}{a}.
\end{equation}
Here, an overdot represents a derivative with respect to cosmic time $t$. At the present time $t=t_0$, the value of the Hubble parameter is denoted as $H_0$, which is commonly known as the Hubble constant.

The first Friedmann equation \eqref{feq1} is a fundamental equation in cosmology that relates the expansion rate of the Universe to its energy density and curvature. It takes the form
\begin{equation}
\label{feq1.1}
H^2=\frac{8\pi}{3}(\rho_m+\rho_r+\rho_\Lambda)-\frac{k}{a^2},
\end{equation}
where $\rho_m$, $\rho_r$, and $\rho_\Lambda$ represent the matter, radiation, and dark energy density, respectively. The term $k/a^2$ accounts for the curvature of spatial hypersurfaces, where $k$ takes the values $-1$, $0$, or $1$ corresponding to open, flat, or closed spatial geometries, respectively.
Equation \eqref{feq1.1} can be rewritten in terms of the Hubble constant $H_0$ as
\begin{equation}
\label{feq1.2}
H^2=H_0^2\left[\frac{8\pi}{3H_0^2}(\rho_m+\rho_r+\rho_\Lambda)-\frac{k}{a^2H_0^2}\right].
\end{equation}
We can define the cosmological parameters as follows:
\begin{itemize}
    \item The non-relativistic matter density parameter
    \begin{equation}
    \label{Omegam}
        \Omega_m\coloneqq\frac{8\pi}{3H_0^2}\rho_m=\frac{8\pi}{3H_0^2}\rho_{m0}\left(\frac{a_0}{a}\right)^3=\Omega_{m0}\left(\frac{a_0}{a}\right)^3,
    \end{equation}
    where the subscript ``0'' denotes the current value of the given quantities.
    \item The radiation density parameter
    \begin{equation}
    \label{Omegar}
        \Omega_r\coloneqq\frac{8\pi}{3H_0^2}\rho_r=\frac{8\pi}{3H_0^2}\rho_{r0}\left(\frac{a_0}{a}\right)^4=\Omega_{r0}\left(\frac{a_0}{a}\right)^4.
    \end{equation}
    \item The cosmological constant dark energy density parameter
    \begin{equation}
    \label{Omegal}
        \Omega_\Lambda\coloneqq\frac{8\pi}{3H_0^2}\rho_\Lambda.
    \end{equation}
    \item The curvature density parameter
    \begin{equation}
    \label{Omegak}
        \Omega_k\coloneqq-\frac{k}{a^2H_0^2}=\Omega_{k0}\left(\frac{a_0}{a}\right)^2.
    \end{equation}
\end{itemize}
Using these parameters, equation \eqref{feq1.2} becomes
\begin{equation}
\label{feq1.3}
H^2=H_0^2(\Omega_m+\Omega_r+\Omega_k+\Omega_\Lambda)=H_0^2\left[\Omega_{m0}\left(\frac{a_0}{a}\right)^3+\Omega_{r0}\left(\frac{a_0}{a}\right)^4+\Omega_{k0}\left(\frac{a_0}{a}\right)^2+\Omega_\Lambda\right].
\end{equation}

\section{Distance measures in cosmology}
\label{makereference1.3}

Light travels along null geodesics, which can be expressed as $ds^2=0$. For an isotropic observer in the FLRW metric, it is always possible to find a new set of $(\theta, \phi)$ coordinates such that the null geodesic is radial, leading to the following line element:
\begin{equation}
    ds^2=-dt^2+a^2\frac{dr^2}{1-kr^2}=0,
\end{equation}
or equivalently,
\begin{equation}
\label{lp}
    \frac{dt}{a}=\frac{dr}{\sqrt{1-kr^2}}.
\end{equation}
Consider a light source located at $r=R$ that emits two light beams separated by a time interval equal to the wavelength of light at the source, $\lambda_{\rm S}$. The light beams are emitted at times $t_{\rm S}$ and $t_{\rm S}+\lambda_{\rm S}$. An observer located at $r=0$ detects the light beams with a wavelength of light at the observer, $\lambda_{\rm O}$, at two different times: $t_{\rm O}$ and $t_{\rm O}+\lambda_{\rm O}$, respectively. Integrating the light path \eqref{lp} over $t$ and $r$, we obtain
\begin{equation}
\label{los1}
    \int^{t_{\rm O}}_{{t_{\rm S}}} \frac{dt}{a}\equiv\int^{t_{\rm S}+\lambda_{\rm S}}_{{t_{\rm S}}} \frac{dt}{a}+\int^{{t_{\rm O}}}_{t_{\rm S}+\lambda_{\rm S}} \frac{dt}{a}=\int^0_R\frac{dr}{\sqrt{1-kr^2}}=\int^{t_{\rm O}+\lambda_{\rm O}}_{{t_{\rm S}+\lambda_{\rm S}}} \frac{dt}{a},
\end{equation}
which yields
\begin{equation}
    \int^{t_{\rm S}+\lambda_{\rm S}}_{{t_{\rm S}}} \frac{dt}{a}=\int^{t_{\rm O}+\lambda_{\rm O}}_{{t_{\rm O}}} \frac{dt}{a}.
\end{equation}
For small time intervals $\lambda_{\rm S}$ and $\lambda_{\rm O}$, we can consider scale factors at the source ($a_{\rm S}$) and at the observer ($a_{\rm O}$) to be constants. Therefore, the above equation becomes
\begin{equation}
\label{red1}
    \frac{\lambda_{\rm S}}{a_{\rm S}}=\frac{\lambda_{\rm O}}{a_{\rm O}}\quad \text{or}\quad \frac{\lambda_{\rm S}}{\lambda_{\rm O}}=\frac{a_{\rm S}}{a_{\rm O}},
\end{equation}
We define the redshift $z$ as
\begin{equation}
    z=\frac{\lambda_{\rm O}-\lambda_{\rm S}}{\lambda_{\rm S}}=\frac{\lambda_{\rm O}}{\lambda_{\rm S}}-1,
\end{equation}
which, combined with equation \eqref{red1}, gives
\begin{equation}
\label{red2}
    1+z=\frac{a_{\rm O}}{a_{\rm S}}.
\end{equation}
We can simplify equation \eqref{red2} by considering ourselves as the observer at the present time $t_0$ when the scale factor is $a_0$, with the source at time $t$ when the scale factor is $a$. Then we have
\begin{equation}
\label{redshift}
1+z=\frac{a_{0}}{a}.
\end{equation}
By combining equation \eqref{redshift} with equation \eqref{feq1.3}, we can express the Hubble parameter as a function of redshift $z$
\begin{equation}
\label{HzLCDM}
    H(z)=H_0\sqrt{\Omega_{m0}(1+z)^3+\Omega_{r0}(1+z)^4+\Omega_{k0}(1+z)^2+\Omega_\Lambda}\equiv H_0E(z),
\end{equation}
where $E(z)$ is the expansion rate function. 

Several key distance measures are used in cosmology to quantify the expansion of the Universe. The transverse comoving distance $D_M(z)$, also known as the radial distance along the line of sight, can be calculated from equation \eqref{los1} and is given by
\begin{equation}
    \label{eq:DM}
    D_M(z) = 
    \begin{cases}
    \frac{1}{H_0\sqrt{\Omega_{k0}}}\sinh\left[H_0\sqrt{\Omega_{k0}}D_C(z)\right] & \text{if}\ \Omega_{k0} > 0, \\
    D_C(z) & \text{if}\ \Omega_{k0} = 0,\\
    \frac{1}{H_0\sqrt{|\Omega_{k0}|}}\sin\left[H_0\sqrt{|\Omega_{k0}|}D_C(z)\right] & \text{if}\ \Omega_{k0} < 0,
    \end{cases}
\end{equation}
where $D_C(z)$ is the comoving distance to the source, given by
\begin{equation}
\label{eq:DC}
   D_C(z) = \int^z_0 \frac{dz'}{H(z')}.
\end{equation}
$D_M(z)$ is related to other important distance measures, including the luminosity distance $D_L(z)$ and the angular diameter distance $D_A(z)$, through
\begin{equation}
    \label{DM-DL-DA}
    D_M(z)=\frac{D_L(z)}{1+z}=(1+z)D_A(z).
\end{equation}
In addition, the Hubble distance is defined as
\begin{equation}
\label{eq:DH}
    D_H(z)=\frac{1}{H(z)},
\end{equation}
and the volume averaged distance is given by
\begin{equation}
\label{eq:DV}
    D_V(z)=[zD_H(z)D^2_M(z)]^{1/3}.
\end{equation}
Another important quantity is the sound horizon at the drag epoch $z_d$, which is defined as
\begin{equation}
\label{rs}
    r_s=\int^{\infty}_{z_d}\frac{c_s(z)}{H(z)}dz,
\end{equation}
where the sound speed in the photon-baryon fluid is given by
\begin{equation}
    c_s(z)=\frac{1}{\sqrt{3\left(1+\frac{3}{4}\frac{\rho_b}{\rho_r}\right)}},
\end{equation}
with $\rho_b$ being the baryon density. In practice, an approximation for $r_s$ calibrated from numerical simulations is often used:
\begin{equation}
\label{eq:sh}
    r_s=\frac{55.154\exp{[-72.3(\Omega_{\nu}h^2+0.0006)^2]}}{(\Omega_{b}h^2)^{0.12807}(\och+\obh)^{0.25351}} \mathrm{Mpc},
\end{equation}
where \onhs, \obhs, \ochs\ are the physical energy density parameters for non-relativistic neutrinos, baryons, and cold dark matter respectively, and $h$ is the Hubble constant in units of 100 \hunit. This approximation is accurate to 0.021\% for a standard radiation background with effective number of neutrino species $N_{\rm eff}= 3.046$, total neutrino mass $\sum m_\nu < 0.6\ \rm eV$, and values of \obhs\ and \ochs\ within $3\sigma$ of values derived using Planck CMB anisotropy data \citep{PhysRevD.92.123516}.

Note that the sound horizon $r_s$ can be numerically computed using the Cosmic Linear Anisotropy Solving System (\textsc{class}) code \citep{class}, without relying on CMB anisotropy data. This is the approach used in our latest analyses, \cite{CaoRyanRatra2022,CaoKhadkaRatra2022,CaoDainottiRatra2022,CaoRatra2022,CaoDainottiRatra2022b,Cao:2022pdv}.

The distance modulus is defined as
\begin{equation}
\label{dismod}
    \mu=m-M=5\log\left(\frac{D_L}{10\ \rm pc}\right)=5\log D_L +25,
\end{equation}
where $m$ is the apparent magnitude of the object, $M$ is its absolute magnitude, and $D_L$ is the luminosity distance in units of Mpc. Throughout the thesis, unless otherwise specified, the symbol ``$\log$'' represents the base-10 logarithm. The distance modulus is commonly used in observational cosmology as a measure of the distance to distant objects, such as type Ia supernovae, and can be inferred from the observed apparent magnitudes and the known absolute magnitudes of such objects.


\cleardoublepage


\chapter{Dark energy cosmological models}
\label{sec:models}

Various measurements indicate that the Universe is currently expanding at an accelerating rate. If general relativity is valid this acceleration is due to the presence of dark energy, a hypothetical substance that has negative pressure. In the spatially-flat $\Lambda$CDM model \citep{peeb84}, dark energy is represented by a cosmological constant and accounts for around 70\% of the total energy content of the Universe. However, recent observations might indicate potential discrepancies with this model \citep{DiValentinoetal2021b,PerivolaropoulosSkara2021,Morescoetal2022,Abdallaetal2022,Hu:2023jqc}, leading to investigations into alternative models that allow for non-zero spatial curvature and/or dynamical dark energy. In our analyses we also explore some of these alternatives, described next, to better understand the underlying physics of the Universe.

\section{$\Lambda$CDM models}
\label{lcdm}

The Hubble parameter in flat and non-flat \lcdm\ models are already derived in equation \eqref{HzLCDM}, in which the expansion rate function is
\be
\label{eq:EzL}
    E(z) = \sqrt{\Om\left(1 + z\right)^3 + \Ok\left(1 + z\right)^2 + \Omega_{\Lambda}},
\ee
where \om\ and \ok\ are the current values of the cosmological non-relativistic matter density parameter and the spatial curvature energy density parameter, respectively, the cosmological constant dark energy density parameter $\Omega_{\Lambda} = 1 - \Om - \Ok$, and $z$ is the redshift. In the flat \lcdm\ model, $\Ok = 0$. For most of the data sets we used, the cosmological parameters being constrained are $\{H_0, \obh\!, \och\}$ (or $\{H_0, \Om\}$) and $\{H_0, \obh\!, \och\!, \Ok\}$ (or $\{H_0, \Om, \Ok\}$) in the flat and non-flat \lcdm\ models, respectively. In some cases, certain data sets may not provide constraints on $H_0$ and \obhs, in which case fixed values of $\Omega_b$ and $H_0$ are assumed. For example, a commonly used set is $\Omega_b=0.05$ and $H_0=70$ \hunit. As noted in section \ref{makereference1.2}, the equation of state parameter of the cosmological constant dark energy is $w_\Lambda=-1$.

It is worth noting that in equation \eqref{eq:EzL} the contribution of radiation to the cosmological energy budget is neglected. This is justified for the low-redshift data sets that we used in our analyses, as the contribution of radiation becomes negligible at late times.

\section{XCDM parametrizations}
\label{xcdm}

In the XCDM parametrizations, dark energy is modeled as an X-fluid with a generalized constant equation of state parameter \wx, which distinguishes it from the \lcdm\ models where $w_\Lambda=-1$. The cosmological X-fluid dynamical dark energy parameter is defined as
\begin{equation}
\label{OmegaX}
    \Omega_{\rm X}\coloneqq\frac{8\pi}{3H_0^2}\rho_{\rm X}=\frac{8\pi}{3H_0^2}\rho_{\rm X0}\left(\frac{a_0}{a}\right)^{3(1+\wX)}=\Omega_{\rm X0}(1+z)^{3(1+\wX)},
\end{equation}
where $\rho_{\rm X}$ ($a$)  and $\rho_{\rm X0}$ ($a_0$) are the X-fluid dynamical dark energy density (the scale factor) at a given time and at the present time, respectively. By replacing $\Omega_\Lambda$ in equation \eqref{eq:EzL} with $\Omega_{\rm X}$, we obtain the expansion rate
\be
\label{eq:EzX}
    E(z) = \sqrt{\Om\left(1 + z\right)^3 + \Ok\left(1 + z\right)^2 + \Omega_{{\rm X}0}\left(1 + z\right)^{3\left(1 + \wX\right)}},
\ee
where the current value of the X-fluid dynamical dark energy density parameter $\Omega_{{\rm X}0} = 1 - \Om - \Ok$. The cosmological parameters being constrained are $\{H_0, \obh\!, \och\!, \wX\}$ (or $\{H_0, \Om, \wX\}$) and $\{H_0, \obh\!, \och\!, \wX, \Ok\}$ (or $\{H_0, \Om, \wX, \Ok\}$) in the flat and non-flat XCDM parametrizations, respectively. Note that when $\wX=-1$ XCDM reduces to \lcdm.

\section{$\phi$CDM models}
\label{pcdm}

A dynamical scalar field can play an important cosmological role. In particular, scalar fields are often used to model the behavior of the inflaton field, which is thought to have driven the very rapid expansion of the Universe during an early inflationary epoch. During inflation, the scalar field energy dominates the cosmological energy budget. The dynamics of the scalar field during this period determines the rate and duration of inflation, as well as the properties of the resulting density fluctuations that seed the formation of large-scale cosmic structures such as galaxies.

After inflation, scalar fields may also play a role in the late-time acceleration of the expansion of the Universe, and are referred to as dynamical dark energy scalar fields. The study of scalar field dynamics in cosmology involves solving the equations of motion that describe the behavior of the scalar field in the expanding Universe, and determining its effects on the evolution of other cosmological parameters.

In this work we explore a dynamical scalar field dark energy model, also known as Quintessence, where the dynamical dark energy is represented by a scalar field $\phi$ instead of a cosmological constant. The Lagrangian of the scalar field $\phi$ is given by 
\begin{equation}
\label{l_phi}
    \mathcal{L}_\phi=-\frac{1}{16\pi}\left[\frac{1}{2}g^{\mu\nu}\partial_\mu\phi\partial_\nu\phi+V(\phi)\right],
\end{equation}
where $V(\phi)$ is the potential energy density of the scalar field. Following the analogy of the stress-energy tensor of matter, we can define the stress-energy tensor of the scalar field $\phi$ as
\begin{equation}
\label{SET2}
    T_{\mu\nu}\coloneqq -\frac{2}{\sqrt{-g}}\frac{\delta(\mathcal{L}_{\phi}\sqrt{-g})}{\delta g^{\mu\nu}}=-2\frac{\delta\mathcal{L}_{\phi}}{\delta g^{\mu\nu}}+g_{\mu\nu}\mathcal{L}_{\phi}.
\end{equation}
The first term in the right-hand-side of equation \eqref{SET2} can be simplified as
\begin{equation}
-2\frac{\delta\mathcal{L}_{\phi}}{\delta g^{\mu\nu}}=\frac{1}{8\pi}\frac{\delta}{\delta g^{\mu\nu}}\left[\frac{1}{2}g^{\alpha\beta}\partial_\alpha\phi\partial_\beta\phi+V(\phi)\right]=\frac{1}{16\pi}\partial_\mu\phi\partial_\nu\phi.
\end{equation}
Therefore, the stress-energy tensor of $\phi$ in equation \eqref{SET2} becomes
\begin{equation}
\label{SET3}
    T_{\mu\nu}=\frac{1}{16\pi}\left\{\partial_\mu\phi\partial_\nu\phi-g_{\mu\nu}\left[\frac{1}{2}g^{\alpha\beta}\partial_\alpha\phi\partial_\beta\phi+V(\phi)\right]\right\}.
\end{equation}
Consequently, we can obtain the components of the stress-energy tensor in the FLRW metric as
\begin{align}
\nonumber
    &T_{00}=\frac{1}{16\pi}\left\{\partial_0\phi\partial_0\phi-g_{00}\left[\frac{1}{2}g^{00}\partial_0\phi\partial_0\phi+V(\phi)\right]\right\}=\frac{1}{16\pi}\left[\frac{1}{2}\dot{\phi}^2+V(\phi)\right], \\
    &T_{ij}=\frac{1}{16\pi}\left\{\partial_i\phi\partial_j\phi-g_{ij}\left[\frac{1}{2}g^{00}\partial_0\phi\partial_0\phi+V(\phi)\right]\right\}=\frac{1}{16\pi}\left[\frac{1}{2}\dot{\phi}^2-V(\phi)\right]g_{ij},
\end{align}
where $\phi$ is only a function of $t$ due to the homogeneous and isotropic properties of the FLRW metric. Similarly, following the same procedure as for the stress-energy tensor in equation \eqref{Tflrw}, we can identify the energy density and pressure of the scalar field $\phi$ as
\begin{equation}
\label{rho_phi}
    \rho_\phi=\frac{1}{16\pi}\left[\frac{1}{2}\dot{\phi}^2+V(\phi)\right],
\end{equation}
and
\begin{equation}
\label{p_phi}
    p_\phi=\frac{1}{16\pi}\left[\frac{1}{2}\dot{\phi}^2-V(\phi)\right],
\end{equation}
respectively. Therefore, we can express the equation of state of the scalar field dark energy as follows:
\begin{equation}
    w_\phi=\frac{\frac{1}{2}\dot{\phi}^2-V(\phi)}{\frac{1}{2}\dot{\phi}^2+V(\phi)}.
\end{equation}
Therefore, the Friedmann equation can be expressed in terms of the Hubble parameter
\be\label{Hz3}
H(z)=H_0\sqrt{\Om(1+z)^3+\Ok(1+z)^2+\Omega_{\phi}(z,\alpha)},
\ee
where the scalar field dynamical dark energy density parameter
\be\label{op}
\Omega_{\phi}(z,\alpha)\coloneqq\frac{8\pi}{3H_0^2}\rho_\phi.
\ee

The action for the scalar field $\phi$ is given by
\begin{equation}
    S_\phi=\int\mathcal{L}_\phi\sqrt{-g}d^4x=-\frac{1}{16\pi}\int\left[\frac{1}{2}g^{\mu\nu}\partial_\mu\phi\partial_\nu\phi+V(\phi)\right]\sqrt{-g}d^4x.
\end{equation}
The equation of motion for $\phi$ can be obtained using the principle of least action, which states that the variation of the action with respect to $\phi$ is zero:
\begin{align}
\nonumber
    0&=\delta S_\phi \\
    \nonumber
    &=-\frac{1}{16\pi}\delta\int\left[\frac{1}{2}g^{\mu\nu}\partial_\mu\phi\partial_\nu\phi+V(\phi)\right]\sqrt{-g}d^4x \\
    \nonumber
    &=-\frac{1}{16\pi}\int\left[\frac{1}{2}g^{\mu\nu}\delta(\partial_\mu\phi\partial_\nu\phi)+V^{\prime}(\phi)\delta\phi\right]\sqrt{-g}d^4x \\
    \nonumber
    &=-\frac{1}{16\pi}\int\left[g^{\mu\nu}\partial_\mu\phi\delta(\partial_\nu\phi)+V^{\prime}(\phi)\delta\phi\right]\sqrt{-g}d^4x \\
    \nonumber
    &=-\frac{1}{16\pi}\int\left[g^{\mu\nu}\partial_\mu\phi\partial_\nu(\delta\phi)+V^{\prime}(\phi)\delta\phi\right]\sqrt{-g}d^4x \\
    \nonumber
    &=-\frac{1}{16\pi}\int\left(\int^{\infty}_{-\infty} g^{\mu\nu}\partial_\mu\phi\delta\phi\sqrt{-g}dx^\nu\right) d^3x-\frac{1}{16\pi}\int\left[\partial_\nu(\sqrt{-g}g^{\mu\nu}\partial_\mu\phi)-\sqrt{-g}V^{\prime}(\phi)\right]\delta\phi d^4x \\
    &=-\frac{1}{16\pi}\int\left[\partial_\nu(\sqrt{-g}g^{\mu\nu}\partial_\mu\phi)-\sqrt{-g}V^{\prime}(\phi)\right]\delta\phi d^4x.
\label{pla_phi}
\end{align}
where we used the fact that $\delta\phi$ vanishes at boundaries and a prime denotes a derivative with respect to $\phi$. If equation \eqref{pla_phi} is valid for arbitrary $\delta\phi$, we obtain the equation of motion for $\phi$:
\begin{equation}
   \partial_\nu(\sqrt{-g}g^{\mu\nu}\partial_\mu\phi)=\partial_\nu(\sqrt{-g})g^{\mu\nu}\partial_\mu\phi+\sqrt{-g}\partial_\nu(g^{\mu\nu}\partial_\mu\phi)=\sqrt{-g}V^{\prime}(\phi).
\end{equation}
In the FLRW metric, the above equation of motion becomes
\begin{align}
\nonumber
   & &\partial_0(\sqrt{-g})g^{00}\partial_0\phi+\sqrt{-g}\partial_0(g^{00}\partial_0\phi)=\sqrt{-g}V^{\prime}(\phi) \\
    \nonumber
    &\Longrightarrow&\partial_0(\sqrt{-g})(-\dot\phi)+\sqrt{-g}(-\ddot\phi)=\sqrt{-g}V^{\prime}(\phi) \\
    \nonumber
    &\Longrightarrow&\partial_0(a^3)\frac{\sqrt{-g}}{a^3}(-\dot\phi)+\sqrt{-g}(-\ddot\phi)=\sqrt{-g}V^{\prime}(\phi) \\
    \nonumber
    &\Longrightarrow&-3\frac{\dot{a}}{a}\dot\phi-\ddot\phi=V^{\prime}(\phi) \\
    &\Longrightarrow&\ddot\phi+3\frac{\dot{a}}{a}\dot\phi-V^{\prime}(\phi)=0 \Longleftrightarrow \ddot\phi+3H\dot\phi-V^{\prime}(\phi)=0.
\label{eom}
\end{align}

In this work, we use the \pcdm\ models with an inverse power-law scalar field potential energy density \citep{peebrat88,ratpeeb88,pavlov13}
\be
\label{PE}
V(\phi)=\frac{1}{2}\kappa m_p^2\phi^{-\alpha},
\ee
where $m_p$ is the Planck mass, $\alpha$ is a positive constant (with $\alpha=0$ corresponding to a cosmological constant), and $\kappa$ is a constant that determines the strength of the coupling between the scalar field and gravity. We obtain $\kappa$ either by using the shooting method in the Cosmic Linear Anisotropy Solving System (\textsc{class}) code \citep{class} or by adopting the matter-dominated initial condition form \footnote{The detailed derivation of this form can be found in subsection 3.6.3 of \cite{5}.} given by
\be\label{kp}
\kappa=\frac{8}{3m_p^2}\bigg(\frac{\alpha+4}{\alpha+2}\bigg)
\bigg[\frac{2}{3}\alpha(\alpha+2)\bigg]^{\alpha/2}.
\ee

Using the numerical solutions of the Friedmann equation \eqref{Hz3} and the equation of motion \eqref{eom}, we can obtain constraints on the cosmological parameters in the \pcdm\ models. Specifically, we can obtain constraints on the cosmological parameters $\{H_0, \obh\!, \och\!, \alpha\}$ (or $\{H_0, \Om, \alpha\}$) and $\{H_0, \obh\!, \och\!, \alpha, \Ok\}$ (or $\{H_0, \Om, \alpha, \Ok\}$) in the flat and non-flat \pcdm\ models, respectively. For recent studies on constraints on \pcdm\ see Refs.\ \citep{Zhaietal2017, ooba_etal_2018b, ooba_etal_2019, park_ratra_2018, park_ratra_2019b, park_ratra_2020, SolaPercaulaetal2019, Singhetal2019, UrenaLopezRoy2020, SinhaBanerjee2021, Xuetal2021, deCruzetal2021, Jesusetal2022, Adiletal2022} and related references within these papers.

\cleardoublepage


\chapter{Cosmological constraints from HII starburst galaxy apparent magnitude and other cosmological measurements}
\label{makereference3}

This chapter is based on \cite{CaoRyanRatra2020}. Figures and tables by Shulei Cao, from analyses
conducted independently by Shulei Cao and Joseph Ryan \citep{Ryan:2021eiw}.

\section{Introduction}
\label{makereference3.1}

The accelerated expansion of the current universe is now well-established observationally and is usually credited to a dark energy whose origins remain murky (see e.g. \citealp{Ratra_Vogeley,Martin,Coley_Ellis}). The standard \lcdm\ model of cosmology \citep{peeb84} describes a universe with flat spatial hypersurfaces predominantly filled with dark energy in the form of a cosmological constant $\Lambda$ and cold dark matter (CDM) together comprising $\sim95$\% of the total energy budget. While spatially-flat \lcdm\ is mostly consistent with cosmological observations (see e.g. \citealp{Alam_et_al_2017,Farooq_Ranjeet_Crandall_Ratra_2017,scolnic_et_al_2018,planck2018b}), there are indications of some (mild) discrepances between standard \lcdm\ model predictions and cosmological measurements. In addition, the quality and quantity of cosmological data continue to grow, making it possible to consider and constrain additional cosmological parameters beyond those that characterize the standard \lcdm\ model.

Given the uncertainty surrounding the origin of the cosmological constant, many workers have investigated the possibility that the cosmological ``constant'' is not really constant, but rather evolves in time, either by positing an equation of state parameter $w \neq -1$ (thereby introducing a redshift dependence into the dark energy density) or by replacing the constant $\Lambda$ in the Einstein-Hilbert action with a dynamical scalar field $\phi$ (\citealp{peebrat88, ratpeeb88}). Non-flat spatial geometry also introduces a time-dependent source term in the Friedmann equations. In this paper we study the standard spatially-flat \lcdm\ model as well as dynamical dark energy and spatially non-flat extensions of this model.

One major goal of this paper is to use measurements of the redshift, apparent luminosity, and gas velocity dispersion of HII starburst galaxies to constrain cosmological parameters.\footnote{For early attempts see \cite{Siegel_2005}, \cite{Plionis_2009,Plionis_2010,Plionis_2011} and \cite{Mania_2012}. For more recent discussions see \cite{Chavez_2016}, \cite{Wei_2016}, \cite{Yennapureddy_2017}, \cite{Zheng_2019}, \cite{ruan_etal_2019}, \cite{GonzalezMoran2019}, \cite{Wan_2019}, and \cite{Wu_2020}.}
An HII starburst galaxy (hereinafter ``\hiig'') is one that contains a large HII region, an emission nebula sourced by the UV radiation from an O- or B-type star. There is a correlation between the measured luminosity ($L$) and the inferred velocity dispersion ($\sigma$) of the ionized gases within these \hiig, referred to as the $L$-$\sigma$ relation (see Section \ref{makereference3.2}) which has been shown to be a useful cosmological tracer (see \citealp{Melnick_2000,Siegel_2005,Plionis_2011,Chavez_2012,Chavez_2014,Chavez_2016,Terlevich_2015,GonzalezMoran2019}, and references therein). This relation has been used to constrain the Hubble constant $H_0$ (\citealp{Chavez_2012,FernandezArenas}), and it can also be used to put constraints on the dark energy equation of state parameter $w$ (\citealp{Terlevich_2015,Chavez_2016,GonzalezMoran2019}).

\hiig\ data reach to redshift $z\sim2.4$, a little beyond that of the highest redshift baryon acoustic oscillation (BAO) data which reach to $z\sim2.3$. \hiig\ data are among a handful of cosmological observations that probe the largely unexplored part of redshift space from $z\sim2$ to $z\sim1100$. Other data that probe this region include quasar angular size measurements that reach to $z\sim2.7$ (\citealp{gurvits_kellermann_frey_1999,Chen_Ratra_2003,Cao_et_al2017b,Ryanetal2019}, and references therein), quasar flux measurements that reach to  $z\sim5$ (\citealp{RisalitiLusso2015,RisalitiLusso2019,Yangetal2020,KhadkaRatra2020a,KhadkaRatra2020b,Zheng_2020}, and references therein), and gamma ray burst data that reach to $z\sim8$ (\citealp{Lamb2000,samushia_ratra_2010,Demianskietal_2021}, and references therein). In this paper we also use quasar angular size measurements (hereinafter ``QSO'') to constrain cosmological model parameters.

While \hiig\ and QSO data probe the largely unexplored $z\sim2.3$--2.7 part of the universe, current \hiig\ and QSO measurements provide relatively weaker constraints on cosmological parameters than those provided by more widely used measurements, such as BAO peak length scale observations or Hubble parameter (hereinafter ``$H(z)$'') observations (with these latter data being at lower redshift but of better quality than \hiig\ or QSO data). However, we find that the \hiig\ and QSO constraints are consistent with those that follow from BAO and $H(z)$ data, and so we use all four sets of data together to constrain cosmological parameters. We find that the \hiig\ and QSO data tighten parameter constraints relative to the $H(z)$ + BAO only case.

Using six different cosmological models to constrain cosmological parameters allows us to determine which of our results are less model-dependent. In all models, the \hiig\ data favor those parts of cosmological parameter space for which the current cosmological expansion is accelerating.\footnote{This result could weaken, however, as the \hiig\ data constraint contours could broaden when \hiig\ data systematic uncertainties are taken into account. We do not incorporate any \hiig\ systematic uncertainties into our analysis; see below.} The joint analysis of the \hiig\, QSO, BAO and $H(z)$ data results in relatively model-independent and fairly tight determination of the Hubble constant $H_0$ and the current non-relativistic matter density parameter \om.\footnote{The BAO and $H(z)$ data play a more significant role than do the \hiig\ and QSO data in setting these and other limits, but the \hiig\ and QSO data tighten the BAO + $H(z)$ constraints. We note, however, that the $H(z)$ and QSO data, by themselves, give lower central values of $H_0$ but with larger error bars. Also, because we calibrate the distance scale of the BAO measurements listed in Table \ref{tab:BAO} via the sound horizon scale at the drag epoch ($r_s$, about which see below), a quantity that depends on early-Universe physics, we would expect these measurements to push the best-fitting values $H_0$ lower when they are combined with late-Universe measurements like \hiig\ (whose distance scale is not set by the physics of the early Universe).} Depending on the model, \om\ ranges from a low of $0.309^{+0.015}_{-0.014}$ to a high of $0.319 \pm 0.013$, being consistent with most other estimates of this parameter (unless indicated otherwise, uncertainties given in this paper are $\pm 1\sigma$). The best-fitting values of $H_0$, ranging from $68.18^{+0.97}_{-0.75}$ \hunit to $69.90 \pm 1.48$ \hunit, are, from the quadrature sum of the error bars, 2.01$\sigma$ to 3.40$\sigma$ lower than the local $H_0 = 74.03 \pm 1.42$ \hunit measurement of \cite{riess_etal_2019} and only 0.06$\sigma$ to 0.60$\sigma$ higher than the median statistics $H_0 = 68 \pm 2.8$ \hunit estimate of \cite{chenratmed}. These combined measurements are consistent with the spatially-flat \lcdm\ model, but also do not strongly disallow some mild dark energy dynamics, as well as a little non-zero spatial curvature energy density.

This paper is organized as follows. In Section \ref{makereference3.2} we introduce the data we use. The models we analyze are described in Chapter \ref{sec:models}, with a description of our analysis method in Section \ref{makereference3.3}. Our results are in Section \ref{makereference3.4}, and we provide our conclusions in Section \ref{makereference3.6}.

\section{Data}
\label{makereference3.2}
We use a combination of $H(z)$, BAO, QSO, and \hiig\ data to obtain constraints on our cosmological models. The $H(z)$ data, spanning the redshift range $0.070 \leq z \leq 1.965$, are identical to the $H(z)$ data used in \cite{Ryan_1, Ryanetal2019} and compiled in Table 2 of \cite{Ryan_1}; see that paper for description. The QSO data compiled by \cite{Cao_et_al2017b} (listed in Table 1 of that paper) and spanning the redshift range $0.462 \leq z \leq 2.73$, are identical to that used in \cite{Ryanetal2019}; see these papers for descriptions. Our BAO data (see Table \ref{tab:BAO}) have been updated relative to \cite{Ryanetal2019} and span the redshift range $0.38 \leq z \leq 2.34$. Our \hiig\ data are new, comprising 107 low redshift ($0.0088 \leq z \leq 0.16417$) \hiig\ measurements, used in \cite{Chavez_2014}, and 46 high redshift ($0.636427 \leq z \leq 2.42935$) \hiig\ measurements, used in \cite{GonzalezMoran2019}.\footnote{15 from \cite{GonzalezMoran2019}, 25 from \cite{Erb_2006}, \cite{Masters_2014}, and \cite{Maseda_2014}, and 6 from \cite{Terlevich_2015}.} These extinction-corrected measurements (see below for a discussion of extinction correction) were very kindly provided to us by Ana Luisa Gonz\'{a}lez-Mor\'{a}n (private communications, 2019 and 2020).

In order to use BAO measurements to constrain cosmological model parameters, knowledge of the sound horizon scale at the drag epoch ($r_s$) is required. We compute this scale more accurately than in \cite{Ryanetal2019} by using the approximate formula \eqref{eq:sh} \citep{PhysRevD.92.123516}. Here $\Omega_{\rm cb_0} = \Omega_{\rm c_0} + \Omega_{\rm b_0} = \Omega_{\rm m_0} - \Omega_{\rm \nu_0}$ with $\Omega_{\rm cb_0}$, $\Omega_{\rm c_0}$, $\Omega_{\rm b_0}$, and $\Omega_{\nu_0} = 0.0014$ (following \citealp{Carter_2018}) being the current values of the CDM + baryonic matter, CDM, baryonic matter, and neutrino energy density parameters, respectively, and the Hubble constant $H_0 = 100\ h$ \hunit. Here and in what follows, a subscript of `0' on a given quantity denotes the current value of that quantity. Additionally, $\Omega_{\rm b_0}h^2$ is slightly model-dependent; the values of this parameter that we use in this paper are the same as those originally computed in \cite{park_ratra_2018, park_ratra_2019a, ParkRatra2019a} and listed in Table 2 of \cite{Ryanetal2019}.

\begin{table}
\centering
\begin{threeparttable}
\caption{BAO data.}\label{tab:BAO}
\setlength{\tabcolsep}{3.5pt}
\begin{tabular}{lccc}
\hline
$z$ & Measurement\tnote{a} & Value & Ref.\\
\hline
$0.38$ & $D_M\left(r_{s,{\rm fid}}/r_s\right)$ & 1512.39 & \cite{Alam_et_al_2017}\tnote{b}\\
$0.38$ & $H(z)\left(r_s/r_{s,{\rm fid}}\right)$ & 81.2087 & \cite{Alam_et_al_2017}\tnote{b}\\
$0.51$ & $D_M\left(r_{s,{\rm fid}}/r_s\right)$ & 1975.22 & \cite{Alam_et_al_2017}\tnote{b}\\
$0.51$ & $H(z)\left(r_s/r_{s,{\rm fid}}\right)$ & 90.9029 & \cite{Alam_et_al_2017}\tnote{b}\\
$0.61$ & $D_M\left(r_{s,{\rm fid}}/r_s\right)$ & 2306.68 & \cite{Alam_et_al_2017}\tnote{b}\\
$0.61$ & $H(z)\left(r_s/r_{s,{\rm fid}}\right)$ & 98.9647 & \cite{Alam_et_al_2017}\tnote{b}\\
$0.122$ & $D_V\left(r_{s,{\rm fid}}/r_s\right)$ & $539\pm17$ & \cite{Carter_2018}\\
$0.81$ & $D_A/r_s$ & $10.75\pm0.43$ & \cite{DES_2019b}\\
$1.52$ & $D_V\left(r_{s,{\rm fid}}/r_s\right)$ & $3843\pm147$ & \cite{3}\\
$2.34$ & $D_H/r_s$ & 8.86 & \cite{Agathe}\tnote{c}\\
$2.34$ & $D_M/r_s$ & 37.41 & \cite{Agathe}\tnote{c}\\
\hline
\end{tabular}
\begin{tablenotes}
\item[a] $D_M \left(r_{s,{\rm fid}}/r_s\right)$, $D_V \left(r_{s,{\rm fid}}/r_s\right)$, $r_s$, and $r_{s, {\rm fid}}$ have units of Mpc, while $H(z)\left(r_s/r_{s,{\rm fid}}\right)$ has units of \hunit, and $D_A/r_s$ is dimensionless.
\item[b] The six measurements from \cite{Alam_et_al_2017} are correlated; see eq. (20) of \cite{Ryanetal2019} for their correlation matrix.
\item[c] The two measurements from \cite{Agathe} are correlated; see eq. \eqref{CovM} below for their correlation matrix.
\end{tablenotes}
\end{threeparttable}
\end{table}

As mentioned in Section \ref{makereference3.1}, \hiig\ can be used as cosmological probes because they exhibit a tight correlation between the observed luminosity ($L$) of their Balmer emission lines and the velocity dispersion ($\sigma$) of their ionized gas (as measured from the widths of the emission lines). That correlation can be expressed in the form
\begin{equation}
\label{eq:logL}
    \log L = \beta \log \sigma + \gamma,
\end{equation}
where $\gamma$ and $\beta$ are the intercept and slope, respectively, and $\log = \log_{10}$ here and in what follows. In order to determine the values of $\beta$ and $\gamma$, it is necessary to establish the extent to which light from an \hiig\ is extinguished as it propagates through space. A correction must be made to the observed flux so as to account for the effect of this extinction. As mentioned above, the data we received from Ana Luisa Gonz\'{a}lez-Mor\'{a}n have been corrected for extinction. In \cite{GonzalezMoran2019}, the authors used the \cite{Gordon_2003} extinction law, and in so doing found
\begin{equation}
    \label{eq:Gordon_beta}
    \beta = 5.022 \pm 0.058,
\end{equation}
and
\begin{equation}
    \label{eq:Gordon_gamma}
    \gamma = 33.268 \pm 0.083,
\end{equation}
respectively. These are the values of $\beta$ and $\gamma$ that we use in the $L$-$\sigma$ relation, eq. \eqref{eq:logL}.

Once the luminosity of an \hiig\ has been established through eq. \eqref{eq:logL}, this luminosity can be used, in conjunction with a measurement of the flux ($f$) emitted by the \hiig, to determine the distance modulus of the \hiig\ via
\begin{equation}
    \mu_{\rm obs} = 2.5\log L - 2.5\log f - 100.2
\end{equation}
(see e.g. \citealp{Terlevich_2015}, \citealp{GonzalezMoran2019}, and references therein).\footnote{For each \hiig\ in our sample we have the measured values and uncertainties of $\log f$\!, $\log \sigma$\!, and $z$.} This quantity can then be compared to the value of the distance modulus predicted within a given cosmological model
\be
\label{eq:mu_th}
    \mu_{\rm th}\left(\textbf{p}, z\right) = 5\log D_{\rm L}\left(\textbf{p}, z\right) + 25,
\ee
where the luminosity distance $D_L(\textbf{p}, z)$ is related to the transverse comoving distance $D_M(\textbf{p}, z)$ and the angular size distance $D_A(\textbf{p}, z)$ through equation \eqref{DM-DL-DA}. These are functions of the redshift $z$ and the parameters $\textbf{p}$ of the model in question, and $D_M$ is defined in equation \eqref{eq:DM}.

As the precision of cosmological observations has grown over the last few years, a tension between measurements of the Hubble constant made with early-Universe probes and measurements made with late-Universe probes has revealed itself (for a review, see \citealp{riess_2019}). Whether a given cosmological observation supports a lower value of $H_0$ (i.e. one that is closer to the early-Universe \textit{Planck} measurement) or a higher value of $H_0$ (i.e. one that is closer to the late-Universe value measured by \citealp{riess_etal_2019}) may depend on whether the distance scale associated with this observation has been set by early- or late-Universe physics. It is therefore important to know what distance scale cosmological observations have been calibrated to, so that the extent to which measurements of $H_0$ are pushed higher or lower by these different distance calibrations can be clearly identified.

The $H_0$ values we measure from the combined $H(z)$, BAO, QSO, and \hiig\ data are based on a combination of both early- and late-Universe distance calibrations. As mentioned above, the distance scale of our BAO measurements is set by the size of the sound horizon at the drag epoch $r_s$. The sound horizon, in turn, depends on $\Omega_{\rm b_0}h^2$, which was computed by \cite{park_ratra_2018, park_ratra_2019a, ParkRatra2019a} using early-Universe data. Our \hiig\ measurements, on the other hand, have been calibrated using cosmological model independent distance ladder measurements of the distances to nearby giant HII regions (see \citealp{GonzalezMoran2019} and references therein), so these data qualify as late-Universe probes. The distance scale of our QSO measurements is set by the intrinsic linear size ($l_m$) of the QSOs themselves, which is a late-Universe measurement (see \citealp{Cao_et_al2017b}). Finally, our $H(z)$ data depend on late-Universe astrophysics through the modeling of the star formation histories of the galaxies whose ages are measured to obtain the Hubble parameter (although the differences between different models are not thought to have a significant effect on measurements of $H(z)$ from these galaxies; see \citealp{moresco_et_al_2018, moresco_et_al_2020}).

\section{Data Analysis Methodology}
\label{makereference3.3}

We perform a Markov chain Monte Carlo (MCMC) analysis with the Python module emcee \citep{emcee} and maximize the likelihood function, $\mathcal{L}$, 
to determine the best-fitting values of the parameters $\textbf{p}$ of the models. We use flat priors for all parameters $\textbf{p}$. For all models, the priors on $\Omega_{\rm m_0}$ and $h$ are non-zero over the ranges $0.1 \leq \Omega_{\rm m_0} \leq 0.7$ and $0.50 \leq h \leq 0.85$. In the non-flat \lcdm\ model the $\Omega_{\Lambda}$ prior is non-zero over $0.2 \leq \Omega_{\Lambda} \leq 1$. In the flat and non-flat XCDM parametrizations the prior range on $w_{\rm X}$ is $-2 \leq w_{\rm X} \leq 0$, and the prior range on $\Omega_{\rm k_0}$ in the non-flat XCDM parametrization is $-0.7 \leq \Omega_{\rm k_0} \leq 0.7$. In the flat and non-flat \pcdm\ models the prior range on $\alpha$ is $0.01 \leq \alpha \leq 3$ and $0.01 \leq \alpha \leq 5$, respectively, and the prior range on $\Omega_{\rm k_0}$ is also $-0.7 \leq \Omega_{\rm k_0} \leq 0.7$.

For \hiig, the likelihood function is
\be
\label{eq:LH1}
    \mathcal{L}_{\rm \hiig}= e^{-\chi^2_{\rm \hiig}/2},
\ee
where
\be
\label{eq:chi2_HIIG}
    \chi^2_{\rm \hiig}(\textbf{p}) = \sum^{153}_{i = 1} \frac{[\mu_{\rm th}(\textbf{p}, z_i) - \mu_{\rm obs}(z_i)]^2}{\epsilon_i^2},
\ee
and $\epsilon_i$ is the uncertainty of the $i_{\rm th}$ measurement. Following \cite{GonzalezMoran2019}, $\epsilon$ has the form
\be
\label{eq:err_HIIG}
    \epsilon=\sqrt{\epsilon^2_{\rm stat}+\epsilon^2_{\rm sys}},
\ee
where the statistical uncertainties are
\be
\label{eq:stat_err_HIIG}
    \epsilon^2_{\rm stat}=6.25\left[\epsilon^2_{\log f}+\beta^2\epsilon^2_{\log\sigma}+\epsilon^2_{\beta}(\log\sigma)^2+\epsilon^2_{\gamma}\right]+\left(\frac{\partial{\mu_{\rm th}}}{\partial{z}}\right)^2\epsilon^2_{z}.
\ee
Following \cite{GonzalezMoran2019} we do not account for systematic uncertainties in our analysis, so the uncertainty on the \hiig\ measurements consists entirely of the statistical uncertainty (so that $\epsilon = \epsilon_{\rm stat}$).\footnote{A systematic error budget for \hiig\ data is available in the literature, however; see \cite{Chavez_2016}.} The reader should also note here that although the theoretical statistical uncertainty depends our cosmological model parameters (through the theoretical distance modulus $\mu_{\rm th} = \mu_{\rm th}\left(\textbf{p}, z\right)$), the effect of this model-dependence on the parameter constraints is negligible for the current data.\footnote{In contrast to our definition of $\chi^2$ in eq. \eqref{eq:chi2_HIIG}, \cite{GonzalezMoran2019} defined an $H_0$-independent $\chi^2$ function in their eq. (27) and weighted this $\chi^2$ function by $1/\epsilon^2_{\rm stat}$ (where $\epsilon^2_{\rm stat}$ is given by their eq. (15)) which we do not do. This procedure is discussed in the literature \citep{Melnick_2017,FernandezArenas}, and when we use it we find that it leads to a reduced $\chi^2$ identical to that given in \cite{GonzalezMoran2019} (being less than 2 but greater than 1) without having a noticeable effect on the shapes or peak locations of our posterior likelihoods (hence providing very similar best-fitting values and error bars of the cosmological model parameters). As discussed below, with our $\chi^2$ definition we find reduced $\chi^2$ values $\sim2.75$. \cite{GonzalezMoran2019} note that an accounting of systematic uncertainties could decrease the reduced $\chi^2$ values towards unity.\label{fn5}}

For $H(z)$, the likelihood function is
\be
\label{eq:LH2}
    \mathcal{L}_{\rm H}= e^{-\chi^2_{\rm H}/2},
\ee
where
\be
\label{eq:chi2_Hz}
    \chi^2_{\rm H}(\textbf{p}) = \sum^{31}_{i = 1} \frac{[H_{\rm th}(\textbf{p}, z_i) - H_{\rm obs}(z_i)]^2}{\epsilon_i^2},
\ee
and $\epsilon_i$ is the uncertainty of $H_{\rm obs}(z_i)$.

For the BAO data, the likelihood function is
\be
\label{eq:LH3}
    \mathcal{L}_{\rm BAO}= e^{-\chi^2_{\rm BAO}/2},
\ee
and for the uncorrelated BAO data (lines 7-9 in Table \ref{tab:BAO}) the $\chi^2$ function takes the form
\be
\label{eq:chi2_BAO1}
    \chi^2_{\rm BAO}(\textbf{p}) = \sum^{3}_{i = 1} \frac{[A_{\rm th}(\textbf{p}, z_i) - A_{\rm obs}(z_i)]^2}{\epsilon_i^2},
\ee
where $A_{\rm th}$ and $A_{\rm obs}$ are, respectively, the theoretical and observational quantities as listed in Table \ref{tab:BAO}, and $\epsilon_{i}$ corresponds to the uncertainty of $A_{\rm obs}(z_i)$. For the correlated BAO data, the $\chi^2$ function takes the form
\be
\label{eq:chi2_BAO2}
    \chi^2_{\rm BAO}(\textbf{p}) = [A_{\rm th}(\textbf{p}) - A_{\rm obs}(z_i)]^T\textbf{C}^{-1}[A_{\rm th}(\textbf{p}) - A_{\rm obs}(z_i)],
\ee
where superscripts $T$ and $-1$ denote the transpose and inverse of the matrices, respectively. The covariance matrix $\textbf{C}$ for the BAO data, taken from \cite{Alam_et_al_2017}, is given in eq. (20) of \cite{Ryanetal2019}, while for the BAO data from \cite{Agathe},
\be
\label{CovM}
    \textbf{C}=
    \begin{bmatrix}
    0.0841 & -0.183396 \\
    -0.183396 & 3.4596
    \end{bmatrix}.
\ee

For QSO, the likelihood function is
\be
\label{eq:LH4}
    \mathcal{L}_{\rm QSO}= e^{-\chi^2_{\rm QSO}/2},
\ee
and the $\chi^2$ function takes the form
\be
\label{eq:chi2_QSO}
    \chi^2_{\rm QSO}(\textbf{p}) = \sum^{120}_{i = 1} \left[\frac{\theta_{\rm th}(\textbf{p}, z_i) - \theta_{\rm obs}(z_i)}{\epsilon_i+0.1\theta_{\rm obs}(z_i)}\right]^2,
\ee
where $\theta_{\rm th}(\textbf{p}, z_i)$ and $\theta_{\rm obs}(z_i)$ are theoretical and observed values of the angular size at redshift $z_i$, respectively, and $\epsilon_i$ is the uncertainty of $\theta_{\rm obs}(z_i)$ (see \citealp{Ryanetal2019} for more details).

For the joint analysis of these data, the total likelihood function is obtained by multiplying the individual likelihood functions (that is, eqs. \eqref{eq:LH1}, \eqref{eq:LH2}, \eqref{eq:LH3}, and \eqref{eq:LH4}) together in various combinations. For example, for $H(z)$, BAO, and QSO data, we have
\be
\label{LH}
    \mathcal{L}=\mathcal{L}_{\rm H}\mathcal{L}_{\rm BAO}\mathcal{L}_{\rm QSO}.
\ee

We also use the Akaike Information
Criterion ($AIC$) and Bayesian Information Criterion ($BIC$) to compare the goodness of fit of models with different numbers of parameters, where
\be
\label{AIC}
    AIC=-2\ln \mathcal{L}_{\rm max} + 2n\equiv\chi^2_{\rm min}+2n,
\ee
and
\be
\label{BIC}
    BIC=-2\ln \mathcal{L}_{\rm max} + n\ln N\equiv\chi^2_{\rm min}+n\ln N.
\ee
In these two equations, $\mathcal{L}_{\rm max}$ refers to the maximum value of the given likelihood function, $\chi^2_{\rm min}$ refers to the corresponding minimum $\chi^2$ value, $n$ is the number of parameters of the given model, and $N$ is the number of data points (for example for \hiig\ we have $N=153$, etc.).

\section{Results}
\label{makereference3.4}
\subsection{\hiig\ constraints}
\label{subsec:HIIG}

We present the posterior one-dimensional (1D) probability distributions and two-dimensional (2D) confidence regions of the cosmological parameters for the six flat and non-flat models in Figs. \ref{fig01}--\ref{fig06}, in gray. The unmarginalized best-fitting parameter values are listed in Table \ref{tab:BFP}, along with the corresponding $\chi^2$, $AIC$, $BIC$, and degrees of freedom $\nu$ (where $\nu \equiv N - n$). The marginalized best-fitting parameter values and uncertainties ($\pm 1\sigma$ error bars or $2\sigma$ limits) are given in Table \ref{tab:1d_BFP}.\footnote{We plot these figures by using the Python package GetDist \citep{Lewis_2019}, which we also use to compute the central values (posterior means) and uncertainties of the cosmological parameters listed in Table \ref{tab:1d_BFP}.}

From the fit to the \hiig\ data, we see that most of the probability lies in the part of the parameter space corresponding to currently-accelerating cosmological expansion (see the gray contours in Figs. \ref{fig01}--\ref{fig06}). This means that the \hiig\ data favor currently-accelerating cosmological expansion,\footnote{Although a full accounting of the systematic uncertainties in the \hiig\ data could weaken this conclusion.} in agreement with supernova Type Ia, BAO, $H(z)$, and other cosmological data.

From the \hiig\ data, we find that the constraints on the non-relativistic matter density parameter \om\ are consistent with other estimates, ranging between a high of $0.300^{+0.106}_{-0.083}$ (flat XCDM) and a low of $\Omega_{\rm m_0} = 0.210^{+0.043}_{-0.092}$ (flat \pcdm).

The \hiig\ data constraints on $H_0$ in Table \ref{tab:1d_BFP} are consistent with the estimate of $H_0 = 71.0 \pm 2.8 ({\rm stat.}) \pm 2.1 ({\rm sys.})$ \hunit made by \cite{FernandezArenas} based on a compilation of \hiig\ measurements that differs from what we have used here. The \hiig\ $H_0$ constraints listed in Table \ref{tab:1d_BFP} are also consistent with other recent measurements of $H_0$, being between $0.90\sigma$ (flat XCDM) and $1.56\sigma$ (non-flat \pcdm) lower than the recent local expansion rate measurement of $H_0 = 74.03 \pm 1.42$ \hunit \citep{riess_etal_2019},\footnote{Note that other local expansion rate measurements are slightly lower with slightly larger error bars \citep{rigault_etal_2015,zhangetal2017,Dhawan,FernandezArenas,freedman_etal_2019,freedman_etal_2020,rameez_sarkar_2021}.} and between $0.78\sigma$ (non-flat \pcdm) and $1.13\sigma$ (flat XCDM) higher than the median statistics estimate of $H_0=68 \pm 2.8$ \hunit \citep{chenratmed},\footnote{This is consistent with earlier median statistics estimates \citep{gott_etal_2001,chen_etal_2003} and also with a number of recent $H_0$ measurements \citep{chen_etal_2017,DES_2018,Gomez-ValentAmendola2018,haridasu_etal_2018,planck2018b,zhang_2018,dominguez_etal_2019,martinelli_tutusaus_2019,Cuceu_2019,zeng_yan_2019,schoneberg_etal_2019,lin_ishak_2021,zhang_huang_2019}.} with our measurements ranging from a low of $H_0=70.60^{+1.68}_{-1.84}$ \hunit (non-flat \pcdm) to a high of $H_0=71.85 \pm 1.96$ \hunit (flat XCDM).

As for spatial curvature, from the marginalized 1D likelihoods in Table \ref{tab:1d_BFP}, for non-flat \lcdm, non-flat XCDM, and non-flat \pcdm, we measure $\Omega_{\rm k_0}=0.094^{+0.237}_{-0.363}$,\footnote{Since $\Omega_{\rm k_0}=1-\Omega_{\rm m_0}-\Omega_{\Lambda}$, in the non-flat \lcdm\ model analysis we replace $\Omega_{\Lambda}$ with $\Omega_{\rm k_0}$ in the MCMC chains of $\{H_0, \Omega_{\rm m_0}, \Omega_{\Lambda}\}$ to obtain new chains of $\{H_0, \Omega_{\rm m_0}, \Omega_{\rm k_0}\}$ and so measure $\Omega_{\rm k_0}$ central values and uncertainties. A similar procedure, based on $\Omega_{\Lambda}=1-\Omega_{\rm m_0}$, is used to measure $\Omega_{\Lambda}$ in the flat \lcdm\ model.} $\Omega_{\rm k_0}=0.011^{+0.457}_{-0.460}$, and $\Omega_{\rm k_0}=0.291^{+0.348}_{-0.113}$, respectively. From the marginalized likelihoods, we see that non-flat \lcdm\ and XCDM models are consistent with all three spatial geometries, while non-flat \pcdm\ favors the open case at 2.58$\sigma$. However, this seems to be a little odd, especially for non-flat \pcdm, considering their unmarginalized best-fitting $\Omega_{\rm k_0}$\!'s are all negative (see Table \ref{tab:BFP}).

The fits to the \hiig\ data are consistent with dark energy being a cosmological constant but don't rule out dark energy dynamics. For flat (non-flat) XCDM, $w_{\rm X}=-1.180^{+0.560}_{-0.330}$ ($w_{\rm X}=-1.125^{+0.537}_{-0.321}$), which are both within 1$\sigma$ of $w_{\rm X}=-1$. For flat (non-flat) \pcdm, $2\sigma$ upper limits of $\alpha$ are $\alpha<2.784$ ($\alpha<4.590$), with the 1D likelihood functions, in both cases, peaking at $\alpha=0$.

Current \hiig\ data do not provide very restrictive constraints on cosmological model parameters, but when used in conjunction with other cosmological data they can help tighten the constraints.

\subsection{$H(z)$, BAO, and \hiig\ (HzBH) constraints}
\label{subsec:HzBH}

The \hiig\ constraints discussed in the previous subsection are consistent with constraints from most other cosmological data, so it is appropriate to use the \hiig\ data in conjunction with other data to constrain parameters. In this subsection we perform a full analysis of $H(z)$, BAO, and \hiig\ (HzBH) data and derive tighter constraints on cosmological parameters.

\begin{figure*}
\centering
  \subfloat[Full parameter range]{%
    \includegraphics[width=3.25in,height=3.25in]{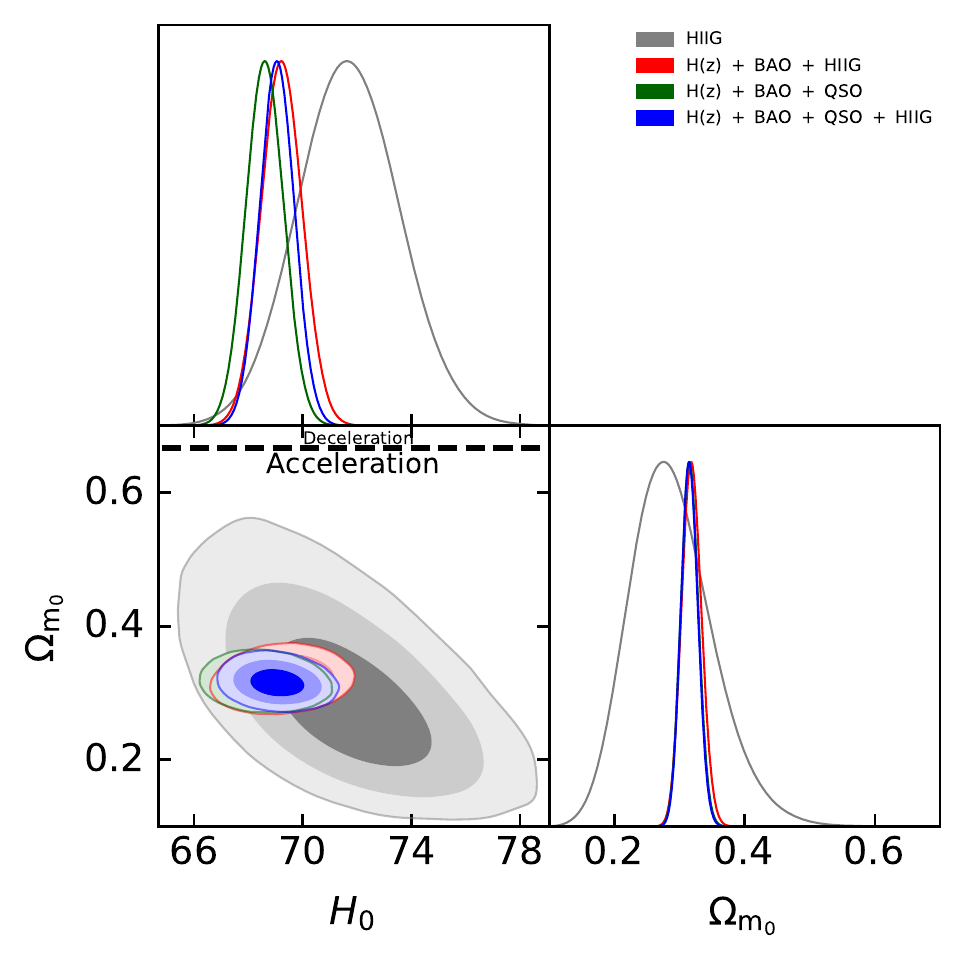}}
  \subfloat[Zoom in]{%
    \includegraphics[width=3.25in,height=3.25in]{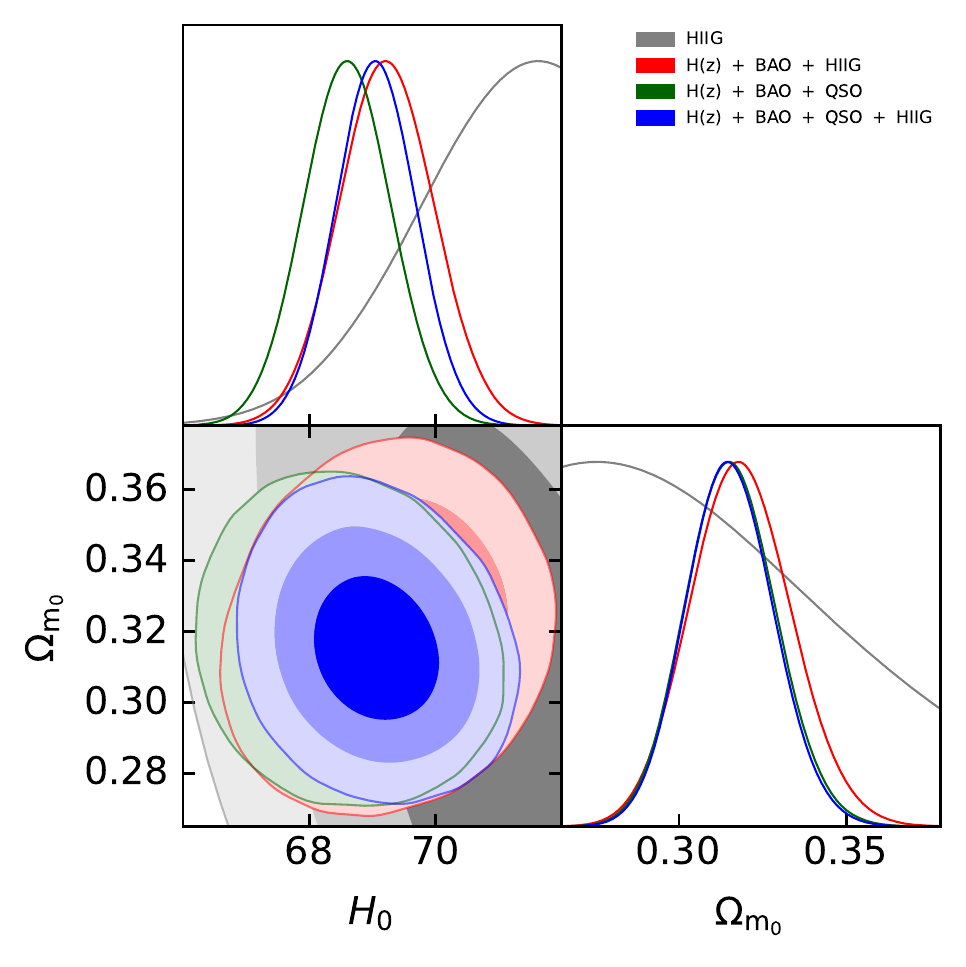}}\\
\caption{1$\sigma$, 2$\sigma$, and 3$\sigma$ confidence contours for flat \lcdm, where the right panel is the comparison zoomed in. The black dotted line is the zero-acceleration line, which divides the parameter space into regions associated with currently accelerated (below) and currently decelerated (above) cosmological expansion.}
\label{fig01}
\end{figure*}

\begin{figure*}
\centering
  \subfloat[Full parameter range]{%
    \includegraphics[width=3.25in,height=3.25in]{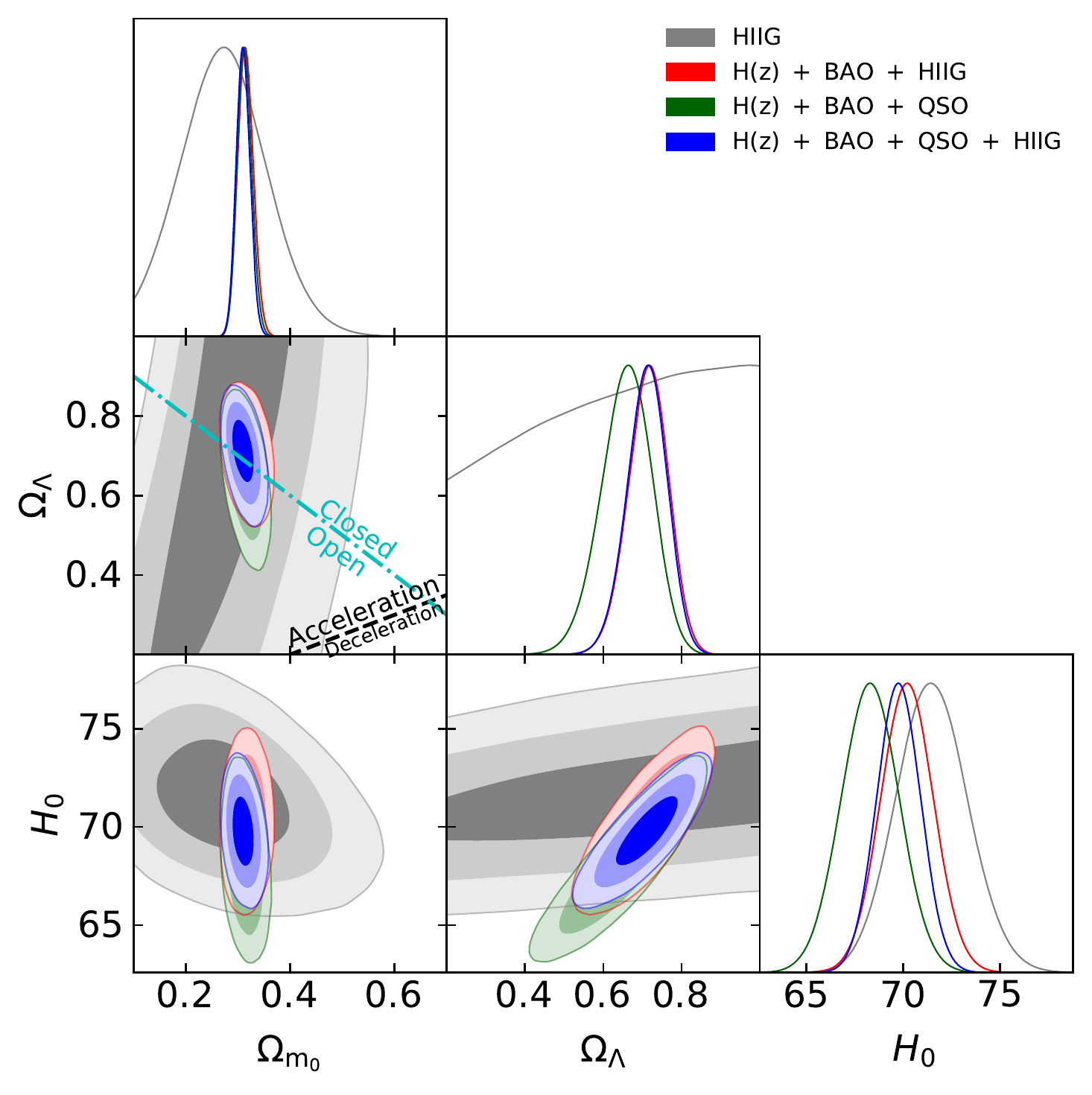}}
  \subfloat[Zoom in]{%
    \includegraphics[width=3.25in,height=3.25in]{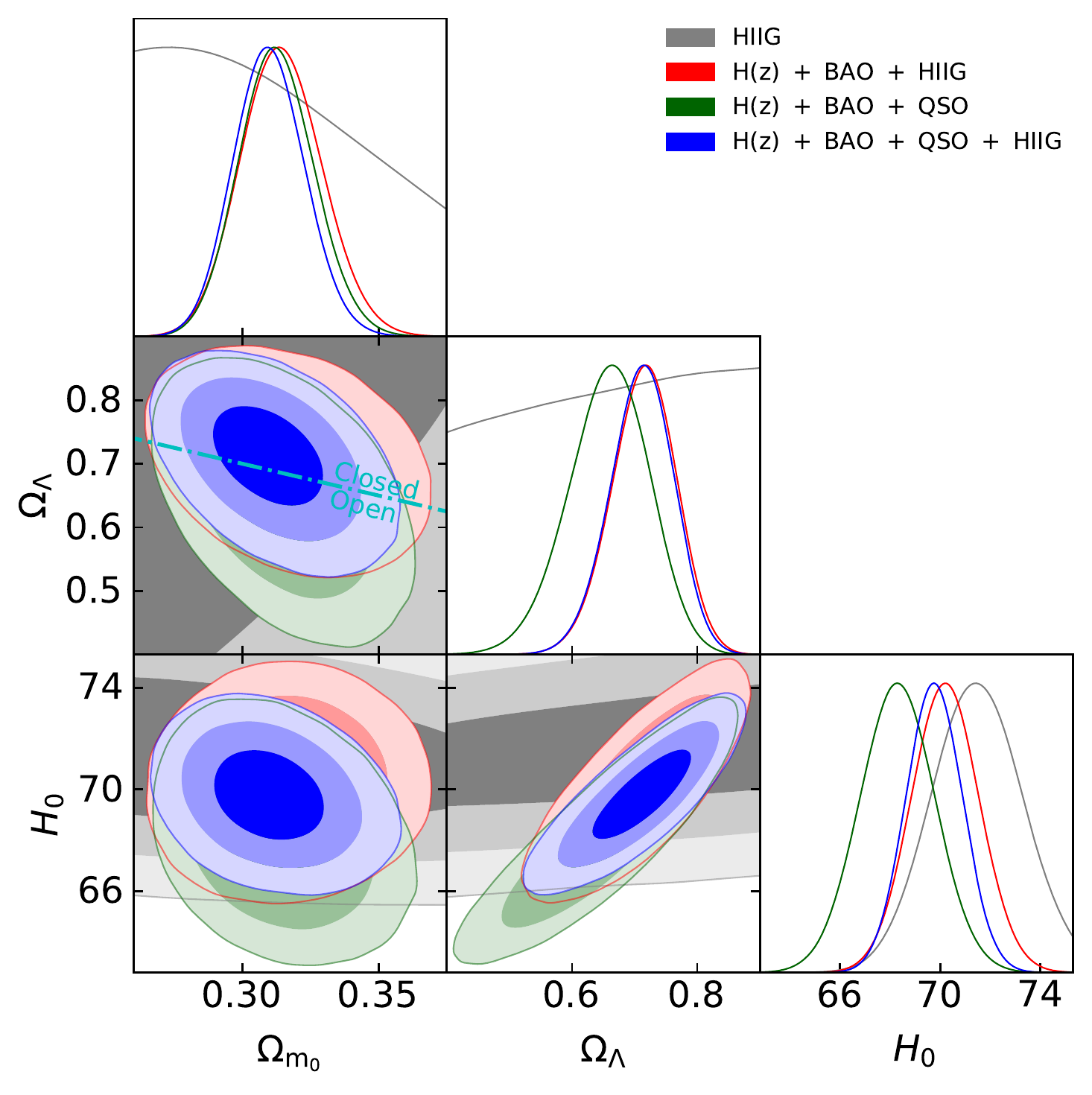}}\\
\caption{Same as Fig. \ref{fig01} but for non-flat \lcdm. The cyan dash-dot line represents the flat case, with closed spatial hypersurfaces to the upper right. The black dotted line is the zero-acceleration line, which divides the parameter space into regions associated with currently accelerated (above left) and currently decelerated (below right) cosmological expansion.}
\label{fig02}
\end{figure*}

\begin{figure*}
\centering
  \subfloat[Full parameter range]{%
    \includegraphics[width=3.25in,height=3.25in]{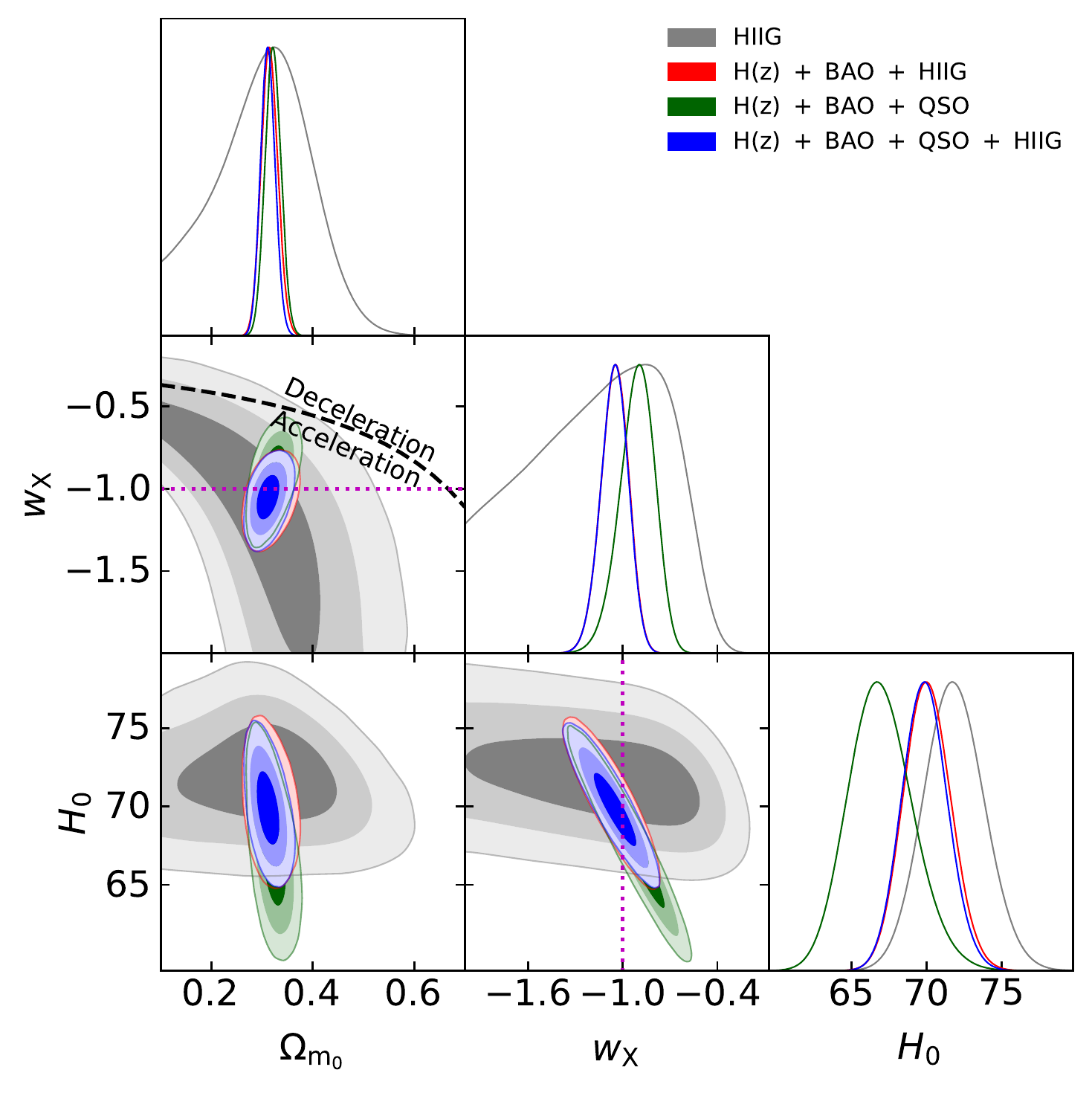}}
  \subfloat[Zoom in]{%
    \includegraphics[width=3.25in,height=3.25in]{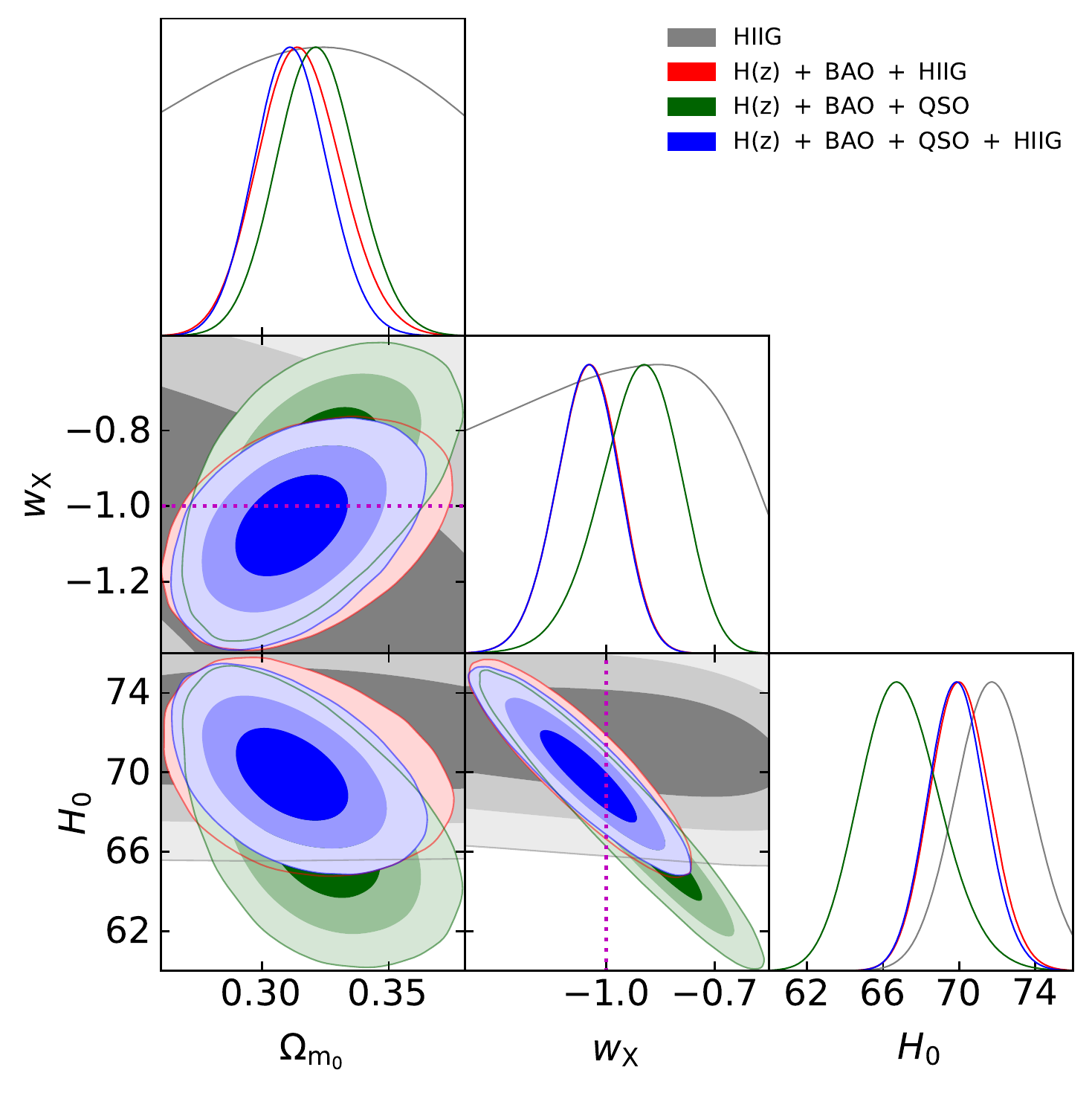}}\\
\caption{1$\sigma$, 2$\sigma$, and 3$\sigma$ confidence contours for flat XCDM. The black dotted line is the zero-acceleration line, which divides the parameter space into regions associated with currently accelerated (below left) and currently decelerated (above right) cosmological expansion. The magenta lines denote $w_{\rm X}=-1$, i.e. the flat \lcdm\ model.}
\label{fig03}
\end{figure*}

\begin{figure*}
\centering
  \subfloat[Full parameter range]{%
    \includegraphics[width=3.25in,height=3.25in]{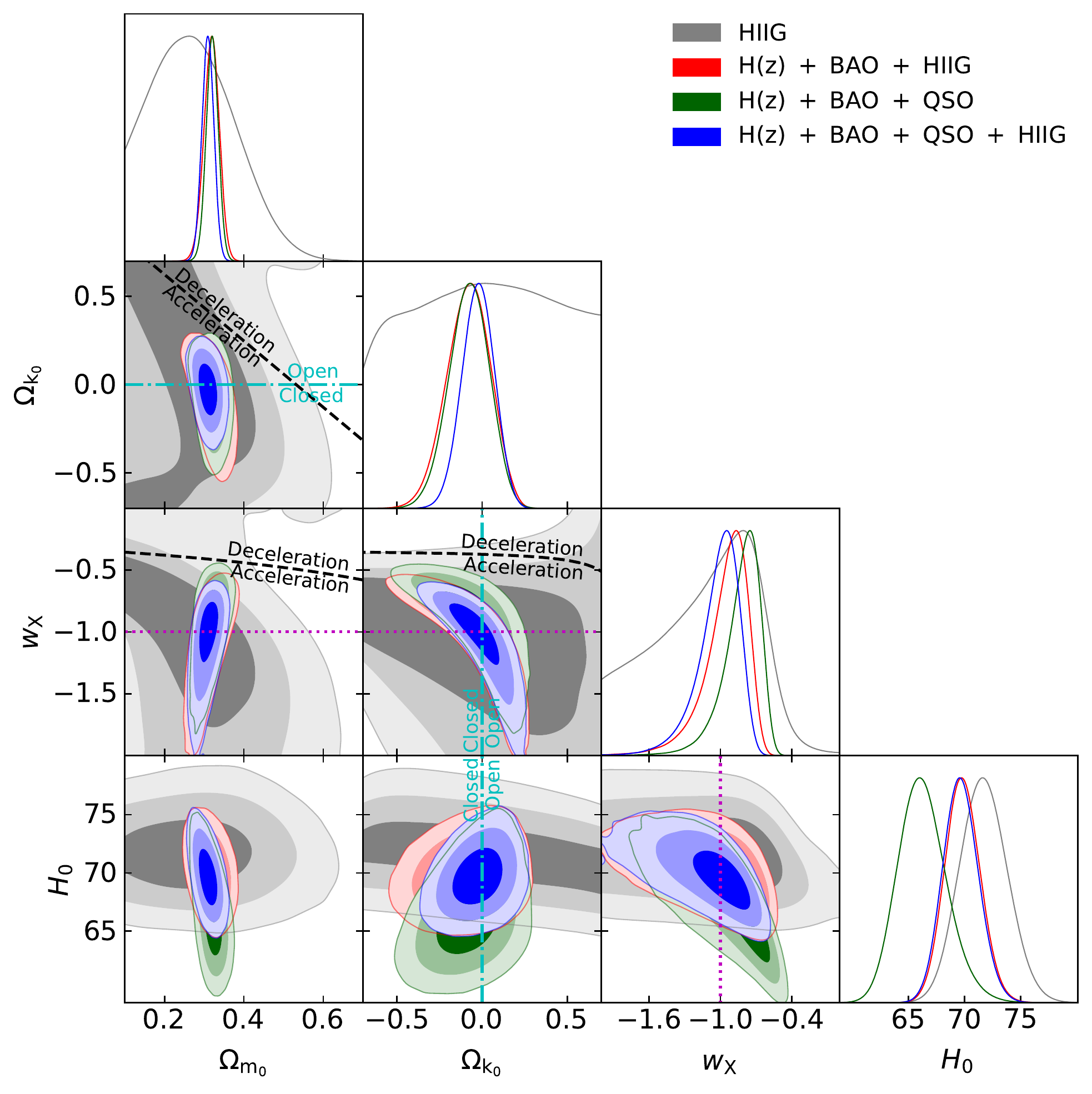}}
  \subfloat[Zoom in]{%
    \includegraphics[width=3.25in,height=3.25in]{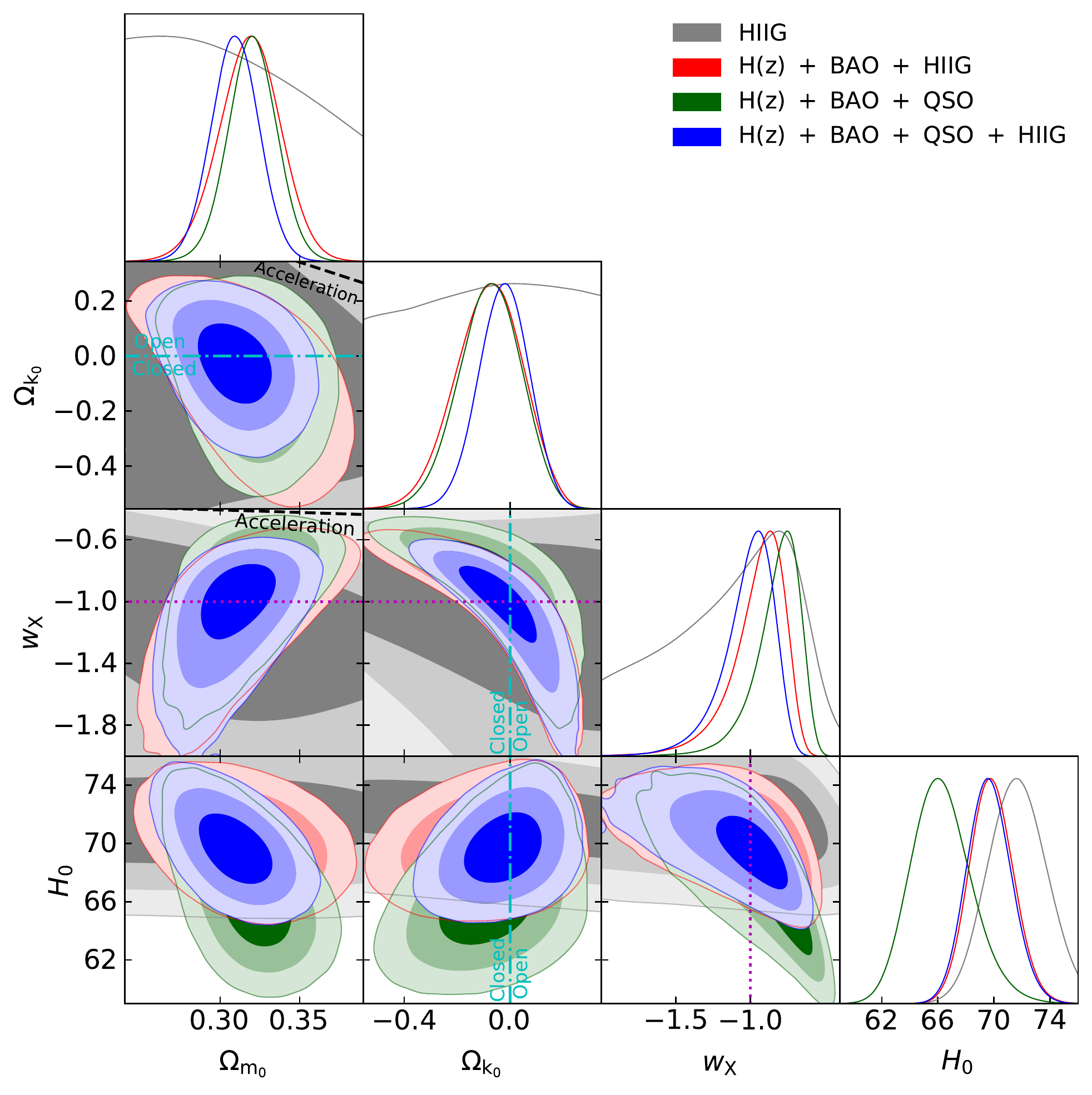}}\\
\caption{Same as Fig. \ref{fig03} but for non-flat XCDM, where the zero acceleration lines in each of the three subpanels are computed for the third cosmological parameter set to the \hiig\ data only best-fitting values listed in Table \ref{tab:BFP}. Currently-accelerated cosmological expansion occurs below these lines. The cyan dash-dot lines represent the flat case, with closed spatial hypersurfaces either below or to the left. The magenta lines indicate $w_{\rm X} = -1$, i.e. the non-flat \lcdm\ model.}
\label{fig04}
\end{figure*}

\begin{figure*}
\centering
  \subfloat[Full parameter range]{%
    \includegraphics[width=3.25in,height=3.25in]{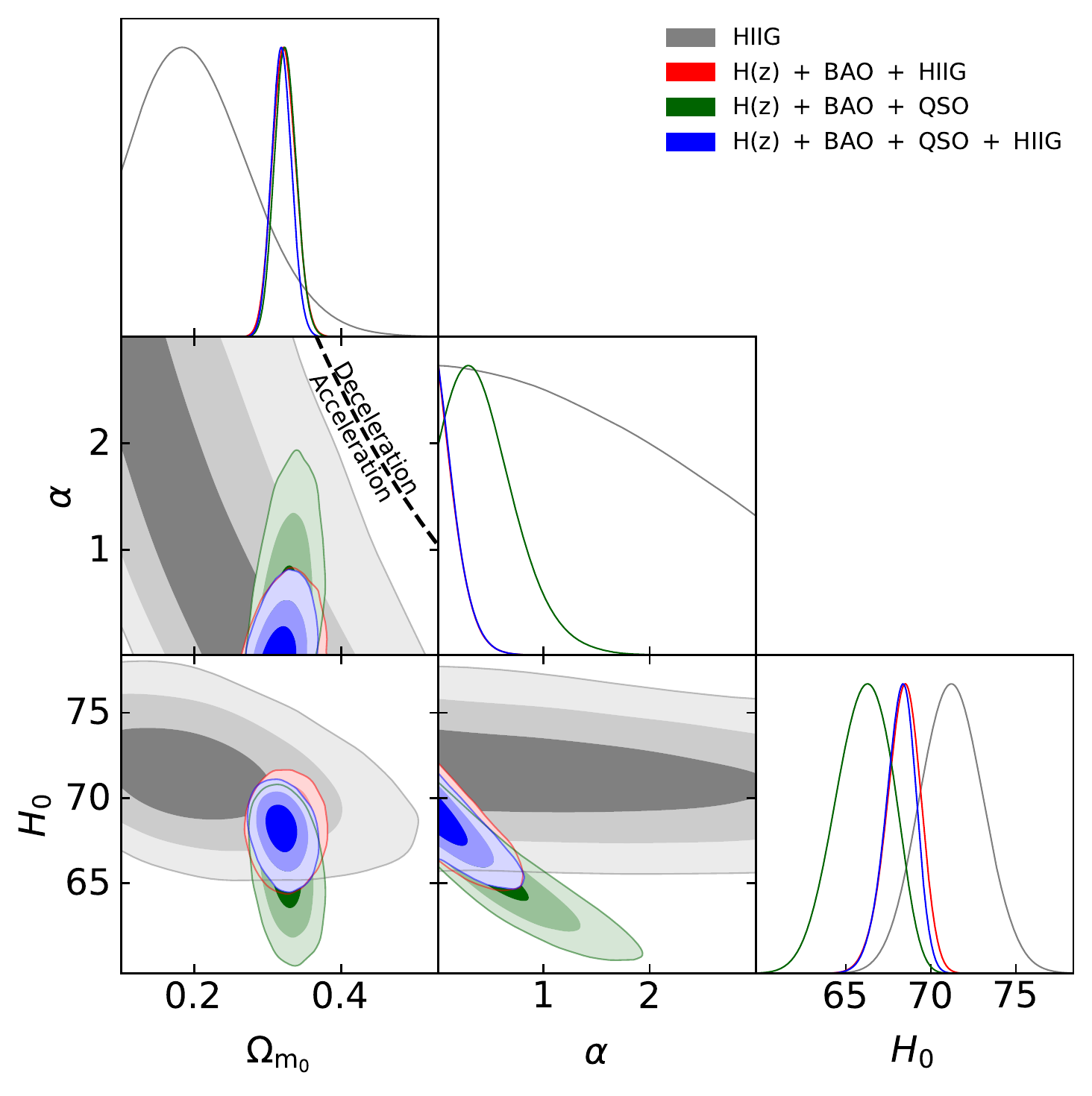}}
  \subfloat[Zoom in]{%
    \includegraphics[width=3.25in,height=3.25in]{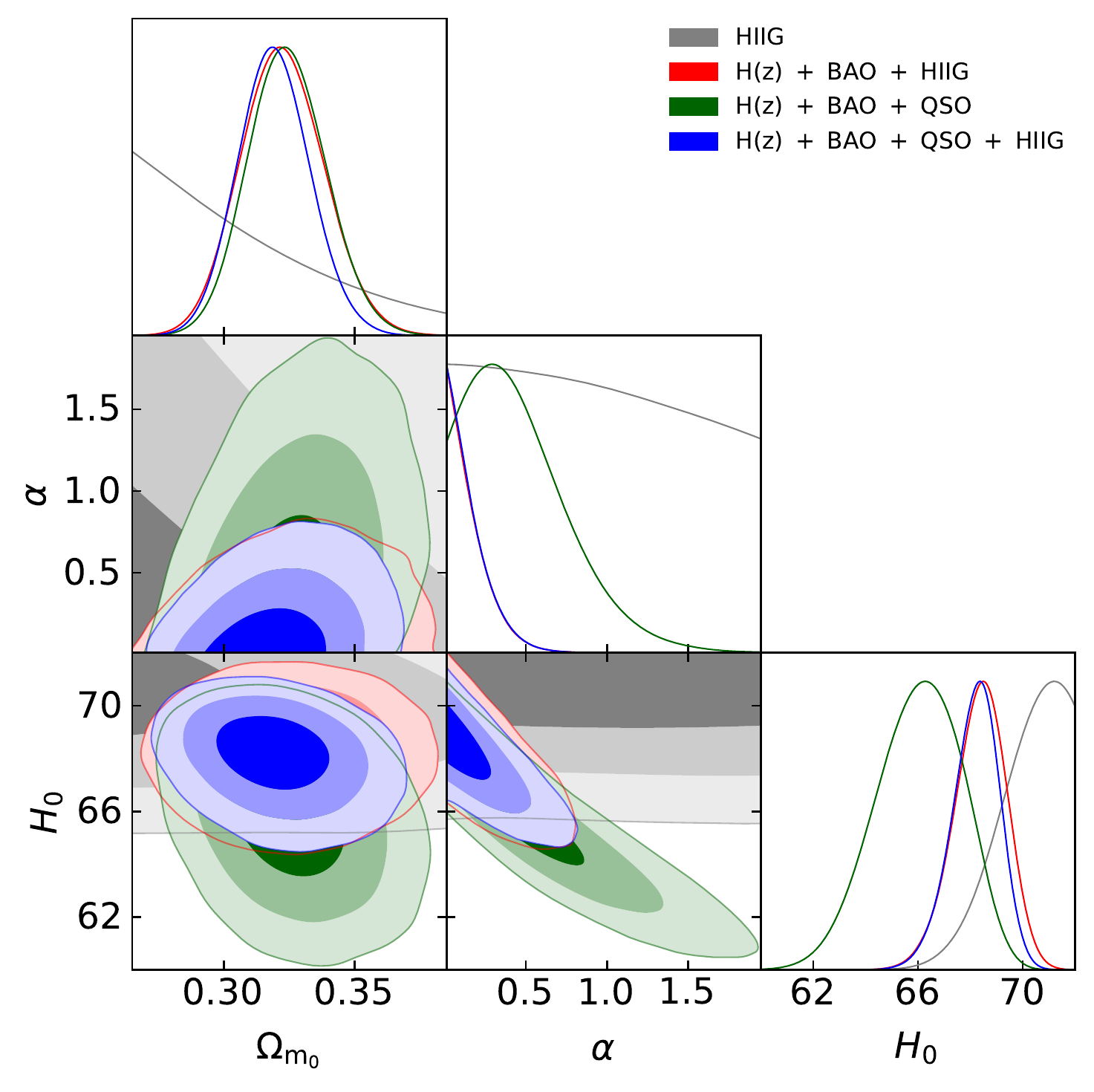}}\\
\caption{1$\sigma$, 2$\sigma$, and 3$\sigma$ confidence contours for flat $\phi$CDM. The black dotted zero-acceleration line splits the parameter space into regions of currently accelerated (below left) and currently decelerated (above right) cosmological expansion. The $\alpha = 0$ axis is the flat \lcdm\ model.}
\label{fig05}
\end{figure*}

\begin{figure*}
\centering
  \subfloat[Full parameter range]{%
    \includegraphics[width=3.25in,height=3.25in]{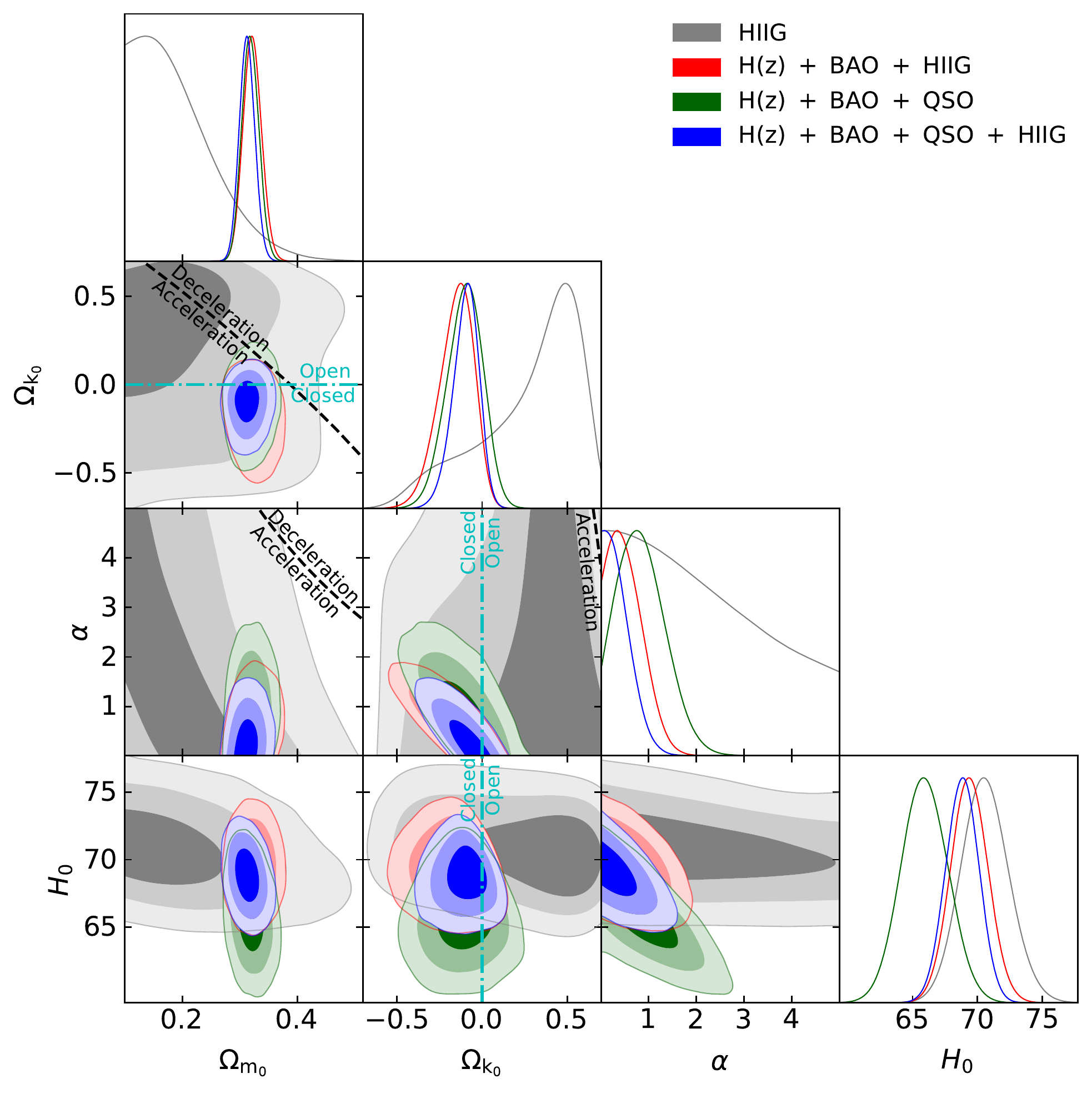}}
  \subfloat[Zoom in]{%
    \includegraphics[width=3.25in,height=3.25in]{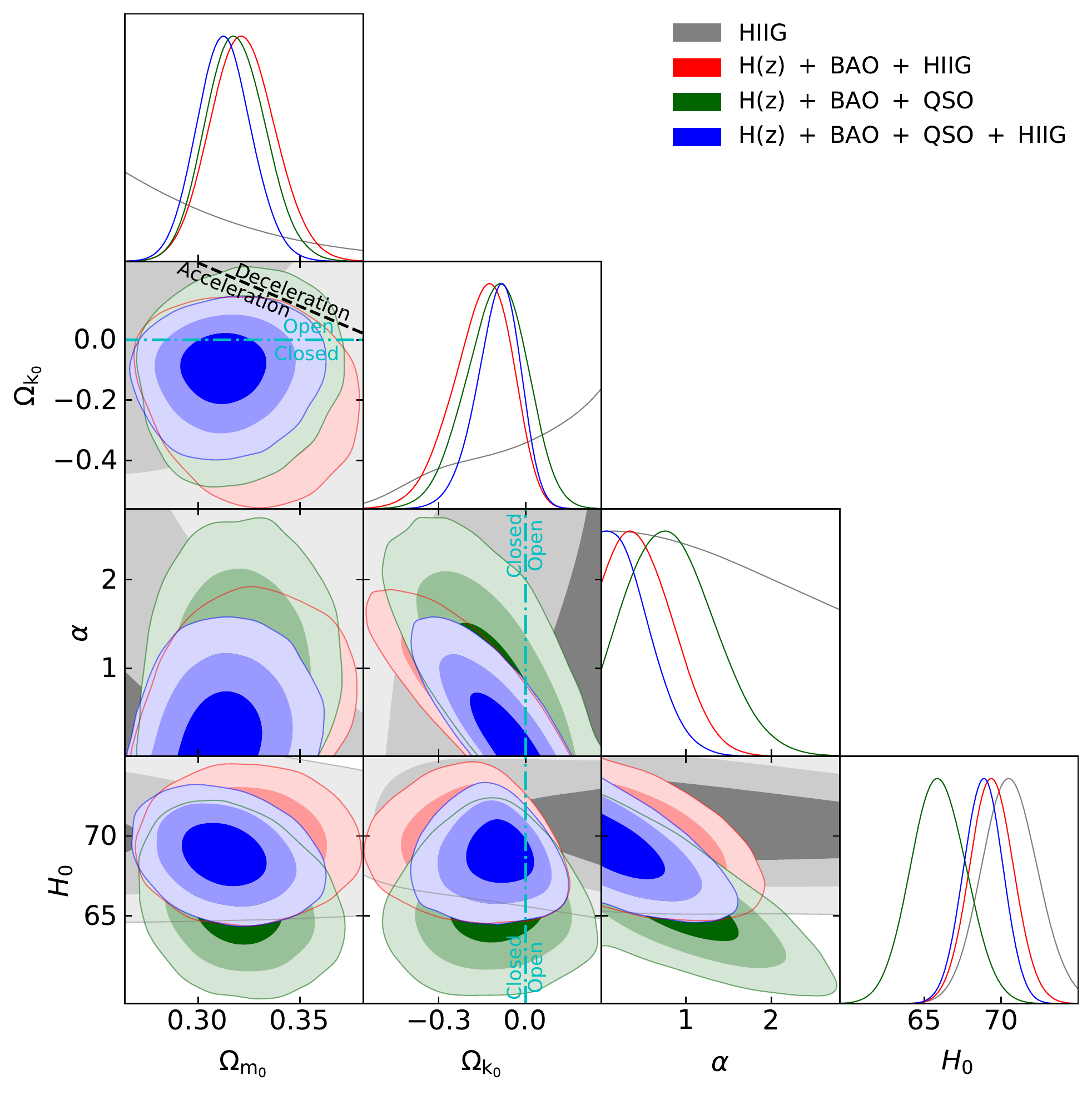}}\\
\caption{Same as Fig. \ref{fig05} but for non-flat \pcdm, where the zero-acceleration lines in each of the subpanels are computed for the third cosmological parameter set to the \hiig\ data only best-fitting values listed in Table \ref{tab:BFP}. Currently-accelerating cosmological expansion occurs below these lines. The cyan dash-dot lines represent the flat case, with closed spatial geometry either below or to the left. The $\alpha = 0$ axis is the non-flat \lcdm\ model.}
\label{fig06}
\end{figure*}

The 1D probability distributions and 2D confidence regions of the cosmological parameters for the six flat and non-flat models are shown in Figs. \ref{fig01}--\ref{fig06}, in red. The best-fitting results and uncertainties are listed in Tables \ref{tab:BFP} and \ref{tab:1d_BFP}.

When we fit our cosmological models to the HzBH data we find that the measured values of the matter density parameter \om\ fall within a narrower range in comparison to the \hiig\ only case, being between $0.314 \pm 0.015$ (non-flat \lcdm) and $0.323^{+0.014}_{-0.016}$ (flat \pcdm).

\begin{table*}
\centering
\resizebox*{1\columnwidth}{1.2\columnwidth}{%
\begin{threeparttable}
\caption{Unmarginalized best-fitting parameter values for all models from various combinations of data.}\label{tab:BFP}
\begin{tabular}{lccccccccccc}
\hline
 Model & Data set & $\Omega_{\mathrm m_0}$ & $\Omega_{\Lambda}$ & $\Omega_{\mathrm k_0}$ & $w_{\mathrm X}$ & $\alpha$ & $H_0$\tnote{a} & $\chi^2$ & $AIC$ & $BIC$ & $\nu$\\
\hline
Flat \lcdm & \hiig\ & 0.276 & 0.724 & --- & --- & --- & 71.81 & 410.75 & 414.75 & 420.81 & 151\\
 & $H(z)$ + BAO + \hiig\ & 0.318 & 0.682 & --- & --- & --- & 69.22 & 434.29 & 438.29 & 444.84 & 193 \\
 & $H(z)$ + BAO + QSO & 0.315 & 0.685 & --- & --- & --- & 68.61 & 372.88 & 376.88 & 383.06 & 160\\
 & $H(z)$ + BAO + QSO + \hiig\ & 0.315 & 0.685 & --- & --- & --- & 69.06 & 786.50 & 790.50 & 798.01 & 313\\
\\
Non-flat \lcdm & \hiig\ & 0.312 & 0.998 & $-0.310$ & --- & --- & 72.35 & 410.44 & 416.44 & 425.53 & 150\\
 & $H(z)$ + BAO + \hiig\ & 0.313 & 0.718 & $-0.031$ & --- & --- & 70.24 & 433.38 & 439.38 & 449.19 & 192\\
 & $H(z)$ + BAO + QSO & 0.311 & 0.665 & 0.024 & --- & --- & 68.37 & 372.82 & 378.82 & 388.08 & 159\\
 & $H(z)$ + BAO + QSO + \hiig\ & 0.309 & 0.716 & $-0.025$ & --- & --- & 69.82 & 785.79 & 791.79 & 803.05 & 312\\
\\
Flat XCDM & \hiig\ & 0.249 & --- & --- & $-0.892$ & --- & 71.65 & 410.72 & 416.72 & 425.82 & 150\\
 & $H(z)$ + BAO + \hiig\ & 0.314 & --- & --- & $-1.044$ & --- & 69.94 & 433.99 & 439.99 & 449.81 & 192\\
 & $H(z)$ + BAO + QSO & 0.322 & --- & --- & $-0.890$ & --- & 66.62 & 371.95 & 377.95 & 387.21 & 159\\
 & $H(z)$ + BAO + QSO + \hiig\ & 0.311 & --- & --- & $-1.045$ & --- & 69.80 & 786.19 & 792.19 & 803.45 & 312\\
\\
Non-flat XCDM & \hiig\ & 0.104 & --- & $-0.646$ & $-0.712$ & --- & 72.61 & 407.69 & 415.69 & 427.81 & 149\\
 & $H(z)$ + BAO + \hiig\ & 0.322 & --- & $-0.117$ & $-0.878$ & --- & 66.67 & 432.85 & 440.85 & 453.94 & 191\\
 & $H(z)$ + BAO + QSO & 0.322 & --- & $-0.112$ & $-0.759$ & --- & 65.80 & 370.68 & 378.68 & 391.03 & 158\\
 & $H(z)$ + BAO + QSO + \hiig\ & 0.310 & --- & $-0.048$ & $-0.957$ & --- & 69.53 & 785.70 & 793.70 & 808.71 & 311\\
\\
Flat $\phi$CDM & \hiig\ & 0.255 & --- & --- & --- & 0.261 & 71.70 & 410.70 & 416.70 & 425.80 & 150\\
 & $H(z)$ + BAO + \hiig\ & 0.318 & --- & --- & --- & 0.011 & 69.09 & 434.36 & 440.36 & 450.18 & 192\\
 & $H(z)$ + BAO + QSO & 0.321 & --- & --- & --- & 0.281 & 66.82 & 372.05 & 378.05 & 387.31 & 159\\
 & $H(z)$ + BAO + QSO + \hiig\ & 0.315 & --- & --- & --- & 0.012 & 68.95 & 786.58 & 792.58 & 803.84 & 312\\
\\
Non-flat $\phi$CDM & \hiig\ & 0.114 & --- & $-0.437$ & --- & 2.680 & 72.14 & 409.91 & 417.91 & 430.03 & 149\\
 & $H(z)$ + BAO + \hiig\ & 0.321 & --- & $-0.132$ & --- & 0.412 & 69.69 & 432.75 & 440.75 & 453.84 & 191\\
 & $H(z)$ + BAO + QSO & 0.317 & --- & $-0.106$ & --- & 0.778 & 66.27 & 370.83 & 378.83 & 391.18 & 158\\
 & $H(z)$ + BAO + QSO + \hiig\ & 0.310 & --- & $-0.054$ & --- & 0.150 & 69.40 & 785.65 & 793.65 & 808.66 & 311\\
\hline
\end{tabular}
\begin{tablenotes}
\item [a] \hunit.
\end{tablenotes}
\end{threeparttable}%
}
\end{table*}

\begin{table*}
\centering
\resizebox*{1\columnwidth}{1.2\columnwidth}{%
\begin{threeparttable}
\caption{One-dimensional marginalized best-fitting parameter values and uncertainties ($\pm 1\sigma$ error bars or $2\sigma$ limits) for all models from various combinations of data.}\label{tab:1d_BFP}
\begin{tabular}{lccccccc}
\hline
 Model & Data set & $\Omega_{\mathrm m_0}$ & $\Omega_{\Lambda}$ & $\Omega_{\mathrm k_0}$ & $w_{\mathrm X}$ & $\alpha$ & $H_0$\tnote{a} \\
\hline
Flat \lcdm & \hiig\ & $0.289^{+0.053}_{-0.071}$ & --- & --- & --- & --- & $71.70\pm1.83$ \\
 & $H(z)$ + BAO + \hiig\ & $0.319^{+0.014}_{-0.015}$ & --- & --- & --- & --- & $69.23\pm0.74$ \\
 & $H(z)$ + BAO + QSO & $0.316^{+0.013}_{-0.014}$ & --- & --- & --- & --- & $68.60\pm0.68$ \\
 & $H(z)$ + BAO + QSO + \hiig\ & $0.315^{+0.013}_{-0.012}$ & --- & --- & --- & --- & $69.06^{+0.63}_{-0.62}$ \\
\\
Non-flat \lcdm & \hiig\ & $0.275^{+0.081}_{-0.078}$ & $>0.501$\tnote{b} & $0.094^{+0.237}_{-0.363}$ & --- & --- & $71.50^{+1.80}_{-1.81}$ \\
 & $H(z)$ + BAO + \hiig\ & $0.314\pm0.015$ & $0.714^{+0.054}_{-0.049}$ & $-0.029^{+0.049}_{-0.048}$ & --- & --- & $70.21\pm1.33$ \\
 & $H(z)$ + BAO + QSO & $0.313^{+0.013}_{-0.015}$ & $0.658^{+0.069}_{-0.060}$ & $0.029^{+0.056}_{-0.063}$ & --- & --- & $68.29\pm1.47$ \\
 & $H(z)$ + BAO + QSO + \hiig\ & $0.310\pm0.013$ & $0.711^{+0.053}_{-0.048}$ & $-0.021^{+0.044}_{-0.048}$ & --- & --- & $69.76^{+1.12}_{-1.11}$ \\
\\
Flat XCDM & \hiig\ & $0.300^{+0.106}_{-0.083}$ & --- & --- & $-1.180^{+0.560}_{-0.330}$ & --- & $71.85\pm1.96$ \\
 & $H(z)$ + BAO + \hiig\ & $0.315^{+0.016}_{-0.017}$ & --- & --- & $-1.052^{+0.092}_{-0.082}$ & --- & $70.05\pm1.54$ \\
 & $H(z)$ + BAO + QSO & $0.322^{+0.015}_{-0.016}$ & --- & --- & $-0.911^{+0.122}_{-0.098}$ & --- & $66.98^{+1.95}_{-2.30}$ \\
 & $H(z)$ + BAO + QSO + \hiig\ & $0.312\pm0.014$ & --- & --- & $-1.053^{+0.091}_{-0.082}$ & --- & $69.90\pm1.48$ \\
\\
Non-flat XCDM & \hiig\ & $0.275^{+0.084}_{-0.125}$ & --- & $0.011^{+0.457}_{-0.460}$ & $-1.125^{+0.537}_{-0.321}$ & --- & $71.71^{+2.07}_{-2.08}$ \\
 & $H(z)$ + BAO + \hiig\ & $0.318\pm0.019$ & --- & $-0.082^{+0.135}_{-0.119}$ & $-0.958^{+0.219}_{-0.098}$ & --- & $69.83^{+1.50}_{-1.62}$ \\
 & $H(z)$ + BAO + QSO & $0.320\pm0.015$ & --- & $-0.078^{+0.124}_{-0.112}$ & $-0.826^{+0.185}_{-0.088}$ & --- & $66.29^{+1.90}_{-2.35}$ \\
 & $H(z)$ + BAO + QSO + \hiig\ & $0.309^{+0.015}_{-0.014}$ & --- & $-0.025\pm0.092$ & $-1.022^{+0.208}_{-0.104}$ & --- & $69.68^{+1.49}_{-1.64}$ \\
\\
Flat $\phi$CDM & \hiig\ & $0.210^{+0.043}_{-0.092}$ & --- & --- & --- & $<2.784$ & $71.23^{+1.79}_{-1.80}$ \\
 & $H(z)$ + BAO + \hiig\ & $0.323^{+0.014}_{-0.016}$ & --- & --- & --- & $<0.411$ & $68.36^{+1.05}_{-0.86}$ \\
 & $H(z)$ + BAO + QSO & $0.324^{+0.014}_{-0.015}$ & --- & --- & --- & $0.460^{+0.116}_{-0.440}$ & $66.03^{+1.79}_{-1.42}$ \\
 & $H(z)$ + BAO + QSO + \hiig\ & $0.319\pm0.013$ & --- & --- & --- & $<0.411$ & $68.18^{+0.97}_{-0.75}$\\
\\
Non-flat $\phi$CDM & \hiig\ & $<0.321$ & --- & $0.291^{+0.348}_{-0.113}$ & --- & $<4.590$ & $70.60^{+1.68}_{-1.84}$ \\
 & $H(z)$ + BAO + \hiig\ & $0.322^{+0.015}_{-0.016}$ & --- & $-0.153^{+0.114}_{-0.079}$ & --- & $0.538^{+0.151}_{-0.519}$ & $69.39\pm1.37$ \\
 & $H(z)$ + BAO + QSO & $0.319^{+0.013}_{-0.015}$ & --- & $-0.103^{+0.111}_{-0.091}$ & --- & $0.854^{+0.379}_{-0.594}$ & $65.94^{+1.75}_{-1.73}$ \\
 & $H(z)$ + BAO + QSO + \hiig\ & $0.313^{+0.012}_{-0.014}$ & --- & $-0.098^{+0.082}_{-0.061}$ & --- & $<0.926$ & $68.83\pm1.23$ \\
\hline
\end{tabular}
\begin{tablenotes}
\item [a] \hunit.
\item [b] This is the 1$\sigma$ lower limit. The $2\sigma$ lower limit is set by the prior, and is not shown here.
\end{tablenotes}
\end{threeparttable}%
}
\end{table*}

Similarly, the measured values of $H_0$ also fall within a narrower range when our models are fit to the HzBH data combination (and are in better agreement with the median statistics estimate of $H_0$ from \citealp{chenratmed} than with the local measurement carried out by \citealp{riess_etal_2019}; this is because the $H(z)$ and BAO data favor a lower $H_0$ value) being between $H_0=68.36^{+1.05}_{-0.86}$ \hunit (flat \pcdm) and $70.21 \pm 1.33$ \hunit (non-flat \lcdm). We assume that the tension between early- and late-Universe measurements of $H_0$ is not a major issue here, because the 2D and 1D contours in Fig. \ref{fig01} overlap, and so we compute a combined $H_0$ value (but if one is concerned about the early- vs late-Universe $H_0$ tension then one should not compare our combined-data $H_0$\!'s here, and in Secs. \ref{subsec:HzBQ} and \ref{subsec:HzBQH}, directly to the measurements of \citealp{riess_etal_2019} or of \citealp{planck2018b}).

In contrast to the \hiig\ only cases, when fit to the HzBH data combination the non-flat models mildly favor closed spatial hypersurfaces. This is because the $H(z)$ and BAO data mildly favor closed spatial hypersurfaces; see, e.g. \cite{park_ratra_2019b} and \cite{Ryanetal2019}. For non-flat \lcdm, non-flat XCDM, and non-flat \pcdm, we find $\Omega_{\rm k_0}=-0.029^{+0.049}_{-0.048}$, $\Omega_{\rm k_0}=-0.082^{+0.135}_{-0.119}$, and $\Omega_{\rm k_0}=-0.153^{+0.114}_{-0.079}$, respectively, with the non-flat \pcdm\ model favoring closed spatial hypersurfaces at 1.34$\sigma$.

The fit to the HzBH data combination produces weaker evidence for dark energy dynamics (in comparison to the \hiig\ only case) with tighter error bars on the measured values of $w_{\rm X}$ and $\alpha$. For flat (non-flat) XCDM, $w_{\rm X}=-1.052^{+0.092}_{-0.082}$ ($w_{\rm X}=-0.958^{+0.219}_{-0.098}$), with $w_{\rm X}=-1$ still being within the 1$\sigma$ range. For flat (non-flat) \pcdm, $\alpha<0.411$ ($\alpha=0.538^{+0.151}_{-0.519}$), where the former is peaked at $\alpha=0$ but for the latter, $\alpha=0$ is just out of the 1$\sigma$ range.

\subsection{$H(z)$, BAO, and QSO (HzBQ) constraints}
\label{subsec:HzBQ}

The $H(z)$, BAO, and QSO (HzBQ) data combination has previously been studied \citep{Ryanetal2019}. Relative to that analysis, we use an updated BAO data compilation, a more accurate formula for $r_s$, and the MCMC formalism (instead of the grid-based $\chi^2$ approach); consequently the parameter constraints derived here slightly differ from those of \cite{Ryanetal2019}.

The 1D probability distributions and 2D confidence regions of the cosmological parameters for all models are presented in Figs. \ref{fig01}--\ref{fig06}, in green. The corresponding best-fitting results and uncertainties are listed in Tables \ref{tab:BFP} and \ref{tab:1d_BFP}.

The measured values of \om\ here fall within a similar range to the range quoted in the last subsection, being between $0.313^{+0.013}_{-0.015}$ (non-flat \lcdm) and $0.324^{+0.014}_{-0.015}$ (flat \pcdm). This range is larger than, but still consistent with, the range of \om\ reported in \cite{Ryanetal2019}, where the same models are fit to the HzBQ data combination.

The $H_0$ measurements in this case fall within a broader range than in the HzBH case, being between $65.94^{+1.75}_{-1.73}$ \hunit (non-flat \pcdm) and $68.60 \pm 0.68$ \hunit (flat \lcdm). In addition, they are lower than the corresponding measurements in the HzBH cases, and are in better agreement with the median statistics \citep{chenratmed} estimate of $H_0$ than with what is measured from the local expansion rate \citep{riess_etal_2019}. Compared with \cite{Ryanetal2019}, the central values are lower except for the non-flat XCDM model.

For non-flat \lcdm, non-flat XCDM, and non-flat \pcdm, we measure $\Omega_{\rm k_0}=0.029^{+0.056}_{-0.063}$, $\Omega_{\rm k_0}=-0.078^{+0.124}_{-0.112}$, and $\Omega_{\rm k_0}=-0.103^{+0.111}_{-0.091}$, respectively. These results are consistent with their unmarginalized best-fittings (see Table \ref{tab:BFP}), where the best-fitting to the non-flat \lcdm\ model favors open spatial hypersurfaces, and the best-fittings to the non-flat XCDM parametrization and the non-flat \pcdm\ model both favor closed spatial hypersurfaces. Note that the central values are larger than those of \cite{Ryanetal2019}, especially for non-flat \lcdm\ (positive instead of negative). In all three models the constraints are consistent with flat spatial hyperfurfaces.

The fit to the HzBQ data combination provides slightly stronger evidence for dark energy dynamics than does the fit to the HzBH data combination. For flat (non-flat) XCDM, $w_{\rm X}=-0.911^{+0.122}_{-0.098}$ ($w_{\rm X}=-0.826^{+0.185}_{-0.088}$), with the former barely within 1$\sigma$ of $w_{\rm X}=-1$ and the latter almost 2$\sigma$ away from $w_{\rm X}=-1$. For flat (non-flat) \pcdm, $\alpha=0.460^{+0.116}_{-0.440}$ ($\alpha=0.854^{+0.379}_{-0.594}$), with the former 1.05$\sigma$ and the latter 1.44$\sigma$ away from the $\alpha=0$ cosmological constant. In comparison with \cite{Ryanetal2019}, central values of $w_{\rm X}$ are larger and smaller for flat and non-flat XCDM models, respectively, and that of $\alpha$ are larger for both flat and non-flat \pcdm\ models.

\subsection{$H(z)$, BAO, QSO, and \hiig\ (HzBQH) constraints}
\label{subsec:HzBQH}

Comparing the results of the previous two subsections, we see that when used in conjunction with $H(z)$ and BAO data, the QSO data result in tighter constraints on $\Omega_{\rm m_0}$, $\Omega_{\rm k_0}$ (in non-flat XCDM), $w_{\rm X}$ (in non-flat XCDM), and $H_0$ (in flat \lcdm), while the \hiig\ data result in tighter constraints on $H_0$ (except for flat \lcdm), $\Omega_{\Lambda}$, $\Omega_{\rm k_0}$(in non-flat \lcdm\ and \pcdm), $w_{\rm X}$ (in flat XCDM), and $\alpha$. Consequently, it is useful to derive constraints from an analysis of the combined $H(z)$, BAO, QSO, and \hiig\ (HzBQH) data. We present the results of such an analysis in this subsection.

In Figs. \ref{fig01}--\ref{fig06}, we present the 1D probability distributions and 2D confidence constraints for the HzBQH cases in blue. Tables \ref{tab:BFP} and \ref{tab:1d_BFP} list the best-fitting results and uncertainties.

It is interesting that the best-fitting values of $\Omega_{\rm m_0}$ in this case are lower compared with both the HzBQ and the HzBH results, being between $0.309^{+0.015}_{-0.014}$ (non-flat XCDM) and $0.319 \pm 0.013$ (flat \pcdm). The best-fitting values of $H_0$ are higher than the HzBQ cases and have central values that are closer to those of the HzBH cases, but are still in better agreement with the lower median statistics estimate of $H_0$ \citep{chenratmed} than the higher local expansion rate measurement of $H_0$ \citep{riess_etal_2019}, being between $68.18^{+0.97}_{-0.75}$ \hunit (flat \pcdm) and $69.90 \pm 1.48$ \hunit (flat XCDM). 

For non-flat \lcdm, non-flat XCDM, and non-flat \pcdm, we measure $\Omega_{\rm k_0}=-0.021^{+0.044}_{-0.048}$, $\Omega_{\rm k_0}=-0.025 \pm 0.092$, and $\Omega_{\rm k_0}=-0.098^{+0.082}_{-0.061}$, respectively. For non-flat \lcdm\ and XCDM, the measured values of the curvature energy density parameter are within 0.48$\sigma$ and 0.27$\sigma$ of $\Omega_{\rm k_0} = 0$, respectively, while the non-flat \pcdm\ model favors a closed geometry with an $\Omega_{\rm k_0}$ that is 1.20$\sigma$ away from zero.

There is not much evidence in support of dark energy dynamics in the HzBQH case, with $\Lambda$ consistent with this data combination. For flat (non-flat) XCDM, $w_{\rm X}=-1.053^{+0.091}_{-0.082}$ ($w_{\rm X}=-1.022^{+0.208}_{-0.104}$). For flat (non-flat) \pcdm, the $2\sigma$ upper limits are $\alpha<0.411$ ($\alpha<0.926$), which indicates that $\alpha = 0$ or $\Lambda$ is consistent with these data.

\subsection{Model comparison}
\label{makereference3.5}

From Table \ref{tab:cab}, we see that the reduced $\chi^2$ for all models is relatively large (being between 2.25 and 2.75). This could probably be attributed to underestimated systematic uncertainties in the \hiig\ data.\footnote{Underestimated systematic uncertainties might also explain the large reduced $\chi^2$ of QSO data \citep{Ryanetal2019}.} This is suggested by \cite{GonzalezMoran2019}, who also found relatively large values of $\chi^2/\nu$ in their cosmological model fits to the \hiig\ data (though not as large as ours, because they compute a different $\chi^2$, as explained in footnote \ref{fn5} in Section \ref{makereference3.3}). They note that an additional systematic uncertainty of $\sim0.22$ could bring their $\chi^2/\nu$ down to $\sim1$. As mentioned previously, we do not account for \hiig\ systematic uncertainties in our analysis.

\begin{table*}
\centering
\caption{$\Delta \chi^2$, $\Delta AIC$, $\Delta BIC$, and $\chi^2_{\mathrm{min}}/\nu$ values.}\label{tab:cab}
\resizebox*{1\columnwidth}{0.45\columnwidth}{%
\begin{tabular}{lccccccc}
\hline
 Quantity & Data set & Flat \lcdm & Non-flat \lcdm & Flat XCDM & Non-flat XCDM & Flat \pcdm & Non-flat \pcdm\\
\hline
 & \hiig\ & 3.06 & 2.75 & 3.03 & 0.00 & 3.01 & 2.22\\
$\Delta \chi^2$ & $H(z)$ + BAO + \hiig\ & 1.54 & 0.63 & 1.24 & 0.10 & 1.61 & 0.00 \\
 & $H(z)$ + BAO + QSO & 2.20 & 2.14 & 1.27 & 0.00 & 1.37 & 0.15\\
 & $H(z)$ + BAO + QSO + \hiig\ & 0.85 & 0.14 & 0.54 & 0.05 & 0.93 & 0.00\\
 \\
 & \hiig\ & 0.00 & 1.69 & 1.97 & 0.94 & 1.95 & 3.16\\
$\Delta AIC$ & $H(z)$ + BAO + \hiig\ & 0.00 & 1.09 & 1.70 & 2.56 & 2.07 & 2.46\\
 & $H(z)$ + BAO + QSO & 0.00 & 1.94 & 1.07 & 1.80 & 1.17 & 1.95\\
 & $H(z)$ + BAO + QSO + \hiig\ & 0.00 & 1.29 & 1.69 & 3.20 & 2.08 & 3.15\\
 \\
 & \hiig\ & 0.00 & 4.72 & 5.01 & 7.00 & 4.99 & 9.22\\
$\Delta BIC$ & $H(z)$ + BAO + \hiig\ & 0.00 & 4.35 & 4.97 & 9.10 & 5.34 & 9.00\\
 & $H(z)$ + BAO + QSO & 0.00 & 5.02 & 4.15 & 7.97 & 4.25 & 8.12\\
 & $H(z)$ + BAO + QSO + \hiig\ & 0.00 & 5.04 & 5.44 & 10.70 & 5.83 & 10.65\\
 \\
 & \hiig\ & 2.72 & 2.74 & 2.74 & 2.74 & 2.74 & 2.75\\
$\chi^2_{\mathrm{min}}/\nu$ & $H(z)$ + BAO + \hiig\ & 2.25 & 2.26 & 2.26 & 2.27 & 2.26 & 2.27\\
 & $H(z)$ + BAO + QSO & 2.33 & 2.34 & 2.34 & 2.35 & 2.34 & 2.35\\
 & $H(z)$ + BAO + QSO + \hiig\ & 2.51 & 2.52 & 2.52 & 2.53 & 2.52 & 2.53\\
\hline
\end{tabular}}
\end{table*}

One thing that is clear, regardless of the absolute size of \hiig\ or QSO systematics (and ignoring the large values of $\chi^2/\nu$), is that the flat \lcdm\ model remains the most favored model among the six models we studied, based on the $AIC$ and $BIC$ criteria (see Table \ref{tab:cab}).\footnote{Note that based on the $\Delta \chi^2$ results of Table \ref{tab:cab} non-flat XCDM has the minimum $\chi^2$ in the \hiig\ and HzBQ cases, whereas non-flat \pcdm\ has the minimum $\chi^2$ for the HzBH and HzBQH cases. The $\Delta \chi^2$ values do not, however, penalize a model for having more parameters.} In Table \ref{tab:cab} we define $\Delta \chi^2$, $\Delta AIC$, and $\Delta BIC$, respectively, as the differences between the values of the $\chi^2$, $AIC$, and $BIC$ associated with a given model and their corresponding minimum values among all models.

From the \hiig\ results for $\Delta AIC$ and $\Delta BIC$ listed in Table \ref{tab:cab}, we see that the evidence against non-flat \lcdm, flat XCDM, and flat \pcdm\ is weak (according to $\Delta AIC$) and positive (according to $\Delta BIC$) where, among these three models, the flat XCDM model is the least favored. The evidence against the non-flat XCDM model is weak regarding $\Delta AIC$ but strong based on $\Delta BIC$, while the evidence against non-flat \pcdm\ in this case is positive ($\Delta AIC$) and strong ($\Delta BIC$), respectively, with it being the least favored model overall.

Largely similar conclusions result from $\Delta AIC$ and $\Delta BIC$ values for the \hiig\ and HzBQ data. The exception is that the HzBQ $\Delta AIC$ value gives only weak evidence against non-flat \pcdm, instead of the positive evidence against it from the \hiig\ $\Delta AIC$ value.

The HzBH and HzBQH values of $\Delta AIC$ and $\Delta BIC$ result in the following conclusions:

1) the evidence against both non-flat \lcdm\ and flat XCDM is weak (HzBH) and positive (HzBQH) for $\Delta AIC$ and $\Delta BIC$;

2) the evidence against flat \pcdm\ is positive; 

3) non-flat XCDM is the least favored model with non-flat \pcdm\ doing almost as badly. $\Delta AIC$ gives positive evidence against non-flat XCDM and non-flat \pcdm, while $\Delta BIC$ strongly disfavors (HzBH) and very strongly disfavors (HzBQH) both of these non-flat models.

\section{Conclusions}
\label{makereference3.6}
In this paper, we have constrained cosmological parameters in six flat and non-flat cosmological models by analyzing a total of 315 observations, comprising 31 $H(z)$, 11 BAO, 120 QSO, and 153 \hiig\ measurements. The QSO angular size and \hiig\ apparent magnitude measurements are particularly noteworthy, as they reach to $z\sim2.7$ and $z\sim2.4$ respectively (somewhat beyond the highest $z\sim2.3$ reached by BAO data) and into a much less studied area of redshift space. While the current \hiig\ and QSO data do not provide very restrictive constraints, they do tighten the limits when they are used in conjunction with BAO + $H(z)$ data.

By measuring cosmological parameters in a variety of cosmological models, we are able to draw some relatively model-independent conclusions (i.e. conclusions that do not differ significantly between the different models). Specifically, for the full data set (i.e the HzBQH data), we find quite restrictive constraints on \om, a reasonable summary perhaps being $\Omega_{\rm m_0}=0.310 \pm 0.013$, in good agreement with many other recent estimates. $H_0$ is also fairly tightly constrained, with a reasonable summary perhaps being $H_0=69.5 \pm 1.2$ \hunit, which is in better agreement with the results of \cite{chenratmed} and \cite{planck2018b} than that of \cite{riess_etal_2019}. The HzBQH measurements are consistent with the standard spatially-flat \lcdm\ model, but do not strongly rule out mild dark energy dynamics or a little spatial curvature energy density. More and better-quality \hiig, QSO, and other data at $z \sim 2$--4 will significantly help to test these extensions.

\cleardoublepage


\chapter{Cosmological constraints from higher-redshift gamma-ray burst, \hii\ starburst galaxy, and quasar (and other) data}
\label{makereference4}

This chapter is based on \cite{Caoetal_2021}. Figures and tables by Shulei Cao, from analyses
conducted independently by Shulei Cao, Joseph Ryan, and Narayan Khadka.

\section{Introduction}
\label{makereference4.1}
There is a large body of evidence indicating that the Universe recently transitioned from a decelerated to an accelerated phase of expansion (at redshift $z \sim 3/4$; see e.g. \citealp{Farooq_Ranjeet_Crandall_Ratra_2017}) and has been undergoing accelerated expansion ever since (for reviews, see e.g. \citealp{Ratra_Vogeley,Martin,Coley_Ellis}). In the standard model of cosmology, called the \lcdm\ model \citep{peeb84}, the accelerated expansion is powered by a constant dark energy density (the cosmological constant, $\Lambda$). This model also assumes that spatial hypersurfaces are flat on cosmological scales, and that the majority of non-relativistic matter in the Universe consists of cold dark matter (CDM). 

Out of all the models that have been devised to explain the observed accelerated expansion of the Universe, the \lcdm\ model is currently the most highly favored in terms of both observational data and theoretical parsimony (see e.g. \citealp{Farooq_Ranjeet_Crandall_Ratra_2017,scolnic_et_al_2018,planck2018b,eBOSS_2020}). In spite of these virtues, however, there are some indications that the \lcdm\ model may not tell the whole story. On the observational side, some workers have found evidence of discrepancies between the \lcdm\ model and cosmological observations (\citealp{riess_2019, martinelli_tutusaus_2019}) and on the theoretical side, the origin of $\Lambda$ has yet to be explained in fundamental terms (e.g., \citealp{Martin}). One way to pin down the nature of dark energy is by studying its dynamics phenomenologically. It is possible that the dark energy density may evolve in time (\citealp{peebrat88}), and many dark energy models exhibiting this behavior have been proposed.

Cosmological models have largely been tested in the redshift range $0 \lesssim z \lesssim 2.3$, with baryon acoustic oscillation (BAO\footnote{In our BAO data analyses in this paper the sound horizon computation assumes a value for the current baryonic matter physical density parameter $\Omega_{\rm b_0} h^2$, appropriate for the model under study, computed from Planck CMB anisotropy data.}) measurements probing the upper end of this range, and at $z\sim1100$, using cosmic microwave background (CMB) anisotropy data. To determine the accuracy of our cosmological models, we also need to test them in the redshift range $2.3 \lesssim z \lesssim 1100$. Quasar angular size (QSO-AS), \hii\ starburst galaxy (\hiig), quasar X-ray and UV flux (QSO-Flux), and gamma-ray burst (GRB) measurements are some of the handful of data available in this range. The main goal of this paper is, therefore, to examine the effect that QSO-AS, \hiig, and GRB data have on cosmological model parameter constraints, in combination with each other, and in combination with more well-known probes.\footnote{We relegate the analysis of QSO-Flux data to Appendix \ref{AppendixA}, the reasons for which are discussed there.} 

Gamma-ray bursts are promising cosmological probes for two reasons. First, it is believed that they can be used as standardizable candles \citep{Lamb2000, Lamb2001, Amati2002, Amati2008, Amati2009, Ghirlanda2004, Demianski2011, Wangetal2015}. Second, they cover a redshift range that is wider than most other commonly-used cosmological probes, having been observed up to $z \sim 8.2$ \citep{Amati2008, Amati2009, Amati2019, samushia_ratra_2010, Demianski2011, Wang_2016, Demianski_2017a, Demianskietal_2021, Dirirsa2019, KhadkaRatra2020c}. In particular, the $z\sim 2.7$--8.2 part of the Universe is primarily accessed by GRBs,\footnote{Though QSO-Flux measurements can reach up to $z \sim 5.1$.} so if GRBs can be standardized, they could provide useful information about a large, mostly unexplored, part of the Universe.

QSO-AS data currently reach to $z\sim 2.7$. These data, consisting of measurements of the angular size of astrophysical radio sources, furnish a standard ruler that is independent of that provided by the BAO sound horizon scale. The intrinsic linear size $l_m$ of intermediate luminosity QSOs has recently been accurately determined by \cite{Cao_et_al2017b}, opening the way for QSOs to, like GRBs, test cosmological models in a little-explored region of redshift space.\footnote{The use of QSO-AS measurements to constrain cosmological models dates back to near the turn of the century (e.g. \citealp{gurvits_kellermann_frey_1999, vishwakarma_2001, lima_alcaniz_2002, zhu_fujimoto_2002, Chen_Ratra_2003}), but, as discussed in \cite{Ryanetal2019}, these earlier results are suspect, because they are based on an inaccurate determination of $l_m$.}

\hiig\ data reach to $z\sim 2.4$, just beyond the range of current BAO data. Measurements of the luminosities of the Balmer lines in \hii\ galaxies can be correlated with the velocity dispersion of the radiating gas, making \hii\ galaxies a standard candle that can complement both GRBs and lower-redshift standard candles like supernovae (\citealp{Siegel_2005,Plionis_2009,Mania_2012,Chavez_2014,GonzalezMoran2019}).

Current QSO-Flux measurements reach to $z\sim 5.1$, but they favor a higher value of the current (denoted by the subscript ``0'') non-relativistic matter density parameter ($\Omega_{\rm m_0}$) than what is currently thought to be reasonable. The $\Omega_{\rm m_0}$ values obtained using QSO-Flux data, in a number of cosmological models, are in nearly 2$\sigma$ tension with the values obtained by using other well-established cosmological probes like CMB, BAO, and Type Ia supernovae (\citealp{RisalitiLusso2019, Yangetal2020, Wei_Melia_2020, KhadkaRatra2020b}). Techniques for standardizing QSO-Flux measurements are still under development, so it might be too early to draw strong conclusions about the cosmological constraints obtained from QSO-Flux measurements. Therefore, in this paper, we use QSO-Flux data alone and in combination with other data to constrain cosmological parameters in four different models, and record these results in Appendix \ref{AppendixA}.

We find that the GRB, \hiig, and QSO-AS constraints are largely mutually consistent, and that their joint constraints are consistent with those from more widely used, and more restrictive, BAO and Hubble parameter ($H(z)$) data. When used jointly with the $H(z)$ + BAO data, these higher-$z$ data tighten the $H(z)$ + BAO constraints.

This paper is organized as follows. In Section \ref{makereference4.2} we introduce the data we use. The models we analyze are described in Chapter \ref{sec:models}, with a description of our analysis method in Section \ref{makereference4.3}. Our results are in Section \ref{makereference4.4}. We provide our conclusions in Section \ref{makereference4.6}. Additionally, we discuss our results for QSO-Flux measurements in Appendix \ref{AppendixA}.

\section{Data}
\label{makereference4.2}
We use QSO-AS, \hiig, QSO-Flux, and GRB data to obtain constraints on the cosmological models we study. The QSO-AS data, comprising 120 measurements compiled by \cite{Cao_et_al2017b} (listed in Table 1 of that paper) and spanning the redshift range $0.462 \leq z \leq 2.73$, are also used in \cite{Ryanetal2019}; see these papers for descriptions. The \hiig\ data, comprising 107 low redshift ($0.0088 \leq z \leq 0.16417$) \hiig\ measurements, used in \cite{Chavez_2014} (recalibrated by \citealp{GonzalezMoran2019}), and 46 high redshift ($0.636427 \leq z \leq 2.42935$) \hiig\ measurements, used in \cite{GonzalezMoran2019}, are also used in \cite{CaoRyanRatra2020}. The GRB data, spanning the redshift range $0.48 \leq z \leq 8.2$, are collected from \cite{Dirirsa2019} (25 from Table 2 of that paper (F10), and the remaining 94 from Table 5 of the same, which are a subset of those compiled by \citealp{Wang_2016}) and also used in \cite{KhadkaRatra2020c}. Note that in our analyses here we did not use the correct value of $E_{\rm p}=871\pm123$ keV for GRB081121, as discussed in Ref.\ \cite{Liuetal2022}, although the effects on the parameter constraints are small. We also add 1598 QSO-Flux measurements spanning the redshift range $0.036 \leq z \leq 5.1003$, from \cite{RisalitiLusso2019}. These data are used in \cite{KhadkaRatra2020b}; see that paper for details. Results related to these QSO-Flux data are discussed in Appendix \ref{AppendixA}. 

In order to be useful as cosmological probes, GRBs need to be standardized, and many phenomenological relations have been proposed for this purpose (\citealp{Amati2002}, \citealp{Ghirlanda2004}, \citealp{Liang2005}, \citealp{Muccino_2020}, and references therein). As in \cite{KhadkaRatra2020c}, we use the Amati relation (\citealp{Amati2002}), which is an observed correlation between the peak photon energy $E_{\rm p}$ and the isotropic-equivalent radiated energy $E_{\rm iso}$ of long-duration GRBs, to standardize GRB measurements. There have been many attempts to standardize GRBs using the Amati relation. Some analyses assume a fixed value of $\Omega_{\rm m_0}$ to calibrate the Amati relation, so they favor a relatively reasonable value of $\Omega_{\rm m_0}$. Others use supernovae data to calibrate the Amati relation, while some use $H(z)$ data to calibrate it. This means that most previous GRB analyses are affected by some non-GRB external factors. In some cases this leads to a circularity problem, in which the models to be constrained by using the Amati relation are also used to calibrate the Amati relation itself (\citealp{Liu_Wei_2015, Demianski_2017a, Demianskietal_2021, Dirirsa2019}). In other cases, the data used in the calibration process dominate the analysis results. To overcome these problems, we fit the parameters of the Amati relation simultaneously with the parameters of the cosmological models we study (as done in \citealp{KhadkaRatra2020c}; also see \citealp{Wang_2016}).

The isotropic radiated energy $E_{\rm iso}$ of a source in its rest frame at a luminosity distance $D_L$ is
\be
\label{Eiso}
    E_{\rm iso}=\frac{4\pi D_L^2}{1+z}S_{\rm bolo},
\ee
where $S_{\rm bolo}$ is the bolometric fluence, and $D_L$ (defined below) depends on $z$ and on the parameters of our cosmological models. $E_{\rm iso}$ is connected to the source's peak energy output $E_{\rm p}$ via the Amati relation \citep{Amati2008, Amati2009}
\begin{equation}
    \label{eq:Amati}
    \log E_{\rm iso} = a  + b\log E_{\rm p},
\end{equation}
where $a$ and $b$ are free parameters that we vary in our model fits.\footnote{$\log=\log_{10}$ is implied hereinafter.} Note here that the peak energy $E_{\rm p} = (1+z)E_{\rm p, obs}$ where $E_{\rm p, obs}$ is the observed peak energy.

The correlation between \hiig\ luminosity ($L$) and velocity dispersion ($\sigma$) is shown in equation \eqref{eq:logL}. One can test a cosmological model with parameters $\textbf{p}$ by using it to compute a theoretical distance modulus \eqref{eq:mu_th} and comparing this prediction to the distance modulus computed from observational \hiig\ luminosity and flux ($f$) data
\begin{equation}
\label{eq:mu_obs}
    \mu_{\rm obs} = 2.5\log L - 2.5\log f - 100.2,
\end{equation}
(\citealp{Terlevich_2015, GonzalezMoran2019}).

QSO-AS data can be used to test cosmological models by comparing the theoretical angular size of the QSO
\be
\label{eq:theta_th}
\theta_{\rm th} = \frac{l_m}{D_{A}}
\ee
with its observed angular size $\theta_{\rm obs}$. In equation \eqref{eq:theta_th}, $l_m$ is the characteristic linear size of the QSO,\footnote{For the data sample we use, this quantity is equal to $11.03 \pm 0.25$ pc; see \cite{Cao_et_al2017b}.} and $D_{A}$ (defined below) is its angular size distance.

Underestimated systematic uncertainties for both \hiig\ and QSO-AS data might be responsible for the large reduced $\chi^2$ (described in Sec. \ref{subsec:comparison}).

The transverse comoving distance $D_M(\textbf{p}, z)$ is related to the luminosity distance $D_L(\textbf{p}, z)$ and the angular size distance $D_A(\textbf{p}, z)$ through equation \eqref{DM-DL-DA}, and is a function of $z$ and the parameters $\textbf{p}$ as shown in equation \eqref{eq:DM}.

We also use $H(z)$ and BAO measurements to constrain cosmological parameters. The $H(z)$ data, 31 measurements spanning the redshift range $0.070 \leq z \leq 1.965$, are compiled in Table 2 of \cite{Ryan_1}. The BAO data, 11 measurements spanning the redshift range $0.38 \leq z \leq 2.34$, are listed in Table 1 of \cite{CaoRyanRatra2020}. 

Systematic errors that affect $H(z)$ measurements include assumptions about the stellar metallicity of the galaxies in which cosmic chronometers are found, progenitor bias, the presence of a population of young stars in these galaxies, and assumptions about stellar population synthesis models. These effects were studied in \cite{70, moresco_et_al_2016, moresco_et_al_2018, moresco_et_al_2020}. \cite{moresco_et_al_2020} found that the dominant contribution to the systematic error budget comes from the choice of stellar population synthesis model, which introduces an average systematic error of $\sim8.9$\% (though the authors say that this can be reduced to $\sim4.5$\% by removing an outlier model from the analysis). The impacts of a population of young stars and of the progenitor bias were found to be negligible in \cite{moresco_et_al_2018, 70}, and \cite{moresco_et_al_2020} found that the impact of a $\sim$ 5--10\% uncertainty in the metallicity estimates produces a $\sim$ 4--9\% systematic error in the $H(z)$ measurements.

The systematic uncertainties of BAO from \cite{Alam_et_al_2017} (described in Sec. 7) are included in their covariance matrix. The BAO data from \cite{Carter_2018} is the combined result of the 6dF Galaxy Survey1 (6dFGS) and the SDSS DR7 MGS, where the systematic effects are described in detail in \cite{Jones_et_al_2009} and \cite{Ross_et_al_2015} (negligible), respectively. As described in \cite{DES_2019b}, the BAO systematic uncertainty is 15\% of their statistical uncertainty and thus negligible. The same negligible systematic effect applies to the BAO measurement from \cite{3}. \cite{Agathe} added polynomial terms to the correlation function, so as to test the sensitivity of the slowly-varying part of the correlation function to systematic effects. They found that this shifted the BAO peak position by less than $1\sigma$ relative to its position in their fiducial model.

\section{Data Analysis Methodology}
\label{makereference4.3}

By using the \textsc{python} module \textsc{emcee} \citep{emcee}, we perform a Markov chain Monte Carlo (MCMC) analysis to maximize the likelihood function, $\mathcal{L}$, and thereby determine the best-fitting values of the free parameters. The flat cosmological parameter priors are the same as those used in \cite{CaoRyanRatra2020} and the flat priors of the parameters of the Amati relation are non-zero over $0\leq\sigma_{\rm ext}\leq10$ (described below), $40\leq a\leq60$, and $0\leq b\leq5$. 

The likelihood functions associated with $H(z)$, BAO, \hiig, and QSO-AS data are described in \cite{CaoRyanRatra2020}. For GRB data, the natural log of its likelihood function \citep{D'Agostini_2005} is
\be
\label{eq:LH_GRBC4}
    \ln\mathcal{L}_{\rm GRB}= -\frac{1}{2}\Bigg[\chi^2_{\rm GRB}+\sum^{119}_{i=1}\ln\left(2\pi(\sigma_{\rm ext}^2+\sigma_{{y_i}}^2+b^2\sigma_{{x_i}}^2)\right)\Bigg],
\ee
where
\be
\label{eq:chi2_GRBC4}
    \chi^2_{\rm GRB} = \sum^{119}_{i=1}\bigg[\frac{(y_i-b x_i-a)^2}{(\sigma_{\rm ext}^2+\sigma_{{y_i}}^2+b^2\sigma_{{x_i}}^2)}\bigg],
\ee
$x=\log\frac{E_{\rm p}}{\rm keV}$, $\sigma_{x}=\frac{\sigma_{E_{\rm p}}}{E_{\rm p}\ln 10}$, $y=\log\frac{E_{\rm iso}}{\rm erg}$, and $\sigma_{\rm ext}$ is the extrinsic scatter parameter, which contains the unknown systematic uncertainty. For the GRB with $\sigma_z$ uncertainty in $z$,
\be
\label{eq:err_y2}
    \sigma^2_{y}=\left(\frac{\sigma_{S_{\rm bolo}}}{S_{\rm bolo}\ln 10}\right)^2+\left(\frac{2(1+z)\frac{\partial D_M}{\partial z}+D_M}{(1+z)D_M\ln 10}\sigma_z\right)^2,
\ee
and for those without $z$ uncertainties $\sigma_z=0$ (the non-zero $\sigma_z$ has a negligible effect on our results).

The Akaike Information Criterion ($AIC$) and the Bayesian Information Criterion ($BIC$) are used to compare the goodness of fit of models with different numbers of parameters, where
\be
\label{AIC1}
    AIC=-2\ln \mathcal{L}_{\rm max} + 2n,
\ee
and
\be
\label{BIC1}
    BIC=-2\ln \mathcal{L}_{\rm max} + n\ln N.
\ee
In these equations, $\mathcal{L}_{\rm max}$ is the maximum value of the relevant likelihood function, $n$ is the number of free parameters of the model under consideration, and $N$ is the number of data points (e.g., for GRB $N=119$).

\section{Results}
\label{makereference4.4}
\subsection{\hiig, QSO-AS, and GRB constraints, individually}
\label{subsec:GRB}

\begin{figure*}
\centering
  \subfloat[All parameters]{%
    \includegraphics[width=3.25in,height=3.25in]{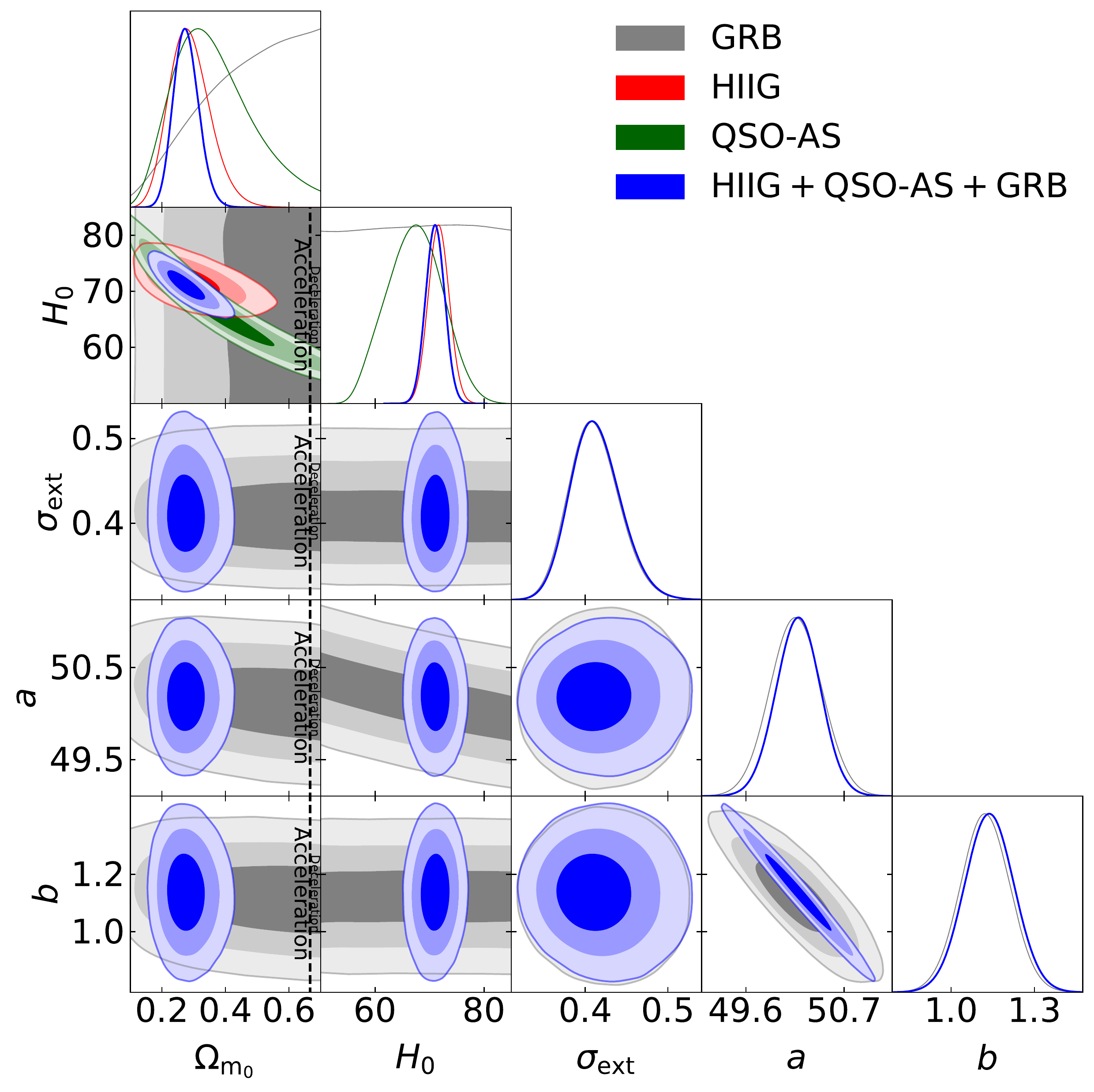}}
  \subfloat[Cosmological parameters zoom in]{%
    \includegraphics[width=3.25in,height=3.25in]{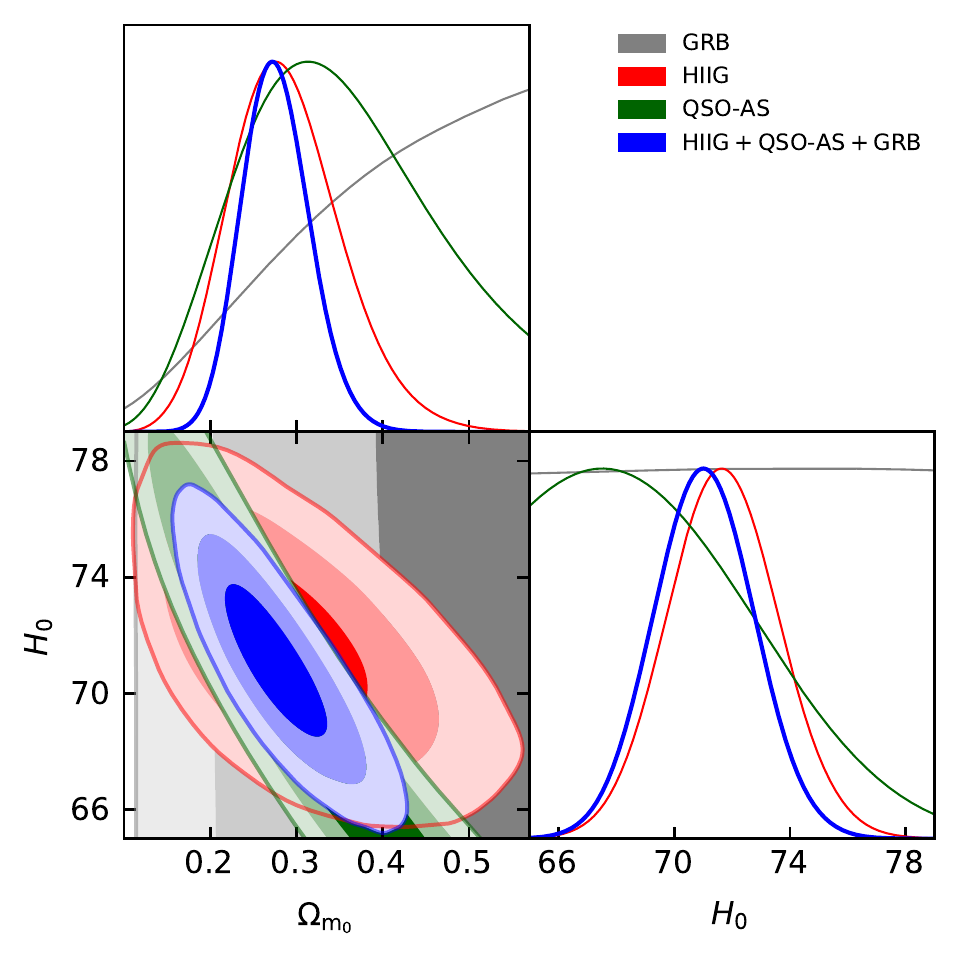}}\\
\caption{1$\sigma$, 2$\sigma$, and 3$\sigma$ confidence contours for flat \lcdm, where the right panel is the cosmological parameters comparison zoomed in. The black dotted lines in the left sub-panels of the left panel are the zero-acceleration lines, which divide the parameter space into regions associated with currently-accelerating (left) and currently-decelerating (right) cosmological expansion.}
\label{fig1}
\end{figure*}

\begin{figure*}
\centering
  \subfloat[All parameters]{%
    \includegraphics[width=3.25in,height=3.25in]{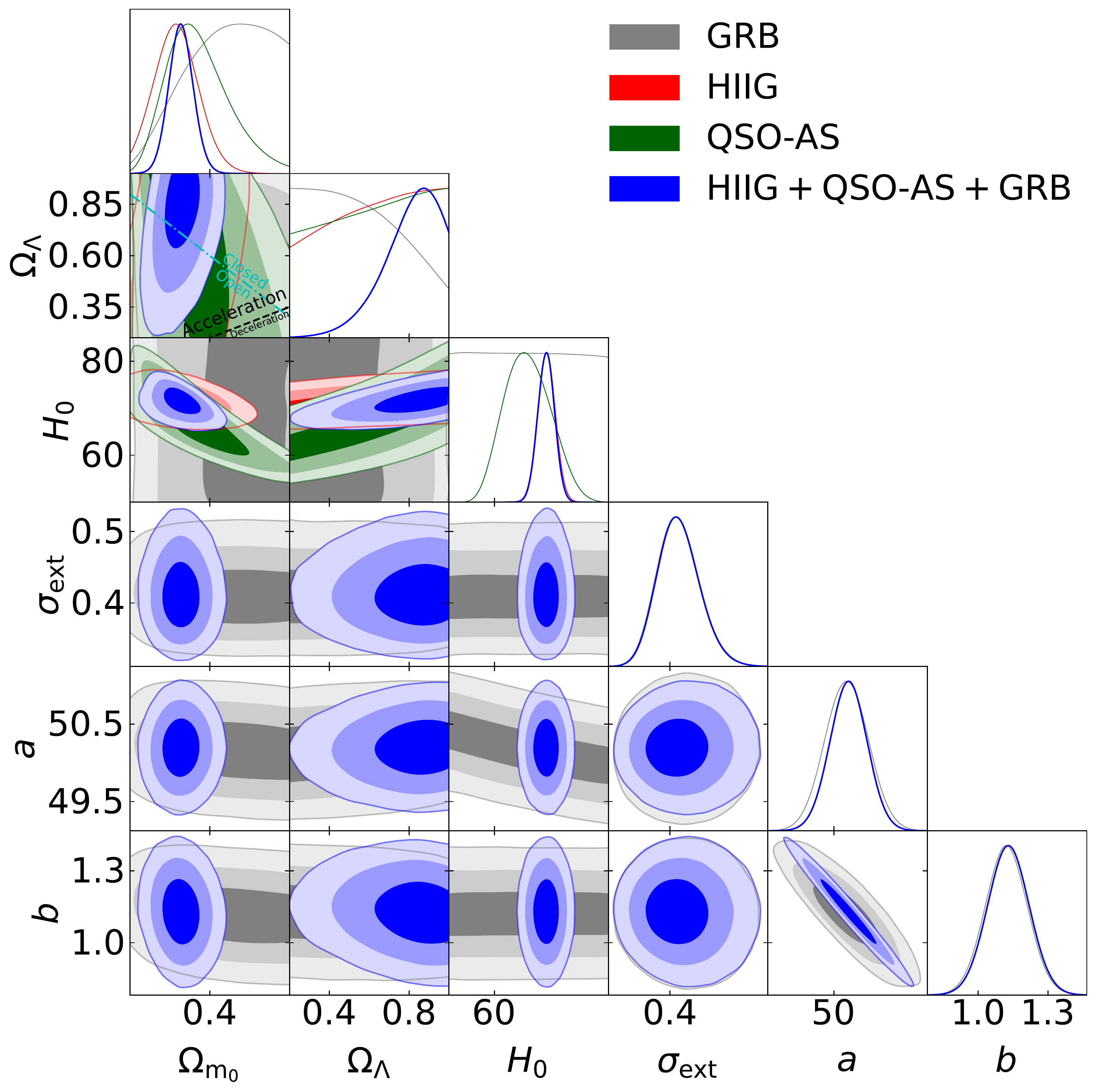}}
  \subfloat[Cosmological parameters zoom in]{%
    \includegraphics[width=3.25in,height=3.25in]{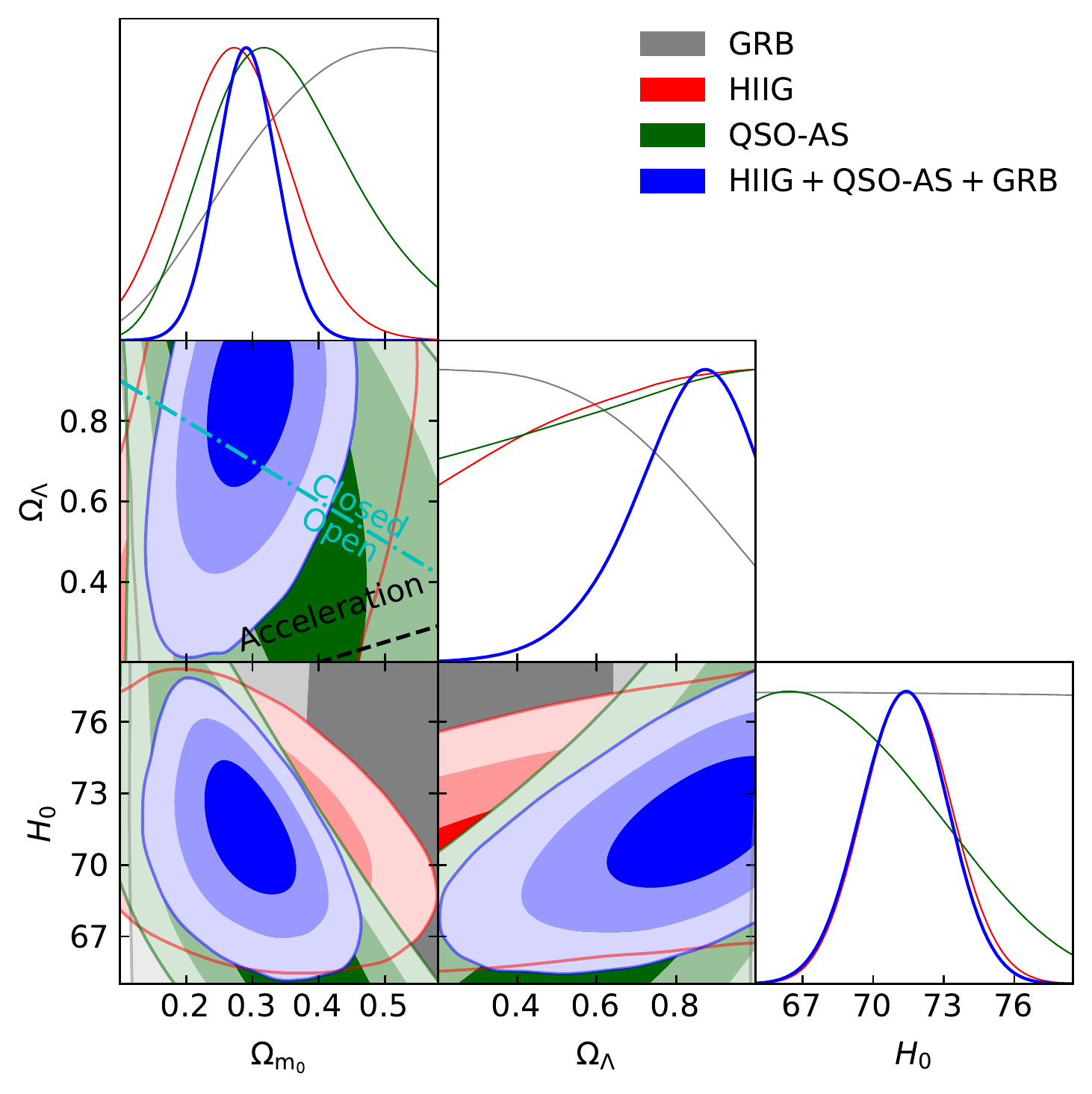}}\\
\caption{Same as Fig. \ref{fig1} but for non-flat \lcdm. The cyan dash-dot line represents the flat \lcdm\ case, with closed spatial hypersurfaces to the upper right. The black dotted line is the zero-acceleration line, which divides the parameter space into regions associated with currently-accelerating (above left) and currently-decelerating (below right) cosmological expansion.}
\label{fig2}
\end{figure*}

\begin{figure*}
\centering
  \subfloat[All parameters]{%
    \includegraphics[width=3.25in,height=3.25in]{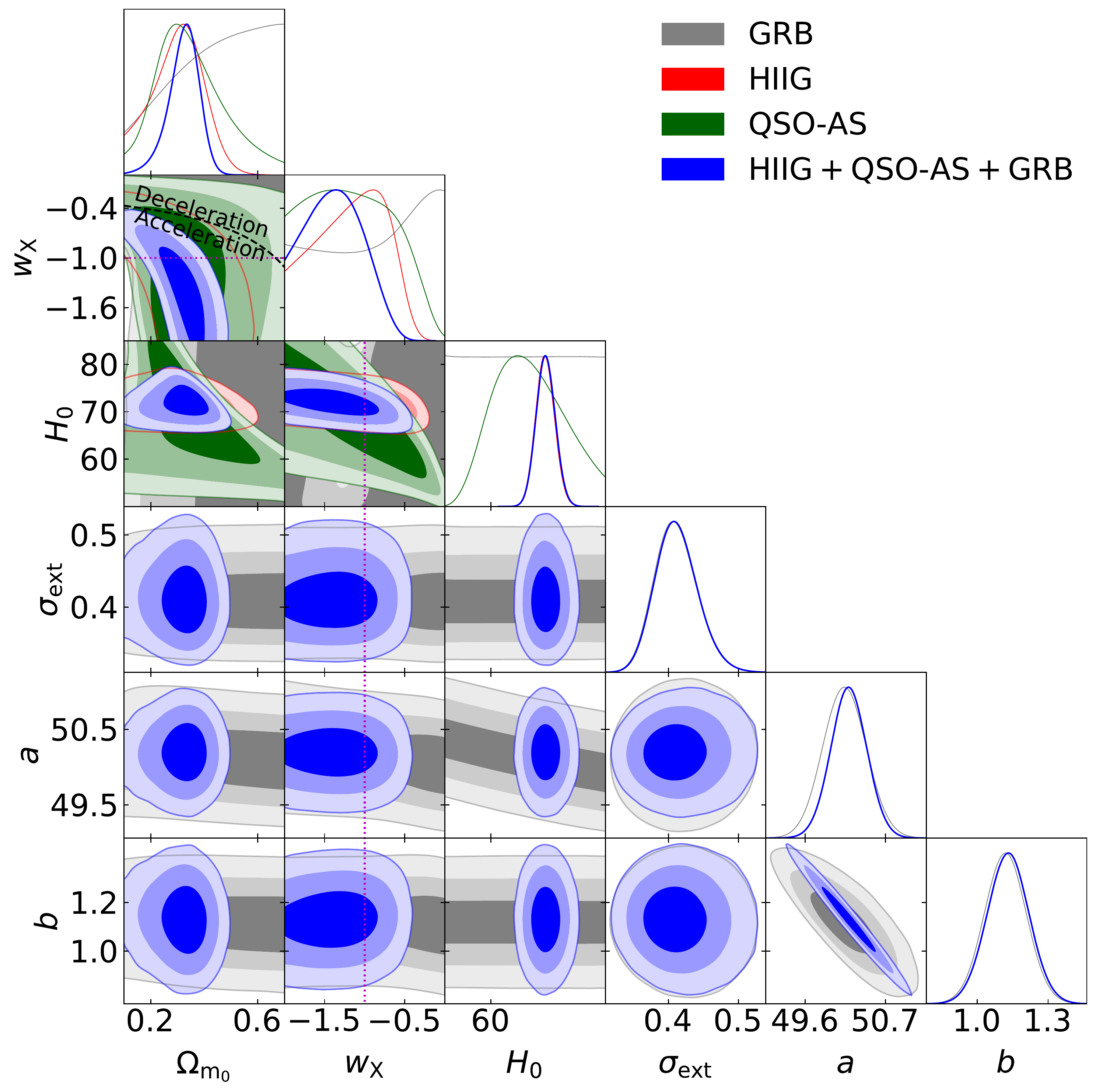}}
  \subfloat[Cosmological parameters zoom in]{%
    \includegraphics[width=3.25in,height=3.25in]{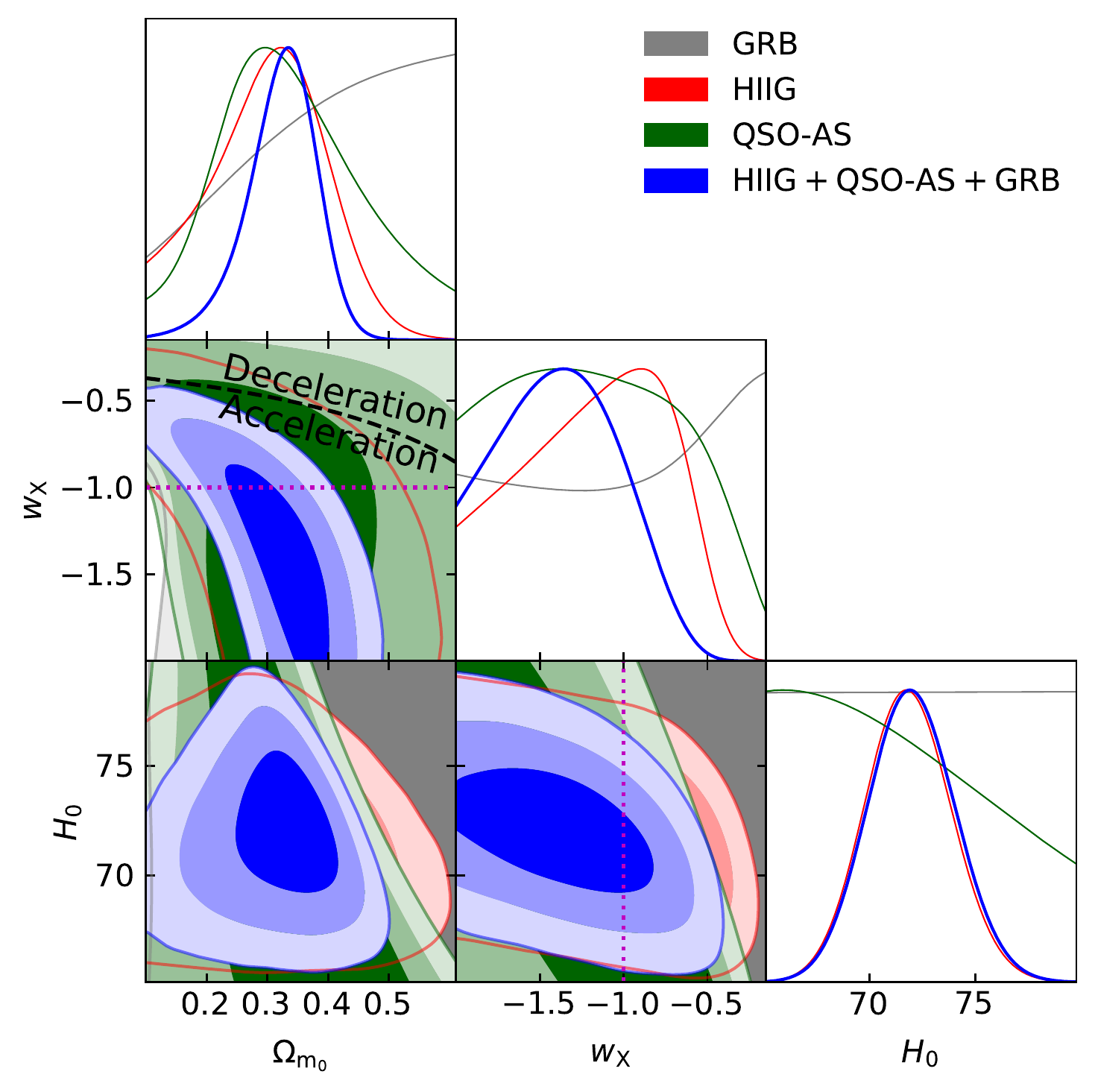}}\\
\caption{1$\sigma$, 2$\sigma$, and 3$\sigma$ confidence contours for flat XCDM. The black dotted line is the zero-acceleration line, which divides the parameter space into regions associated with currently-accelerating (below left) and currently-decelerating (above right) cosmological expansion. The magenta lines denote $w_{\rm X}=-1$, i.e. the flat \lcdm\ model.}
\label{fig3}
\end{figure*}

\begin{figure*}
\centering
  \subfloat[All parameters]{%
    \includegraphics[width=3.25in,height=3.25in]{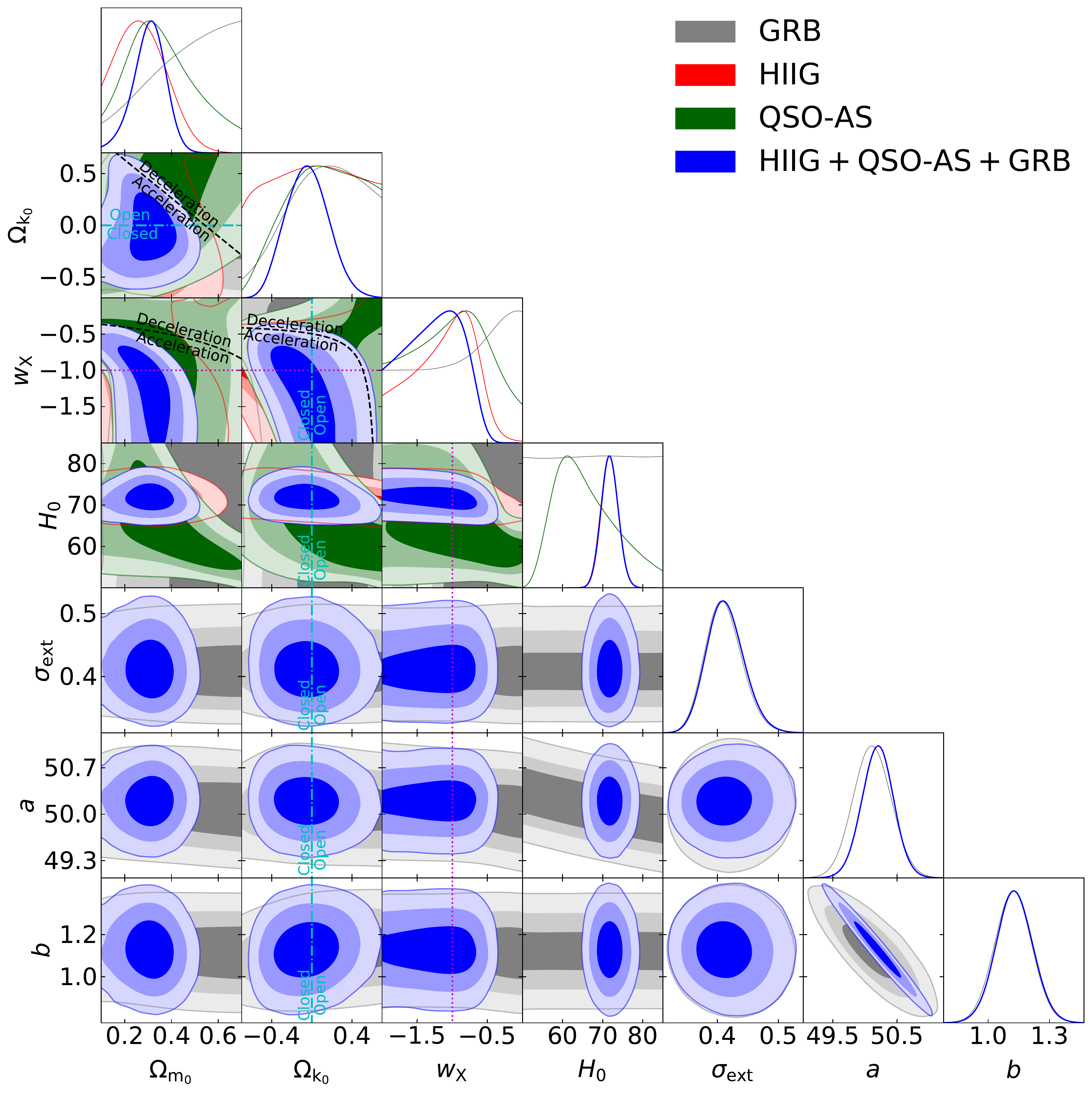}}
  \subfloat[Cosmological parameters zoom in]{%
    \includegraphics[width=3.25in,height=3.25in]{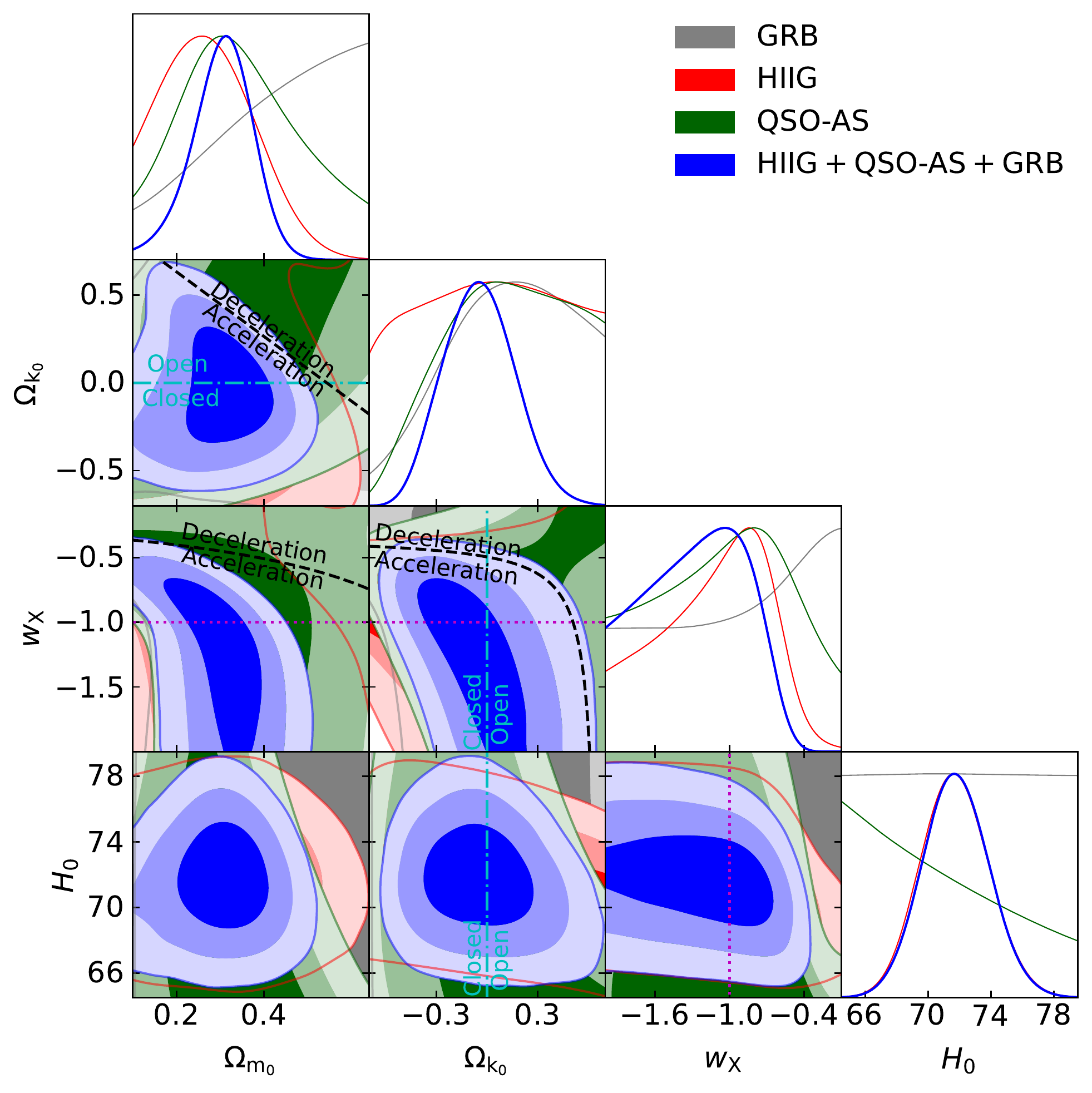}}\\
\caption{Same as Fig. \ref{fig3} but for non-flat XCDM, where the zero acceleration lines in each of the three subpanels are computed for the third cosmological parameter set to the $H(z)$ + BAO data best-fitting values listed in Table \ref{tab:BFPC4}. Currently-accelerating cosmological expansion occurs below these lines. The cyan dash-dot lines represent the flat XCDM case, with closed spatial hypersurfaces either below or to the left. The magenta lines indicate $w_{\rm X} = -1$, i.e. the non-flat \lcdm\ model.}
\label{fig4}
\end{figure*}

We present the posterior one-dimensional (1D) probability distributions and two-dimensional (2D) confidence regions of the cosmological and Amati relation parameters for the six flat and non-flat models in Figs. \ref{fig1}--\ref{fig6}, in gray (GRB), red (\hiig), and green (QSO-AS). The unmarginalized best-fitting parameter values are listed in Table \ref{tab:BFPC4}, along with the corresponding $\chi^2$, $-2\ln\mathcal{L}_{\rm max}$, $AIC$, $BIC$, and degrees of freedom $\nu$ (where $\nu \equiv N - n$).\footnote{Note that the $\chi^2$ values listed in Tables \ref{tab:BFPC4} and \ref{tab:BFPaA} are computed from the best-fitting parameter values and are not necessarily the minimum (especially when including GRB and QSO-Flux data).} The values of $\Delta\chi^2$, $\Delta AIC$, and $\Delta BIC$ reported in Table \ref{tab:BFPC4} are discussed in Section \ref{subsec:comparison}, where we define $\Delta \chi^2$, $\Delta AIC$, and $\Delta BIC$, respectively, as the differences between the values of the $\chi^2$, $AIC$, and $BIC$ associated with a given model and their corresponding minimum values among all models. The marginalized best-fitting parameter values and uncertainties ($\pm 1\sigma$ error bars or $2\sigma$ limits) are given in Table \ref{tab:1d_BFPC4}.\footnote{We use the \textsc{python} package \textsc{getdist} \citep{Lewis_2019} to plot these figures and compute the central values (posterior means) and uncertainties of the free parameters listed in Table \ref{tab:1d_BFPC4}.}
From Table \ref{tab:1d_BFPC4} we find that the QSO-AS constraints on \om\ are consistent with other results within a 1$\sigma$ range but with large error bars, ranging from a low of $0.329^{+0.086}_{-0.171}$ (flat \pcdm) to a high of $0.364^{+0.083}_{-0.150}$ (flat \lcdm). 

The QSO-AS constraints on $H_0$ are between $H_0=61.91^{+2.83}_{-4.92}$ \hunit (non-flat \pcdm) and $H_0=68.39^{+6.14}_{-8.98}$ \hunit (flat XCDM), with large error bars and relatively low values for non-flat XCDM and the \pcdm\ models. 

The non-flat models mildly favor open geometry, but are also consistent, given the large error bars, with spatially-flat hypersurfaces (except for non-flat \pcdm, where the open case is favored at $2.76\sigma$). For non-flat \lcdm, non-flat XCDM, and non-flat \pcdm, we find $\Omega_{\rm k_0}=0.017^{+0.184}_{-0.277}$, $\Omega_{\rm k_0}=0.115^{+0.466}_{-0.293}$, and $\Omega_{\rm k_0}=0.254^{+0.304}_{-0.092}$, respectively.\footnote{From Table \ref{tab:1d_BFPC4} we see that GRB data are also consistent with flat spatial geometry in the non-flat \lcdm\ and XCDM cases, but also favor, at $2.92\sigma$, open spatial geometry in the case of non-flat \pcdm.}

The fits to the QSO-AS data favor dark energy being a cosmological constant but do not strongly disfavor dark energy dynamics. For flat (non-flat) XCDM, $w_{\rm X}=-1.161^{+0.430}_{-0.679}$ ($w_{\rm X}=-1.030^{+0.593}_{-0.548}$), and for flat (non-flat) \pcdm, $2\sigma$ upper limits of $\alpha$ are $\alpha<2.841$ ($\alpha<4.752$). In the former case, both results are within 1$\sigma$ of $w_{\rm X}=-1$, and in the latter case, both 1D likelihoods peak at $\alpha=0$. 

Constraints on cosmological model parameters derived solely from \hiig\ data are discussed in Sec. 5.1 of \cite{CaoRyanRatra2020}, while those derived from GRB data are described in Sec. 5.1 of \cite{KhadkaRatra2020a} (though there are slight differences coming from the different treatments of $H_0$ and the different ranges of flat priors used there and here); both are listed in Table \ref{tab:1d_BFPC4} here. In contrast to the \hiig\ and QSO-AS data sets, the GRB data alone cannot constrain $H_0$ because there is a degeneracy between the intercept parameter ($a$) of the Amati relation and $H_0$; for consistency with the analyses of the \hiig\ and QSO-AS data, we treat $H_0$ as a free parameter in the GRB data analysis here.

Cosmological constraints obtained using the \hiig, QSO-AS, and GRB data sets are mutually consistent, and are also consistent with those obtained from most other cosmological probes. This is partially a consequence of the larger \hiig, QSO-AS, and GRB data error bars, which lead to relatively weaker constraints on cosmological parameters when each of these data sets is used alone (see Table \ref{tab:1d_BFPC4}). However, because the \hiig, QSO-AS, and GRB constraints are mutually consistent, we may jointly analyze these data. Their combined cosmological constraints will therefore be more restrictive than when they are analyzed individually.

We note, from Figs. \ref{fig1}--\ref{fig6}, that a significant part of the likelihood of each of these three data sets lies in the parameter space part with currently-accelerating cosmological expansion.

\subsection{\hiig, QSO-AS, and GRB (HQASG) joint constraints}
\label{subsec:HQASG}

Because the \hiig, QSO-AS, and GRB contours are mutually consistent for all six of the models we study, we jointly analyze these data to obtain HQASG constraints.

\begin{figure*}
\centering
  \subfloat[All parameters]{%
    \includegraphics[width=3.25in,height=3.25in]{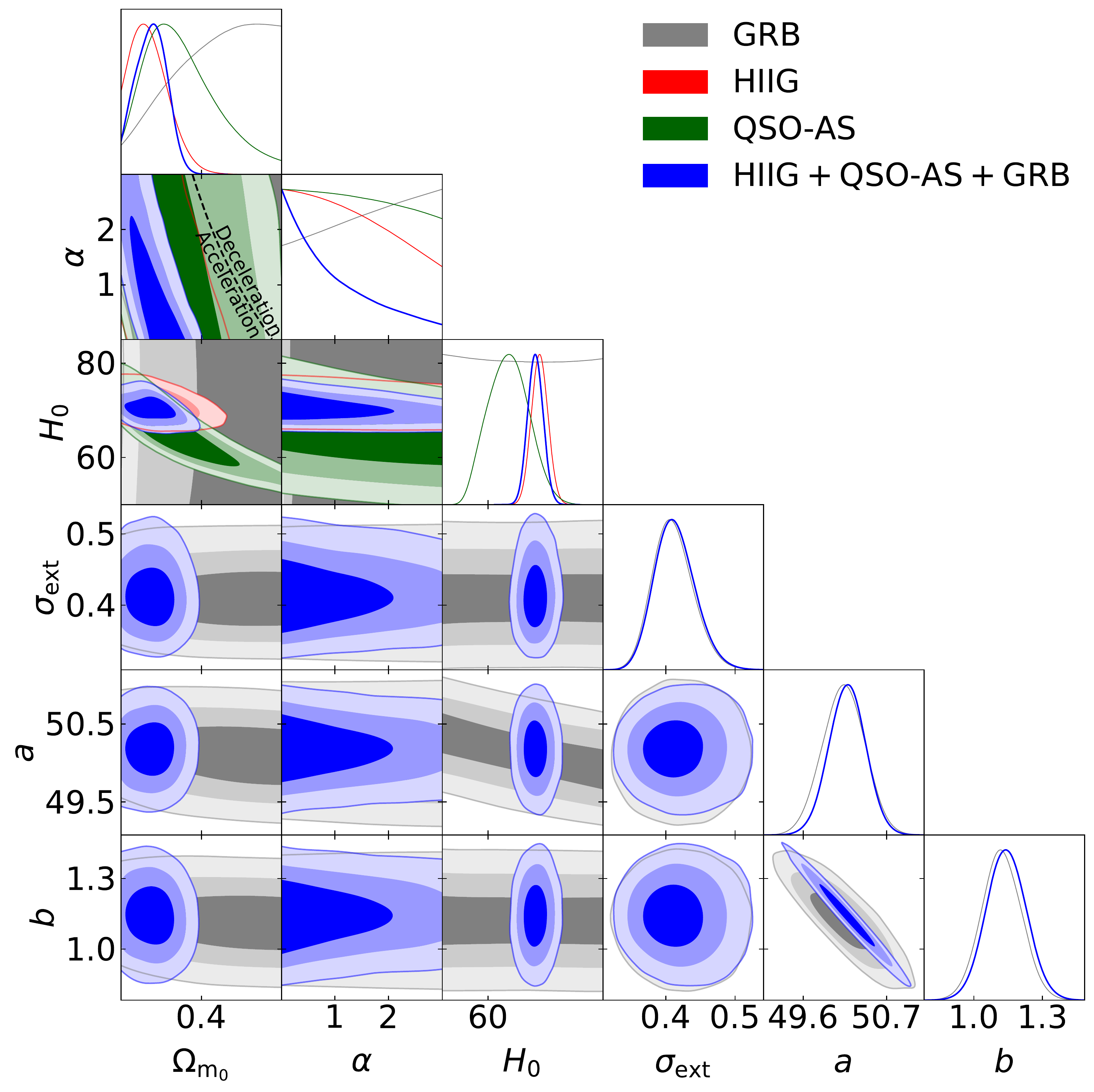}}
  \subfloat[Cosmological parameters zoom in]{%
    \includegraphics[width=3.25in,height=3.25in]{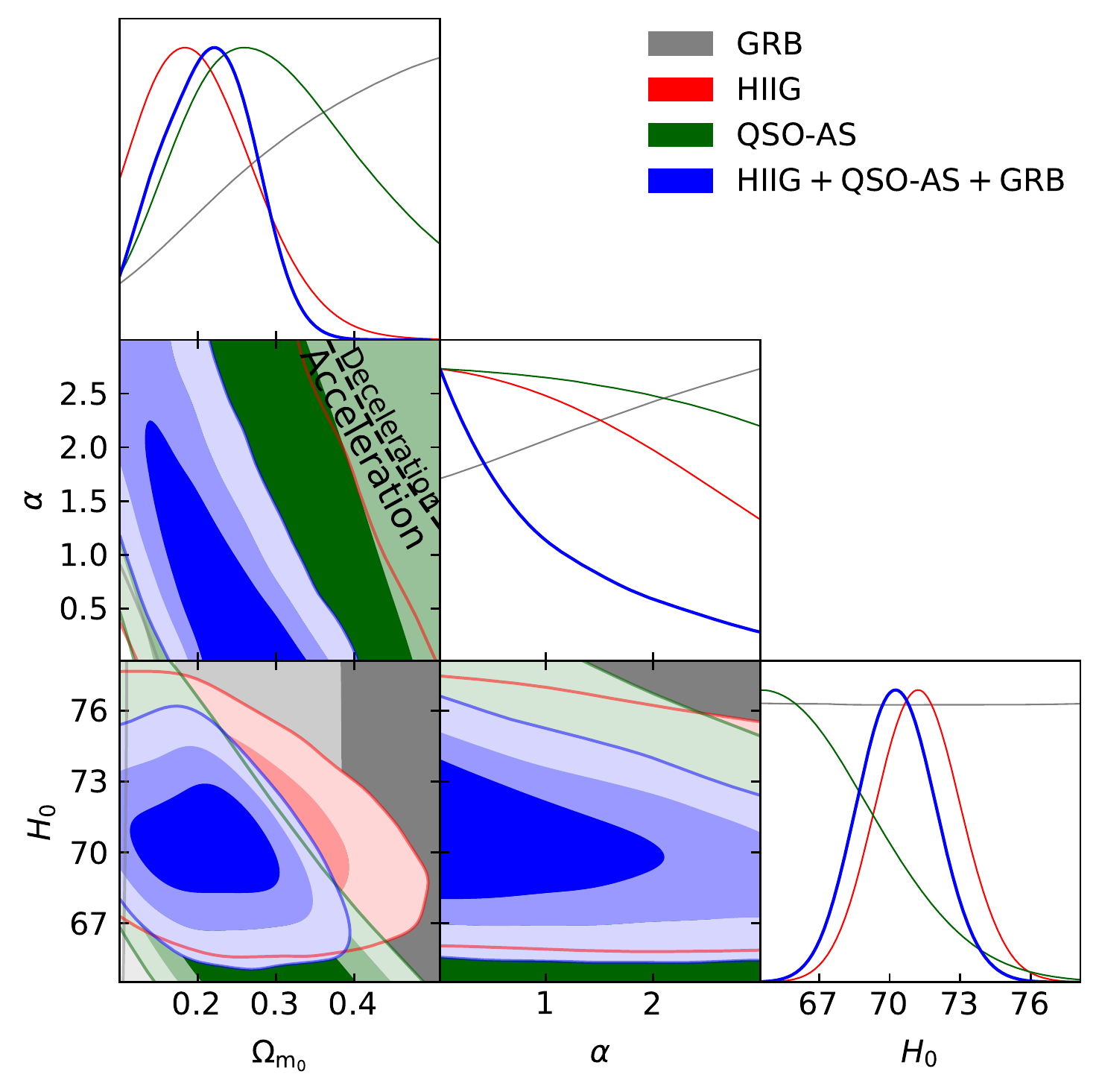}}\\
\caption{1$\sigma$, 2$\sigma$, and 3$\sigma$ confidence contours for flat \pcdm. The black dotted zero-acceleration line splits the parameter space into regions of currently-accelerating (below left) and currently-decelerating (above right) cosmological expansion. The $\alpha = 0$ axis is the flat \lcdm\ model.}
\label{fig5}
\end{figure*}

\begin{figure*}
\centering
  \subfloat[All parameters]{%
    \includegraphics[width=3.25in,height=3.25in]{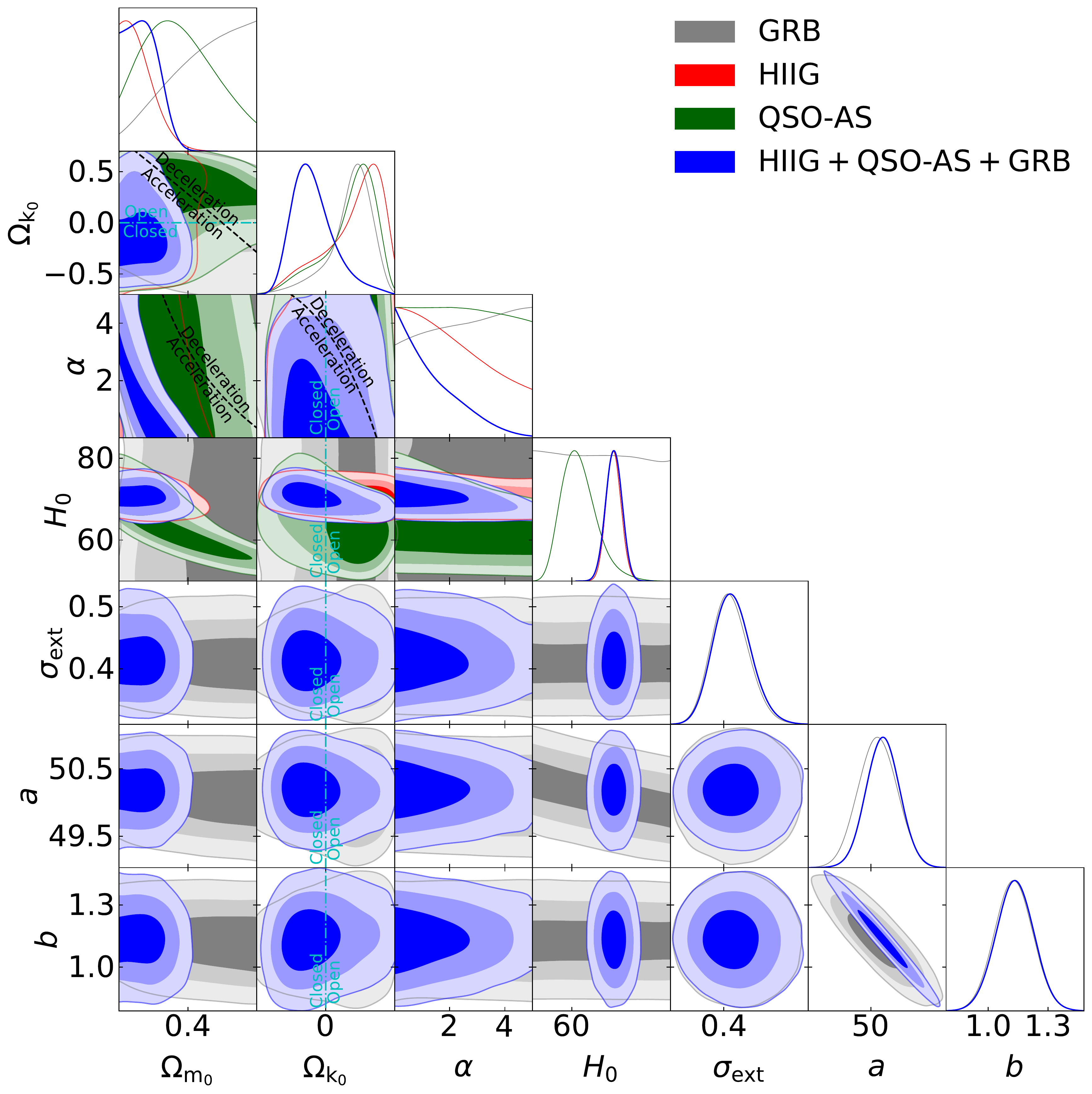}}
  \subfloat[Cosmological parameters zoom in]{%
    \includegraphics[width=3.25in,height=3.25in]{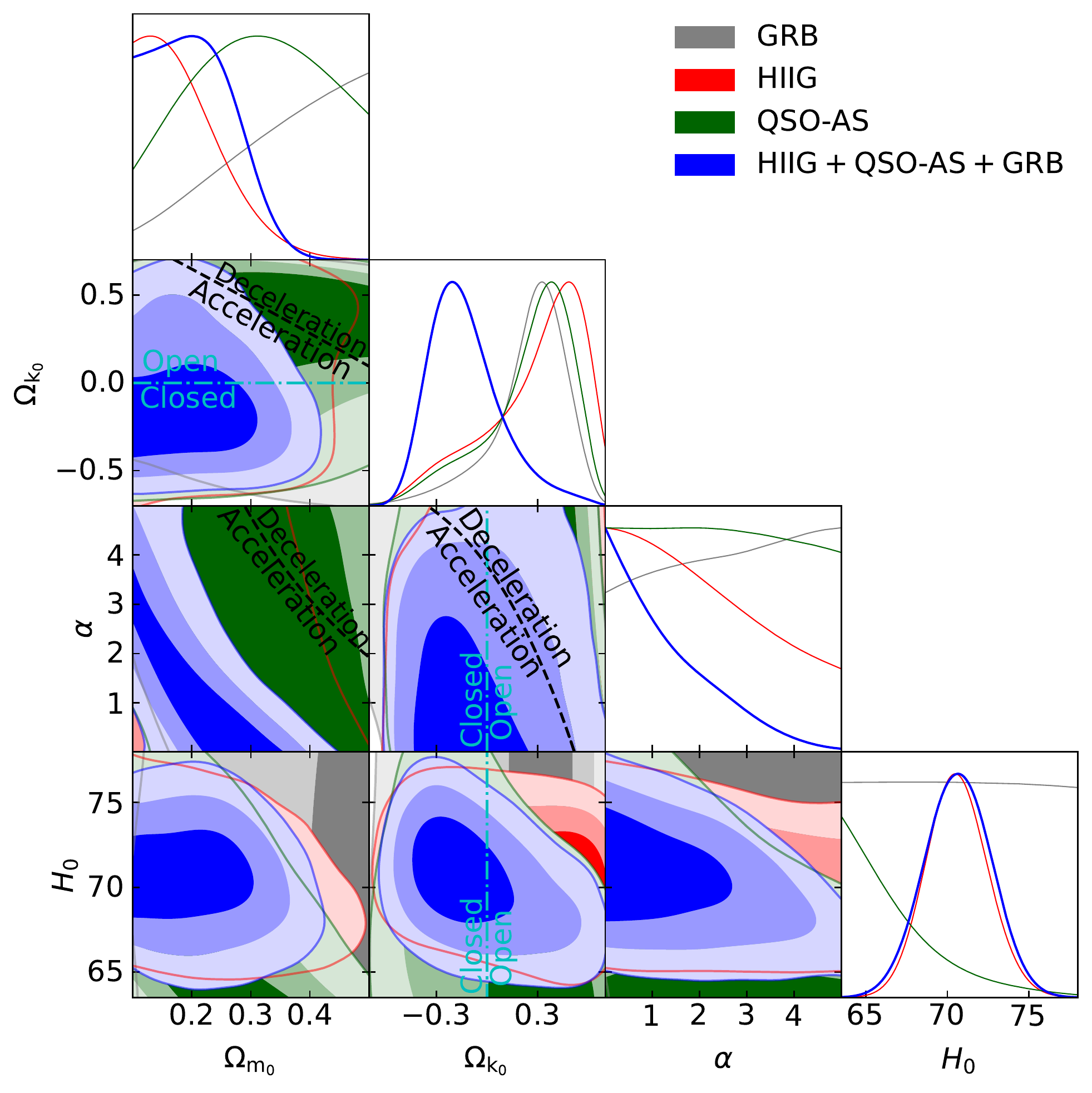}}\\
\caption{Same as Fig. \ref{fig5} but for non-flat \pcdm, where the zero-acceleration lines in each of the sub-panels are computed for the third cosmological parameter set to the $H(z)$ + BAO data best-fitting values listed in Table \ref{tab:BFPC4}. Currently-accelerating cosmological expansion occurs below these lines. The cyan dash-dot lines represent the flat \pcdm\ case, with closed spatial geometry either below or to the left. The $\alpha = 0$ axis is the non-flat \lcdm\ model.}
\label{fig6}
\end{figure*}

\begin{figure*}
\centering
  \subfloat[All parameters]{%
    \includegraphics[width=3.25in,height=3.25in]{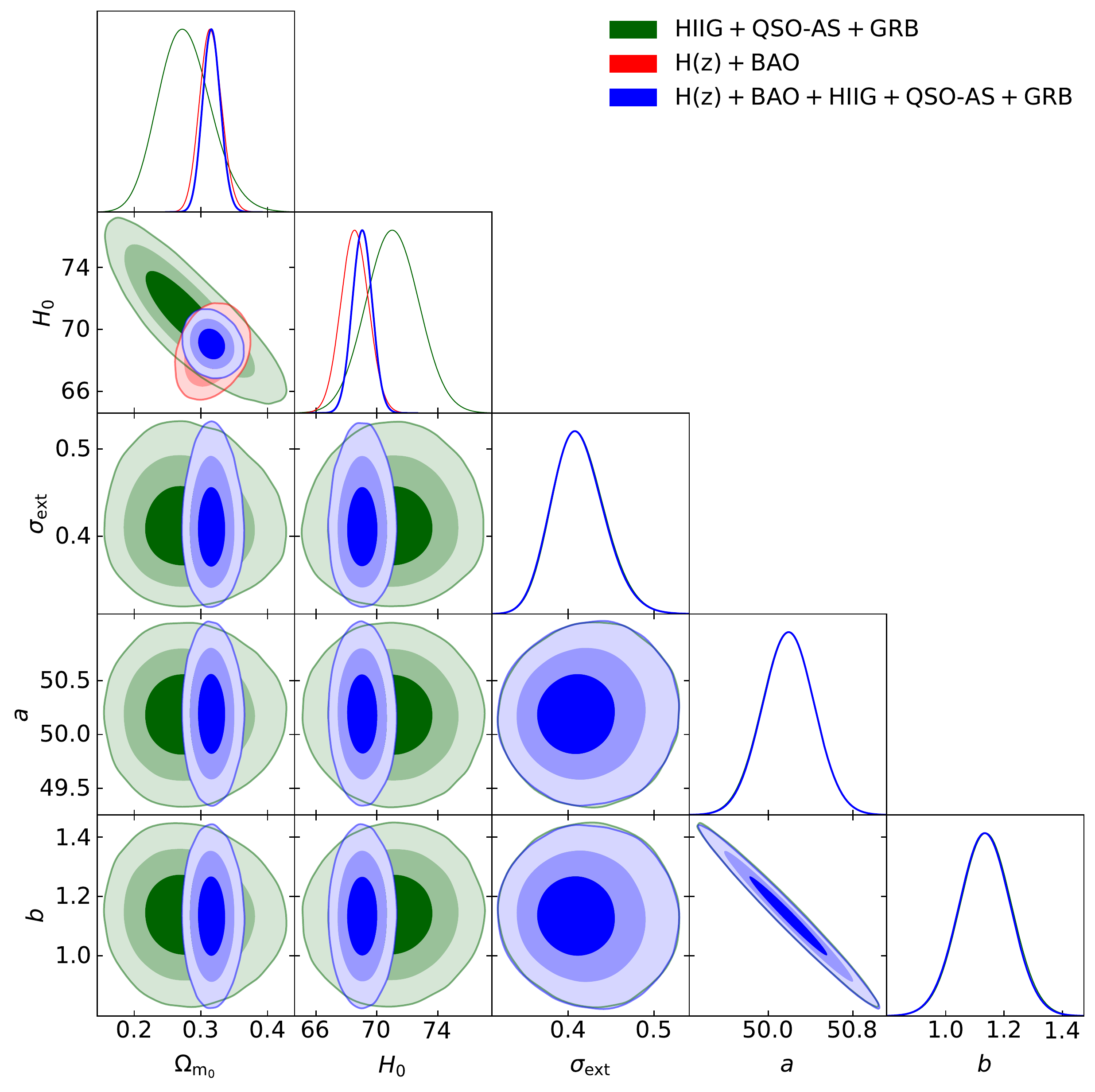}}
  \subfloat[Cosmological parameters zoom in]{%
    \includegraphics[width=3.25in,height=3.25in]{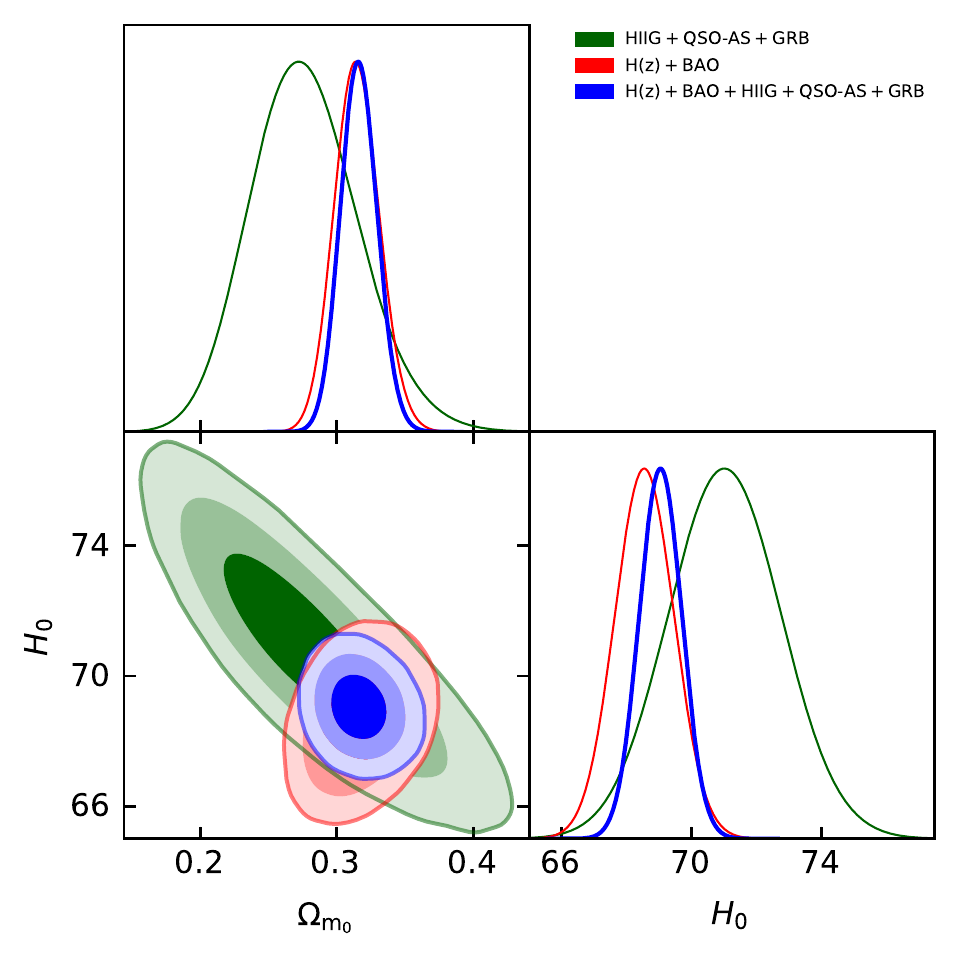}}\\
\caption{Same as Fig. \ref{fig1} (flat \lcdm) but for different combinations of data.}
\label{fig7}
\end{figure*}

\begin{figure*}
\centering
  \subfloat[All parameters]{%
    \includegraphics[width=3.25in,height=3.25in]{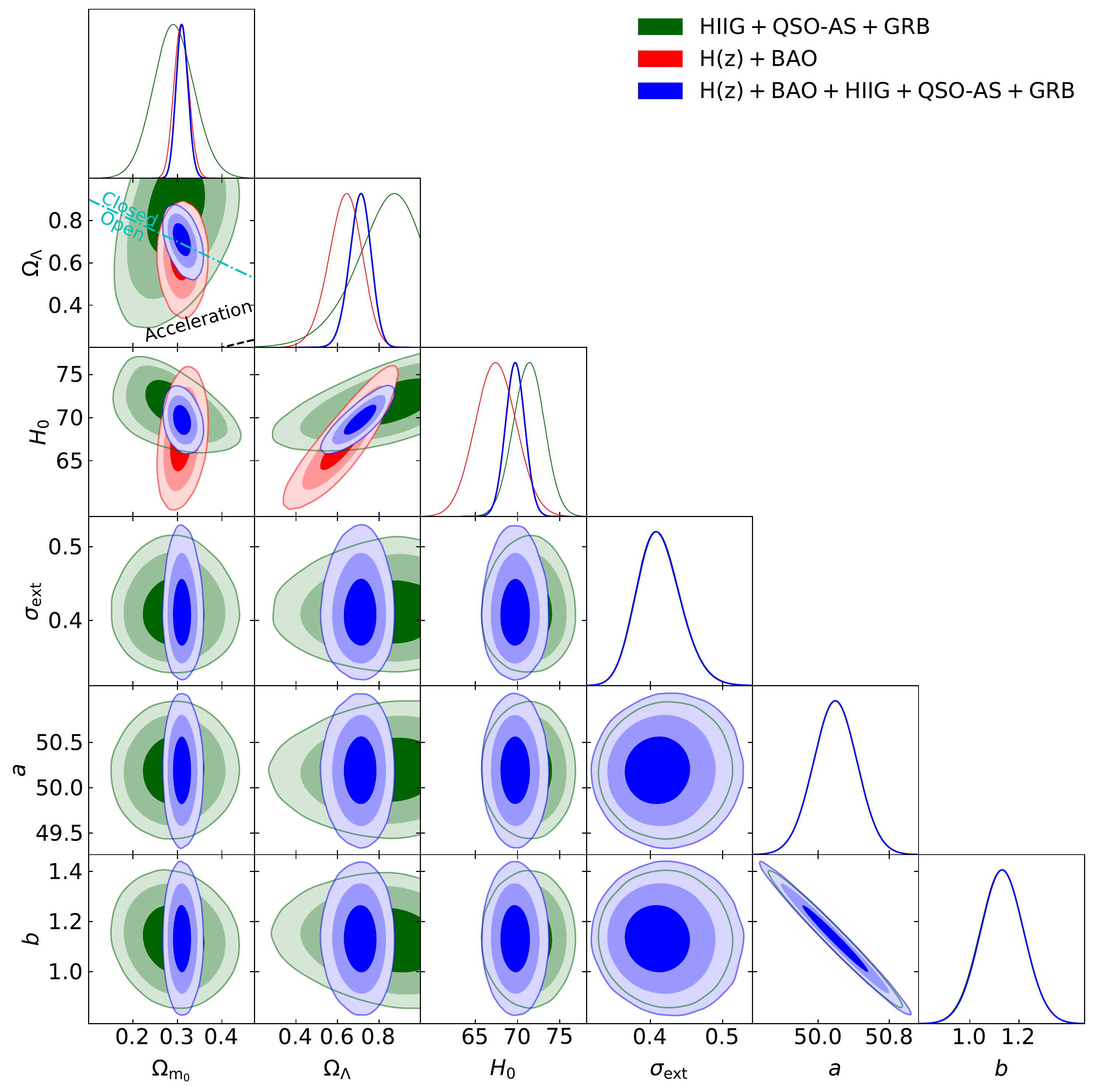}}
  \subfloat[Cosmological parameters zoom in]{%
    \includegraphics[width=3.25in,height=3.25in]{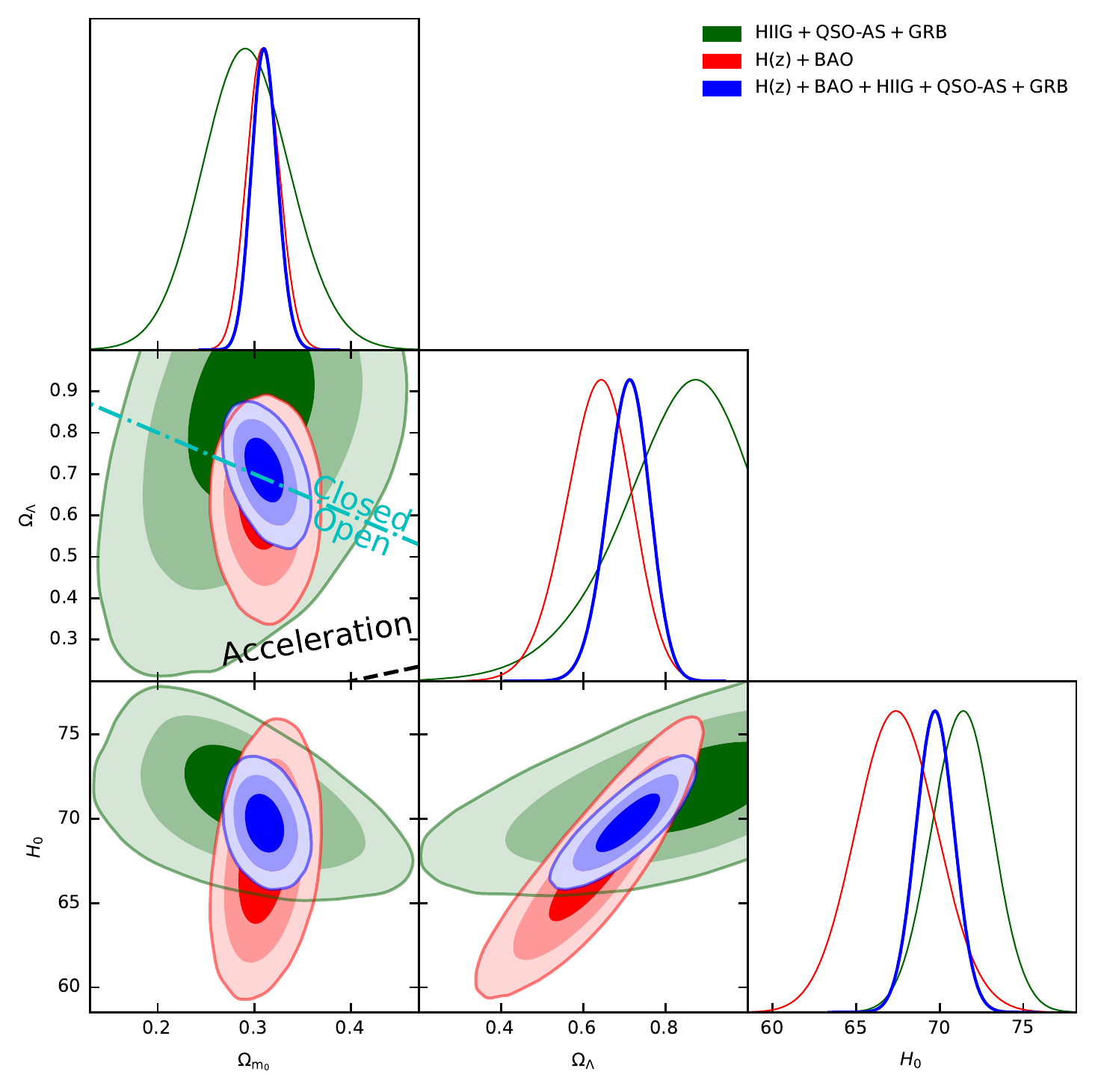}}\\
\caption{Same as Fig. \ref{fig2} (non-flat \lcdm) but for different combinations of data.}
\label{fig8}
\end{figure*}

\begin{figure*}
\centering
  \subfloat[All parameters]{%
    \includegraphics[width=3.25in,height=3.25in]{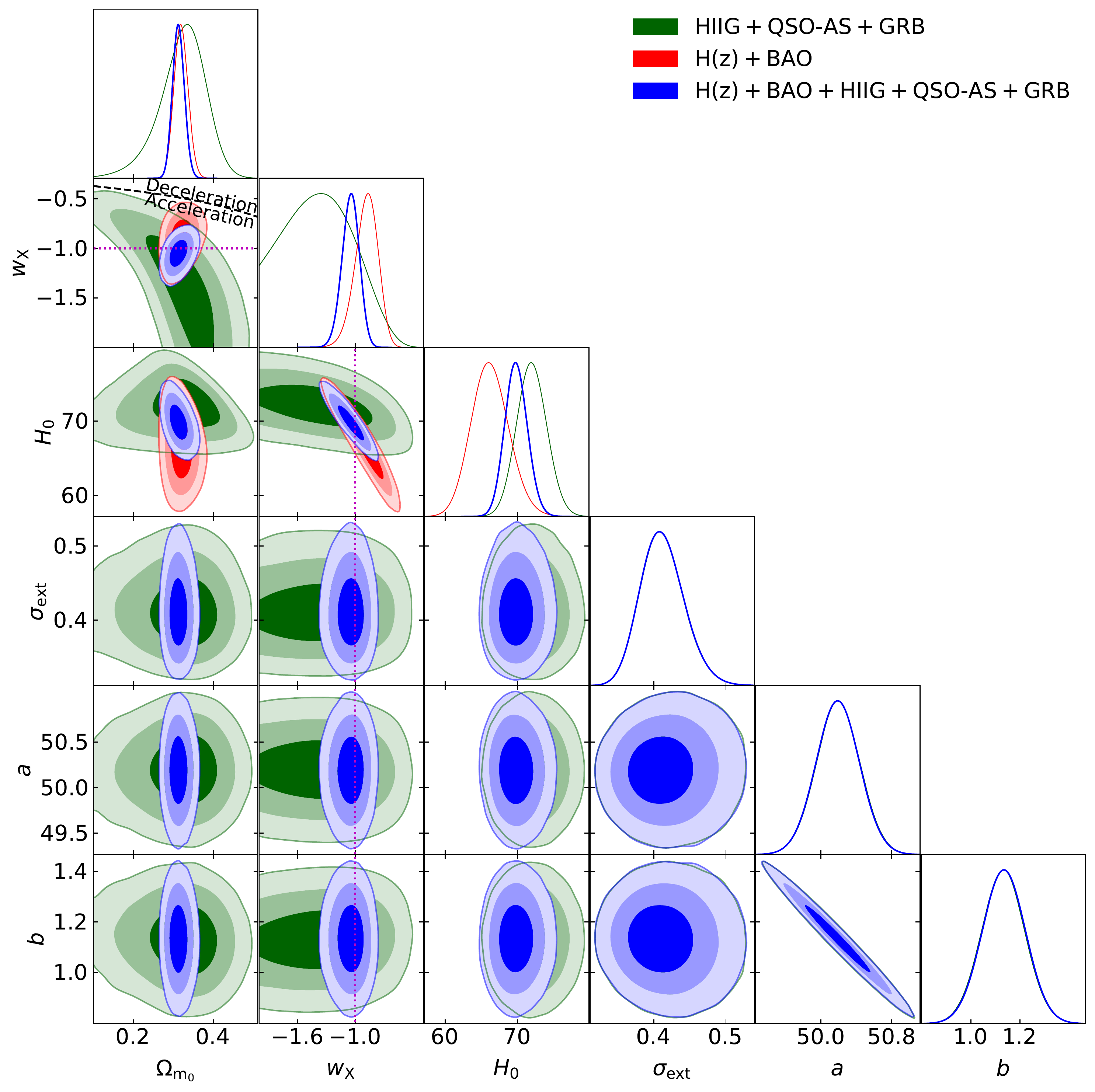}}
  \subfloat[Cosmological parameters zoom in]{%
    \includegraphics[width=3.25in,height=3.25in]{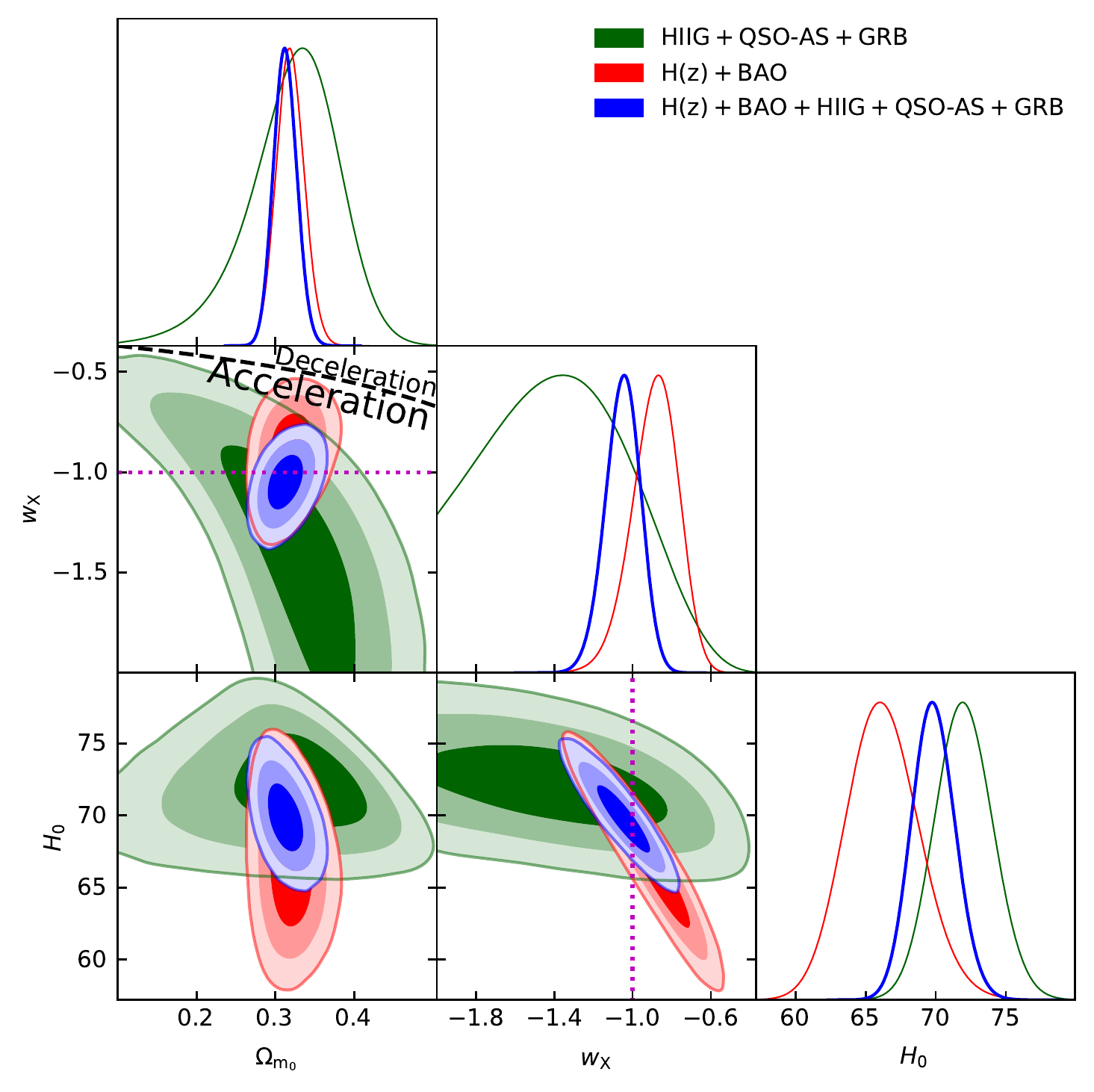}}\\
\caption{Same as Fig. \ref{fig3} (flat XCDM) but for different combinations of data.}
\label{fig9}
\end{figure*}

\begin{figure*}
\centering
  \subfloat[All parameters]{%
    \includegraphics[width=3.25in,height=3.25in]{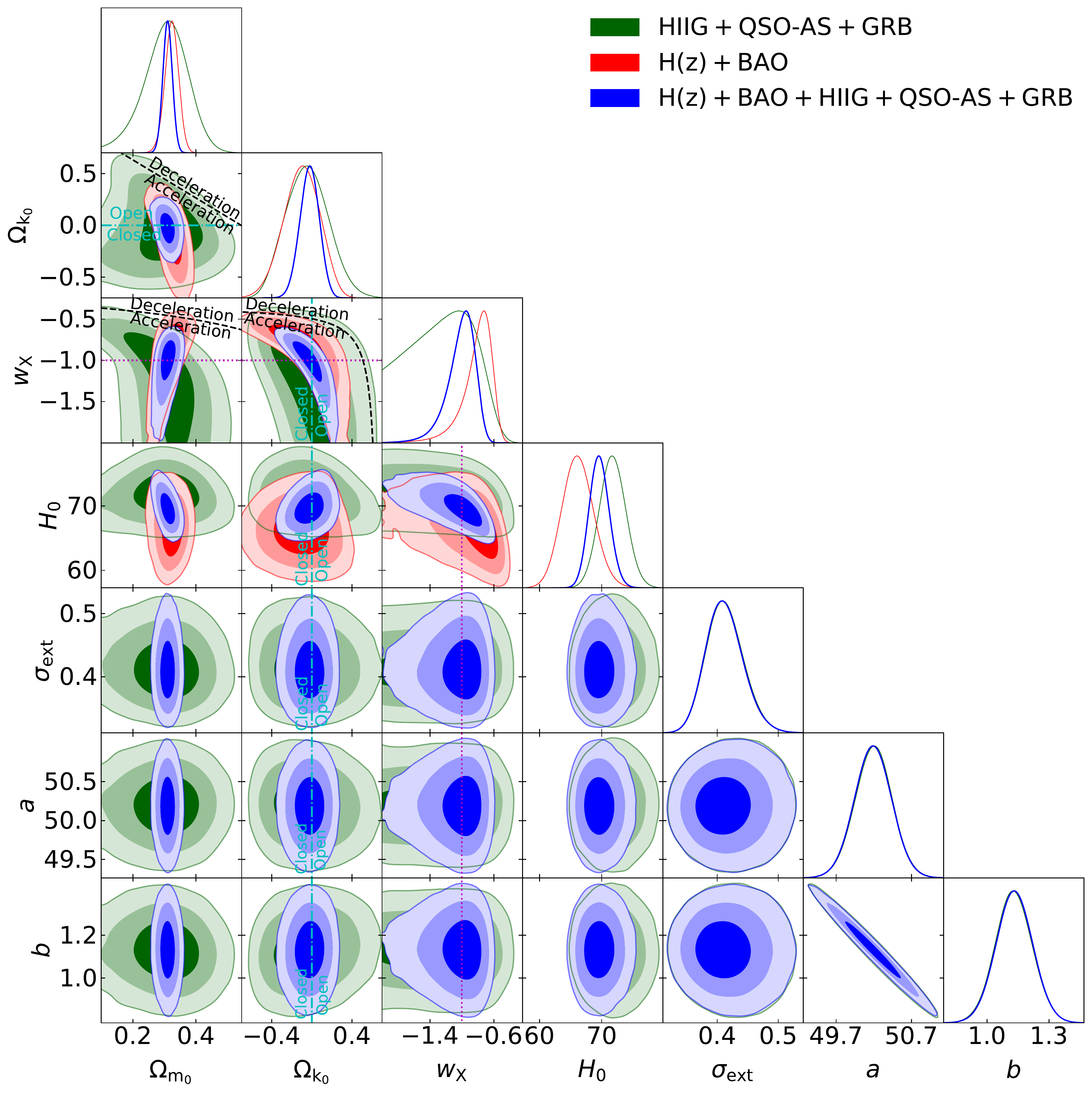}}
  \subfloat[Cosmological parameters zoom in]{%
    \includegraphics[width=3.25in,height=3.25in]{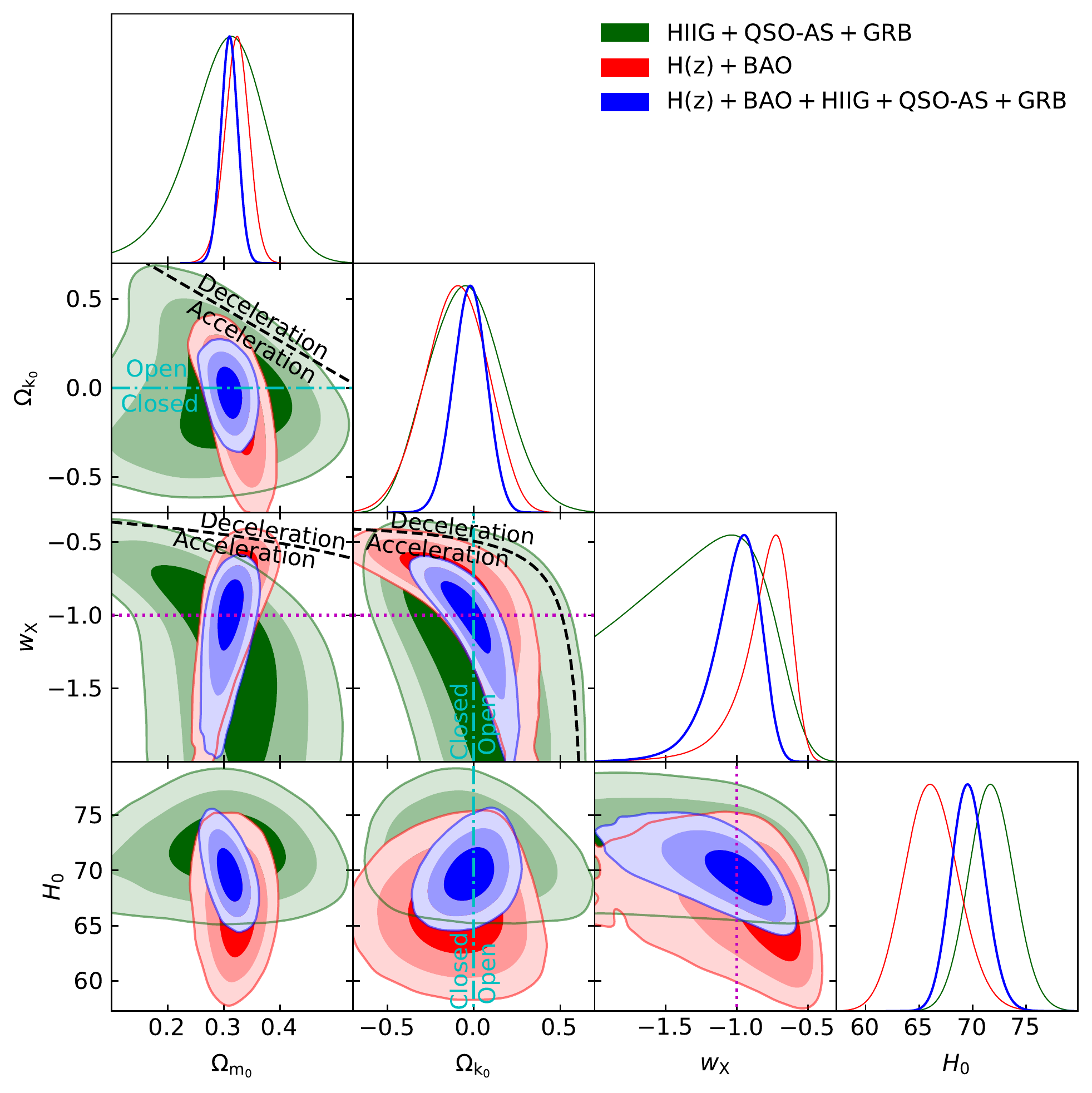}}\\
\caption{Same as Fig. \ref{fig4} (non-flat XCDM) but for different combinations of data.}
\label{fig10}
\end{figure*}

\begin{figure*}
\centering
  \subfloat[All parameters]{%
    \includegraphics[width=3.25in,height=3.25in]{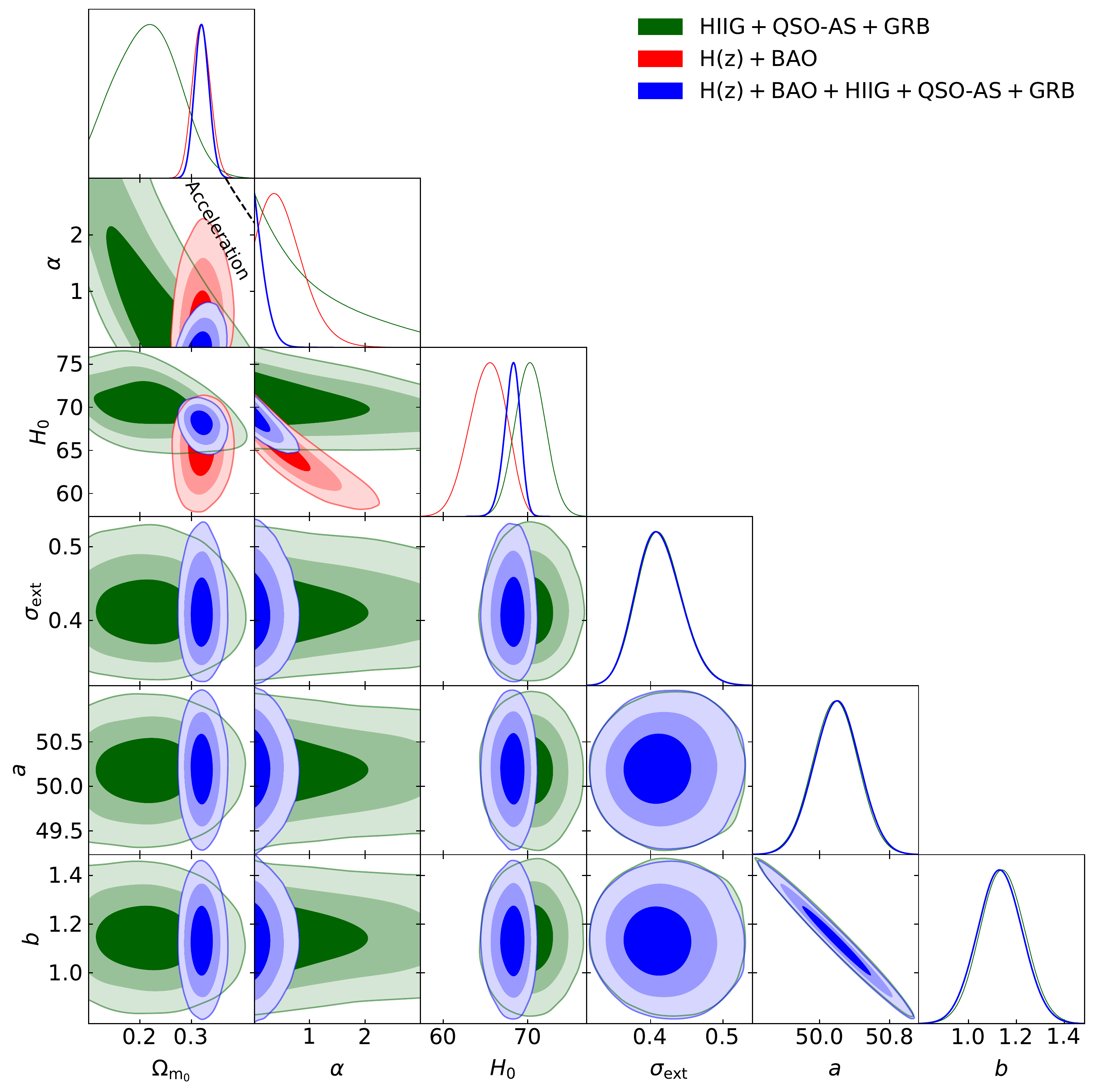}}
  \subfloat[Cosmological parameters zoom in]{%
    \includegraphics[width=3.25in,height=3.25in]{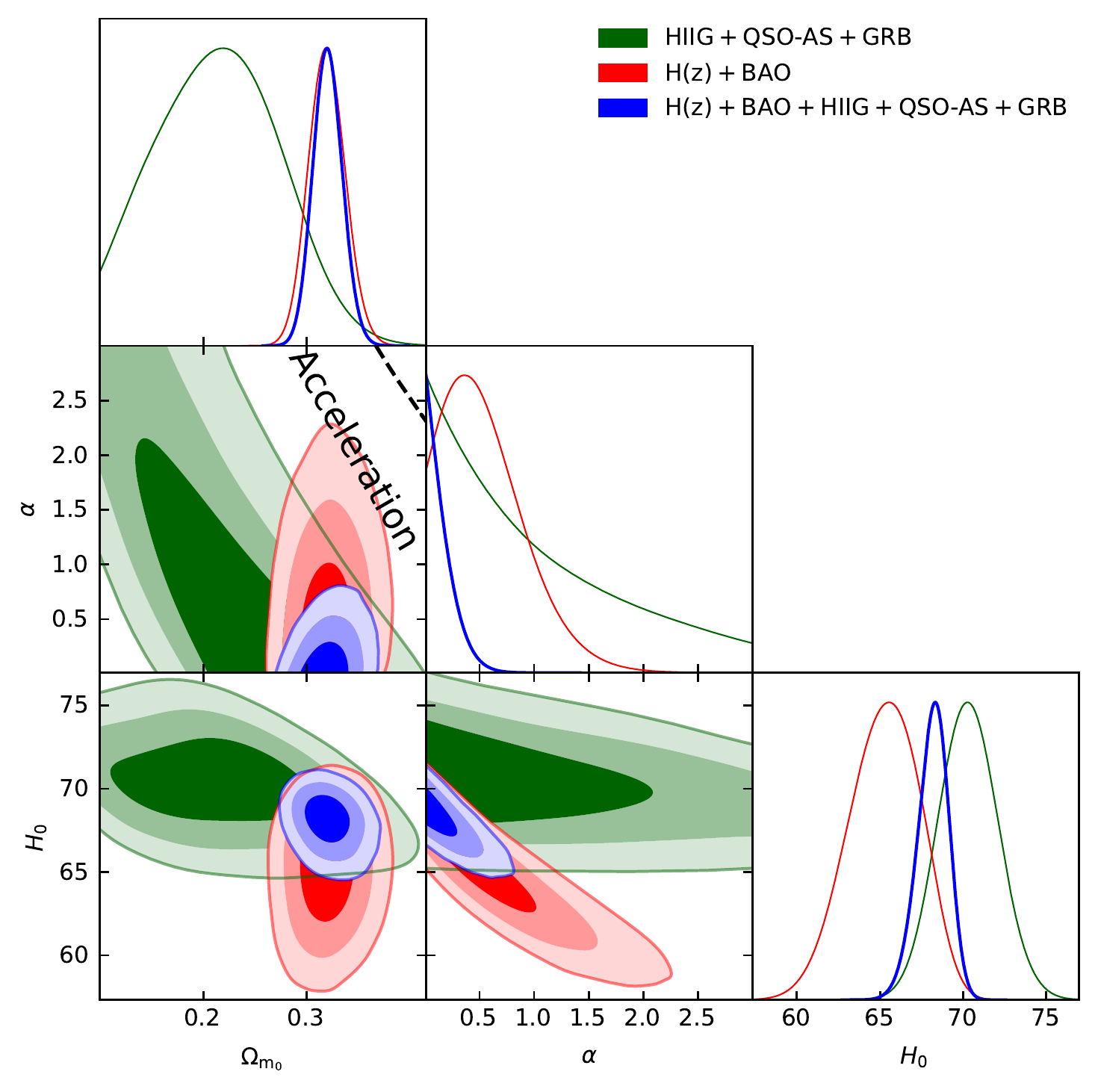}}\\
\caption{Same as Fig. \ref{fig5} (flat \pcdm) but for different combinations of data.}
\label{fig11}
\end{figure*}

\begin{figure*}
\centering
  \subfloat[All parameters]{%
    \includegraphics[width=3.25in,height=3.25in]{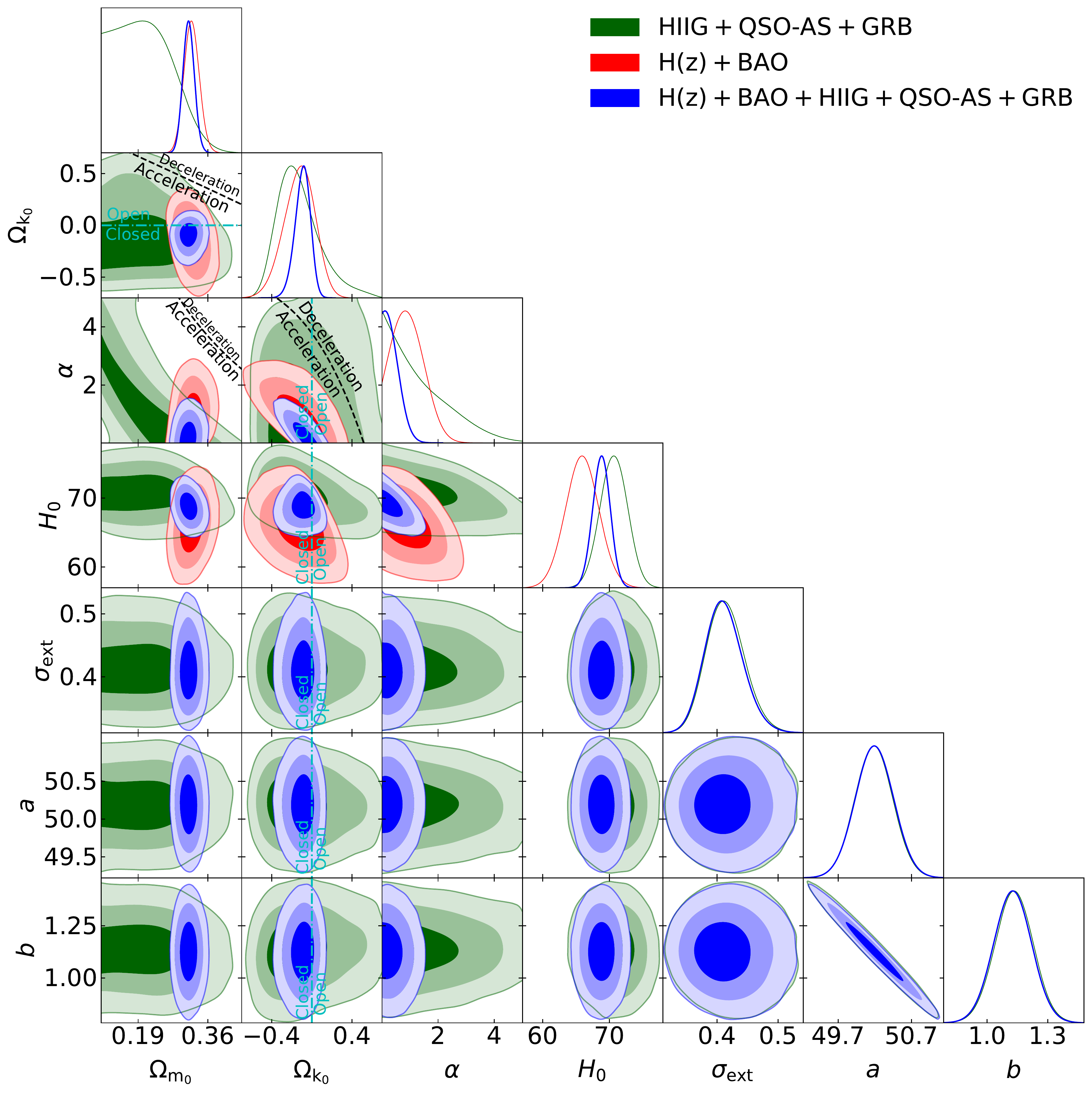}}
  \subfloat[Cosmological parameters zoom in]{%
    \includegraphics[width=3.25in,height=3.25in]{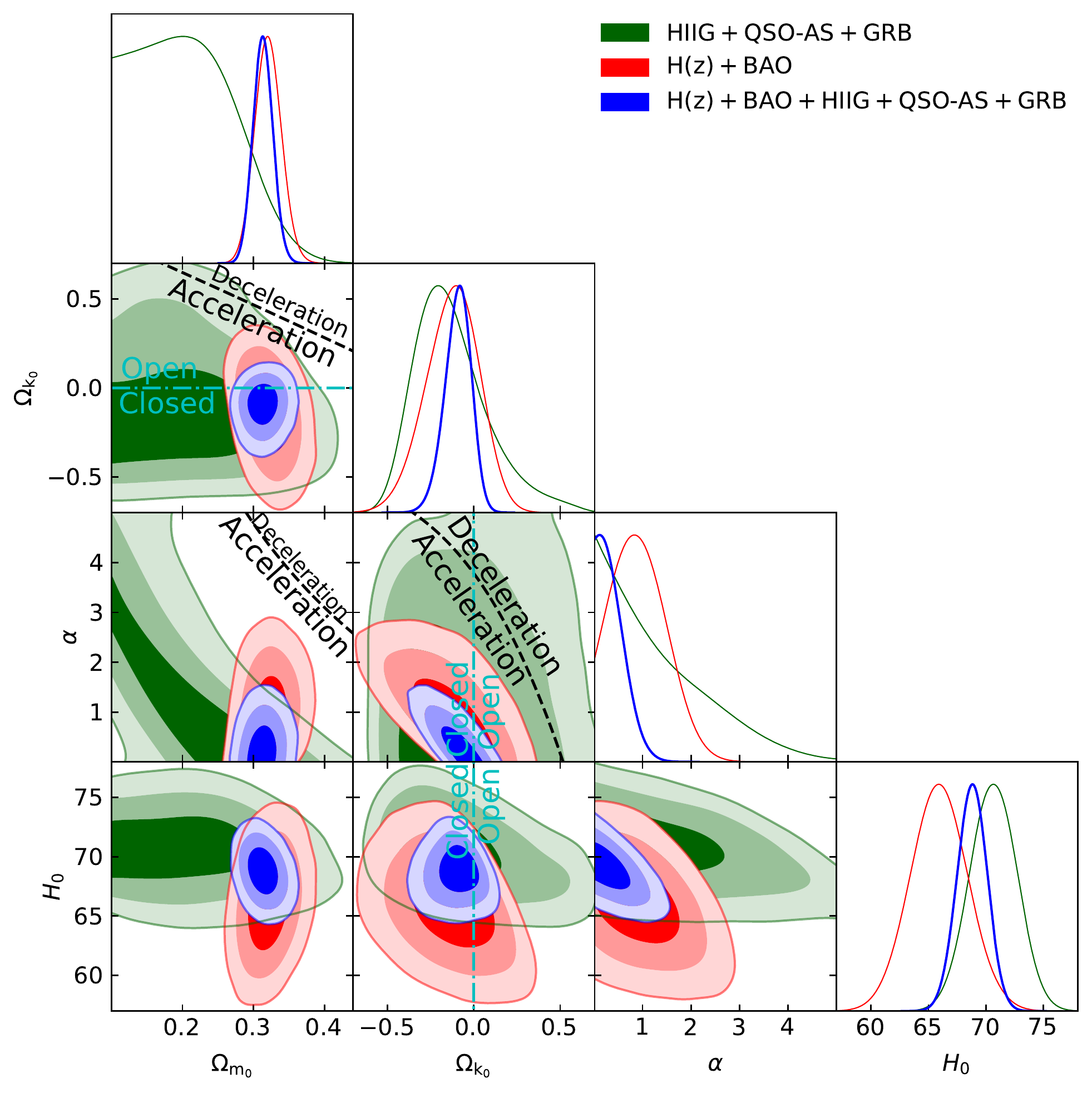}}\\
\caption{Same as Fig. \ref{fig6} (non-flat \pcdm) but for different combinations of data.}
\label{fig12}
\end{figure*}

The 1D probability distributions and 2D confidence regions of the cosmological and Amati relation parameters from the HQASG data are in Figs. \ref{fig1}--\ref{fig6}, in blue, Figs. \ref{fig7}--\ref{fig12}, in green, and panels (a) of Figs. \ref{fig04aA}--\ref{fig04aA}, in red. The best-fitting results and uncertainties are in Tables \ref{tab:BFPC4} and \ref{tab:1d_BFPC4}.

We find that the HQASG data combination favors currently-accelerating cosmological expansion.

The fit to the HQASG data produces best-fitting values of \om\ that lie between $0.205^{+0.044}_{-0.094}$ (non-flat \pcdm) at the low end, and $0.322^{+0.062}_{-0.044}$ (flat XCDM) at the high end. This range is smaller than the ranges within which \om\ falls when it is determined from the \hiig, QSO-AS, and GRB data individually, but the low and high ends of the range are still somewhat mutually inconsistent, being 2.66$\sigma$ away from each other. This is a consequence of the low \om\ value for non-flat \pcdm; the \om\ values for \lcdm\ and XCDM are quite consistent with the recent estimate of \cite{planck2018b}. In contrast, the best-fitting values of $H_0$ that we measure from the HQASG data are mutually very consistent (within $0.65\sigma$), with $H_0=70.30\pm1.68$ \hunit (flat \pcdm) at the low end of the range and $H_0=72.00^{+1.99}_{-1.98}$ \hunit (flat XCDM) at the high end of the range. These measurements are $0.83\sigma$ (flat XCDM) and $1.70\sigma$ (flat \pcdm) lower than the local Hubble constant measurement of $H_0 = 74.03 \pm 1.42$ \hunit \citep{riess_etal_2019}, and $0.70\sigma$ (flat \pcdm) and $1.16\sigma$ (flat XCDM) higher than the median statistics estimate of $H_0=68 \pm 2.8$ \hunit \citep{chenratmed}.\footnote{Other local expansion rate determinations have slightly lower central values with slightly larger error bars \citep{rigault_etal_2015,zhangetal2017,Dhawan,FernandezArenas,freedman_etal_2019,freedman_etal_2020,rameez_sarkar_2019,Breuvaletal_2020, Efstathiou_2020, Khetan_et_al_2021}. Our $H_0$ measurements are consistent with earlier median statistics estimates \citep{gott_etal_2001,chen_etal_2003} and with other recent $H_0$ determinations \citep{chen_etal_2017,DES_2018,Gomez-ValentAmendola2018, planck2018b, zhang_2018,dominguez_etal_2019,martinelli_tutusaus_2019,Cuceu_2019,zeng_yan_2019,schoneberg_etal_2019,lin_ishak_2021, Blum_et_al_2020, Lyu_et_al_2020, Philcox_et_al_2020, Zhang_Huang_2021, Birrer_et_al_2020, Denzel_et_al_2020}.}

In contrast to the \hiig, QSO-AS, and GRB only cases, when fitted to the HQASG data combination the non-flat models mildly favor closed spatial hypersurfaces. For non-flat \lcdm, non-flat XCDM, and non-flat \pcdm, we find $\Omega_{\rm k_0}=-0.093^{+0.092}_{-0.190}$, $\Omega_{\rm k_0}=-0.044^{+0.193}_{-0.217}$, and $\Omega_{\rm k_0}=-0.124^{+0.127}_{-0.253}$, respectively, with the non-flat \lcdm\ model favoring closed spatial hypersurfaces at 1.01$\sigma$.

The fit to the HQASG data combination produces stronger evidence for dark energy dynamics in the flat and non-flat XCDM parametrizations but weaker evidence in the flat and non-flat \pcdm\ models (in comparison to the \hiig\ and QSO-AS only cases) with tighter error bars on the measured values of $w_{\rm X}$ and $\alpha$. For flat (non-flat) XCDM, $w_{\rm X}=-1.379^{+0.361}_{-0.375}$ ($w_{\rm X}=-1.273^{+0.501}_{-0.321}$), with $w_{\rm X}=-1$ being within the 1$\sigma$ range for non-flat XCDM and being 1.05$\sigma$ larger for flat XCDM. For flat (non-flat) \pcdm, $\alpha<2.584$ ($\alpha<3.414$), where both likelihoods peak at $\alpha=0$.

The constraints on the Amati relation parameters from the HQASG data are also model-independent, but with slightly larger central values and smaller error bars for the parameter $a$. A reasonable summary is $\sigma_{\rm ext}=0.413^{+0.026}_{-0.032}$, $a=50.19\pm0.24$, and $b=1.133\pm0.086$.

The HQASG cosmological constraints are largely consistent with those from other data, like the constraints from the $H(z)$ + BAO data used in \cite{CaoRyanRatra2020} and \cite{KhadkaRatra2020c}, that are shown in red in Figs. \ref{fig7}--\ref{fig12}. We note, however, that there is some mild tension between \pcdm\ \om\ values, and between XCDM and \pcdm\ $H_0$ values determined from $H(z)$ + BAO and HQASG data, with the $2.46\sigma$ difference between \om\ values estimated from the two different data combinations in the non-flat \pcdm\ model being the only somewhat troubling difference (see Table \ref{tab:1d_BFPC4}).

\begin{table*}
\centering
\resizebox*{1\columnwidth}{1.2\columnwidth}{%
\begin{threeparttable} 
\caption{Unmarginalized best-fitting parameter values for all models from various combinations of data.}\label{tab:BFPC4}
\begin{tabular}{lcccccccccccccccccc}
\toprule
 Model & Data set & $\Omega_{\mathrm{m_0}}$ & $\Omega_{\Lambda}$ & $\Omega_{\mathrm{k_0}}$ & $w_{\mathrm{X}}$ & $\alpha$ & $H_0$\tnote{c} & $\sigma_{\mathrm{ext}}$ & $a$ & $b$ & $\chi^2$ & $\nu$ & $-2\ln\mathcal{L}_{\mathrm{max}}$ & $AIC$ & $BIC$ & $\Delta\chi^2$ & $\Delta AIC$ & $\Delta BIC$ \\
\midrule
Flat \lcdm & GRB & 0.698 & 0.302 & -- & -- & -- & 80.36 & 0.404 & 49.92& 1.113 & 117.98 & 114 & 130.12 & 140.12 & 154.01 & 1.08 & 0.00 & 0.00\\
 & \hiig\ & 0.276 & 0.724 & -- & -- & -- & 71.81 & -- & -- & -- & 410.75 & 151 & 410.75 & 414.75 & 420.81 & 3.15 & 0.00 & 0.00\\
 & QSO-AS & 0.315 & 0.685 & -- & -- & -- & 68.69 & -- & -- & -- & 352.05 & 118 & 352.05 & 356.05 & 361.62 & 1.76 & 0.00 & 0.00\\
 & HQASG\tnote{d} & 0.271 & 0.729 & -- & -- & -- & 71.13 & 0.407 & 50.18 & 1.138 & 879.42 & 387 & 895.05 & 905.05 & 924.91 & 0.12 & 0.00 & 0.00\\
 & $H(z)$ + BAO & 0.314 & 0.686 & -- & -- & -- & 68.53 & -- & -- & -- & 20.82 & 40 & 20.82 & 24.82 & 28.29 & 2.39 & 0.00 & 0.00\\
 & HzBHQASG\tnote{e} & 0.317 & 0.683 & -- & -- & -- & 69.06 & 0.404 & 50.19 & 1.134 & 903.61 & 429 & 917.79 & 927.79 & 948.16 & 4.05 & 0.00 & 0.00\\
\\
Non-flat \lcdm & GRB & 0.691 & 0.203 & 0.106 & -- & -- & 77.03 & 0.402 & 49.96 & 1.115 & 117.37 & 113 & 129.96 & 141.96 & 158.64 & 0.47 & 1.84 & 4.63\\
 & \hiig\ & 0.311 & 1.000 & $-0.311$ & -- & -- & 72.41 & -- & -- & -- & 410.44 & 150 & 410.44 & 416.44 & 425.53 & 2.84 & 1.69 & 4.72\\
 & QSO-AS & 0.266 & 1.000 & $-0.268$ & -- & -- & 74.73 & -- & -- & -- & 351.30 & 117 & 351.30 & 357.30 & 365.66 & 1.01 & 1.25 & 4.04\\
 & HQASG\tnote{d} & 0.291 & 0.876 & $-0.167$ & -- & -- & 72.00 & 0.406 & 50.22 & 1.120 & 879.30 & 386 & 894.02 & 906.02 & 929.85 & 0.00 & 0.97 & 4.94\\
 & $H(z)$ + BAO & 0.308 & 0.643 & 0.049 & -- & -- & 67.52 & -- & -- & -- & 20.52 & 39 & 20.52 & 26.52 & 31.73 & 2.09 & 1.70 & 3.44\\
 & HzBHQASG\tnote{e} & 0.309 & 0.716 & $-0.025$ & -- & -- & 69.77 & 0.402 & 50.17 & 1.141 & 904.47 & 428 & 917.17 & 929.17 & 953.61 & 4.91 & 1.38 & 5.45\\
\\
Flat XCDM & GRB & 0.102 & -- & -- & $-0.148$ & -- & 55.30 & 0.400 & 50.22 & 1.117 & 118.28 & 113 & 129.79 & 141.79 & 158.47 & 1.38 & 1.67 & 4.46\\
 & \hiig\ & 0.251 & -- & -- & $-0.899$ & -- & 71.66 & -- & -- & -- & 410.72 & 150 & 410.72 & 416.72 & 425.82 & 3.12 & 1.97 & 5.01\\
 & QSO-AS & 0.267 & -- & -- & $-2.000$ & -- & 81.70 & -- & -- & -- & 351.84 & 117 & 351.84 & 357.84 & 366.20 & 1.55 & 1.79 & 4.58\\
 & HQASG\tnote{d} & 0.320 & -- & -- & $-1.306$ & -- & 72.03 & 0.404 & 50.20 & 1.131 & 880.47 & 386 & 894.27 & 906.27 & 930.10 & 1.17 & 1.22 & 5.19\\
 & $H(z)$ + BAO & 0.319 & -- & -- & $-0.865$ & -- & 65.83 & -- & -- & -- & 19.54 & 39 & 19.54 & 25.54 & 30.76 & 1.11 & 0.72 & 2.47\\
 & HzBHQASG\tnote{e} & 0.313 & -- & -- & $-1.052$ & -- & 69.90 & 0.407 & 50.19 & 1.132 & 902.09 & 428 & 917.55 & 929.55 & 953.99 & 2.53 & 1.76 & 5.83\\
\\
Non-flat XCDM & GRB & 0.695 & -- & 0.556 & $-1.095$ & -- & 57.64 & 0.399 & 50.13 & 1.133 & 118.43 & 112 & 129.73 & 143.73 & 163.19 & 1.53 & 3.61 & 9.18\\
 & \hiig\ & 0.100 & -- & $-0.702$ & $-0.655$ & -- & 72.57 & -- & -- & -- & 407.60 & 149 & 407.60 & 415.60 & 427.72 & 0.00 & 0.85 & 6.91\\
 & QSO-AS & 0.100 & -- & $-0.548$ & $-0.670$ & -- & 74.04 & -- & -- & -- & 350.29 & 116 & 350.29 & 358.29 & 369.44 & 0.00 & 2.24 & 7.82\\
 & HQASG\tnote{d} & 0.300 & -- & $-0.161$ & $-1.027$ & -- & 80.36 & 0.405 & 50.21 & 1.122 & 879.48 & 385 & 894.01 & 908.01 & 935.81 & 0.18 & 2.96 & 10.90\\
 & $H(z)$ + BAO & 0.327 & -- & $-0.159$ & $-0.730$ & -- & 65.97 & -- & -- & -- & 18.43 & 38 & 18.43 & 26.43 & 33.38 & 0.00 & 1.61 & 5.09\\
 & HzBHQASG\tnote{e} & 0.312 & -- & $-0.045$ & $-0.959$ & -- & 69.46 & 0.402 & 50.23 & 1.117 & 904.17 & 427 & 917.07 & 931.07 & 959.58 & 4.61 & 3.28 & 11.42\\
\\
Flat $\phi$CDM & GRB & 0.674 & -- & -- & -- & 2.535 & 84.00 & 0.399 & 49.88 & 1.104 & 119.15 & 113 & 130.14 & 142.14 & 158.82 & 2.25 & 2.02 & 4.81\\
 & \hiig\ & 0.255 & -- & -- & -- & 0.260 & 71.70 & -- & -- & -- & 410.70 & 150 & 410.70 & 416.70 & 425.80 & 3.10 & 1.95 & 4.99\\
 & QSO-AS & 0.319 & -- & -- & -- & 0.012 & 68.47 & -- & -- & -- & 352.05 & 117 & 352.05 & 358.05 & 366.41 & 1.76 & 2.00 & 4.79\\
 & HQASG\tnote{d} & 0.282 & -- & -- & -- & 0.012 & 70.81 & 0.402 & 50.19 & 1.135 & 882.56 & 386 & 895.28 & 907.28 & 931.11 & 3.26 & 2.23 & 6.20\\
 & $H(z)$ + BAO & 0.318 & -- & -- & -- & 0.364 & 66.04 & -- & -- & -- & 19.65 & 39 & 19.65 & 25.65 & 30.86 & 1.22 & 0.83 & 2.57\\
 & HzBHQASG\tnote{e} & 0.316 & -- & -- & -- & 0.013 & 69.15 & 0.405 & 50.24 & 1.114 & 903.52 & 428 & 918.12 & 930.12 & 954.56 & 3.96 & 2.33 & 6.40\\
\\
Non-flat $\phi$CDM & GRB & 0.664 & -- & 0.188 & -- & 4.269 & 59.65 & 0.403 & 50.17 & 1.111 & 116.90 & 112 & 129.93 & 143.93 & 163.39 & 0.00 & 3.81 & 9.38\\
 & \hiig\ & 0.114 & -- & $-0.437$ & -- & 2.680 & 72.14 & -- & -- & -- & 409.91 & 149 & 409.91 & 417.91 & 430.03 & 2.31 & 3.16 & 9.22\\
 & QSO-AS & 0.100 & -- & $-0.433$ & -- & 2.948 & 72.37 & -- & -- & -- & 350.98 & 116 & 350.98 & 358.98 & 370.13 & 0.69 & 2.93 & 8.51\\
 & HQASG\tnote{d} & 0.276 & -- & $-0.185$ & -- & $0.145$ & 72.11 & 0.402 & 50.16 & 1.142 & 881.09 & 385 & 894.24 & 908.24 & 936.03 & 1.79 & 3.19 & 11.12\\
 & $H(z)$ + BAO & 0.321 & -- & $-0.137$ & -- & 0.887 & 66.41 & -- & -- & -- & 18.61 & 39 & 18.61 & 26.61 & 33.56 & 0.18 & 1.79 & 5.27\\
 & HzBHQASG\tnote{e} & 0.310 & -- & $-0.052$ & -- & 0.193 & 69.06 & 0.411 & 50.21 & 1.126 & 899.56 & 427 & 917.26 & 931.26 & 959.77 & 0.00 & 3.47 & 11.61\\
\bottomrule
\end{tabular}
\begin{tablenotes}[flushleft]
\item [c] \hunit.
\item [d] \hiig\ + QSO-AS + GRB.
\item [e] $H(z)$ + BAO + \hiig\ + QSO-AS + GRB.
\end{tablenotes}
\end{threeparttable}%
}
\end{table*}

\subsection{$H(z)$, BAO, \hiig, QSO-AS, and GRB (HzBHQASG) constraints}
\label{subsec:HzBHQASG}

Given the good mutual consistency between constraints derived from $H(z)$ + BAO data and those derived from HQASG data, in this subsection we determine more restrictive joint constraints from the combined $H(z)$, BAO, \hiig, QSO-AS, and GRB (HzBHQASG) data on the parameters of our six cosmological models.

The 1D probability distributions and 2D confidence regions of the cosmological and Amati relation parameters for all models from the HzBHQASG data are in blue in Figs. \ref{fig7}--\ref{fig12}, and in red in panels (b) of Figs. \ref{fig04aA}--\ref{fig04aA}. The best-fitting results and uncertainties are in Tables \ref{tab:BFPC4} and \ref{tab:1d_BFPC4}.

The measured values of \om\ here are a little larger, and significantly more restrictively constrained, than the ones in the HQASG cases (except for flat XCDM), being between $0.310\pm0.014$ (non-flat XCDM) and $0.320\pm0.013$ (flat \pcdm). The $H_0$ measurements are a little lower, and more tightly constrained, than in the HQASG cases, and are in better agreement with the lower median statistics estimate of $H_0$ \citep{chenratmed} than the higher local expansion rate measurement of $H_0$ \citep{riess_etal_2019}, being between $68.16^{+1.01}_{-0.80}$ \hunit (flat \pcdm) and $69.85^{+1.42}_{-1.55}$ \hunit (flat XCDM).

For non-flat \lcdm, non-flat XCDM, and non-flat \pcdm, we measure $\Omega_{\rm k_0}=-0.019^{+0.043}_{-0.048}$, $\Omega_{\rm k_0}=-0.024^{+0.092}_{-0.093}$, and $\Omega_{\rm k_0}=-0.094^{+0.082}_{-0.064}$, respectively, where the central values are a little higher (closer to 0) than what was measured in the HQASG cases. The joint constraints are more restrictive, with non-flat \lcdm\ and XCDM within 0.44$\sigma$ and 0.26$\sigma$ of $\Omega_{\rm k_0} = 0$, respectively. The non-flat \pcdm\ model, on the other hand, still favors a closed geometry with an $\Omega_{\rm k_0}$ that is 1.15$\sigma$ away from zero.

The HzBHQASG case has slightly larger measured values and tighter error bars for $w_{\rm X}$ and $\alpha$ than the HQASG case, so there is also not much evidence in support of dark energy dynamics. For flat (non-flat) XCDM, $w_{\rm X}=-1.050^{+0.090}_{-0.081}$ ($w_{\rm X}=-1.019^{+0.202}_{-0.099}$). For flat (non-flat) \pcdm, the $2\sigma$ upper limits are $\alpha<0.418$ ($\alpha<0.905$).

The cosmological model-independent constraints from the HzBHQASG data combination on the parameters of the Amati relation can be summarized as $\sigma_{\rm ext}=0.412^{+0.026}_{-0.032}$, $a=50.19\pm0.24$, and $b=1.132\pm0.085$.

\subsection{Model comparison}
\label{subsec:comparison}

From Table \ref{tab:BFPC4}, we see that the reduced $\chi^2$ values determined from GRB data alone are around unity for all models (being between 1.03 and 1.06) while those values determined from the $H(z)$ + BAO data combination range from 0.48 to 0.53, with the lower reduced $\chi^2$ here being due to the $H(z)$ data (that probably have overestimated error bars). As discussed in \cite{Ryanetal2019} and \cite{CaoRyanRatra2020}, the cases that involve \hiig\ and QSO-AS data have a larger reduced $\chi^2$ (between 2.11 and 3.02), which is probably due to underestimated systematic uncertainties in both cases.

Based on the $AIC$ and the $BIC$ (see Table \ref{tab:BFPC4}), the flat \lcdm\ model remains the most favored model, across all data combinations, among the six models we study.\footnote{Note that based on the $\Delta \chi^2$ results of Table \ref{tab:BFPC4} non-flat \lcdm\ has the minimum $\chi^2$ in the HQASG case and non-flat XCDM has the minimum $\chi^2$ in the \hiig, QSO-AS, and $H(z)$ + BAO cases, whereas non-flat \pcdm\ has the minimum $\chi^2$ for the GRB and HzBHQASG cases. The $\Delta \chi^2$ values do not, however, penalize a model for having more parameters.} From $\Delta AIC$ and $\Delta BIC$, we find mostly weak or positive evidence against the models we considered, and only in a few cases do we find strong evidence against them. According to $\Delta BIC$, the evidence against non-flat XCDM is strong for the \hiig, QSO-AS, and GRB only cases, and very strong for the HQASG and HzBHQASG cases. Similarly, the evidence against flat \pcdm\ is strong for the HQASG and HzBHQASG cases, and the evidence against non-flat \pcdm\ is strong for the \hiig, QSO-AS, and GRB only cases, and very strong for the HQASG and HzBHQASG cases.

Among these six models, a comparison of the $\Delta BIC$ values from Table \ref{tab:BFPC4} shows that the most disfavored model is non-flat \pcdm, and that the second most disfavored model is non-flat XCDM. This is especially true when these models are fitted to the HQASG and HzBHQASG data combinations, in which cases non-flat \pcdm\ and non-flat XCDM are very strongly disfavored. These models aren't as strongly disfavored by the $AIC$, however; from a comparison of the $\Delta AIC$ values in Table \ref{tab:BFPC4}, we see that the evidence against the most disfavored model (non-flat \pcdm) is only positive.

\begin{table*}
\centering
\resizebox*{1\columnwidth}{1.2\columnwidth}{%
\begin{threeparttable}
\caption{One-dimensional marginalized best-fitting parameter values and uncertainties ($\pm 1\sigma$ error bars or $2\sigma$ limits) for all models from various combinations of data.}\label{tab:1d_BFPC4}
\begin{tabular}{lcccccccccc}
\toprule
 Model & Data set & $\Omega_{\mathrm{m_0}}$ & $\Omega_{\Lambda}$ & $\Omega_{\mathrm{k_0}}$ & $w_{\mathrm{X}}$ & $\alpha$ & $H_0$\tnote{c} & $\sigma_{\mathrm{ext}}$ & $a$ & $b$ \\
\midrule
Flat \lcdm & GRB & $>0.208$ & -- & -- & -- & -- & -- & $0.411^{+0.026}_{-0.032}$ & $50.16\pm0.27$ & $1.123\pm0.085$\\
 & \hiig\ & $0.289^{+0.053}_{-0.071}$ & -- & -- & -- & -- & $71.70\pm1.83$ & -- & -- & -- \\
 & QSO-AS & $0.364^{+0.083}_{-0.150}$ & -- & -- & -- & -- & $67.29^{+4.93}_{-5.07}$ & -- & -- & -- \\
 & HQASG\tnote{e} & $0.277^{+0.034}_{-0.041}$ & -- & -- & -- & -- & $71.03\pm1.67$ & $0.413^{+0.026}_{-0.032}$ & $50.19\pm0.24$ & $1.138\pm0.085$\\
 & $H(z)$ + BAO & $0.315^{+0.015}_{-0.017}$ & -- & -- & -- & -- & $68.55\pm0.87$ & -- & -- & -- \\
 & HzBHQASG\tnote{f} & $0.316\pm0.013$ & -- & -- & -- & -- & $69.05^{+0.62}_{-0.63}$ & $0.412^{+0.026}_{-0.032}$ & $50.19\pm0.23$ & $1.133\pm0.085$\\
\\
Non-flat \lcdm & GRB & $0.463^{+0.226}_{-0.084}$ & $<0.658$\tnote{d} & $-0.007^{+0.251}_{-0.234}$ & -- & -- & -- & $0.412^{+0.026}_{-0.032}$ & $50.17\pm0.28$ & $1.121\pm0.086$\\
 & \hiig\ & $0.275^{+0.081}_{-0.078}$ & $>0.501$\tnote{d} & $0.094^{+0.237}_{-0.363}$ & -- & -- & $71.50^{+1.80}_{-1.81}$ & -- & -- & -- \\
 & QSO-AS & $0.357^{+0.082}_{-0.135}$ & -- & $0.017^{+0.184}_{-0.277}$ & -- & -- & $67.32^{+4.49}_{-5.44}$ & -- & -- & -- \\
 & HQASG\tnote{e} & $0.292\pm0.044$ & $0.801^{+0.191}_{-0.055}$ & $-0.093^{+0.092}_{-0.190}$ & -- & -- & $71.33^{+1.75}_{-1.77}$ & $0.413^{+0.026}_{-0.032}$ & $50.19\pm0.24$ & $1.130\pm0.086$\\
 & $H(z)$ + BAO & $0.309\pm0.016$ & $0.636^{+0.081}_{-0.072}$ & $0.055^{+0.082}_{-0.074}$ & -- & -- & $67.44\pm2.33$ & -- & -- & -- \\
 & HzBHQASG\tnote{f} & $0.311^{+0.012}_{-0.014}$ & $0.708^{+0.053}_{-0.046}$ & $-0.019^{+0.043}_{-0.048}$ & -- & -- & $69.72\pm1.10$ & $0.412^{+0.026}_{-0.032}$ & $50.19\pm0.23$ & $1.132\pm0.085$ \\
\\
Flat XCDM & GRB & $>0.366$\tnote{d} & -- & -- & -- & -- & -- & $0.411^{+0.025}_{-0.032}$ & $50.14\pm0.28$ & $1.119\pm0.085$\\
 & \hiig\ & $0.300^{+0.106}_{-0.083}$ & -- & -- & $-1.180^{+0.560}_{-0.330}$ & -- & $71.85\pm1.96$ & -- & -- & -- \\
 & QSO-AS & $0.349^{+0.090}_{-0.143}$ & -- & -- & $-1.161^{+0.430}_{-0.679}$ & -- & $68.39^{+6.14}_{-8.98}$ & -- & -- & -- \\
 & HQASG\tnote{e} & $0.322^{+0.062}_{-0.044}$ & -- & -- & $-1.379^{+0.361}_{-0.375}$ & -- & $72.00^{+1.99}_{-1.98}$ & $0.412^{+0.026}_{-0.032}$ & $50.20\pm0.24$ & $1.130\pm0.085$\\
 & $H(z)$ + BAO & $0.319^{+0.016}_{-0.017}$ & -- & -- & $-0.888^{+0.126}_{-0.098}$ & -- & $66.26^{+2.32}_{-2.63}$ & -- & -- & -- \\
 & HzBHQASG\tnote{f} & $0.313^{+0.014}_{-0.015}$ & -- & -- & $-1.050^{+0.090}_{-0.081}$ & -- & $69.85^{+1.42}_{-1.55}$ & $0.412^{+0.026}_{-0.032}$ & $50.19\pm0.24$ & $1.134\pm0.085$ \\
\\
Non-flat XCDM & GRB & $>0.386$\tnote{d} & -- & $0.121^{+0.464}_{-0.275}$ & $>-1.218$\tnote{d} & -- & -- & $0.411^{+0.026}_{-0.032}$ & $50.12\pm0.28$ & $1.122\pm0.087$\\
 & \hiig\ & $0.275^{+0.084}_{-0.125}$ & -- & $0.011^{+0.457}_{-0.460}$ & $-1.125^{+0.537}_{-0.321}$ & -- & $71.71^{+2.07}_{-2.08}$ & -- & -- & -- \\
 & QSO-AS & $0.359^{+0.111}_{-0.174}$ & -- & $0.115^{+0.466}_{-0.293}$ & $-1.030^{+0.593}_{-0.548}$ & -- & $65.92^{+4.54}_{-9.54}$ & -- & -- & -- \\
 & HQASG\tnote{e} & $0.303^{+0.073}_{-0.058}$ & -- & $-0.044^{+0.193}_{-0.217}$ & $-1.273^{+0.501}_{-0.321}$ & -- & $71.77\pm2.02$ & $0.413^{+0.026}_{-0.031}$ & $50.20\pm0.24$ & $1.129\pm0.085$\\
 & $H(z)$ + BAO & $0.323^{+0.021}_{-0.020}$ & -- & $-0.105^{+0.187}_{-0.162}$ & $-0.818^{+0.212}_{-0.071}$ & -- & $66.20^{+2.29}_{-2.55}$ & -- & -- & -- \\
 & HzBHQASG\tnote{f} & $0.310\pm0.014$ & -- & $-0.024^{+0.092}_{-0.093}$ & $-1.019^{+0.202}_{-0.099}$ & -- & $69.63^{+1.45}_{-1.62}$ & $0.412^{+0.026}_{-0.031}$ & $50.19\pm0.23$ & $1.132\pm0.085$ \\
\\
Flat $\phi$CDM & GRB & $>0.376$\tnote{d} & -- & -- & -- & -- & -- & $0.411^{+0.025}_{-0.032}$ & $50.13\pm0.28$ & $1.121\pm0.087$\\
 & \hiig\ & $0.210^{+0.043}_{-0.092}$ & -- & -- & -- & $<2.784$ & $71.23^{+1.79}_{-1.80}$ & -- & -- & -- \\
 & QSO-AS & $0.329^{+0.086}_{-0.171}$ & -- & -- & -- & $<2.841$ & $64.42^{+4.47}_{-4.62}$ & -- & -- & -- \\
 & HQASG\tnote{e} & $0.214^{+0.057}_{-0.061}$ & -- & -- & -- & $<2.584$ & $70.30\pm1.68$ & $0.413^{+0.026}_{-0.032}$ & $50.18\pm0.24$ & $1.142\pm0.087$\\
 & $H(z)$ + BAO & $0.319^{+0.016}_{-0.017}$ & -- & -- & -- & $0.550^{+0.169}_{-0.494}$ & $65.25^{+2.25}_{-1.82}$ & -- & -- & -- \\
 & HzBHQASG\tnote{f} & $0.320\pm0.013$ & -- & -- & -- & $<0.418$ & $68.16^{+1.01}_{-0.80}$ & $0.412^{+0.027}_{-0.033}$ & $50.20\pm0.24$ & $1.131\pm0.088$ \\
\\
Non-flat $\phi$CDM & GRB & $>0.189$ & -- & $0.251^{+0.247}_{-0.086}$ & -- & -- & -- & $0.411^{+0.026}_{-0.032}$ & $50.11\pm0.28$ & $1.128\pm0.089$\\
 & \hiig\ & $<0.321$ & -- & $0.291^{+0.348}_{-0.113}$ & -- & $<4.590$ & $70.60^{+1.68}_{-1.84}$ & -- & -- & -- \\
 & QSO-AS & $0.362^{+0.117}_{-0.193}$ & -- & $0.254^{+0.304}_{-0.092}$ & -- & $<4.752$ & $61.91^{+2.83}_{-4.92}$ & -- & -- & -- \\
 & HQASG\tnote{e} & $0.205^{+0.044}_{-0.094}$ & -- & $-0.124^{+0.127}_{-0.253}$ & -- & $<3.414$ & $70.66\pm1.90$ & $0.414^{+0.027}_{-0.033}$ & $50.19\pm0.24$ & $1.134\pm0.088$\\
 & $H(z)$ + BAO & $0.321\pm0.017$ & -- & $-0.126^{+0.157}_{-0.130}$ & -- & $0.938^{+0.439}_{-0.644}$ & $65.93\pm2.33$ & -- & -- & -- \\
 & HzBHQASG\tnote{f} & $0.313\pm0.013$ & -- & $-0.094^{+0.082}_{-0.064}$ & -- & $<0.905$ & $68.79\pm1.22$ & $0.412^{+0.027}_{-0.033}$ & $50.20\pm0.24$ & $1.126\pm0.087$ \\
\bottomrule
\end{tabular}
\begin{tablenotes}[flushleft]
\item [c] \hunit.
\item [d] This is the 1$\sigma$ limit. The $2\sigma$ limit is set by the prior, and is not shown here.
\item [e] \hiig\ + QSO-AS + GRB.
\item [f] $H(z)$ + BAO + \hiig\ + QSO-AS + GRB.
\end{tablenotes}
\end{threeparttable}%
}
\end{table*}

\section{Conclusion}
\label{makereference4.6}
We find that cosmological constraints determined from higher-$z$ GRB, \hiig, and QSO-AS data are mutually consistent. It is both reassuring and noteworthy that these higher-$z$ data jointly favor currently-accelerating cosmological expansion, and that their constraints are consistent with the constraints imposed by more widely used and more restrictive $H(z)$ and BAO data. Using a data set consisting of 31 $H(z)$, 11 BAO, 120 QSO-AS, 153 \hiig, and 119 GRB measurements, we jointly constrain the parameters of the GRB Amati relation and of six cosmological models. 

The GRB measurements are of special interest because they reach to $z\sim8.2$ (far beyond the highest $z\sim2.3$ reached by BAO data) and into a much less studied area of redshift space. Current GRB data do not provide very restrictive constraints on cosmological model parameters, but in the near future we expect there to be more GRB observations \citep{Shirokov2020} which should improve the GRB data and provide more restrictive cosmological constraints.

Some of our conclusions do not differ significantly between models and so are model-independent. In particular, for the HzBHQASG data (the full data set excluding QSO-Flux data), we find a fairly restrictive summary value of $\Omega_{\rm m_0}=0.313 \pm 0.013$ that agrees well with many other recent measurements. From these data we also find a fairly restrictive summary value of $H_0=69.3 \pm 1.2$ \hunit\ that is in better agreement with the results of \cite{chenratmed} and \cite{planck2018b} than with the result of \cite{riess_etal_2019}; note that we do not take the $H_0$ tension issue into account (for a review, see \citealp{riess_2019}). The HzBHQASG measurements are consistent with flat \lcdm, but do not rule out mild dark energy dynamics or a little spatial curvature energy density. More and better-quality higher-$z$ GRB, \hiig, QSO, and other data will significantly help to test these extensions of flat \lcdm.


\cleardoublepage


\chapter{Using Pantheon and DES supernova, baryon acoustic oscillation, and Hubble parameter data to constrain the Hubble constant, dark energy dynamics, and spatial curvature}
\label{makereference5}

This chapter is based on \cite{CaoRyanRatra2021}. Figures and tables by Shulei Cao, from analyses
conducted independently by Shulei Cao and Joseph Ryan.

\section{Introduction} 
\label{makereference5.1}

That the Universe is currently in a phase of accelerated expansion is well-supported by observations but not fully explained by fundamental theory (see e.g. \citealp{Ratra_Vogeley,Martin,Coley_Ellis}). The standard spatially flat \lcdm\ model \citep{peeb84} interprets this phenomenon as a consequence of dark energy with negative pressure (a cosmological constant, $\Lambda$) and requires the major part of the energy budget of the Universe to consist of time-independent dark energy and cold dark matter (CDM). Although flat \lcdm\ is consistent with many observations (see e.g. \citealp{Farooq_Ranjeet_Crandall_Ratra_2017,scolnic_et_al_2018,planck2018b,eBOSS_2020}),\footnote{Note that the \textit{Planck} TT,TE,EE+lowE+lensing data favor positive spatial curvature \citep{planck2018b} but are consistent with a spatially flat model at $1.63\sigma$.} there exist some potential observational discrepancies \citep{riess_2019, martinelli_tutusaus_2019} and theoretical puzzles (e.g., \citealp{Martin}), which leaves room for other cosmological models, including non-flat \lcdm. As the quality and quantity of observational data grow in time, constraining these models is within reach. Many workers have investigated the merits of the flat and non-flat XCDM parametrizations and \pcdm\ models, where dark energy dynamics and spatial curvature come into play.\footnote{For observational constraints on spatial curvature see \cite{Farooq_Mania_Ratra_2015}, \cite{Chen_et_al_2016}, \cite{rana_jain_mahajan_mukherjee_2017}, \cite{ooba_etal_2018a, ooba_etal_2018b, ooba_etal_2018c}, \cite{Yuetal2018}, \cite{park_ratra_2018, park_ratra_2019a, ParkRatra2019a, park_ratra_2020}, \cite{wei_2018}, \cite{DES_2019}, \cite{handley_2019a}, \cite{jesus_etal_2019}, \cite{li_etal_2020}, \cite{geng_etal_2020}, \cite{kumar_etal_2020}, \cite{efstathiou_gratton_2020}, \cite{DiValentinoetal2021a}, \cite{DiValentinoetal2021b}, \cite{gao_etal_2020}, \cite{Abbassi_2020}, \cite{Yang_2020}, \cite{Agudelo_Ruiz_2020}, \cite{Velasquez-Toribio_2020}, \cite{Vagnozzi_2020a,Vagnozzi_2020b}, and references therein.}\footnote{For observational constraints on the \pcdm\ model see \cite{yashar_et_al_2009}, \cite{Samushia_2010}, \cite{Campanelli_etal_2012}, \cite{Avsajanishvili_2015}, \cite{Sola_etal_2017}, \cite{Sola_perez_gomez_2018,SolaPercaulaetal2019}, \cite{Zhaietal2017}, \cite{ooba_etal_2018b,ooba_etal_2019}, \cite{sangwan_tripathi_jassal_2018}, \cite{Singhetal2019}, \cite{KhadkaRatra2020a,KhadkaRatra2020b,KhadkaRatra2020c,KhadkaRatra2021}, \cite{UrenaLopezRoy2020}, and references therein.}

Many observational data sets have been used to place constraints on the parameters of cosmological models, such as the equation of state parameter ($w$) of dark energy. Most recently, in \cite{CaoRyanRatra2020}, we used Hubble parameter ($H(z)$), baryon acoustic oscillation (BAO), quasar angular size (QSO), quasar X-ray and UV flux, \hii\ starburst galaxy (\hiig), and gamma-ray burst (GRB) data to constrain this parameter (among others). The tightest constraints on $w$, we found, come from low-redshift $H(z)$ (cosmic chronometer) and BAO (standard ruler) data, with the standard candle data (\hiig\ and GRB) giving very broad constraints. In this paper we combine measurements of the distances to 1255 Type Ia supernovae (SNe Ia) with our set of $H(z)$ and BAO data (along with QSO and \hiig\ observations) to obtain tight cosmological parameter constraints.

The usefulness of SN Ia data to cosmology is well-known. SN Ia measurements revealed the accelerated expansion of the Universe over twenty years ago, and they are employed today to place constraints on cosmological parameters and to break parameter degeneracies. Over this time period, the sample size of SN Ia distance measurements has grown considerably, and the analysis and mitigation of systematic uncertainties has improved \citep{DES_2019c, DES_2019d}. Supernovae are therefore a reasonably empirically well-understood cosmological probe\footnote{Though the relatively simpler physics underlying cosmic microwave background (CMB) anisotropies and BAO makes those probes better understood than SNe Ia.}, and so can be used to obtain reliable constraints on cosmological model parameters.

In our earlier studies that made use of BAO data (e.g., \citealp{Ryanetal2019, CaoRyanRatra2020}), we relied on CMB-derived values of the baryon density\footnote{Here $\Omega_{\rm b_0}$ is the baryon density parameter and $h=H_0/(100\ \rm{km \ s^{-1} \ Mpc^{-1}})$.} $\obh$ in order to compute the size of the sound horizon $r_{s}$. The size of the sound horizon is needed to calibrate the BAO scale (see Table \ref{tab:BAOC5}), so the constraints we derived from our BAO measurements were indirectly dependent on CMB physics. \cite{park_ratra_2018, park_ratra_2019a, ParkRatra2019a} computed $\obh$ within each of the six models we study (namely flat/non-flat \lcdm, flat/non-flat XCDM, and flat/non-flat \pcdm) from CMB data using primordial energy density fluctuation power spectra $P(k)$ appropriate for flat and curved geometries \citep{Lucchin_1985, ratra_1989,ratra_2017,ratra_peebles_1995}. Other power spectra have been considered in the non-flat case \citep{Lesgourgues_2014,Bonga_2016,handley_2019b,Thavanesan_2021}. Since we do not make use of $P(k)$, the controversy associated with $P(k)$ in non-flat models is avoided in our analyses here.

The constraints from $H(z)$ + BAO data and from SN Ia data are not inconsistent, and so these data can be jointly used to constrain cosmological parameters. \cite{park_ratra_2019b} used $H(z)$, BAO, and Pantheon SN Ia apparent magnitude (SN-Pantheon) measurements in such a joint analysis. Here we use a more recent BAO data compilation and new DES-3yr binned SN Ia apparent magnitude (SN-DES) data. We find for all combinations of data we study here that all or almost all of the favored parameter space corresponds to currently accelerating cosmological expansion. The most reliable constraints come from the $H(z)$ + BAO + SN-Pantheon + SN-DES (HzBSNPD) data combination, with fairly model-independent determinations of the Hubble constant, $H_0=68.8\pm1.8\ \rm{km \ s^{-1} \ Mpc^{-1}}$, and the non-relativistic matter density parameter, $\Omega_{\rm m_0}=0.294\pm0.020$. The estimate of $H_0$ is in better agreement with the median statistics $H_0 = 68 \pm 2.8$ \hunit\ estimate of \cite{chenratmed} and the \cite{planck2018b} estimate of $H_0 = 67.4 \pm 0.5$ \hunit\ than with the local $H_0 = 74.03 \pm 1.42$ \hunit\ measurement of \cite{riess_etal_2019}. The combined measurements are consistent with the spatially flat \lcdm\ model, but also favor some dark energy dynamics, as well as a little non-zero spatial curvature energy density. More restrictive constraints are derived when these data are combined with QSO and \hiig\ data.

This paper is organized as follows. We use the cosmological models described in Chapter \ref{sec:models}. In Section \ref{makereference5.2} the data used are introduced and our method of analyzing these data is described in Section \ref{makereference5.3}. We present our results in Section \ref{makereference5.4}, and our conclusions in Section \ref{makereference5.5}.

\section{Data}
\label{makereference5.2}

In this paper, we use a combination of $H(z)$, BAO, SN-Pantheon, SN-DES, QSO, and \hiig\ data to constrain the cosmological models we study. 

The $H(z)$ data, compiled in Table 2 of \cite{Ryan_1}, consist of 31 measurements spanning the redshift range $0.070 \leq z \leq 1.965$. The BAO data, which have been updated relative to \cite{CaoRyanRatra2020}, consist of 11 measurements spanning the redshift range $0.38 \leq z \leq 2.334$, listed in Table \ref{tab:BAOC5}. 

The SN-Pantheon data, compiled by \cite{scolnic_et_al_2018}, consist of 1048 SN Ia measurements spanning the redshift range $0.01<z<2.3$. The SN-DES data, compiled by \cite{DES_2019d}, consist of 20 binned measurements of 207 SN Ia measurements spanning the redshift range $0.015 \leq z \leq 0.7026$.

The QSO data, listed in Table 1 of \cite{Cao_et_al2017b}, consist of 120 measurements of the angular size
\begin{equation}
    \theta(z) = \frac{l_{\rm m}}{D_{A}(z)},
\end{equation}
spanning the redshift range $0.462 \leq z \leq 2.73$. $l_{\rm m}$ is the characteristic linear size of the quasars in the sample. This quantity is determined by using the Gaussian Process method to reconstruct the expansion history of the Universe from 24 cosmic chronometer measurements over $z < 1.2$. This $H(z)$ function is used to reconstruct the angular size distance $D_{A}(z)$, which can then be used to compute $l_{\rm m}$ given measurements $(\theta_{\rm obs}(z)$) of quasar angular sizes. QSO and $H(z)$ data are therefore somewhat correlated, but the error bars on the constraints derived from QSO data are so large that we do not believe this correlation to be an issue. 

The \hiig\ data consist of 107 low redshift ($0.0088 \leq z \leq 0.16417$) measurements, used in \cite{Chavez_2014} (recalibrated by \citealp{GonzalezMoran2019}), and 46 high redshift ($0.636427 \leq z \leq 2.42935$) measurements.

\begin{table}
\centering
\begin{threeparttable}
\caption{BAO data.}\label{tab:BAOC5}
\setlength{\tabcolsep}{3.5pt}
\begin{tabular}{lccc}
\toprule
$z$ & Measurement\tnote{a} & Value & Ref.\\
\midrule
$0.38$ & $D_M\left(r_{s,{\rm fid}}/r_s\right)$ & 1512.39 & \cite{Alam_et_al_2017}\tnote{b}\\
$0.38$ & $H(z)\left(r_s/r_{s,{\rm fid}}\right)$ & 81.2087 & \cite{Alam_et_al_2017}\tnote{b}\\
$0.51$ & $D_M\left(r_{s,{\rm fid}}/r_s\right)$ & 1975.22 & \cite{Alam_et_al_2017}\tnote{b}\\
$0.51$ & $H(z)\left(r_s/r_{s,{\rm fid}}\right)$ & 90.9029 & \cite{Alam_et_al_2017}\tnote{b}\\
$0.61$ & $D_M\left(r_{s,{\rm fid}}/r_s\right)$ & 2306.68 & \cite{Alam_et_al_2017}\tnote{b}\\
$0.61$ & $H(z)\left(r_s/r_{s,{\rm fid}}\right)$ & 98.9647 & \cite{Alam_et_al_2017}\tnote{b}\\
$0.122$ & $D_V\left(r_{s,{\rm fid}}/r_s\right)$ & $539\pm17$ & \cite{Carter_2018}\\
$0.81$ & $D_A/r_s$ & $10.75\pm0.43$ & \cite{DES_2019b}\\
$1.52$ & $D_V\left(r_{s,{\rm fid}}/r_s\right)$ & $3843\pm147$ & \cite{3}\\
$2.334$ & $D_M/r_s$ & 37.5 & \cite{duMas2020}\tnote{c}\\
$2.334$ & $D_H/r_s$ & 8.99 & \cite{duMas2020}\tnote{c}\\
\bottomrule
\end{tabular}
\begin{tablenotes}[flushleft]
\item[a] $D_M$, $D_V$, $r_s$, $r_{s, {\rm fid}}$, $D_A$, and $D_H$ have units of Mpc, while $H(z)$ has units of \hunit.
\item[b] The six measurements from \cite{Alam_et_al_2017} are correlated; see equation (20) of \cite{Ryanetal2019} for their correlation matrix.
\item[c] The two measurements from \cite{duMas2020} are correlated; see equation \eqref{CovM1} below for their correlation matrix.
\end{tablenotes}
\end{threeparttable}
\end{table}

The covariance matrix $\textbf{C}$ for the BAO data, taken from \cite{Alam_et_al_2017}, is given in equation (20) of \cite{Ryanetal2019}. For the BAO data from \cite{duMas2020}, the covariance matrix is
\be
\label{CovM1}
    \textbf{C}=
    \begin{bmatrix}
    1.3225 & -0.1009 \\
    -0.1009 & 0.0380
    \end{bmatrix}.
\ee
The scale of BAO measurements is set by the sound horizon ($r_{s}$) during the epoch of radiation drag. To compute this quantity, we use the approximate formula \eqref{eq:sh} \citep{PhysRevD.92.123516}.

In our previous studies we did not vary \obhs\ as a free parameter. Instead we used CMB-derived, model-dependent values of \obhs\ to compute $r_{s}$. Because we vary \obhs\ as a free parameter in this paper, our computations of the sound horizon (and therefore our calibration of the scale of our BAO measurements) are fully independent of CMB physics (at the cost of enlarging the parameter space and so somewhat weakening the constraints).

Following \cite{Conley_et_al_2011} and \cite{Deng_Wei_2018}, we define the theoretical magnitude of a supernova to be
\begin{equation}
\label{eq:m_th}
    m_{\rm th} = 5\log\mathcal{D}_{L}(z) + \mathcal{M},
\end{equation}
where $\mathcal{M}$ is a nuisance parameter to be marginalized over, and $\mathcal{D}_{L}(z)$ is
\begin{equation}
\label{eq:D_L}
    \mathcal{D}_{L}(z) \equiv \left(1 + z_{\rm hel}\right) \int_0^{z_{\rm cmb}} \frac{d\tilde{z}}{E\left(\tilde{z}\right)}.
\end{equation}
In this equation, $z_{\rm hel}$ is the heliocentric redshift, and $z_{\rm cmb}$ is the CMB-frame redshift. In \cite{Conley_et_al_2011}, equation \eqref{eq:D_L} is called the ``Hubble-constant free luminosity distance'', because $E(z)$ does not contain $H_0$. In our case, because we use $h$, \obhs\!, and \ochs\ as free parameters, our expansion rate function (and thus our luminosity distance) depends on the Hubble constant. We therefore obtain weak constraints on $H_0$ from the supernova data, unlike \cite{Conley_et_al_2011} and \cite{Deng_Wei_2018} (see Section \ref{makereference5.4}, below).

\section{Data Analysis Methodology}
\label{makereference5.3}

We use the \textsc{python} module \textsc{emcee} \citep{emcee} to maximize the likelihood functions, thereby determining the constraints on the free parameters. In our analyses here the priors on the cosmological parameters are different from zero (and flat) over the ranges $0.005 \leq \Omega_{\rm b_0}\!h^2 \leq 0.1$, $0.001 \leq \Omega_{\rm c_0}\!h^2 \leq 0.99$, $0.2 \leq h \leq 1.0$, $-3 \leq w_{\rm X} \leq 0.2$, $-0.7 \leq \Omega_{\rm k_0} \leq 0.7$, and $0 < \alpha \leq 10$. \om\ is a derived parameter and depends on $h$. Our approach here differs from that of those earlier papers in that, instead of varying the non-relativistic matter density parameter \om\ as a free parameter, we vary the baryonic (\obhs) and cold dark matter (\ochs) densities as free parameters, treating \om\ as a derived parameter.\footnote{We do this to eliminate the dependence of the BAO data on CMB physics; see Section \ref{makereference5.2} for details.} 

The likelihood functions of $H(z)$, BAO, \hiig, and QSO data are described in \cite{CaoRyanRatra2020} and \cite{Caoetal_2021}. For the SN Ia (SN-Pantheon and SN-DES) data, the likelihood function is
\be
\label{eq:LH_SN}
    \mathcal{L}_{\rm SN}= e^{-\chi^2_{\rm SN}/2},
\ee
where, as in \cite{park_ratra_2019b}, $\chi^2_{\rm SN}$ takes the form of equation (C1) in Appendix C of \cite{Conley_et_al_2011} with $\mathcal{M}$ being marginalized. The covariance matrices of the SN Ia data, $\textbf{C}_{\rm SN}$ are the sum of the diagonal statistical uncertainty covariance matrices, $\textbf{C}_{\rm stat} = \rm diag(\sigma^2_{\rm SN})$, and the systematic uncertainty covariance matrices, $\textbf{C}_{\rm sys}$: $\textbf{C}_{\rm SN} = \textbf{C}_{\rm stat} + \textbf{C}_{\rm sys}$.\footnote{Note that the covariance matrices for the SN-DES data are the ones described in equation (18) of \cite{DES_2019d}.} $\sigma_{\rm SN}$ are the SN Ia statistical uncertainties.

As in \cite{Caoetal_2021}, we use the Akaike Information Criterion ($AIC$) in equation \eqref{AIC1} and the Bayesian Information Criterion ($BIC$) in equation \eqref{BIC1} to compare the quality of models with different numbers of parameters.

\section{Results}
\label{makereference5.4}

\begin{figure*}
\centering
 \subfloat[]{%
    \includegraphics[width=3.25in,height=3.25in]{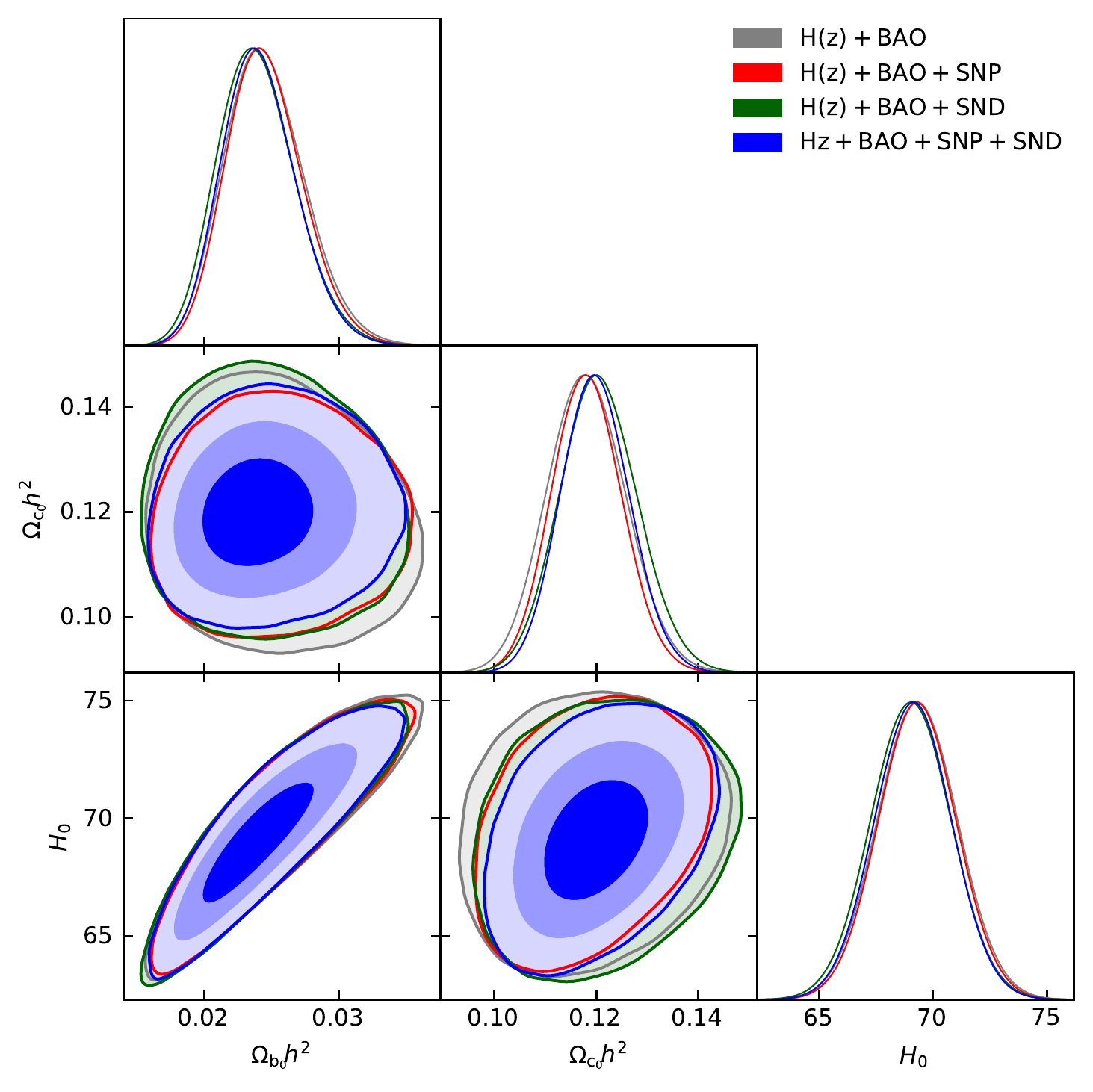}}
 \subfloat[]{%
    \includegraphics[width=3.25in,height=3.25in]{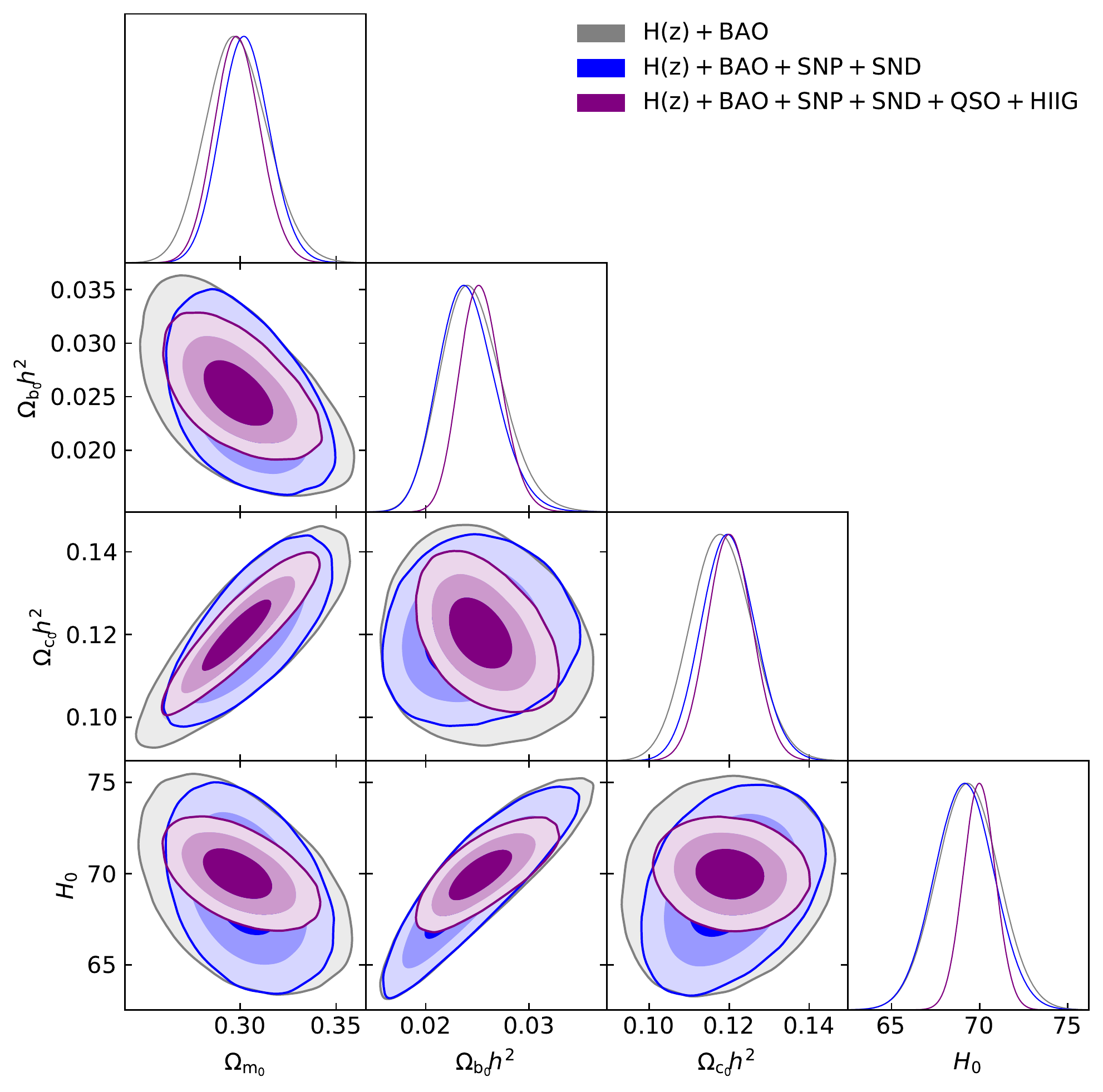}}\\
\caption{1$\sigma$, 2$\sigma$, and 3$\sigma$ confidence contours for flat \lcdm, where the right panel is the comparison including derived cosmological matter density parameter \om. In all cases, the favored parameter space is associated with currently-accelerating cosmological expansion.}
\label{fig1C5}
\end{figure*}

\begin{figure*}
\centering
 \subfloat[]{%
    \includegraphics[width=3.25in,height=3.25in]{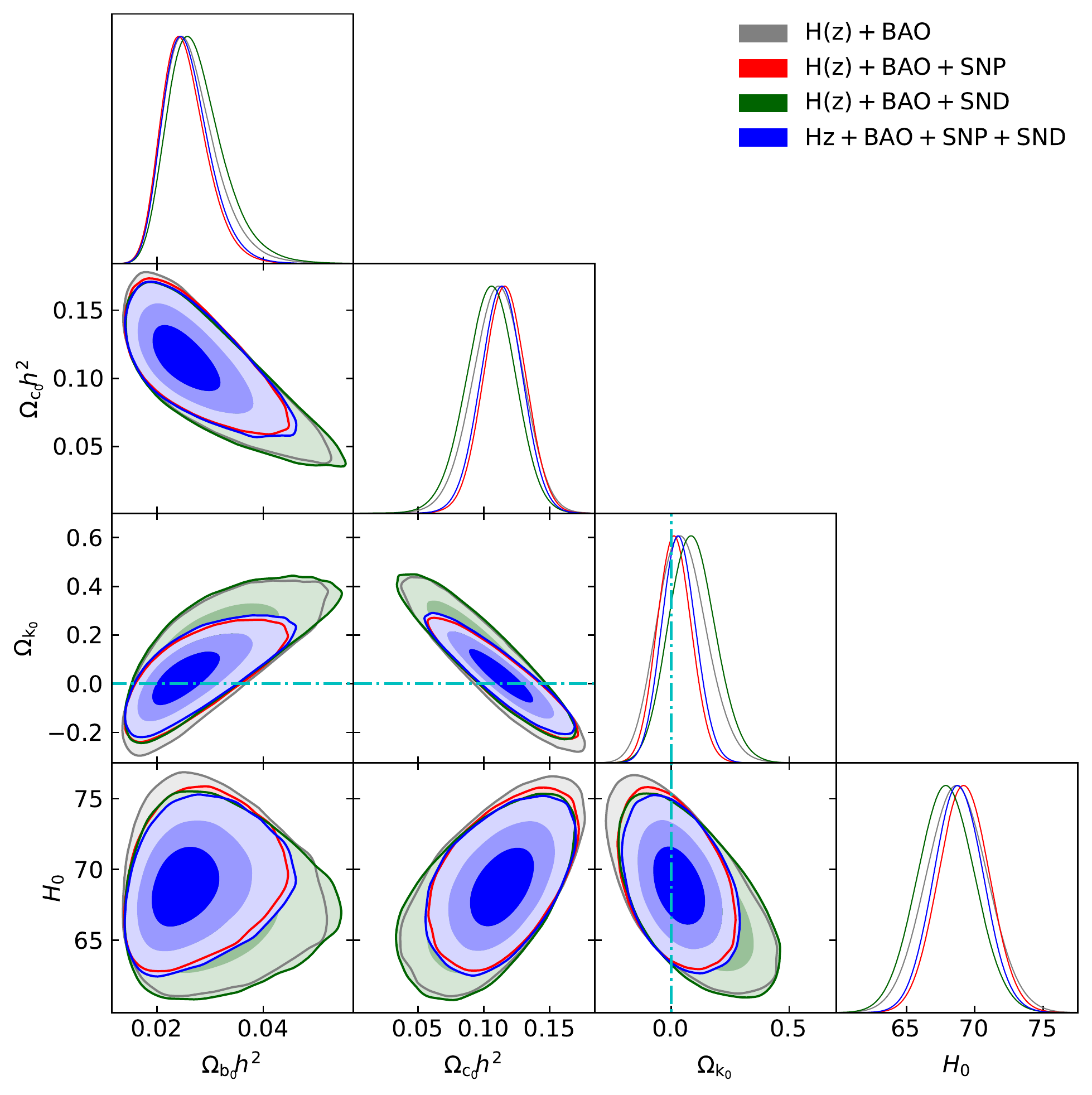}}
 \subfloat[]{%
    \includegraphics[width=3.25in,height=3.25in]{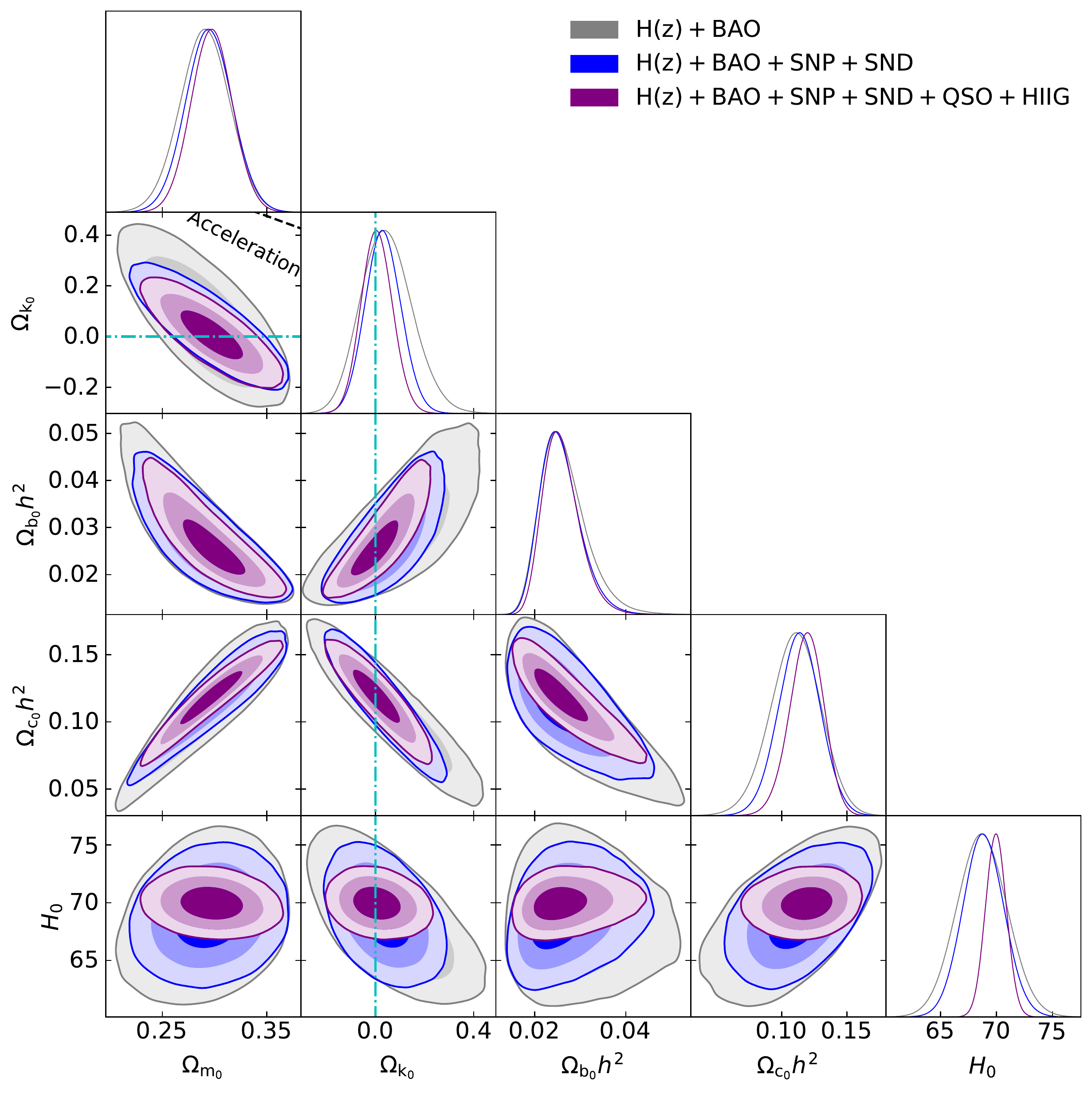}}\\
\caption{Same as Fig. \ref{fig1C5} but for non-flat \lcdm, where the cyan dash-dot lines represent the flat \lcdm\ case, with closed spatial hypersurfaces either below or to the left. The black dotted line in the right subpanel is the zero-acceleration line, which divides the parameter space into regions associated with currently-accelerating (below left) and currently-decelerating (above right) cosmological expansion. In all cases, the favored parameter space is associated with currently-accelerating cosmological expansion.}
\label{fig2C5}
\end{figure*}

\begin{figure*}
\centering
 \subfloat[]{%
    \includegraphics[width=3.25in,height=3.25in]{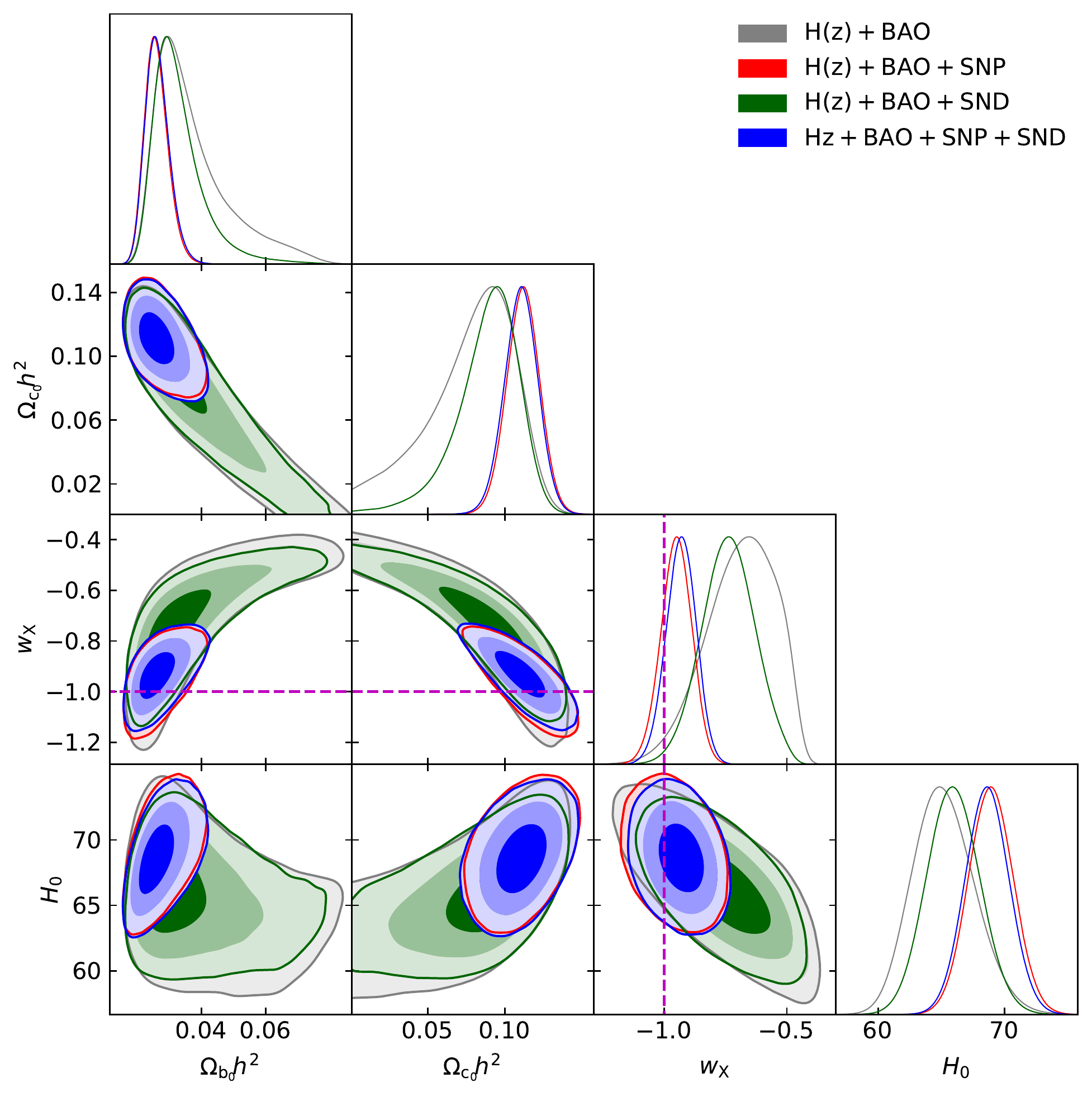}}
 \subfloat[]{%
    \includegraphics[width=3.25in,height=3.25in]{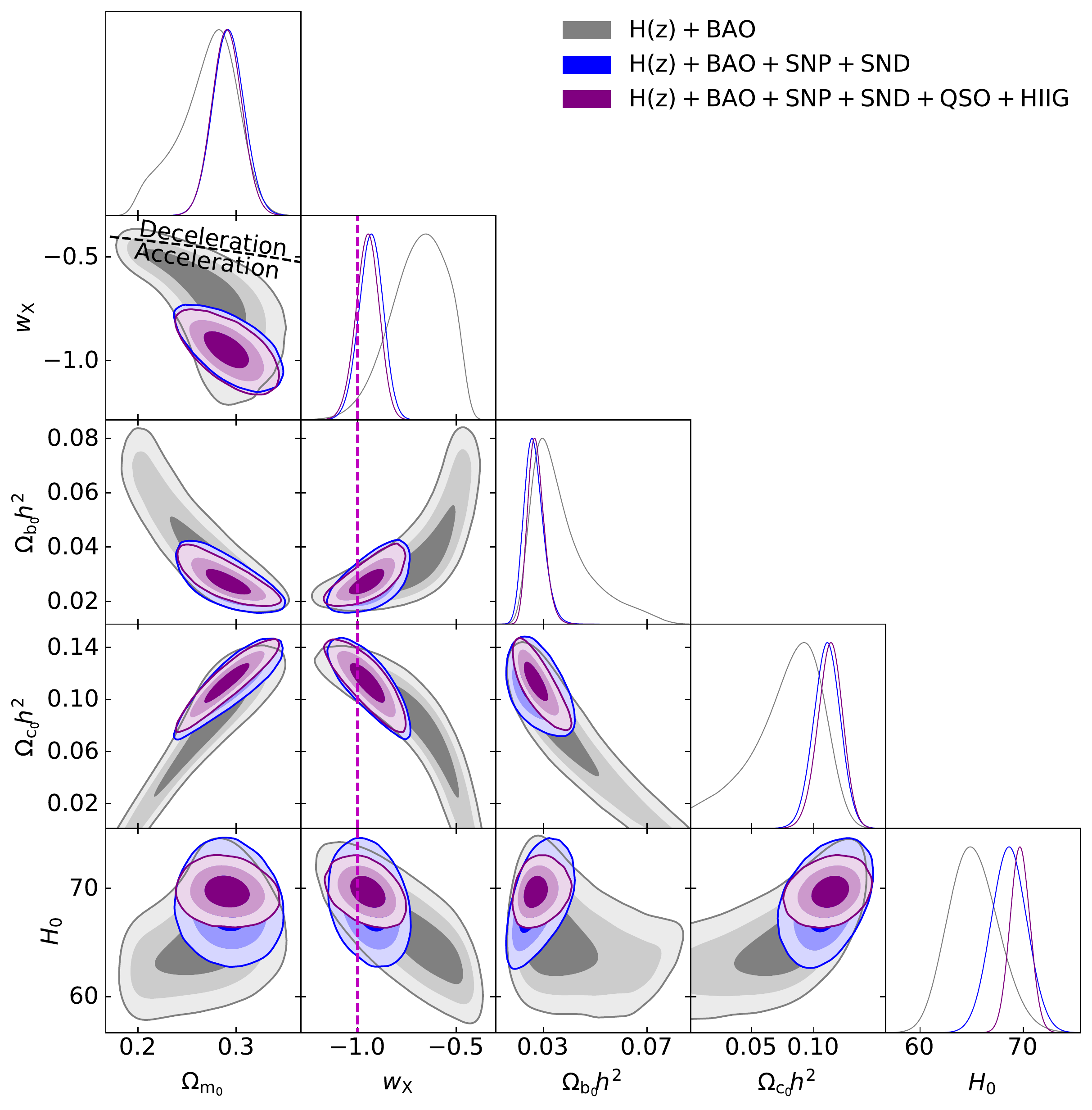}}\\
\caption{1$\sigma$, 2$\sigma$, and 3$\sigma$ confidence contours for flat XCDM. The black dotted line in the right panel is the zero-acceleration line, which divides the parameter space into regions associated with currently-accelerating (below) and currently-decelerating (above) cosmological expansion. In all cases, almost all of the favored parameter space is associated with currently-accelerating cosmological expansion. The magenta lines denote $w_{\rm X}=-1$, i.e. the flat \lcdm\ model.}
\label{fig3C5}
\end{figure*}

\begin{figure*}
\centering
 \subfloat[]{%
    \includegraphics[width=3.25in,height=3.25in]{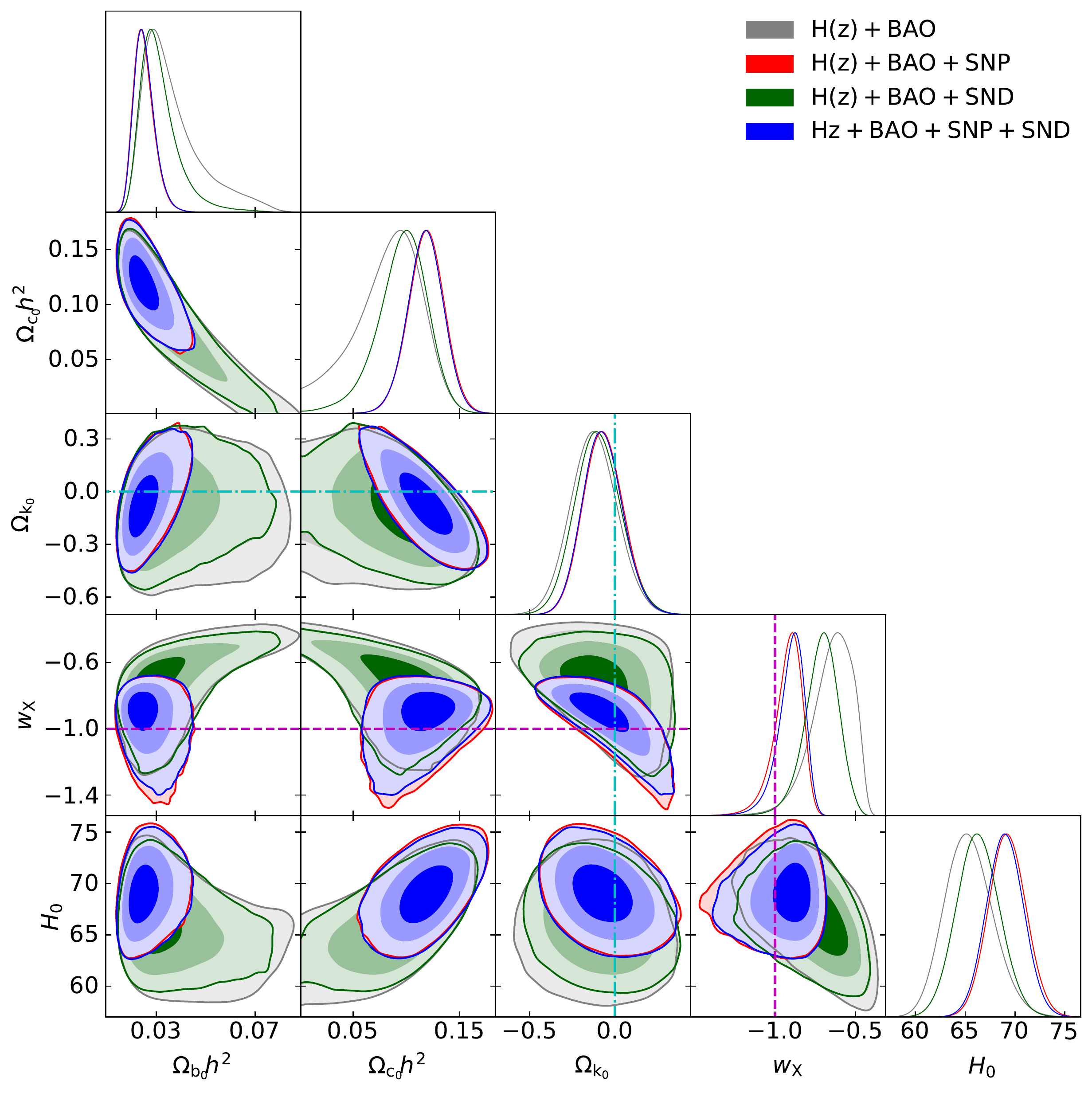}}
 \subfloat[]{%
    \includegraphics[width=3.25in,height=3.25in]{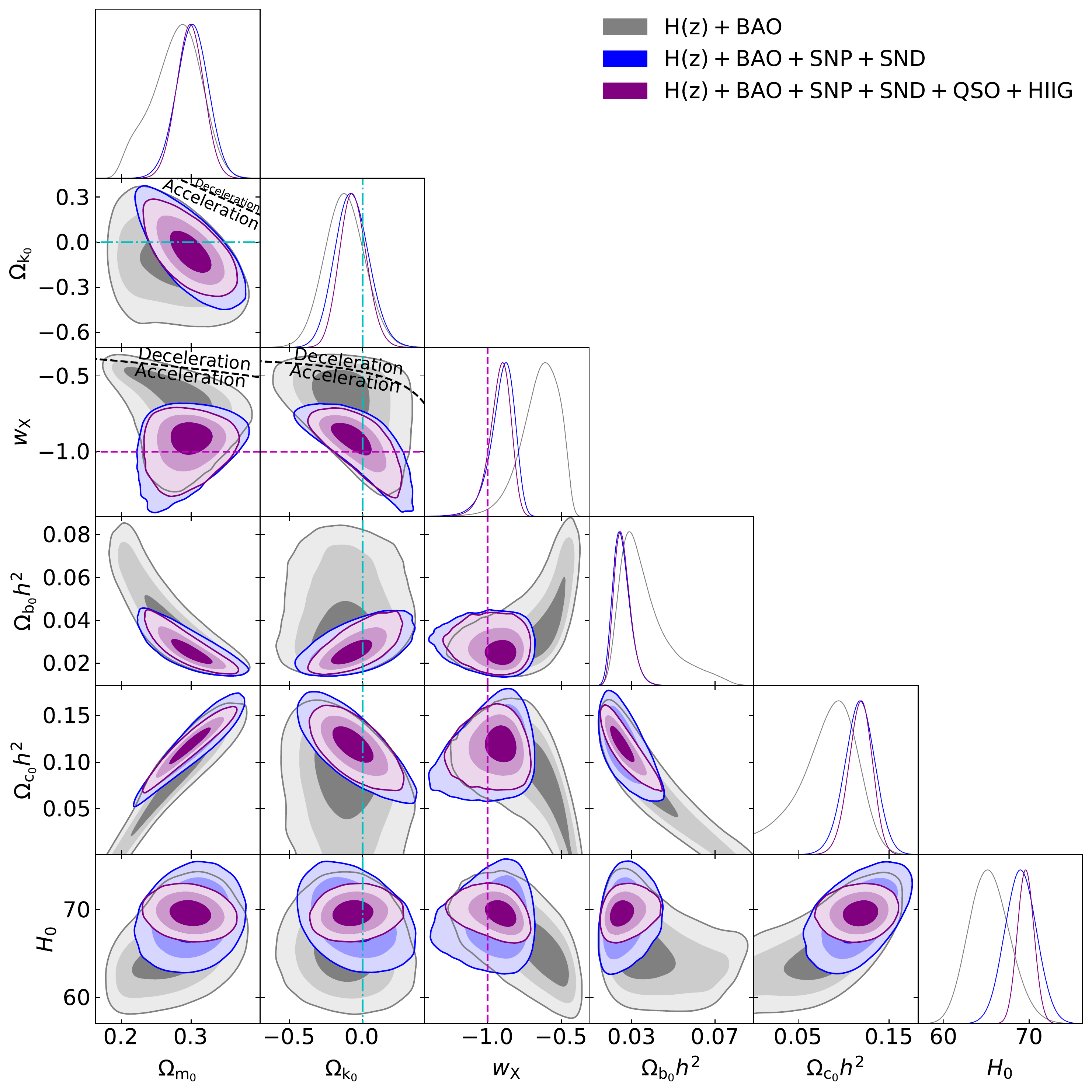}}\\
\caption{Same as Fig. \ref{fig3C5} but for non-flat XCDM, where the zero acceleration lines in each of the three subpanels of the right panel are computed for the third cosmological parameter set to the $H(z)$ + BAO data best-fitting values listed in Table \ref{tab:BFPC5}. Currently-accelerating cosmological expansion occurs below these lines. The cyan dash-dot lines represent the flat XCDM case, with closed spatial hypersurfaces either below or to the left. In all cases, almost all of the favored parameter space is associated with currently-accelerating cosmological expansion. The magenta lines indicate $w_{\rm X} = -1$, i.e. the non-flat \lcdm\ model.}
\label{fig4C5}
\end{figure*}

\begin{figure*}
\centering
 \subfloat[]{%
    \includegraphics[width=3.25in,height=3.25in]{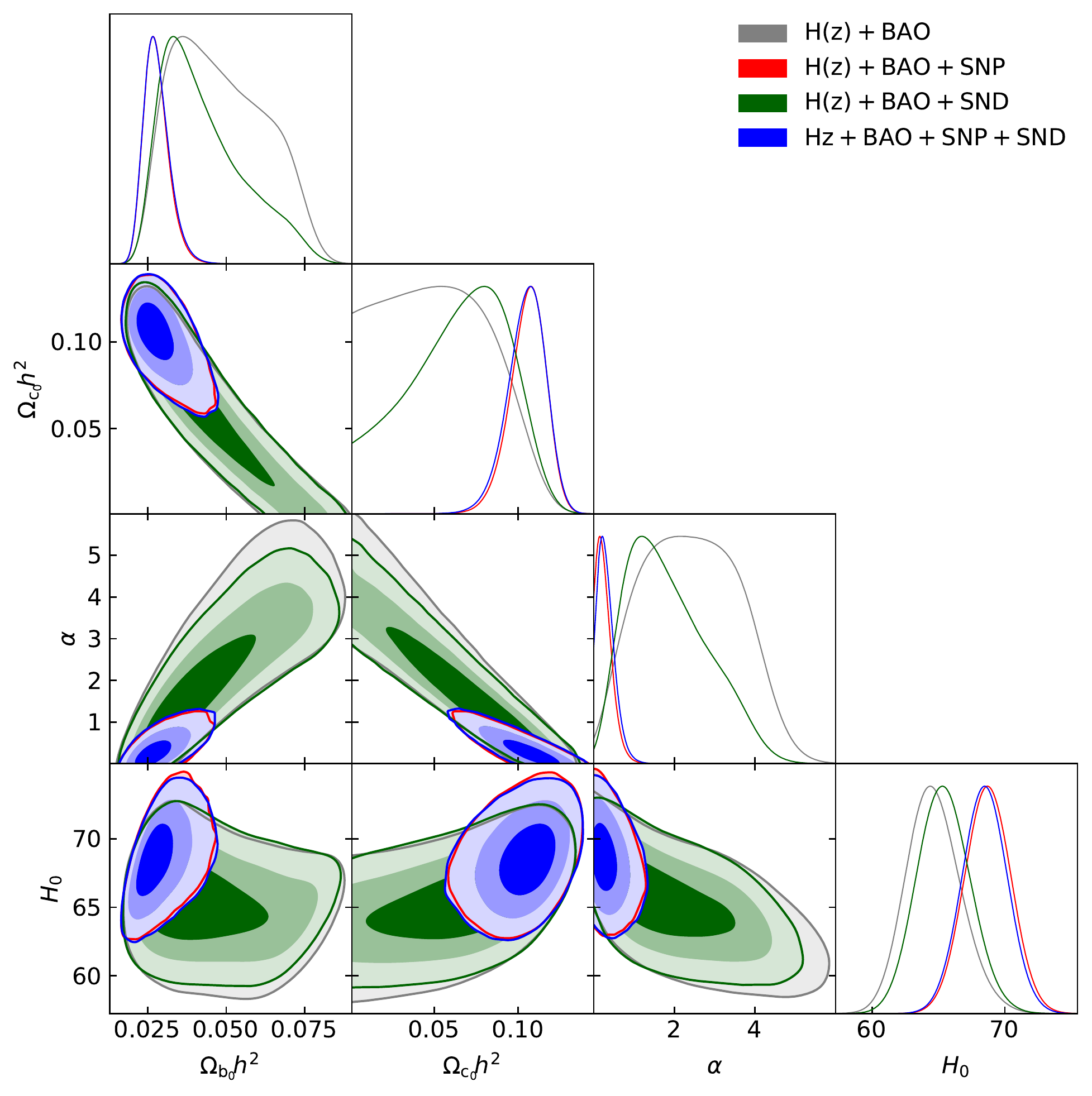}}
 \subfloat[]{%
    \includegraphics[width=3.25in,height=3.25in]{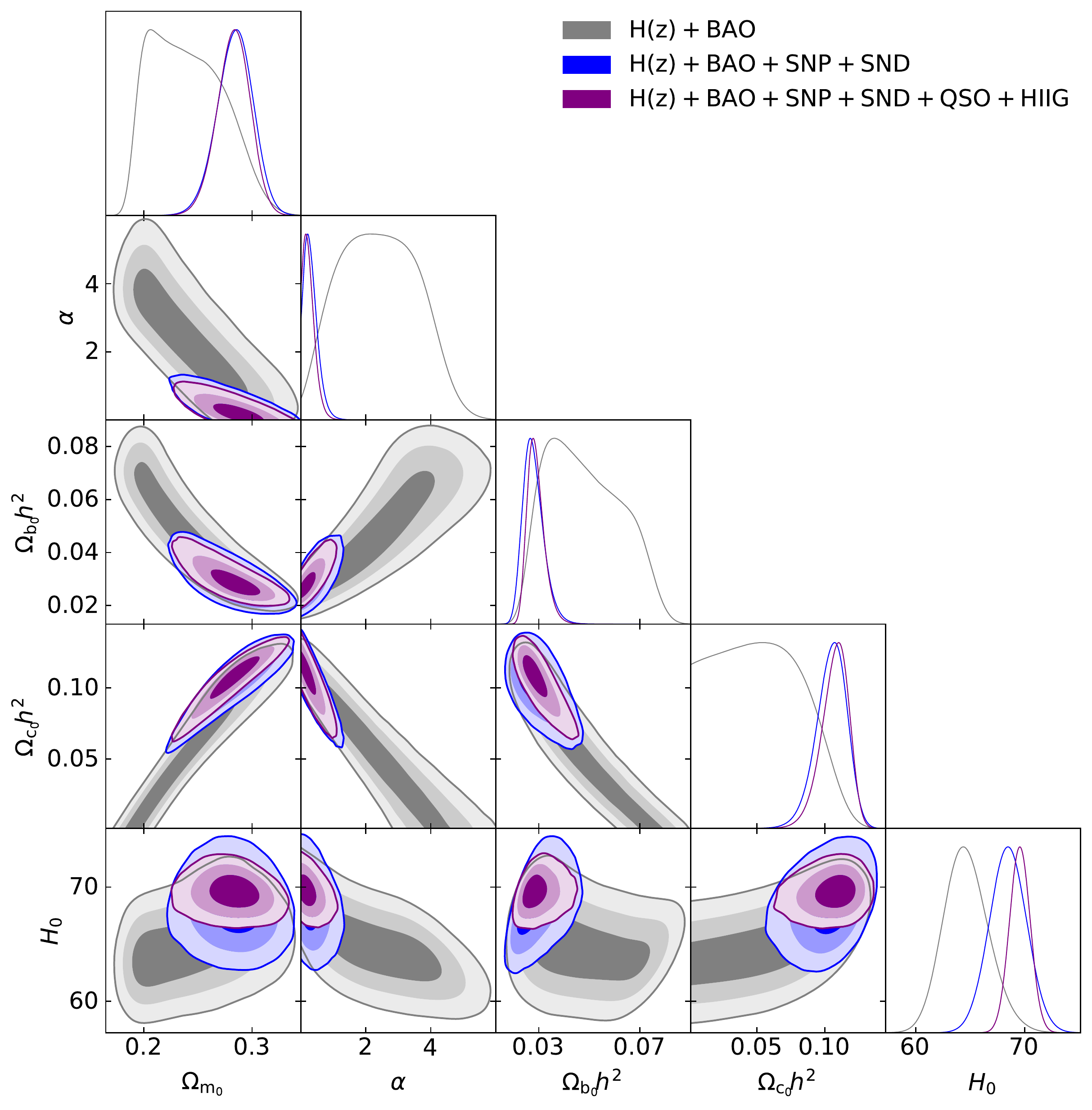}}\\
\caption{1$\sigma$, 2$\sigma$, and 3$\sigma$ confidence contours for flat \pcdm. In all cases, the favored parameter space is associated with currently-accelerating cosmological expansion. The $\alpha = 0$ axis is the flat \lcdm\ model.}
\label{fig5C5}
\end{figure*}

\begin{figure*}
\centering
 \subfloat[]{%
    \includegraphics[width=3.25in,height=3.25in]{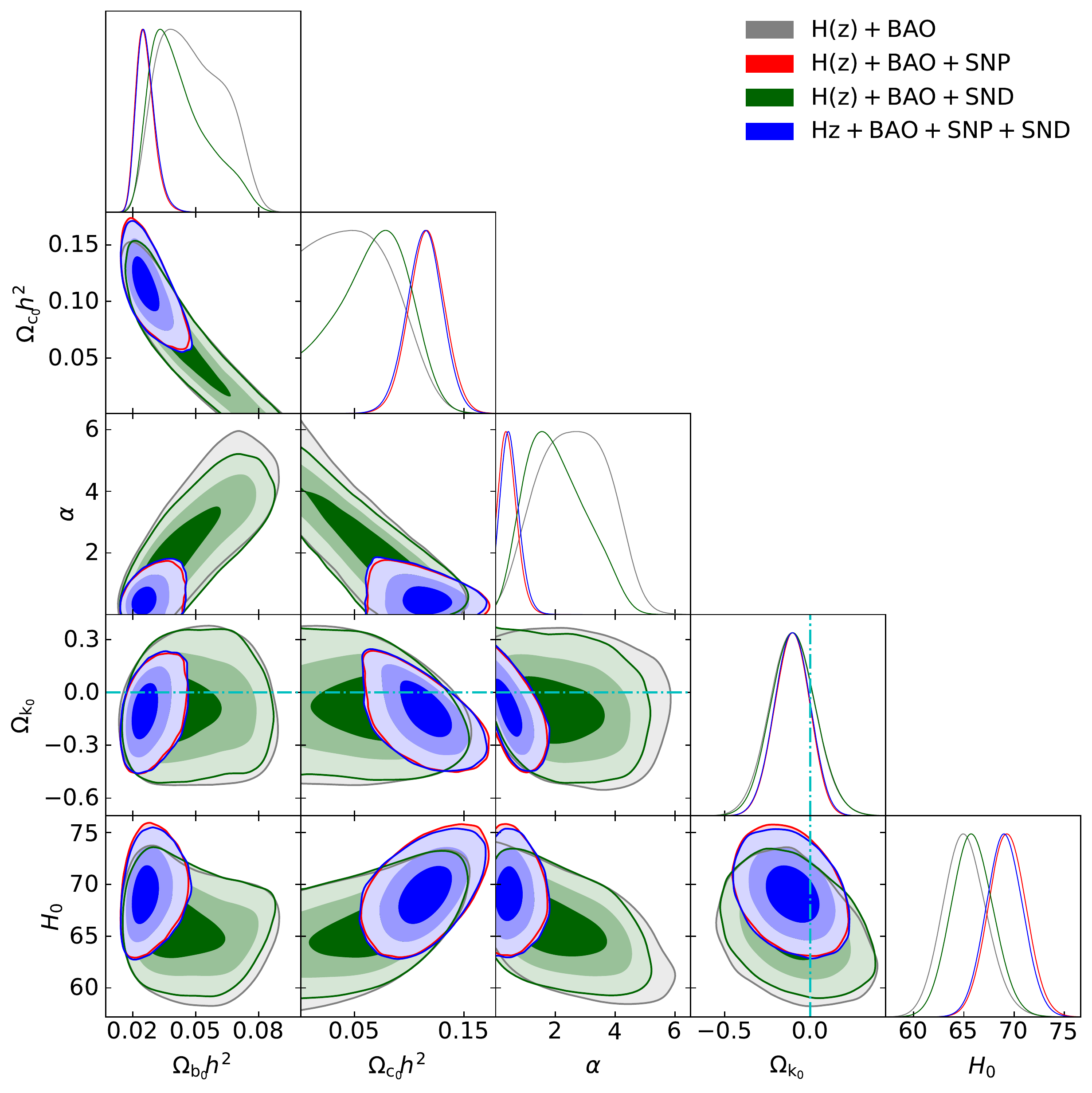}}
 \subfloat[]{%
    \includegraphics[width=3.25in,height=3.25in]{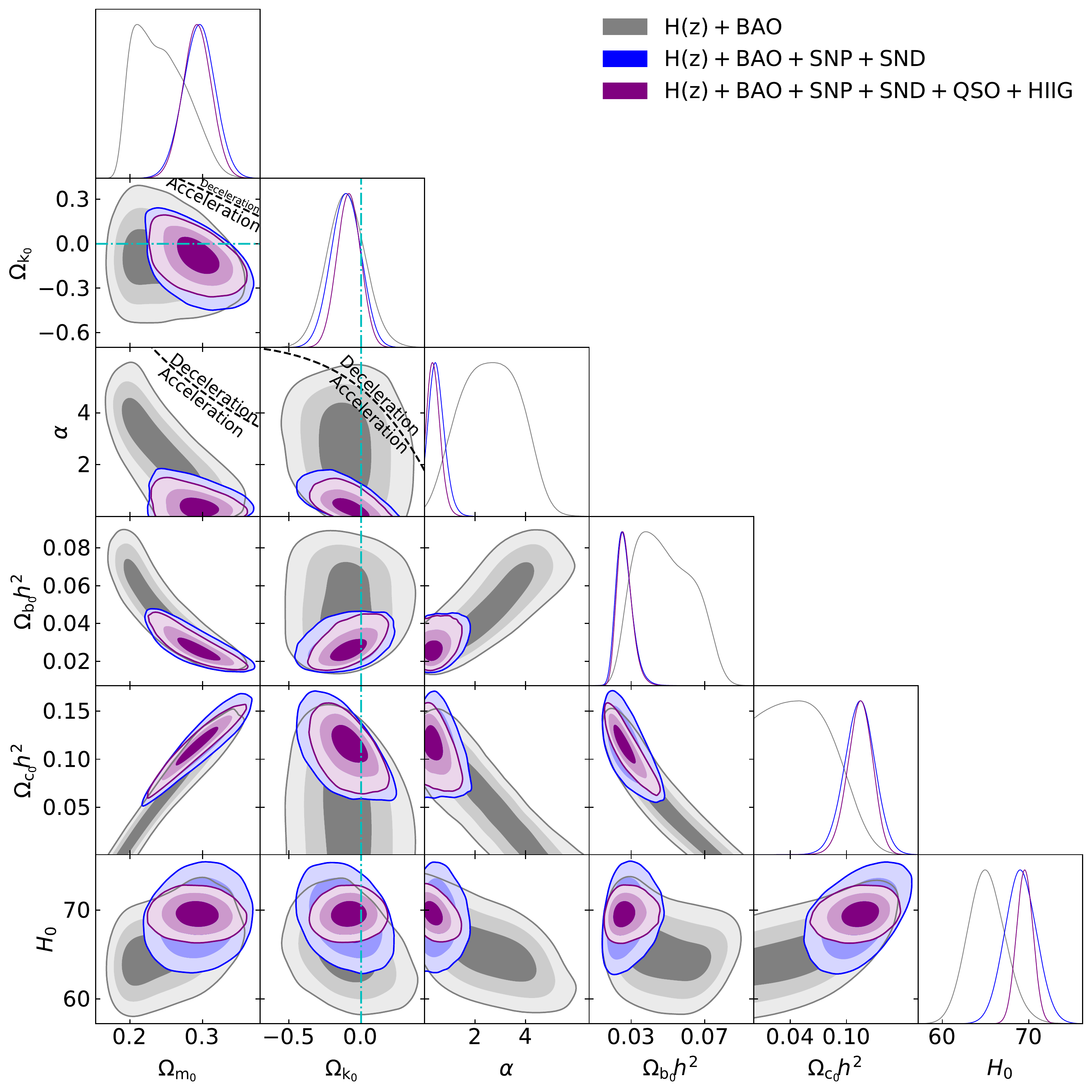}}\\
\caption{Same as Fig. \ref{fig5C5} but for non-flat \pcdm, where the zero-acceleration lines in each of the sub-panels of the right panel are computed for the third cosmological parameter set to the $H(z)$ + BAO data best-fitting values listed in Table \ref{tab:BFPC5}. Currently-accelerating cosmological expansion occurs below these lines. In all cases, almost all of the favored parameter space is associated with currently-accelerating cosmological expansion. The cyan dash-dot lines represent the flat \pcdm\ case, with closed spatial geometry either below or to the left. The $\alpha = 0$ axis is the non-flat \lcdm\ model.}
\label{fig6C5}
\end{figure*}

The posterior one-dimensional (1D) probability distributions and two-dimensional (2D) confidence regions of the cosmological parameters for the six flat and non-flat models are shown in Figs. \ref{fig1C5}--\ref{fig6C5}, in gray ($H(z)$+BAO), red ($H(z)$ + BAO + SN-Pantheon, HzBSNP), green ($H(z)$ + BAO + SN-DES, HzBSND), blue ($H(z)$ + BAO + SN-Pantheon + SN-DES, HzBSNPD), and purple ($H(z)$ + BAO + SN-Pantheon + SN-DES + QSO + \hiig, HzBSNPDQH). We list the unmarginalized best-fitting parameter values, as well as the corresponding $\chi^2$, $AIC$, $BIC$, and degrees of freedom $\nu$ ($\nu \equiv N - n$) for all models and data combinations, in Table \ref{tab:BFPC5}. The marginalized best-fitting parameter values and uncertainties ($\pm 1\sigma$ error bars or $2\sigma$ limits), for all models and data combinations, are listed in Table \ref{tab:1d_BFPC5}.\footnote{The \textsc{python} package \textsc{getdist} \citep{Lewis_2019} is used to analyze the samples.}

\subsection{$H(z)$ + BAO, HzBSNP, and HzBSND constraints}
\label{subsec:HzB}

The 1D marginalized $H(z)$ + BAO constraints on the cosmological parameters are listed in Table \ref{tab:1d_BFPC5}. These are (slightly) different from the ones obtained by \cite{KhadkaRatra2021}, because of the different treatments of both the prior ranges and the coefficient $\kappa$ in the \pcdm\ models.\footnote{We treated $\kappa$ as a derived constant determined from the parameter $\alpha$ (see e.g. eq. (14) of \citealp{Caoetal_2021}), while \cite{KhadkaRatra2021} treated it as a constant derived from the energy budget equation.}

The $H(z)$, BAO, and SN-Pantheon data combinations have previously been studied \citep{park_ratra_2019b}. Relative to that analysis, we use the updated BAO data, shown in Table \ref{tab:BAOC5}, in our analysis here. In the HzBSNP case, we find that the determinations of \ok\ are more consistent with flat spatial hypersurfaces than what \cite{park_ratra_2019b} found and dark energy dynamics favors less deviation from a cosmological constant in the XCDM cases, while favoring a somewhat stronger deviation from $\alpha=0$ in the non-flat \pcdm\ case.

Because the $H(z)$, BAO, and SN-DES constraints are consistent across all six of the models we study, we also perform a joint analysis of these data to determine HzBSND constraints. Relative to the HzBSNP constraints, the measured values of $\Omega_{\rm b_0}\!h^2$, $\Omega_{\rm c_0}\!h^2$, and \om\ are a little higher, lower, and lower (except for flat \lcdm) than those values measured from the HzBSNP case, respectively. Given the error bars, these differences are not statistically significant. The measured values of $H_0$ are lower than those for the HzBSNP case. The non-flat XCDM and \pcdm\ models favor more and less closed geometry than in the HzBSNP case. The non-flat \lcdm\ model favors more open geometry than in the HzBSNP case. The constraints for all three non-flat models are consistent with spatially flat hypersurfaces. The fits to the HzBSND data produce stronger evidence for dark energy dynamics than the fits to the HzBSNP data.

\begin{table*}
\centering
\resizebox*{1\columnwidth}{1.2\columnwidth}{%
\begin{threeparttable}
\caption{Unmarginalized best-fitting parameter values for all models from various combinations of data.}\label{tab:BFPC5}
\begin{tabular}{lcccccccccccc}
\toprule
Model & Data set & $\Omega_{\mathrm{b_0}}\!h^2$ & $\Omega_{\mathrm{c_0}}\!h^2$ & $\Omega_{\mathrm{m_0}}$ & $\Omega_{\mathrm{k_0}}$ & $w_{\mathrm{X}}$ & $\alpha$ & $H_0$\tnote{a} & $\chi^2$ & $\nu$ & $AIC$ & $BIC$ \\
\midrule
Flat \lcdm & $H(z)$ + BAO & 0.0240 & 0.1179 & 0.299 & -- & -- & -- & 69.11 & 23.64 & 39 & 29.64 & 34.86\\
 & HzBSNP\tnote{b} & 0.0240 & 0.1180 & 0.299 & -- & -- & -- & 69.10 & 1053.22 & 1087 & 1059.22 & 1074.21\\
 & HzBSND\tnote{c} & 0.0234 & 0.1203 & 0.305 & -- & -- & -- & 68.82 & 50.83 & 59 & 56.83 & 63.21\\
 & HzBSNPD\tnote{d} & 0.0236 & 0.1196 & 0.303 & -- & -- & -- & 68.91 & 1080.46 & 1107 & 1086.46 & 1101.50\\
 & HzBSNPDQH\tnote{e} & 0.0251 & 0.1203 & 0.299 & -- & -- & -- & 69.92 & 1844.99 & 1380 & 1850.99 & 1866.69\\
\\
Non-flat \lcdm & $H(z)$ + BAO & 0.0248 & 0.1136 & 0.294 & 0.026 & -- & -- & 68.75 & 23.58 & 38 & 31.58 & 38.53\\
 & HzBSNP\tnote{b} & 0.0241 & 0.1172 & 0.298 & 0.004 & -- & -- & 69.06 & 1053.22 & 1086 & 1061.22 & 1081.20\\
 & HzBSND\tnote{c} & 0.0258 & 0.1081 & 0.292 & 0.071 & -- & -- & 67.92 & 50.28 & 58 & 58.28 & 66.79\\
 & HzBSNPD\tnote{d} & 0.0245 & 0.1150 & 0.297 & 0.023 & -- & -- & 68.68 & 1080.35 & 1106 & 1088.35 & 1108.40\\
 & HzBSNPDQH\tnote{e} & 0.0249 & 0.1209 & 0.300 & $-0.004$ & -- & -- & 69.93 & 1844.99 & 1379 & 1852.99 & 1873.92\\
\\
Flat XCDM & $H(z)$ + BAO & 0.0323 & 0.0860 & 0.280 & -- & $-0.696$ & -- & 65.12 & 19.65 & 38 & 27.65 & 34.60\\
 & HzBSNP\tnote{b} & 0.0254 & 0.1120 & 0.292 & -- & $-0.951$ & -- & 68.72 & 1052.63 & 1086 & 1060.63 & 1080.61\\
 & HzBSND\tnote{c} & 0.0300 & 0.0934 & 0.286 & -- & $-0.752$ & -- & 65.90 & 45.46 & 58 & 53.46 & 61.97\\
 & HzBSNPD\tnote{d} & 0.0256 & 0.1107 & 0.293 & -- & $-0.932$ & -- & 68.43 & 1079.23 & 1106 & 1087.23 & 1107.28\\
 & HzBSNPDQH\tnote{e} & 0.0268 & 0.1136 & 0.291 & -- & $-0.949$ & -- & 69.63 & 1844.27 & 1379 & 1852.27 & 1873.20\\
\\
Non-flat XCDM & $H(z)$ + BAO & 0.0302 & 0.0956 & 0.294 & $-0.155$ & $-0.650$ & -- & 65.55 & 18.31 & 37 & 28.31 & 37.00\\
 & HzBSNP\tnote{b} & 0.0234 & 0.1231 & 0.307 & $-0.103$ & $-0.895$ & -- & 69.25 & 1051.82 & 1085 & 1061.82 & 1086.79\\
 & HzBSND\tnote{c} & 0.0277 & 0.1046 & 0.301 & $-0.136$ & $-0.711$ & -- & 66.45 & 44.34 & 57 & 54.34 & 64.98\\
 & HzBSNPD\tnote{d} & 0.0236 & 0.1220 & 0.307 & $-0.107$ & $-0.877$ & -- & 68.98 & 1078.36 & 1105 & 1088.36 & 1113.42\\
 & HzBSNPDQH\tnote{e} & 0.0242 & 0.1217 & 0.303 & $-0.092$ & $-0.900$ & -- & 69.54 & 1843.25 & 1378 & 1853.25 & 1879.41\\
\\
Flat $\phi$CDM & $H(z)$ + BAO & 0.0361 & 0.0758 & 0.264 & -- & -- & 1.484 & 65.30 & 19.48 & 38 & 27.48 & 34.43\\
 & HzBSNP\tnote{b} & 0.0260 & 0.1145 & 0.292 & -- & -- & 0.101 & 69.51 & 1051.46 & 1086 & 1059.46 & 1079.44\\
 & HzBSND\tnote{c} & 0.0328 & 0.0860 & 0.273 & -- & -- & 1.061 & 66.16 & 45.17 & 58 & 53.17 & 61.68\\
 & HzBSNPD\tnote{d} & 0.0254 & 0.1102 & 0.292 & -- & -- & 0.168 & 68.35 & 1078.18 & 1106 & 1086.18 & 1106.22\\
 & HzBSNPDQH\tnote{e} & 0.0264 & 0.1135 & 0.290 & -- & -- & 0.132 & 69.57 & 1842.95 & 1379 & 1850.95 & 1871.88\\
\\
Non-flat $\phi$CDM & $H(z)$ + BAO & 0.0354 & 0.0811 & 0.269 & $-0.148$ & -- & 1.819 & 66.06 & 18.16 & 37 & 28.16 & 36.85\\
 & HzBSNP\tnote{b} & 0.0234 & 0.1225 & 0.305 & $-0.133$ & -- & 0.393 & 69.32 & 1050.31 & 1085 & 1060.31 & 1085.28\\
 & HzBSND\tnote{c} & 0.0319 & 0.0933 & 0.282 & $-0.140$ & -- & 1.411 & 66.84 & 44.09 & 57 & 54.09 & 64.72\\
 & HzBSNPD\tnote{d} & 0.0256 & 0.1159 & 0.298 & $-0.080$ & -- & 0.377 & 69.09 & 1077.13 & 1105 & 1087.13 & 1112.19\\
 & HzBSNPDQH\tnote{e} & 0.0258 & 0.1155 & 0.293 & $-0.078$ & -- & 0.354 & 69.55 & 1842.00 & 1378 & 1852.00 & 1878.16\\
\bottomrule
\end{tabular}
\begin{tablenotes}[flushleft]
\item [a] \hunit.
\item [b] $H(z)$ + BAO + SN-Pantheon.
\item [c] $H(z)$ + BAO + SN-DES.
\item [d] $H(z)$ + BAO + SN-Pantheon + SN-DES.
\item [e] $H(z)$ + BAO + SN-Pantheon + SN-DES + QSO + \hiig.
\end{tablenotes}
\end{threeparttable}%
}
\end{table*}

\subsection{HzBSNPD constraints}
\label{subsec:HzBSNPD}

The results of the previous three subsections show that, when combined with $H(z)$ + BAO data, SN-Pantheon data produce tighter constraints on almost all cosmological parameters, than do SN-DES data (with a few exceptions including $\Omega_{\rm b_0}\!h^2$ for non-flat \lcdm, $\Omega_{\rm c_0}\!h^2$ for non-flat \pcdm, and \om\ and $H_0$ for flat and non-flat \pcdm). Since the $H(z)$ + BAO, SN-Pantheon, and SN-DES data constraints are not inconsistent, it is useful to derive constraints from an analysis of the combined $H(z)$, BAO, SN-Pantheon, and SN-DES (HzBSNPD) data. The results of such an analysis are presented in this subsection. We discuss these results in some detail here because, as discussed in Sec. \ref{makereference5.4.1}, we believe that the constraints we obtain from the HzBSNPD data combination are more reliable than the constraints we obtain from the other data combinations we study.

The measured values of $\Omega_{\rm b_0}\!h^2$ range from a low of $0.0241^{+0.0024}_{-0.0030}$ (flat \lcdm) to a high of $0.0279^{+0.0031}_{-0.0048}$ (flat \pcdm) and those of $\Omega_{\rm c_0}\!h^2$ range from a low of $0.1047^{+0.0125}_{-0.0096}$ (flat \pcdm) to a high of $0.1199\pm0.0067$ (flat \lcdm). The derived constraints on \om\ range from a low of $0.284^{+0.017}_{-0.016}$ (flat \pcdm) to a high of $0.303\pm0.013$ (flat \lcdm). These measurements are consistent with what is measured by \cite{planck2018b}. In particular, for flat \lcdm, comparing to the TT,TE,EE+lowE+lensing results in Table 2 of \cite{planck2018b} the error bars we find here for \obhs\!, \ochs\!, and \om\ are a factor of 18, 5.6, and 1.8, respectively, larger than the \textit{Planck} error bars, and our estimates here for the quantities differ from the \textit{Planck} estimates by 0.58$\sigma$, 0.015$\sigma$, and 0.82$\sigma$, respectively.

The constraints on $H_0$ are between $H_0=68.48^{+1.71}_{-1.70}$ \hunit\ (flat \pcdm) and $H_0=69.14\pm1.68$ \hunit\ (flat \lcdm), which are $0.35\sigma$ (flat \lcdm) and $0.15\sigma$ (flat \pcdm) higher than the median statistics estimate of $H_0=68 \pm 2.8$ \hunit\ \citep{chenratmed}, and $2.22\sigma$ (flat \lcdm) and $2.50\sigma$ (flat \pcdm) lower than the local Hubble constant measurement of $H_0 = 74.03 \pm 1.42$ \hunit\ \citep{riess_etal_2019}.\footnote{Other local expansion rate $H_0$ measurements result in slightly lower central values with slightly larger error bars \citep{rigault_etal_2015,zhangetal2017,Dhawan,FernandezArenas,freedman_etal_2020,rameez_sarkar_2019,Breuvaletal_2020, Efstathiou_2020, Khetan_et_al_2021}. Our $H_0$ determinations are consistent with earlier median statistics determinations \citep{gott_etal_2001,chen_etal_2003} as well as with other recent $H_0$ measurements \citep{chen_etal_2017,DES_2018,Gomez-ValentAmendola2018, planck2018b,dominguez_etal_2019,Cuceu_2019,zeng_yan_2019,schoneberg_etal_2019,lin_ishak_2021, Blum_et_al_2020, Lyu_et_al_2020, Philcox_et_al_2020, Zhang_Huang_2021, Birrer_et_al_2020, Denzel_et_al_2020,Pogosianetal_2020,Boruahetal_2021,Kimetal_2020,Harvey_2020}.} For flat \lcdm\ our $H_0$ error bar is a factor of 3.1 larger than that from the \textit{Planck} data and our $H_0$ estimate is 1.01$\sigma$ higher than that of \textit{Planck}.

For non-flat \lcdm, non-flat XCDM, and non-flat \pcdm, we find $\Omega_{\rm k_0}=0.032\pm0.072$, $\Omega_{\rm k_0}=-0.071^{+0.110}_{-0.123}$, and $\Omega_{\rm k_0}=-0.105\pm0.104$, respectively, with non-flat \pcdm\ favoring closed geometry at 1.01$\sigma$. The non-flat XCDM and \pcdm\ models favor closed geometry, while the non-flat \lcdm\ model favors open geometry. The constraints for non-flat \lcdm\ and XCDM models are consistent with spatially flat hypersurfaces.

The fits to the HzBSNPD data favor dark energy dynamics, where for flat (non-flat) XCDM, $w_{\rm X}=-0.932\pm0.061$ ($w_{\rm X}=-0.904^{+0.098}_{-0.058}$), with best-fitting value being 1.11$\sigma$ (1.66$\sigma$) away from $w_{\rm X}=-1$; and for flat (non-flat) \pcdm, $\alpha=0.320^{+0.108}_{-0.277}$ ($\alpha=0.509^{+0.212}_{-0.370}$), with best-fitting value being 1.16$\sigma$ (1.38$\sigma$) away from $\alpha=0$.

\begin{table*}
\centering
\resizebox*{1\columnwidth}{1.2\columnwidth}{%
\begin{threeparttable}
\caption{One-dimensional marginalized best-fitting parameter values and uncertainties ($\pm 1\sigma$ error bars or $2\sigma$ limits) for all models from various combinations of data.}\label{tab:1d_BFPC5}
\begin{tabular}{lcccccccc}
\toprule
Model & Data set & $\Omega_{\mathrm{b_0}}\!h^2$ & $\Omega_{\mathrm{c_0}}\!h^2$ & $\Omega_{\mathrm{m_0}}$ & $\Omega_{\mathrm{k_0}}$ & $w_{\mathrm{X}}$ & $\alpha$ & $H_0$\tnote{a}\\
\midrule
Flat \lcdm & $H(z)$ + BAO & $0.0245^{+0.0026}_{-0.0032}$ & $0.1182\pm0.0077$ & $0.298^{+0.015}_{-0.017}$ & -- & -- & -- & $69.33\pm1.75$ \\
 & HzBSNP\tnote{b} & $0.0245^{+0.0025}_{-0.0031}$ & $0.1182\pm0.0068$ & $0.298\pm0.013$ & -- & -- & -- & $69.32\pm1.70$ \\
 & HzBSND\tnote{c} & $0.0239^{+0.0025}_{-0.0032}$ & $0.1206\pm0.0076$ & $0.305^{+0.015}_{-0.017}$ & -- & -- & -- & $69.04\pm1.74$ \\
 & HzBSNPD\tnote{d} & $0.0241^{+0.0024}_{-0.0030}$ & $0.1199\pm0.0067$ & $0.303\pm0.013$ & -- & -- & -- & $69.14\pm1.68$ \\
 & HzBSNPDQH\tnote{e} & $0.0253^{+0.0019}_{-0.0022}$ & $0.1202\pm0.0057$ & $0.299\pm0.012$ & -- & -- & -- & $69.98\pm0.91$ \\
\\
Non-flat \lcdm & $H(z)$ + BAO & $0.0265^{+0.0035}_{-0.0059}$ & $0.1104\pm0.0192$ & $0.291\pm0.024$ & $0.047^{+0.095}_{-0.112}$ & -- & -- & $68.71\pm2.24$ \\
 & HzBSNP\tnote{b} & $0.0253^{+0.0033}_{-0.0049}$ & $0.1158^{+0.0161}_{-0.0160}$ & $0.296\pm0.022$ & $0.013\pm0.073$ & -- & -- & $69.22\pm1.86$ \\
 & HzBSND\tnote{c} & $0.0276^{+0.0038}_{-0.0062}$ & $0.1049^{+0.0188}_{-0.0187}$ & $0.288\pm0.024$ & $0.090^{+0.093}_{-0.106}$ & -- & -- & $67.92\pm2.10$ \\
 & HzBSNPD\tnote{d} & $0.0257^{+0.0033}_{-0.0050}$ & $0.1133\pm0.0160$ & $0.295\pm0.022$ & $0.032\pm0.072$ & -- & -- & $68.83\pm1.82$ \\
 & HzBSNPDQH\tnote{e} & $0.0260^{+0.0031}_{-0.0046}$ & $0.1188^{+0.0138}_{-0.0123}$ & $0.297\pm0.020$ & $0.007\pm0.063$ & -- & -- & $69.95\pm0.93$ \\
\\
Flat XCDM & $H(z)$ + BAO & $0.0372^{+0.0045}_{-0.0138}$ & $0.0777^{+0.0351}_{-0.0182}$ & $0.270^{+0.036}_{-0.022}$ & -- & $-0.688^{+0.174}_{-0.109}$ & -- & $65.22^{+2.21}_{-2.64}$ \\
 & HzBSNP\tnote{b} & $0.0261^{+0.0030}_{-0.0041}$ & $0.1118\pm0.0105$ & $0.292\pm0.016$ & -- & $-0.951\pm0.063$ & -- & $68.91\pm1.76$ \\
 & HzBSND\tnote{c} & $0.0331^{+0.0038}_{-0.0091}$ & $0.0881^{+0.0235}_{-0.0137}$ & $0.279^{+0.027}_{-0.019}$ & -- & $-0.739^{+0.110}_{-0.108}$ & -- & $65.95\pm2.08$ \\
 & HzBSNPD\tnote{d} & $0.0264^{+0.0031}_{-0.0042}$ & $0.1105\pm0.0107$ & $0.292\pm0.016$ & -- & $-0.932\pm0.061$ & -- & $68.62\pm1.73$ \\
 & HzBSNPDQH\tnote{e} & $0.0273^{+0.0026}_{-0.0035}$ & $0.1131^{+0.0104}_{-0.0095}$ & $0.291\pm0.015$ & -- & $-0.949\pm0.059$ & -- & $69.67^{+0.97}_{-0.96}$ \\
\\
Non-flat XCDM & $H(z)$ + BAO & $0.0367^{+0.0049}_{-0.0145}$ & $0.0822^{+0.0376}_{-0.0233}$ & $0.278^{+0.041}_{-0.030}$ & $-0.122^{+0.137}_{-0.136}$ & $-0.647^{+0.159}_{-0.084}$ & -- & $65.39^{+2.18}_{-2.59}$ \\
 & HzBSNP\tnote{b} & $0.0251^{+0.0031}_{-0.0049}$ & $0.1186\pm0.0167$ & $0.301\pm0.023$ & $-0.066^{+0.111}_{-0.124}$ & $-0.923^{+0.104}_{-0.060}$ & -- & $69.24\pm1.87$ \\
 & HzBSND\tnote{c} & $0.0315^{+0.0039}_{-0.0091}$ & $0.0956^{+0.0260}_{-0.0190}$ & $0.290^{+0.031}_{-0.026}$ & $-0.099\pm0.133$ & $-0.714^{+0.116}_{-0.089}$ & -- & $66.30\pm2.14$ \\
 & HzBSNPD\tnote{d} & $0.0253^{+0.0032}_{-0.0048}$ & $0.1178^{+0.0166}_{-0.0165}$ & $0.301\pm0.023$ & $-0.071^{+0.110}_{-0.123}$ & $-0.904^{+0.098}_{-0.058}$ & -- & $69.00\pm1.85$ \\
 & HzBSNPDQH\tnote{e} & $0.0256^{+0.0030}_{-0.0046}$ & $0.1182^{+0.0136}_{-0.0121}$ & $0.299\pm0.020$ & $-0.063^{+0.087}_{-0.097}$ & $-0.919^{+0.085}_{-0.056}$ & -- & $69.59\pm0.97$ \\
\\
Flat $\phi$CDM & $H(z)$ + BAO & $0.0480^{+0.0113}_{-0.0195}$ & $0.0524^{+0.0246}_{-0.0427}$ & $0.240^{+0.024}_{-0.044}$ & -- & -- & $2.418^{+1.197}_{-1.331}$ & $64.67^{+1.86}_{-2.22}$ \\
 & HzBSNP\tnote{b} & $0.0278^{+0.0030}_{-0.0046}$ & $0.1055^{+0.0119}_{-0.0091}$ & $0.284\pm0.016$ & -- & -- & $<0.666$ & $68.71^{+1.73}_{-1.74}$ \\
 & HzBSND\tnote{c} & $0.0429^{+0.0071}_{-0.0170}$ & $0.0641^{+0.0371}_{-0.0235}$ & $0.251^{+0.038}_{-0.031}$ & -- & -- & $1.863^{+0.674}_{-1.316}$ & $65.41^{+1.91}_{-2.08}$ \\
 & HzBSNPD\tnote{d} & $0.0279^{+0.0031}_{-0.0048}$ & $0.1047^{+0.0125}_{-0.0096}$ & $0.284^{+0.017}_{-0.016}$ & -- & -- & $0.320^{+0.108}_{-0.277}$ & $68.48^{+1.71}_{-1.70}$ \\
 & HzBSNPDQH\tnote{e} & $0.0289^{+0.0025}_{-0.0040}$ & $0.1073^{+0.0116}_{-0.0081}$ & $0.283^{+0.016}_{-0.014}$ & -- & -- & $0.261^{+0.067}_{-0.254}$ & $69.57\pm0.94$ \\
\\
Non-flat $\phi$CDM & $H(z)$ + BAO & $0.0482^{+0.0126}_{-0.0190}$ & $0.0544^{+0.0194}_{-0.0497}$ & $0.242^{+0.024}_{-0.046}$ & $-0.103\pm0.132$ & -- & $2.618^{+1.213}_{-1.226}$ & $65.14^{+2.02}_{-2.29}$\\
 & HzBSNP\tnote{b} & $0.0260^{+0.0033}_{-0.0051}$ & $0.1159^{+0.0163}_{-0.0161}$ & $0.296\pm0.022$ & $-0.106\pm0.102$ & -- & $0.454^{+0.174}_{-0.372}$ & $69.33\pm1.86$ \\
 & HzBSND\tnote{c} & $0.0427^{+0.0076}_{-0.0177}$ & $0.0670^{+0.0379}_{-0.0282}$ & $0.253^{+0.037}_{-0.039}$ & $-0.097\pm0.130$ & -- & $2.058^{+0.779}_{-1.269}$ & $65.86\pm2.09$ \\
 & HzBSNPD\tnote{d} & $0.0264^{+0.0034}_{-0.0052}$ & $0.1139\pm0.0161$ & $0.295\pm0.022$ & $-0.105\pm0.104$ & -- & $0.509^{+0.212}_{-0.370}$ & $69.06^{+1.84}_{-1.83}$ \\
 & HzBSNPDQH\tnote{e} & $0.0265^{+0.0031}_{-0.0048}$ & $0.1142^{+0.0141}_{-0.0123}$ & $0.293\pm0.020$ & $-0.085\pm0.081$ & -- & $0.399^{+0.159}_{-0.313}$ & $69.53\pm0.95$ \\
\bottomrule
\end{tabular}
\begin{tablenotes}[flushleft]
\item [a] \hunit.
\item [b] $H(z)$ + BAO + SN-Pantheon.
\item [c] $H(z)$ + BAO + SN-DES.
\item [d] $H(z)$ + BAO + SN-Pantheon + SN-DES.
\item [e] $H(z)$ + BAO + SN-Pantheon + SN-DES + QSO + \hiig.
\end{tablenotes}
\end{threeparttable}%
}
\end{table*}

\subsection{HzBSNPDQH constraints}
\label{subsec:HzBSNPDQH}
Since the constraints derived from $H(z)$, BAO, SN-Pantheon, SN-DES, QSO, and \hiig\ data are not inconsistent, in this subsection we jointly analyze HzBSNPDQH data to determine more restrictive constraints on the cosmological parameters (though as discussed in Sec. \ref{makereference5.4.1}, we believe these constraints to be less reliable than those that stem from the HzSNPD combination, so we only describe the broad outlines here).

For flat \lcdm, the error bars we derive for \obhs\!, \ochs\!, and \om\ are larger than the \textit{Planck} error bars, though our central estimates of these quantities are broadly consistent with those derived from \textit{Planck}. In a similar fashion, we find larger error bars on $H_0$ in flat \lcdm\ than does \textit{Planck}, though our central estimate is higher than theirs. Generally, the constraints we derive on $H_0$ are more consistent with the median statistics estimate of $H_0=68 \pm 2.8$ \hunit\ \citep{chenratmed}, than with the local Hubble constant measurement of $H_0 = 74.03 \pm 1.42$ \hunit\ \citep{riess_etal_2019}.

We find mild evidence for spatial curvature, with non-flat XCDM and \pcdm\ favoring closed geometry, and non-flat \lcdm\ mildly favoring open geometry. The constraints from non-flat \lcdm\ and XCDM are consistent with spatially flat hypersurfaces to within less than 1$\sigma$. Additionally, we find mild evidence for dark energy dynamics, with the best-fitting value of $w_{\rm X}$ being 0.86$\sigma$ (1.45$\sigma$) away from $w_{\rm X}=-1$ in flat (non-flat) XCDM, and the best-fitting value of $\alpha$ being 1.03$\sigma$ (1.27$\sigma$) away from $\alpha=0$ in flat (non-flat) \pcdm.

\subsection{Model comparison}
\label{makereference5.4.1}

The values of $\Delta\chi^2$, $\Delta AIC$, $\Delta BIC$, and the reduced $\chi^2$ ($\chi^2/\nu$) are reported in Table \ref{tab:cabC5}, where $\Delta \chi^2$, $\Delta AIC$, and $\Delta BIC$, respectively, are defined as the differences between the values of the $\chi^2$, $AIC$, and $BIC$ for a given model and their corresponding minimum values among all models. From Table \ref{tab:cabC5}, we see that the reduced $\chi^2$ values determined from the $H(z)$ + BAO data combination range from 0.49 to 0.62, which is probably due to the $H(z)$ data having overestimated error bars (see \citealp{Caoetal_2021} for discussions of the systematic errors of these data). As discussed in \cite{Ryanetal2019} and \cite{CaoRyanRatra2020}, the underestimated systematic uncertainties in QSO and \hiig\ data\footnote{Roberto Terlevich and his colleagues are currently investigating the systematic uncertainties of the \hiig\ data, the results of which they plan to publish in a future paper (Roberto Terlevich, private communication, 2021).} result in larger reduced $\chi^2$ ($\sim1.34$) for the models in the HzBSNPDQH case. The reduced $\chi^2$ values for the HzBSNP and HzBSNPD cases are around unity for all models and for the HzBSND case range from 0.77 to 0.87. Of the combinations we study here, on the basis of these reduced $\chi^2$ values, the HzBSNPD constraints should be viewed as the most reliable ones.

We find that based on the $AIC$ and $BIC$, flat \lcdm\ and flat \pcdm\ are the most favored models in different data combination cases. The $\Delta AIC$ results show that the most favored model is flat \lcdm\ in the HzBSNP case, while the most favored model is flat \pcdm\ in the rest of the data combinations. The $\Delta BIC$ results show that the most favored model is flat \pcdm\ in the $H(z)$ + BAO and HzBSND cases, and is flat \lcdm\ in the remaining cases. For both $\Delta AIC$ and $\Delta BIC$ results, the most disfavored model is non-flat \lcdm\ in the $H(z)$ + BAO and HzBSND cases, and is non-flat XCDM in all other cases, with positive evidence against non-flat \lcdm\ and either positive or very strong evidence (depending on the data combination) against non-flat XCDM.

Overall, the $\Delta AIC$ results show no strong evidence against any model, and neither do the $\Delta BIC$ results for the $H(z)$ + BAO and HzBSND cases. However, in the HzBSNP and HzBSNPDQH cases, the $\Delta BIC$ results show strong evidence against the non-flat \lcdm\ and flat XCDM models, and very strong evidence against the non-flat \pcdm\ and XCDM models. In the HzBSNPD case, the evidence against flat XCDM and flat \pcdm\ is positive, the evidence against non-flat \lcdm\ is strong, and the evidence against non-flat \pcdm\ and non-flat XCDM is very strong. Based on the $\Delta \chi^2$ results, non-flat \pcdm\ has the minimum $\chi^2$ in all cases. 

In summary, the HzBSNPD data favor flat \pcdm\ ($AIC$) or flat \lcdm\ ($BIC$) among the six models we study here.

\begin{table*}
\centering
\resizebox*{1\columnwidth}{0.7\columnwidth}{%
\begin{threeparttable}
\caption{$\Delta \chi^2$, $\Delta AIC$, $\Delta BIC$, and $\chi^2_{\mathrm{min}}/\nu$ values.}\label{tab:cabC5}
\begin{tabular}{lccccccc}
\toprule
 Quantity & Data set & Flat \lcdm & Non-flat \lcdm & Flat XCDM & Non-flat XCDM & Flat \pcdm & Non-flat \pcdm\\
\hline
 & $H(z)$ + BAO & 5.48 & 5.42 & 1.49 & 0.15 & 1.32 & 0.00\\
 & HzBSNP\tnote{a} & 2.91 & 2.91 & 2.32 & 1.51 & 1.15 & 0.00 \\
$\Delta \chi^2$ & HzBSND\tnote{b} & 6.74 & 6.19 & 1.37 & 0.25 & 1.08 & 0.00\\
 & HzBSNPD\tnote{c} & 3.33 & 3.22 & 2.10 & 1.23 & 1.05 & 0.00\\
 & HzBSNPDQH\tnote{d} & 2.99 & 2.99 & 2.27 & 1.25 & 0.95 & 0.00\\
 \\
 & $H(z)$ + BAO & 2.16 & 4.10 & 0.17 & 0.83 & 0.00 & 0.68\\
 & HzBSNP\tnote{a} & 0.00 & 2.00 & 1.41 & 2.60 & 0.24 & 1.09 \\
$\Delta AIC$ & HzBSND\tnote{b} & 3.66 & 5.11 & 0.29 & 1.17 & 0.00 & 0.92\\
 & HzBSNPD\tnote{c} & 0.28 & 2.17 & 1.05 & 2.18 & 0.00 & 0.95\\
 & HzBSNPDQH\tnote{d} & 0.04 & 2.04 & 1.32 & 2.30 & 0.00 & 1.05\\
 \\
 & $H(z)$ + BAO & 0.43 & 4.10 & 0.17 & 2.57 & 0.00 & 2.42\\
 & HzBSNP\tnote{a} & 0.00 & 6.99 & 6.40 & 12.58 & 5.23 & 11.07 \\
$\Delta BIC$ & HzBSND\tnote{b} & 1.53 & 5.11 & 0.29 & 3.30 & 0.00 & 3.04\\
 & HzBSNPD\tnote{c} & 0.00 & 6.90 & 5.78 & 11.92 & 4.72 & 10.69\\
 & HzBSNPDQH\tnote{d} & 0.00 & 7.23 & 6.51 & 12.72 & 5.19 & 11.47\\
 \\
 & $H(z)$ + BAO & 0.61 & 0.62 & 0.52 & 0.49 & 0.51 & 0.49\\
 & HzBSNP\tnote{a} & 0.97 & 0.97 & 0.97 & 0.97 & 0.97 & 0.97 \\
$\chi^2_{\mathrm{min}}/\nu$ & HzBSND\tnote{b} & 0.86 & 0.87 & 0.78 & 0.78 & 0.78 & 0.77\\
 & HzBSNPD\tnote{c} & 0.98 & 0.98 & 0.98 & 0.98 & 0.97 & 0.97\\
 & HzBSNPDQH\tnote{d} & 1.34 & 1.34 & 1.34 & 1.34 & 1.34 & 1.34\\
\bottomrule
\end{tabular}
\begin{tablenotes}[flushleft]
\item [a] $H(z)$ + BAO + SN-Pantheon.
\item [b] $H(z)$ + BAO + SN-DES.
\item [c] $H(z)$ + BAO + SN-Pantheon + SN-DES.
\item [d] $H(z)$ + BAO + SN-Pantheon + SN-DES + QSO + \hiig.
\end{tablenotes}
\end{threeparttable}%
}
\end{table*}

\section{Conclusion}
\label{makereference5.5}

By analyzing a total of 1383 measurements, consisting of 31 $H(z)$, 11 BAO, 1048 SN-Pantheon, 20 SN-DES, 120 QSO, and 153 \hiig\ data points, we jointly constrain cosmological parameters in six flat and non-flat cosmological models.

From the constraints derived using the cosmological models, we can identify some relatively model-independent features. As discussed in Sec. \ref{makereference5.4.1}, the $H(z)$ + BAO + SN-Pantheon + SN-DES (HzBSNPD) data combination produces the most reliable constraints. In particular, for the HzBSNPD data combination, we find a reasonable and fairly restrictive summary value of $\Omega_{\rm m_0}=0.294 \pm 0.020$,\footnote{Here we take the summary central value to be the mean of the two of six central-most values. As for the uncertainty, we call the difference between the two central-most values twice the systematic uncertainty and the average of the two central-most error bars the statistical uncertainty, and compute the summary error bar as the quadrature sum of the two uncertainties.} which is in good agreement with many other recent measurements (e.g. $0.315\pm0.007$ from \citealp{planck2018b}). A fairly restrictive summary value of $H_0=68.8\pm1.8$ \hunit\ is found to be in better agreement with the estimates of \cite{chenratmed} and \cite{planck2018b} than with the measurement of \cite{riess_etal_2019}; note that the constraints from BAO data do not depend on physics of the early Universe (with $\Omega_{\rm b_0}\!h^2$ being a free parameter that is fitted to the data used here). There is some room for dark energy dynamics or a little spatial curvature energy density in the HzBSNPD constraints, but based on $AIC$ and $BIC$ criteria, flat \pcdm\ or flat \lcdm\ are the best candidate models.


\cleardoublepage


\chapter{Cosmological constraints from H\,\textsc{ii} starburst galaxy, quasar angular size, and other measurements}
\label{makereference6}

This chapter is based on \cite{CaoRyanRatra2022}.

\section{Introduction} 
\label{makereference6.1}

Many observations indicate that the Universe is currently in a phase of accelerated expansion, however, the theory behind this is not yet well-established. Although the spatially flat \lcdm\ model\footnote{The flat \lcdm\ model has flat spatial hypersurfaces and a time-independent dark energy, a cosmological constant $\Lambda$, that provides approximately 70\% of the current cosmological energy budget. Non-relativistic cold dark matter (CDM) accounts for approximately 25\% and non-relativistic baryonic matter accounts for almost all of the remaining $\sim 5\%$ of the energy budget.} \citep{peeb84} is consistent with most observations (see e.g.\ \citealp{Farooq_Ranjeet_Crandall_Ratra_2017,scolnic_et_al_2018,planck2018b,eBOSS_2020}), some potential observational discrepancies and theoretical puzzles (see e.g.\ \citealp{DiValentinoetal2021b, PerivolaropoulosSkara2021}) suggest that there still is room for other cosmological models, including, for example, non-flat \lcdm\ (the \citealp{planck2018b} cosmic microwave background (CMB) anisotropy TT,TE,EE+lowE+lensing data favor positive spatial curvature) as well as dynamical dark energy. These discrepancies and puzzles motivate us to also study dynamical dark energy models and spatially non-flat models in this paper.

In this paper we use the new \cite{GonzalezMoranetal2021} \hii\ starburst galaxy (\hiig) measured fluxes and inferred absolute luminosities (from their correlation with their measured ionized gas velocity dispersions) as standard candles to constrain cosmological models.\footnote{Our analyses of the earlier \cite{GonzalezMoran2019} data are described in \cite{CaoRyanRatra2020} (also see \citealp{Caoetal_2021,CaoRyanRatra2021,Johnsonetal2022}). For analyses of earlier \hiig\ data, see  \cite{Melnick_2000,Siegel_2005,Plionis_2011,Mania_2012,Chavez_2014,Terlevich_2015,Chavez_2016}, and references therein. For recent analyses of the \cite{GonzalezMoranetal2021} \hiig\ measurements, see \cite{Tsiapietal2021, Mehrabietal2021}.} These new \hiig\ data reach to a slightly higher redshift $z \sim 2.5$, somewhat higher than the baryon acoustic oscillation (BAO) standard ruler data that reach to $z \sim 2.3$, that we also use in this paper. In order to determine the expansion rate and geometry of the Universe, it is vital to measure distances using either standard candles or standard rules. More data sets probing wider redshift regions would provide more information and make more contributions to a better understanding of our Universe, so it is worthwhile to seek additional standard rulers. The angular sizes of quasars (QSOs) provide one such additional probe, reaching to $z \sim 2.7$, which we have explored in previous work \citep{Ryanetal2019, CaoRyanRatra2020, Caoetal_2021, CaoRyanRatra2021}. As described in those papers and in Sec. \ref{makereference6.2} below, intermediate luminosity QSOs have, over a fairly wide range of redshifts ($0.46 \lesssim z \lesssim 2.7$), very similar intrinsic lengths $l_{\rm m} = 11.03 \pm 0.25$ pc \citep{Cao_et_al2017b}. A knowledge of this intrinsic length scale, combined with measurements of the angular sizes of these QSOs allows one to determine the angular diameter distance out to the redshifts of the QSOs.

The QSO data that we have used in the past (from \citealp{Cao_et_al2017b}) have the following drawback, however: $l_{\rm m}$ was determined with a Gaussian process interpolation \citep{Seikel_Clarkson_Smith_2012} of the Hubble parameter from Hubble parameter ($H(z)$) data (as described in \citealp{Cao_et_al2017b}), many of which we have used in our previous analyses and also use in this paper. We have discussed this correlation between our QSO data and our $H(z)$ data in the past \citep{Ryan_powerlaw, CaoRyanRatra2021}, although we made the assumption in our earlier analyses that the correlation is not significant enough to have a strong effect on our results (owing to the weakness of the constraints from QSO data). Here we sidestep this problem by treating $l_{\rm m}$ as a free (nuisance) parameter, thereby constraining its value directly from our analysis. As discussed in Sec. \ref{makereference6.4}, we find that the value of $l_{\rm m}$ is almost independent of cosmological model, and is consistent with the value $l_{\rm m} = 11.03 \pm 0.25$ pc from \cite{Cao_et_al2017b} that we used in our earlier work. This finding suggests that these QSOs are close to being standard rulers, and it validates the result of \cite{Cao_et_al2017b}, independently of their method.

Significant constraints on cosmology now largely come from only a few data sets, at low $z \lesssim 2.3$, including BAO, Type Ia supernova (SN Ia), and $H(z)$ measurements, and at $z \sim 1100$ from CMB anisotropy observations. As mentioned above, it is useful and important to develop new probes, especially in the intermediate $ 2.3 \lesssim z \lesssim 1100$ redshift range. \hiig\ is an example, as are QSO angular sizes that have been under discussion for a longer time (see e.g.\  \citealp{gurvits_kellermann_frey_1999, vishwakarma_2001, lima_alcaniz_2002, zhu_fujimoto_2002, Chen_Ratra_2003}) with the compilation of \cite{Cao_et_al2017b} being a significant step forward. Other probes under development now include reverberation-measured Mg\,\textsc{ii} time-lag radius-luminosity relation QSOs that reach to $z \sim 1.9$ (\citealp{Mehrabietal2021, Czernyetal2021, Zajaceketal2021, Yuetal2021, Khadkaetal_2021a}). High redshift options include QSO X-ray and UV flux measurements which extend to $z \sim 7.5$ (\citealp{RisalitiLusso2015, RisalitiLusso2019, KhadkaRatra2020a, KhadkaRatra2020b, KhadkaRatra2022, KhadkaRatra2021, Yangetal2020, Lussoetal2020, Lietal2021, Lian_etal_2021}),\footnote{However the current QSO compilation is standardizable up to only $z \sim 1.5$--1.7 (\citealp{KhadkaRatra2021, KhadkaRatra2022}).} and gamma-ray burst (GRB) data that extend to $z \sim 8.2$ (\citealp{Amati2008, Amati2019, samushia_ratra_2010, Wang_2016, Demianskietal_2021, Dirirsa2019, KhadkaRatra2020c, Khadkaetal_2021b, Wangetal_2021, Huetal2021}).\footnote{Only a smaller sample of 118 GRBs is reliable enough to be used for cosmological purposes, but include GRBs that probe to $z \sim 8.2$ (\citealp{KhadkaRatra2020c, Khadkaetal_2021b}).} As of now, all five of these probes provide mostly only weak cosmological constraints, but new data should yield tighter constraints that have the potential to soon usefully probe the largely unexplored $2 \lesssim z \lesssim 8$ part of cosmological redshift space.

Our comparisons here between the constraints from the new \cite{GonzalezMoranetal2021} data and the old \cite{GonzalezMoran2019} data show that the new data provide more restrictive constraints on most cosmological parameters. As noted above, QSO angular size data provide relatively cosmological model-independent estimates of $l_{\rm m}$. We find that the cosmological constraints from $H(z)$, BAO, SN Ia, QSO, and the new \hiig\ measurements are not mutually inconsistent, thus we combine them to provide more restrictive constraints on the cosmological and nuisance parameters. The almost model-independent summary constraints from this data combination are measurements of the Hubble constant, $H_0=69.7\pm1.2$ \hunit, the non-relativistic matter density parameter, $\Omega_{\rm m_0}=0.293\pm0.021$, and the QSO characteristic linear size, $l_{\rm m}=10.93\pm0.25$ pc. The estimate of $H_0$ is in better agreement with the median statistics estimate of \cite{chenratmed} ($H_0 = 68 \pm 2.8$ \hunit) than with the measurements of \cite{planck2018b} ($H_0 = 67.4 \pm 0.5$ \hunit) and \cite{Riess_2021} ($H_0 = 73.2 \pm 1.3$ \hunit). Although the most-favored model is the spatially-flat \lcdm\ model, there is room for some mild dark energy dynamics and a little non-zero spatial curvature energy density. We also find that currently accelerating cosmological expansion is favored by most of the data combinations we study (except for QSO data alone).

This paper is organized as follows. The models we study are described in Chapter \ref{sec:models}. The data we used are introduced in Section \ref{makereference6.2} with the data analysis method presented in Section \ref{makereference6.3}. We summarize our results and conclusions in Sections \ref{makereference6.4} and \ref{makereference6.5}.

\section{Data}
\label{makereference6.2}

In this paper our main focus is on a new set of \hiig\ data (\citealp{GonzalezMoranetal2021}, which we dub ``\hiig-2021''). We compare cosmological constraints from these \hiig-2021 data to those from earlier \hiig\ data. We also use these \hiig-2021 data and BAO, $H(z)$, SN Ia, and QSO angular size measurements to constrain cosmological parameters in the models we study.

The 31 $H(z)$ measurements we use, that span the redshift range $0.070 \leq z \leq 1.965$, are given in Table 2 of \cite{Ryan_1}.\footnote{These measurements were taken from \cite{69}, \cite{71}, \cite{70}, \cite{73}, \cite{72}, \cite{moresco_et_al_2016}, and \cite{15}.} The 11 BAO measurements we use, that span the redshift range $0.38 \leq z \leq 2.334$, are listed in Table 1 of \cite{CaoRyanRatra2021}.\footnote{These measurements were taken from \cite{Alam_et_al_2017}, \cite{3}, \cite{Carter_2018}, \cite{DES_2019b}, and \cite{duMas2020}.} Information on systematic errors of these data can be found in \cite{Caoetal_2021}.

The SN-Pantheon data we use consist of 1048 SN Ia measurements, spanning the redshift range $0.01<z<2.3$, compiled in \cite{scolnic_et_al_2018}. The SN-DES data we use consist of 20 binned measurements (of 207 SN Ia measurements), spanning the redshift range $0.015 \leq z \leq 0.7026$, compiled in \cite{DES_2019d}. See \cite{CaoRyanRatra2021} for a description of how we use these SN Ia data.

The QSO data we use, that span the redshift range $0.462 \leq z \leq 2.73$, are listed in Table 1 of \cite{Cao_et_al2017b}. These consist of 120 measurements of the angular size
\begin{equation}
    \theta(z) = \frac{l_{\rm m}}{D_{A}(z)}.
\end{equation}
Here $l_{\rm m}$ is the characteristic linear size of QSOs in the sample and $D_{A}$ (defined below) is the angular size distance. Here we improve on the approach of \cite{Cao_et_al2017b}, \cite{Ryanetal2019}, and \cite{CaoRyanRatra2020, Caoetal_2021, CaoRyanRatra2021}, by treating $l_{\rm m}$ as a nuisance parameter to be determined from these measurements so that these QSO data are independent of $H(z)$ data.

The old \hiig\ data (which we dub ``\hiig-2019'') consist of 107 low redshift measurements that span $0.0088 \leq z \leq 0.16417$, used in \cite{Chavez_2014} (recalibrated by \citealp{GonzalezMoran2019}), and 46 high redshift measurements that span $0.636427 \leq z \leq 2.42935$. The new \hiig-2021 data, comprising the original 107 low redshift measurements and 74 updated high redshift measurements (that now span $0.636427 \leq z \leq 2.545$), are listed in Table A3 of \cite{GonzalezMoranetal2021}. 

The correlation between \hiig\ luminosity ($L$) and velocity dispersion ($\sigma$) is in equation \eqref{eq:logL}. Note that the systematic uncertainties of both \hiig\ and QSO data are not considered so that the reduced $\chi^2$'s are relatively large.

The transverse comoving distance $D_M(z)$ in equation \eqref{eq:DM}, the luminosity distance $D_L(z)$, and the angular size distance $D_A(z)$ are related through equation \eqref{DM-DL-DA}.

\section{Data Analysis Methodology}
\label{makereference6.3}

In this paper we use the \textsc{class} code to compute cosmological model predictions as a function of the cosmological model and other parameters. These predictions are compared to observational data using the Markov chain Monte Carlo (MCMC) code \textsc{MontePython} \citep{Audrenetal2013} to maximize the likelihood function, $\mathcal{L}$, and thereby determine the best-fitting values of the free parameters. The priors on the cosmological parameters are flat and nonzero over the same ranges as used in \cite{CaoRyanRatra2021}, except that now $\obh\in[0.00499, 0.03993]$.\footnote{The value of primordial Helium abundance $Y_{\rm p}$ is set using a standard big-bang nucleosynthesis prediction by interpolation on a grid of values computed using version 1.2 of the PArthENoPE BBN code for a neutron lifetime of 880.2 s. Since we choose the effective number of relativistic neutrino species $N_{\rm eff}=3.046$, \obhs\ is therefore limited to the range of [0.00499, 0.03993] by the correlated predictions of $Y_{\rm p}$.} The prior range of the QSO nuisance parameter $l_{\rm m}$ is not bounded. We assume a minimal neutrino sector, with three massless neutrino species, with the effective number of relativistic neutrino species $N_{\rm eff} = 3.046$. We neglect the late-time contribution of non-relativistic neutrinos and treat the baryonic (\obhs) and cold dark matter (\ochs) energy density parameters as free cosmological parameters to be determined from the data. The non-relativistic matter density parameter $\Om = (\obh + \och)/{h^2}$ is a derived parameter.

The computation of the likelihood functions of $H(z)$, BAO, \hiig, and QSO data are described in \cite{CaoRyanRatra2020} and \cite{Caoetal_2021}, whereas that of the likelihood functions of SN Ia measurements can be found in \cite{CaoRyanRatra2021}. One can also find the definitions of the Akaike Information Criterion ($AIC$) and the Bayesian Information Criterion ($BIC$) in those papers.

\section{Results}
\label{makereference6.4}

\begin{figure*}
\centering
 \subfloat[]{%
    \includegraphics[width=3.25in,height=3.25in]{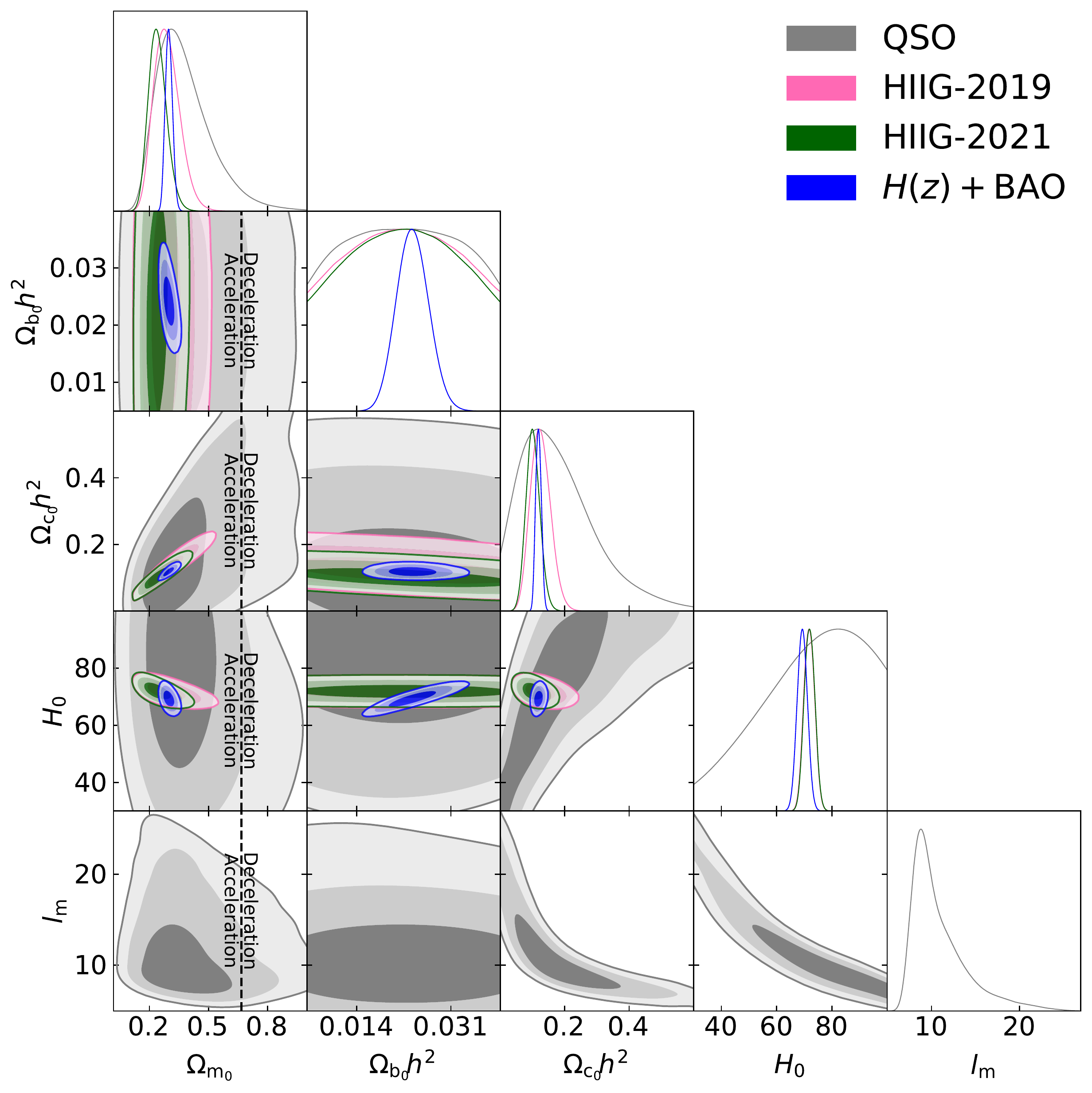}}
 \subfloat[]{%
    \includegraphics[width=3.25in,height=3.25in]{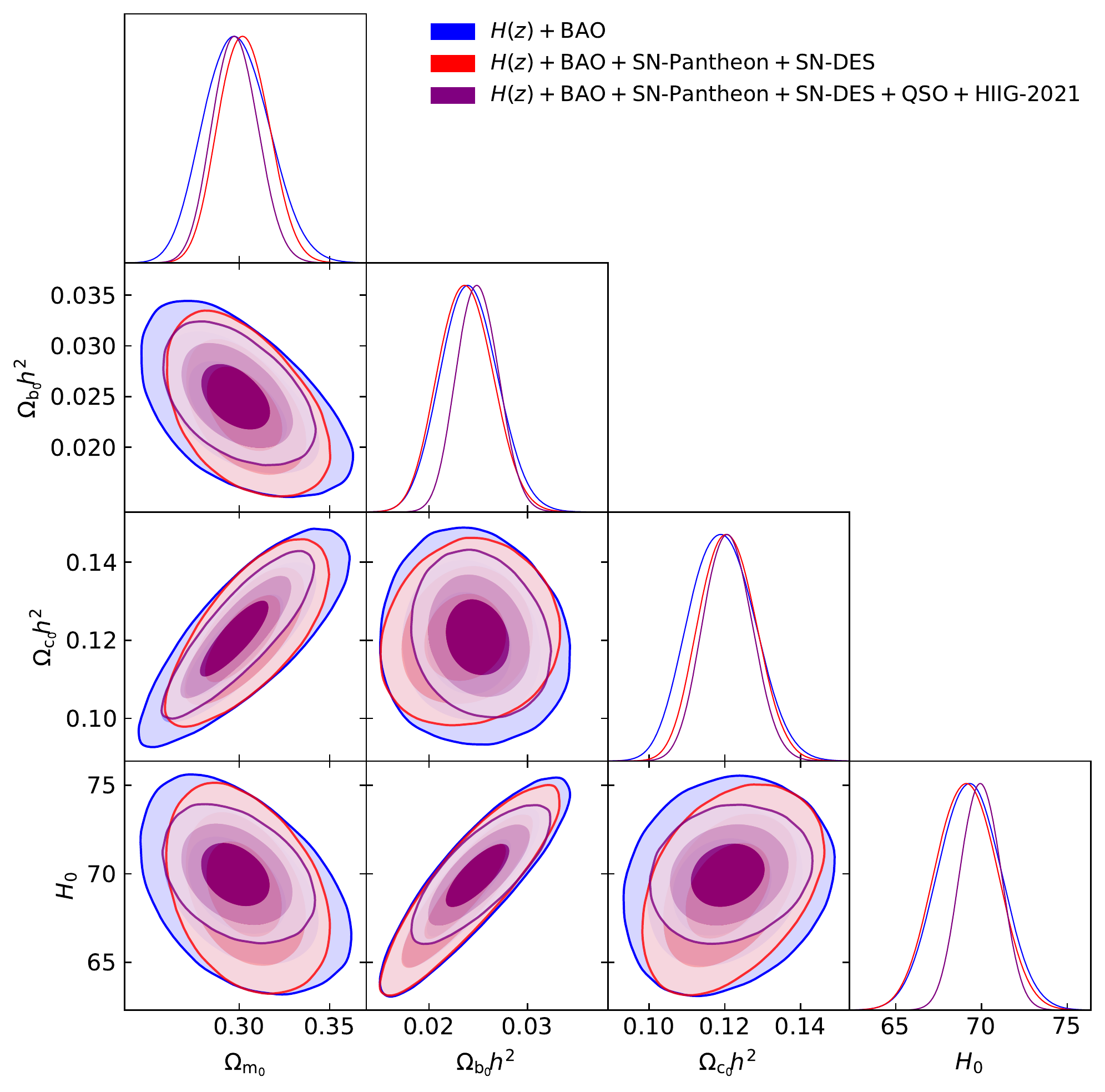}}\\
\caption{One-dimensional likelihoods and 1$\sigma$, 2$\sigma$, and 3$\sigma$ two-dimensional likelihood confidence contours for flat \lcdm. Left panel shows individual data set and $H(z)$ + BAO results and the right panel shows joint data sets constraints. The zero-acceleration lines are shown as black dashed lines in the left panel, which divide the parameter space into regions associated with currently-accelerating (left) and currently-decelerating (right) cosmological expansion, while in the right panel the joint analyses favor currently-accelerating expansion.}
\label{fig1C6}
\end{figure*}

\begin{figure*}
\centering
 \subfloat[]{%
    \includegraphics[width=3.25in,height=3.25in]{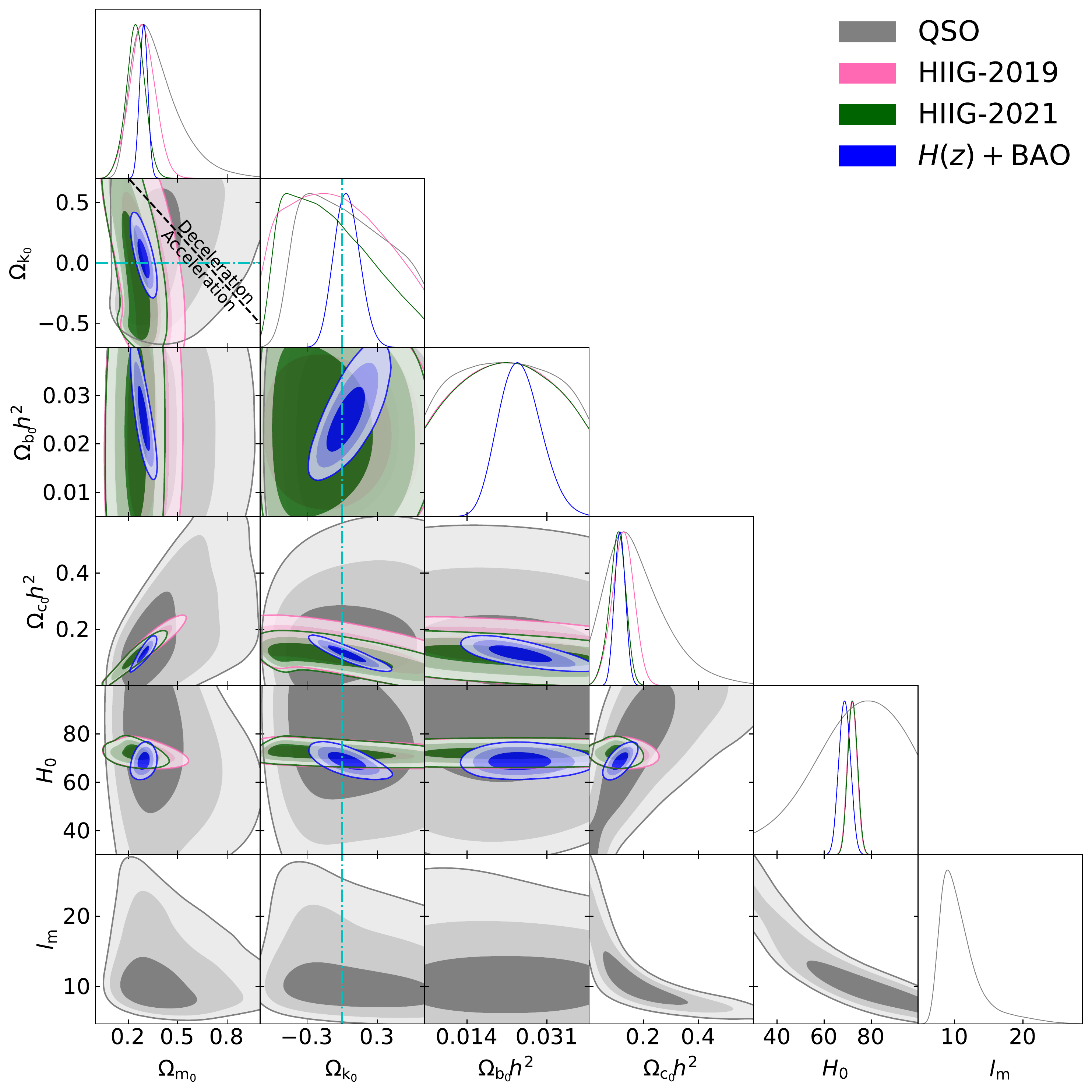}}
 \subfloat[]{%
    \includegraphics[width=3.25in,height=3.25in]{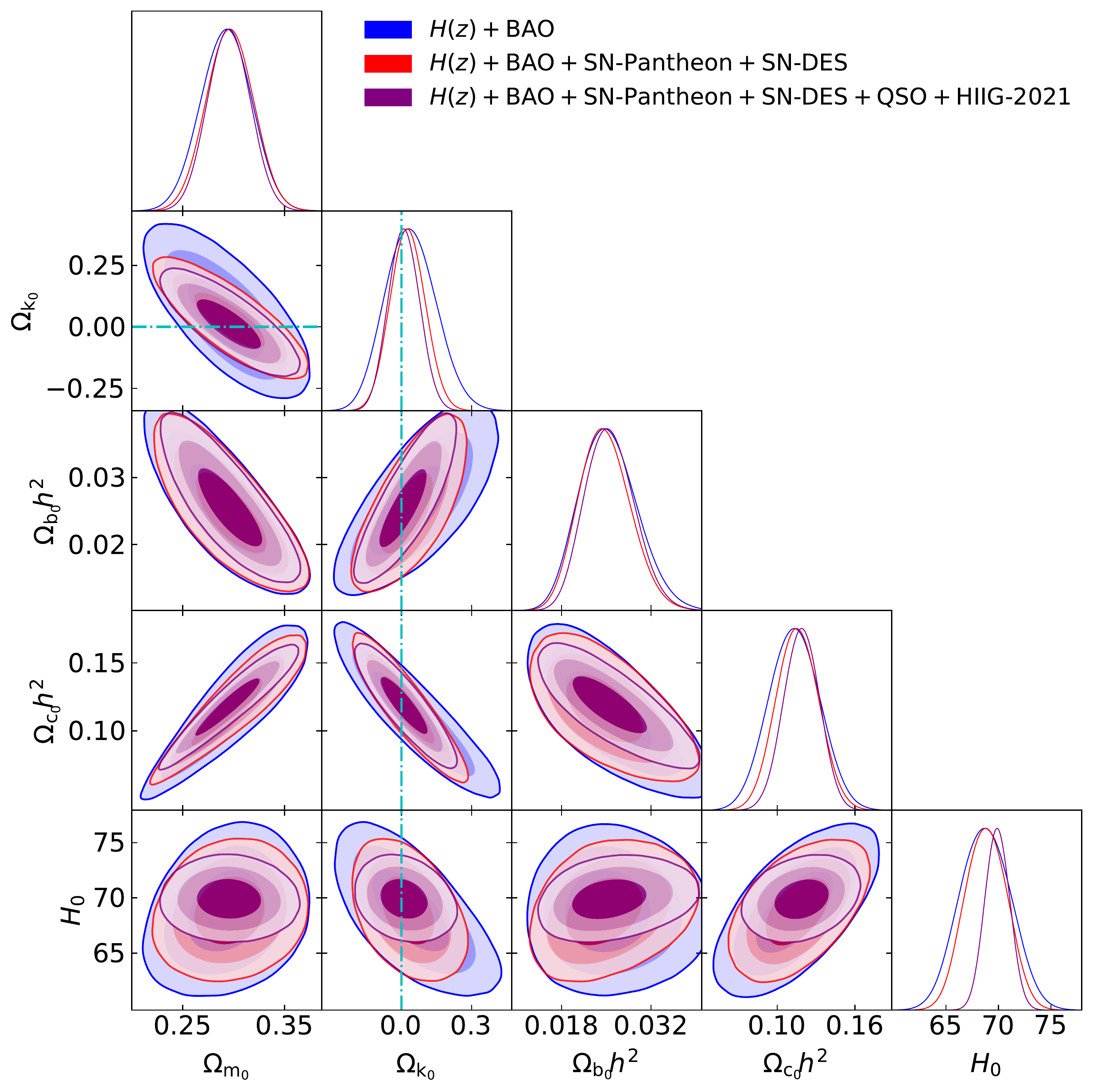}}\\
\caption{Same as Fig. \ref{fig1C6} but for non-flat \lcdm. The flat \lcdm\ case is shown as the cyan dash-dot lines, with closed spatial hypersurfaces either below or to the left. The black dashed line in the left panel is the zero-acceleration line, which divides the parameter space into regions associated with currently-accelerating (below left) and currently-decelerating (above right) cosmological expansion. In the right panel, the joint analyses favor currently-accelerating expansion.}
\label{fig2C6}
\end{figure*}

\begin{figure*}
\centering
 \subfloat[]{%
    \includegraphics[width=3.25in,height=3.25in]{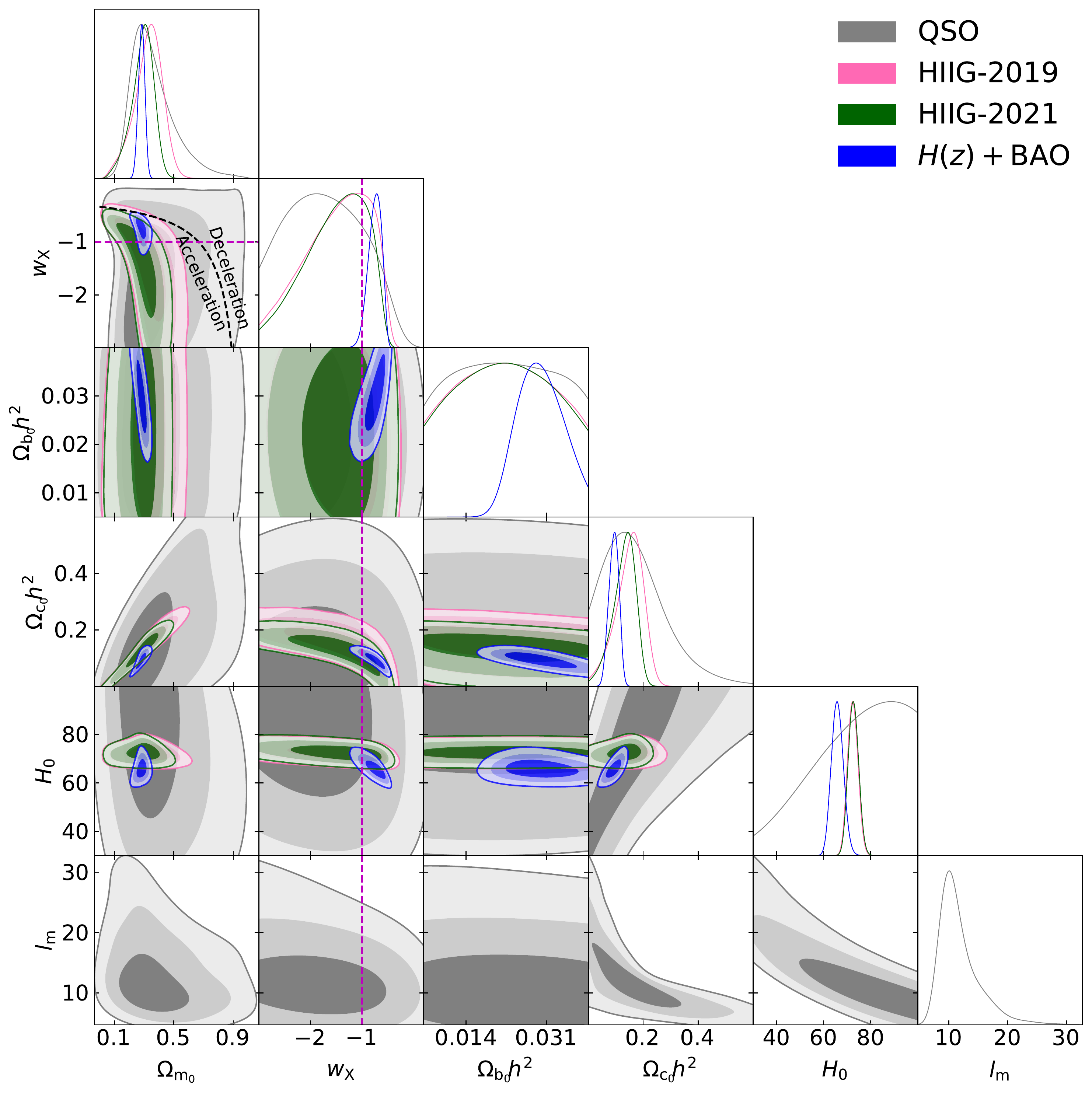}}
 \subfloat[]{%
    \includegraphics[width=3.25in,height=3.25in]{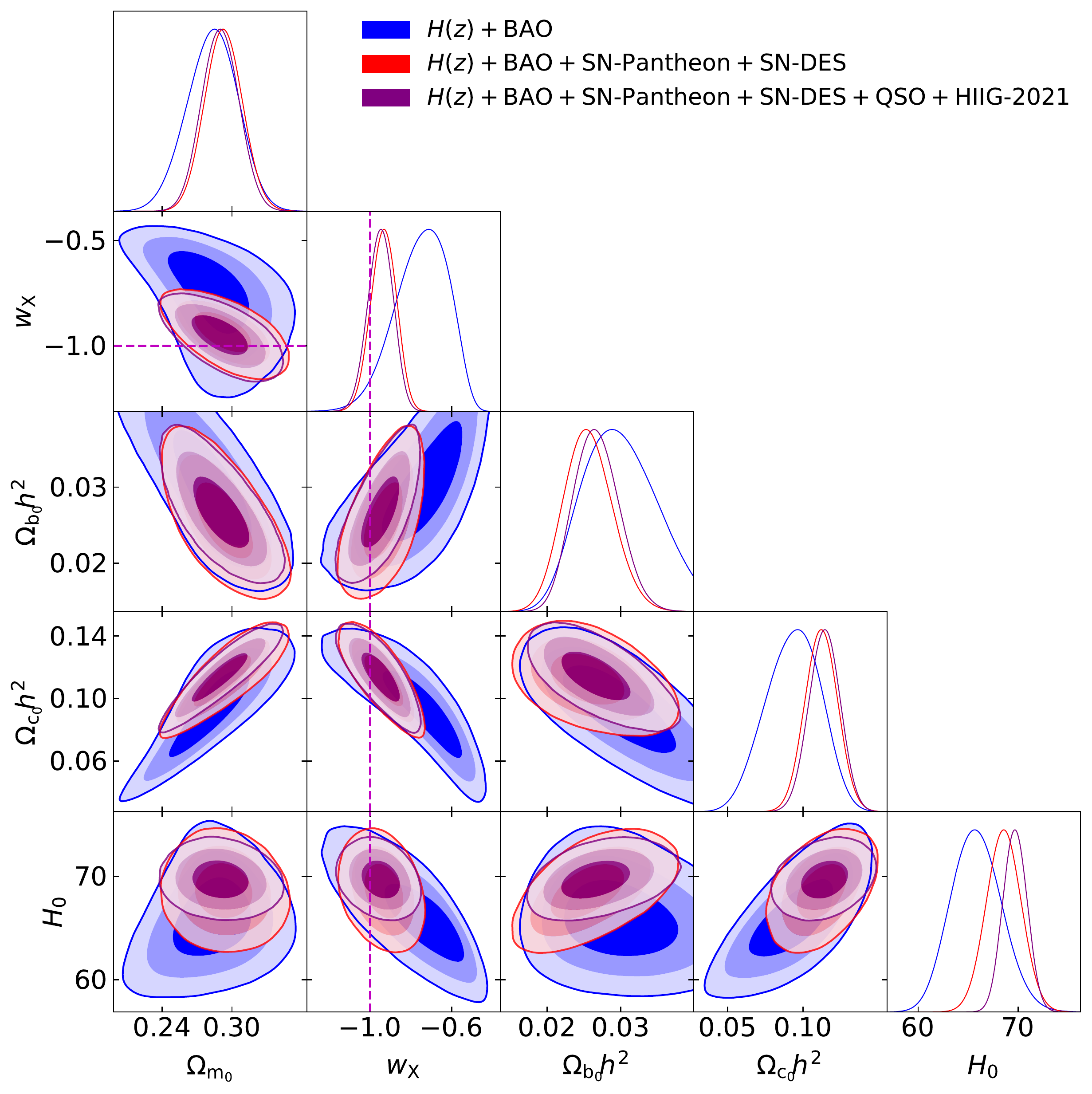}}\\
\caption{One-dimensional likelihoods and 1$\sigma$, 2$\sigma$, and 3$\sigma$ two-dimensional likelihood confidence contours for flat XCDM. The black dashed line in the left panel is the zero-acceleration line, which divides the parameter space into regions associated with currently-accelerating (below left) and currently-decelerating (above right) cosmological expansion. In the right panel, the joint analyses favor currently-accelerating expansion. The magenta lines represent $w_{\rm X}=-1$, i.e. the flat \lcdm\ model.}
\label{fig3C6}
\end{figure*}

\begin{figure*}
\centering
 \subfloat[]{%
    \includegraphics[width=3.25in,height=3.25in]{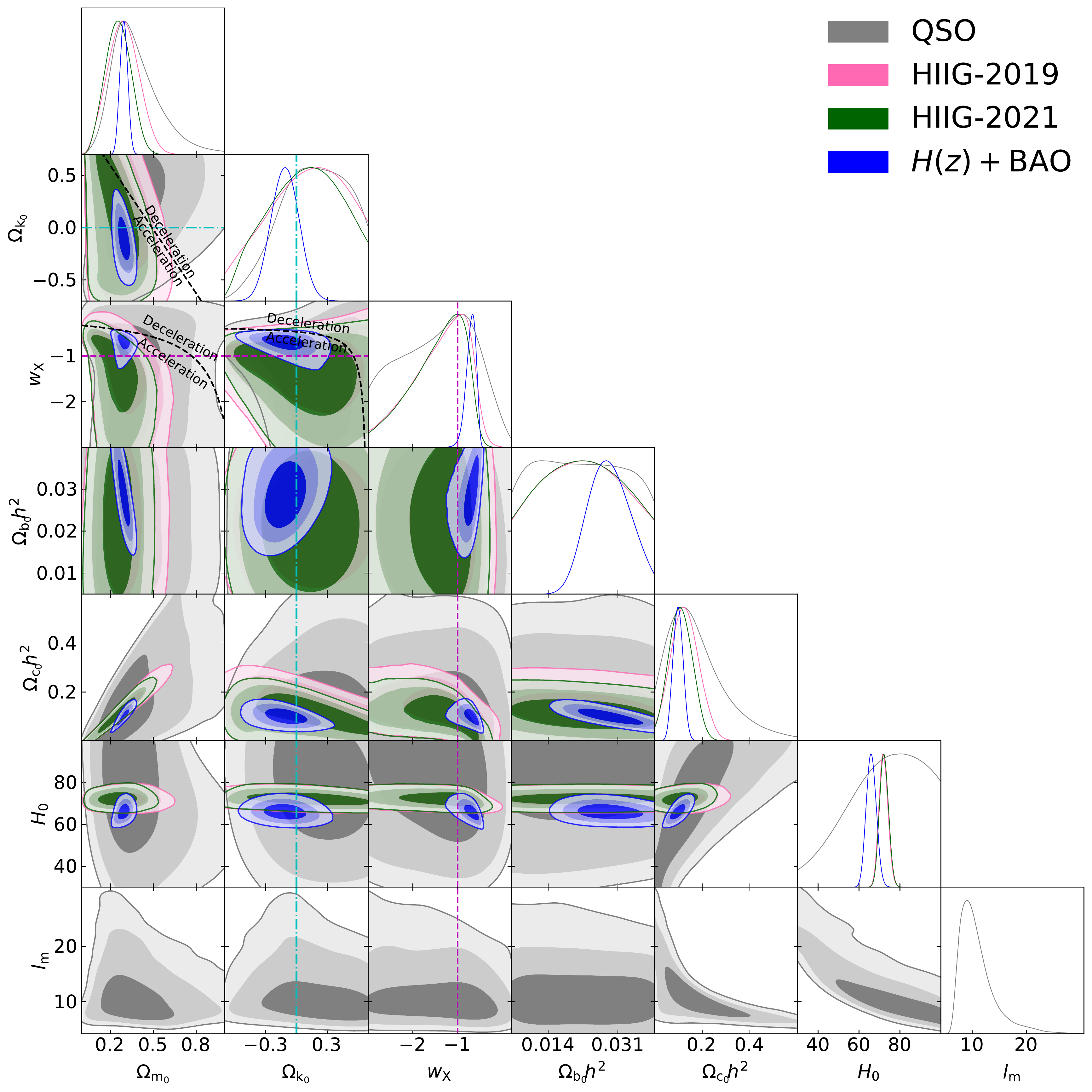}}
 \subfloat[]{%
    \includegraphics[width=3.25in,height=3.25in]{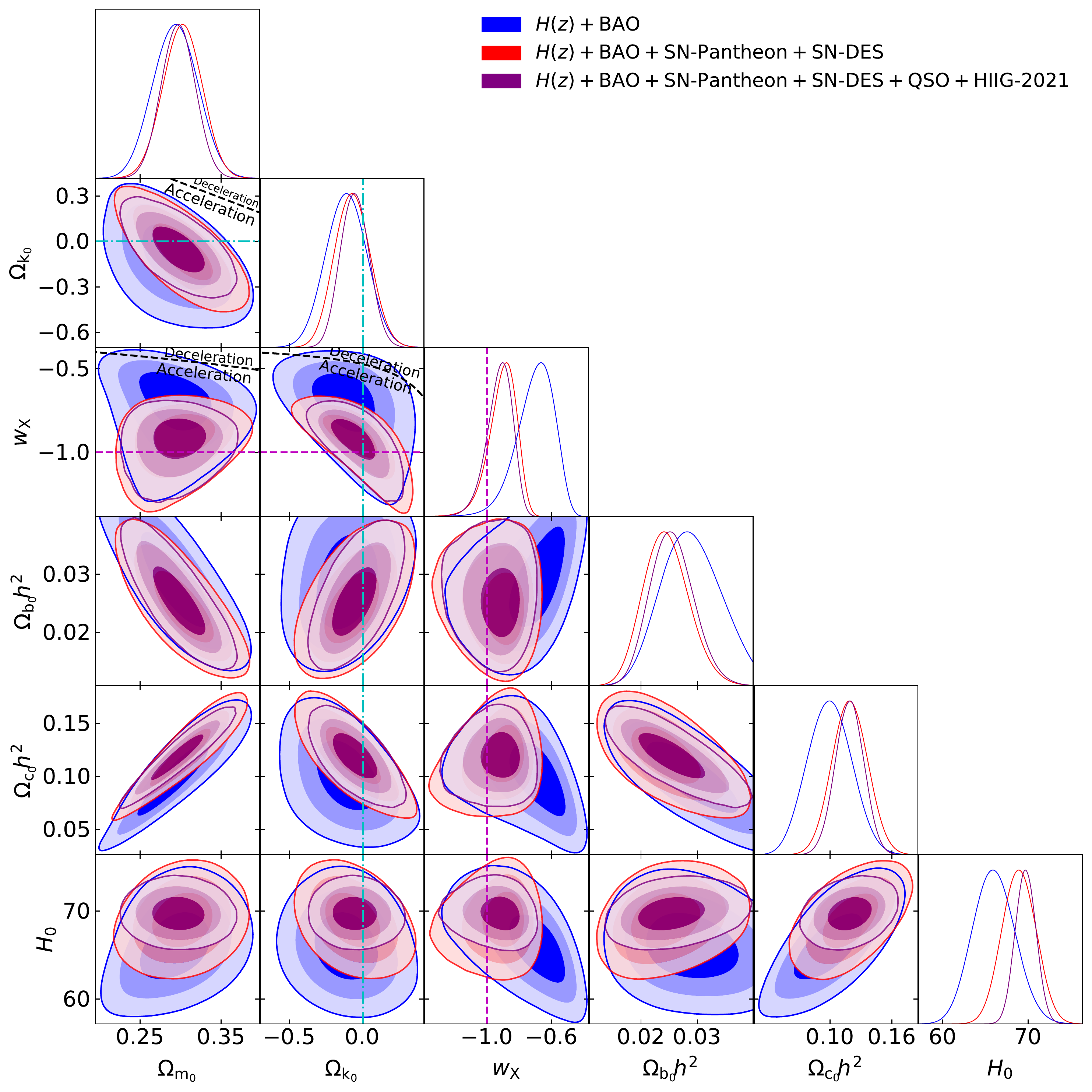}}\\
\caption{Same as Fig. \ref{fig3C6} but for non-flat XCDM, where the black dashed zero-acceleration lines are computed for the third cosmological parameter set to the $H(z)$ + BAO data best-fitting values listed in Table \ref{tab:BFPC6}, with currently-accelerating cosmological expansion residing below left. The flat XCDM case is denoted as the cyan dash-dot lines, with closed spatial hypersurfaces either below or to the left. The magenta lines represent $w_{\rm X} = -1$, i.e. the non-flat \lcdm\ model. In all cases except for the QSO only case, almost all of the favored parameter space is associated with currently-accelerating cosmological expansion.}
\label{fig4C6}
\end{figure*}

\begin{figure*}
\centering
 \subfloat[]{%
    \includegraphics[width=3.25in,height=3.25in]{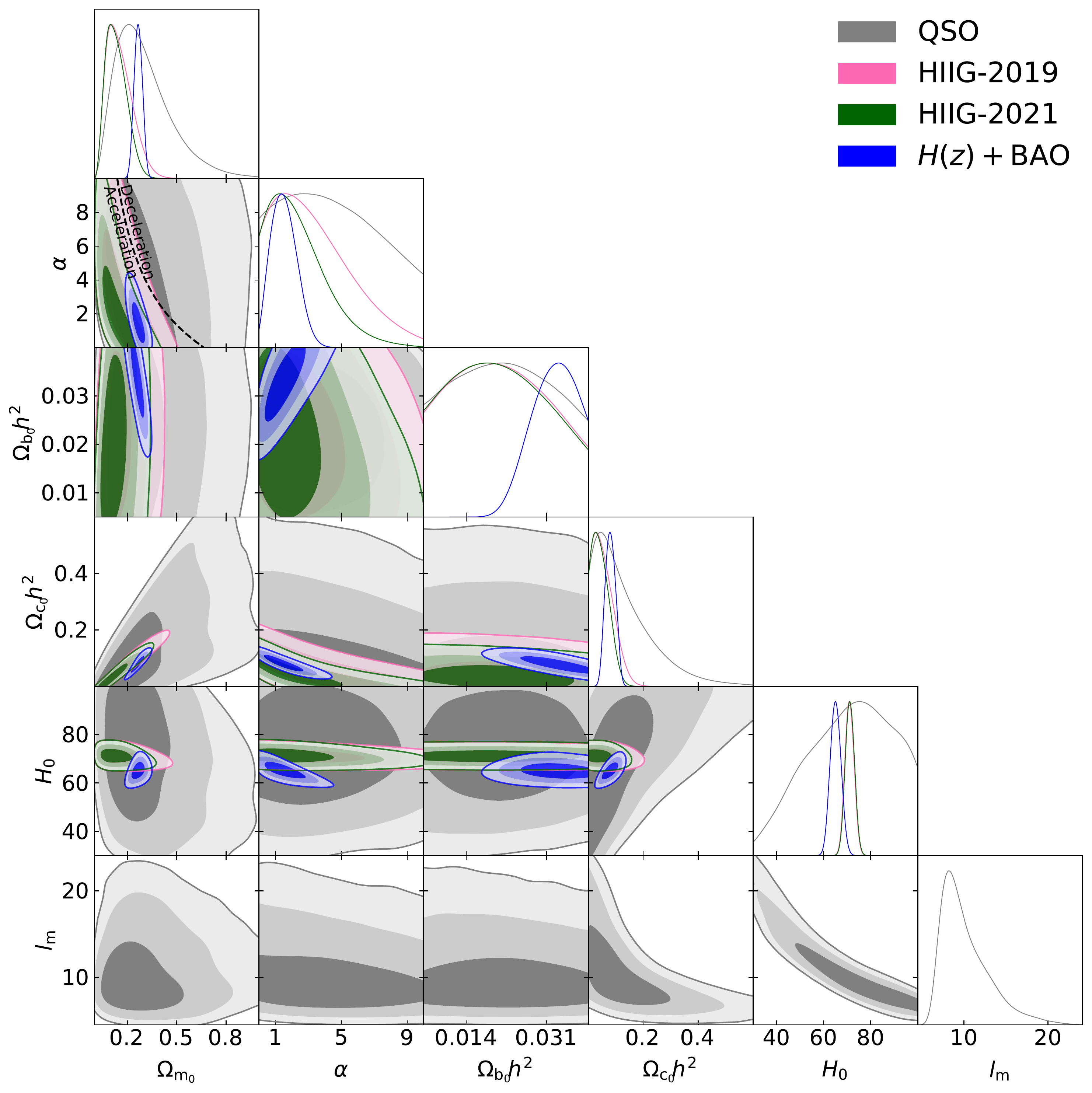}}
 \subfloat[]{%
    \includegraphics[width=3.25in,height=3.25in]{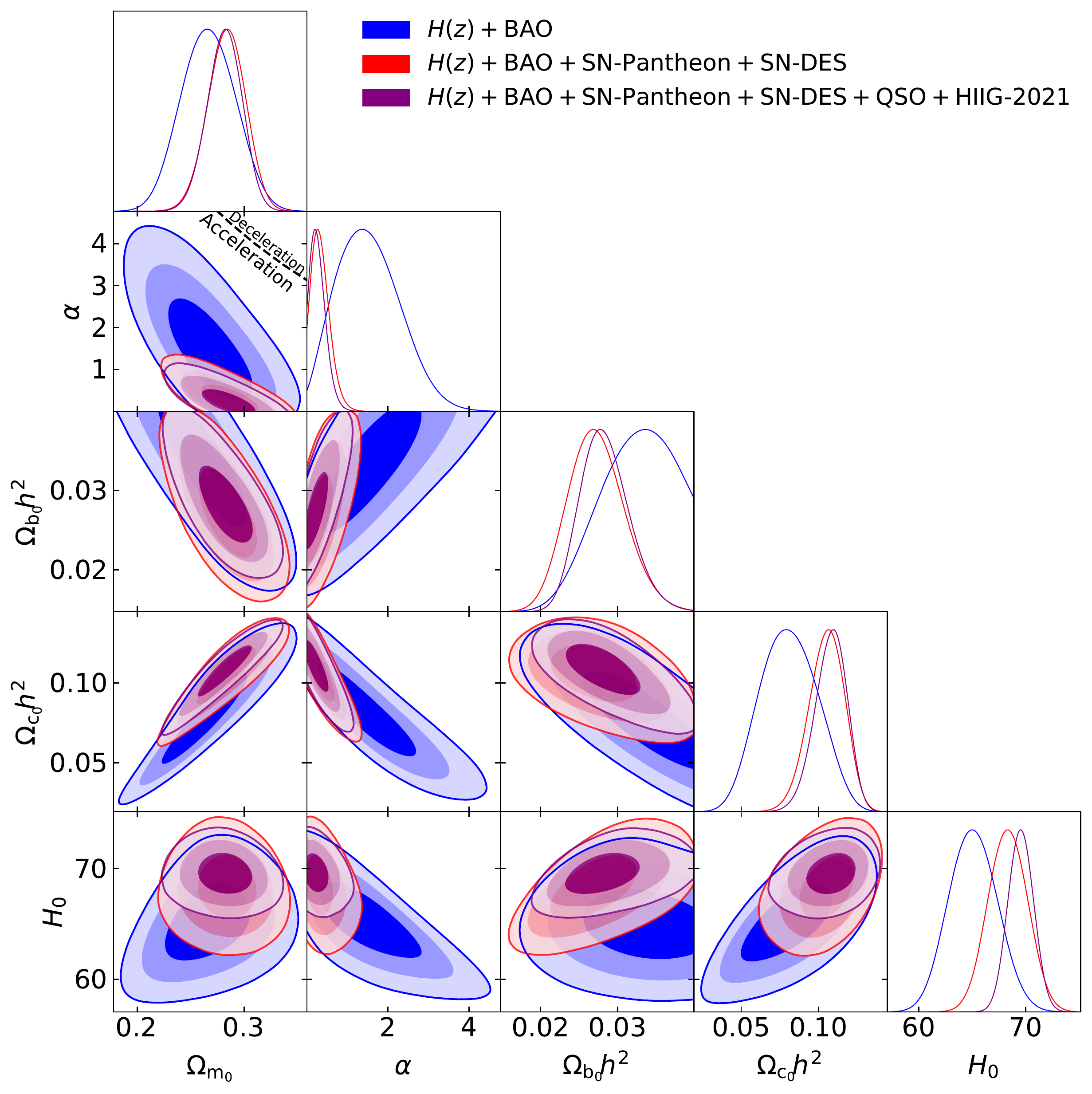}}\\
\caption{One-dimensional likelihoods and 1$\sigma$, 2$\sigma$, and 3$\sigma$ two-dimensional likelihood confidence contours for flat \pcdm. The black dashed lines are the zero-acceleration lines, which divides the parameter space into regions associated with currently-accelerating (below left) and currently-decelerating (above right) cosmological expansion. The $\alpha = 0$ axis is the flat \lcdm\ model. In all cases except for the QSO only case, almost all of the favored parameter space is associated with currently-accelerating cosmological expansion.}
\label{fig5C6}
\end{figure*}

\begin{figure*}
\centering
 \subfloat[]{%
    \includegraphics[width=3.25in,height=3.25in]{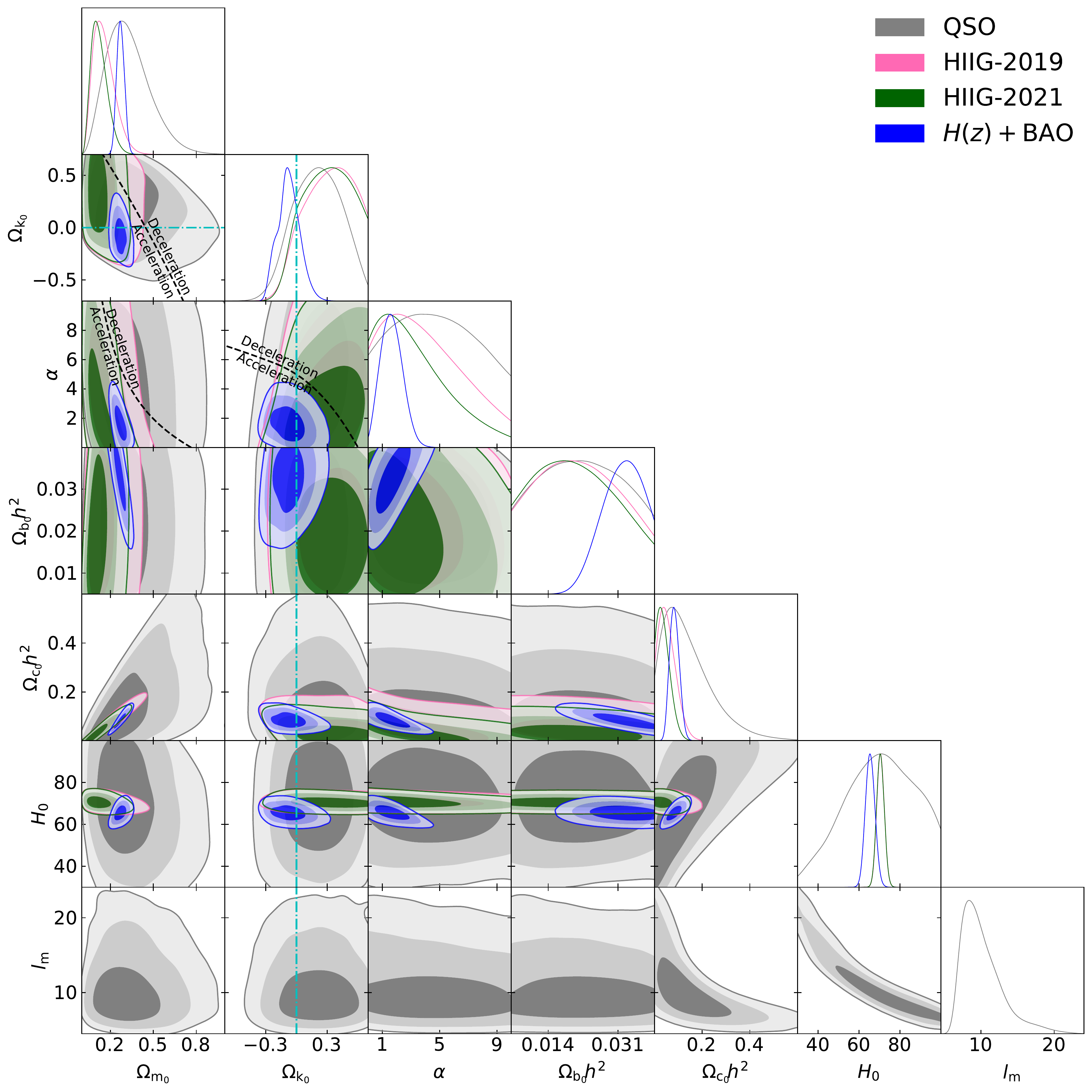}}
 \subfloat[]{%
    \includegraphics[width=3.25in,height=3.25in]{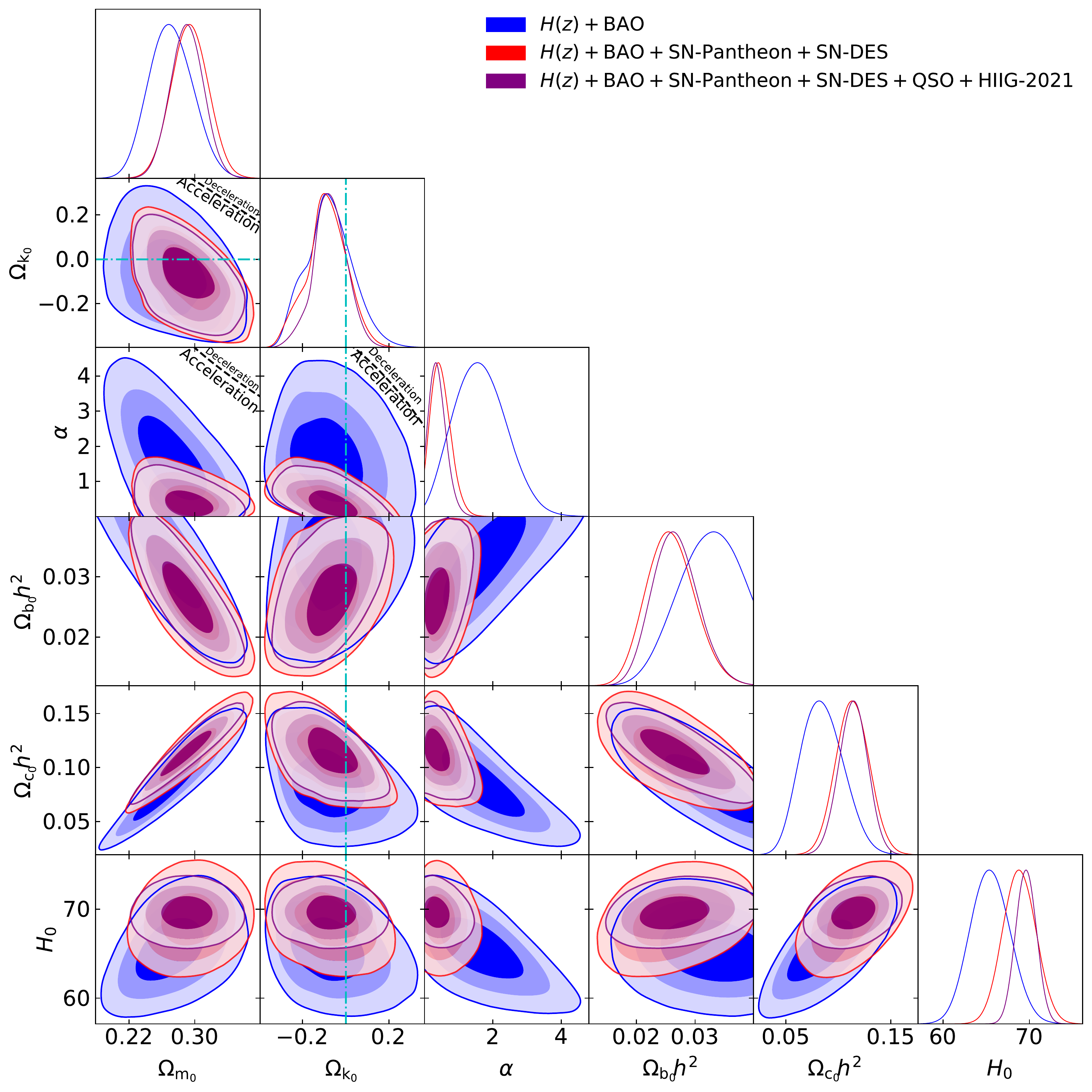}}\\
\caption{Same as Fig. \ref{fig5C6} but for non-flat \pcdm, where the black dashed zero-acceleration lines are computed for the third cosmological parameter set to the $H(z)$ + BAO data best-fitting values listed in Table \ref{tab:BFPC6}. Currently-accelerating cosmological expansion occurs below left of these lines. The cyan dash-dot lines represent the flat \pcdm\ case, with closed spatial geometry either below or to the left. The $\alpha = 0$ axis is the non-flat \lcdm\ model. In the right panel, the joint analyses favor currently-accelerating expansion. }
\label{fig6C6}
\end{figure*}

The posterior one-dimensional (1D) probability distributions and two-dimensional (2D) confidence regions of the cosmological and nuisance parameters for the six flat and non-flat models are shown in Figs. \ref{fig1C6}--\ref{fig6C6}, in gray (QSO), pink (\hiig-2019), green (\hiig-2021), blue ($H(z)$ + BAO), red ($H(z)$ + BAO + SN-Pantheon + SN-DES, HzBSNPD), and purple ($H(z)$ + BAO + SN-Pantheon + SN-DES + QSO + \hiig-2021, HzBSNPDQH). We list the unmarginalized best-fitting parameter values, as well as the corresponding $\chi^2$, $AIC$, $BIC$, degrees of freedom $\nu$ ($\nu \equiv N - n$), reduced $\chi^2$ ($\chi^2/\nu$), $\Delta \chi^2$, $\Delta AIC$, and $\Delta BIC$ for all models and data combinations, in Table \ref{tab:BFPC6}. The marginalized posterior mean parameter values and uncertainties ($\pm 1\sigma$ error bars or $2\sigma$ limits), for all models and data combinations, are listed in Table \ref{tab:1d_BFPC6}.\footnote{We use the \textsc{python} package \textsc{getdist} \citep{Lewis_2019} to determine the posterior means and uncertainties and to generate the marginalized likelihood contours.}

\begin{table*}
\centering
\resizebox*{1.0\columnwidth}{1.2\columnwidth}{%
\begin{threeparttable}
\caption{Unmarginalized best-fitting parameter values for all models from various combinations of data.}\label{tab:BFPC6}
\begin{tabular}{lccccccccccccccccc}
\toprule
Model & Data set & $\Omega_{\mathrm{b_0}}\!h^2$ & $\Omega_{\mathrm{c_0}}\!h^2$ & $\Omega_{\mathrm{m_0}}$ & $\Omega_{\mathrm{k_0}}$ & $w_{\mathrm{X}}$ & $\alpha$ & $H_0$\tnote{a} & $l_{\mathrm{m}}$\tnote{b} & $\chi^2$ & $\nu$ & $AIC$ & $BIC$ & $\chi^2/\nu$ & $\Delta \chi^2$ & $\Delta AIC$ & $\Delta BIC$ \\
\midrule
Flat \lcdm & $H(z)$ + BAO & 0.0239 & 0.1187 & 0.298 & -- & -- & -- & 69.13 & -- & 23.66 & 39 & 29.66 & 34.87 & 0.61 & 0.00 & 0.00 & 0.00\\
 & \hiig-2019 & 0.0200 & 0.1215 & 0.274 & -- & -- & -- & 71.86 & -- & 410.75 & 150 & 416.75 & 425.84 & 2.74 & 0.00 & 0.00 & 0.00\\
 & \hiig-2021 & 0.0156 & 0.1058 & 0.235 & -- & -- & -- & 71.89 & -- & 433.86 & 178 & 439.86 & 449.45 & 2.44 & 0.00 & 0.00 & 0.00\\
 & QSO & 0.0095 & 0.0172 & 0.315 & -- & -- & -- & 29.14 & 25.99 & 352.04 & 116 & 360.04 & 371.19 & 3.03 & 0.00 & 0.00 & 0.00\\
 & HzBSNPD\tnote{c} & 0.0235 & 0.1200 & 0.302 & -- & -- & -- & 68.90 & -- & 1080.48 & 1107 & 1086.48 & 1101.52 & 0.98 & 0.00 & 0.00 & 0.00\\
 & HzBSNPDQH\tnote{d} & 0.0248 & 0.1205 & 0.298 & -- & -- & -- & 69.83 & 10.98 & 1868.64 & 1407 & 1876.64 & 1897.64 & 1.33 & 0.00 & 0.00 & 0.00\\
\\
Non-flat \lcdm & $H(z)$ + BAO & 0.0247 & 0.1140 & 0.294 & 0.029 & -- & -- & 68.68 & -- & 23.60 & 38 & 31.60 & 38.55 & 0.62 & $-0.06$ & 1.94 & 3.68\\
 & \hiig-2019 & 0.0075 & 0.1585 & 0.314 & $-0.424$ & -- & -- & 72.70 & -- & 410.40 & 149 & 418.40 & 430.52 & 2.75 & $-0.35$ & 1.65 & 4.68\\
 & \hiig-2021 & 0.0114 & 0.1282 & 0.260 & $-0.490$ & -- & -- & 73.20 & -- & 432.79 & 177 & 440.79 & 453.58 & 2.45 & $-1.07$ & 0.93 & 4.13\\
 & QSO & 0.0320 & 0.1277 & 0.236 & $-0.363$ & -- & -- & 82.31 & 10.52 & 351.12 & 115 & 361.12 & 375.06 & 3.05 & $-0.92$ & 1.08 & 3.87\\
 & HzBSNPD\tnote{c} & 0.0243 & 0.1153 & 0.296 & 0.025 & -- & -- & 68.61 & -- & 1080.38 & 1106 & 1088.38 & 1108.42 & 0.98 & $-0.10$ & 1.90 & 6.90\\
 & HzBSNPDQH\tnote{d} & 0.0249 & 0.1192 & 0.296 & 0.005 & -- & -- & 69.76 & 10.99 & 1868.63 & 1406 & 1878.63 & 1904.89 & 1.33 & $-0.01$ & 1.99 & 7.25\\
\\
Flat XCDM & $H(z)$ + BAO & 0.0304 & 0.0891 & 0.281 & -- & $-0.701$ & -- & 65.18 & -- & 19.65 & 38 & 27.65 & 34.60 & 0.52 & $-4.01$ & $-2.01$ & $-0.27$\\
 & \hiig-2019 & 0.0107 & 0.1180 & 0.251 & -- & $-0.896$ & -- & 71.62 & -- & 410.72 & 149 & 418.72 & 430.85 & 2.76 & $-0.03$ & 1.97 & 5.01\\
 & \hiig-2021 & 0.0291 & 0.0942 & 0.239 & -- & $-1.013$ & -- & 71.88 & -- & 433.86 & 177 & 441.86 & 454.65 & 2.45 & 0.00 & 2.00 & 5.20\\
 & QSO & 0.0173 & 0.0050 & 0.253 & -- & $-2.137$ & -- & 29.68 & 31.24 & 351.83 & 115 & 361.83 & 375.77 & 3.06 & $-0.21$ & 1.79 & 4.58\\
 & HzBSNPD\tnote{c} & 0.0254 & 0.1119 & 0.293 & -- & $-0.934$ & -- & 68.51 & -- & 1079.24 & 1106 & 1087.24 & 1107.29 & 0.98 & $-1.24$ & 0.76 & 5.77\\
 & HzBSNPDQH\tnote{d} & 0.0266 & 0.1135 & 0.289 & -- & $-0.948$ & -- & 69.63 & 10.95 & 1867.85 & 1406 & 1877.85 & 1904.11 & 1.33 & $-0.79$ & 1.21 & 6.47\\
\\
Non-flat XCDM & $H(z)$ + BAO & 0.0290 & 0.0980 & 0.295 & $-0.152$ & $-0.655$ & -- & 65.59 & -- & 18.31 & 37 & 28.31 & 37.00 & 0.49 & $-5.35$ & $-1.35$ & 2.13\\
 & \hiig-2019 & 0.0128 & 0.0011 & 0.027 & $-0.625$ & $-0.630$ & -- & 72.24 & -- & 405.56 & 148 & 415.56 & 430.72 & 2.74 & $-5.19$ & $-1.19$ & 4.88\\
 & \hiig-2021 & 0.0260 & 0.0066 & 0.062 & $-0.573$ & $-0.680$ & -- & 72.43 & -- & 430.07 & 176 & 440.07 & 456.06 & 2.44 & $-3.79$ & 0.21 & 6.61\\
 & QSO & 0.0153 & 0.0122 & 0.060 & $-0.570$ & $-0.617$ & -- & 67.92 & 11.82 & 350.22 & 114 & 362.22 & 378.95 & 3.07 & $-1.82$ & 2.18 & 7.76\\
 & HzBSNPD\tnote{c} & 0.0234 & 0.1232 & 0.309 & $-0.111$ & $-0.876$ & -- & 68.94 & -- & 1078.38 & 1105 & 1088.38 & 1113.44 & 0.98 & $-2.10$ & 1.90 & 11.92\\
 & HzBSNPDQH\tnote{d} & 0.0246 & 0.1208 & 0.299 & $-0.083$ & $-0.900$ & -- & 69.72 & 10.89 & 1867.11 & 1405 & 1879.11 & 1910.63 & 1.33 & $-1.53$ & 2.47 & 12.99\\
\\
Flat $\phi$CDM & $H(z)$ + BAO & 0.0333 & 0.0788 & 0.264 & -- & -- & 1.504 & 65.20 & -- & 19.49 & 38 & 27.49 & 34.44 & 0.51 & $-4.17$ & $-2.17$ & $-0.43$\\
 & \hiig-2019 & 0.0304 & 0.1038 & 0.261 & -- & -- & 0.174 & 71.75 & -- & 410.74 & 149 & 418.74 & 430.86 & 2.76 & $-0.01$ & 1.99 & 5.02\\
 & \hiig-2021 & 0.0334 & 0.0876 & 0.234 & -- & -- & 0.001 & 71.94 & -- & 433.86 & 177 & 441.86 & 454.65 & 2.45 & 0.00 & 2.00 & 5.20\\
 & QSO & 0.0370 & 0.0960 & 0.316 & -- & -- & 0.001 & 64.91 & 11.66 & 352.05 & 115 & 362.05 & 375.98 & 3.06 & 0.01 & 2.01 & 4.79\\
 & HzBSNPD\tnote{c} & 0.0257 & 0.1097 & 0.290 & -- & -- & 0.226 & 68.38 & -- & 1079.09 & 1106 & 1087.09 & 1107.14 & 0.98 & $-1.39$ & 0.61 & 5.62\\
 & HzBSNPDQH\tnote{d} & 0.0270 & 0.1125 & 0.288 & -- & -- & 0.174 & 69.63 & 10.94 & 1867.75 & 1406 & 1877.75 & 1904.01 & 1.33 & $-0.89$ & 1.11 & 6.37\\
\\
Non-flat $\phi$CDM & $H(z)$ + BAO & 0.0334 & 0.0816 & 0.266 & $-0.147$ & -- & 1.915 & 65.70 & -- & 18.15 & 37 & 28.15 & 36.84 & 0.49 & $-5.51$ & $-1.51$ & 1.97\\
 & \hiig-2019 & 0.0251 & 0.1126 & 0.265 & $-0.265$ & -- & 0.433 & 72.04 & -- & 410.37 & 148 & 420.37 & 435.52 & 2.77 & $-0.38$ & 3.62 & 9.68\\
 & \hiig-2021 & 0.0146 & 0.1014 & 0.249 & $-0.246$ & -- & 0.101 & 72.48 & -- & 433.19 & 176 & 443.19 & 459.19 & 2.46 & $-0.67$ & 3.33 & 9.74\\
 & QSO & 0.0282 & 0.0469 & 0.261 & $-0.261$ & -- & 0.008 & 53.61 & 15.39 & 351.32 & 114 & 363.32 & 380.04 & 3.08 & $-0.72$ & 3.28 & 8.85\\
 & HzBSNPD\tnote{c} & 0.0243 & 0.1197 & 0.302 & $-0.110$ & -- & 0.442 & 69.05 & -- & 1078.07 & 1105 & 1088.07 & 1113.13 & 0.98 & $-2.41$ & 1.59 & 11.61\\
 & HzBSNPDQH\tnote{d} & 0.0250 & 0.1192 & 0.297 & $-0.092$ & -- & 0.348 & 69.69 & 10.87 & 1866.93 & 1405 & 1878.93 & 1910.45 & 1.33 & $-1.71$ & 2.29 & 12.81\\
\bottomrule
\end{tabular}
\begin{tablenotes}[flushleft]
\item [a] \hunit.
\item [b] pc.
\item [c] $H(z)$ + BAO + SN-Pantheon + SN-DES.
\item [d] $H(z)$ + BAO + SN-Pantheon + SN-DES + QSO + \hiig-2021.
\end{tablenotes}
\end{threeparttable}%
}
\end{table*}

\subsection{Constraints from \hiig-2021 versus \hiig-2019}
\label{submakereference6.4_HIIG_comparison}

Here we compare constraints from the current compilation of \hiig\ data (\hiig-2021) with the constraints from the older compilation (\hiig-2019). 

For both data sets, from the figures, most of the probability lies in the part of parameter space that corresponds to currently accelerating cosmological expansion.

For both the flat and non-flat \lcdm\ models, \hiig-2021 data favor lower values of \ochs\ and \om. For the non-flat case, the \hiig-2021 data prefer a lower (more negative) value of \ok\ than the \hiig-2019 data. It is worth noting that $H_0$ is not particularly sensitive to the change in the \hiig\ data. It is also worth noting that the \hiig-2021 compilation provides constraints on \obhs\ in the flat case, in contrast to \hiig-2019, although these constraints are not as tight as those obtained from the other data combinations. \hiig-2021 data also more tightly constrain \ok\ compared to the \hiig-2019 data. In comparison with the flat \lcdm\ \hiig-2021 constraints given in \cite{GonzalezMoranetal2021}, $h=0.717\pm0.018$ and $\Om=0.243^{+0.039}_{-0.050}$, our constraints, $h=0.7191\pm0.0192$ and $\Om=0.243^{+0.039}_{-0.051}$, are a bit less restrictive (due to our models having more free parameters) but are consistent with theirs.

In the flat and non-flat XCDM parametrizations, the \hiig-2021 data favor lower values of \ochs\!, \om\!, \wx\!, and \ok\ than those favored by the \hiig-2019 data, while the constraints on \obhs\ and $H_0$ are consistent with each other. The flat XCDM \hiig-2021 constraints are $\{h, \Om, w_{\rm X}\}=\{0.719\pm0.020, 0.250^{+0.10}_{-0.061}, -1.19^{+0.46}_{-0.38}\}$ in \cite{GonzalezMoranetal2021}, whereas our results are $\{h, \Om, w_{\rm X}\}=\{0.7266\pm0.0219, 0.288^{+0.087}_{-0.058}, -1.527^{+0.786}_{-0.391}\}$ and consistent with theirs within 1$\sigma$, although due to different prior ranges the posterior means deviate more than those for the flat \lcdm\ model.

In the flat \pcdm\ model, the \hiig-2021 data prefer 
lower values of \ochs\!, \om\!, and $\alpha$, with consistent constraints on \obhs\ and $H_0$. The \hiig-2021 data constrain $\alpha$ more tightly than the \hiig-2019 data, leading to $\alpha$ being consistent with zero to within a little more than 1$\sigma$. In the non-flat case, the \hiig-2021 data prefer lower values for all parameters. It is worth noting that both \hiig-2021 and \hiig-2019 data in the flat and non-flat \pcdm\ models determine lower values of \om, and \hiig-2021 data prefer the lowest \om\ value in the non-flat \pcdm\ model among all models.

\hiig-2021 and \hiig-2019 data result in higher values of $H_0$ than the other probes we study in this paper. The highest $H_0$ values are in the flat XCDM parametrization and are $72.66\pm2.19$ \hunit and $72.37^{+2.18}_{-2.20}$ \hunit, respectively, which are $1.31\sigma$ and $1.23\sigma$ higher than the median statistics estimate of $H_0=68 \pm 2.8$ \hunit\ \citep{chenratmed}, and $0.21\sigma$ and $0.33\sigma$ lower than the local Hubble constant measurement of $H_0 = 73.2 \pm 1.3$ \hunit\ \citep{Riess_2021}. The lowest $H_0$ estimates are in the non-flat \pcdm\ model and are $70.49\pm1.81$ \hunit and $70.53\pm1.79$ \hunit, which are $0.75\sigma$ and $0.76\sigma$ higher than the median statistics estimate of $H_0=68 \pm 2.8$ \hunit, and $1.22\sigma$ and $1.21\sigma$ lower than the local Hubble constant measurement of $H_0 = 73.2 \pm 1.3$ \hunit.

In the non-flat \lcdm\ model, \hiig-2021 and \hiig-2019 data favor closed spatial hypersurfaces, while in the non-flat XCDM parametrization and the non-flat \pcdm\ model, they favor open spatial hypersurfaces. Only in the non-flat \pcdm\ model, however, is \ok\ more than 1$\sigma$ away from spatial flatness. Dark energy dynamics is favored by both data sets, but dark energy being a cosmological constant is not disfavored (it is within 1$\sigma$ or just a little bit more away).

\subsection{QSO constraints alone and in comparison to those from other probes}
\label{submakereference6.4_QSO_constraints}

In this subsection we discuss the constraints we obtain solely from the QSO data. As mentioned in Sec. \ref{makereference6.2}, in this paper we improve on earlier analyses of the QSO angular size data by now treating $l_{\rm m}$, the characteristic linear size of QSOs, as a nuisance parameter to be determined from the observational data. From QSO data alone, in Table \ref{tab:1d_BFPC6}, $l_{\rm m}$ ranges from a low of $10.26^{+1.24}_{-3.42}$ pc for the non-flat \pcdm\ model to a high of $11.90^{+1.52}_{-4.17}$ pc for the flat XCDM parametrization, differing by just 0.38$\sigma$. These values are consistent from model to model, largely justifying the use of QSOs as standard rulers, with $l_{\rm m}=11.03$ pc, the value we used in our previous studies (taken from \citealp{Cao_et_al2017b}). However, ignoring the dependence on cosmological model and the $l_{\rm m}$ errors, as we and others have previously done, results in mildly biased and somewhat more restrictive QSO angular size constraints than is warranted by data. These deficiencies are corrected in our improved analyses here.\footnote{When QSO data are combined with other probes, as in the HzBSNPDQH combination, the model-independence of $l_{\rm m}$ is evident and the determination here is consistent with $l_{\rm m}=11.03 \pm 0.25$ pc found by \cite{Cao_et_al2017b}.}  Additionally, we note that in Table \ref{tab:BFPC6}, the best-fitting values of $H_0$ in flat \lcdm\ and flat XCDM appear to be unreasonably low. This strange behavior is caused by the large values of $l_{\rm m}$, which push the $H_0$ values lower to obtain locally minimized $\chi^2$ values. Specifically, from the form of the model-predicted angular size of a quasar,
\begin{equation}
    \theta\left(z\right) \propto \frac{l_{\rm m} H_0}{d_{\rm M}(z)}
\end{equation}
(where $d_{\rm M}(z) := \frac{H_0}{c}D_{\rm M}(z)$ and suppressing irrelevant parameters), we can see that a large value of $l_{\rm m}$ requires a small value of $H_0$ in order to keep $\theta\left(z\right)$ constant. Since $l_{\rm m}$ has an unbounded prior range, it can roam over a larger region of parameter space than $H_0$. It therefore has the freedom to move into regions of parameter space where its value is unusually large; if this happens, then $H_0$ must be made small to compensate. This is only a partial answer, since it does not account for the variation of \ok\ (the effect of which is more complex, as \ok\ is coupled to the redshift $z$ through the function $d_{\rm M}(z)$), and so does not fully capture the behavior of $\theta(z)$ across all models, but it does give some insight into the apparently anomalously low values of $H_0$ that appear in some cases.\footnote{The relatively higher values of $H_0$ seen in the $\phi$CDM models pose an apparent challenge to this explanation, but here the best-fitting values of \obhs\ and \ochs\ need to be taken into account. In comparing, for example, the flat \lcdm\ model to the flat \pcdm\ model (both of which have nearly identical best-fitting values of \om), we can see that the flat \pcdm\ model has larger best-fitting values of both \obhs\ and \ochs. From the defining relationship $\Om = (\obh + \och)/{h^2}$, keeping \om\ constant requires \obhs\ + \ochs\ and $H_0$ to vary in tandem. If \obhs\ and \ochs\ both increase, as they do in going from flat \lcdm\ to flat \pcdm, then $H_0$ must also increase. This then has the effect of lowering $l_{\rm m}$ (all other parameters being held fixed).}

From the results listed in Table \ref{tab:1d_BFPC6}, we can draw the following conclusions. First, QSO data alone can only constrain the values of \obhs\ in the flat and non-flat \pcdm\ models. Second, QSO data alone prefer higher values of \om, which are consistent with almost all other probes except for the non-flat \pcdm\ \hiig-2021 case (the posterior mean values being 1.1$\sigma$ away from each other in this case). Furthermore, QSO data alone do not give tight constraints on $H_0$ or \ok. Although in each non-flat model open geometry is favored, given the large error bars, flat geometry is within 1$\sigma$.

QSO data favor higher central values of \ochs\ and \om, in both flat and non-flat \lcdm, compared to the central values favored by the other probes (although QSO constraints have wider error bars than the other constraints). QSO data only very weakly constrain the value of $H_0$ in the flat \lcdm\ model, while the fit of QSO data to the non-flat \lcdm\ model produces a tighter constraint whose central value is closer to that of the \hiig\ data and the local value favored by \cite{Riess_2021} (with wide error bars, however). In both the flat and non-flat cases, the marginalized values of $l_{\rm m}$ are close to the value obtained by \cite{Cao_et_al2017b}, with the central value in the flat \lcdm\ model here being only 0.02 pc away from that of \cite{Cao_et_al2017b} (with wider error bars than what they found). QSO data do not provide strong evidence for non-zero spatial curvature in the non-flat \lcdm\ model, as the marginalized posterior mean value of \ok\ is consistent with \ok $= 0$ to within 1$\sigma$.

When we look at the flat and non-flat XCDM parametrizations, we find that QSO data again favor somewhat large values of \ochs\ and \om\ (but, as with flat and non-flat \lcdm, these have wide error bars) and weak constraints on $H_0$. The central value of $H_0$ in the non-flat case is more consistent with \cite{Riess_2021} and with the values derived from the \hiig\ data. In both cases we find that the marginalized values of $l_{\rm m}$ are consistent with that of \cite{Cao_et_al2017b}. We also find that QSO data favor values of \wx\ that are in the phantom regime (consistent with the findings from the \hiig\ data). In the non-flat case, QSO data favor a relatively large and positive central value (0.170) for \ok, corresponding to a spatially open universe, but the error bars are wide enough that this result is still consistent with spatial flatness.

Both the flat and non-flat \pcdm\ models have central values of \obhs\ from QSO data that are similar to earlier findings (specifically, they are close to the values of \obhs\ obtained for the flat \lcdm\ and \pcdm\ models by \citealp{park_ratra_2018, ParkRatra2019a}). Both flat and non-flat \pcdm\ have relatively high central values of \ochs\ and \om\ (compared to the other probes), both favor similar large values of $\alpha$ (consistent with $\alpha = 0$, however, to within 1.15$\sigma$ and 1.29$\sigma$ in the flat and non-flat cases, respectively), and both show weak constraints on $H_0$. Both flat and non-flat \pcdm\ favor posterior mean values of $l_{\rm m}$ that are consistent to within 1$\sigma$ with the central value obtained by \cite{Cao_et_al2017b}. Like non-flat XCDM, non-flat \pcdm\ favors a relatively large and positive value of \ok, that is nevertheless consistent with spatial flatness to within 1$\sigma$.

\begin{table*}
\centering
\resizebox*{1.0\columnwidth}{1.2\columnwidth}{%
\begin{threeparttable}
\caption{One-dimensional posterior mean parameter values and uncertainties ($\pm 1\sigma$ error bars or $2\sigma$ limits) for all models from various combinations of data.}\label{tab:1d_BFPC6}
\begin{tabular}{lccccccccc}
\toprule
Model & Data set & $\Omega_{\mathrm{b_0}}\!h^2$ & $\Omega_{\mathrm{c_0}}\!h^2$ & $\Omega_{\mathrm{m_0}}$ & $\Omega_{\mathrm{k_0}}$ & $w_{\mathrm{X}}$ & $\alpha$ & $H_0$\tnote{a} & $l_{\mathrm{m}}$\tnote{b}\\
\midrule
Flat \lcdm & $H(z)$ + BAO & $0.0241\pm0.0029$ & $0.1193^{+0.0082}_{-0.0090}$ & $0.299^{+0.017}_{-0.019}$ & -- & -- & -- & $69.29^{+1.84}_{-1.85}$ & -- \\
 & \hiig-2019 & -- & $0.1258^{+0.0278}_{-0.0335}$ & $0.289^{+0.054}_{-0.074}$ & -- & -- & -- & $71.80\pm1.94$ & -- \\
 & \hiig-2021 & $0.0225\pm0.0108$ & $0.1023^{+0.0197}_{-0.0229}$ & $0.243^{+0.039}_{-0.051}$ & -- & -- & -- & $71.91\pm1.92$ & -- \\
 & QSO & -- & $0.1874^{+0.0592}_{-0.1595}$ & $0.387^{+0.078}_{-0.177}$ & -- & -- & -- & $>38.09$ & $11.05^{+1.10}_{-3.85}$ \\
 & HzBSNPD\tnote{c} & $0.0237\pm0.0028$ & $0.1208\pm0.0074$ & $0.303^{+0.013}_{-0.014}$ & -- & -- & -- & $69.10\pm1.80$ & -- \\
 & HzBSNPDQH\tnote{d} & $0.0250\pm0.0021$ & $0.1208\pm0.0064$ & $0.298\pm0.013$ & -- & -- & -- & $69.95\pm1.18$ & $10.96\pm0.26$ \\
\\
Non-flat \lcdm & $H(z)$ + BAO & $0.0253^{+0.0040}_{-0.0049}$ & $0.1134^{+0.0196}_{-0.0197}$ & $0.293\pm0.025$ & $0.040^{+0.102}_{-0.115}$ & -- & -- & $68.75\pm2.45$ & -- \\
 & \hiig-2019 & $0.0224\pm0.0108$ & $0.1245^{+0.0413}_{-0.0380}$ & $0.285\pm0.077$ & $-0.052^{+0.289}_{-0.530}$ & -- & -- & $71.95\pm2.04$ & -- \\
 & \hiig-2021 & $0.0225\pm0.0108$ & $0.1035^{+0.0328}_{-0.0268}$ & $0.243^{+0.060}_{-0.055}$ & $-0.100^{+0.216}_{-0.484}$ & -- & -- & $72.15^{+2.05}_{-2.04}$ & -- \\
 & QSO & -- & $0.1797^{+0.0610}_{-0.1489}$ & $0.385^{+0.074}_{-0.191}$ & $0.043^{+0.265}_{-0.464}$ & -- & -- & $71.33^{+26.04}_{-9.82}$ & $11.27^{+0.91}_{-3.92}$ \\
 & HzBSNPD\tnote{c} & $0.0248^{+0.0036}_{-0.0043}$ & $0.1157\pm0.0164$ & $0.296\pm0.023$ & $0.027\pm0.074$ & -- & -- & $68.82\pm2.02$ & -- \\
 & HzBSNPDQH\tnote{d} & $0.0256^{+0.0035}_{-0.0042}$ & $0.1188\pm0.0138$ & $0.295\pm0.021$ & $0.011\pm0.067$ & -- & -- & $69.90\pm1.18$ & $10.96\pm0.25$ \\
\\
Flat XCDM & $H(z)$ + BAO & $0.0297^{+0.0046}_{-0.0053}$ & $0.0934^{+0.0195}_{-0.0169}$ & $0.283^{+0.023}_{-0.021}$ & -- & $-0.751^{+0.152}_{-0.106}$ & -- & $65.85^{+2.38}_{-2.65}$ & -- \\
 & \hiig-2019 & $0.0224\pm0.0109$ & $0.1491^{+0.0596}_{-0.0390}$ & $0.327^{+0.108}_{-0.077}$ & -- & $-1.494^{+0.858}_{-0.411}$ & -- & $72.37^{+2.18}_{-2.20}$ & -- \\
 & \hiig-2021 & $0.0223\pm0.0108$ & $0.1297^{+0.0493}_{-0.0308}$ & $0.288^{+0.087}_{-0.058}$ & -- & $-1.527^{+0.786}_{-0.391}$ & -- & $72.66\pm2.19$ & -- \\
 & QSO & -- & $0.1785^{+0.0536}_{-0.1603}$ & $0.373^{+0.072}_{-0.187}$ & -- & $-1.709^{+0.696}_{-0.882}$ & -- & $>39.72$ & $11.90^{+1.52}_{-4.17}$ \\
 & HzBSNPD\tnote{c} & $0.0256^{+0.0031}_{-0.0035}$ & $0.1121^{+0.0107}_{-0.0108}$ & $0.293\pm0.016$ & -- & $-0.935\pm0.063$ & -- & $68.57\pm1.74$ & -- \\
 & HzBSNPDQH\tnote{d} & $0.0267^{+0.0029}_{-0.0033}$ & $0.1142\pm0.0103$ & $0.290\pm0.016$ & -- & $-0.950\pm0.062$ & -- & $69.69\pm1.20$ & $10.94\pm0.25$ \\
\\
Non-flat XCDM & $H(z)$ + BAO & $0.0288^{+0.0049}_{-0.0054}$ & $0.0997^{+0.0210}_{-0.0211}$ & $0.294\pm0.027$ & $-0.112^{+0.136}_{-0.137}$ & $-0.706^{+0.135}_{-0.084}$ & -- & $66.01\pm2.43$ & -- \\
 & \hiig-2019 & $0.0223\pm0.0107$ & $0.1288^{+0.0548}_{-0.0620}$ & $0.291^{+0.109}_{-0.110}$ & $0.089^{+0.484}_{-0.283}$ & $-1.409^{+0.810}_{-0.402}$ & -- & $71.96^{+2.06}_{-2.29}$ & -- \\
 & \hiig-2021 & $0.0224\pm0.0107$ & $0.1109^{+0.0472}_{-0.0514}$ & $0.255\pm0.090$ & $0.078^{+0.417}_{-0.322}$ & $-1.461^{+0.777}_{-0.361}$ & -- & $72.23^{+2.08}_{-2.28}$ & -- \\
 & QSO & -- & $0.1905^{+0.0461}_{-0.1731}$ & $0.403^{+0.093}_{-0.225}$ & $0.170^{+0.403}_{-0.256}$ & $-1.455^{+1.014}_{-0.776}$ & -- & $>38.43$ & $11.08^{+1.19}_{-4.05}$ \\
 & HzBSNPD\tnote{c} & $0.0245^{+0.0036}_{-0.0043}$ & $0.1193\pm0.0170$ & $0.302\pm0.024$ & $-0.069\pm0.119$ & $-0.907^{+0.099}_{-0.062}$ & -- & $68.95\pm1.96$ & -- \\
 & HzBSNPDQH\tnote{d} & $0.0255^{+0.0035}_{-0.0041}$ & $0.1189^{+0.0136}_{-0.0135}$ & $0.297\pm0.020$ & $-0.054\pm0.096$ & $-0.926^{+0.091}_{-0.062}$ & -- & $69.73^{+1.19}_{-1.20}$ & $10.89\pm0.26$ \\
\\
Flat $\phi$CDM & $H(z)$ + BAO & $0.0323^{+0.0060}_{-0.0034}$ & $0.0810^{+0.0188}_{-0.0185}$ & $0.267\pm0.025$ & -- & -- & $1.530^{+0.644}_{-0.904}$ & $65.09^{+2.23}_{-2.24}$ & -- \\
 & \hiig-2019 & $0.0214^{+0.0084}_{-0.0128}$ & $0.0561^{+0.0157}_{-0.0541}$ & $0.155^{+0.047}_{-0.097}$ & -- & -- & $<7.803$ & $70.97^{+1.91}_{-1.89}$ & -- \\
 & \hiig-2021 & $0.0213^{+0.0080}_{-0.0130}$ & $0.0468^{+0.0152}_{-0.0431}$ & $0.135^{+0.043}_{-0.077}$ & -- & -- & $2.454^{+0.587}_{-2.434}$ & $71.15^{+1.90}_{-1.89}$ & -- \\
 & QSO & $0.0221^{+0.0179}_{-0.0171}$ & $<0.3644$ & $0.305^{+0.081}_{-0.212}$ & -- & -- & $4.526^{+1.546}_{-4.464}$ & $70.79^{+24.27}_{-12.67}$ & $10.32^{+1.08}_{-3.54}$ \\
 & HzBSNPD\tnote{c} & $0.0273^{+0.0032}_{-0.0039}$ & $0.1051^{+0.0122}_{-0.0103}$ & $0.284\pm0.017$ & -- & -- & $0.351^{+0.132}_{-0.284}$ & $68.33\pm1.81$ & -- \\
 & HzBSNPDQH\tnote{d} & $0.0284^{+0.0027}_{-0.0035}$ & $0.1078^{+0.0112}_{-0.0090}$ & $0.282\pm0.016$ & -- & -- & $0.288^{+0.098}_{-0.252}$ & $69.54\pm1.17$ & $10.92\pm0.25$ \\
\\
Non-flat $\phi$CDM & $H(z)$ + BAO & $0.0319^{+0.0061}_{-0.0037}$ & $0.0849^{+0.0178}_{-0.0217}$ & $0.271^{+0.025}_{-0.028}$ & $-0.074^{+0.104}_{-0.111}$ & -- & $1.646^{+0.680}_{-0.840}$ & $65.48\pm2.26$ & -- \\
 & \hiig-2019 & $0.0216^{+0.0089}_{-0.0121}$ & $0.0560^{+0.0198}_{-0.0476}$ & $0.157^{+0.047}_{-0.091}$ & $0.314^{+0.310}_{-0.182}$ & -- & $3.983^{+1.228}_{-3.932}$ & $70.53\pm1.79$ & -- \\
 & \hiig-2021 & $0.0210^{+0.0076}_{-0.0129}$ & $0.0407^{+0.0116}_{-0.0381}$ & $0.125^{+0.037}_{-0.069}$ & $0.302^{+0.278}_{-0.221}$ & -- & $<8.046$ & $70.49\pm1.81$ & -- \\
 & QSO & $0.0222\pm0.0109$ & $0.1444^{+0.0391}_{-0.1352}$ & $0.330^{+0.108}_{-0.183}$ & $0.207^{+0.244}_{-0.241}$ & -- & $4.773^{+2.487}_{-3.699}$ & $70.15^{+20.58}_{-15.20}$ & $10.26^{+1.24}_{-3.42}$ \\
 & HzBSNPD\tnote{c} & $0.0259^{+0.0037}_{-0.0043}$ & $0.1136\pm0.0156$ & $0.294\pm0.022$ & $-0.088^{+0.105}_{-0.075}$ & -- & $0.495^{+0.208}_{-0.344}$ & $68.84\pm1.88$ & -- \\
 & HzBSNPDQH\tnote{d} & $0.0267^{+0.0035}_{-0.0041}$ & $0.1141\pm0.0133$ & $0.290\pm0.020$ & $-0.072^{+0.074}_{-0.073}$ & -- & $0.405^{+0.165}_{-0.304}$ & $69.62\pm1.17$ & $10.87\pm0.26$ \\
\bottomrule
\end{tabular}
\begin{tablenotes}[flushleft]
\item [a] \hunit.
\item [b] pc.
\item [c] $H(z)$ + BAO + SN-Pantheon + SN-DES.
\item [d] $H(z)$ + BAO + SN-Pantheon + SN-DES + QSO + \hiig-2021.
\end{tablenotes}
\end{threeparttable}%
}
\end{table*}

\subsection{Joint analyses results}
\label{submakereference6.4_joint}

Since the constraints derived from $H(z)$, BAO, SN-Pantheon, SN-DES, QSO, \hiig-2019, and \hiig-2021 data are not mutually inconsistent, we jointly analyze combinations of these data and summarize these results in this subsection.

The $H(z)$ + BAO and HzBSNPD results are different from, but consistent with, what we obtained in \cite{CaoRyanRatra2021}. The differences arise from the different codes that we used to analyze the data; in \cite{CaoRyanRatra2021} we used \textsc{emcee}, whereas here we used \textsc{class} and \textsc{MontePython}. It is worth recalling here that, as mentioned above, \textsc{class} constrains \obhs\ in the range $0.00499 \leq \obh \leq 0.03993$. Therefore the parameter constraints differ more when the model and data prefer higher values of \obhs; this is especially true of the \pcdm\ model when it is fitted to the $H(z)$ + BAO data combination. As a result, the present constraints on \om\ and $\alpha$ in \pcdm\ with $H(z)$ + BAO data are higher and lower than the ones given in \cite{CaoRyanRatra2021}. The HzBSNPD results are, however, consistent.

The fit to the HzBSNPDQH data combination produces, for all models, the most interesting results. By adding QSO and \hiig-2021 data to HzBSNPD combination, the constraints are slightly tightened. Although the posterior means of \obhs\ and \ochs\ are relatively higher, those of \om\ are lower than the constraints from HzBSNPD. The \om\ constraints range from a low of $0.282\pm0.016$ (flat \pcdm) to a high of $0.298\pm0.013$ (flat \lcdm), a difference of only 0.78$\sigma$.

The constraints on $H_0$ are between $H_0=69.54\pm1.17$ \hunit\ (flat \pcdm) and $H_0=69.95\pm1.18$ \hunit\ (flat \lcdm) --- a difference of only 0.25$\sigma$ --- which are $0.64\sigma$ (flat \lcdm) and $0.51\sigma$ (flat \pcdm) higher than the median statistics estimate of $H_0=68 \pm 2.8$ \hunit\ \citep{chenratmed}, and $1.85\sigma$ (flat \lcdm) and $2.09\sigma$ (flat \pcdm) lower than the local Hubble constant measurement of $H_0 = 73.2 \pm 1.3$ \hunit\ \citep{Riess_2021}.\footnote{Other local expansion rate $H_0$ measurements result in slightly lower central values with slightly larger error bars \citep{rigault_etal_2015,zhangetal2017,Dhawan,FernandezArenas,Breuvaletal_2020, Efstathiou_2020, Khetan_et_al_2021,rameez_sarkar_2021, Freedman2021}. Our $H_0$ estimates are consistent with earlier median statistics determinations \citep{gott_etal_2001, Calabreseetal2012} as well as with other recent $H_0$ measurements \citep{chen_etal_2017,DES_2018,Gomez-ValentAmendola2018, planck2018b,dominguez_etal_2019,Cuceu_2019,zeng_yan_2019,schoneberg_etal_2019, Blum_et_al_2020, Lyu_et_al_2020, Philcox_et_al_2020, Birrer_et_al_2020, Denzel_et_al_2020,Pogosianetal_2020,Boruahetal_2021,Kimetal_2020,Harvey_2020, Zhang_Huang_2021,lin_ishak_2021}.}

For non-flat \lcdm, non-flat XCDM, and non-flat \pcdm, we find $\Omega_{\rm k_0}=0.011\pm0.067$, $\Omega_{\rm k_0}=-0.054\pm0.096$, and $\Omega_{\rm k_0}=-0.072^{+0.074}_{-0.073}$, respectively. The non-flat XCDM and \pcdm\ models favor closed geometry, while the non-flat \lcdm\ model favors open geometry. Note, however, that these results are all consistent with spatially flat hypersurfaces to within 1$\sigma$.

Our results show a slight preference for dark energy dynamics. For flat (non-flat) XCDM, $w_{\rm X}=-0.950\pm0.062$ ($w_{\rm X}=-0.926^{+0.091}_{-0.062}$), with central values being 0.81$\sigma$ (1.19$\sigma$) away from $w_{\rm X}=-1$; and for flat (non-flat) \pcdm, $\alpha=0.288^{+0.098}_{-0.252}$ ($\alpha=0.405^{+0.165}_{-0.304}$), with central values being 1.14$\sigma$ (1.33$\sigma$) away from $\alpha=0$.

The constraints on the nuisance parameter $l_{\rm m}$ are between $l_{\rm m}=10.87\pm0.26$ pc (non-flat \pcdm) and $l_{\rm m}=10.96\pm0.26$ pc (flat \lcdm), which differ by 0.24$\sigma$ and so are effectively model-independent, and consistent with $l_{\rm m}=11.03\pm0.25$ pc \citep{Cao_et_al2017b}.

\subsection{Model comparison}
\label{subsec:comparisonC6}

The values of the reduced $\chi^2$ ($\chi^2/\nu$), $\Delta \chi^2$, $\Delta AIC$, and $\Delta BIC$ are reported in Table \ref{tab:BFPC6}, where $\Delta \chi^2$, $\Delta AIC$, and $\Delta BIC$, are the differences between the values of the $\chi^2$, $AIC$, and $BIC$ for a given model and the ones for flat \lcdm. Here a negative (positive) value of $\Delta \chi^2$, $\Delta AIC$, or $\Delta BIC$ means that the given statistic favors (disfavors) the model under consideration relative to flat \lcdm. We find that, except for a few of the $H(z)$ + BAO and \hiig-2019 cases, the flat \lcdm\ model is the most favored model among all six models we study. The $AIC$ does not show strong evidence against any of the models.\footnote{There is weak evidence for the reference model when $\Delta AIC(BIC) \in [0,2]$, positive evidence when $\Delta AIC(BIC) \in (2,6]$, strong evidence when $\Delta AIC(BIC) \in (6,10]$, and very strong evidence when $\Delta AIC(BIC) > 10$ \citep{Kass_Raftery}.} However, we find that some data combinations show strong evidence against the models we study, when these models are analyzed using the BIC, as follows. First, the HzBSNPD combination strongly disfavors non-flat \lcdm\ and very strongly disfavors non-flat XCDM and non-flat \pcdm. Second, the HzBSNPDQH combination strongly disfavors non-flat \lcdm, flat XCDM, and flat \pcdm, and very strongly disfavors non-flat XCDM and non-flat \pcdm. Furthermore, strong evidence against non-flat XCDM as well as non-flat \pcdm\ are provided by the \hiig-2021 and QSO data.

\section{Conclusion}
\label{makereference6.5}

We find that the new \hiig-2021 data provide more restrictive cosmological parameter constraints and also prefer lower values of \om\!, \wx\!, and \ok\ than those favored by the \hiig-2019 data. 
 
We find that the QSO characteristic linear size $l_{\rm m}$ is relatively model-independent, so QSOs can be treated as approximate standard rulers but the uncertainty in $l_{\rm m}$ must be accounted for in the analysis.

We also jointly analyzed a total of 1411 measurements, consisting of 31 $H(z)$, 11 BAO, 1048 SN-Pantheon, 20 SN-DES, 120 QSO, and 181 \hiig-2021 data points to constrain cosmological and nuisance parameters in six flat and non-flat cosmological models. We can describe the relatively model-independent summary features of the constraints obtained from this $H(z)$ + BAO + SN-Pantheon + SN-DES + QSO + \hiig-2021 (HzBSNPDQH) data combination as follows.\footnote{The following summary values are obtained with the same method used in \cite{CaoRyanRatra2021}, where we take the summary central value to be the mean of the two of six central-most values. As for the uncertainty, we call the difference between the two central-most values twice the systematic uncertainty and the average of the two central-most error bars the statistical uncertainty, and compute the summary error bar as the quadrature sum of the two uncertainties.} First, the constraint on $l_{\rm m}$ is $l_{\rm m}=10.93\pm0.25$ pc, which is consistent with the $l_{\rm m}=11.03\pm0.25$ pc of  \cite{Cao_et_al2017b}. Second, the constraint on \om\ is $\Omega_{\rm m_0}=0.293 \pm 0.021$, which is in good agreement with many other recent measurements (e.g. $0.315\pm0.007$ from TT,TE,EE+lowE+lensing CMB anisotropy data in the flat \lcdm\ model of  \citealp{planck2018b}). Third, the determination of $H_0$ is $H_0=69.7\pm1.2$ \hunit, which is in better agreement with the estimate of \cite{chenratmed} than with the measurements of \cite{planck2018b} and \cite{Riess_2021}. There is some room for dark energy dynamics or a little spatial curvature energy density, but overall the flat \lcdm\ model is the best candidate model.


\cleardoublepage


\chapter{Standardizing Dainotti-correlated gamma-ray bursts, and using them with standardized Amati-correlated gamma-ray bursts to constrain cosmological model parameters}
\label{makereference7}

This chapter is based on \cite{CaoKhadkaRatra2022}, from analyses
conducted independently by Shulei Cao and Narayan Khadka.

\section{Introduction} 
\label{makereference7.1}

The observed currently accelerated cosmological expansion indicates that --- if general relativity provides an accurate description of gravitation on cosmological scales --- dark energy must contribute significantly to the current cosmological energy budget. The simpler spatially-flat $\Lambda$CDM model \citep{peeb84} is consistent with this and other observations. Fits of this model to most better-established cosmological data suggest that a time-independent cosmological constant $(\Lambda)$ provides $\sim 70\%$ of the current cosmological energy budget, non-relativistic cold dark matter (CDM) provides $\sim 25\%$, and non-relativistic baryonic matter provides most of the remaining $\sim 5\%$ \citep[see, e.g.][]{Farooq_Ranjeet_Crandall_Ratra_2017, scolnic_et_al_2018, planck2018b, eBOSS_2020}. While the spatially-flat $\Lambda$CDM model is consistent with most observations \citep[see, e.g.][]{DiValentinoetal2021a, PerivolaropoulosSkara2021}, observational data do not strongly rule out a little spatial curvature or dynamical dark energy. In this paper, in addition to the spatially-flat $\Lambda$CDM model, we also study spatially non-flat and dynamical dark energy models.  

Observational astronomy now provides many measurements that can be used to test cosmological models. Largely, these data are either at low or at high redshift. So cosmological models are mostly tested at low and high redshifts, remaining poorly tested in the intermediate redshift regime. The highest redshift of the better-established low-redshift data, $\sim 2.3$, is reached through baryon acoustic oscillation (BAO) observations; the high redshift region, $z \sim 1100$, is probed by better-established cosmic microwave background anisotropy data. Fits of these better-established data to cosmological models provide mostly mutually consistent results. However, for a better understanding of our Universe, it is necessary to also test cosmological models in the intermediate redshift range of $2.3 \lesssim z \lesssim 1100$. Some progress has been achieved: methods that test cosmological models in the intermediate redshift region include the use of \hii\ starburst galaxy measurements which reach to $z \sim 2.4$ \citep{Mania_2012, Chavez_2014, GonzalezMoran2019, GonzalezMoranetal2021, CaoRyanRatra2020, CaoRyanRatra2021, Caoetal_2021, Johnsonetal2022}, quasar angular size measurements which reach to $\sim 2.7$ \citep{Cao_et_al2017a, Ryanetal2019, CaoRyanRatra2020, Caoetal_2021, Zhengetal2021, Lian_etal_2021}, and quasar flux measurements which reach to $\sim 7.5$ \citep{RisalitiLusso2015, RisalitiLusso2019, KhadkaRatra2020a, KhadkaRatra2020b, KhadkaRatra2022, KhadkaRatra2021, Yangetal2020, Lussoetal2020, ZhaoXia2021, Lietal2021, Lian_etal_2021, Rezaeietal2022, Luongoetal2021}.\footnote{Note that in the latest \cite{Lussoetal2020} quasar flux compilation, their assumed UV--X-ray correlation is valid only to a much lower redshift, $z \sim 1.5-1.7$, and so these quasars can be used to derive only lower-$z$ cosmological constraints \citep{KhadkaRatra2022, KhadkaRatra2021}.}  

Gamma-ray burst (GRB) measurements are another high redshift probe and reach to $z \sim 8.2$ \citep{Amati2008, Amati2019, Salvaterraetal2009, Tanviretal2009, samushia_ratra_2010, Cardoneetal2010, Dainottietal2013a, Wangetal2015, Wang_2016, DainottiaDelVecchio2017, Dirirsa2019, KhadkaRatra2020c, Demianskietal_2021, Khadkaetal_2021b, Luongoetal2021, LuongoMuccino2021}. While there are quite a few Amati correlation long GRBs that have been used to constrain cosmological parameters, currently only a smaller fraction of 118 such GRBs (hereafter A118) that cover the redshift range $0.3399 \leq z \leq 8.2$ \citep{KhadkaRatra2020c, Khadkaetal_2021a} are reliable enough to be used to constrain cosmological parameters. To date, this is the lower-$z$ data set used to constrain cosmological parameters that spans the widest range of redshifts. These A118 data provide cosmological constraints which are consistent with those obtained from the better-established cosmological probes but the GRB constraints are significantly less restrictive. To obtain tighter cosmological constraints using GRB data, we need to make use of more GRBs.

Recently \cite{Wangetal_2021} and \cite{Huetal2021} have compiled smaller GRB data sets that together probe the redshift range $0.35 \leq z \leq 5.91$. These are GRBs whose plateau phase luminosity $L_0$ and spin-down characteristic time $t_b$ are correlated through the Dainotti ($L_0-t_b$) correlation \citep{Dainottietal2013b,Dainottietal2017}. This correlation between $L_0$ and $t_b$ allows one to use these GRBs for cosmological purposes. These GRBs can be classified in two categories depending on whether the plateau phase is dominated by magnetic dipole (MD) radiation or gravitational wave (GW) emission \citep{Wangetal_2021, Huetal2021}. In this paper we use long and short GRBs whose plateau phase is dominated by MD radiation (hereafter MD-LGRBs and MD-SGRBs) and long GRBs whose plateau phase is dominated by GW emission (hereafter GW-LGRBs). All three sets of GRBs obey the Dainotti correlation but each set can have different correlation parameters. We use the three individual GRB data sets, as well as some combinations of them, to constrain cosmological model parameters and Dainotti correlation parameters simultaneously.\footnote{The advantage of fitting cosmological and GRB correlation parameters simultaneously is that the fitting process is free from the circularity problem. More specifically, this procedure allows us to determine whether the GRB correlation parameters depend on the assumed cosmological model and so determine whether the GRBs are standardizable.} We find that these GRBs are standardizable, as was assumed in \cite{Wangetal_2021} and \cite{Huetal2021}. However, cosmological constraints obtained from these Dainotti correlation GRB data sets are very weak. 

When we combine the MD-LGRB or GW-LGRB data sets with the 115 non-overlapping Amati correlation GRBs from the A118 data set, they slightly tighten the constraints from the 115 Amati correlation GRBs, but not significantly so. Each of the individual Amati or Dainotti correlation GRB data sets, as well as combinations of these GRB data sets, mostly provide only lower limits on the current value of the non-relativistic matter energy density parameter $\Omega_{\rm m0}$ and the resulting cosmological parameter constraints are mostly consistent with those obtained from better-established cosmological data. 

In this paper, we use a combination of Hubble parameter ($H(z)$) and BAO data, $H(z)$ + BAO, results as a proxy for better-established data results, to compare with our GRB data results. Qualitatively, results from the individual GRB data sets, as well as those from combinations of GRB data sets, are consistent with those from the $H(z)$ + BAO data which favor $\Omega_{\rm m0} \sim 0.3$, but there are a few combinations of GRB data sets with constraints on \om\ being more than $2\sigma$ away from 0.3 in the \lcdm\ models.

This paper is structured as follows. We use the cosmological models described in Chapter \ref{sec:models}. In Sec.\ \ref{makereference7.2} we describe the data sets we analyze. In Sec.\ \ref{makereference7.3} we summarize our analyses techniques. In Sec.\ \ref{makereference7.4} we present our results. We conclude in Sec.\ \ref{makereference7.5}.

\section{Data}
\label{makereference7.2}

In this paper, we analyze four different GRB data sets as well as some combinations of these data sets. We also use a joint $H(z)$ + BAO data set. These data sets are summarized in Table \ref{tab:dataC7} and described in what follows.\footnote{In this table and elsewhere, for compactness, we sometimes use ML, MS, and GL as abbreviations for the MD-LGRB, MD-SGRB, and GW-LGRB data sets compiled by \cite{Wangetal_2021} and \cite{Huetal2021}.}

\begin{itemize}

\item[]{\bf MD-LGRB sample}. This includes 31 long GRBs, with burst duration longer than 2 seconds, listed in Table 1 of \cite{Wangetal_2021}. For this data set, measured quantities for a GRB are redshift $z$, X-ray flux $F_0$, characteristic time scale $t_b$, and spectral index during the plateau phase $\beta^{\prime}$.\footnote{ML, MS, and GL data error bars on $F_0$ and $t_b$ are mostly asymmetric. We symmetrize these error bars using the method applied in \cite{Wangetal_2021} and \cite{Huetal2021}, with the symmetrized error bar $\sigma = \sqrt{(\sigma_u^2 + \sigma_d^2)/2}$, where $\sigma_u$ and $\sigma_d$ are the asymmetric upper and lower error bars.} This sample probes the redshift range $1.45 \leq z \leq 5.91$.

\item[]{\bf MD-SGRB sample}. This includes 5 short GRBs, with burst duration shorter than 2 seconds, listed in Table 1 of \cite{Huetal2021}. For this data set, measured quantities for a GRB are the same as those for the MD-LGRB sample. This data set probes the redshift range $0.35 \leq z \leq 2.6$.

\item[]{\bf GW-LGRB sample}. This includes 24 long GRBs listed in Table 1 of \cite{Huetal2021}. For this data set, measured quantities for a GRB are the same as those for the MD-LGRB sample. This sample probes the redshift range $0.55 \leq z \leq 4.81$.

\item[]{\bf A118 sample}. This sample include 118 long GRBs listed in Table 7 of \cite{Khadkaetal_2021a}. For this data set, measured quantities for a GRB are $z$, rest-frame spectral peak energy $E_{\rm p}$, and measured bolometric fluence $S_{\rm bolo}$, computed in the standard rest-frame energy band $1-10^4$ keV. This sample probes the redshift range $0.3399 \leq z \leq 8.2$. Note that in our analyses here we did not use the correct value of $E_{\rm p}=871\pm123$ keV for GRB081121, as discussed in Ref.\ \cite{Liuetal2022}, although the effects on the parameter constraints are small.

The A118 data and the MD-LGRB data sets have 3 common GRBs, GRB060526, GRB081008, and GRB090516. We exclude these common GRBs from the A118 data set to form the A115 data set for joint analyses with the MD-LGRB data set. There are also 3 common GRBs between the A118 data set and the GW-LGRB data set, GRB060206, GRB091029, and GRB131105A. We exclude these common GRBs from the A118 data set to form the A115$^{\prime}$ data set for joint analyses with the GW-LGRB data set.

\item[]{$\textbf{ \emph{H(z)}}$ \bf and BAO data}. In addition to the GRB data, we also use 31 $H(z)$ and 11 BAO measurements. These $H(z)$ and BAO measurements probe the redshift range $0.07 \leq z \leq 1.965$ and $0.0106 \leq z \leq 2.33$, respectively. The $H(z)$ data are in Table 2 of \cite{Ryan_1} and the BAO data are in Table 1 of \cite{KhadkaRatra2020c}. We use cosmological constraints from the better-established $H(z)$ + BAO data to compare with those obtained from the GRB data sets.

\end{itemize}

\section{Data Analysis Methodology}
\label{makereference7.3}

\begin{table}
\centering
\begin{threeparttable}
\caption{Summary of data sets used.}
\setlength{\tabcolsep}{3.5pt}
\label{tab:dataC7}
\begin{tabular}{lcc}
\toprule
Data set & $N$ (Number of points) & Redshift range\\
\midrule
ML & 31 & $1.45 \leq z \leq 5.91$ \\
MS & 5 & $0.35 \leq z \leq 2.6$ \\
GL & 24 & $0.55 \leq z \leq 4.81$ \\
MS + GL & 29 & $0.35 \leq z \leq 4.81$ \\
A118 & 118 & $0.3399 \leq z \leq 8.2$ \\
A115\tnote{a} & 115 & $0.3399 \leq z \leq 8.2$ \\
A115$^{\prime}$\tnote{b} & 115 & $0.3399 \leq z \leq 8.2$ \\
\midrule
$H(z)$ & 31 & $0.070 \leq z \leq 1.965$ \\
BAO & 11 & $0.38 \leq z \leq 2.334$ \\
\bottomrule
\end{tabular}
\begin{tablenotes}[flushleft]
\item [a] Excluding from A118 those GRBs in common with MD-LGRB (GRB060526, GRB081008, and GRB090516).
\item [b] Excluding from A118 those GRBs in common with GW-LGRB (GRB060206, GRB091029, and GRB131105A).
\end{tablenotes}
\end{threeparttable}%
\end{table}

\begin{table}
\centering
\begin{threeparttable}
\caption{Flat priors of the constrained parameters.}
\label{tab:priorsC7}
\setlength{\tabcolsep}{3.5pt}
\begin{tabular}{lcc}
\toprule
Parameter & & Prior\\
\midrule
 & Cosmological Parameters & \\
\midrule
$H_0$\,\tnote{a} &  & [None, None]\\
\obhs\,\tnote{b} &  & [0, 1]\\
\ochs\,\tnote{c} &  & [0, 1]\\
\ok &  & [-2, 2]\\
$\alpha$ &  & [0, 10]\\
\wx &  & [-5, 0.33]\\
\midrule
 & GRB Nuisance Parameters\tnote{d} & \\
\midrule
$k$ &  & [-10, 10]\\
$b$\,\tnote{e} &  & [0, 10]\\
$\sigma_{\rm int}$ &  & [0, 5]\\
$\beta$ &  & [0, 5]\\
$\gamma$ &  & [0, 300]\\
\bottomrule
\end{tabular}
\begin{tablenotes}[flushleft]
\item [a] \hunit. In the GRB alone cases, $H_0$ is set to be 70 \hunit, while in the $H(z)$ + BAO case, the prior range is irrelevant (unbounded).
\item [b] In the GRB alone cases, \obhs\ is set to be 0.0245, i.e. $\Omega_{b}=0.05$.
\item [c] In the GRB alone cases, $\Omega_{c}\in[-0.05,0.95]$ to ensure $\Om\in[0,1]$.
\item [d] Note that $k$, $b$, and $\sigma_{\rm int}$ of MD-LGRBs are different from those of MD-SGRBs/GW-LGRBs, but with the same prior ranges.
\item [e] $b<0$ values are possible for MD-SGRBs (due to fewer data points) but, as discussed below, requiring $b\geq 0$ does not have significant consequences.
\end{tablenotes}
\end{threeparttable}%
\end{table}

For GRBs which obey the Dainotti correlation the luminosity of the plateau phase is \citep{Dainottietal2008, Dainottietal2010, Dainottietal2011}
\be
\label{eq:L0}
    L_0=\frac{4\pi D_L^2F_0}{(1+z)^{1-\beta^{\prime}}},
\ee
where $F_0$ is the GRB X-ray flux, $\beta^{\prime}$ is the spectral index in the plateau phase, and $D_L$ is the luminosity distance. 

$D_L$, as a function of redshift $z$ and cosmological parameters $\textbf{\emph{p}}$, is given by
\begin{equation}
  \label{eq:DL}
  \frac{H_0\sqrt{\left|\Omega_{\rm k0}\right|}D_L(z, \textbf{\emph{p}})}{c(1+z)} = 
    \begin{cases}
    {\rm sinh}\left[g(z, \textbf{\emph{p}})\right] & \text{if}\ \Omega_{\rm k0} > 0, \\
    \vspace{1mm}
    g(z, \textbf{\emph{p}}) & \text{if}\ \Omega_{\rm k0} = 0,\\
    \vspace{1mm}
    {\rm sin}\left[g(z, \textbf{\emph{p}})\right] & \text{if}\ \Omega_{\rm k0} < 0,
    \end{cases}   
\end{equation}
where
\begin{equation}
\label{eq:gz}
   g(z, \textbf{\emph{p}}) = H_0\sqrt{\left|\Omega_{\rm k0}\right|}\int^z_0 \frac{dz'}{H(z', \textbf{\emph{p}})},
\end{equation}
$c$ is the speed of light, and $H(z, \textbf{\emph{p}})$ is the Hubble parameter that is described in Chapter \ref{sec:models} for each cosmological model.

For these GRBs the luminosity of the plateau phase $L_0$ and the characteristic time scale $t_b$ are correlated through the Dainotti or luminosity-time relation
\begin{equation}
    \label{eq:dcorr}
    y\equiv\log \left(\frac{L_0}{10^{47}\ \mathrm{erg/s}}\right) = k\log \frac{t_b}{10^3(1+z)\ \mathrm{s}} + b\equiv kx + b,
\end{equation}
where $\log=\log_{10}$ and the slope $k$ and the intercept $b$ are free parameters to be determined from the data.

We predict $L_0$ as a function of cosmological parameters $\textbf{\emph{p}}$ at the redshift of each GRB by using eqs.\ \eqref{eq:L0}, \eqref{eq:DL}, and \eqref{eq:dcorr}. We then compare predicted and measured values of $L_0$ by using the natural log of the likelihood function \citep{D'Agostini_2005}
\be
\label{eq:LH_MD-LGRB}
    \ln\mathcal{L}_{\rm GRB}= -\frac{1}{2}\Bigg[\chi^2_{\rm GRB}+\sum^{N}_{i=1}\ln\left(2\pi(\sigma_{\rm int}^2+\sigma_{{y_i}}^2+k^2\sigma_{{x_i}}^2)\right)\Bigg],
\ee
where
\be
\label{eq:chi2_MD-LGRB}
    \chi^2_{\rm GRB} = \sum^{N}_{i=1}\bigg[\frac{(y_i-k x_i-b)^2}{(\sigma_{\rm int}^2+\sigma_{{y_i}}^2+k^2\sigma_{{x_i}}^2)}\bigg].
\ee
Here $N$ is the number of data points (e.g., for MD-LGRB $N=31$), and $\sigma_{\rm int}$ is the intrinsic scatter parameter (which also contains the unknown systematic uncertainty).

For GRBs which obey the Amati correlation the rest frame isotropic radiated energy $E_{\rm iso}$ is
\be
\label{eq:Eiso}
    E_{\rm iso}=\frac{4\pi D_L^2}{1+z}S_{\rm bolo},
\ee
where $S_{\rm bolo}$ is the bolometric fluence. 

For these GRBs the rest frame peak photon energy $E_{\rm p}$ and $E_{\rm iso}$ are correlated through the Amati (or $E_{\rm p}-E_{\rm iso}$) relation \eqref{eq:Amati} \citep{Amati2008, Amati2009}.

We predict $E_{\rm iso}$ as a function of cosmological parameters $\textbf{\emph{p}}$ at the redshift of each GRB by using eqs.\ \eqref{eq:DL}, \eqref{eq:Eiso}, and \eqref{eq:Amati}. We then compare predicted and measured values of $E_{\rm iso}$ by using the natural log of the likelihood function \citep{D'Agostini_2005}
\be
\label{eq:LH_GRBC7}
    \ln\mathcal{L}_{\rm A118}= -\frac{1}{2}\Bigg[\chi^2_{\rm A118}+\sum^{N}_{i=1}\ln\left(2\pi(\sigma_{\rm int}^2+\sigma_{{y^{\prime}_i}}^2+\beta^2\sigma_{{x^{\prime}_i}}^2)\right)\Bigg],
\ee
where
\be
\label{eq:chi2_GRBC7}
    \chi^2_{\rm A118} = \sum^{N}_{i=1}\bigg[\frac{(y^{\prime}_i-\beta x^{\prime}_i-\gamma)^2}{(\sigma_{\rm int}^2+\sigma_{{y^{\prime}_i}}^2+\beta^2\sigma_{{x^{\prime}_i}}^2)}\bigg].
\ee
Here $x^{\prime}=\log(E_{\rm p}/{\rm keV})$, $\sigma_{x^{\prime}}=\sigma_{E_{\rm p}}/(E_{\rm p}\ln 10)$, $y^{\prime}=\log(E_{\rm iso}/{\rm erg})$, and $\sigma_{\rm int}$ is the intrinsic scatter parameter, which also contains the unknown systematic uncertainty.

The $H(z)$ + BAO data analyses follow the method described in Sec.\ 4 of \cite{KhadkaRatra2022}. 

We maximize the likelihood function using the Markov chain Monte Carlo (MCMC) method as implemented in the \textsc{MontePython} code \citep{Brinckmann2019} and determine the best-fitting and posterior mean values and the corresponding uncertainties for all free parameters. We assure convergence of the MCMC chains for each free parameter from the Gelman-Rubin criterion ($R-1 < 0.05$). Flat priors used for the free parameters are given in Table \ref{tab:priorsC7}.

The Akaike Information Criterion ($AIC$) in equation \eqref{AIC1} and the Bayesian Information Criterion ($BIC$) in equation \eqref{BIC1} are used to compare the goodness of fit of models with different numbers of parameters.

\section{Results}
\label{makereference7.4}

\subsection{Constraints from ML, MS, and GL data}
 \label{subsec:MLSGL}
 
\begin{table}
\centering
\begin{threeparttable}
\caption{One-dimensional marginalized posterior means and 68.27\% limits of the Dainotti correlation parameters for the ML, GL, and MS data sets using the flat \lcdm\ model with $\Omega_{\rm m0} = 0.3$ and $H_0 = 70$ \hunit, and comparison with the results given in \protect\cite{Wangetal_2021} and \protect\cite{Huetal2021}.}
\label{tab:fix}
\setlength{\tabcolsep}{3.5pt}
\begin{tabular}{lcccc}
\toprule
Data set & Source & $k$ & $b$ & $\sigma_{\rm int}$\\
\midrule
 & {}\tnote{a} & $-1.02^{+0.09}_{-0.08}$ & $1.72^{+0.07}_{-0.07}$ & --\\
ML & {}\tnote{b} & $-1.026\pm0.085$ & $1.726\pm0.074$ & $0.303^{+0.032}_{-0.050}$\\
 & {}\tnote{c} & $-1.026\pm0.086$ & $1.726\pm0.074$ & $0.303^{+0.032}_{-0.050}$\\
\midrule 
 & {}\tnote{d} & $-1.77^{+0.20}_{-0.20}$ & $0.66^{+0.01}_{-0.01}$ & $0.42^{+0.08}_{-0.06}$\\
GL & {}\tnote{b} & $-1.753^{+0.187}_{-0.208}$ & $0.642^{+0.100}_{-0.071}$ & $0.428^{+0.053}_{-0.086}$\\
 & {}\tnote{c} & $-1.769\pm0.205$ & $0.656\pm0.094$ & $0.431^{+0.054}_{-0.088}$\\
\midrule
 & {}\tnote{d} & $-1.38^{+0.17}_{-0.19}$ & $0.33^{+0.17}_{-0.16}$ & $0.35^{+0.20}_{-0.12}$\\
MS & {}\tnote{b} & $-1.381^{+0.209}_{-0.213}$ & $0.327^{+0.195}_{-0.189}$ & $0.420^{+0.086}_{-0.242}$\\
 & {}\tnote{c} & $-1.397^{+0.247}_{-0.241}$ & $0.354^{+0.195}_{-0.242}$ & $0.525^{+0.044}_{-0.358}$\\
\bottomrule
\end{tabular}
\begin{tablenotes}[flushleft]
\item [a] Results from \cite{Wangetal_2021} with the prior ranges of the parameters being $k\in(-1.3,-0.75)$, $b\in(1.4,2.0)$, and $\sigma_{\rm int}\in(0.1,0.6)$ for ML.
\item [b] Our results with the same prior ranges of the parameters as \cite{Wangetal_2021} or \cite{Huetal2021}.
\item [c] Our results with wider prior ranges of the parameters $k\in[-10,10]$, $b\in[0,10]$ ($b\in[-0.5,10]$), and $\sigma_{\rm int}\in[0,3]$ for ML and GL (MS).
\item [d] Results from \cite{Huetal2021} with the prior ranges of the parameters being $k\in(-2.2,-1)$, $b\in(0.1,0.8)$, and $\sigma_{\rm int}\in(0.01,0.8)$ for GL and being $k\in(-2.1,-0.55)$, $b\in(-0.5,1.0)$, and $\sigma_{\rm int}\in(0.01,1)$ for MS.
\end{tablenotes}
\end{threeparttable}%
\end{table}

In Table \ref{tab:fix} we list Dainotti correlation parameters computed using the ML, GL, and MS data sets. These are computed in the flat \lcdm\ model with $\Omega_{\rm m0} = 0.3$ and $H_0 = 70$ \hunit, the same model and parameter values used in \cite{Wangetal_2021} and \cite{Huetal2021}. The first line of parameter values in each of the three subpanels of Table \ref{tab:fix} are taken from these papers.\footnote{\cite{Wangetal_2021} do not list a value for $\sigma_{\rm int}$ in the ML case.} To compare to these results, we used \textsc{emcee} (\citealp{emcee}) to compute the values listed in the second and third lines of each subpanel. Comparing the first and second lines in each subpanel, we find that they are consistent, except: i) for the GL case our $b$ uncertainties are larger than those of \cite{Huetal2021}; and, ii) for the MS case we have larger $b$ and $k$ error bars and a larger central value of $\sigma_{\rm int}$ than those of \cite{Huetal2021}, but they agree within 1$\sigma$. In the third line of each subpanel we list results obtained assuming wider prior ranges of the parameters. We find that the ML results do not change, the GL results are shifted closer to those of \cite{Huetal2021}, except for the values of $\sigma_{\rm int}$, and the MS results are shifted away from those of \cite{Huetal2021} with larger error bars, especially for $\sigma_{\rm int}$.

\begin{figure*}
\centering
 \subfloat[MD-LGRB]{%
    \includegraphics[width=3.25in,height=1.84in]{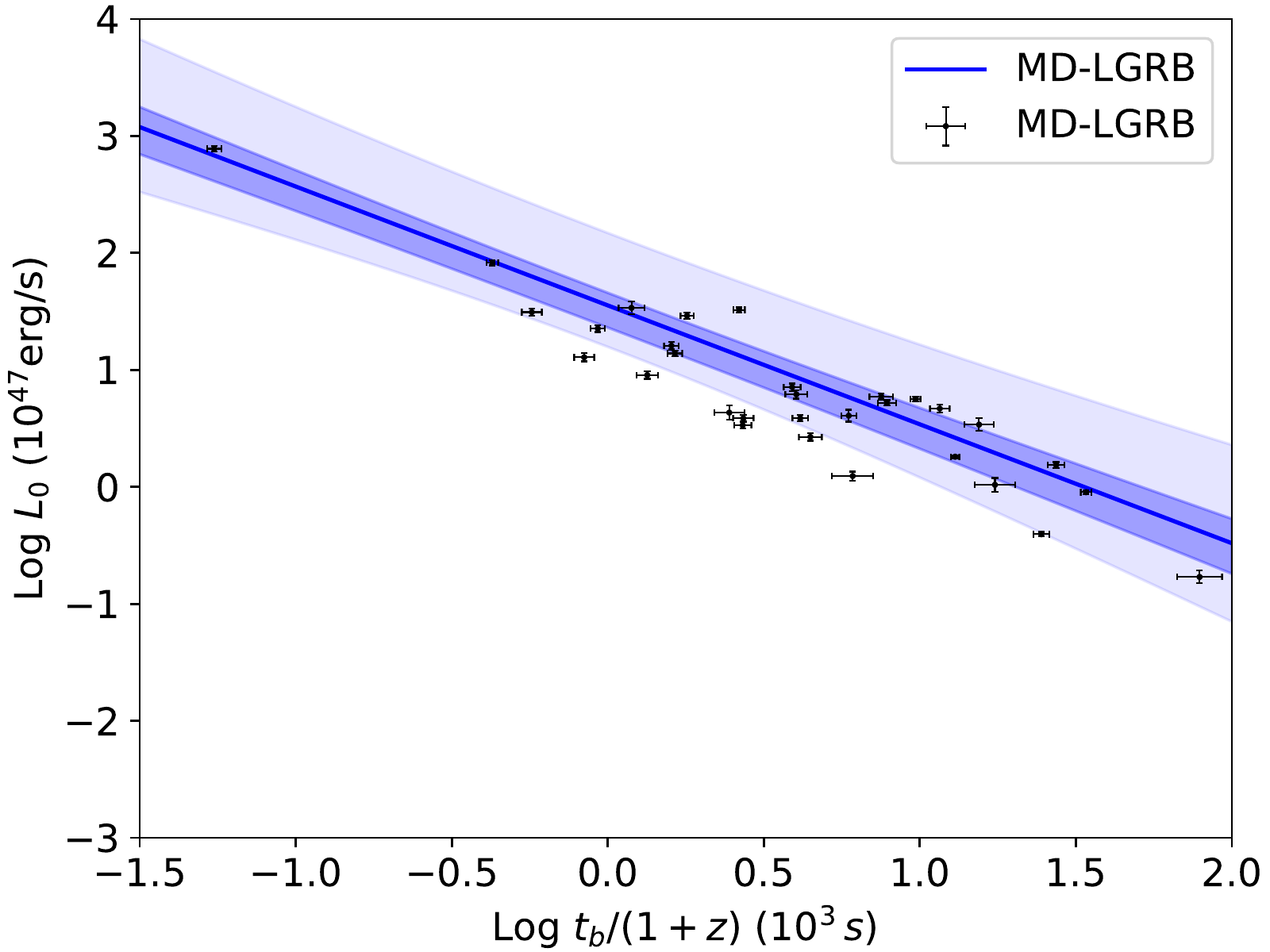}}
 \subfloat[MD-SGRB]{%
    \includegraphics[width=3.25in,height=1.84in]{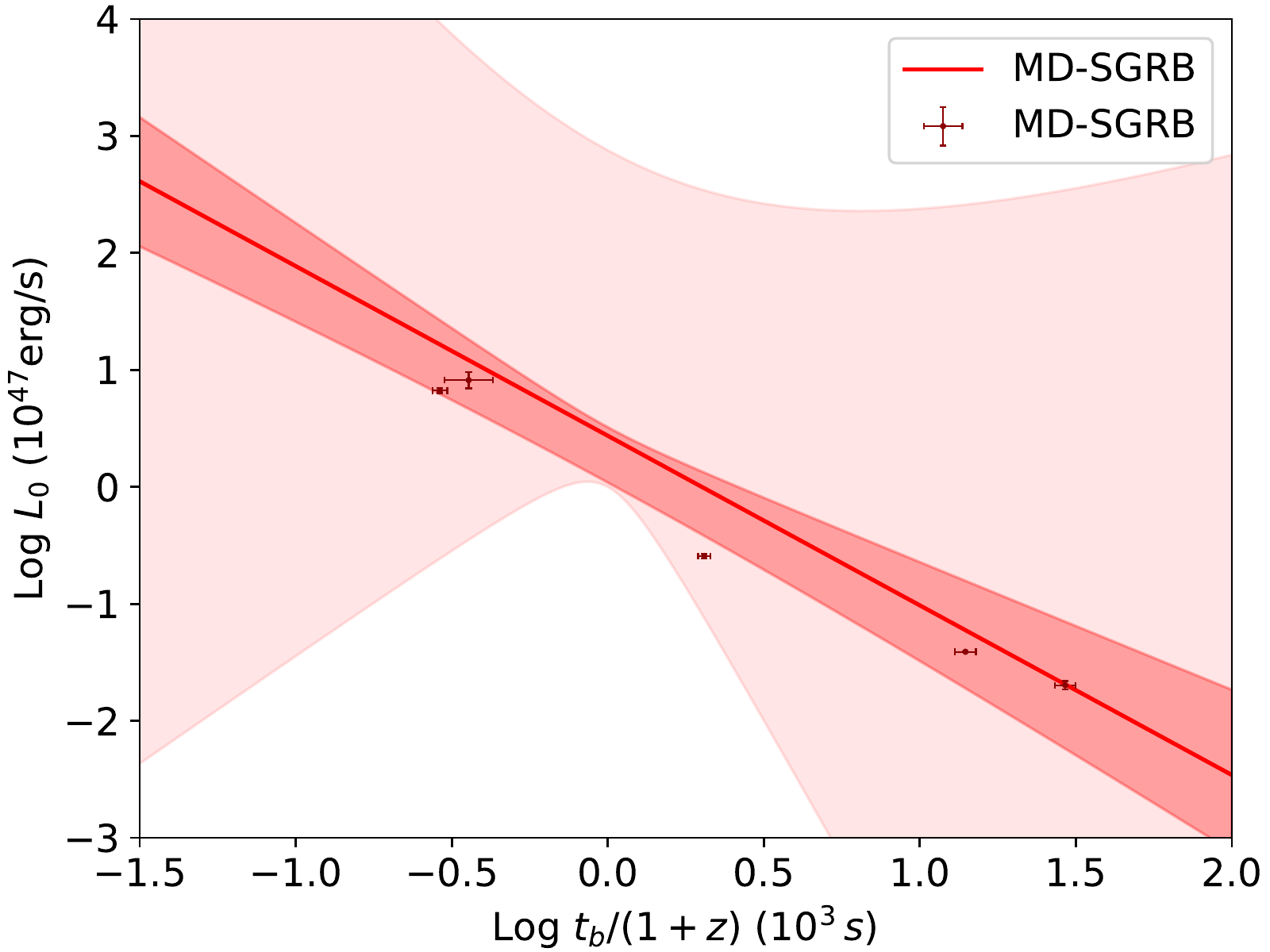}}\\
 \subfloat[GW-LGRB and MD-SGRB]{%
    \includegraphics[width=3.25in,height=1.84in]{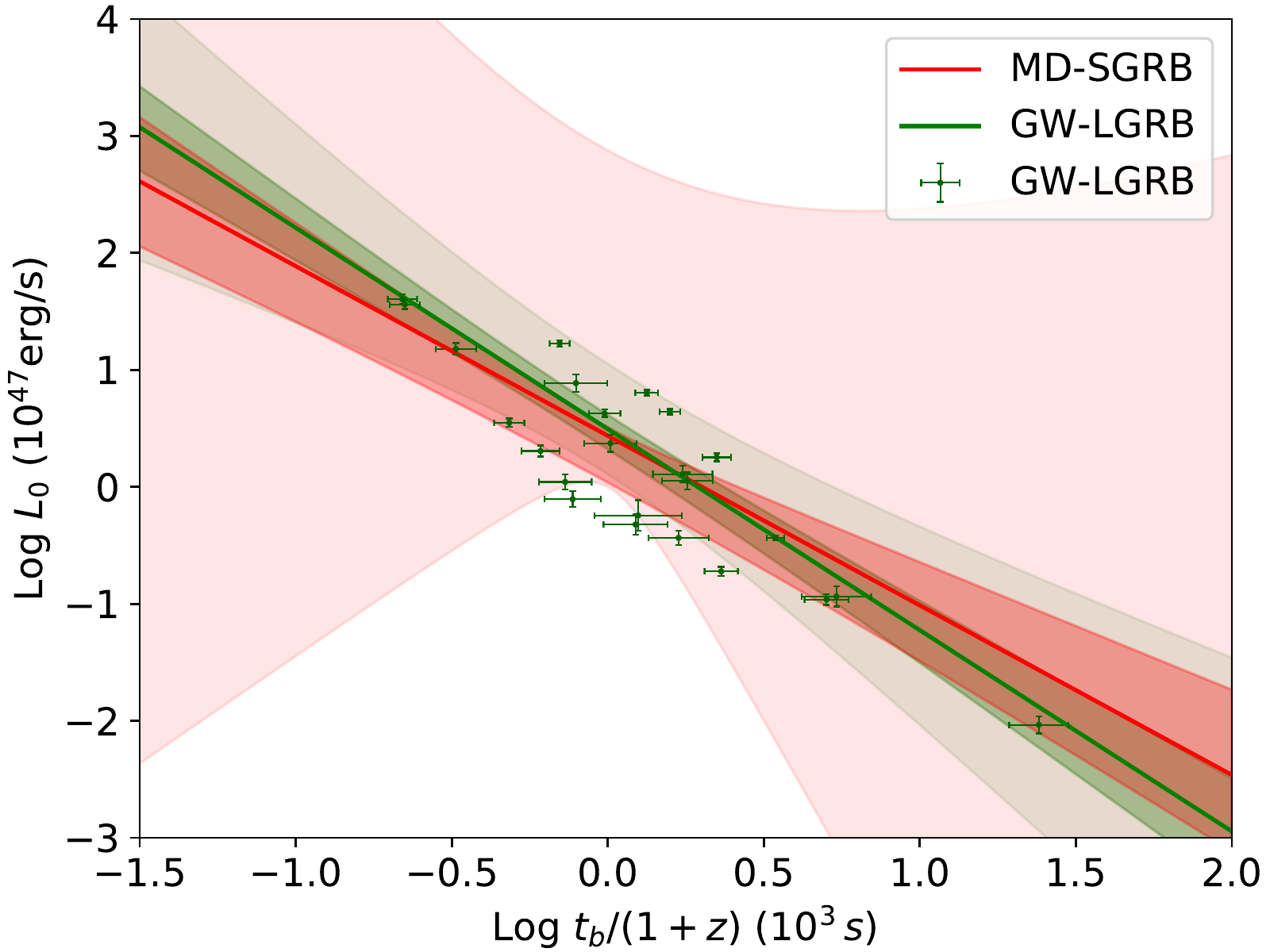}}
 \subfloat[MD-SGRB + GW-LGRB]{%
    \includegraphics[width=3.25in,height=1.84in]{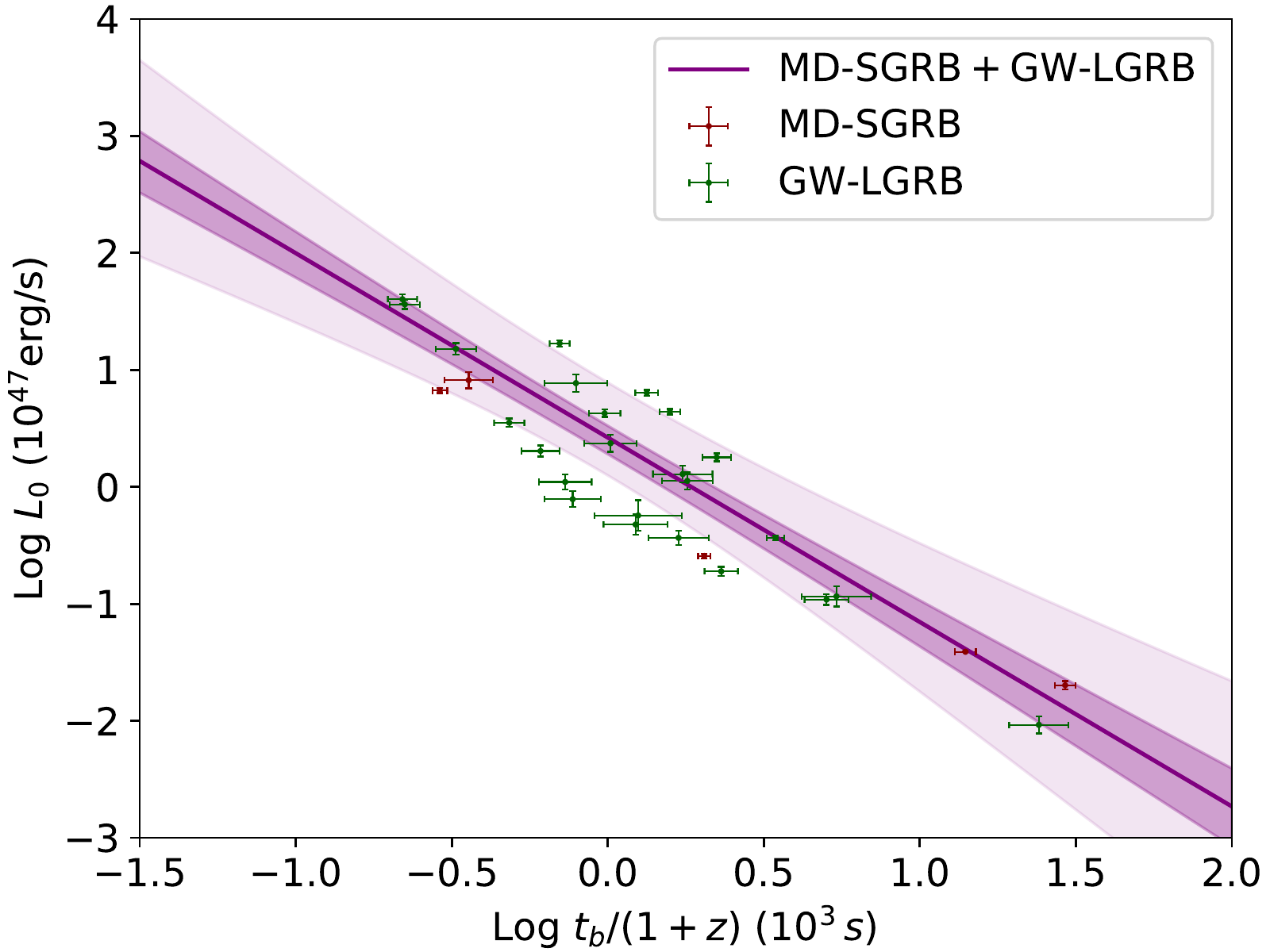}}\\
\caption{$L_0-t_b$ correlations for MD-LGRB, MD-SGRB, GW-LGRB, and MD-SGRB + GW-LGRB data using the flat \lcdm\ model. The MD-LGRB, MD-SGRB, and GW-LGRB data with error bars are shown in black, dark red, and dark green, respectively. The solid lines are the $L_0-t_b$ correlations with posterior mean values for the slopes and intercepts listed in Table \ref{tab:1d_BFPC7}, for MD-LGRB (blue), MD-SGRB (red), GW-LGRB (green), and MD-SGRB + GW-LGRB (purple) data. The $1\sigma$ and $3\sigma$ confidence regions are the dark and light colored shaded regions with the uncertainties propagated from those of $k$ and $b$ (without considering $\sigma_{\rm int}$).}
\label{fig00C7}
\end{figure*}

\begin{figure*}
\centering
 \subfloat[Flat \lcdm]{%
    \includegraphics[width=3.25in,height=1.84in]{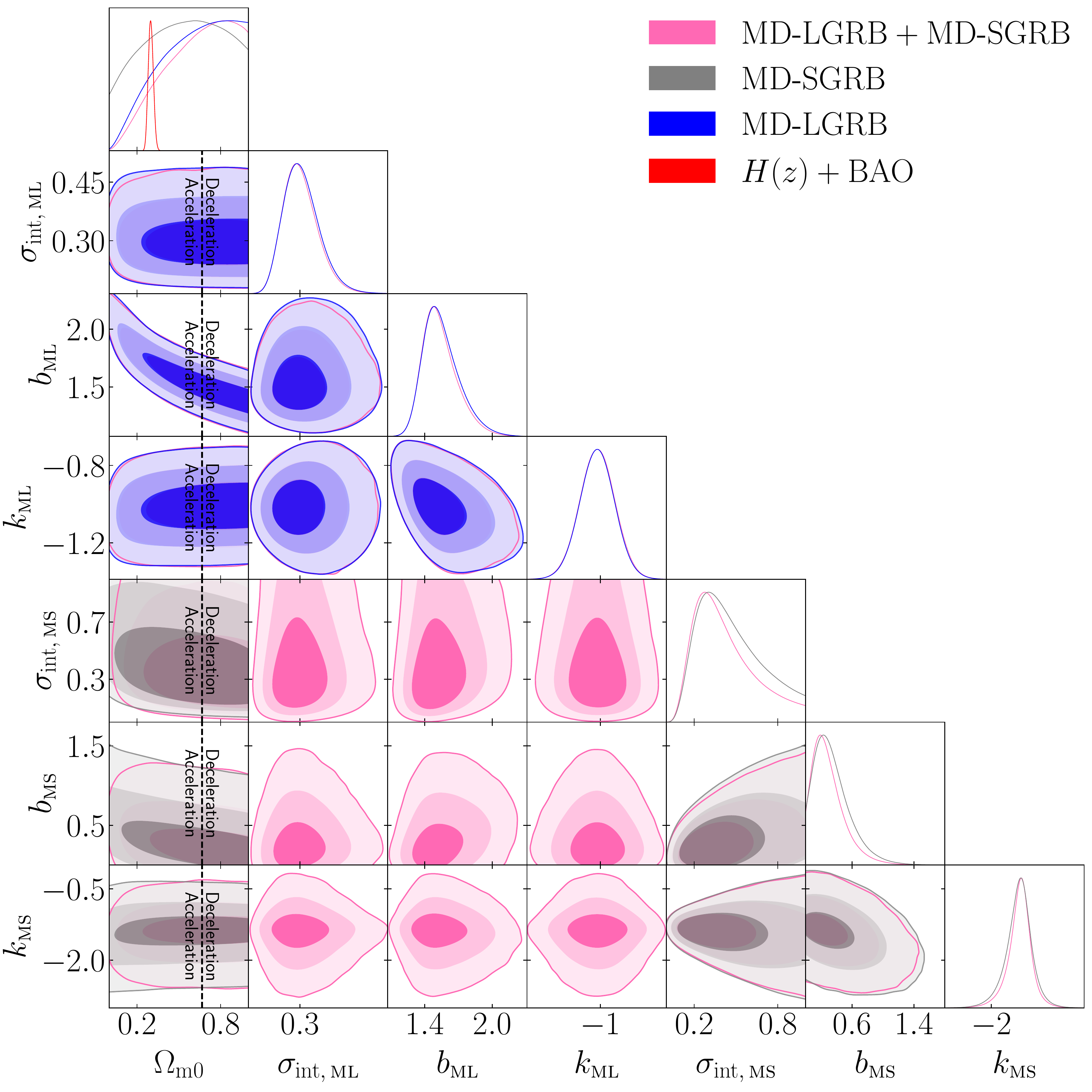}}
 \subfloat[Non-flat \lcdm]{%
    \includegraphics[width=3.25in,height=1.84in]{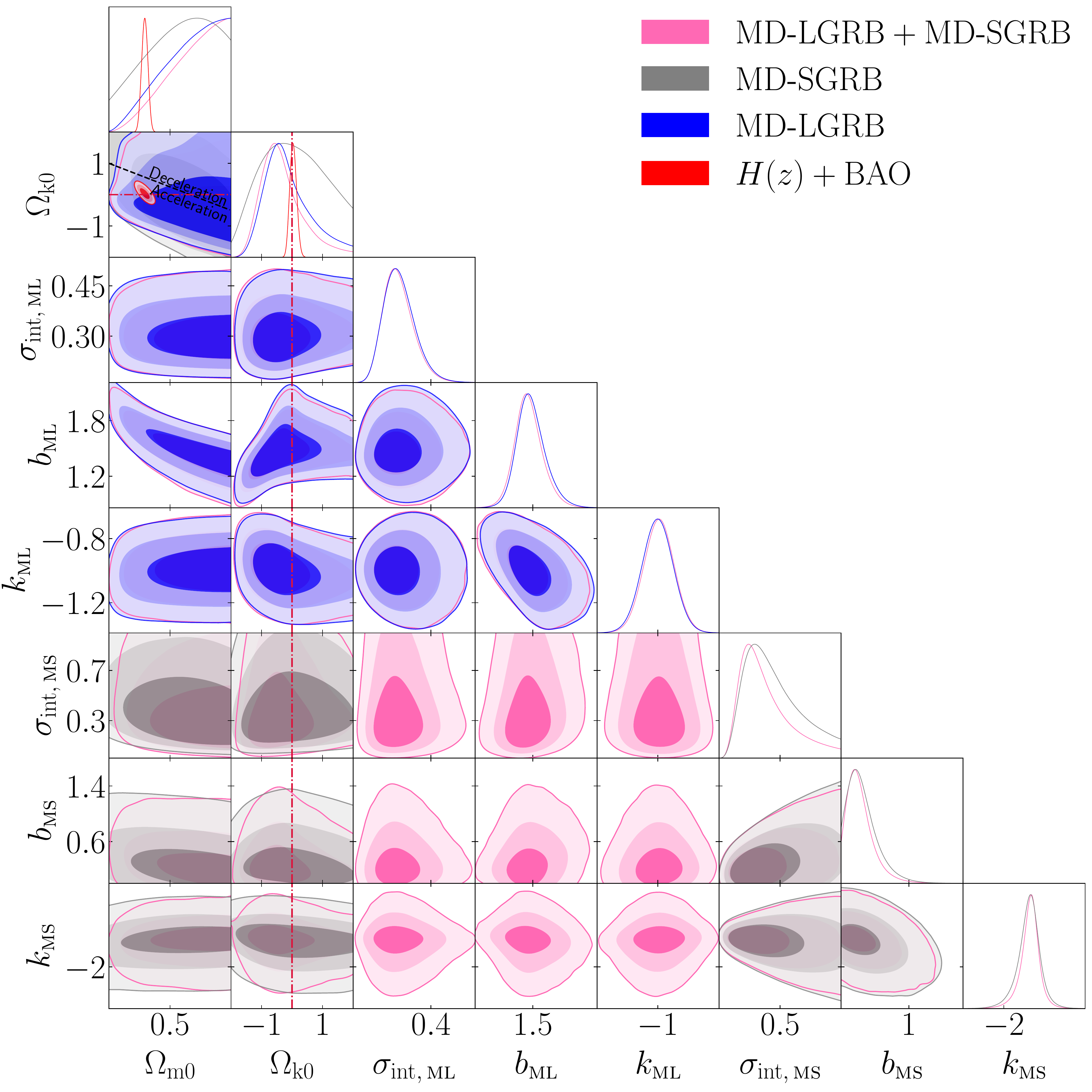}}\\
 \subfloat[Flat XCDM]{%
    \includegraphics[width=3.25in,height=1.84in]{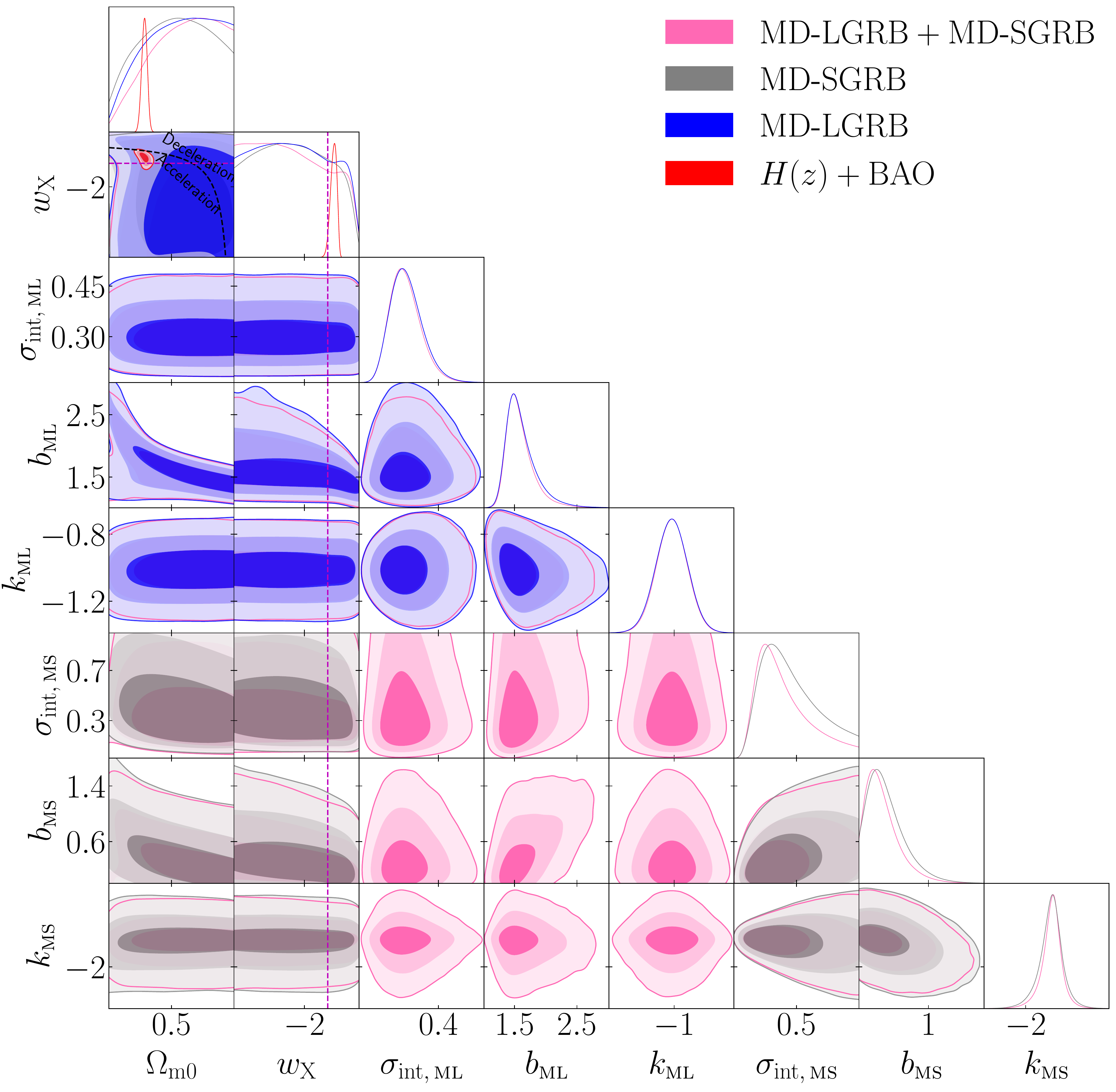}}
 \subfloat[Non-flat XCDM]{%
    \includegraphics[width=3.25in,height=1.84in]{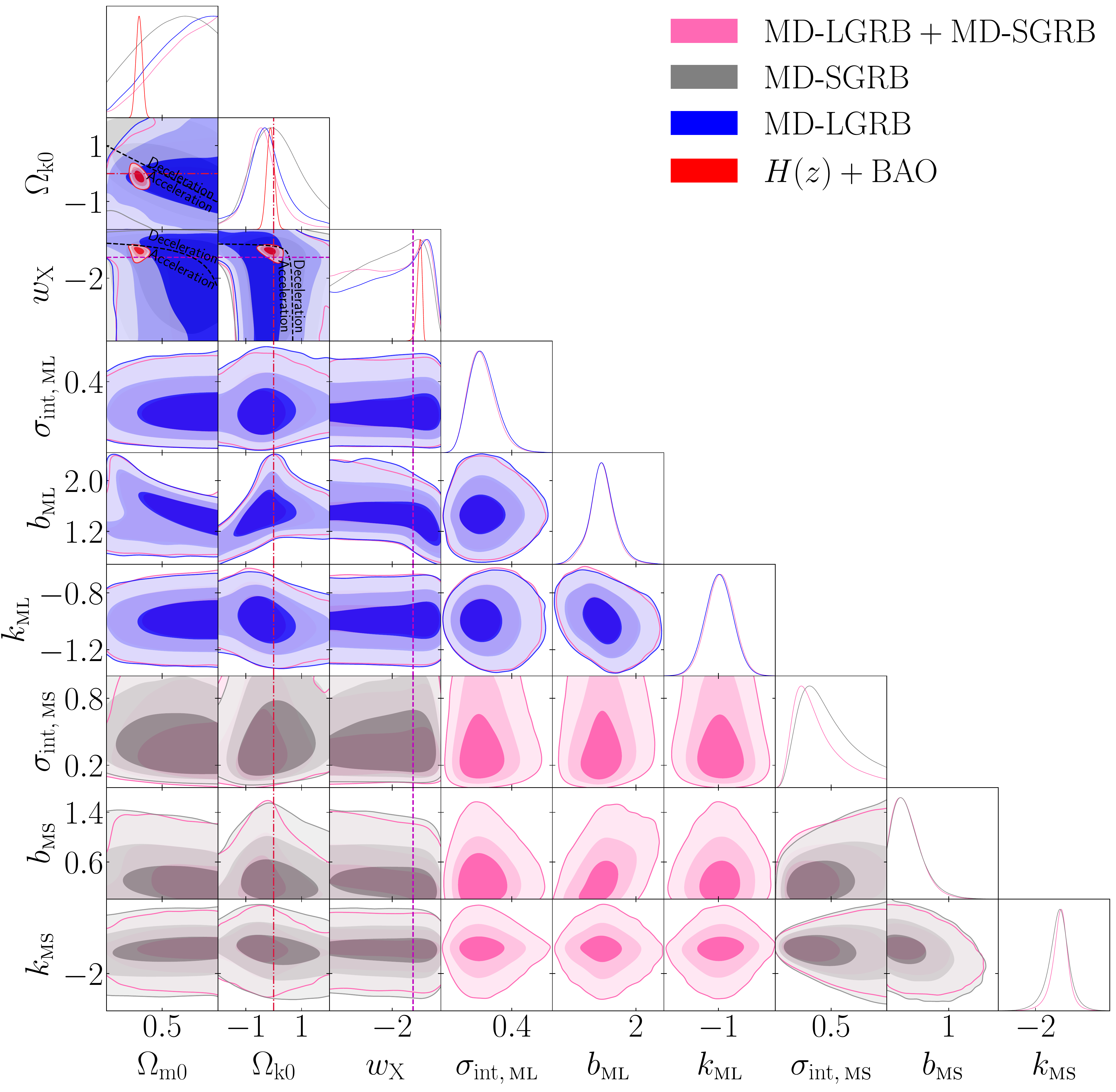}}\\
 \subfloat[Flat \pcdm]{%
    \includegraphics[width=3.25in,height=1.84in]{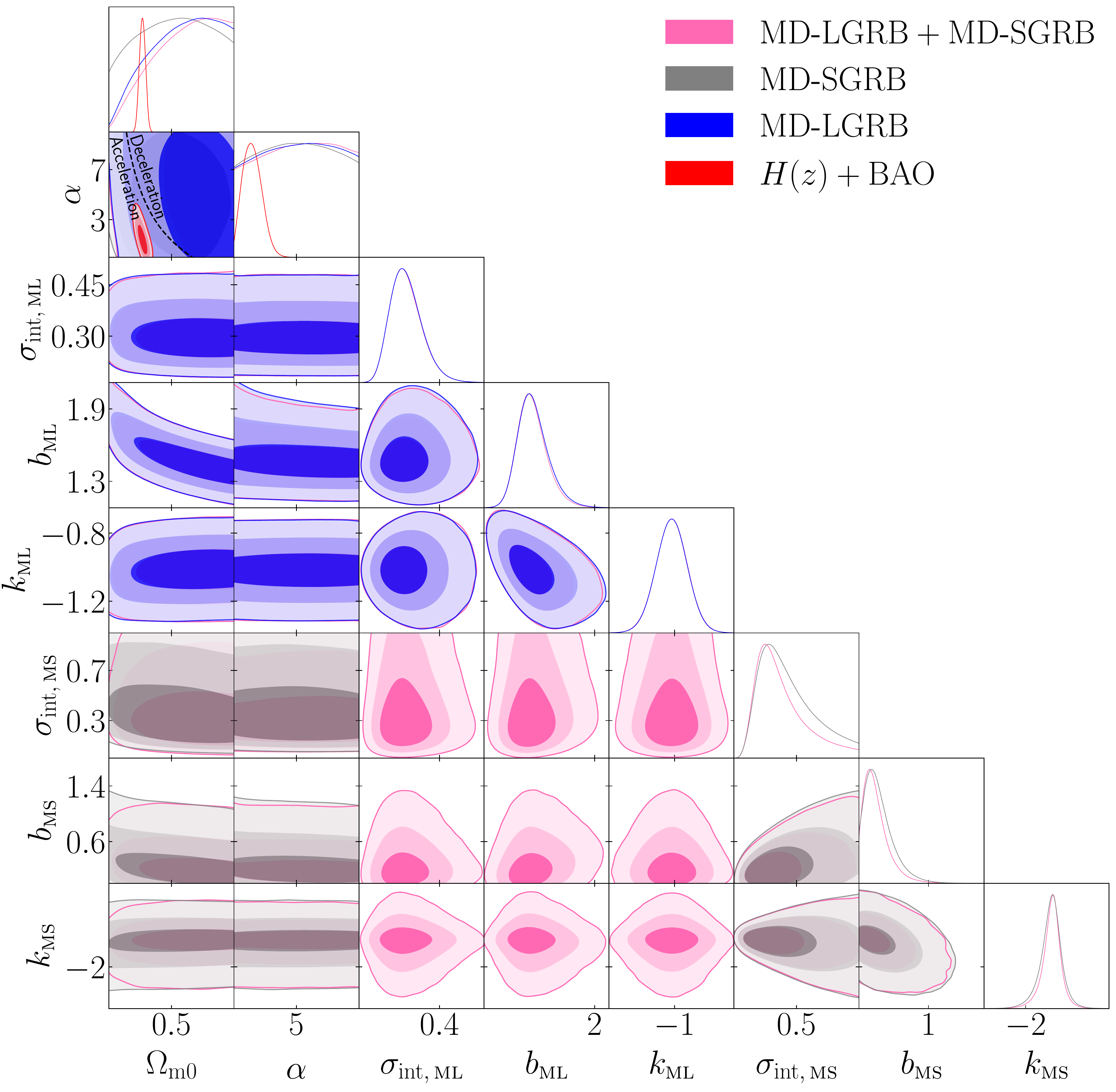}}
  \subfloat[Non-flat \pcdm]{%
     \includegraphics[width=3.25in,height=1.84in]{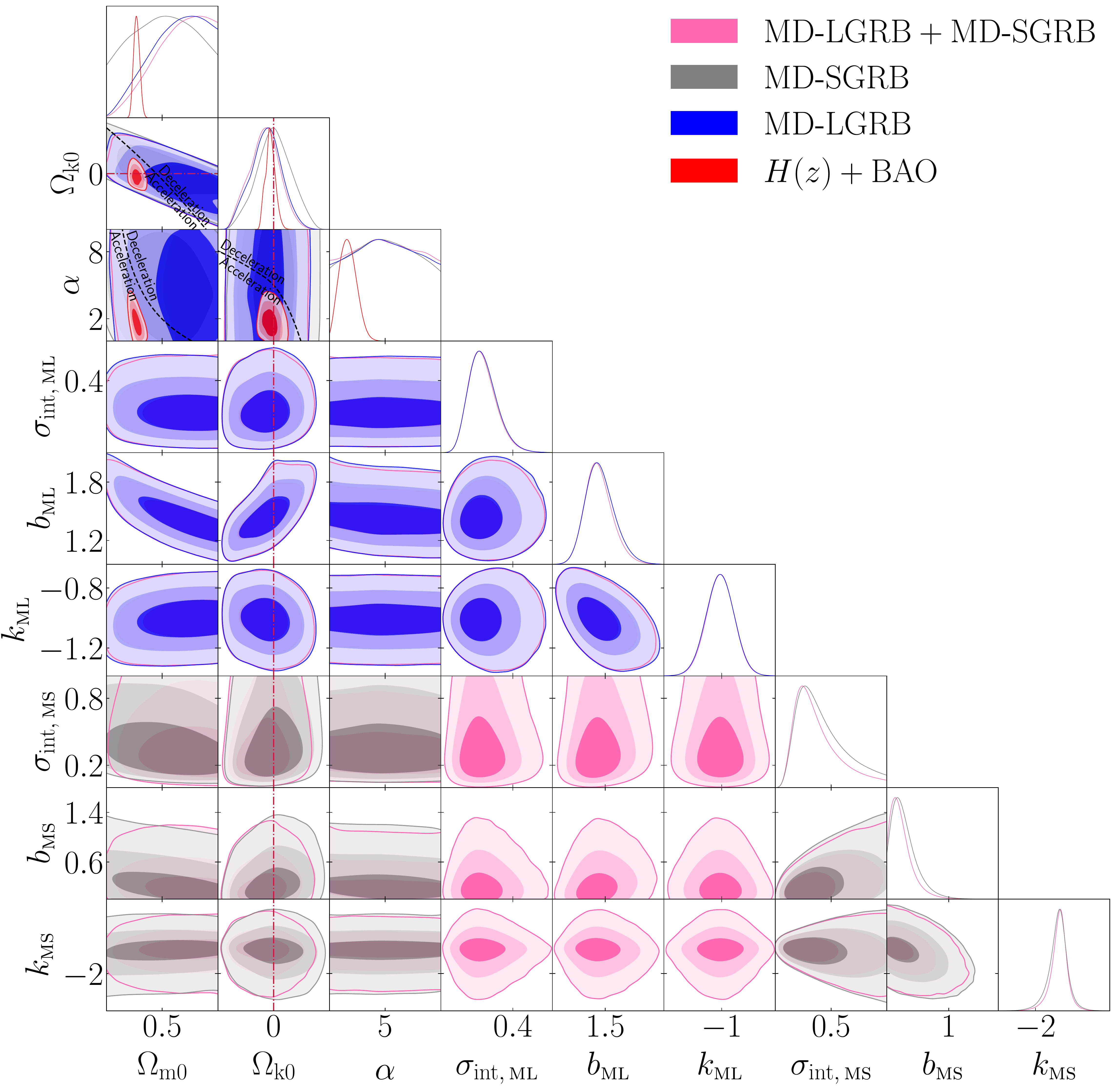}}\\
\caption{One-dimensional likelihoods and 1$\sigma$, 2$\sigma$, and 3$\sigma$ two-dimensional likelihood confidence contours from MD-LGRB (blue), MD-SGRB (gray), MD-LGRB + MD-SGRB (pink), and $H(z)$ + BAO (red) data for all six models. The zero-acceleration lines are shown as black dashed lines, which divide the parameter space into regions associated with currently-accelerating and currently-decelerating cosmological expansion. In the non-flat XCDM and non-flat \pcdm\ cases, the zero-acceleration lines are computed for the third cosmological parameter set to the $H(z)$ + BAO data best-fitting values listed in Table \ref{tab:BFPC7}. The crimson dash-dot lines represent flat hypersurfaces, with closed spatial hypersurfaces either below or to the left. The magenta lines represent $w_{\rm X}=-1$, i.e.\ flat or non-flat \lcdm\ models. The $\alpha = 0$ axes correspond to flat and non-flat \lcdm\ models in panels (e) and (f), respectively.}
\label{fig1C7}
\end{figure*}

\begin{figure*}
\centering
 \subfloat[Flat \lcdm]{%
    \includegraphics[width=3.25in,height=1.84in]{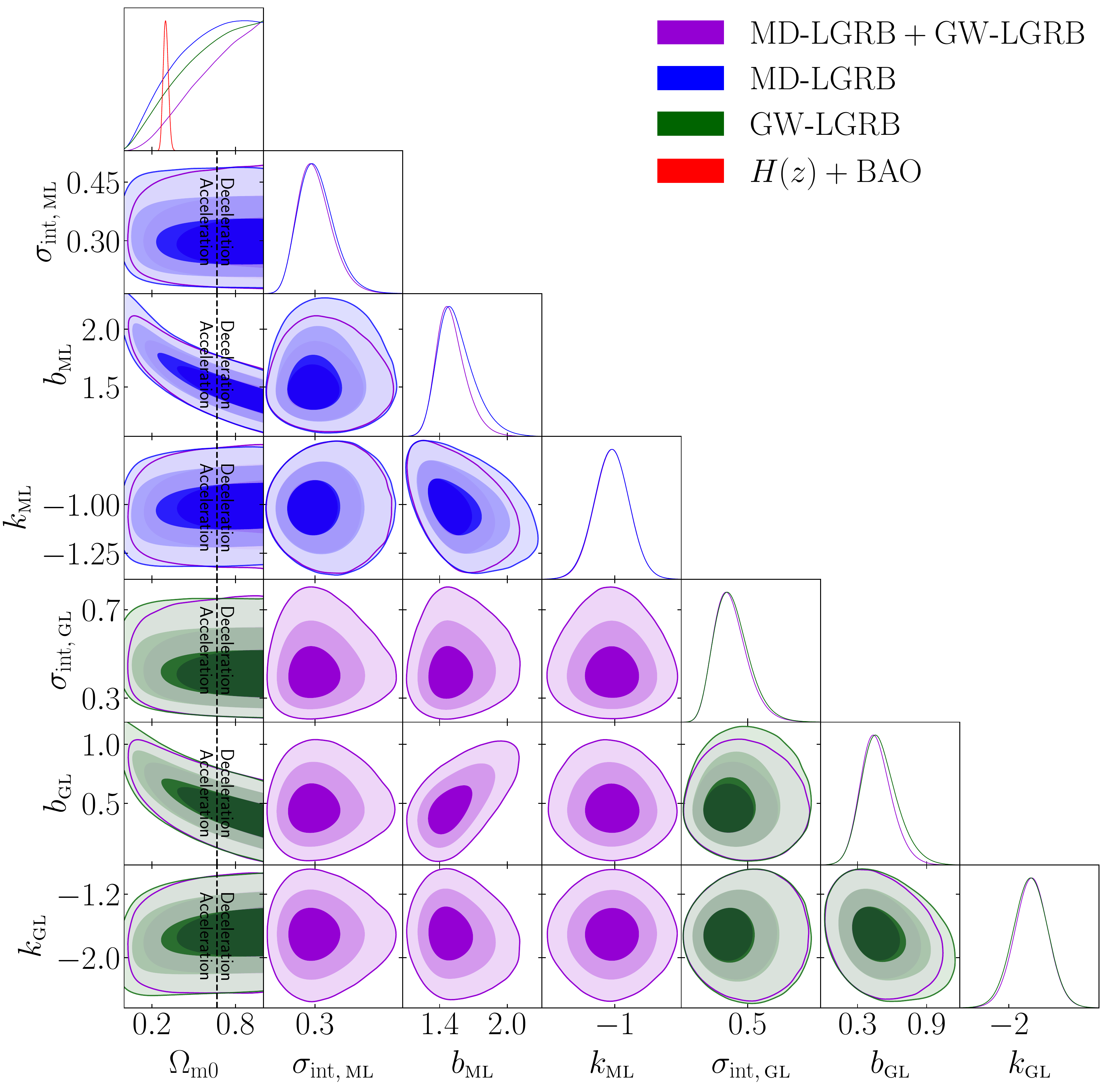}}
 \subfloat[Non-flat \lcdm]{%
    \includegraphics[width=3.25in,height=1.84in]{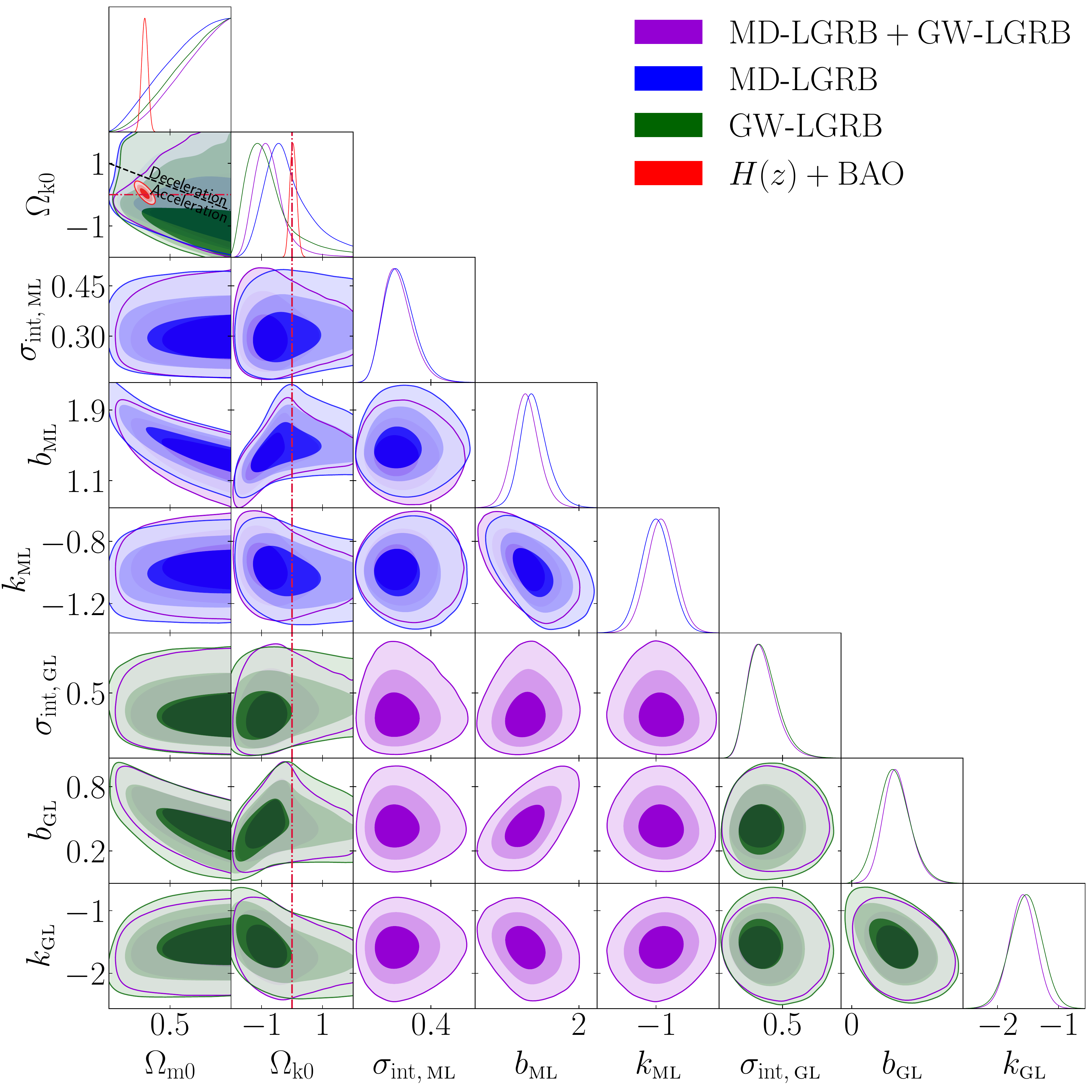}}\\
 \subfloat[Flat XCDM]{%
    \includegraphics[width=3.25in,height=1.84in]{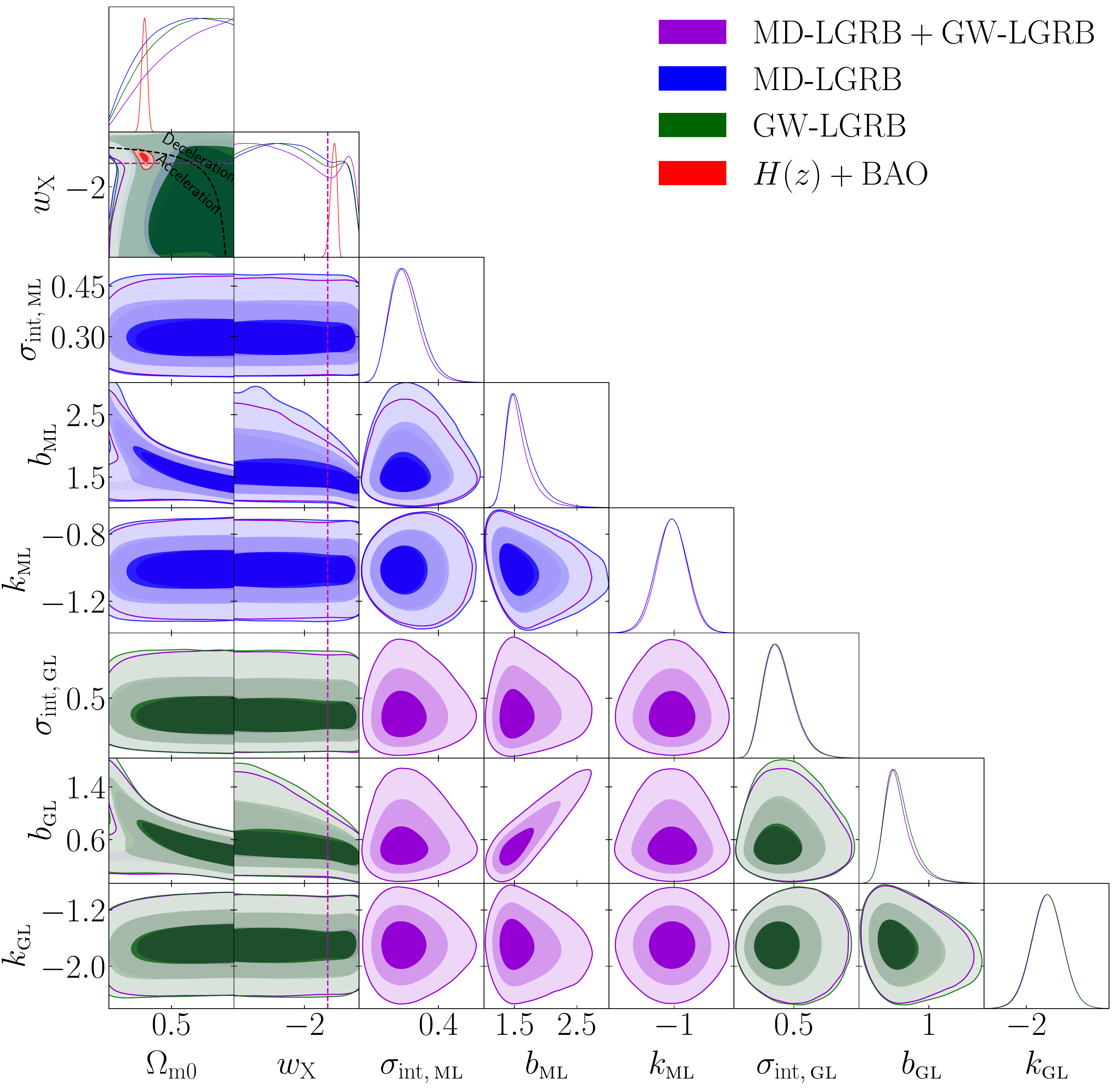}}
 \subfloat[Non-flat XCDM]{%
    \includegraphics[width=3.25in,height=1.84in]{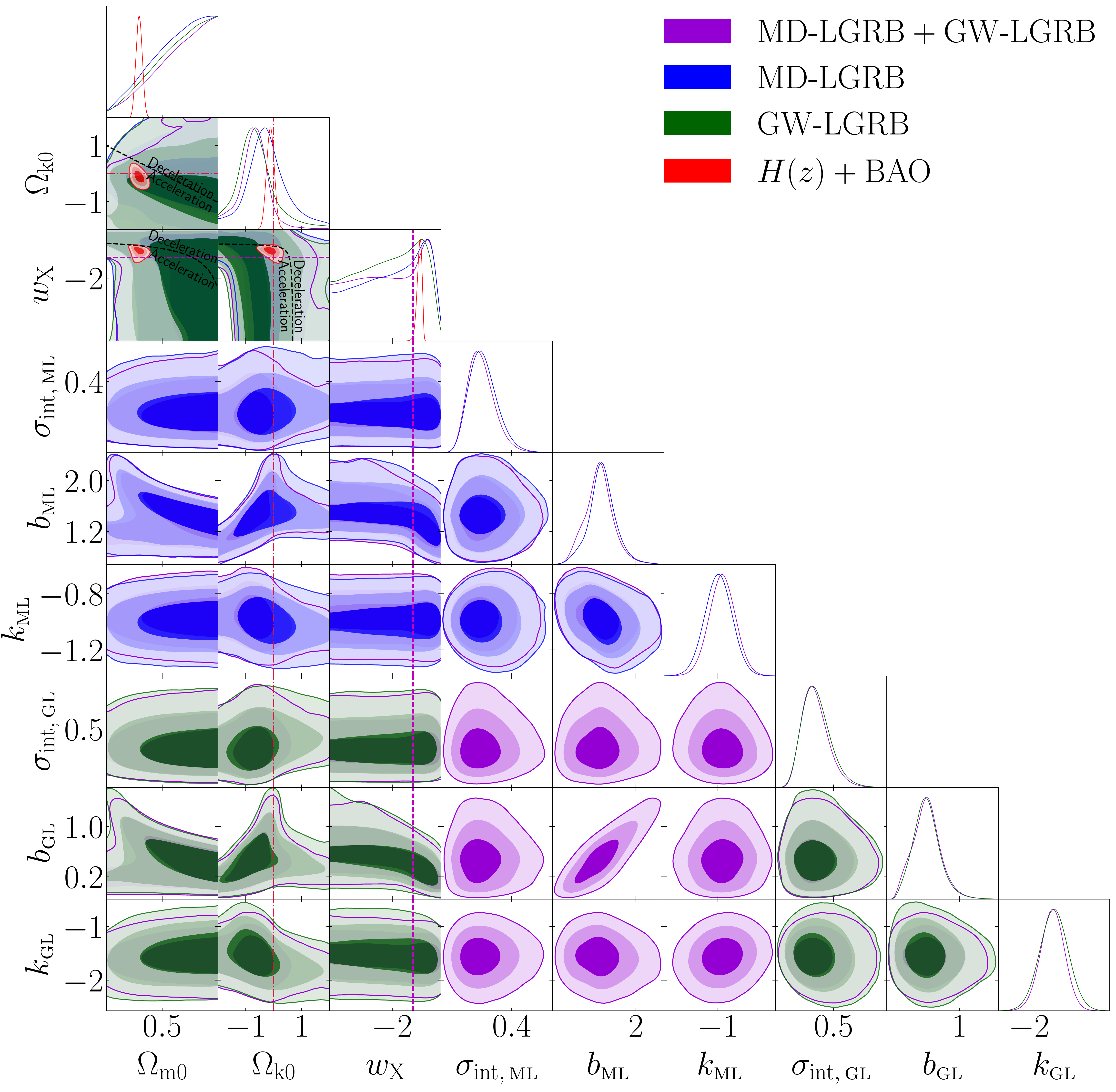}}\\
 \subfloat[Flat \pcdm]{%
    \includegraphics[width=3.25in,height=1.84in]{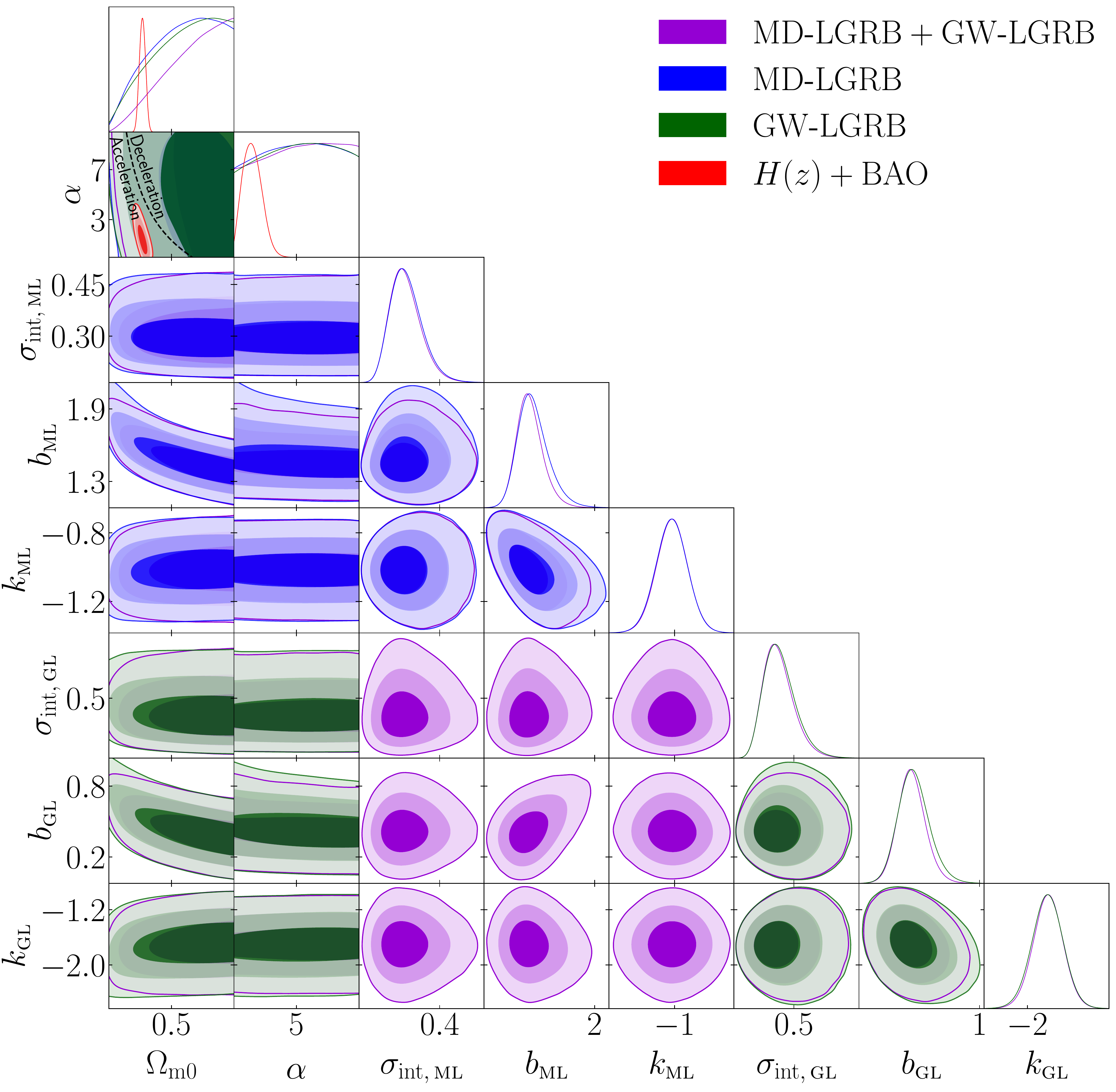}}
  \subfloat[Non-flat \pcdm]{%
     \includegraphics[width=3.25in,height=1.84in]{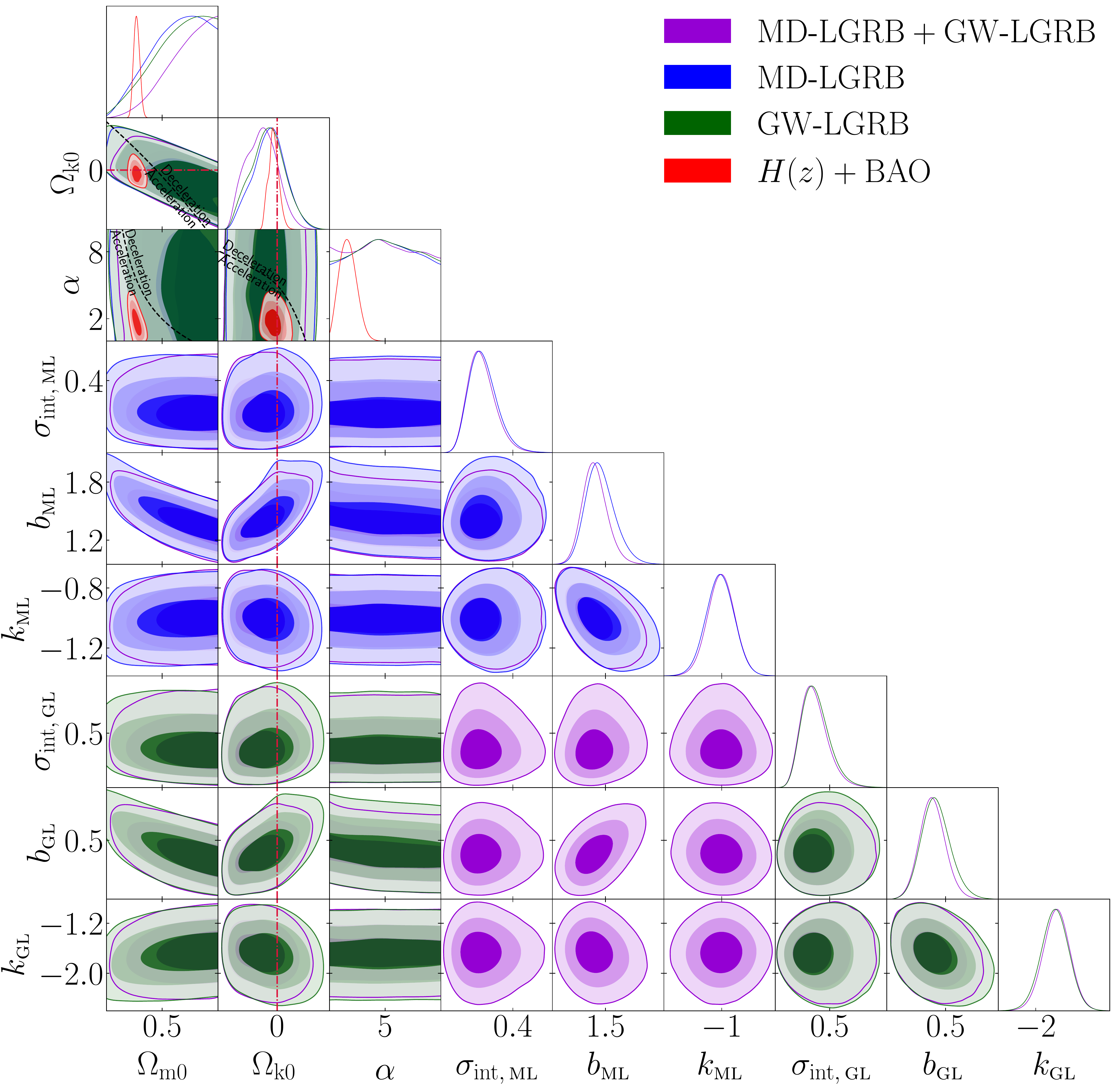}}\\
\caption{One-dimensional likelihoods and 1$\sigma$, 2$\sigma$, and 3$\sigma$ two-dimensional likelihood confidence contours from MD-LGRB (blue), GW-LGRB (green), MD-LGRB + GW-LGRB (violet), and $H(z)$ + BAO (red) data for all six models. The zero-acceleration lines are shown as black dashed lines, which divide the parameter space into regions associated with currently-accelerating and currently-decelerating cosmological expansion. In the non-flat XCDM and non-flat \pcdm\ cases, the zero-acceleration lines are computed for the third cosmological parameter set to the $H(z)$ + BAO data best-fitting values listed in Table \ref{tab:BFPC7}. The crimson dash-dot lines represent flat hypersurfaces, with closed spatial hypersurfaces either below or to the left. The magenta lines represent $w_{\rm X}=-1$, i.e.\ flat or non-flat \lcdm\ models. The $\alpha = 0$ axes correspond to flat and non-flat \lcdm\ models in panels (e) and (f), respectively.}
\label{fig2C7}
\end{figure*}

\begin{figure*}
\centering
 \subfloat[Flat \lcdm]{%
    \includegraphics[width=3.25in,height=1.84in]{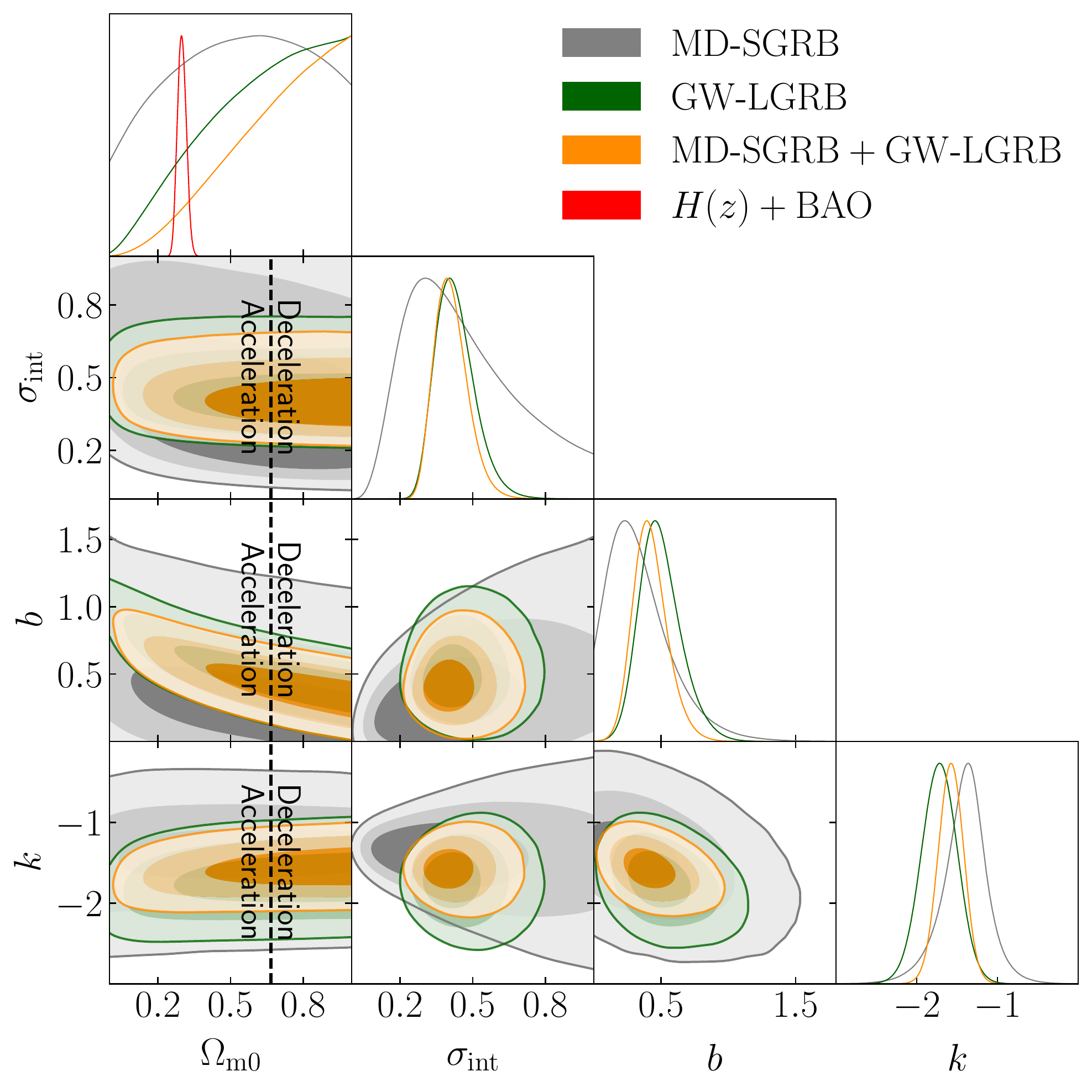}}
 \subfloat[Non-flat \lcdm]{%
    \includegraphics[width=3.25in,height=1.84in]{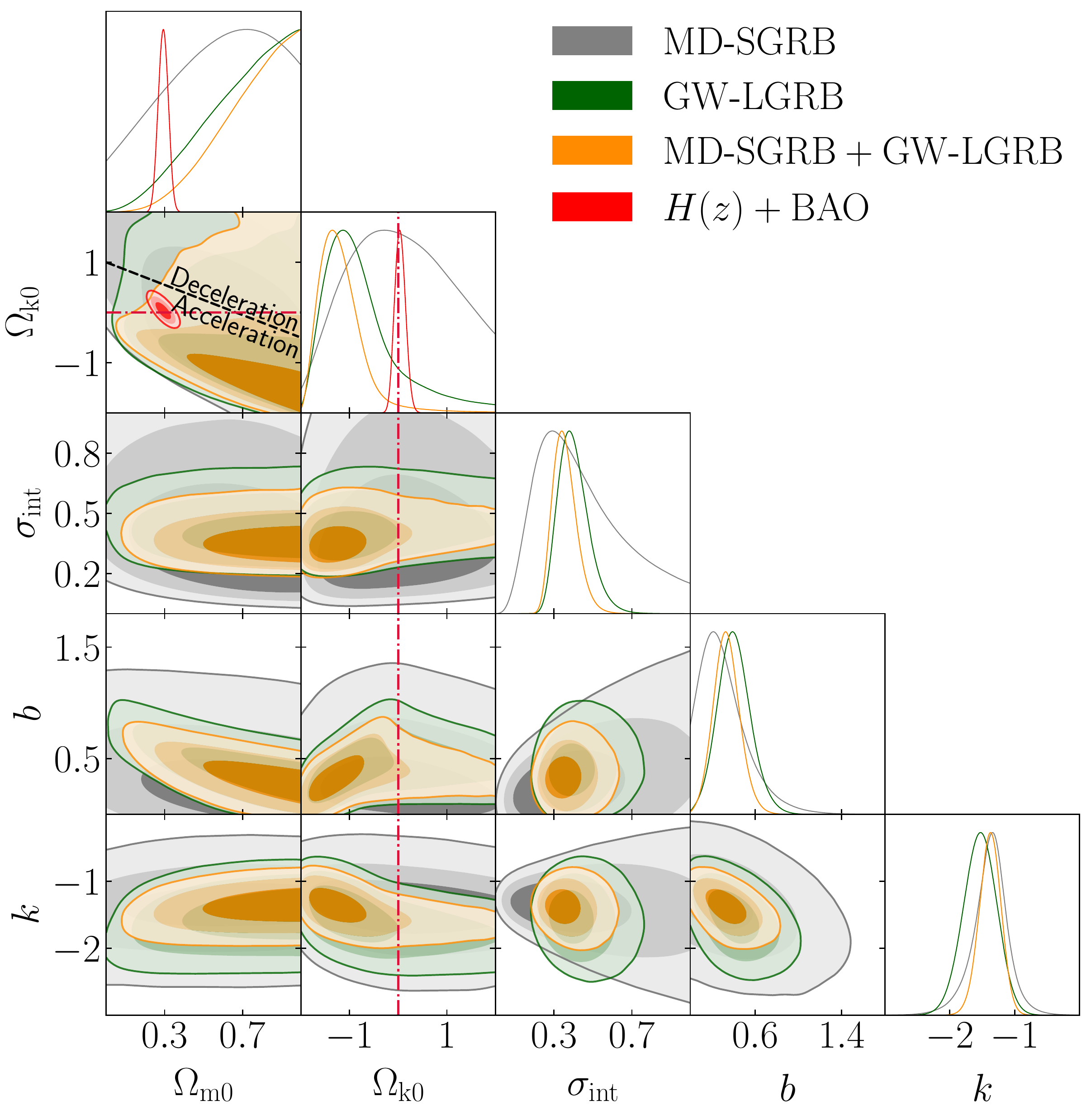}}\\
 \subfloat[Flat XCDM]{%
    \includegraphics[width=3.25in,height=1.84in]{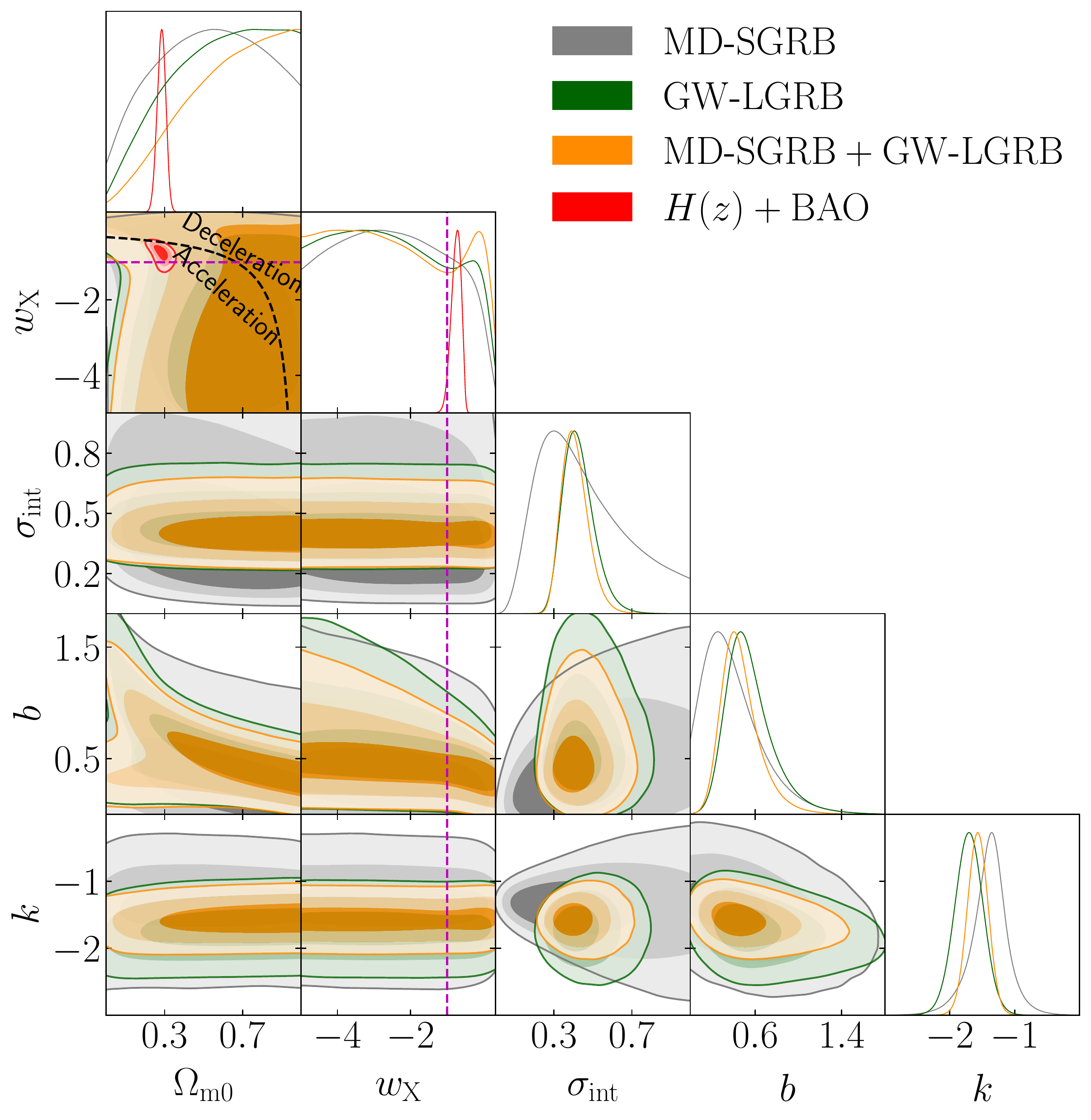}}
 \subfloat[Non-flat XCDM]{%
    \includegraphics[width=3.25in,height=1.84in]{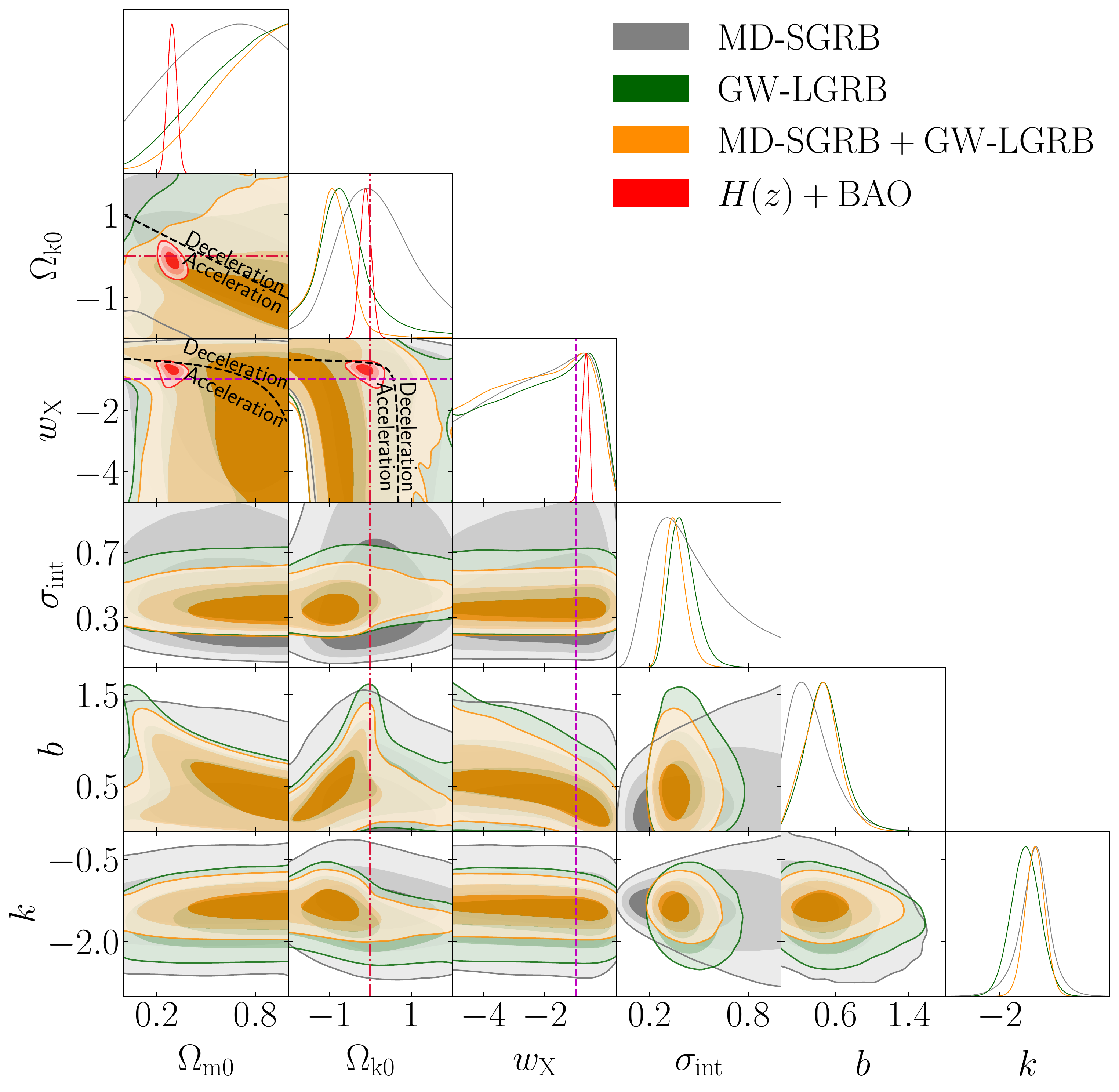}}\\
 \subfloat[Flat \pcdm]{%
    \includegraphics[width=3.25in,height=1.84in]{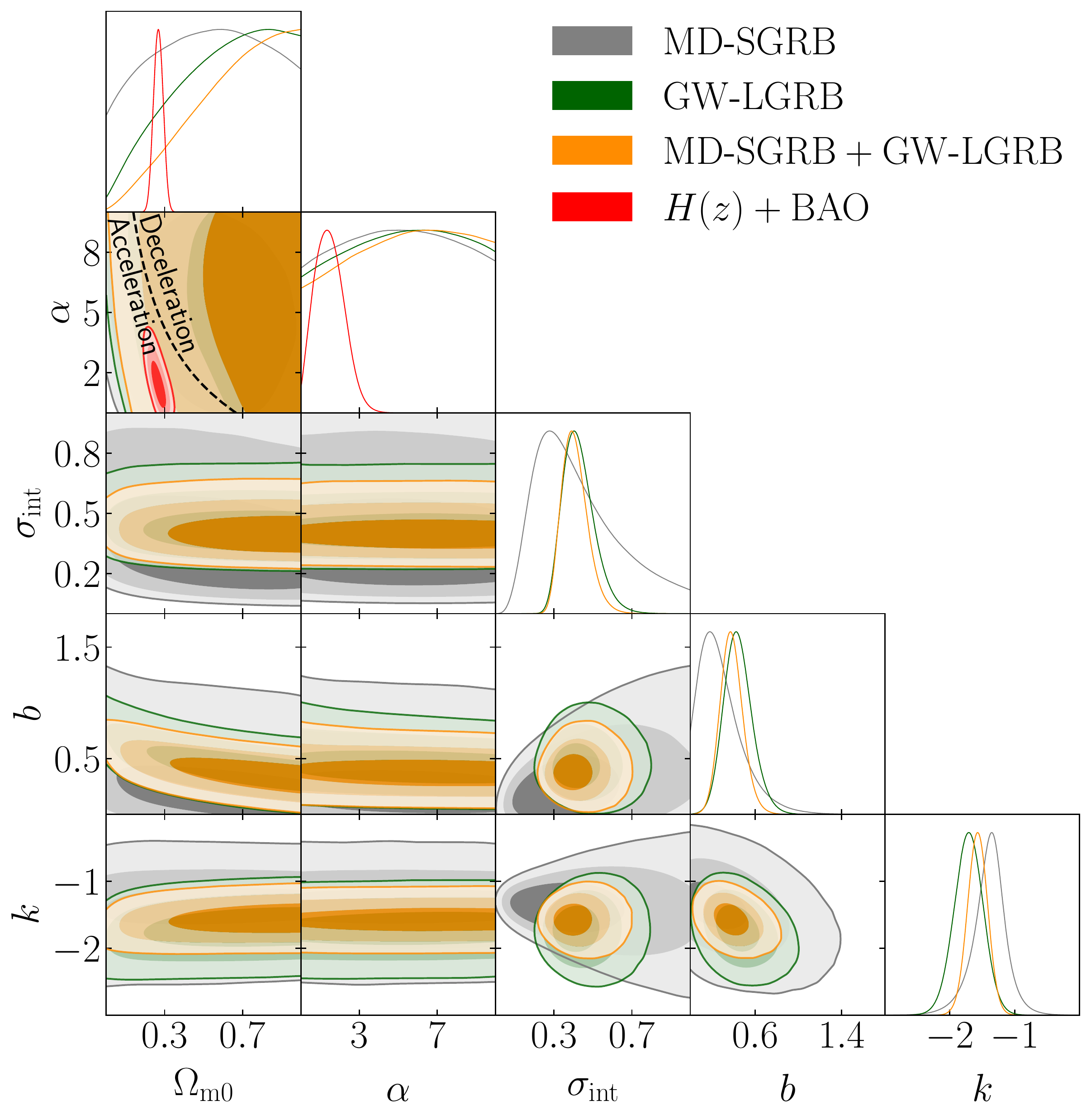}}
 \subfloat[Non-flat \pcdm]{%
    \includegraphics[width=3.25in,height=1.84in]{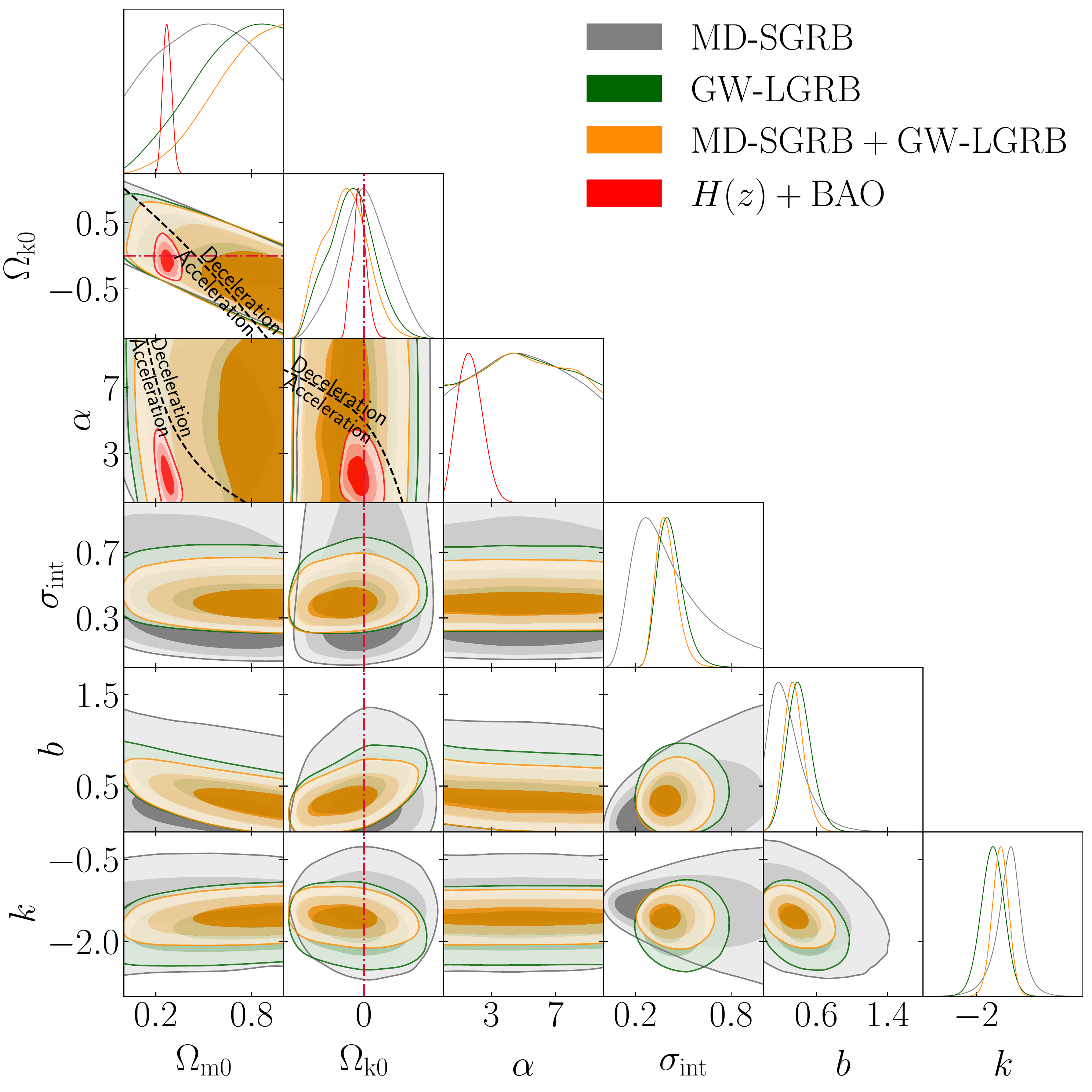}}\\
\caption{One-dimensional likelihoods and 1$\sigma$, 2$\sigma$, and 3$\sigma$ two-dimensional likelihood confidence contours from MD-SGRB (gray), GW-LGRB (green), MD-SGRB + GW-LGRB (orange), and $H(z)$ + BAO (red) data for all six models, without subscripts on $\sigma_{\rm int}$, $k$, and $b$. The zero-acceleration lines are shown as black dashed lines, which divide the parameter space into regions associated with currently-accelerating and currently-decelerating cosmological expansion. In the non-flat XCDM and non-flat \pcdm\ cases, the zero-acceleration lines are computed for the third cosmological parameter set to the $H(z)$ + BAO data best-fitting values listed in Table \ref{tab:BFPC7}. The crimson dash-dot lines represent flat hypersurfaces, with closed spatial hypersurfaces either below or to the left. The magenta lines represent $w_{\rm X}=-1$, i.e.\ flat or non-flat \lcdm\ models. The $\alpha = 0$ axes correspond to flat and non-flat \lcdm\ models in panels (e) and (f), respectively.}
\label{fig3C7}
\end{figure*}

We now record and discuss results when these data sets are used to jointly constrain the Dainotti parameters and the cosmological parameters of the six spatially-flat and non-flat dark energy cosmological models. Figure \ref{fig00C7} shows the flat $\Lambda$CDM Dainotti correlations for the ML, MS, GL, and MS + GL data sets. The unmarginalized best-fitting results and the one-dimensional (1D) posterior mean values and uncertainties are reported in Tables \ref{tab:BFPC7} and \ref{tab:1d_BFPC7}, respectively. The corresponding posterior 1D probability distributions and two-dimensional (2D) confidence regions of these parameters are shown in Figs.\ \ref{fig1C7}--\ref{fig3C7}, in blue (ML), gray (MS), green (GL), pink (ML + MS), violet (ML + GL), orange (MS + GL), and red ($H(z)$ + BAO, as a baseline). Note that $H_0=70$ \hunit\ and $\Omega_{b}=0.05$ are applied in the GRB cases.

\begin{sidewaystable*}
\centering
\resizebox*{\columnwidth}{0.65\columnwidth}{%
\begin{threeparttable}
\caption{Unmarginalized best-fitting parameter values for all models from various combinations of data.}\label{tab:BFPC7}
\begin{tabular}{lccccccccccccccccccccccccc}
\toprule
Model & Data set & $\Omega_{b}h^2$ & $\Omega_{c}h^2$ & $\Omega_{\mathrm{m0}}$ & $\Omega_{\mathrm{k0}}$ & $w_{\mathrm{X}}$ & $\alpha$ & $H_0$\tnote{a} & $\sigma_{\mathrm{int,\,\textsc{ml}}}$ & $b_{\mathrm{\textsc{ml}}}$ & $k_{\mathrm{\textsc{ml}}}$ & $\sigma_{\mathrm{int,\,\textsc{ms}}}$ & $b_{\mathrm{\textsc{ms}}}$ & $k_{\mathrm{\textsc{ms}}}$ & $\sigma_{\mathrm{int,\,\textsc{gl}}}$ & $b_{\mathrm{\textsc{gl}}}$ & $k_{\mathrm{\textsc{gl}}}$ & $\sigma_{\mathrm{int,\,\textsc{ms+gl}}}$ & $b_{\mathrm{\textsc{ms+gl}}}$ & $k_{\mathrm{\textsc{ms+gl}}}$ & $-2\ln\mathcal{L}_{\mathrm{max}}$ & $AIC$ & $BIC$ & $\Delta AIC$ & $\Delta BIC$ \\
\midrule
 & $H(z)$ + BAO & 0.0239 & 0.1187 & 0.298 & -- & -- & -- & 69.13 & -- & -- & -- & -- & -- & -- & -- & -- & -- & -- & -- & -- & 23.66 & 29.66 & 34.87 & 0.00 & 0.00\\
 & ML & 0.0245 & 0.4645 & 0.998 & -- & -- & -- & 70 & 0.275 & 1.383 & $-1.010$ & -- & -- & -- & -- & -- & -- & -- & -- & -- & 8.68 & 16.68 & 22.41 & 0.00 & 0.00\\
 & MS & 0.0245 & 0.4651 & 0.999 & -- & -- & -- & 70 & -- & -- & -- & 0.159 & 0.105 & $-1.310$ & -- & -- & -- & -- & -- & -- & $-3.49$ & 4.51 & 2.95 & 0.00 & 0.00\\
Flat \lcdm & GL & 0.0245 & 0.4641 & 0.997 & -- & -- & -- & 70 & -- & -- & -- & -- & -- & -- & 0.370 & 0.359 & $-1.675$ & -- & -- & -- & 22.94 & 30.94 & 35.65 & 0.00 & 0.00\\
 & MS + GL & 0.0245 & 0.4652 & 0.999 & -- & -- & -- & 70 & -- & -- & -- & -- & -- & -- & -- & -- & -- & 0.361 & 0.313 & $-1.545$ & 25.59 & 33.59 & 38.30 & 0.00 & 0.00\\
 & ML + GL & 0.0245 & 0.4642 & 0.997 & -- & -- & -- & 70 & 0.274 & 1.378 & $-1.008$ & -- & -- & -- & 0.362 & 0.358 & $-1.708$ & -- & -- & -- & 31.66 & 45.66 & 59.71 & 0.00 & 0.00\\
 & ML + MS & 0.0245 & 0.463 & 0.995 & -- & -- & -- & 70 & 0.274 & 1.388 & $-1.013$ & 0.148 & 0.080 & $-1.322$ & -- & -- & -- & -- & -- & -- & 5.34 & 19.34 & 30.42 & 0.00 & 0.00\\
\\
 & $H(z)$ + BAO & 0.0247 & 0.1140 & 0.294 & 0.029 & -- & -- & 68.68 & -- & -- & -- & -- & -- & -- & -- & -- & -- & -- & -- & -- & 23.60 & 31.60 & 38.55 & 1.94 & 3.68\\
 & ML & 0.0245 & 0.4410 & 0.950 & $-0.973$ & -- & -- & 70 & 0.268 & 1.316 & $-0.967$ & -- & -- & -- & -- & -- & -- & -- & -- & -- & 7.48 & 17.48 & 24.65 & 0.80 & 2.24\\
 & MS & 0.0245 & 0.2727 & 0.607 & $-1.730$ & -- & -- & 70 & -- & -- & -- & 0.007 & 0.023 & $-0.919$ & -- & -- & -- & -- & -- & -- & $-15.81$ & $-5.81$ & $-7.76$ & $-10.32$ & $-10.71$\\
Non-flat \lcdm & GL & 0.0245 & 0.4640 & 0.997 & $-1.703$ & -- & -- & 70 & -- & -- & -- & -- & -- & -- & 0.329 & 0.238 & $-1.377$ & -- & -- & -- & 17.00 & 27.00 & 32.89 & $-3.94$ & $-2.76$\\
 & MS + GL & 0.0245 & 0.4649 & 0.999 & $-1.738$ & -- & -- & 70 & -- & -- & -- & -- & -- & -- & -- & -- & -- & 0.304 & 0.220 & $-1.303$ & 16.02 & 26.02 & 31.91 & $-7.57$ & $-6.39$\\
 & ML + GL & 0.0245 & 0.4330 & 0.934 & $-1.288$ & -- & -- & 70 & 0.269 & 1.248 & $-0.953$ & -- & -- & -- & 0.336 & 0.330 & $-1.529$ & -- & -- & -- & 26.79 & 42.79 & 58.84 & $-2.87$ & $-0.87$\\
 & ML + MS & 0.0245 & 0.4625 & 0.994 & $-1.130$ & -- & -- & 70 & 0.259 & 1.278 & $-0.966$ & 0.136 & 0.159 & $-1.249$ & -- & -- & -- & -- & -- & -- & 2.91 & 18.91 & 31.58 & $-0.43$ & 1.14\\
\\
 & $H(z)$ + BAO & 0.0304 & 0.0891 & 0.281 & -- & $-0.701$ & -- & 65.18 & -- & -- & -- & -- & -- & -- & -- & -- & -- & -- & -- & -- & 19.65 & 27.65 & 34.60 & $-2.01$ & $-0.27$\\
 & ML & 0.0245 & 0.0327 & 0.117 & -- & 0.133 & -- & 70 & 0.275 & 1.288 & $-0.997$ & -- & -- & -- & -- & -- & -- & -- & -- & -- & 8.14 & 18.14 & 25.31 & 1.46 & 2.90\\
 & MS & 0.0245 & 0.0939 & 0.242 & -- & 0.141 & -- & 70 & -- & -- & -- & 0.160 & 0.054 & $-1.285$ & -- & -- & -- & -- & -- & -- & $-4.23$ & 5.77 & 3.81 & 1.26 & 0.86\\
Flat XCDM & GL & 0.0245 & 0.0035 & 0.057 & -- & 0.139 & -- & 70 & -- & -- & -- & -- & -- & -- & 0.364 & 0.259 & $-1.651$ & -- & -- & -- & 21.97 & 31.97 & 37.86 & 1.03 & 2.21\\ 
 & MS + GL & 0.0245 & 0.0058 & 0.062 & -- & 0.141 & -- & 70 & -- & -- & -- & -- & -- & -- & -- & -- & -- & 0.346 & 0.248 & $-1.518$ & 23.92 & 33.92 & 39.81 & 0.33 & 1.51\\
 & ML + GL & 0.0245 & 0.0735 & 0.200 & -- & 0.143 & -- & 70 & 0.273 & 1.300 & $-1.015$ & -- & -- & -- & 0.359 & 0.273 & $-1.636$ & -- & -- & -- & 30.20 & 46.20 & 62.26 & 0.54 & 6.55\\
 & ML + MS & 0.0245 & $-0.0207$ & 0.008 & -- & 0.137 & -- & 70 & 0.278 & 1.269 & $-0.984$ & 0.175 & 0.031 & $-1.282$ & -- & -- & -- & -- & -- & -- & 3.99 & 19.99 & 32.65 & 0.65 & 2.23\\
\\
 & $H(z)$ + BAO & 0.0290 & 0.0980 & 0.295 & $-0.152$ & $-0.655$ & -- & 65.59 & -- & -- & -- & -- & -- & -- & -- & -- & -- & -- & -- & -- & 18.31 & 28.31 & 37.00 & $-1.35$ & 2.13\\
 & ML & 0.0245 & 0.1525 & 0.361 & $-1.893$ & 0.036 & -- & 70 & 0.269 & 0.949 & $-0.976$ & -- & -- & -- & -- & -- & -- & -- & -- & -- & 7.39 & 19.39 & 27.99 & 2.71 & 5.58\\
 & MS & 0.0245 & 0.3596 & 0.784 & $-1.915$ & $-1.108$ & -- & 70 & -- & -- & -- & 0.057 & 0.098 & $-0.995$ & -- & -- & -- & -- & -- & -- & $-13.61$ & $-1.61$ & $-3.95$ & $-6.12$ & $-6.90$\\
Non-flat XCDM & GL & 0.0245 & 0.0378 & 0.127 & $-0.174$ & $-4.518$ & -- & 70 & -- & -- & -- & -- & -- & -- & 0.327 & 1.237 & $-1.299$ & -- & -- & -- & 16.61 & 28.61 & 35.68 & $-2.33$ & 0.03\\
 & MS + GL & 0.0245 & 0.4212 & 0.910 & $-1.218$ & $-2.308$ & -- & 70 & -- & -- & -- & -- & -- & -- & -- & -- & -- & 0.298 & 0.403 & $-1.248$ & 15.65 & 27.65 & 34.72 & $-5.94$ & $-3.58$\\
 & ML + GL & 0.0245 & 0.3594 & 0.783 & $-0.789$ & $-4.432$ & -- & 70 & 0.269 & 1.449 & $-0.927$ & -- & -- & -- & 0.338 & 0.576 & $-1.429$ & -- & -- & -- & 26.04 & 44.04 & 62.11 & $-1.62$ & 2.40\\
 & ML + MS & 0.0245 & 0.1499 & 0.306 & $-1.993$ & $0.130$ & -- & 70 & 0.281 & 0.881 & $-0.978$ & 0.095 & $-0.164$ & $-1.206$ & -- & -- & -- & -- & -- & -- & $-0.04$ & 17.96 & 32.21 & $-1.38$ & 1.79\\
\\
 & $H(z)$ + BAO & 0.0333 & 0.0788 & 0.264 & -- & -- & 1.504 & 65.20 & -- & -- & -- & -- & -- & -- & -- & -- & -- & -- & -- & -- & 19.49 & 27.49 & 34.44 & $-2.17$ & $-0.43$\\
 & ML & 0.0245 & 0.4651 & 0.999 & -- & -- & 5.225 & 70 & 0.275 & 1.383 & $-1.011$ & -- & -- & -- & -- & -- & -- & -- & -- & -- & 8.68 & 18.68 & 25.85 & 2.00 & 3.44\\
 & MS & 0.0245 & 0.4649 & 0.999 & -- & -- & 8.046 & 70 & -- & -- & -- & 0.160 & 0.099 & $-1.306$ & -- & -- & -- & -- & -- & -- & $-3.49$ & 6.51 & 4.56 & 2.00 & 1.61\\
Flat $\phi$CDM & GL & 0.0245 & 0.4641 & 0.997 & -- & -- & 4.299 & 70 & -- & -- & -- & -- & -- & -- & 0.372 & 0.360 & $-1.674$ & -- & -- & -- & 22.94 & 32.94 & 38.83 & 2.00 & 3.18\\
 & MS + GL & 0.0245 & 0.4653 & 1.000 & -- & -- & 6.323 & 70 & -- & -- & -- & -- & -- & -- & -- & -- & -- & 0.359 & 0.314 & $-1.545$ & 25.59 & 35.59 & 41.48 & 2.00 & 3.18\\
 & ML + GL & 0.0245 & 0.4648 & 0.999 & -- & -- & 7.886 & 70 & 0.274 & 1.375 & $-1.013$ & -- & -- & -- & 0.371 & 0.333 & $-1.647$ & -- & -- & -- & 31.72 & 47.72 & 63.78 & 2.06 & 4.07\\
 & ML + MS & 0.0245 & 0.4611 & 0.991 & -- & -- & 6.029 & 70 & 0.268 & 1.371 & $-0.997$ & 0.148 & 0.113 & $-1.317$ & -- & -- & -- & -- & -- & -- & 5.32 & 21.32 & 33.99 & 1.98 & 3.57\\
\\
 & $H(z)$ + BAO & 0.0334 & 0.0816 & 0.266 & $-0.147$ & -- & 1.915 & 65.70 & -- & -- & -- & -- & -- & -- & -- & -- & -- & -- & -- & -- & 18.15 & 28.15 & 36.84 & $-1.51$ & 1.97\\
 & ML & 0.0245 & 0.4558 & 0.980 & $-0.980$ & -- & 0.423 & 70 & 0.266 & 1.296 & $-0.973$ & -- & -- & -- & -- & -- & -- & -- & -- & -- & 7.48 & 19.48 & 28.09 & 2.80 & 5.68\\
 & MS & 0.0245 & 0.4482 & 0.965 & $-0.964$ & -- & 8.262 & 70 & -- & -- & -- & 0.132 & 0.003 & $-1.261$ & -- & -- & -- & -- & -- & -- & $-5.25$ & 6.75 & 4.40 & 2.24 & 1.45\\
Non-flat $\phi$CDM & GL & 0.0245 & 0.4644 & 0.998 & $-0.993$ & -- & 0.173 & 70 & -- & -- & -- & -- & -- & -- & 0.340 & 0.337 & $-1.547$ & -- & -- & -- & 20.15 & 32.15 & 39.22 & 1.21 & 3.57\\
 & MS + GL & 0.0245 & 0.4601 & 0.989 & $-0.982$ & -- & 0.011 & 70 & -- & -- & -- & -- & -- & -- & -- & -- & -- & 0.315 & 0.310 & $-1.484$ & 21.46 & 33.46 & 40.53 & $-0.13$ & 2.23\\
 & ML + GL & 0.0245 & 0.4465 & 0.961 & $-0.950$ & -- & 0.232 & 70 & 0.255 & 1.290 & $-0.969$ & -- & -- & -- & 0.348 & 0.356 & $-1.600$ & -- & -- & -- & 28.04 & 46.04 & 64.11 & 0.38 & 4.40\\
 & ML + MS & 0.0245 & 0.4628 & 0.995 & $-0.936$ & -- & 8.517 & 70 & 0.281 & 1.202 & $-0.994$ & 0.152 & 0.027 & $-1.291$ & -- & -- & -- & -- & -- & -- & 2.71 & 20.71 & 34.96 & 1.36 & 4.53\\
\bottomrule
\end{tabular}
\begin{tablenotes}[flushleft]
\item [a] \hunit. In the GRB only cases, $H_0$ is set to be 70 \hunit.
\end{tablenotes}
\end{threeparttable}%
}
\end{sidewaystable*}

\begin{sidewaystable*}
\centering
\resizebox*{\columnwidth}{0.65\columnwidth}{%
\begin{threeparttable}
\caption{One-dimensional marginalized posterior mean values and uncertainties ($\pm 1\sigma$ error bars or $2\sigma$ limits) of the parameters for all models from various combinations of data.}\label{tab:1d_BFPC7}
\begin{tabular}{lcccccccccccccccccccc}
\toprule
Model & Data set & $\Omega_{b}h^2$ & $\Omega_{c}h^2$ & $\Omega_{\mathrm{m0}}$ & $\Omega_{\mathrm{k0}}$ & $w_{\mathrm{X}}$ & $\alpha$ & $H_0$\tnote{a} & $\sigma_{\mathrm{int,\,\textsc{ml}}}$ & $b_{\mathrm{\textsc{ml}}}$ & $k_{\mathrm{\textsc{ml}}}$ & $\sigma_{\mathrm{int,\,\textsc{ms}}}$ & $b_{\mathrm{\textsc{ms}}}$ & $k_{\mathrm{\textsc{ms}}}$ & $\sigma_{\mathrm{int,\,\textsc{gl}}}$ & $b_{\mathrm{\textsc{gl}}}$ & $k_{\mathrm{\textsc{gl}}}$ & $\sigma_{\mathrm{int,\,\textsc{ms+gl}}}$ & $b_{\mathrm{\textsc{ms+gl}}}$ & $k_{\mathrm{\textsc{ms+gl}}}$ \\
\midrule
 & $H(z)$ + BAO & $0.0241\pm0.0029$ & $0.1193^{+0.0082}_{-0.0090}$ & $0.299^{+0.017}_{-0.019}$ & -- & -- & -- & $69.30\pm1.84$ & -- & -- & -- & -- & -- & -- & -- & -- & -- & -- & -- & -- \\
 & ML & -- & -- & $>0.188$ & -- & -- & -- & -- & $0.305^{+0.035}_{-0.053}$ & $1.552^{+0.108}_{-0.189}$ & $-1.017\pm0.090$ & -- & -- & -- & -- & -- & -- & -- & -- & -- \\
 & MS & -- & -- & $0.520^{+0.379}_{-0.253}$ & -- & -- & -- & -- & -- & -- & -- & $0.695^{+0.044}_{-0.550}$ & $0.437^{+0.073}_{-0.400}$ & $-1.450^{+0.362}_{-0.258}$ & -- & -- & -- & -- & -- & -- \\
Flat \lcdm & GL & -- & -- & $>0.202$ & -- & -- & -- & -- & -- & -- & -- & -- & -- & -- & $0.429^{+0.059}_{-0.094}$ & $0.495^{+0.120}_{-0.173}$ & $-1.720\pm0.219$ & -- & -- & -- \\
 & MS + GL & -- & -- & $>0.293$ & -- & -- & -- & -- & -- & -- & -- & -- & -- & -- & -- & -- & -- & $0.412^{+0.052}_{-0.079}$ & $0.421^{+0.101}_{-0.141}$ & $-1.577\pm0.155$ \\
 & ML + GL & -- & -- & $>0.294$ & -- & -- & -- & -- & $0.301^{+0.033}_{-0.051}$ & $1.507^{+0.095}_{-0.151}$ & $-1.015\pm0.089$ & -- & -- & -- & $0.424^{+0.056}_{-0.090}$ & $0.465^{+0.113}_{-0.149}$ & $-1.708\pm0.210$ & -- & -- & -- \\
 & ML + MS & -- & -- & $>0.206$ & -- & -- & -- & -- & $0.302^{+0.034}_{-0.052}$ & $1.542^{+0.104}_{-0.184}$ & $-1.016\pm0.091$ & $0.613^{+0.010}_{-0.479}$ & $0.379^{+0.049}_{-0.350}$ & $-1.426^{+0.312}_{-0.210}$ & -- & -- & -- & -- & -- & -- \\
\\
 & $H(z)$ + BAO & $0.0253^{+0.0041}_{-0.0050}$ & $0.1135^{+0.0196}_{-0.0197}$ & $0.293\pm0.025$ & $0.039^{+0.102}_{-0.115}$ & -- & -- & $68.75^{+2.37}_{-2.36}$ & -- & -- & -- & -- & -- & -- & -- & -- & -- & -- & -- & -- \\
 & ML & -- & -- & $>0.241$ &  $-0.131^{+0.450}_{-0.919}$ & -- & -- & -- & $0.304^{+0.035}_{-0.053}$ & $1.478^{+0.123}_{-0.166}$ & $-1.000\pm0.096$ & -- & -- & -- & -- & -- & -- & -- & -- & -- \\
 & MS & -- & -- & $0.564^{+0.426}_{-0.149}$ &  $0.066^{+1.002}_{-1.199}$ & -- & -- & -- & -- & -- & -- & $0.670^{+0.024}_{-0.533}$ & $0.413^{+0.047}_{-0.393}$ & $-1.430^{+0.360}_{-0.239}$ & -- & -- & -- & -- & -- & -- \\
Non-flat \lcdm & GL & -- & -- & $>0.290$ &  $-0.762^{+0.271}_{-0.888}$ & -- & -- & -- & -- & -- & -- & -- & -- & -- & $0.402^{+0.057}_{-0.090}$ & $0.407^{+0.136}_{-0.160}$ & $-1.536\pm0.252$ & -- & -- & -- \\
 & MS + GL & -- & -- & $>0.391$ &  $-1.165^{+0.225}_{-0.519}$ & -- & -- & -- & -- & -- & -- & -- & -- & -- & -- & -- & -- & $0.357^{+0.046}_{-0.070}$ & $0.337^{+0.110}_{-0.127}$ & $-1.382^{+0.164}_{-0.163}$ \\
 & ML + GL & -- & -- & $>0.338$ &  $-0.737^{+0.299}_{-0.547}$ & -- & -- & -- & $0.300^{+0.033}_{-0.051}$ & $1.386^{+0.138}_{-0.154}$ & $-0.966\pm0.093$ & -- & -- & -- & $0.397^{+0.053}_{-0.086}$ & $0.437^{+0.110}_{-0.142}$ & $-1.588^{+0.215}_{-0.214}$ & -- & -- & -- \\
 & ML + MS & -- & -- & $>0.270$ &  $-0.300^{+0.380}_{-0.836}$ & -- & -- & -- & $0.302^{+0.034}_{-0.052}$ & $1.457^{+0.128}_{-0.163}$ & $-0.993\pm0.095$ & $0.518^{+0.009}_{-0.393}$ & $0.339^{+0.056}_{-0.301}$ & $-1.385^{+0.271}_{-0.186}$ & -- & -- & -- & -- & -- & -- \\
\\
 & $H(z)$ + BAO & $0.0296^{+0.0046}_{-0.0052}$ & $0.0939^{+0.0194}_{-0.0171}$ & $0.284^{+0.023}_{-0.021}$ & -- & $-0.754^{+0.155}_{-0.107}$ & -- & $65.89^{+2.41}_{-2.71}$ & -- & -- & -- & -- & -- & -- & -- & -- & -- & -- & -- & -- \\
 & ML & -- & -- & $>0.123$ & -- & $-2.456^{+2.567}_{-2.180}$ & -- & -- & $0.306^{+0.036}_{-0.054}$ & $1.611^{+0.113}_{-0.277}$ & $-1.014\pm0.092$ & -- & -- & -- & -- & -- & -- & -- & -- & -- \\
 & MS & -- & -- & $0.520^{+0.340}_{-0.276}$ & -- & $-2.494^{+1.264}_{-2.050}$ & -- & -- & -- & -- & -- & $0.704^{+0.035}_{-0.564}$ & $0.497^{+0.086}_{-0.458}$ & $-1.441^{+0.366}_{-0.259}$ & -- & -- & -- & -- & -- & -- \\
Flat XCDM & GL & -- & -- & $>0.141$ & -- & $<-0.046$ & -- & -- & -- & -- & -- & -- & -- & -- & $0.428^{+0.058}_{-0.092}$ & $0.556^{+0.127}_{-0.256}$ & $-1.706\pm0.215$ & -- & -- & -- \\
 & MS + GL & -- & -- & $>0.192$ & -- & $<0.028$ & -- & -- & -- & -- & -- & -- & -- & -- & -- & -- & -- & $0.409^{+0.052}_{-0.078}$ & $0.470^{+0.108}_{-0.206}$ & $-1.570\pm0.155$ \\
 & ML + GL & -- & -- & $>0.164$ & -- & $<0.022$ & -- & -- & $0.300^{+0.033}_{-0.051}$ & $1.566^{+0.099}_{-0.239}$ & $-1.012\pm0.087$ & -- & -- & -- & $0.424^{+0.057}_{-0.092}$ & $0.531^{+0.116}_{-0.240}$ & $-1.700\pm0.213$ & -- & -- & -- \\
 & ML + MS & -- & -- & $>0.142$ & -- & $<-0.072$ & -- & -- & $0.303^{+0.034}_{-0.052}$ & $1.597^{+0.108}_{-0.255}$ & $-1.013\pm0.089$ & $0.562^{+0.010}_{-0.431}$ & $0.409^{+0.066}_{-0.377}$ & $-1.408^{+0.288}_{-0.196}$ & -- & -- & -- & -- & -- & -- \\
\\
 & $H(z)$ + BAO & $0.0290^{+0.0052}_{-0.0055}$ & $0.0990^{+0.0214}_{-0.0215}$ & $0.293\pm0.028$ & $-0.116\pm0.134$ & $-0.700^{+0.138}_{-0.083}$ & -- & $65.96^{+2.32}_{-2.55}$ & -- & -- & -- & -- & -- & -- & -- & -- & -- & -- & -- & -- \\
 & ML & -- & -- & $>0.174$ & $-0.262^{+0.580}_{-0.724}$ & $-2.000^{+2.117}_{-1.264}$ & -- & -- & $0.305^{+0.036}_{-0.054}$ & $1.462^{+0.194}_{-0.196}$ & $-0.996\pm0.097$ & -- & -- & -- & -- & -- & -- & -- & -- & -- \\
 & MS & -- & -- & $0.552^{+0.442}_{-0.152}$ & $0.134^{+0.793}_{-0.987}$ & $-2.234^{+2.159}_{-0.969}$ & -- & -- & -- & -- & -- & $0.733^{+0.031}_{-0.596}$ & $0.464^{+0.052}_{-0.444}$ & $-1.433^{+0.394}_{-0.270}$ & -- & -- & -- & -- & -- & -- \\
Non-flat XCDM & GL & -- & -- & $>0.194$ & $-0.615^{+0.470}_{-0.685}$ & $-2.212^{+2.186}_{-0.962}$ & -- & -- & -- & -- & -- & -- & -- & -- & $0.403^{+0.058}_{-0.092}$ & $0.480^{+0.177}_{-0.223}$ & $-1.532^{+0.259}_{-0.260}$ & -- & -- & -- \\
 & MS + GL & -- & -- & $>0.268$ & $-0.920^{+0.460}_{-0.386}$ & $-2.323^{+2.085}_{-1.095}$ & -- & -- & -- & -- & -- & -- & -- & -- & -- & -- & -- & $0.358^{+0.047}_{-0.072}$ & $0.452^{+0.185}_{-0.203}$ & $-1.363\pm0.175$ \\
 & ML + GL & -- & -- & $>0.196$ & $-0.696^{+0.484}_{-0.408}$ & $-2.158^{+2.254}_{-1.424}$ & -- & -- & $0.298^{+0.033}_{-0.051}$ & $1.422^{+0.201}_{-0.221}$ & $-0.968\pm0.092$ & -- & -- & -- & $0.397^{+0.054}_{-0.087}$ & $0.480^{+0.205}_{-0.224}$ & $-1.561^{+0.220}_{-0.219}$ & -- & -- & -- \\
 & ML + MS & -- & -- & $>0.198$ & $-0.375^{+0.458}_{-0.611}$ & $-2.235^{+2.289}_{-2.007}$ & -- & -- & $0.301^{+0.034}_{-0.052}$ & $1.470^{+0.184}_{-0.183}$ & $-0.988^{+0.095}_{-0.094}$ & $0.525^{+0.007}_{-0.409}$ & $0.397^{+0.079}_{-0.364}$ & $-1.367^{+0.275}_{-0.193}$ & -- & -- & -- & -- & -- & -- \\
\\
 & $H(z)$ + BAO & $0.0321^{+0.0056}_{-0.0039}$ & $0.0823^{+0.0186}_{-0.0183}$ & $0.268\pm0.024$ & -- & -- & $1.467^{+0.637}_{-0.866}$ & $65.24^{+2.15}_{-2.35}$ & -- & -- & -- & -- & -- & -- & -- & -- & -- & -- & -- & -- \\
 & ML & -- & -- & $>0.148$ & -- & -- & -- & -- & $0.304^{+0.035}_{-0.053}$ & $1.493^{+0.093}_{-0.143}$ & $-1.017\pm0.089$ & -- & -- & -- & -- & -- & -- & -- & -- & -- \\
 & MS & -- & -- & $0.514^{+0.365}_{-0.275}$ & -- & -- & -- & -- & -- & -- & -- & $0.597^{+0.029}_{-0.457}$ & $0.355^{+0.052}_{-0.331}$ & $-1.425^{+0.311}_{-0.221}$ & -- & -- & -- & -- & -- & -- \\
Flat $\phi$CDM & GL & -- & -- & $>0.148$ & -- & -- & -- & -- & -- & -- & -- & -- & -- & -- & $0.428^{+0.059}_{-0.094}$ & $0.444^{+0.112}_{-0.141}$ & $-1.710\pm0.218$ & -- & -- & -- \\
 & MS + GL & -- & -- & $>0.222$ & -- & -- & -- & -- & -- & -- & -- & -- & -- & -- & -- & -- & -- & $0.408^{+0.050}_{-0.076}$ & $0.384^{+0.092}_{-0.111}$ & $-1.573\pm0.151$ \\
 & ML + GL & -- & -- & $>0.235$ & -- & -- & -- & -- & $0.301^{+0.033}_{-0.051}$ & $1.463^{+0.086}_{-0.119}$ & $-1.015\pm0.088$ & -- & -- & -- & $0.423^{+0.056}_{-0.091}$ & $0.423^{+0.104}_{-0.121}$ & $-1.701\pm0.209$ & -- & -- & -- \\
 & ML + MS & -- & -- & $>0.166$ & -- & -- & -- & -- & $0.303^{+0.034}_{-0.053}$ & $1.485^{+0.092}_{-0.137}$ & $-1.017\pm0.090$ & $0.517^{+0.001}_{-0.385}$ & $0.304^{+0.033}_{-0.283}$ & $-1.410^{+0.268}_{-0.173}$ & -- & -- & -- & -- & -- & -- \\
\\
 & $H(z)$ + BAO & $0.0319^{+0.0062}_{-0.0037}$ & $0.0848^{+0.0181}_{-0.0220}$ & $0.271^{+0.025}_{-0.028}$ & $-0.074^{+0.104}_{-0.110}$ & -- & $1.653^{+0.685}_{-0.856}$ & $65.46^{+2.31}_{-2.29}$ & -- & -- & -- & -- & -- & -- & -- & -- & -- & -- & -- & -- \\
 & ML & -- & -- & $>0.207$ & $-0.163^{+0.355}_{-0.317}$ & -- & -- & -- & $0.303^{+0.035}_{-0.053}$ & $1.448^{+0.120}_{-0.165}$ & $-1.011\pm0.091$ & -- & -- & -- & -- & -- & -- & -- & -- & -- \\
 & MS & -- & -- & $0.505^{+0.310}_{-0.313}$ & $0.017^{+0.387}_{-0.375}$ & -- & -- & -- & -- & -- & -- & $0.589^{+0.023}_{-0.455}$ & $0.352^{+0.048}_{-0.331}$ & $-1.432^{+0.314}_{-0.207}$ & -- & -- & -- & -- & -- & -- \\
Non-flat $\phi$CDM & GL & -- & -- & $>0.207$ & $-0.193^{+0.364}_{-0.347}$ & -- & -- & -- & -- & -- & -- & -- & -- & -- & $0.422^{+0.057}_{-0.092}$ & $0.408^{+0.121}_{-0.149}$ & $-1.693^{+0.215}_{-0.214}$ & -- & -- & -- \\
 & MS + GL & -- & -- & $>0.313$ & $-0.293^{+0.330}_{-0.359}$ & -- & -- & -- & -- & -- & -- & -- & -- & -- & -- & -- & -- & $0.397^{+0.050}_{-0.076}$ & $0.343^{+0.101}_{-0.120}$ & $-1.546\pm0.150$ \\
 & ML + GL & -- & -- & $>0.340$ & $-0.327^{+0.321}_{-0.345}$ & -- & -- & -- & $0.299^{+0.033}_{-0.051}$ & $1.391^{+0.107}_{-0.136}$ & $-1.004\pm0.088$ & -- & -- & -- & $0.414^{+0.055}_{-0.088}$ & $0.374^{+0.113}_{-0.128}$ & $-1.668^{+0.207}_{-0.208}$ & -- & -- & -- \\
 & ML + MS & -- & -- & $>0.238$ & $-0.198^{+0.351}_{-0.321}$ & -- & -- & -- & $0.301^{+0.034}_{-0.052}$ & $1.435^{+0.114}_{-0.156}$ & $-1.010\pm0.089$ & $0.507^{+0.003}_{-0.379}$ & $0.292^{+0.025}_{-0.281}$ & $-1.393^{+0.258}_{-0.173}$ & -- & -- & -- & -- & -- & -- \\
\bottomrule
\end{tabular}
\begin{tablenotes}[flushleft]
\item [a] \hunit. In the GRB only cases, $H_0$ is set to be 70 \hunit.
\end{tablenotes}
\end{threeparttable}%
}
\end{sidewaystable*}

ML, MS, and GL GRB data have almost cosmological-model independent Dainotti parameters. This means that it is not unreasonable to treat the ML, MS, and GL GRBs as standardizable candles, as was assumed in \cite{Wangetal_2021} and \cite{Huetal2021}. 

In the ML case (with subscript ``ML'' in the first line of Tables \ref{tab:BFPC7}, \ref{tab:1d_BFPC7}, \ref{tab:BFP2C7}, and \ref{tab:1d_BFP2C7}), the slope $k$ ranges from a high of $-0.996\pm0.097$ (non-flat XCDM) to a low of $-1.017\pm0.090$ (flat \lcdm), the intercept $b$ ranges from a high of $1.611^{+0.113}_{-0.277}$ (flat XCDM) to a low of $1.448^{+0.120}_{-0.165}$ (non-flat \pcdm), and the intrinsic scatter $\sigma_{\rm int}$ ranges from a high of $0.306^{+0.036}_{-0.054}$ (flat XCDM) to a low of $0.303^{+0.035}_{-0.053}$ (non-flat \pcdm), with central values of each pair being $0.16\sigma$, $0.54\sigma$, and $0.05\sigma$ away from each other, respectively. 

In the MS case (with subscript ``MS'' in the first line of Tables \ref{tab:BFPC7} and \ref{tab:1d_BFPC7}) with prior range of $b\in[0,10]$, the slope $k$ ranges from a high of $-1.425^{+0.311}_{-0.221}$ (flat \pcdm) to a low of $-1.450^{+0.362}_{-0.258}$ (flat \lcdm), the intercept $b$ ranges from a high of $0.497^{+0.086}_{-0.458}$ (flat XCDM) to a low of $0.352^{+0.048}_{-0.331}$ (non-flat \pcdm), and the intrinsic scatter $\sigma_{\rm int}$ ranges from a high of $0.733^{+0.031}_{-0.596}$ (non-flat XCDM) to a low of $0.589^{+0.023}_{-0.455}$ (non-flat \pcdm), with central values of each pair being $0.06\sigma$, $0.54\sigma$, and $0.05\sigma$ away from each other, respectively.\footnote{Note, however, that the lower error bars of $b$ and $\sigma_{\rm int}$ are considerably larger than the upper ones due to cut-off prior ranges of the former and skewed distributions of the latter. Therefore here we also consider the MS case with wider prior range of $b\in[-10,10]$, which are not listed in the tables due to their insignificant differences. Because the lowest and highest values of $k$, $b$, and $\sigma_{\rm int}$ from these two MS cases differ from each other at only $0.22\sigma$, $0.56\sigma$, and $0.37\sigma$, respectively, and the constraints of the cosmological parameters are also within $1\sigma$ range, the prior range of $b\in[0,10]$ is an acceptable choice.}

In the GL case (with subscript ``GL'' in the first line of Tables \ref{tab:BFPC7}, \ref{tab:1d_BFPC7}, \ref{tab:BFP2C7}, and \ref{tab:1d_BFP2C7}), the slope $k$ ranges from a high of $-1.532^{+0.259}_{-0.260}$ (non-flat XCDM) to a low of $-1.720\pm0.219$ (flat \lcdm), the intercept $b$ ranges from a high of $0.556^{+0.127}_{-0.256}$ (flat XCDM) to a low of $0.407^{+0.136}_{-0.160}$ (non-flat \lcdm), and the intrinsic scatter $\sigma_{\rm int}$ ranges from a high of $0.429^{+0.059}_{-0.094}$ (flat \lcdm) to a low of $0.402^{+0.057}_{-0.090}$ (non-flat \pcdm), with central values of each pair being $0.55\sigma$, $0.51\sigma$, and $0.25\sigma$ away from each other, respectively.

Figure \ref{fig00C7}, panel (c), shows that the GL and MS GRBs obey the same Dainotti correlation in the flat \lcdm\ model, within the uncertainties.\footnote{It is unclear if this is more than just a coincidence, as the plateau phases in the two cases are dominated by GW emission (GL) and MD radiation (MS), respectively.} Table \ref{tab:comp} shows that the differences between the GL and MS Dainotti parameters in all six cosmological models are within $1\sigma$. GL and MS GRBs however follow a different Dainotti correlation than the ML GRBs. Given the similarity of the GL and MS Dainotti correlation parameters, it is not unreasonable to use just three (not six) correlation parameters in joint analyses of MS and GL data (with subscript ``MS+GL'' in the first line of Tables \ref{tab:BFPC7} and \ref{tab:1d_BFPC7}). In this case, the slope $k$ ranges from a high of $-1.363\pm0.175$ (non-flat XCDM) to a low of $-1.577\pm0.155$ (flat \lcdm), the intercept $b$ ranges from a high of $0.470^{+0.108}_{-0.206}$ (flat XCDM) to a low of $0.337^{+0.110}_{-0.127}$ (non-flat \lcdm), and the intrinsic scatter $\sigma_{\rm int}$ ranges from a high of $0.412^{+0.052}_{-0.079}$ (flat \lcdm) to a low of $0.357^{+0.046}_{-0.070}$ (non-flat \lcdm), with central values of each pair being $0.92\sigma$, $0.57\sigma$, and $0.60\sigma$ away from each other, respectively. In contrast to the GL case, the MS + GL case tightens the constraints a little bit, with smaller error bars, and prefers lower values of $b$ and $\sigma_{\rm int}$, and higher values of $k$. When we jointly analyze ML + GL and ML + MS, the constraints on the Dainotti parameters follow the same pattern as that of MS + GL against GL.

The constraints on the cosmological parameters are very loose for all of these cases. In the flat \lcdm\ model, the highest $2\sigma$ lower limit of \om\ among these cases is $\Om>0.294$ of the ML + GL data. In the non-flat \lcdm\ model, the highest $2\sigma$ lower limit of \om\ is $\Om>0.391$ of the MS + GL case, which is inconsistent with that of the $H(z)$ + BAO case. The MS data favor open hypersufaces while all other cases favor closed hypersufaces, with the favored spatial geometries for GL, ML + GL, and MS + GL data being more than $1\sigma$ (or even $2\sigma$) away from flat geometry. In the flat XCDM parametrization, the highest $2\sigma$ lower limit of \om\ is $\Om>0.192$ for the MS + GL data, and the constraints on the X-fluid equation of state parameter \wx\ are very loose, with the highest $1\sigma$ upper limit being $0.111$ for the ML case. In the non-flat XCDM parametrization, the highest $2\sigma$ lower limit of \om\ is $\Om>0.268$ for the MS + GL data, and the constraints on \wx\ are very loose, with the highest $1\sigma$ upper limit being $0.238$ for the MS + GL data. The favored spatial geometries for these cases follow the same pattern as that for non-flat \lcdm, but with larger upper limits of \ok\ except for the MS data. In the flat \pcdm\ model, the highest $2\sigma$ lower limit of \om\ is $\Om>0.235$ for the ML + GL data. In the non-flat \pcdm\ model, the highest $2\sigma$ lower limit of \om\ is $\Om>0.340$ for the ML + GL case, which is inconsistent with that of the $H(z)$ + BAO data. Except for the MS case, closed spatial hypersurfaces are favored, but only in the ML + GL case is flat geometry slightly more than $1\sigma$ away. There are no constraints on $\alpha$ from these GRB data.

In the \lcdm\ and XCDM cases, all GRB data combinations more favor currently accelerating cosmological expansion. They however more favor currently decelerating cosmological expansion in the \pcdm\ models, in the $\Om-\alpha$ and $\Om-\Ok$ parameter subspaces.

From the $AIC$ and $BIC$ values we compute $\Delta AIC$ and $\Delta BIC$ values with respect to the flat \lcdm\ model. These are listed in the last two columns of Table \ref{tab:BFPC7}. In the ML case, flat \lcdm\ is the most favored model but there is only weak or positive evidence against any other model. In the MS case non-flat \lcdm\ model is the most favored model and, except for non-flat XCDM (with positive evidence against it), the other models are very strongly disfavored. In the GL case, non-flat \lcdm\ is again the most favored model, while the evidence against the others are mostly positive, except for non-flat \pcdm\ (with strong $BIC$ evidence against it). In the MS + GL case, similar to the MS case, non-flat \lcdm\ is the most favored model and, except for non-flat XCDM (with weak $AIC$ and positive $BIC$ evidence against it), the others are strongly disfavored. In the ML + GL case, non-flat \lcdm\ is the most favored model but, except for flat XCDM (with strong $BIC$ evidence against it), the evidence against the other models is either weak or positive. In the ML + MS case, the best candidates are non-flat XCDM based on $AIC$ and flat \lcdm\ based on $BIC$, while the evidence against the other models is either weak or positive.

\begin{table}
\begin{threeparttable}
\caption{MD-SGRB and GW-LGRB data $L_0-t_b$ correlation parameters (and $\sigma_{\rm int}$) differences.}
\label{tab:comp}
\setlength{\tabcolsep}{14.5pt}
\begin{tabular}{lccc}
\toprule
Model & $\Delta \sigma_{\rm int}$  & $\Delta k$  & $\Delta b$\\
\midrule
Flat \lcdm\  & $0.48\sigma$ & $0.31\sigma$ & $0.80\sigma$\\
Non-flat \lcdm\ & $0.50\sigma$ & $0.31\sigma$ & $0.01\sigma$ \\
Flat XCDM  & $0.49\sigma$ & $0.79\sigma$ & $0.22\sigma$\\
Non-flat XCDM  & $0.55\sigma$ & $0.26\sigma$ & $0.07\sigma$\\
Flat $\phi$CDM  & $0.37\sigma$ & $0.92\sigma$ & $0.59\sigma$\\
Non-flat $\phi$CDM & $0.36\sigma$ & $0.87\sigma$ & $0.36\sigma$\\
\bottomrule
\end{tabular}
\end{threeparttable}
\end{table}

\subsection{Constraints from A118, A115 (and jointly with ML), and A115$^{\prime}$ (and jointly with GL) data}
 \label{subsec:GRB-A}

\begin{sidewaystable*}
\centering
\resizebox*{\columnwidth}{0.65\columnwidth}{%
\begin{threeparttable}
\caption{Unmarginalized best-fitting parameter values for all models from various combinations of data.\tnote{a}}\label{tab:BFP2C7}
\begin{tabular}{lcccccccccccccccccccc}
\toprule
Model & Data set & $\Omega_{c}h^2$ & $\Omega_{\mathrm{m0}}$ & $\Omega_{\mathrm{k0}}$ & $w_{\mathrm{X}}$ & $\alpha$ & $\sigma_{\mathrm{int,\,\textsc{ml}}}$ & $b_{\mathrm{\textsc{ml}}}$ & $k_{\mathrm{\textsc{ml}}}$ & $\sigma_{\mathrm{int}}$ & $\gamma$ & $\beta$ & $\sigma_{\mathrm{int,\,\textsc{gl}}}$ & $b_{\mathrm{\textsc{gl}}}$ & $k_{\mathrm{\textsc{gl}}}$ & $-2\ln\mathcal{L}_{\mathrm{max}}$ & $AIC$ & $BIC$ & $\Delta AIC$ & $\Delta BIC$ \\
\midrule
 & A118 & 0.4089 & 0.884 & -- & -- & -- & -- & -- & -- & 0.401 & 50.02 & 1.099 & -- & -- & -- & 128.72 & 136.72 & 147.81 & 0.00 & 0.00\\
 & ML & 0.4645 & 0.998 & -- & -- & -- & 0.275 & 1.383 & $-1.010$ & -- & -- & -- & -- & -- & -- & 8.68 & 16.68 & 22.41 & 0.00 & 0.00\\
 & A115 & 0.4172 & 0.901 & -- & -- & -- & -- & -- & -- & 0.405 & 50.01 & 1.099 & -- & -- & -- & 127.97 & 135.97 & 146.95 & 0.00 & 0.00\\
Flat \lcdm & ML + A115 & 0.4540 & 0.977 & -- & -- & -- & 0.274 & 1.400 & $-1.019$ & 0.407 & 50.00 & 1.097 & -- & -- & -- & 136.70 & 150.70 & 171.59 & 0.00 & 0.00\\
 & GL & 0.4641 & 0.997 & -- & -- & -- & -- & -- & -- & -- & -- & -- & 0.370 & 0.359 & $-1.675$ & 22.94 & 30.94 & 35.65 & 0.00 & 0.00\\
 & A115$^{\prime}$ & 0.4629 & 0.995 & -- & -- & -- & -- & -- & -- & 0.403 & 50.01 & 1.091 & -- & -- & -- & 126.34 & 134.34 & 145.32 & 0.00 & 0.00\\
 & GL + A115$^{\prime}$ & 0.4652 & 0.999 & -- & -- & -- & -- & -- & -- & 0.402 & 49.96 & 1.110 & 0.370 & 0.363 & $-1.666$ & 149.34 & 163.34 & 183.88 & 0.00 & 0.00\\
\\
 & A118 & 0.4622 & 0.993 & 0.907 & -- & -- & -- & -- & -- & 0.400 & 49.92 & 1.115 & -- & -- & -- & 127.96 & 137.96 & 151.82 & 1.24 & 4.01\\
 & ML & 0.4410 & 0.950 & $-0.973$ & -- & -- & 0.268 & 1.316 & $-0.967$ & -- & -- & -- & -- & -- & -- & 7.48 & 17.48 & 24.65 & 0.80 & 2.24\\
 & A115 & 0.4631 & 0.995 & 1.014 & -- & -- & -- & -- & -- & 0.403 & 49.90 & 1.118 & -- & -- & -- & 127.18 & 137.18 & 150.90 & 1.21 & 3.95\\
Non-flat \lcdm & ML + A115 & 0.4637 & 0.996 & 0.062 & -- & -- & 0.283 & 1.383 & $-1.006$ & 0.410 & 50.01 & 1.088 & -- & -- & -- & 136.74 & 152.74 & 176.61 & 2.04 & 5.02\\
 & GL & 0.4640 & 0.997 & $-1.703$ & -- & -- & -- & -- & -- & -- & -- & -- & 0.329 & 0.238 & $-1.377$ & 17.00 & 27.00 & 32.89 & $-3.94$ & $-2.76$\\
 & A115$^{\prime}$ & 0.4647 & 0.998 & 0.729 & -- & -- & -- & -- & -- & 0.404 & 49.94 & 1.110 & -- & -- & -- & 125.93 & 135.93 & 149.65 & 1.59 & 4.33\\
 & GL + A115$^{\prime}$ & 0.4544 & 0.977 & $-0.244$ & -- & -- & -- & -- & -- & 0.402 & 50.00 & 1.100 & 0.362 & 0.364 & $-1.674$ & 149.29 & 165.29 & 188.77 & 1.95 & 4.89\\
\\
 & A118 & $-0.0115$ & 0.027 & -- & $-0.098$ & -- & -- & -- & -- & 0.399 & 50.04 & 1.102 & -- & -- & -- & 128.43 & 138.43 & 152.29 & 1.71 & 4.48\\
 & ML & 0.0327 & 0.117 & -- & 0.133 & -- & 0.275 & 1.288 & $-0.997$ & -- & -- & -- & -- & -- & -- & 8.14 & 18.14 & 25.31 & 1.46 & 2.90\\
 & A115 & $-0.0197$ & 0.010 & -- & $-0.102$ & -- & -- & -- & -- & 0.407 & 50.01 & 1.115 & -- & -- & -- & 127.66 & 137.66 & 151.38 & 1.69 & 4.43\\
Flat XCDM & ML + A115 & 0.4065 & 0.880 & -- & $-4.386$ & -- & 0.273 & 1.431 & $-1.008$ & 0.403 & 50.05 & 1.093 & -- & -- & -- & 136.70 & 152.70 & 176.57 & 2.00 & 4.98\\
 & GL & 0.0035 & 0.057 & -- & 0.139 & -- & -- & -- & -- & -- & -- & -- & 0.364 & 0.259 & $-1.651$ & 21.97 & 31.97 & 37.86 & 1.03 & 2.21\\ 
 & A115$^{\prime}$ & 0.0119 & 0.074 & -- & $-0.082$ & -- & -- & -- & -- & 0.402 & 50.03 & 1.100 & -- & -- & -- & 126.22 & 136.22 & 149.94 & 1.88 & 4.62\\
 & GL + A115$^{\prime}$ & 0.2817 & 0.625 & -- & 0.122 & -- & -- & -- & -- & 0.401 & 49.98 & 1.084 & 0.376 & 0.330 & $-1.672$ & 149.21 & 165.21 & 188.69 & 1.87 & 4.81\\
\\
 & A118 & 0.4603 & 0.989 & 0.841 & $-1.048$ & -- & -- & -- & -- & 0.399 & 49.93 & 1.112 & -- & -- & -- & 127.99 & 139.99 & 156.61 & 3.27 & 8.80\\
 & ML & 0.1525 & 0.361 & $-1.893$ & 0.036 & -- & 0.269 & 0.949 & $-0.976$ & -- & -- & -- & -- & -- & -- & 7.39 & 19.39 & 27.99 & 2.71 & 5.58\\
 & A115 & 0.4602 & 0.989 & 0.955 & $-1.097$ & -- & -- & -- & -- & 0.404 & 49.91 & 1.115 & -- & -- & -- & 127.18 & 139.18 & 155.65 & 3.21 & 8.70\\
Non-flat XCDM & ML + A115 & 0.3647 & 0.794 & 0.002 & $-4.000$ & -- & 0.269 & 1.478 & $-1.015$ & 0.404 & 50.05 & 1.107 & -- & -- & -- & 136.75 & 154.75 & 181.60 & 4.05 & 10.01\\
 & GL & 0.0378 & 0.127 & $-0.174$ & $-4.518$ & -- & -- & -- & -- & -- & -- & -- & 0.327 & 1.237 & $-1.299$ & 16.61 & 28.61 & 35.68 & $-2.33$ & 0.03\\
 & A115$^{\prime}$ & 0.4643 & 0.997 & 0.783 & $-0.956$ & -- & -- & -- & -- & 0.403 & 49.94 & 1.112 & -- & -- & -- & 125.92 & 137.92 & 154.39 & 3.58 & 9.07\\
 & GL + A115$^{\prime}$ & 0.4349 & 0.938 & $-0.255$ & $-0.043$ & -- & -- & -- & -- & 0.410 & 50.03 & 1.063 & 0.358 & 0.296 & $-1.616$ & 149.16 & 167.16 & 193.57 & 3.82 & 9.69\\
\\
 & A118 & 0.2301 & 0.520 & -- & -- & 9.936 & -- & -- & -- & 0.402 & 50.02 & 1.109 & -- & -- & -- & 128.56 & 138.56 & 152.41 & 1.84 & 4.60\\
 & ML & 0.4651 & 0.999 & -- & -- & 5.225 & 0.275 & 1.383 & $-1.011$ & -- & -- & -- & -- & -- & -- & 8.68 & 18.68 & 25.85 & 2.00 & 3.44\\
 & A115 & 0.2065 & 0.471 & -- & -- & 9.932 & -- & -- & -- & 0.405 & 50.04 & 1.106 & -- & -- & -- & 127.79 & 137.79 & 151.52 & 1.82 & 4.57\\
Flat $\phi$CDM & ML + A115 & 0.4563 & 0.981 & -- & -- & 2.762 & 0.273 & 1.391 & $-1.021$ & 0.405 & 49.99 & 1.099 & -- & -- & -- & 136.70 & 152.70 & 176.57 & 2.00 & 4.98\\
 & GL & 0.4641 & 0.997 & -- & -- & 4.299 & -- & -- & -- & -- & -- & -- & 0.372 & 0.360 & $-1.674$ & 22.94 & 32.94 & 38.83 & 2.00 & 3.18\\
 & A115$^{\prime}$ & 0.3334 & 0.730 & -- & -- & 9.652 & -- & -- & -- & 0.402 & 50.02 & 1.096 & -- & -- & -- & 126.29 & 136.29 & 150.01 & 1.95 & 4.69\\
 & GL + A115$^{\prime}$ & 0.4484 & 0.965 & -- & -- & 8.745 & -- & -- & -- & 0.403 & 50.05 & 1.077 & 0.372 & 0.348 & $-1.663$ & 149.36 & 165.36 & 188.83 & 2.02 & 4.95\\
\\
 & A118 & 0.3245 & 0.712 & 0.245 & -- & 8.862 & -- & -- & -- & 0.400 & 50.01 & 1.116 & -- & -- & -- & 128.42 & 140.42 & 157.04 & 3.70 & 9.23\\
 & ML & 0.4558 & 0.980 & $-0.980$ & -- & 0.423 & 0.266 & 1.296 & $-0.973$ & -- & -- & -- & -- & -- & -- & 7.48 & 19.48 & 28.09 & 2.80 & 5.68\\
 & A115 & 0.3217 & 0.706 & 0.290 & -- & 3.150 & -- & -- & -- & 0.406 & 50.05 & 1.108 & -- & -- & -- & 127.64 & 139.64 & 156.11 & 3.67 & 9.16\\
Non-flat $\phi$CDM & ML + A115 & 0.3044 & 0.671 & $-0.076$ & -- & 8.893 & 0.274 & 1.406 & $-1.001$ & 0.404 & 50.03 & 1.095 & -- & -- & -- & 136.77 & 154.77 & 181.62 & 4.07 & 10.03\\
 & GL & 0.4644 & 0.998 & $-0.993$ & -- & 0.173 & -- & -- & -- & -- & -- & -- & 0.340 & 0.337 & $-1.547$ & 20.15 & 32.15 & 39.22 & 1.21 & 3.57\\
 & A115$^{\prime}$ & 0.3787 & 0.823 & 0.161 & -- & 7.940 & -- & -- & -- & 0.404 & 50.02 & 1.105 & -- & -- & -- & 126.21 & 138.21 & 154.68 & 3.87 & 9.36\\
 & GL + A115$^{\prime}$ & 0.4420 & 0.952 & $-0.266$ & -- & 8.343 & -- & -- & -- & 0.400 & 50.00 & 1.077 & 0.372 & 0.310 & $-1.622$ & 149.18 & 167.18 & 193.59 & 3.84 & 9.71\\
\bottomrule
\end{tabular}
\begin{tablenotes}[flushleft]
\item [a] In these GRB cases, $\Omega_b$ and $H_0$ are set to be 0.05 and 70 \hunit, respectively.
\end{tablenotes}
\end{threeparttable}%
}
\end{sidewaystable*}

\begin{sidewaystable*}
\centering
\resizebox*{\columnwidth}{0.65\columnwidth}{%
\begin{threeparttable}
\caption{One-dimensional marginalized posterior mean values and uncertainties ($\pm 1\sigma$ error bars or $2\sigma$ limits) of the parameters for all models from various combinations of data.\tnote{a}}\label{tab:1d_BFP2C7}
\begin{tabular}{lcccccccccccccc}
\toprule
Model & Data set & $\Omega_{\mathrm{m0}}$ & $\Omega_{\mathrm{k0}}$ & $w_{\mathrm{X}}$ & $\alpha$ & $\sigma_{\mathrm{int,\,\textsc{ml}}}$ & $b_{\mathrm{\textsc{ml}}}$ & $k_{\mathrm{\textsc{ml}}}$ & $\sigma_{\mathrm{int}}$ & $\gamma$ & $\beta$ & $\sigma_{\mathrm{int,\,\textsc{gl}}}$ & $b_{\mathrm{\textsc{gl}}}$ & $k_{\mathrm{\textsc{gl}}}$ \\
\midrule
 & A118 & $>0.247$ & -- & -- & -- & -- & -- & -- & $0.412^{+0.027}_{-0.033}$ & $50.09\pm0.26$ & $1.110\pm0.090$ & -- & -- & -- \\
 & ML & $>0.188$ & -- & -- & -- & $0.305^{+0.035}_{-0.053}$ & $1.552^{+0.108}_{-0.189}$ & $-1.017\pm0.090$ & -- & -- & -- & -- & -- & -- \\
 & A115 & $0.630^{+0.352}_{-0.135}$ & -- & -- & -- & -- & -- & -- & $0.417^{+0.028}_{-0.035}$ & $50.09\pm0.26$ & $1.112\pm0.093$ & -- & -- & -- \\
Flat \lcdm & ML + A115 & $>0.298$ & -- & -- & -- & $0.301^{+0.033}_{-0.051}$ & $1.515^{+0.101}_{-0.151}$ & $-1.015\pm0.088$ & $0.416^{+0.027}_{-0.034}$ & $50.07\pm0.25$ & $1.111\pm0.089$ & -- & -- & -- \\
 & GL & $>0.202$ & -- & -- & -- & -- & -- & -- & -- & -- & -- & $0.429^{+0.059}_{-0.094}$ & $0.495^{+0.120}_{-0.173}$ & $-1.720\pm0.219$ \\
 & A115$^{\prime}$ & $>0.264$ & -- & -- & -- & -- & -- & -- & $0.414^{+0.028}_{-0.034}$ & $50.10\pm0.26$ & $1.107\pm0.090$ & -- & -- & -- \\%
 & GL + A115$^{\prime}$ & $>0.339$ & -- & -- & -- & -- & -- & -- & $0.413^{+0.027}_{-0.034}$ & $50.08\pm0.25$ & $1.104\pm0.089$ & $0.423^{+0.056}_{-0.090}$ & $0.458^{+0.112}_{-0.139}$ & $-1.705\pm0.210$ \\
\\
 & A118 & $>0.287$ &  $0.694^{+0.626}_{-0.848}$ & -- & -- & -- & -- & -- & $0.412^{+0.027}_{-0.034}$ & $50.01\pm0.26$ & $1.121\pm0.090$ & -- & -- & -- \\
 & ML & $>0.241$ &  $-0.131^{+0.450}_{-0.919}$ & -- & -- & $0.304^{+0.035}_{-0.053}$ & $1.478^{+0.123}_{-0.166}$ & $-1.000\pm0.096$ & -- & -- & -- & -- & -- & -- \\
 & A115 & $>0.275$ &  $0.698^{+0.639}_{-0.857}$ & -- & -- & -- & -- & -- & $0.417^{+0.028}_{-0.034}$ & $50.00\pm0.27$ & $1.124\pm0.092$ & -- & -- & -- \\
Non-flat \lcdm & ML + A115 & $>0.346$ &  $0.352^{+0.427}_{-0.830}$ & -- & -- & $0.304^{+0.034}_{-0.052}$ & $1.486^{+0.096}_{-0.136}$ & $-1.019\pm0.091$ & $0.416^{+0.028}_{-0.034}$ & $50.03\pm0.26$ & $1.113\pm0.091$ & -- & -- & -- \\
 & GL & $>0.290$ &  $-0.762^{+0.271}_{-0.888}$ & -- & -- & -- & -- & -- & -- & -- & -- & $0.402^{+0.057}_{-0.090}$ & $0.407^{+0.136}_{-0.160}$ & $-1.536\pm0.252$ \\
 & A115$^{\prime}$ & $>0.299$ & $0.599^{+0.582}_{-0.887}$ & -- & -- & -- & -- & -- & $0.414^{+0.028}_{-0.034}$ & $50.02\pm0.26$ & $1.117\pm0.090$ & -- & -- & -- \\
 & GL + A115$^{\prime}$ & $>0.381$ &  $0.214^{+0.428}_{-0.855}$ & -- & -- & -- & -- & -- & $0.414^{+0.027}_{-0.034}$ & $50.05\pm0.26$ & $1.103\pm0.090$ & $0.423^{+0.056}_{-0.090}$ & $0.432^{+0.111}_{-0.132}$ & $-1.701\pm0.212$ \\
\\
 & A118 & $0.599^{+0.350}_{-0.175}$ & -- & $-2.440^{+1.779}_{-1.715}$ & -- & -- & -- & -- & $0.412^{+0.028}_{-0.034}$ & $50.15^{+0.26}_{-0.30}$ & $1.106\pm0.089$ & -- & -- & -- \\
 & ML & $>0.123$ & -- & $-2.456^{+2.567}_{-2.180}$ & -- & $0.306^{+0.036}_{-0.054}$ & $1.611^{+0.113}_{-0.277}$ & $-1.014\pm0.092$ & -- & -- & -- & -- & -- & -- \\
 & A115 & $0.589^{+0.357}_{-0.184}$ & -- & $-2.411^{+1.797}_{-1.729}$ & -- & -- & -- & -- & $0.417^{+0.028}_{-0.035}$ & $50.14^{+0.27}_{-0.31}$ & $1.109\pm0.092$ & -- & -- & -- \\
Flat XCDM & ML + A115 & $>0.191$ & -- & $<-0.041$ & -- & $0.300^{+0.033}_{-0.051}$ & $1.562^{+0.103}_{-0.220}$ & $-1.012\pm0.088$ & $0.417^{+0.028}_{-0.034}$ & $50.13^{+0.27}_{-0.30}$ & $1.108\pm0.090$ & -- & -- & -- \\
 & GL & $>0.141$ & -- & $<-0.046$ & -- & -- & -- & -- & -- & -- & -- & $0.428^{+0.058}_{-0.092}$ & $0.556^{+0.127}_{-0.256}$ & $-1.706\pm0.215$ \\
 & A115$^{\prime}$ & $0.605^{+0.394}_{-0.126}$ & -- & $-2.391^{+1.826}_{-1.758}$ & -- & -- & -- & -- & $0.414^{+0.028}_{-0.034}$ & $50.15^{+0.27}_{-0.30}$ & $1.103\pm0.092$ & -- & -- & -- \\
 & GL + A115$^{\prime}$ & $>0.205$ & -- & $<-0.017$ & -- & -- & -- & -- & $0.413^{+0.027}_{-0.034}$ & $50.14^{+0.26}_{-0.30}$ & $1.101\pm0.088$ & $0.421^{+0.055}_{-0.090}$ & $0.512^{+0.113}_{-0.210}$ & $-1.698\pm0.208$ \\
\\
 & A118 & $>0.246$ & $0.590^{+0.476}_{-0.796}$ & $-2.358^{+2.032}_{-1.154}$ & -- & -- & -- & -- & $0.412^{+0.027}_{-0.033}$ & $50.01\pm0.28$ & $1.121\pm0.091$ & -- & -- & -- \\
 & ML & $>0.174$ & $-0.262^{+0.580}_{-0.724}$ & $-2.000^{+2.117}_{-1.264}$ & -- & $0.305^{+0.036}_{-0.054}$ & $1.462^{+0.194}_{-0.196}$ & $-0.996\pm0.097$ & -- & -- & -- & -- & -- & -- \\
 & A115 & $>0.240$ & $0.563^{+0.498}_{-0.796}$ & $-2.290^{+2.146}_{-1.032}$ & -- & -- & -- & -- & $0.418^{+0.028}_{-0.034}$ & $50.01\pm0.28$ & $1.122\pm0.093$ & -- & -- & -- \\
Non-flat XCDM & ML + A115 & $>0.231$ & $0.202^{+0.376}_{-0.635}$ & $-2.155^{+2.224}_{-1.156}$ & -- & $0.302^{+0.033}_{-0.051}$ & $1.489^{+0.122}_{-0.167}$ & $-1.016\pm0.089$ & $0.417^{+0.027}_{-0.034}$ & $50.05\pm0.28$ & $1.109\pm0.090$ & -- & -- & -- \\
 & GL & $>0.194$ & $-0.615^{+0.470}_{-0.685}$ & $-2.212^{+2.186}_{-0.962}$ & -- & -- & -- & -- & -- & -- & -- & $0.403^{+0.058}_{-0.092}$ & $0.480^{+0.177}_{-0.223}$ & $-1.532^{+0.259}_{-0.260}$ \\
 & A115$^{\prime}$ & $>0.236$ & $0.451^{+0.469}_{-0.789}$ & $-2.210^{+2.208}_{-0.946}$ & -- & -- & -- & -- & $0.415^{+0.028}_{-0.034}$ & $50.03\pm0.28$ & $1.114\pm0.091$ & -- & -- & -- \\
 & GL + A115$^{\prime}$ & $>0.226$ & $0.014^{+0.408}_{-0.604}$ & $-2.080^{+2.201}_{-1.138}$ & -- & -- & -- & -- & $0.415^{+0.027}_{-0.033}$ & $50.09^{+0.27}_{-0.30}$ & $1.096\pm0.090$ & $0.416^{+0.055}_{-0.088}$ & $0.446^{+0.142}_{-0.183}$ & $-1.682\pm0.208$ \\
\\
 & A118 & $0.568^{+0.332}_{-0.230}$ & -- & -- & -- & -- & -- & -- & $0.411^{+0.027}_{-0.033}$ & $50.05\pm0.25$ & $1.110\pm0.089$ & -- & -- & -- \\
 & ML & $>0.148$ & -- & -- & -- & $0.304^{+0.035}_{-0.053}$ & $1.493^{+0.093}_{-0.143}$ & $-1.017\pm0.089$ & -- & -- & -- & -- & -- & -- \\
 & A115 & $0.565^{+0.339}_{-0.228}$ & -- & -- & -- & -- & -- & -- & $0.416^{+0.028}_{-0.034}$ & $50.04\pm0.25$ & $1.113\pm0.091$ & -- & -- & -- \\
Flat $\phi$CDM & ML + A115 & $>0.198$ & -- & -- & -- & $0.302^{+0.033}_{-0.051}$ & $1.477^{+0.091}_{-0.126}$ & $-1.016\pm0.088$ & $0.416^{+0.027}_{-0.034}$ & $50.03\pm0.25$ & $1.110\pm0.089$ & -- & -- & -- \\
 & GL & $>0.148$ & -- & -- & -- & -- & -- & -- & -- & -- & -- & $0.428^{+0.059}_{-0.094}$ & $0.444^{+0.112}_{-0.141}$ & $-1.710\pm0.218$ \\
 & A115$^{\prime}$ & $0.586^{+0.391}_{-0.156}$ & -- & -- & -- & -- & -- & -- & $0.414^{+0.028}_{-0.034}$ & $50.06\pm0.25$ & $1.106\pm0.089$ & -- & -- & -- \\
 & GL + A115$^{\prime}$ & $>0.231$ & -- & -- & -- & -- & -- & -- & $0.413^{+0.027}_{-0.033}$ & $50.05\pm0.24$ & $1.102\pm0.088$ & $0.423^{+0.055}_{-0.090}$ & $0.426^{+0.105}_{-0.119}$ & $-1.703\pm0.210$ \\
\\
 & A118 & $0.560^{+0.256}_{-0.247}$ & $-0.002^{+0.294}_{-0.286}$ & -- & $5.203^{+3.808}_{-2.497}$ & -- & -- & -- & $0.412^{+0.027}_{-0.033}$ & $50.05\pm0.25$ & $1.111^{+0.089}_{-0.090}$ & -- & -- & -- \\
 & ML & $>0.207$ & $-0.163^{+0.355}_{-0.317}$ & -- & -- & $0.303^{+0.035}_{-0.053}$ & $1.448^{+0.120}_{-0.165}$ & $-1.011\pm0.091$ & -- & -- & -- & -- & -- & -- \\
 & A115 & $0.546^{+0.260}_{-0.253}$ & $0.011^{+0.299}_{-0.291}$ & -- & -- & -- & -- & -- & $0.417^{+0.028}_{-0.034}$ & $50.04\pm0.26$ & $1.115\pm0.092$ & -- & -- & -- \\
Non-flat $\phi$CDM & ML + A115 & $0.630^{+0.286}_{-0.181}$ & $-0.108^{+0.296}_{-0.263}$ & -- & -- & $0.301^{+0.033}_{-0.051}$ & $1.453^{+0.115}_{-0.139}$ & $-1.013\pm0.089$ & $0.416^{+0.027}_{-0.034}$ & $50.03\pm0.25$ & $1.106\pm0.090$ & -- & -- & -- \\
 & GL & $>0.207$ & $-0.193^{+0.364}_{-0.347}$ & -- & -- & -- & -- & -- & -- & -- & -- & $0.422^{+0.057}_{-0.092}$ & $0.408^{+0.121}_{-0.149}$ & $-1.693^{+0.215}_{-0.214}$ \\
 & A115$^{\prime}$ & $0.581^{+0.254}_{-0.247}$ & $-0.033^{+0.296}_{-0.290}$ & -- & $5.215^{+3.853}_{-2.429}$ & -- & -- & -- & $0.414^{+0.028}_{-0.034}$ & $50.05\pm0.25$ & $1.106\pm0.091$ & -- & -- & -- \\
 & GL + A115$^{\prime}$ & $0.666^{+0.327}_{-0.106}$ & $-0.169^{+0.317}_{-0.270}$ & -- & -- & -- & -- & -- &  $0.413^{+0.027}_{-0.034}$ & $50.04\pm0.25$ & $1.096\pm0.089$ & $0.419^{+0.055}_{-0.089}$ & $0.401^{+0.115}_{-0.128}$ & $-1.686\pm0.209$ \\
\bottomrule
\end{tabular}
\begin{tablenotes}[flushleft]
\item [a] In these GRB cases, $\Omega_b$ and $H_0$ are set to be 0.05 and 70 \hunit, respectively.
\end{tablenotes}
\end{threeparttable}%
}
\end{sidewaystable*}

\begin{figure*}
\centering
 \subfloat[Flat \lcdm]{%
    \includegraphics[width=3.25in,height=1.84in]{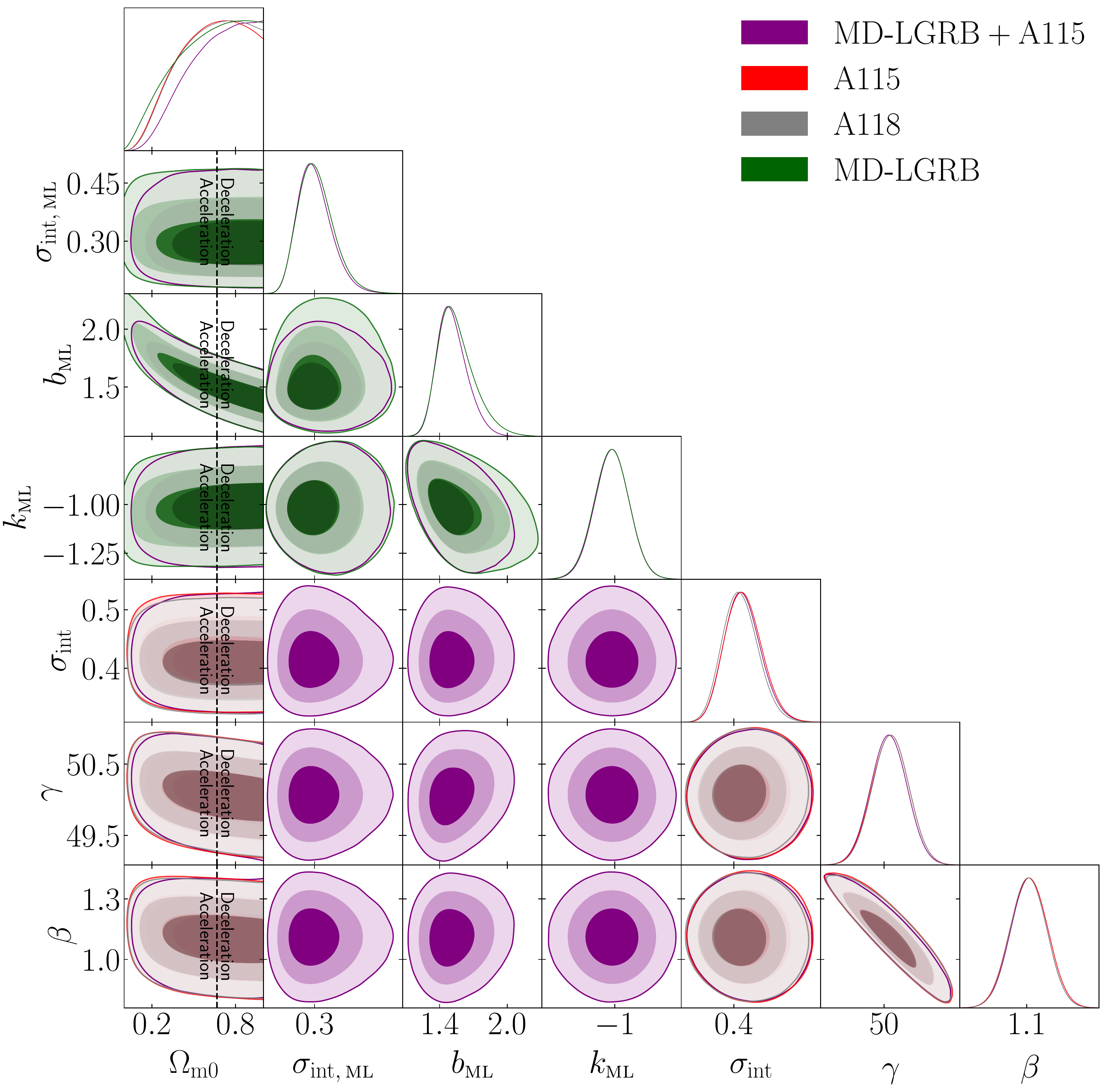}}
 \subfloat[Non-flat \lcdm]{%
    \includegraphics[width=3.25in,height=1.84in]{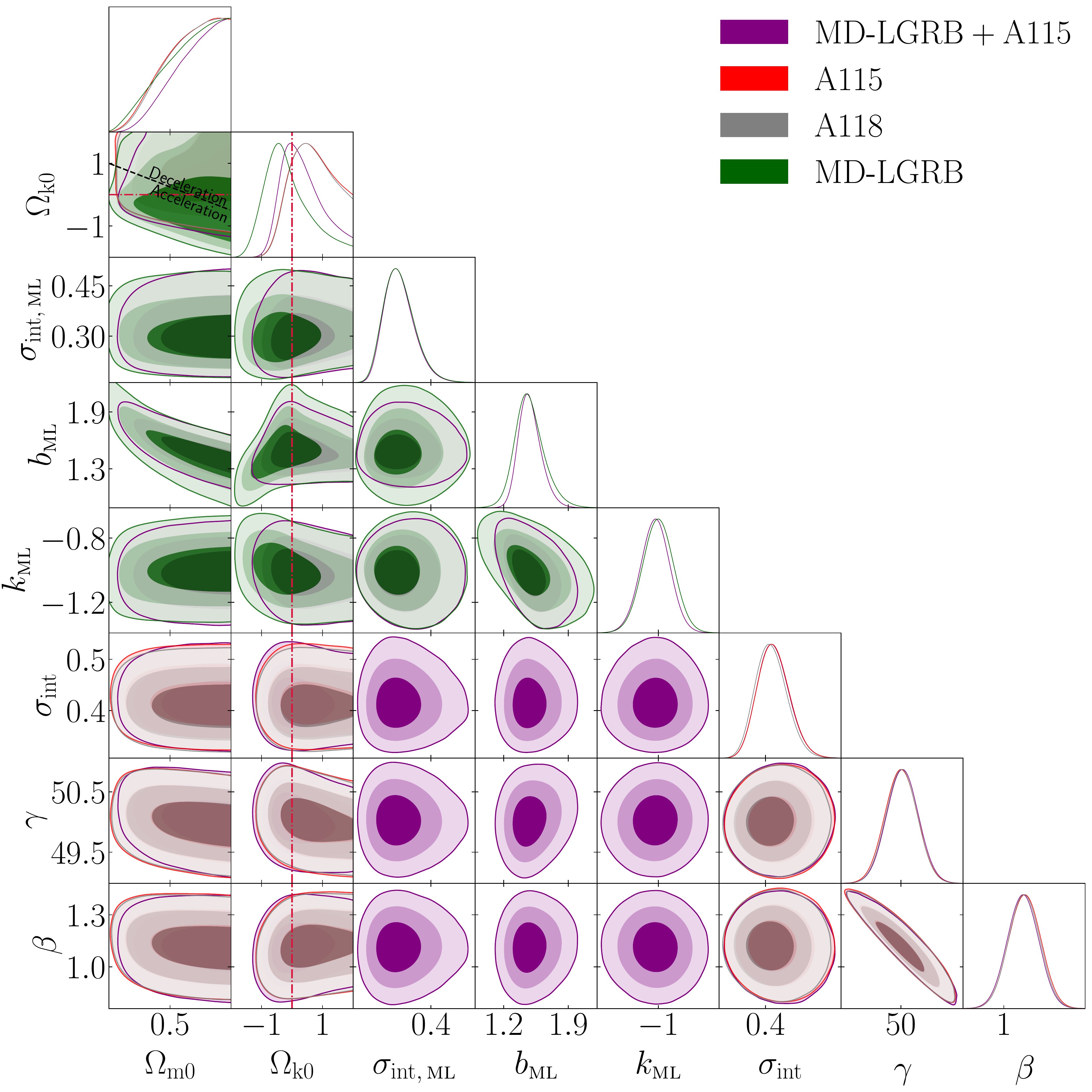}}\\
 \subfloat[Flat XCDM]{%
    \includegraphics[width=3.25in,height=1.84in]{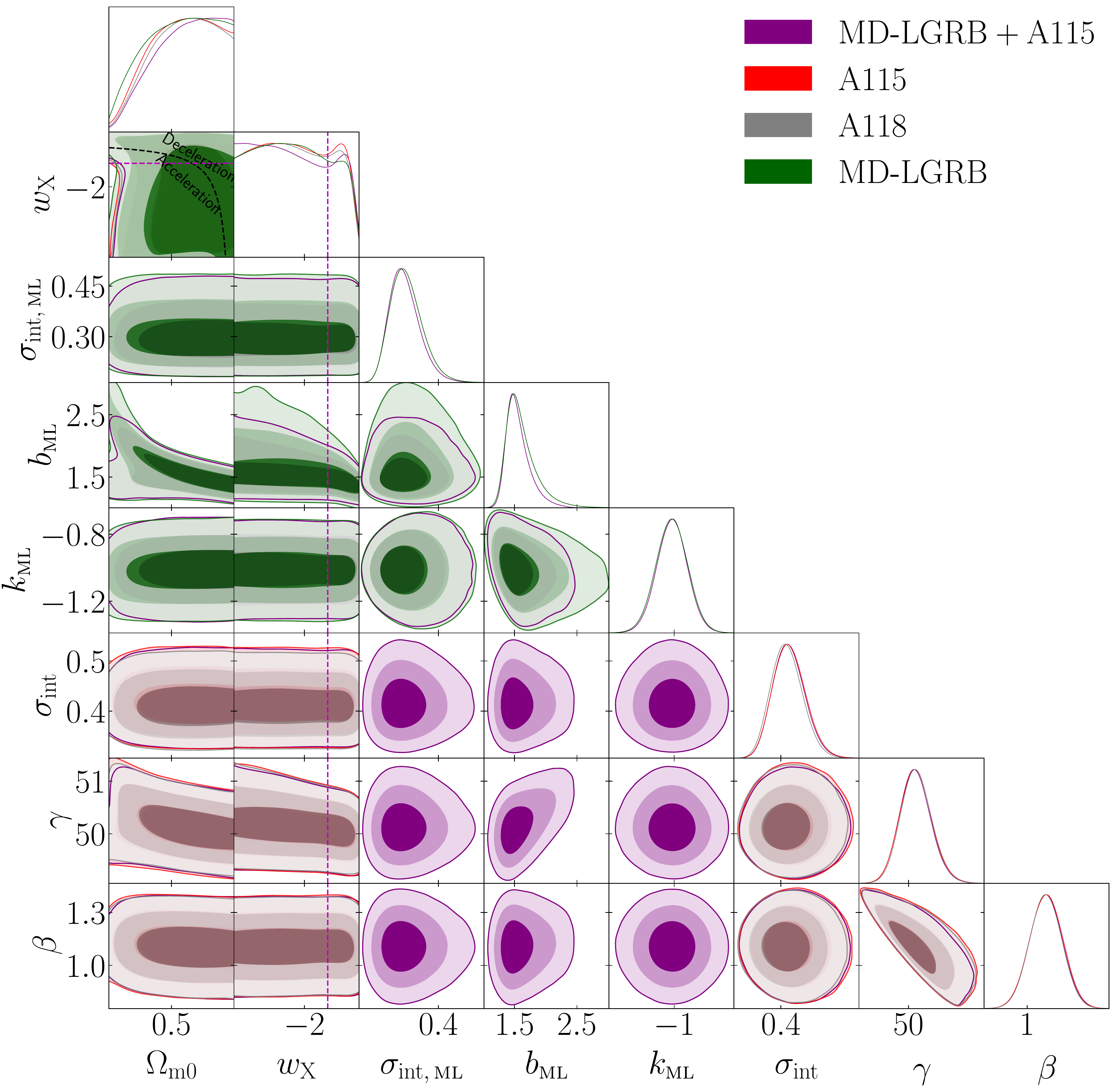}}
 \subfloat[Non-flat XCDM]{%
    \includegraphics[width=3.25in,height=1.84in]{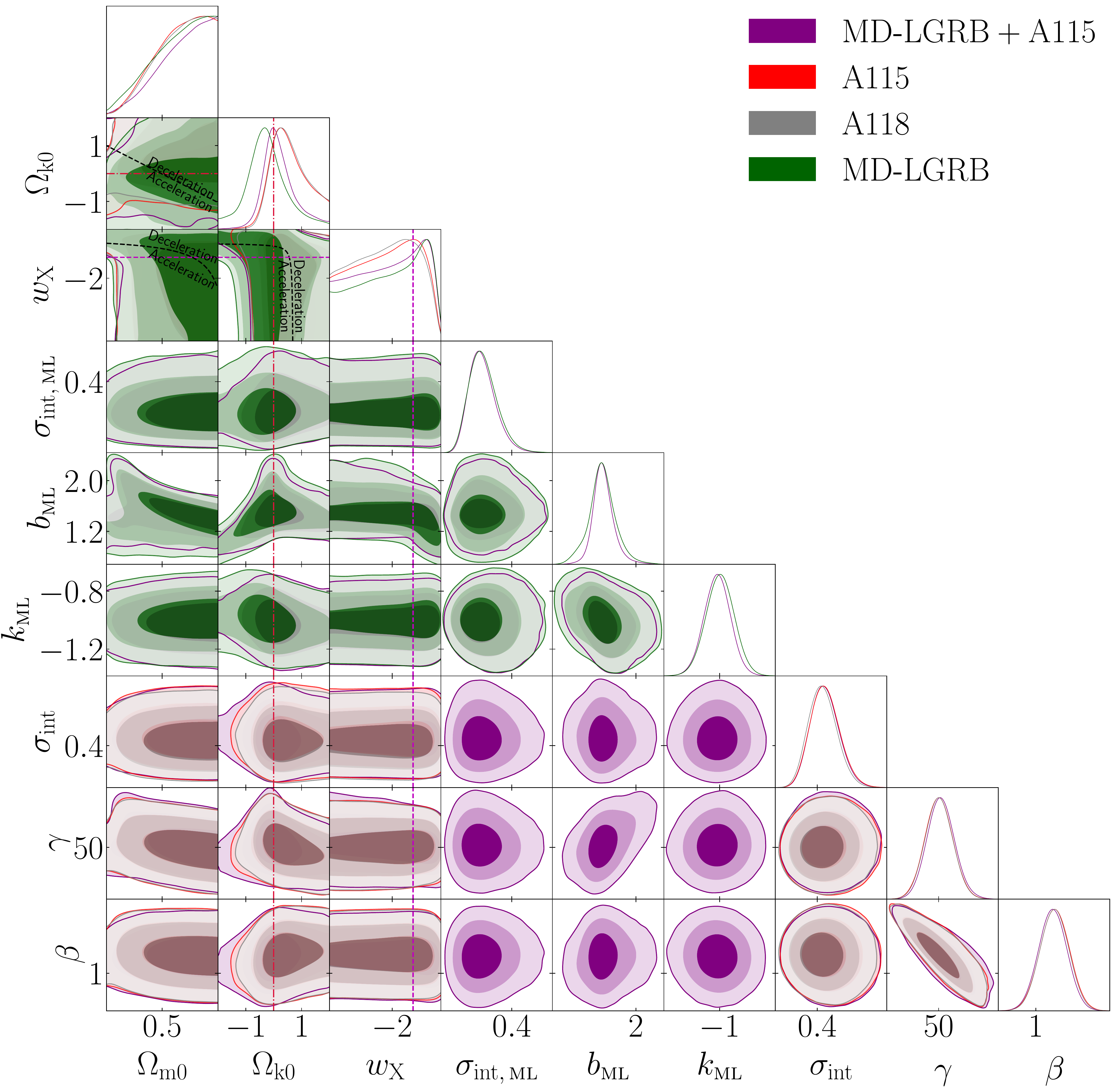}}\\
 \subfloat[Flat \pcdm]{%
    \includegraphics[width=3.25in,height=1.84in]{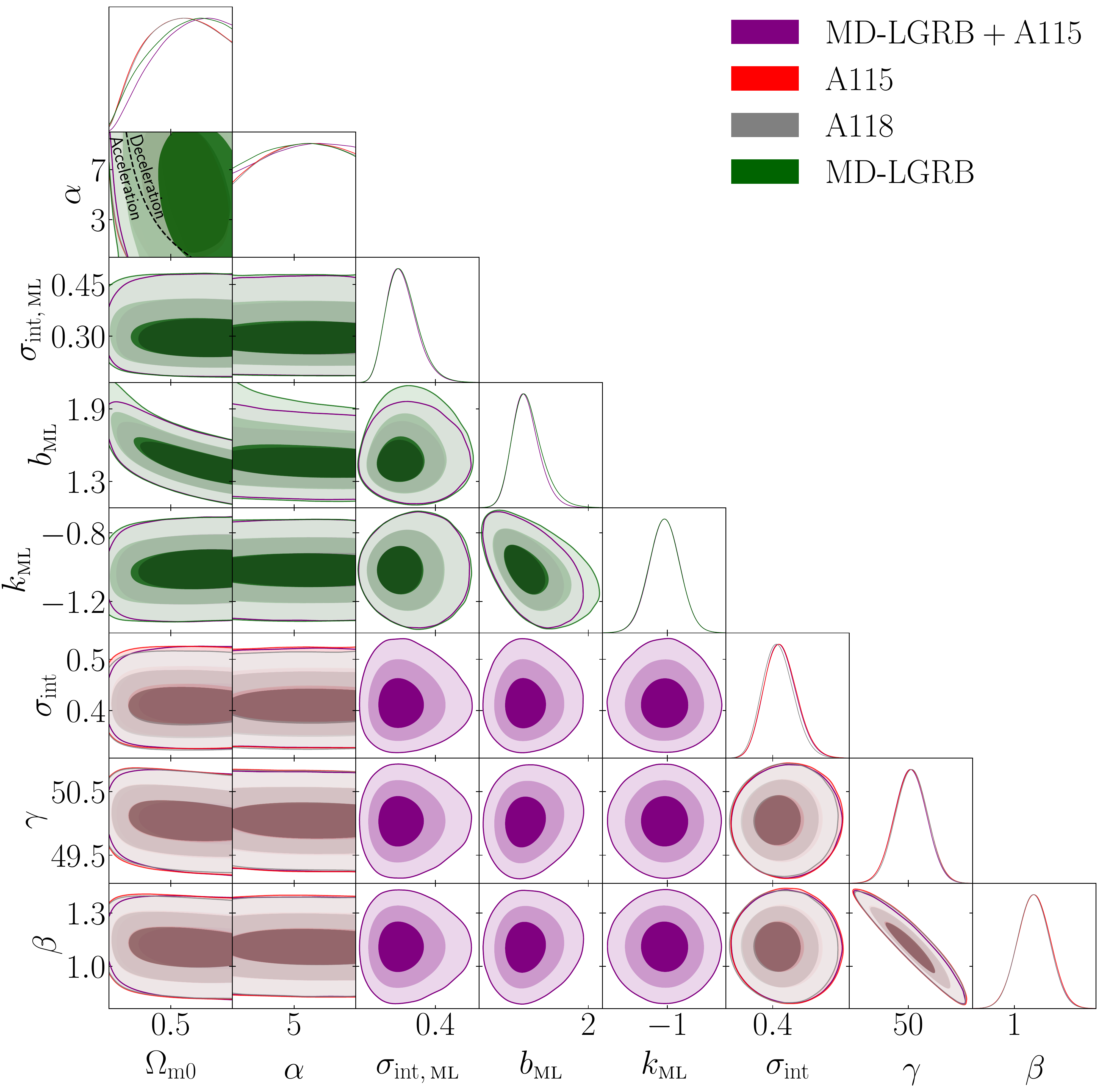}}
  \subfloat[Non-flat \pcdm]{%
     \includegraphics[width=3.25in,height=1.84in]{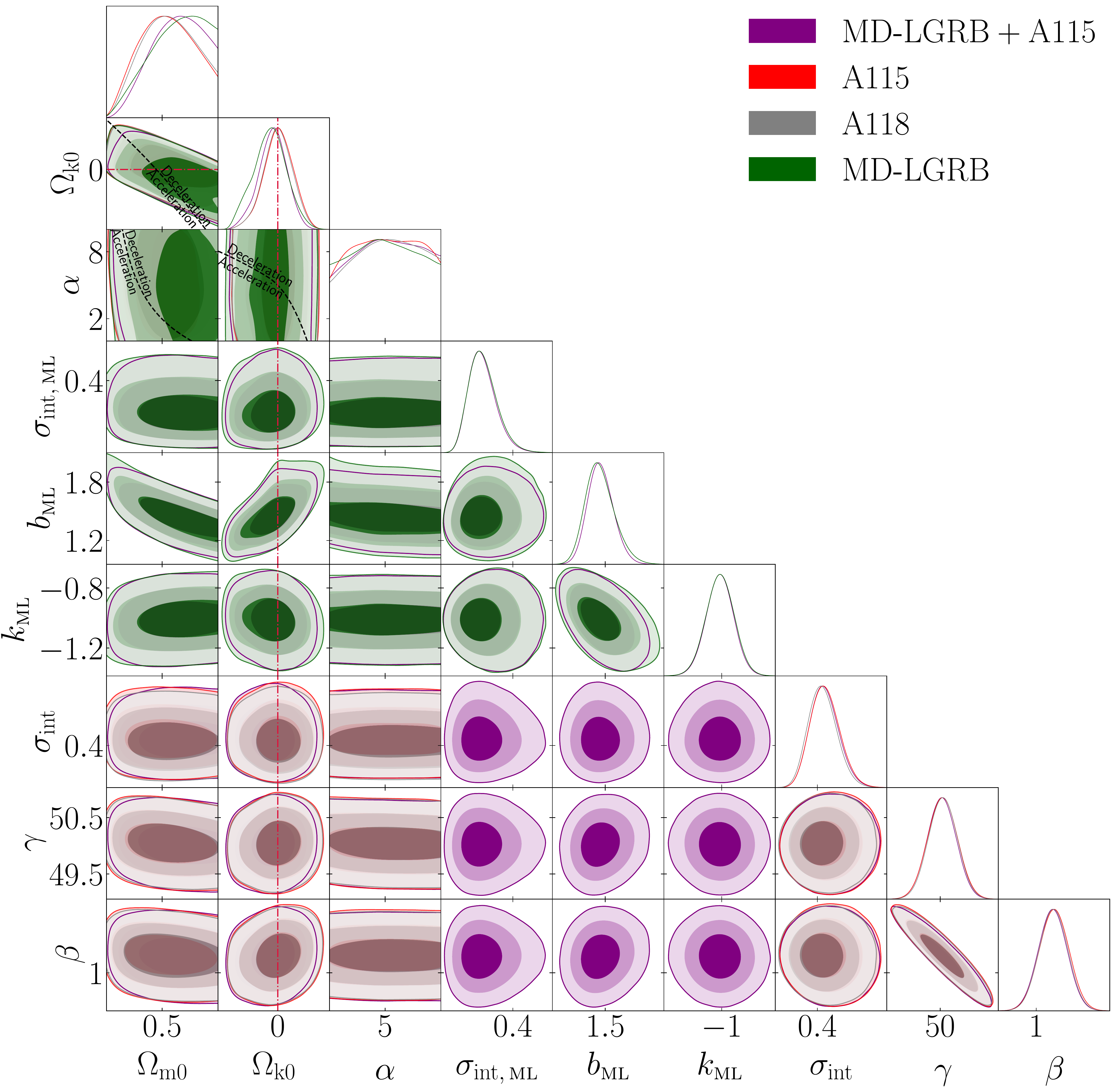}}\\
\caption{One-dimensional likelihoods and 1$\sigma$, 2$\sigma$, and 3$\sigma$ two-dimensional likelihood confidence contours from MD-LGRB (green), A118 (gray), A115 (red), and MD-LGRB + A115 (purple) data for all six models. The zero-acceleration lines are shown as black dashed lines, which divide the parameter space into regions associated with currently-accelerating and currently-decelerating cosmological expansion. In the non-flat XCDM and non-flat \pcdm\ cases, the zero-acceleration lines are computed for the third cosmological parameter set to the $H(z)$ + BAO data best-fitting values listed in Table \ref{tab:BFP2C7}. The crimson dash-dot lines represent flat hypersurfaces, with closed spatial hypersurfaces either below or to the left. The magenta lines represent $w_{\rm X}=-1$, i.e.\ flat or non-flat \lcdm\ models. The $\alpha = 0$ axes correspond to flat and non-flat \lcdm\ models in panels (e) and (f), respectively.}
\label{fig5C7}
\end{figure*}

\begin{figure*}
\centering
 \subfloat[Flat \lcdm]{%
    \includegraphics[width=3.25in,height=1.84in]{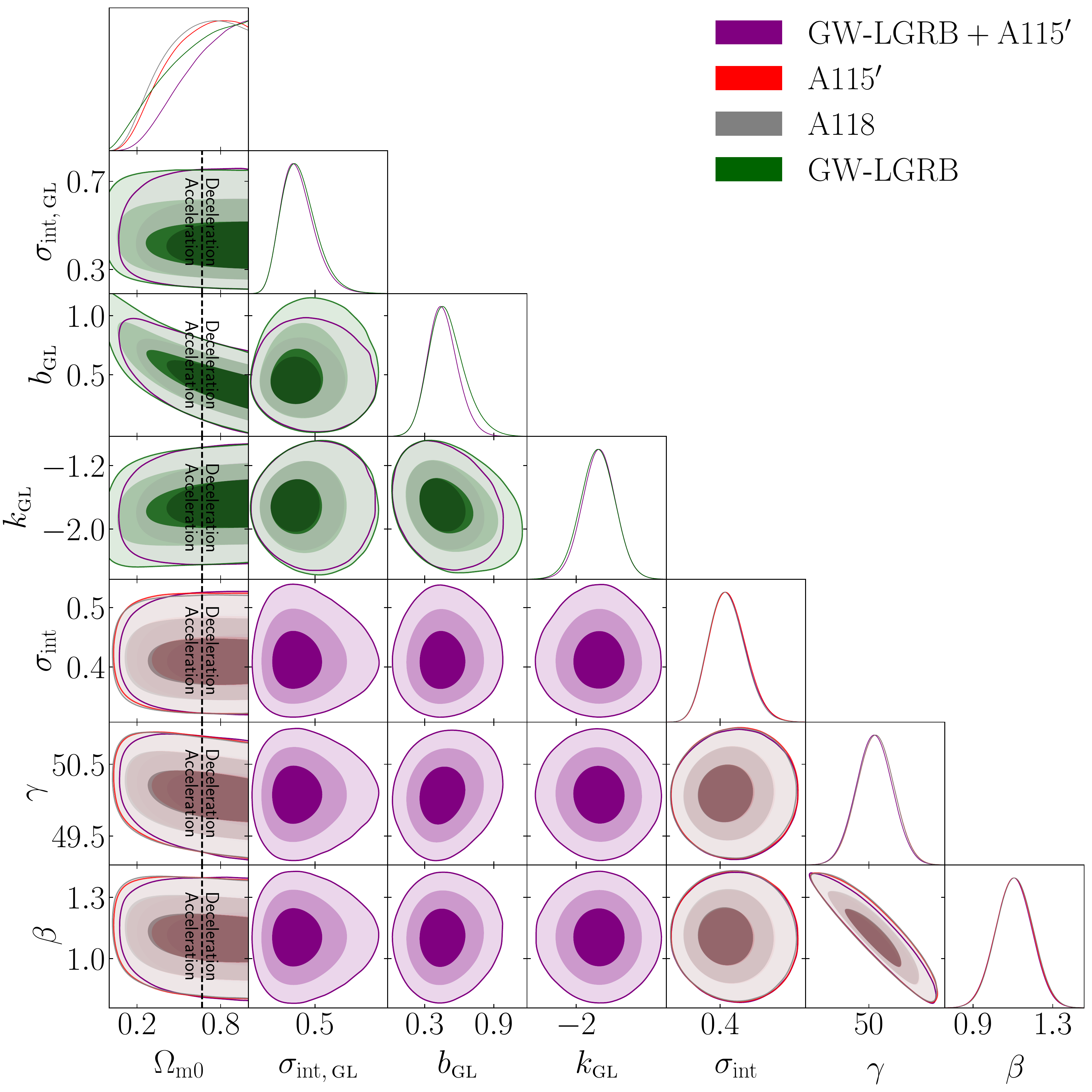}}
 \subfloat[Non-flat \lcdm]{%
    \includegraphics[width=3.25in,height=1.84in]{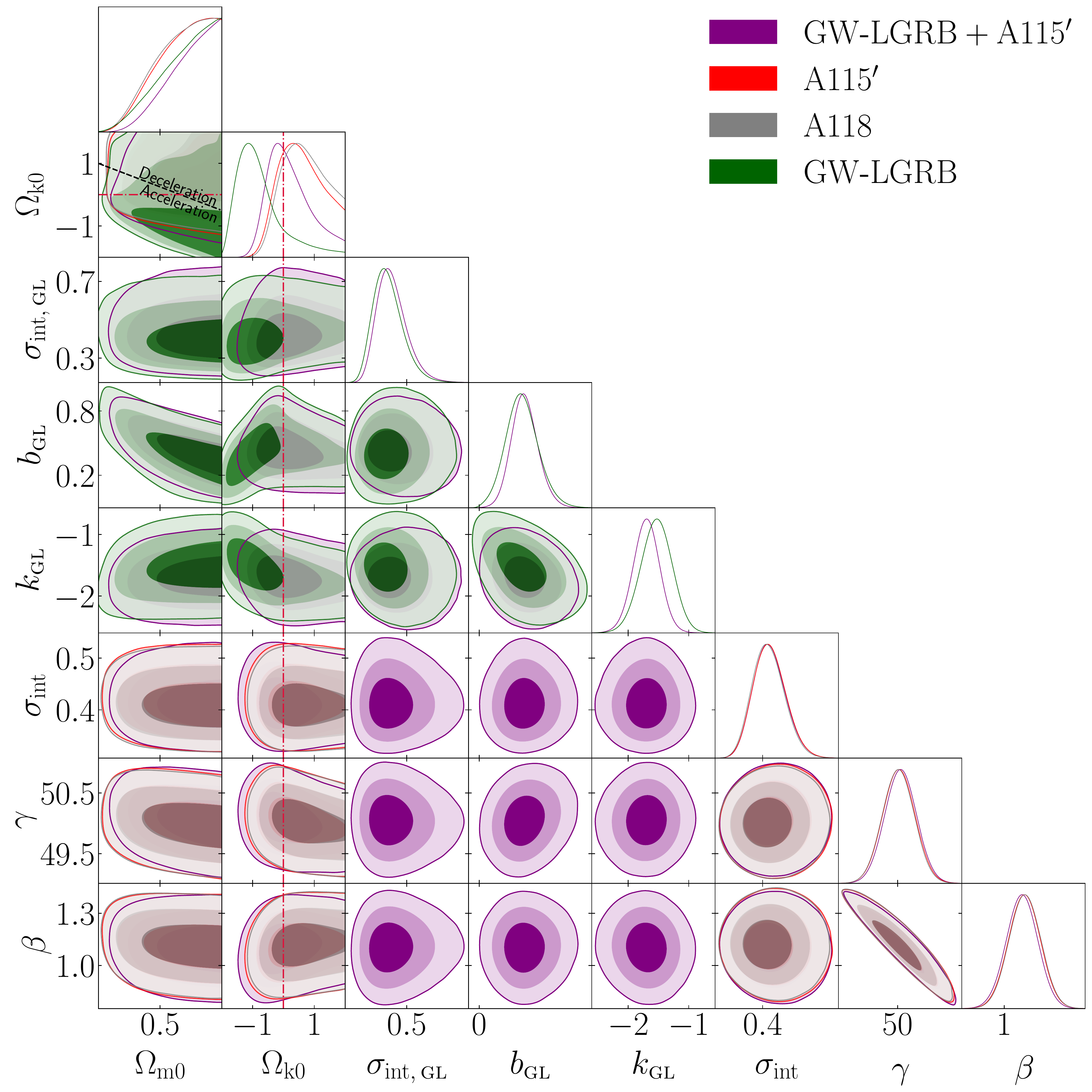}}\\
 \subfloat[Flat XCDM]{%
    \includegraphics[width=3.25in,height=1.84in]{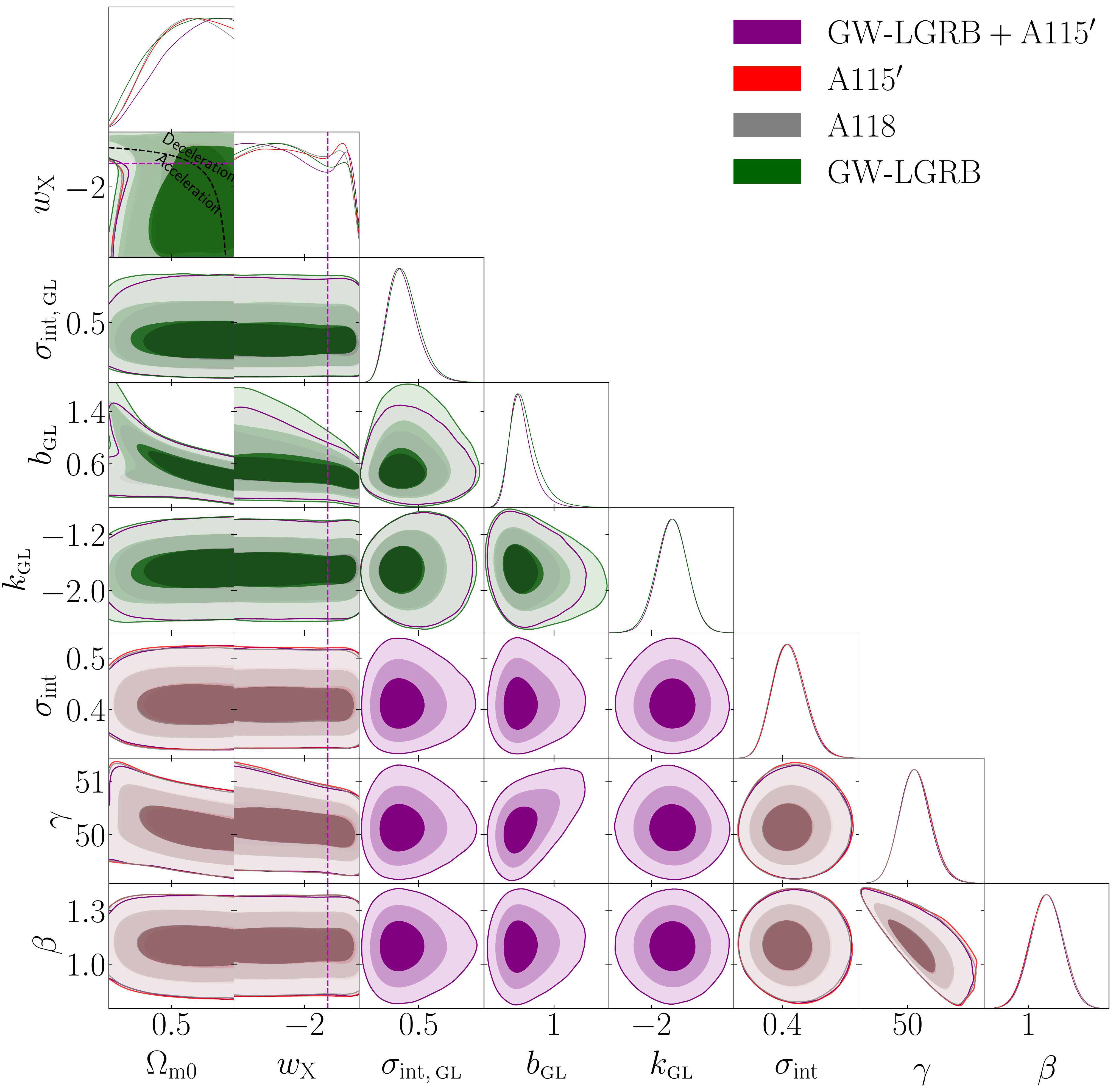}}
 \subfloat[Non-flat XCDM]{%
    \includegraphics[width=3.25in,height=1.84in]{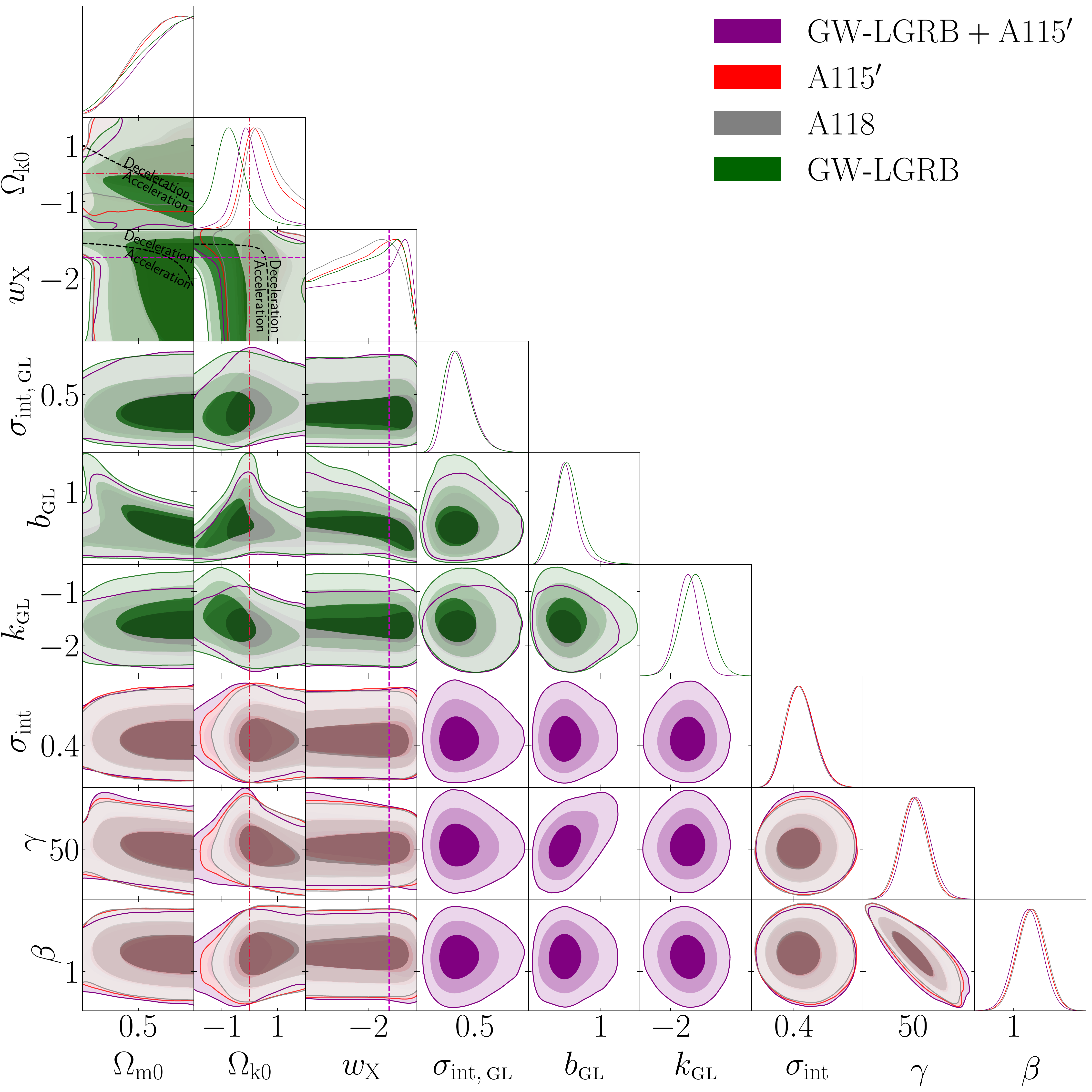}}\\
 \subfloat[Flat \pcdm]{%
    \includegraphics[width=3.25in,height=1.84in]{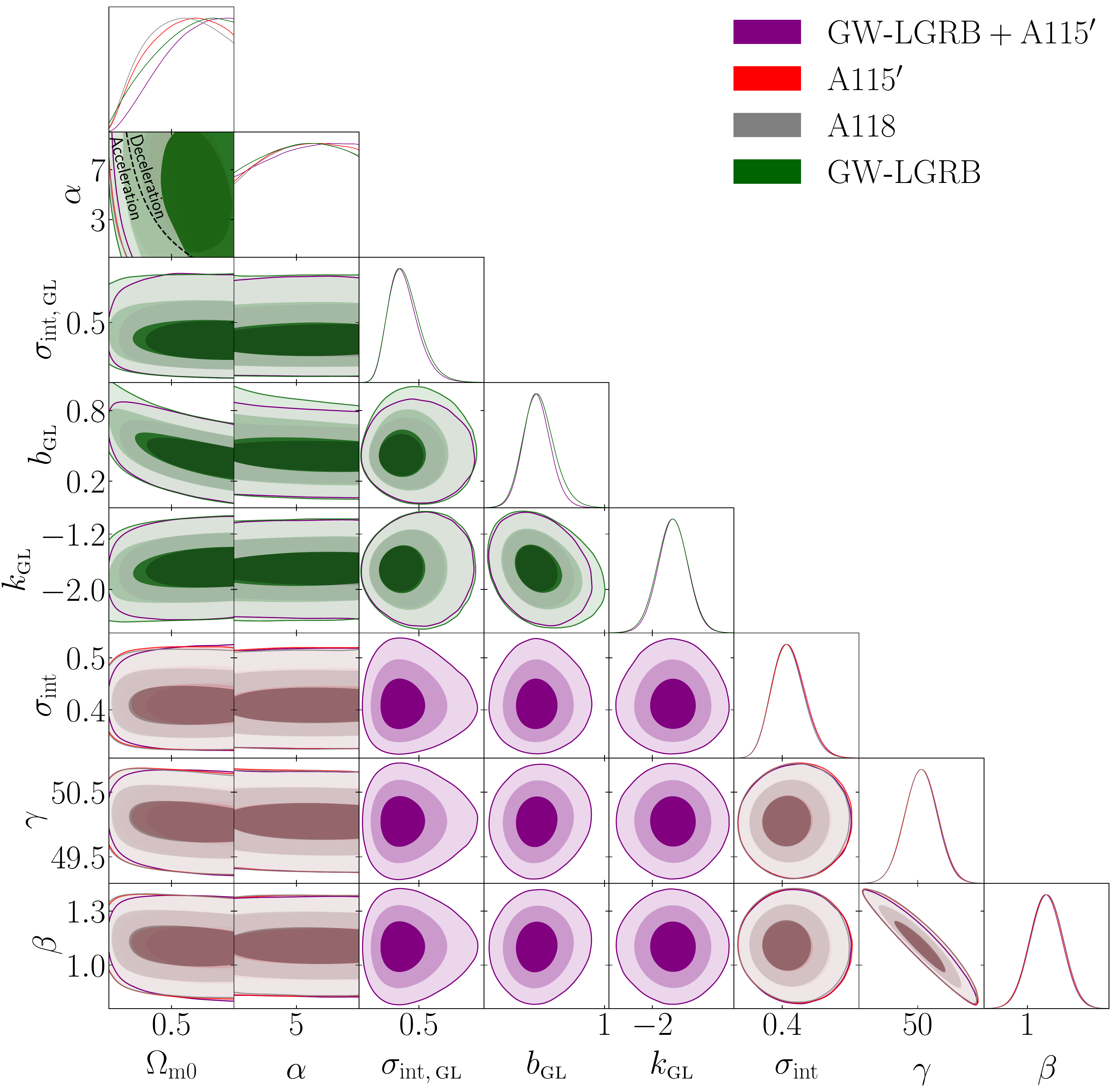}}
  \subfloat[Non-flat \pcdm]{%
     \includegraphics[width=3.25in,height=1.84in]{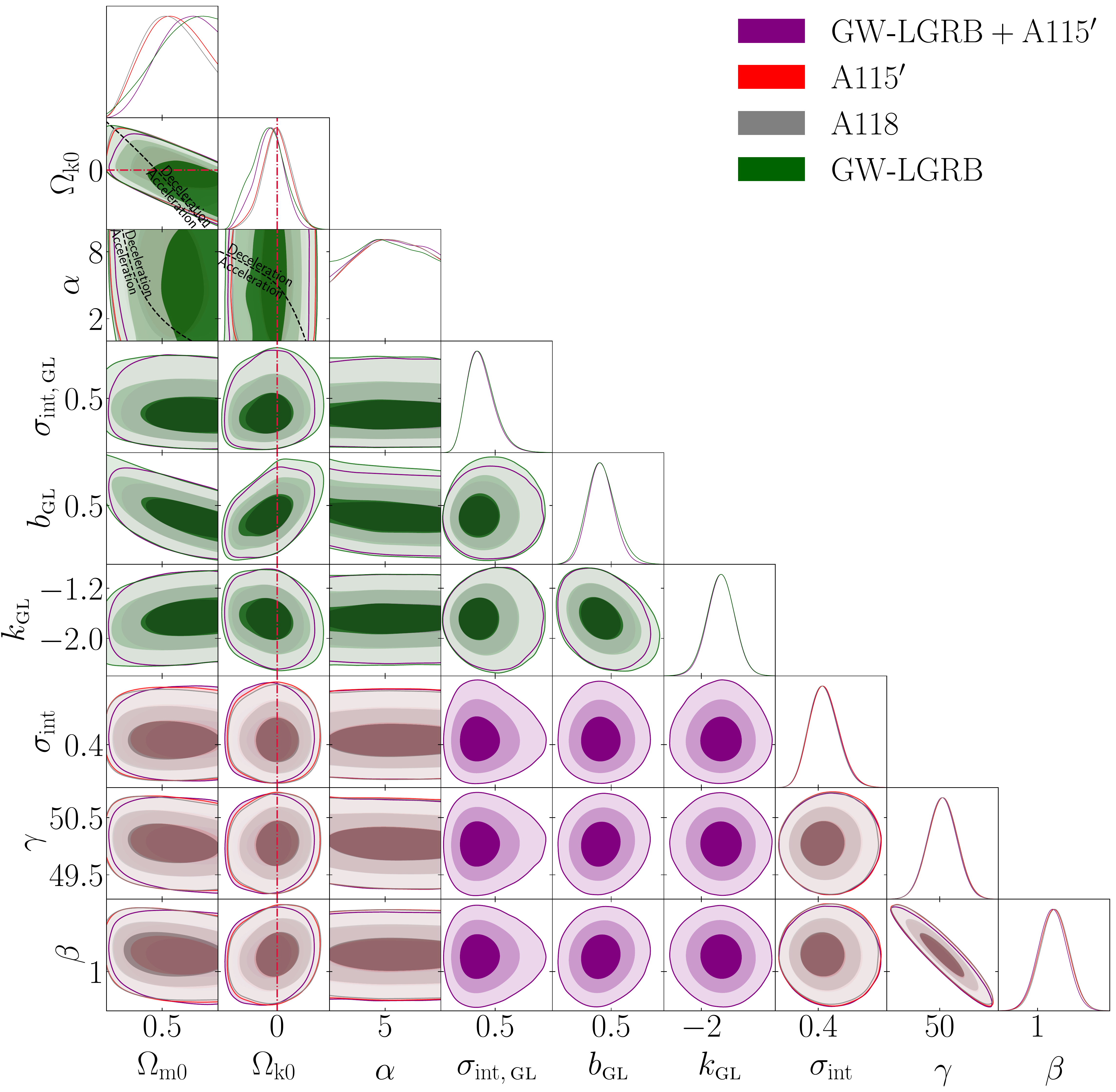}}\\
\caption{One-dimensional likelihoods and 1$\sigma$, 2$\sigma$, and 3$\sigma$ two-dimensional likelihood confidence contours from GW-LGRB (green), A118 (gray), A11$5'$ (red), and GW-LGRB + A11$5'$ (purple) data for all six models. The zero-acceleration lines are shown as black dashed lines, which divide the parameter space into regions associated with currently-accelerating and currently-decelerating cosmological expansion. In the non-flat XCDM and non-flat \pcdm\ cases, the zero-acceleration lines are computed for the third cosmological parameter set to the $H(z)$ + BAO data best-fitting values listed in Table \ref{tab:BFP2C7}. The crimson dash-dot lines represent flat hypersurfaces, with closed spatial hypersurfaces either below or to the left. The magenta lines represent $w_{\rm X}=-1$, i.e.\ flat or non-flat \lcdm\ models. The $\alpha = 0$ axes correspond to flat and non-flat \lcdm\ models in panels (e) and (f), respectively.}
\label{fig6C7}
\end{figure*}

The A118 data set was previously studied by \cite{Khadkaetal_2021b}. Here we analyze it along with the truncated A115 and A115$^{\prime}$ data sets, which are also used in joint analyses with the ML and GL data sets. The constraints from these data sets on the GRB correlation parameters and on the cosmological model parameters are presented in Tables \ref{tab:BFP2C7} and \ref{tab:1d_BFP2C7}. The corresponding posterior 1D probability distributions and 2D confidence regions of these parameters are shown in Figs. \ref{fig5C7} and \ref{fig6C7}, in gray (A118), red (A115 and A115$^{\prime}$), green (ML and GL), and purple (ML + A115 and GL + A115$^{\prime}$). Note that these analyses assume $H_0=70$ \hunit\ and $\Omega_{b}=0.05$.

The constraints from A115 data and from ML data, and from A115$^{\prime}$ data and from GL data, are not mutually inconsistent, so it is not unreasonable to examine joint ML + A115 and GL + A115$^{\prime}$ constraints. The ML data have the smallest intrinsic dispersion, $\sim 0.30-0.31$, with A115, A115$^{\prime}$, and GL having larger intrinsic dispersion, $\sim 0.40-0.43$ 

The constraints on the Amati parameters are quite cosmological-model-independent for these GRB data sets. In the A118 case, the slope $\beta$ ranges from a high of $1.121\pm0.091$ (non-flat XCDM) to a low of $1.106\pm0.089$ (flat XCDM), the intercept $\gamma$ ranges from a high of $50.15^{+0.26}_{-0.30}$ (flat XCDM) to a low of $50.01\pm0.26$ (non-flat \lcdm), and the intrinsic scatter $\sigma_{\rm int}$ ranges from a high of $0.412^{+0.028}_{-0.034}$ (flat XCDM) to a low of $0.411^{+0.027}_{-0.033}$ (flat \pcdm), with central values of each pair being $0.12\sigma$, $0.35\sigma$, and $0.02\sigma$ away from each other, respectively. 

In the A115 case, the slope $\beta$ ranges from a high of $1.124\pm0.092$ (non-flat \lcdm) to a low of $1.109\pm0.092$ (flat XCDM), the intercept $\gamma$ ranges from a high of $50.14^{+0.27}_{-0.31}$ (flat XCDM) to a low of $50.00\pm0.27$ (non-flat \lcdm), and the intrinsic scatter $\sigma_{\rm int}$ ranges from a high of $0.418^{+0.028}_{-0.034}$ (non-flat XCDM) to a low of $0.416^{+0.028}_{-0.034}$ (flat \pcdm), with central values of each pair being $0.12\sigma$, $0.34\sigma$, and $0.05\sigma$ away from each other, respectively.

In the A115$^{\prime}$ case, the slope $\beta$ ranges from a high of $1.117\pm0.090$ (non-flat \lcdm) to a low of $1.103\pm0.092$ (flat XCDM), the intercept $\gamma$ ranges from a high of $50.15^{+0.27}_{-0.30}$ (flat XCDM) to a low of $50.02\pm0.26$ (non-flat \lcdm), and the intrinsic scatter $\sigma_{\rm int}$ ranges from a high of $0.415^{+0.028}_{-0.034}$ (non-flat XCDM) to a low of $0.414^{+0.028}_{-0.034}$ (the others), with central values of each pair being $0.11\sigma$, $0.33\sigma$, and $0.02\sigma$ away from each other, respectively.

The lowest and highest values of $\beta$, $\gamma$, and $\sigma_{\rm int}$ from the A118, A115, and A115$^{\prime}$ cases differ from each other at $0.16\sigma$, $0.37\sigma$, and $0.16\sigma$, respectively. This implies that excluding three GRBs from A118 does not significantly affect the constraints on the Amati parameters. 

In the joint analysis of ML and A115 (ML + A115) data, $\beta$ ranges from a high of $1.113\pm0.091$ (non-flat \lcdm) to a low of $1.106\pm0.090$ (non-flat \pcdm), $\gamma$ ranges from a high of $50.13^{+0.27}_{-0.30}$ (flat XCDM) to a low of $50.03\pm0.25$ (flat and non-flat \pcdm), and $\sigma_{\rm int}$ ranges from a high of $0.417^{+0.028}_{-0.034}$ (flat XCDM) to a low of $0.416^{+0.027}_{-0.034}$ (flat \lcdm, and flat and non-flat \pcdm), with central values of each pair being $0.05\sigma$, $0.26\sigma$, and $0.02\sigma$ away from each other, respectively; also $k$ ranges from a high of $-1.012\pm0.088$ (flat XCDM) to a low of $-1.019\pm0.091$ (non-flat \lcdm), $b$ ranges from a high of $1.562^{+0.103}_{-0.220}$ (flat XCDM) to a low of $1.453^{+0.115}_{-0.139}$ (non-flat \pcdm), and $\sigma_{\mathrm{int,\,\textsc{ml}}}$ ranges from a high of $0.304^{+0.034}_{-0.052}$ (non-flat \lcdm) to a low of $0.300^{+0.033}_{-0.051}$ (flat XCDM), with central values of each pair being $0.06\sigma$, $0.44\sigma$, and $0.06\sigma$ away from each other, respectively. The lowest and highest values of $\beta$, $\gamma$, and $\sigma_{\rm int}$ from the A115 and ML + A115 cases differ from each other at $0.14\sigma$, $0.32\sigma$, and $0.05\sigma$, respectively; also those of $k$, $b$, and $\sigma_{\mathrm{int,\,\textsc{ml}}}$ differ from each other at $0.17\sigma$, $0.53\sigma$, and $0.09\sigma$, respectively.

In the joint analysis of GL and A115$^{\prime}$ (GL + A115$^{\prime}$) data, $\beta$ ranges from a high of $1.104\pm0.089$ (flat \lcdm) to a low of $1.096\pm0.089$ (non-flat \pcdm), $\gamma$ ranges from a high of $50.14^{+0.26}_{-0.30}$ (flat XCDM) to a low of $50.04\pm0.25$ (non-flat \pcdm), and $\sigma_{\rm int}$ ranges from a high of $0.415^{+0.027}_{-0.033}$ (non-flat XCDM) to a low of $0.413^{+0.027}_{-0.034}$ (flat \lcdm, flat XCDM, and non-flat \pcdm), with central values of each pair being $0.06\sigma$, $0.26\sigma$, and $0.05\sigma$ away from each other, respectively; also $k$ ranges from a high of $-1.682\pm0.208$ (non-flat XCDM) to a low of $-1.705\pm0.210$ (flat \lcdm), $b$ ranges from a high of $0.512^{+0.113}_{-0.210}$ (flat XCDM) to a low of $0.401^{+0.115}_{-0.128}$ (non-flat \pcdm), and $\sigma_{\mathrm{int,\,\textsc{gl}}}$ ranges from a high of $0.423^{+0.056}_{-0.090}$ (flat and non-flat \lcdm) to a low of $0.416^{+0.055}_{-0.088}$ (non-flat XCDM), with central values of each pair being $0.08\sigma$, $0.46\sigma$, and $0.07\sigma$ away from each other, respectively. The lowest and highest values of $\beta$, $\gamma$, and $\sigma_{\rm int}$ from the A115$^{\prime}$ and GL + A115$^{\prime}$ cases differ from each other at $0.17\sigma$, $0.30\sigma$, and $0.05\sigma$, respectively; also those of $k$, $b$, and $\sigma_{\mathrm{int,\,\textsc{gl}}}$ differ from each other at $0.52\sigma$, $0.47\sigma$, and $0.20\sigma$, respectively.

Judging from the constraints on the parameters of the Amati and Dainotti correlations, those from the joint analyses do not deviate much from the individual cases. We next focus on constraints on cosmological model parameters.

Similar to GRB data from Sec.\ \ref{subsec:MLSGL}, these data more favor currently accelerating cosmological expansion in the \lcdm\ and XCDM cases, but also more favor currently decelerating cosmological expansion in the \pcdm\ models, in the $\Om-\alpha$ and $\Om-\Ok$ parameter subspaces.

In the flat \lcdm\ model, the A118, A115, and A115$^{\prime}$ $2\sigma$ constraints on \om\ are mutually consistent (A115 with $\Om>0.241$), but they favor higher values of \om\ than do ML and GL while the joint ML + A115 and GL + A115$^{\prime}$ cases favor even higher values of $>0.298$ and $>0.339$, respectively. In fact, this is also true for the ML + GL case ($\Om>0.294$). 

In the non-flat \lcdm\ model, the A118, A115, and A115$^{\prime}$ $2\sigma$ constraints on \om\ are also mutually consistent, but they favor slightly higher values of \om\ than in the flat \lcdm\ case, with $\Om>0.299$ in the A115$^{\prime}$ non-flat \lcdm\ case. The joint ML + A115, GL + A115$^{\prime}$, and ML + GL cases favor higher \om\ $2\sigma$ limits of $>0.346$, $>0.381$, and $>0.338$, respectively. The A118, A115, ML + A115, A115$^{\prime}$, and GL + A115$^{\prime}$ data mildly favor open hypersurfaces in the non-flat \lcdm\ model, being less than $1\sigma$ away from flatness.

In the flat and non-flat XCDM parametrizations, the $2\sigma$ constraints on \om\ are mutually consistent in all cases, where in the flat XCDM parametrization, the $2\sigma$ limits are $\Om>0.181$ (A118), $>0.170$ (A115), and $>0.185$ (A115$^{\prime}$). The constraints on \wx\ are very loose, and thus affected by the \wx\ prior, and consistent with each other in all cases, and mildly favor phantom dark energy (but $\Lambda$ is less than $1\sigma$ away). In the non-flat XCDM parametrization, the A118, A115, ML + A115, A115$^{\prime}$, and GL + A115$^{\prime}$ data also mildly favor open hypersurfaces, with flatness being less than $1\sigma$ away.

In the flat \pcdm\ model, the A118, A115, and A115$^{\prime}$ constraints on \om\ are mutually consistent, with $2\sigma$ limits of $\Om>0.149$ (A118), $>0.145$ (A115), and $>0.159$ (A115$^{\prime}$), which are consistent with the other cases. These GRB data do not provide constraints on $\alpha$ in the flat \pcdm\ model, while in the non-flat \pcdm\ model, A118 and A115$^{\prime}$ provide constraints of $\alpha=5.203^{+3.808}_{-2.497}$ and $\alpha=5.215^{+3.853}_{-2.429}$, respectively. Note that $\alpha=0$ is still within $2\sigma$ for both cases. Similar trends hold for non-flat \pcdm\ \om\ constraints, but ML + A115 and GL + A115$^{\prime}$ data constraints posterior mean values are larger than for the individual data sets. The $2\sigma$ limits are $\Om>0.183$ (A118), $\Om=0.546^{+0.449}_{-0.384}$ (A115), $\Om>0.198$ (A115$^{\prime}$), $>0.251$ (ML + A115), and $>0.286$ (GL + A115$^{\prime}$). Except for the A115 data, non-flat \pcdm\ constraints favor closed hypersurfaces (unlike non-flat \lcdm\ and non-flat XCDM), but with flatness well within $1\sigma$ for all cases.

The $\Delta AIC$ and $\Delta BIC$ values with respect to the flat \lcdm\ model are listed in the last two columns of Table \ref{tab:BFP2C7}. In all cases (except for the GL case, which is discussed in Sec.\ \ref{subsec:MLSGL} above), the flat \lcdm\ model is the most favored model but the evidence against the other models are either weak or positive, except that, based on $BIC$, the evidence against non-flat XCDM and non-flat \pcdm\ are strong. For ML + A115 data the non-flat \pcdm\ model is very strongly disfavored with $\Delta BIC=10.03$.

In summary, while the joint analyses do slightly tighten the constraints, the improvements relative to those from A118 data alone are not significant.

\section{Conclusion}
\label{makereference7.5}

We have used six different cosmological models in analyses of the three (ML, MS, and GL) Dainotti ($L_0-t_b$) correlation GRB data sets compiled by \cite{Wangetal_2021} and \cite{Huetal2021}. We find for each data sets, as well as the MS + GL, ML + GL, and ML + MS combinations, that the GRB correlation parameters are independent of cosmological model. Our results thus indicate that these GRBs are standardizable through the Dainotti correlation and so can be used to constrain cosmological parameters, justifying the assumption made by  \cite{Wangetal_2021} and \cite{Huetal2021}. These results also mean that the circularity problem does not affect cosmological parameter constraints derived from these GRB data.

In contrast to \cite{Wangetal_2021} and \cite{Huetal2021} we do not use $H(z)$ data to calibrate these GRB data, instead we use these data to derive GRB only cosmological constraints. We find that ML, MS, GL, MS + GL, ML + GL, and ML + MS GRBs provide only weak restrictions on cosmological parameters. 

We have also used the more-restrictive  ML and GL Dainotti data sets in joint analyses with the largest available reliable compilation of Amati ($E_{\rm p}-E_{\rm iso}$) correlation A118 GRB data \citep{Khadkaetal_2021b}, but excluding three overlapping GRBs from the A118 data in the joint analyses. While the joint analyses do result in slightly tighter constraints, typically with larger lower limits on \om\ than those from the ML, GL, or A118 data alone, the improvements relative to the A118 data constraints are not significant.

Current GRB data provide quite weak constraints on cosmological parameters but do favor currently accelerated cosmological expansion in the \lcdm\ models and the XCDM parametrizations. We hope that in the near future there will be more and better-quality GRB measurements that will result in more restrictive GRB cosmological constraints. GRBs probe a very wide range of cosmological redshift space, a significant part of which is as yet unprobed, so it is worth putting effort into further developing GRB cosmological constraints. 


\cleardoublepage


\chapter{Standardizing Platinum Dainotti-correlated gamma-ray bursts, and using them with standardized Amati-correlated gamma-ray bursts to constrain cosmological model parameters}
\label{makereference8}

This chapter is based on \cite{CaoDainottiRatra2022}.

\section{Introduction} 
\label{makereference8.1}

Currently accelerated cosmological expansion and other cosmological observations are reasonably well accommodated in the spatially-flat \lcdm\ model \citep{peeb84} with $\sim 70\%$ of the current cosmological energy budget being a time-independent cosmological constant ($\Lambda$), $\sim 25\%$ being non-relativistic cold dark matter (CDM), and most of the remaining $\sim 5\%$ being non-relativistic baryonic matter \citep[see, e.g.][]{Farooq_Ranjeet_Crandall_Ratra_2017, scolnic_et_al_2018, planck2018b, eBOSS_2020}. In this paper we also study cosmological models with a little spatial curvature or dynamical dark energy, since the observations do not rule out such models, and since some sets of measurements seem mutually incompatible when analyzed in the spatially-flat \lcdm\ model \citep[see, e.g.][]{DiValentinoetal2021a, PerivolaropoulosSkara2021}. 

It is still unclear whether this incompatibility is evidence against the spatially-flat \lcdm\ model or is caused by unidentified systematic errors in one of the established cosmological probes or by evolution of the parameters themselves with the redshift \citep{Dainottietal2021a, Dainottietal2022}. Newer, alternate cosmological probes could help alleviate this issue. Recent examples of such probes include reverberation-mapped quasar (QSO) measurements that reach to redshift $z \sim 1.9$ \citep{Czernyetal2021, Zajaceketal2021, Yuetal2021, Khadkaetal_2021a, Khadkaetal2022a}, \hii\ starburst galaxy measurements that reach to $z \sim 2.4$ \citep{Mania_2012, Chavez_2014, GonzalezMoran2019, GonzalezMoranetal2021, CaoRyanRatra2020, CaoRyanRatra2022, Caoetal_2021, Johnsonetal2022, Mehrabietal2022}, QSO angular size measurements that reach to $z \sim 2.7$ \citep{Cao_et_al2017a, Ryanetal2019, CaoRyanRatra2020, Caoetal_2021, Zhengetal2021, Lian_etal_2021}, QSO flux measurements that reach to $z \sim 7.5$ \citep{RisalitiLusso2015, RisalitiLusso2019, KhadkaRatra2020a, KhadkaRatra2020b, KhadkaRatra2021, KhadkaRatra2022, Lussoetal2020, Yangetal2020, ZhaoXia2021, Lietal2021, Lian_etal_2021, Rezaeietal2022, Luongoetal2021},\footnote{The most recent \cite{Lussoetal2020} QSO flux compilation assumes a UV--X-ray correlation model that is invalid above a significantly lower redshift, $z \sim 1.5-1.7$, so these QSOs can only be used to derive lower-$z$ cosmological constraints \citep{KhadkaRatra2021, KhadkaRatra2022}.} and the main subject of this paper, gamma-ray burst (GRB) measurements that reach to $z \sim 8.2$ \citep{Amati2008, Amati2019, CardoneCapozzielloDainotti2009, Cardoneetal2010, samushia_ratra_2010, Dainottietal2013a, Dainottietal2013b, Postnikovetal2014, Wangetal2015, Wang_2016, Wangetal_2021, Dirirsa2019, KhadkaRatra2020c, Huetal2021, Daietal_2021, Demianskietal_2021, Khadkaetal_2021b, Luongoetal2021, LuongoMuccino2021, Caoetal_2021}. Some of these probes might eventually allow for a reliable extension of the Hubble diagram to $z \sim 3-4$, well beyond the reach of Type Ia supernovae. GRBs have been detected to $z \sim 9.4$ \citep{Cucchiaraetal2011}, and might be detectable to $z = 20$ \citep{Lamb2000}, so in principle GRBs could act as a cosmological probe to higher redshifts than 8.2.

\citet{CaoKhadkaRatra2022} recently used the A118 two-parameter Amati-correlated GRB data set \citep{Khadkaetal_2021b} and the long GRB  whose plateau phase is dominated by magnetic dipole radiation (MD-LGRB) and gravational wave emission (GW-LGRB), and short GRB whose plateau phase is dominated by magnetic dipole radiation (MD-SGRB) two-parameter Dainotti-correlated GRB data sets \citep{Wangetal_2021, Huetal2021} to constrain cosmological parameters.\footnote{The Amati correlation relates the rest-frame peak photon energy and the rest-frame isotropic radiated energy \citep{Amati2008} and the two-dimensional Dainotti correlation relates the luminosity at the end of the plateau phase and the rest-frame end-time of plateau emission \citep{Dainottietal2008, Dainottietal2010, Dainottietal2011, Dainottietal2013a, Dainottietal2015, Dainottietal2017}. As discussed below, in this paper we use three-dimensional Dainotti and two-dimensional Amati correlation GRBs.} The circularity problem was circumvented by simultaneously constraining cosmological model parameters and GRB correlation parameters (see \citealp{Dainottietal2013b} for a more extended discussion) and the cosmological-model-independence of the GRB correlation parameters shows that these GRBs are standardizable \citep{KhadkaRatra2020c, CaoKhadkaRatra2022}. Not only does the simultaneous fitting method circumvent the circularity problem, it also allows for the derivation of unbiased GRB-only constraints (unlike the constraints derived from GRBs that have been calibrated by using other data, that are correlated with the calibrating data), that can be straightforwardly used to compare with other constraints derived from other data, such as $H(z)$ + BAO data as we have done here. The A118, MD-LGRB, GW-LGRB, and MD-SGRB GRBs provide cosmological constraints that are mostly compatible with those determined using better-established cosmological probes \citep{CaoKhadkaRatra2022}.

Here, we use the new Platinum compilation of 50 long GRBs, spanning $0.553 \leq z \leq 5.0$, that obey the three-parameter fundamental plane (Dainotti) correlation between the peak prompt luminosity, the luminosity at the end of the plateau emission and its rest frame duration \citep{Dainottietal2016,Dainottietal2017,Dainottietal2020} to constrain cosmological model parameters and GRB correlation parameters. The platinum sample is listed in Table \ref{tab:P50C8}. Note, however, that this table is incorrect. The correct table should be Table \ref{tab:P50aB} in Appendix \ref{AppendixB}, where we reanalyzed the Platinum constraints in \cite{CaoDainottiRatra2022b} and found that the differences are not significant. For this data set, measured quantities for a GRB are redshift $z$, characteristic time scale $T^{*}_{X}$, which marks the end of the plateau emission, the measured $\gamma$-ray energy flux $F_{X}$ at $T^{*}_{X}$ and the prompt peak flux $F_{\rm peak}$ over a 1 s interval, and X-ray spectral index of the plateau phase $\beta^{\prime}$. This sample spans the redshift range $0.553 \leq z \leq 5.0$. We find that the Platinum GRBs are standardizable through the Dainotti correlation and they also provides cosmological parameter constraints compatible with those from better-established cosmological probes, as well as with those derived from A118 GRB data. We also combine the Platinum data set with the 101 non-overlapping Amati-correlated GRBs (A101) from the A118 data set to perform a joint (Platinum + A101) analysis. We find that this joint GRB data set provides slightly more restrictive cosmological constraints (in which the twice-as-large A101 data set dominates the statistics and so plays a more dominant role) that are consistent with those from a combined analysis of baryon acoustic oscillation (BAO) and Hubble parameter [$H(z)$] data. However, the cosmological constraints from Platinum, A118, A101, and Platinum + A101 data are less restrictive (precise) than those from $H(z)$ + BAO data because there are more parameters to be constrained in the GRB cases but not yet enough precise-enough data points to determine more precise GRB constraints. A joint analysis of the $H(z)$ + BAO + Platinum + A101 data results in slightly more restrictive cosmological constraints relative to those from just $H(z)$ + BAO data.

Our paper is organized as follows. We use the cosmological models described in Chapter \ref{sec:models} and outline the data sets adopted in Sec.\ \ref{makereference8.2}. We then describe our analyses methods in Sec.\ \ref{makereference8.3} and discuss results in Sec.\ \ref{makereference8.4}. We summarize our conclusions in Sec.\ \ref{makereference8.5}.

\section{Data}
\label{makereference8.2}

In this paper we analyze two different GRB data sets as well as combinations of them and the joint $H(z)$ + BAO data set. These data sets are summarized in Table \ref{tab:dataC8} and described below.\footnote{In this table and elsewhere, for compactness, we sometimes use Plat. as an abbreviation for the Platinum data set.}

\begin{itemize}

\item[]{\bf Platinum sample}. This includes 50 long GRBs which exhibit a plateau phase with an angle $< 41^\circ$, that do not have a flare during the plateau, and have a plateau with duration longer than 500 s. The first criterion is based on evidence that the plateau angles are Gaussianly distributed and those with angle $> 41^\circ$ are outliers; the second criterion allows one to eliminate cases contaminated by the presence of flaring activity; and, the third criterion allows one to eliminate cases where prompt emission may mask the plateau to the point that the definition of the plateau is uncertain \citep{Willingaleetal2007, Willingaleetal2010}. As discussed below, the Platinum GRBs obey the three-dimensional Dainotti relation. The correct platinum sample is listed in Table \ref{tab:P50aB}. For this data set, measured quantities for a GRB are redshift $z$, characteristic time scale $T^{*}_{X}$, which marks the end of the plateau emission, the measured $\gamma$-ray energy flux $F_{X}$ at $T^{*}_{X}$ and the prompt peak flux $F_{\rm peak}$ over a 1 s interval, and X-ray spectral index of the plateau phase $\beta^{\prime}$. This sample spans the redshift range $0.553 \leq z \leq 5.0$.

\item[]{\bf A118 sample}. This sample include 118 long GRBs, listed in Table 7 of \cite{Khadkaetal_2021b}, that obey the two-dimensional Amati relation. For this data set, measured quantities for a GRB are $z$, rest-frame spectral peak energy $E_{\rm p}$, and measured bolometric fluence $S_{\rm bolo}$, computed in the standard rest-frame energy band $1-10^4$ keV. This sample spans the redshift range $0.3399 \leq z \leq 8.2$. Note that in our analyses here we did not use the correct value of $E_{\rm p}=871\pm123$ keV for GRB081121, as discussed in Ref.\ \cite{Liuetal2022}, although the effects on the parameter constraints are small.

\item[]{\bf A101 sample}. The A118 data and the Platinum data have 17 common GRBs that are listed in the footnote of Table \ref{tab:dataC8}. We exclude these common GRBs from the A118 data set to form the A101 data set for joint analyses with the Platinum data set. This sample spans the redshift range $0.3399 \leq z \leq 8.2$.

\item[]{\bf Platinum + A101 sample}. This combination GRB sample includes 151 GRBs. This sample spans the redshift range $0.3399 \leq z \leq 8.2$.

\item[]{$\textbf{ \emph{H(z)}}$ \bf and BAO data}. There are 31 $H(z)$ and 11 BAO measurements that have a redshift range $0.07 \leq z \leq 1.965$ and $0.0106 \leq z \leq 2.33$, respectively. The $H(z)$ data are in Table 2 of \cite{Ryan_1} and the BAO data are in Table 1 of \cite{CaoRyanRatra2021}. We compare cosmological constraints from $H(z)$ + BAO data with those obtained from the GRB data sets, and also jointly analyze GRB and $H(z)$ + BAO data.

\end{itemize}

\begin{table}
\centering
\begin{threeparttable}
\caption{Summary of data sets used.}
\label{tab:dataC8}
\setlength{\tabcolsep}{3.5pt}
\begin{tabular}{lcc}
\toprule
Data set & $N$ (Number of points) & Redshift range\\
\midrule
Platinum & 50 & $0.553 \leq z \leq 5.0$ \\
A118 & 118 & $0.3399 \leq z \leq 8.2$ \\
A101\tnote{a} & 101 & $0.3399 \leq z \leq 8.2$ \\
Plat. + A101 & 151 & $0.3399 \leq z \leq 8.2$ \\
\midrule
$H(z)$ & 31 & $0.070 \leq z \leq 1.965$ \\
BAO & 11 & $0.38 \leq z \leq 2.334$ \\
\bottomrule
\end{tabular}
\begin{tablenotes}[flushleft]
\item [a] Excluding from A118 those GRBs in common with Platinum [060418, 080721, 081008, 090418(A), 091020, 091029, 110213(A), 110818(A), 111008(A), 120811C, 120922(A), 121128(A), 131030A, 131105A, 140206A, 150314A, and 150403A].
\end{tablenotes}
\end{threeparttable}%
\end{table}

\begin{table*}
\centering
\resizebox*{\columnwidth}{1.3\columnwidth}{%
\begin{threeparttable}
\caption{50 Platinum GRB samples, which are incorrect. The correct ones are listed in Table \ref{tab:P50aB} of Appendix \ref{AppendixB}.}
\label{tab:P50C8}
\begin{tabular}{lccccc}
\toprule
GRB & $z$ & $\log T^{*}_{X}\ (\mathrm{s})$ & $\log F_{X}\ (\mathrm{erg\ cm}^{-2}\ \mathrm{s}^{-1})$ & $\beta^{\prime}$ & $F_{\rm peak}\ (10^{-8}\ \mathrm{erg\ cm}^{-2}\ \mathrm{s}^{-1})$ \\
\midrule

060418 & 1.49 & $3.11887^{+0.05725}_{-0.05724}$ & $-9.79296\pm0.04904$ & 1.98 & $49.9\pm1.63$\\
060605 & 3.8 & $4.04162^{+0.03738}_{-0.03737}$ & $-11.2609^{+0.0531}_{-0.0532}$ & 1.835 & $4.73\pm0.693$\\
060708 & 1.92 & $3.50569^{+0.06273}_{-0.06272}$ & $-10.8688^{+0.0611}_{-0.0612}$ & 2.485 & $6.89\pm0.796$\\
060714 & 2.71 & $3.73449^{+0.04676}_{-0.04677}$ & $-10.8586\pm0.0352$ & 1.87 & $9.13\pm0.549$\\
060814 & 0.84 & $4.22387^{+0.03485}_{-0.03486}$ & $-10.9108\pm0.0353$ & 1.97 & $60.6\pm1.48$\\
060906 & 3.685 & $4.33479^{+0.07214}_{-0.07215}$ & $-11.8833^{+0.0885}_{-0.0884}$ & 2.06 & $12.2\pm1.18$\\
061121 & 1.314 & $3.78753\pm0.01703$ & $-10.0299^{+0.0126}_{-0.0127}$ & 1.9485 & $196\pm2.43$\\
061222A & 2.088 & $3.92091^{+0.02444}_{-0.02443}$ & $-9.95369\pm0.02311$ & 1.86 & $73.3\pm1.51$\\
070110 & 2.352 & $4.26825^{+0.05399}_{-0.05400}$ & $-11.0488^{+0.0547}_{-0.0548}$ & 2.186 & $4.73\pm0.648$\\
070306 & 1.4959 & $4.86926\pm0.02824$ & $-11.2737\pm0.0335$ & 1.831 & $20.1\pm0.899$\\
070508 & 0.82 & $3.02525\pm0.01189$ & $-9.16118^{+0.01080}_{-0.01079}$ & 1.764 & $224\pm3.07$\\
070521 & 0.553 & $3.55243\pm0.04844$ & $-10.01250^{+0.04902}_{-0.04900}$ & 1.98 & $8.71\pm1.2$\\
070529 & 2.4996 & $3.09323^{+0.06805}_{-0.06804}$ & $-10.2468\pm0.0552$ & 1.76 & $11.1\pm1.84$\\
080310 & 2.4266 & $4.326080559^{+0.037986760}_{-0.037986759}$ & $-11.39909559\pm0.04247343$ & 1.878 & $7.52\pm0.893$\\
080430 & 0.767 & $4.217629106^{+0.038388523}_{-0.038388522}$ & $-11.17223461^{+0.02818096}_{-0.02818097}$ & 4.18 & $18.2\pm0.808$\\
080721 & 2.6 & $2.878762860^{+0.007554815}_{-0.007554816}$ & $-8.713485538^{+0.007019298}_{-0.007019299}$ & 1.735 & $193\pm10.3$\\
081008 & 1.967 & $3.855508334\pm0.056285684$ & $-10.79325557\pm0.06052989$ & 1.84 & $10.5\pm0.813$\\
081221 & 2.26 & $2.931957486\pm0.031176651$ & $-9.354590372\pm0.024985500$ & 1.991 & $147\pm2.46$\\
090418A & 1.608 & $3.528136567^{+0.030759021}_{-0.030759020}$ & $-10.04734554^{+0.02865941}_{-0.02865940}$ & 1.98 & $15.8\pm1.66$\\
091018 & 0.971 & $2.896555345^{+0.034497877}_{-0.034497878}$ & $-9.626765539^{+0.024288604}_{-0.024288603}$ & 1.91 & $59.1\pm1.22$\\
091020 & 1.71 & $2.952650108\pm0.045271255$ & $-9.712875280\pm0.034480687$ & 1.895 & $36.4\pm1.71$\\
091029 & 2.752 & $4.303689975^{+0.034224003}_{-0.034224002}$ & $-11.34682089\pm0.02563352$ & 2.064 & $10.9\pm0.786$\\
100219A & 4.7 & $4.721899099\pm0.103773463$ & $-12.13530676^{+0.32639562}_{-0.32639563}$ & 1.46 & $3.15\pm0.721$\\
110213A & 1.46 & $3.84089^{+0.02576}_{-0.02127}$ & $-9.89698^{+0.03865}_{-0.03311}$ & 2.1905 & $7.15\pm1.88$\\
110818A & 3.36 & $3.846274153\pm0.059648416$ & $-11.28292779\pm0.05514905$ & 1.83 & $14.1\pm1.5$\\
111008A & 5 & $3.989933926\pm0.035676696$ & $-10.66765993^{+0.02851108}_{-0.02851109}$ & 1.829 & $54.1\pm3.91$\\
120118B & 2.943 & $3.68307^{+0.15437}_{-0.10916}$ & $-10.9560^{+0.1224}_{-0.1120}$ & 2.01 & $13.8\pm1.32$\\
120404A & 2.88 & $3.79948^{+0.09188}_{-0.04469}$ & $-11.0269^{+0.0927}_{-0.1414}$ & 1.69 & $7.62\pm0.109$\\
120811C & 2.67 & $3.159095907^{+0.084200302}_{-0.084200301}$ & $-10.14322956\pm0.05294686$ & 1.65 & $26\pm1.1$\\
120922A & 3.1 & $3.59799^{+0.21328}_{-0.05347}$ & $-10.5957^{+0.1158}_{-0.2115}$ & 2.17 & $10.6\pm0.757$\\
121128A & 2.2 & $3.233929220^{+0.029795567}_{-0.029795568}$ & $-9.618411717^{+0.028690024}_{-0.028690023}$ & 1.9455 & $104\pm2.1$\\
131030A & 1.29 & $2.84594^{+0.02758}_{-0.02757}$ & $-9.29414^{+0.02517}_{-0.02518}$ & 1.693 & $265\pm4.61$\\
131105A & 1.686 & $3.94880^{+0.04788}_{-0.04789}$ & $-10.9111^{+0.0333}_{-0.0334}$ & 1.94 & $26.8\pm0.204$\\
140206A & 2.7 & $3.59073^{+0.02132}_{-0.02133}$ & $-9.72980^{+0.01300}_{-0.01299}$ & 1.672 & $169\pm2.62$\\
140419A & 3.956 & $3.68383\pm0.02442$ & $-10.00560^{+0.02397}_{-0.02400}$ & 1.678 & $41.8\pm1.28$\\
140506A & 0.889 & $3.30686\pm0.06882$ & $-9.90438\pm0.05516$ & 1.9 & $86.7\pm4.56$\\
140509A & 2.4 & $3.58709^{+0.11327}_{-0.06416}$ & $-11.0977^{+0.0897}_{-0.0782}$ & 1.86 & $11.8\pm1.77$\\
140629A & 2.3 & $2.85608\pm0.06161$ & $-9.84891^{+0.04278}_{-0.04279}$ & 1.815 & $35\pm1.87$\\
150314A & 1.758 & $2.49217^{+0.01586}_{-0.01585}$ & $-8.60990^{+0.01503}_{-0.01502}$ & 1.73 & $381\pm6.06$\\
150403A & 2.06 & $3.19694^{+0.00698}_{-0.00697}$ & $-8.82118^{+0.00510}_{-0.00511}$ & 1.679 & $171\pm3.81$\\
150910A & 1.36 & $3.85419\pm0.02868$ & $-9.97065^{+1.37578}_{-0.03215}$ & 1.6645 & $8.28\pm2.18$\\
151027A & 0.81 & $3.99907\pm0.01731$ & $-9.93519\pm0.02221$ & 1.9235 & $58.2\pm2.88$\\
160121A & 1.96 & $3.82440^{+0.12038}_{-0.12039}$ & $-11.1726\pm0.0694$ & 2.02 & $7.79\pm0.823$\\
160227A & 2.38 & $4.47751\pm0.03885$ & $-10.9957^{+0.0305}_{-0.0306}$ & 1.679 & $4.88\pm0.688$\\
160327A & 4.99 & $3.76407\pm0.06033$ & $-11.2238\pm0.0673$ & 1.78 & $12.5\pm0.877$\\
170202A & 3.645 & $3.77629^{+0.06866}_{-0.06865}$ & $-10.6535^{+0.0449}_{-0.0448}$ & 2.04 & $39.2\pm1.79$\\
170705A & 2.01 & $3.63746\pm0.06603$ & $-10.1867^{+0.0406}_{-0.0405}$ & 1.66 & $113\pm2.41$\\
180329B & 1.998 & $4.02296\pm0.04314$ & $-11.1493^{+0.0483}_{-0.0482}$ & 1.77 & $9.09\pm2.02$\\
190106A & 1.86 & $4.19124^{+0.03109}_{-0.03110}$ & $-10.4325^{+0.0285}_{-0.0286}$ & 2.9365 & $33.6\pm1.3$\\
190114A & 3.37 & $3.74534^{+0.04221}_{-0.04222}$ & $-10.7527^{+0.0336}_{-0.0337}$ & 1.86 & $4.12\pm0.953$\\
\bottomrule
\end{tabular}
\end{threeparttable}%
}
\end{table*}

\section{Data Analysis Methodology}
\label{makereference8.3}

\begin{table}
\centering
\begin{threeparttable}
\caption{Flat priors of the constrained parameters.}
\label{tab:priorsC8}
\setlength{\tabcolsep}{3.5pt}
\begin{tabular}{lcc}
\toprule
Parameter & & Prior\\
\midrule
 & Cosmological Parameters & \\
\midrule
$H_0$\tnote{a} &  & [None, None]\\
\obhs\,\tnote{b} &  & [0, 1]\\
\ochs\,\tnote{c} &  & [0, 1]\\
\ok &  & [-2, 2]\\
$\alpha$ &  & [0, 10]\\
\wx &  & [-5, 0.33]\\
\midrule
 & GRB Correlation Parameters & \\
\midrule
$a$ &  & [-5, 5]\\
$b$ &  & [-5, 5]\\
$C_{o}$ &  & [-50, 50]\\
$\sigma_{\rm int}$ &  & [0, 5]\\
$\beta$ &  & [0, 5]\\
$\gamma$ &  & [0, 300]\\
\bottomrule
\end{tabular}
\begin{tablenotes}[flushleft]
\item [a] \hunit. In all four GRB-only analyses, $H_0$ is set to be 70 \hunit, while in other cases the prior range is irrelevant (unbounded).
\item [b] In all four GRB-only analyses, \obhs\ is set to be 0.0245, i.e. $\Omega_{b}=0.05$.
\item [c] In all four GRB-only analyses, the $\Omega_{c}$ range is adjusted to ensure $\Om\in[0,1]$.
\end{tablenotes}
\end{threeparttable}%
\end{table}

Luminosity distance, $D_L$, as a function of $z$ and cosmological parameters $\textbf{\emph{p}}$, is given by
\begin{equation}
  \label{eq:DLC8}
D_L(z, \textbf{\emph{p}}) = 
    \begin{cases}
    \frac{c(1+z)}{H_0\sqrt{\Omega_{\rm k0}}}\sinh\left[\frac{\sqrt{\Omega_{\rm k0}}H_0}{c}D_C(z, \textbf{\emph{p}})\right] & \text{if}\ \Omega_{\rm k0} > 0, \\
    \vspace{1mm}
    (1+z)D_C(z, \textbf{\emph{p}}) & \text{if}\ \Omega_{\rm k0} = 0,\\
    \vspace{1mm}
    \frac{c(1+z)}{H_0\sqrt{|\Omega_{\rm k0}|}}\sin\left[\frac{H_0\sqrt{|\Omega_{\rm k0}|}}{c}D_C(z, \textbf{\emph{p}})\right] & \text{if}\ \Omega_{\rm k0} < 0,
    \end{cases}
\end{equation}
where the comoving distance is in equation \eqref{eq:DC}, and $H(z, \textbf{\emph{p}})$ is the Hubble parameter that is described in Chapter \ref{sec:models} for each cosmological model.

For Platinum GRBs the X-ray source rest-frame luminosity $L_{X}$, time $T^{*}_{X}$ at the end of the plateau emission, and the peak prompt luminosity $L_{\rm peak}$ are correlated through the three-parameter fundamental plane relation \citep{Dainottietal2016, Dainottietal2017, Dainottietal2020, Dainottietal2021} 
\begin{equation}
    \label{eq:3DC8}
    \log L_{X} = C_{o}  + a\log T^{*}_{X} + b\log L_{\rm peak},
\end{equation}
where 
\be
\label{eq:LxC8}
    L_{X}=\frac{4\pi D_L^2}{(1+z)^{1-\beta^{\prime}}}F_{X},
\ee
\be
\label{eq:LpeakC8}
    L_{\rm peak}=\frac{4\pi D_L^2}{(1+z)^{1-\beta^{\prime}}}F_{\rm peak},
\ee
$C_{o}$ is the intercept parameter, and $a$ and $b$ are the slope parameters, with all three to be determined from data. $F_{X}$ and $F_{\rm peak}$ are the measured $\gamma$-ray energy flux (erg cm$^{-2}$ s$^{-1}$) at $T^{*}_{X}$ and in the peak of the prompt emission over a 1 s interval, respectively, and $\beta^{\prime}$ is the X-ray spectral index of the plateau phase. Note, however, that these expressions are incorrect, where the corrected expressions are described in Chapter \ref{makereference9} from \cite{CaoDainottiRatra2022b}.

We compute $L_{X}$ and $L_{\rm peak}$ as functions of cosmological parameters $\textbf{\emph{p}}$ at the redshift of each GRB by using eqs.\ \eqref{eq:DL}, \eqref{eq:LxC8}, and \eqref{eq:LpeakC8}. We then compute the natural log of the likelihood function \citep{D'Agostini_2005}
\be
\label{eq:LH_GRBC8}
    \ln\mathcal{L}_{\rm GRB}= -\frac{1}{2}\Bigg[\chi^2_{\rm GRB}+\sum^{N}_{i=1}\ln\left(2\pi\sigma^2_{\mathrm{tot},i}\right)\Bigg],
\ee
where
\be
\label{eq:chi2_GRBC8}
    \chi^2_{\rm GRB} = \sum^{N}_{i=1}\bigg[\frac{(\log L_{X,i} - C_{o}  - a\log T^{*}_{X,i} - b\log L_{\mathrm{peak},i})^2}{\sigma^2_{\mathrm{tot},i}}\bigg],
\ee
with
\be
\sigma^2_{\mathrm{tot},i}=\sigma_{\rm int,\,\textsc{p}}^2+\sigma_{{\log L_{X,i}}}^2+a^2\sigma_{{\log T^{*}_{X,i}}}^2+b^2\sigma_{{\log L_{\mathrm{peak},i}}}^2.
\ee
where $\sigma_{\rm int,\,\textsc{p}}$ is the Platinum GRB data intrinsic scatter parameter, that also contains the unknown systematic uncertainty, and $N$ is the number of data points.

For GRBs which obey the Amati correlation, a detailed description of the procedure can be found in Sec.\ 4 of \cite{CaoKhadkaRatra2022}. Here we denote its intrinsic scatter parameter as $\sigma_{\rm int,\,\textsc{a}}$ as opposed to the Platinum one, $\sigma_{\rm int,\,\textsc{p}}$.

Detailed descriptions of the $H(z)$ + BAO data analysis procedure can be found in Sec.\ 4 of \cite{CaoRyanRatra2020} and \cite{CaoRyanRatra2021}.

The flat priors of the free parameters are listed in Table \ref{tab:priorsC8}. The best-fitting values and posterior distributions of all free parameters are obtained through maximizing the likelihood functions using the Markov chain Monte Carlo (MCMC) code \textsc{MontePython} \citep{Brinckmann2019}, with the physics coded in \textsc{class}. The convergence of the MCMC chains for each free parameter is guaranteed by the Gelman-Rubin criterion ($R-1 < 0.05$). The \textsc{python} package \textsc{getdist} \citep{Lewis_2019} is used to compute the posterior means and uncertainties and plot the marginalized likelihood distributions and contours.

We use the Akaike Information Criterion ($AIC$) and the Bayesian Information Criterion ($BIC$) to compare the goodness of fit of models with different numbers of parameters. Their definitions can be found in \cite{CaoKhadkaRatra2022}. We also compare the goodness of fit using the deviance information criterion ($DIC$) \citep{KunzTrottaParkinson2006,Amati2019} defined as
\be
\label{eq:DIC}
DIC=-2\ln \mathcal{L}_{\rm max} + 2n_{\rm eff},
\ee
where $n_{\rm eff}=\langle-2\ln \mathcal{L}\rangle+2\ln \mathcal{L}_{\rm max}$ is the number of effectively constrained parameters with brackets representing the average over the posterior distribution. We compute $\Delta AIC$, $\Delta BIC$, and $\Delta DIC$ differences of the other five cosmological models with respect to the flat \lcdm\ reference model. Positive (negative) values of $\Delta AIC$, $\Delta BIC$, or $\Delta DIC$ indicate that the model under study fits the data worse (better) than does the reference model. Relative to the model with minimum $AIC(BIC/DIC)$, $\Delta AIC(BIC/DIC) \in (0, 2]$ is said to be weak evidence against the candidate model, $\Delta AIC(BIC/DIC) \in (2, 6]$ is positive evidence against the candidate model, while $\Delta AIC(BIC/DIC) \in (6, 10] $ is strong evidence against the candidate model, with $\Delta AIC(BIC/DIC)>10$ being very strong evidence against the candidate model.

Here we consider one massive and two massless neutrino species, with the effective number of relativistic neutrino species $N_{\rm eff} = 3.046$ and the total neutrino mass $\sum m_{\nu}=0.06$ eV. The non-relativistic neutrino physical energy density parameter is $\onh=\sum m_{\nu}/(93.14\ \rm eV)$, where $h$ is the reduced Hubble constant in units of 100 \hunit. With the baryonic (\obhs) and cold dark matter (\ochs) physical energy density parameters as free cosmological parameters to be constrained, the derived non-relativistic matter density parameter is therefore $\Om = (\onh + \obh + \och)/{h^2}$.

\section{Results}
\label{makereference8.4}

\begin{figure*}
\centering
 \subfloat[]{%
    \includegraphics[width=0.5\textwidth,height=0.5\textwidth]{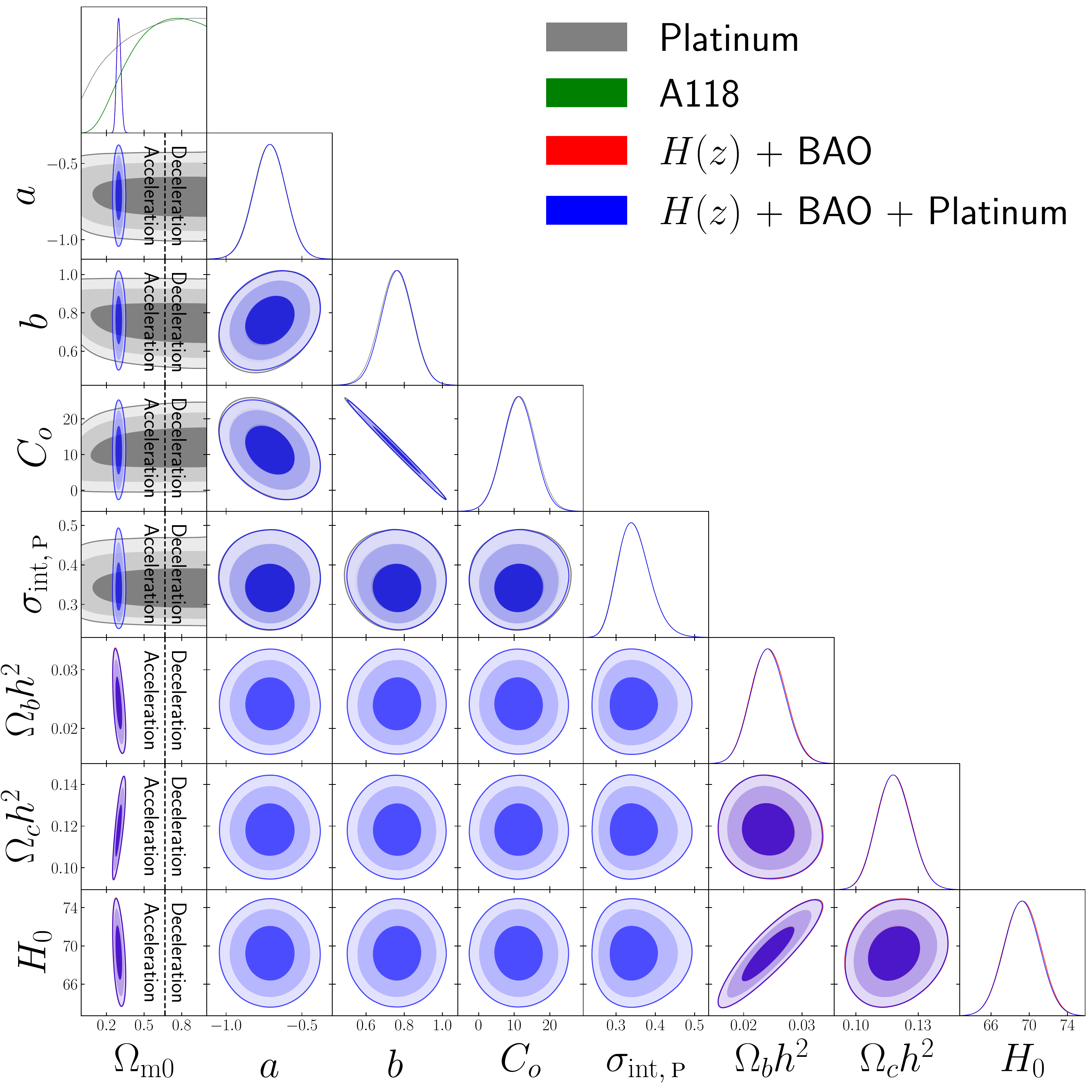}}
 \subfloat[]{%
    \includegraphics[width=0.5\textwidth,height=0.5\textwidth]{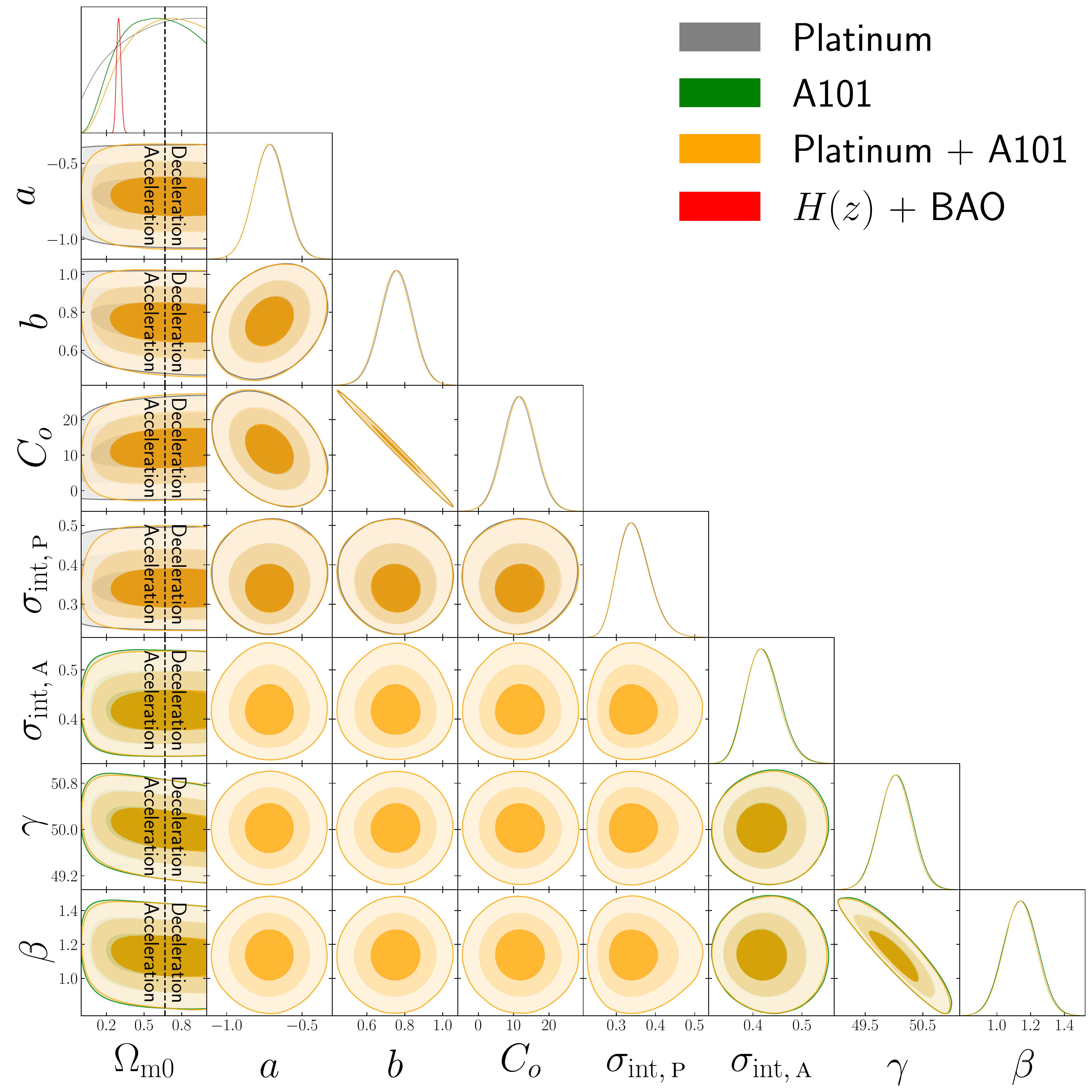}}\\
 \subfloat[]{%
    \includegraphics[width=0.5\textwidth,height=0.5\textwidth]{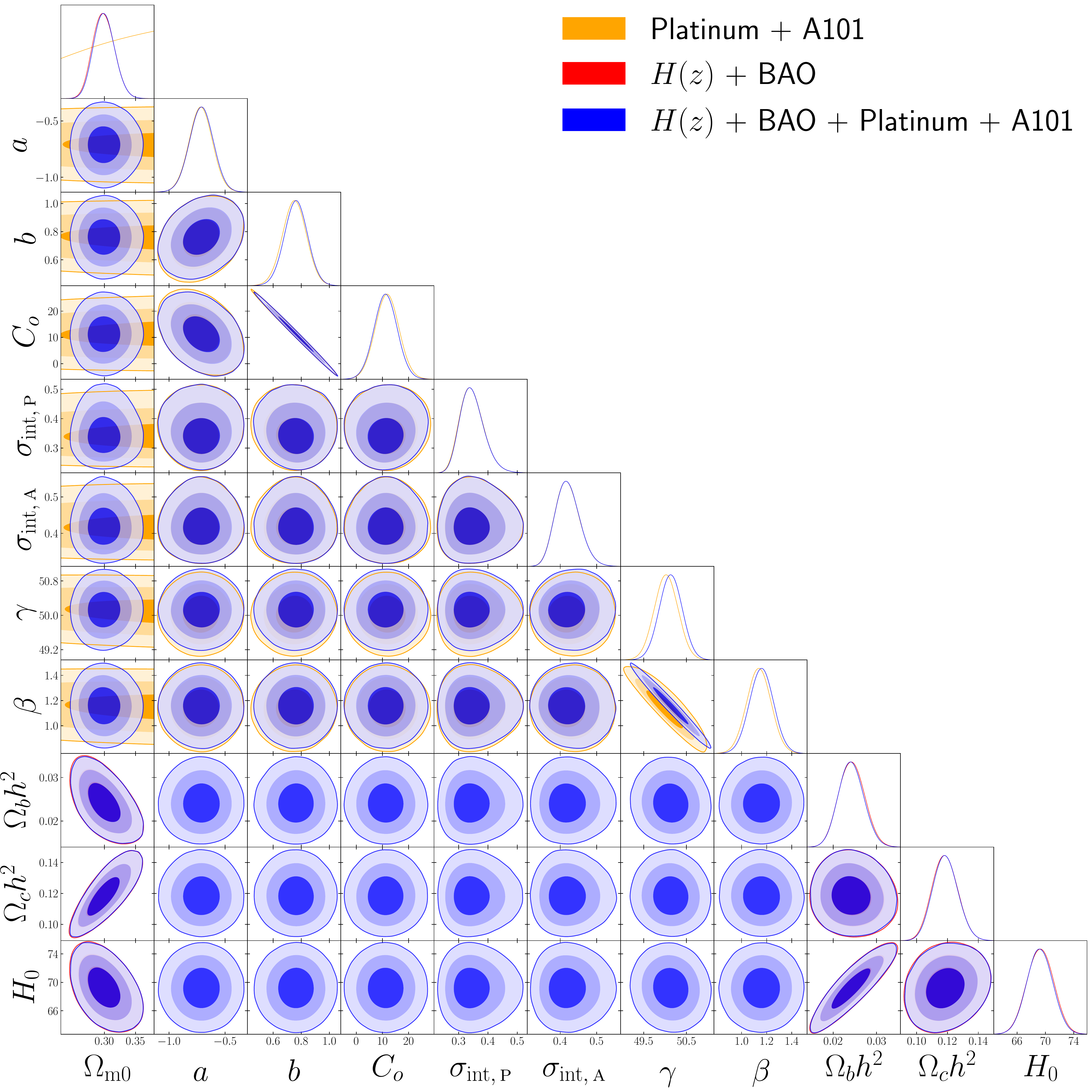}}
 \subfloat[Cosmological parameters zoom in]{%
    \includegraphics[width=0.5\textwidth,height=0.5\textwidth]{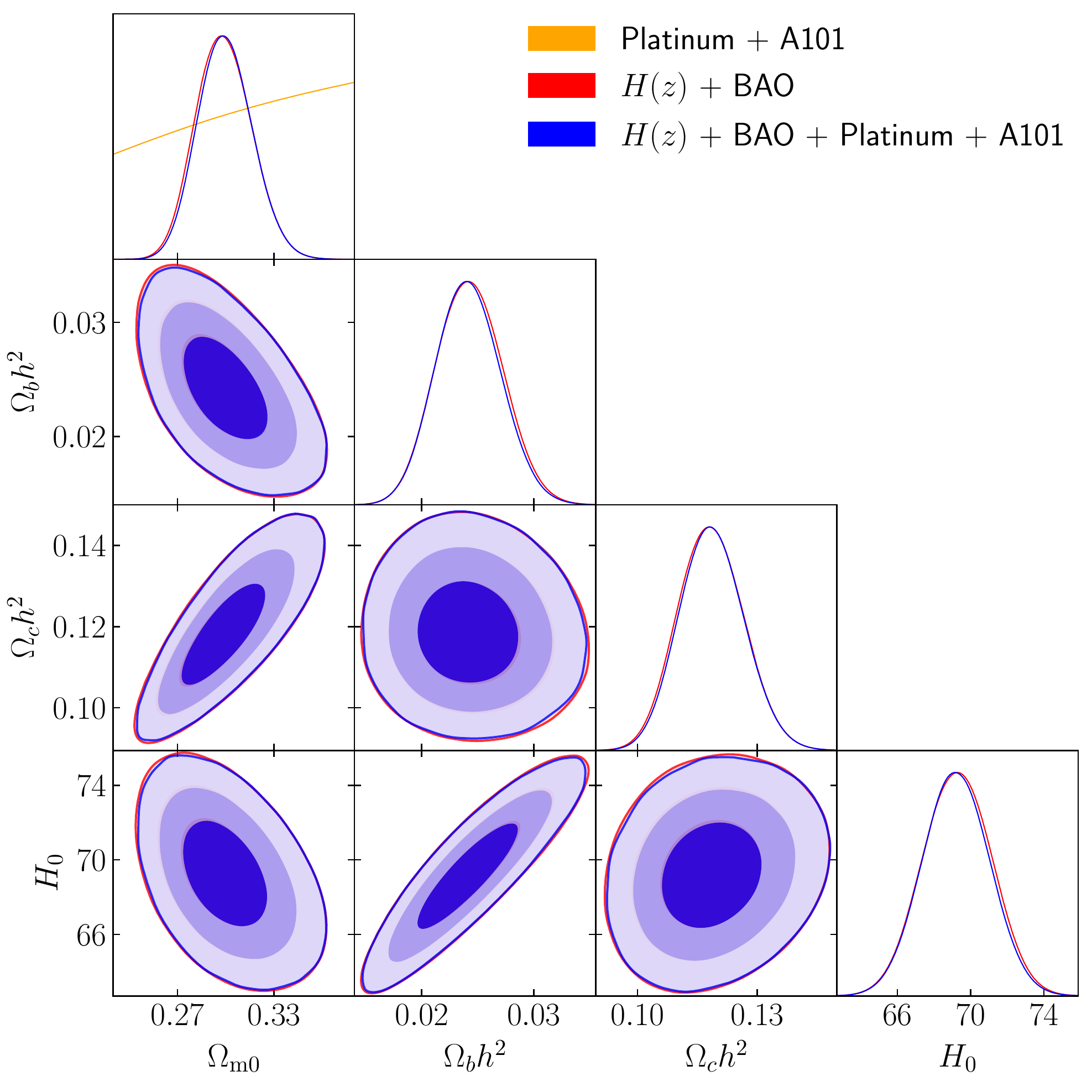}}\\
\caption{One-dimensional likelihood distributions and 1$\sigma$, 2$\sigma$, and 3$\sigma$ two-dimensional likelihood confidence contours for flat \lcdm\ from various combinations of data. The zero-acceleration black dashed lines in panels (a) and (b) divide the parameter space into regions associated with currently-accelerating (left) and currently-decelerating (right) cosmological expansion.}
\label{fig1C8}
\end{figure*}

\begin{figure*}
\centering
 \subfloat[]{%
    \includegraphics[width=0.5\textwidth,height=0.5\textwidth]{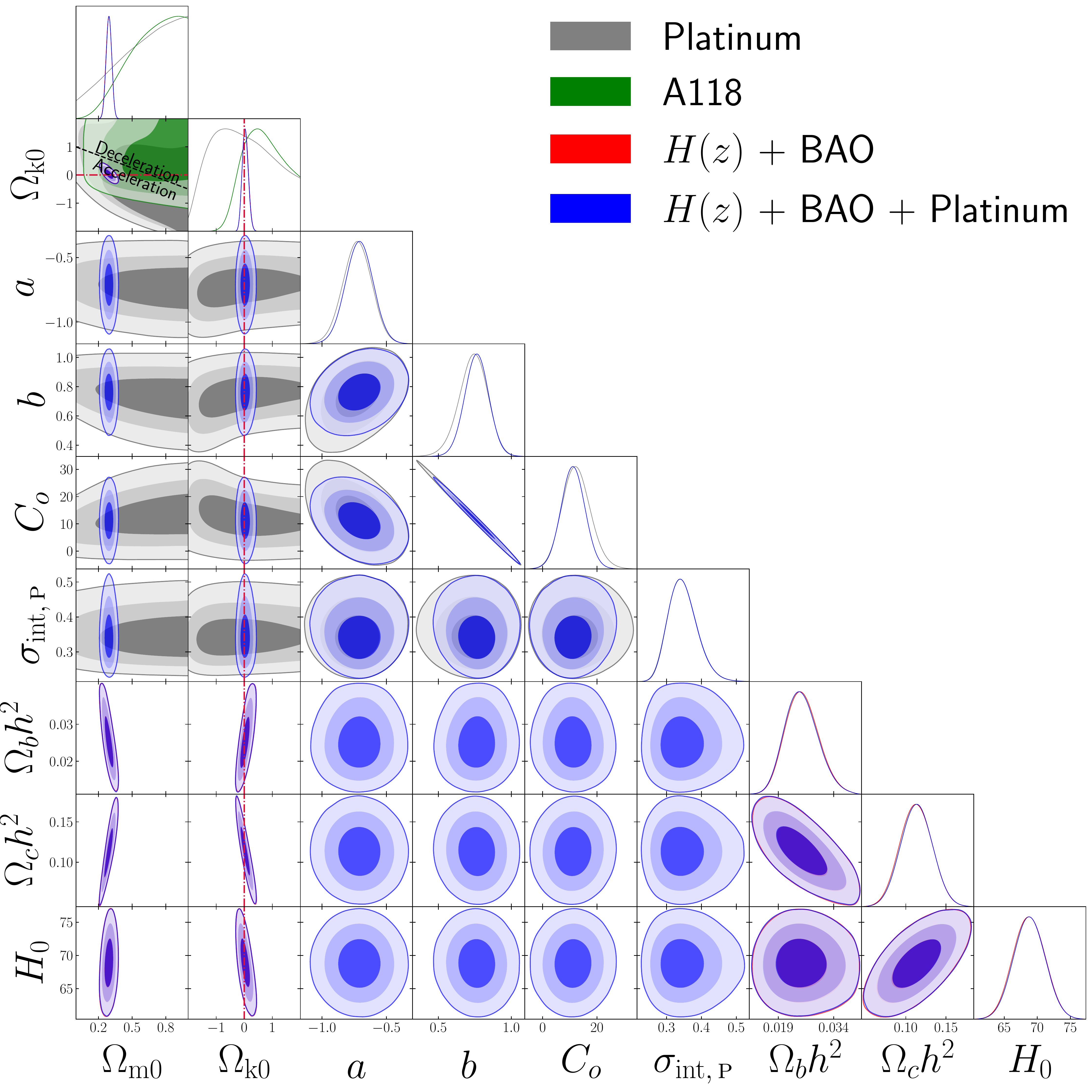}}
 \subfloat[]{%
    \includegraphics[width=0.5\textwidth,height=0.5\textwidth]{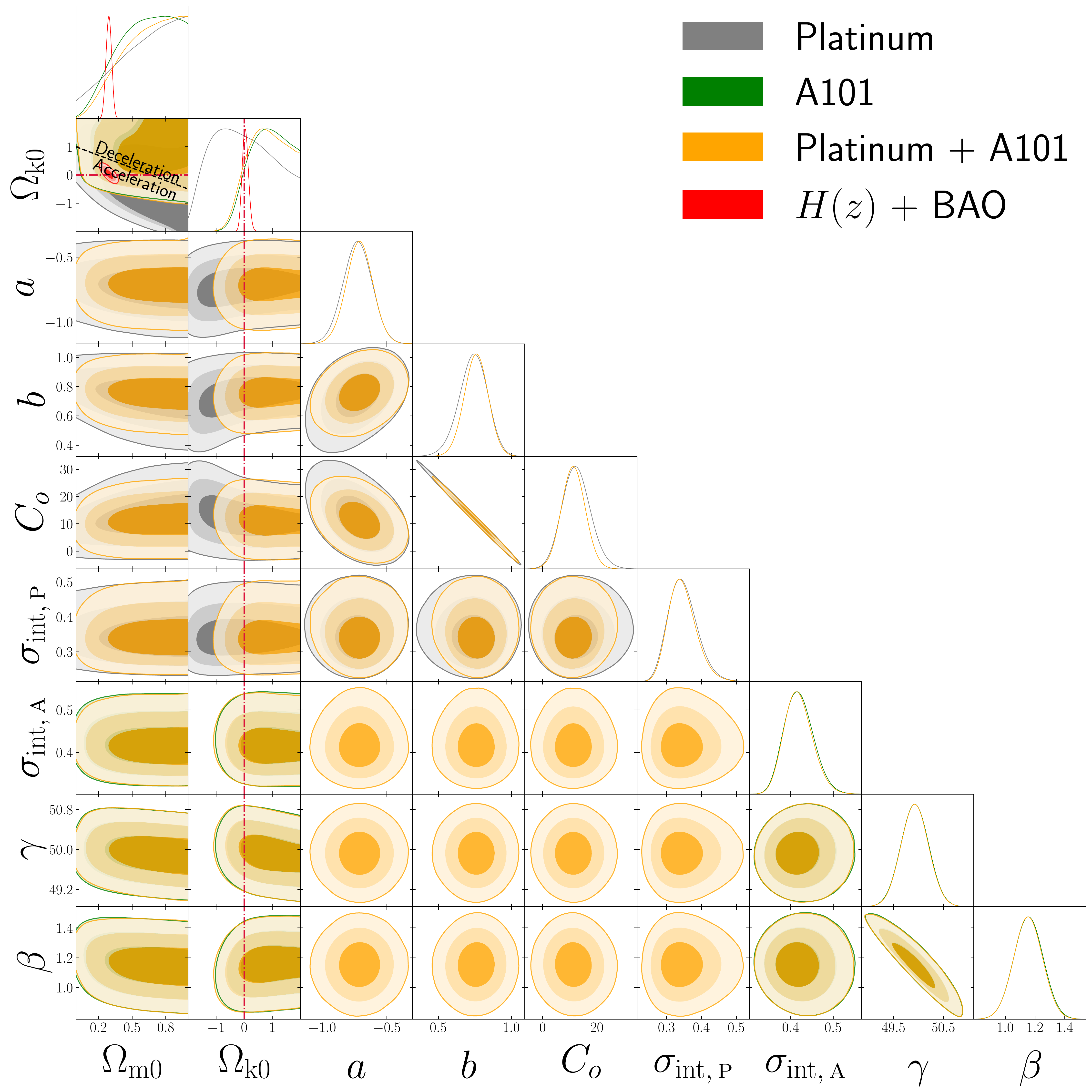}}\\
 \subfloat[]{%
    \includegraphics[width=0.5\textwidth,height=0.5\textwidth]{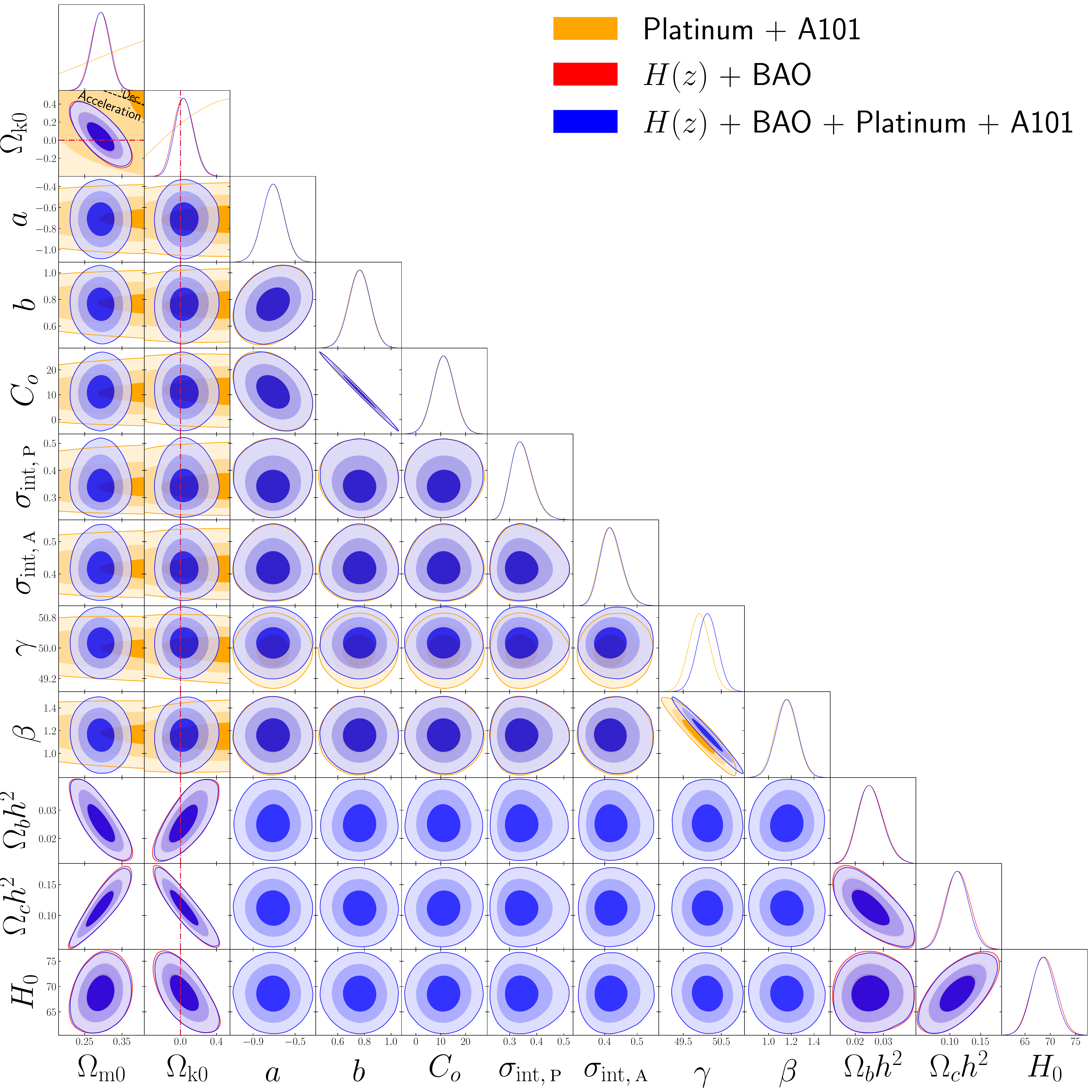}}
 \subfloat[Cosmological parameters zoom in]{%
    \includegraphics[width=0.5\textwidth,height=0.5\textwidth]{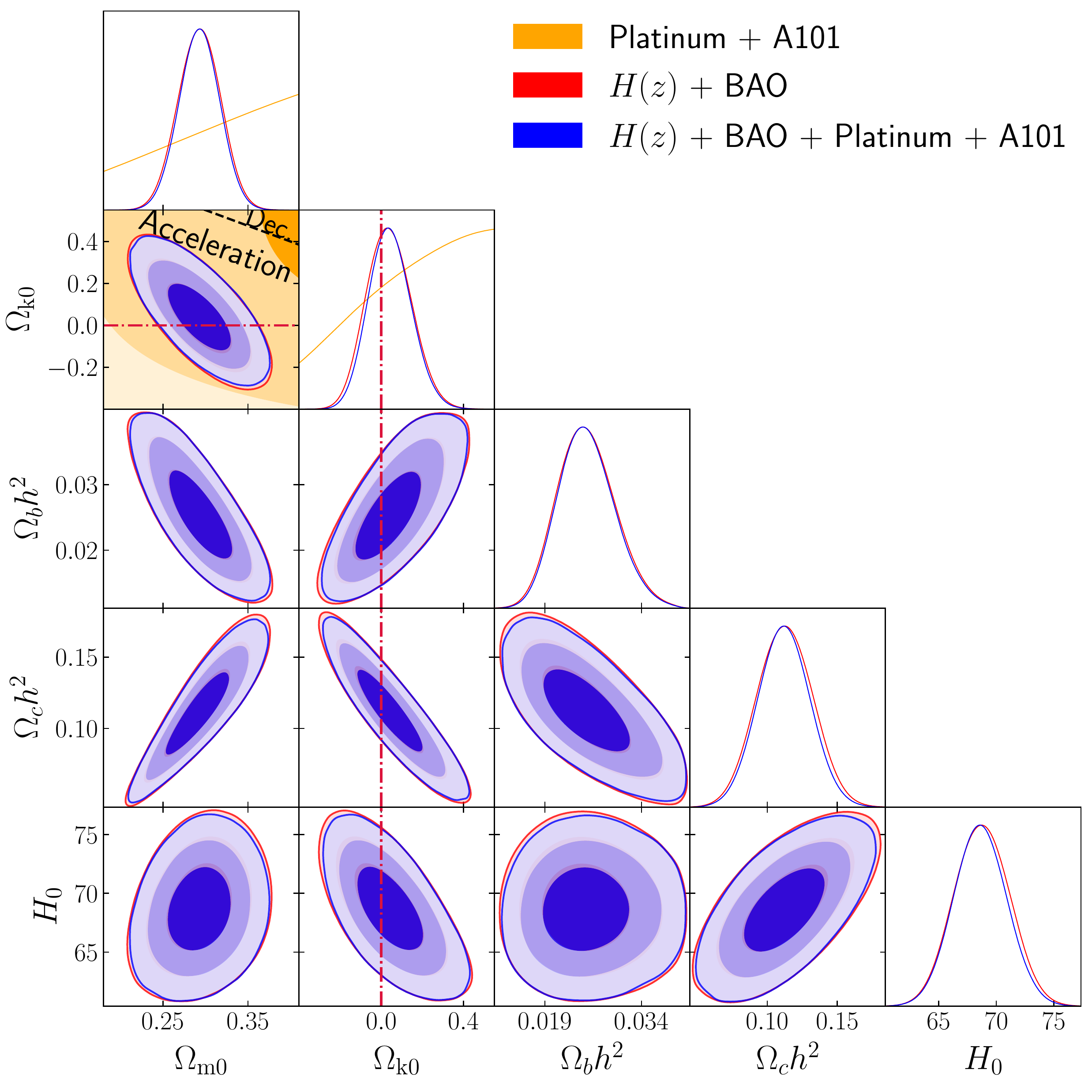}}\\
\caption{Same as Fig.\ \ref{fig1C8} but for non-flat \lcdm. The zero-acceleration black dashed lines divide the parameter space into regions associated with currently-accelerating (below left) and currently-decelerating (above right) cosmological expansion.}
\label{fig2C8}
\end{figure*}

\begin{figure*}
\centering
 \subfloat[]{%
    \includegraphics[width=0.5\textwidth,height=0.5\textwidth]{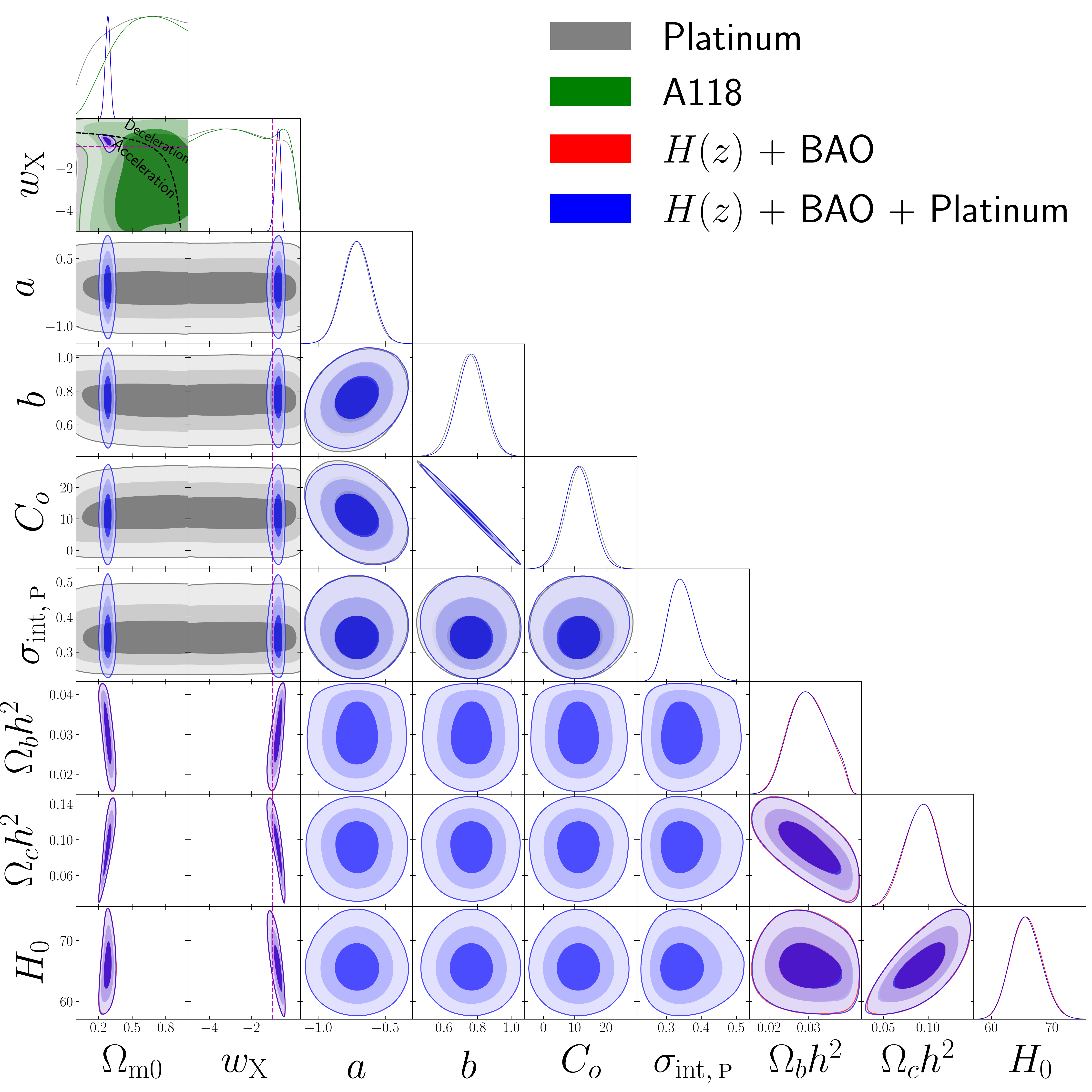}}
 \subfloat[]{%
    \includegraphics[width=0.5\textwidth,height=0.5\textwidth]{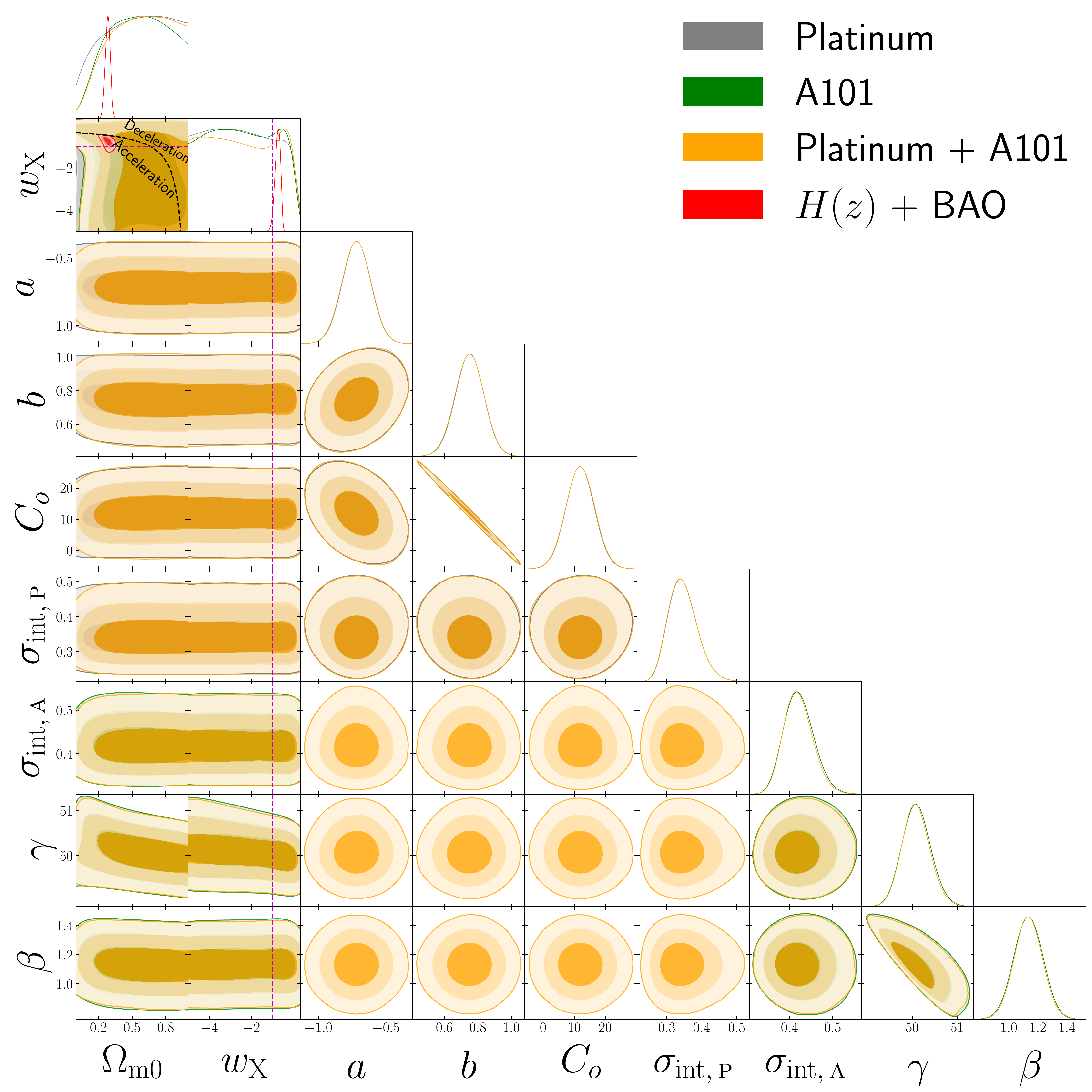}}\\
 \subfloat[]{%
    \includegraphics[width=0.5\textwidth,height=0.5\textwidth]{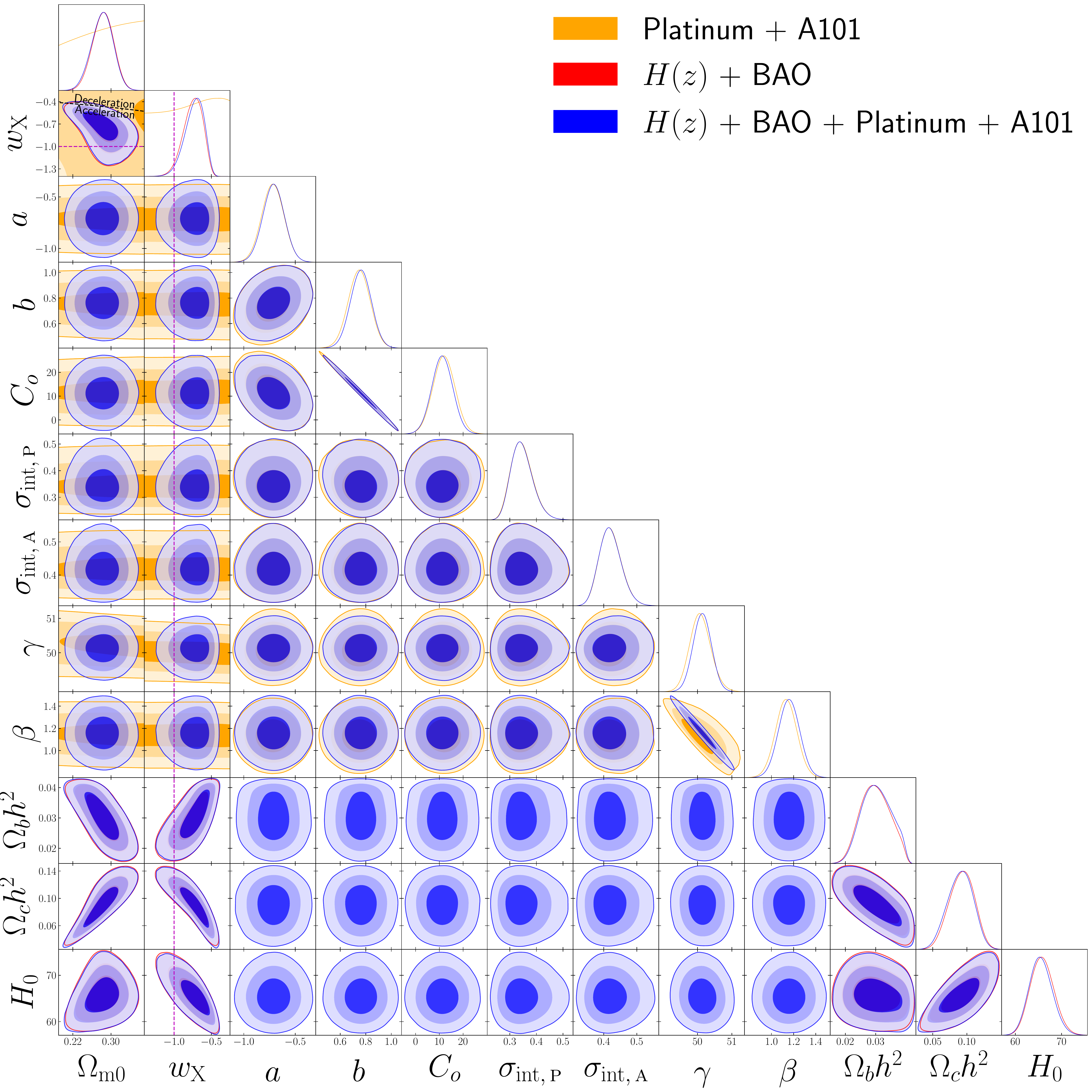}}
 \subfloat[Cosmological parameters zoom in]{%
    \includegraphics[width=0.5\textwidth,height=0.5\textwidth]{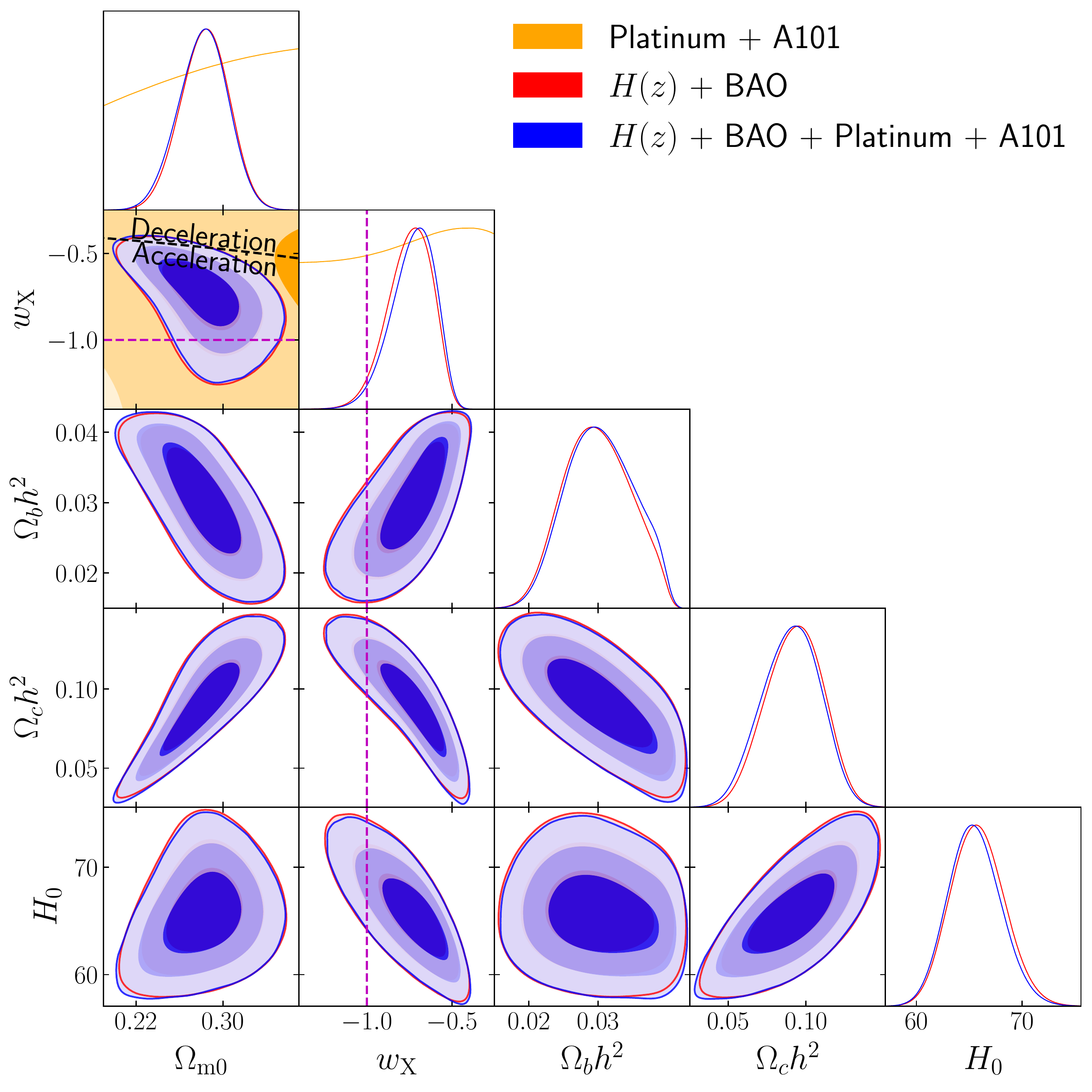}}\\
\caption{One-dimensional likelihood distributions and 1$\sigma$, 2$\sigma$, and 3$\sigma$ two-dimensional likelihood confidence contours for flat XCDM from various combinations of data. The zero-acceleration black dashed lines divide the parameter space into regions associated with currently-accelerating (either below left or below) and currently-decelerating (either above right or above) cosmological expansion. The magenta dashed lines represent $w_{\rm X}=-1$, i.e.\ flat \lcdm.}
\label{fig3C8}
\end{figure*}

\begin{figure*}
\centering
 \subfloat[]{%
    \includegraphics[width=0.5\textwidth,height=0.5\textwidth]{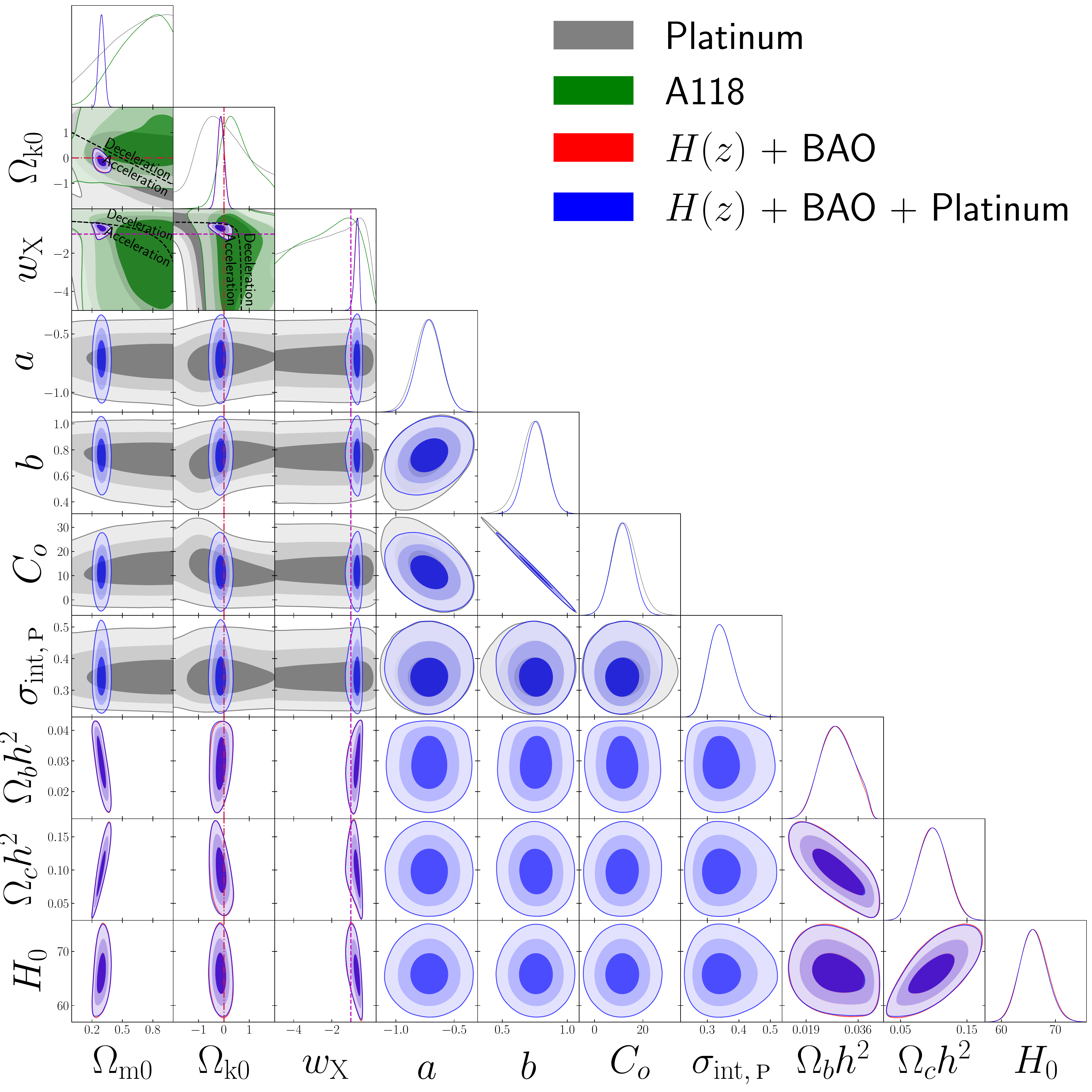}}
 \subfloat[]{%
    \includegraphics[width=0.5\textwidth,height=0.5\textwidth]{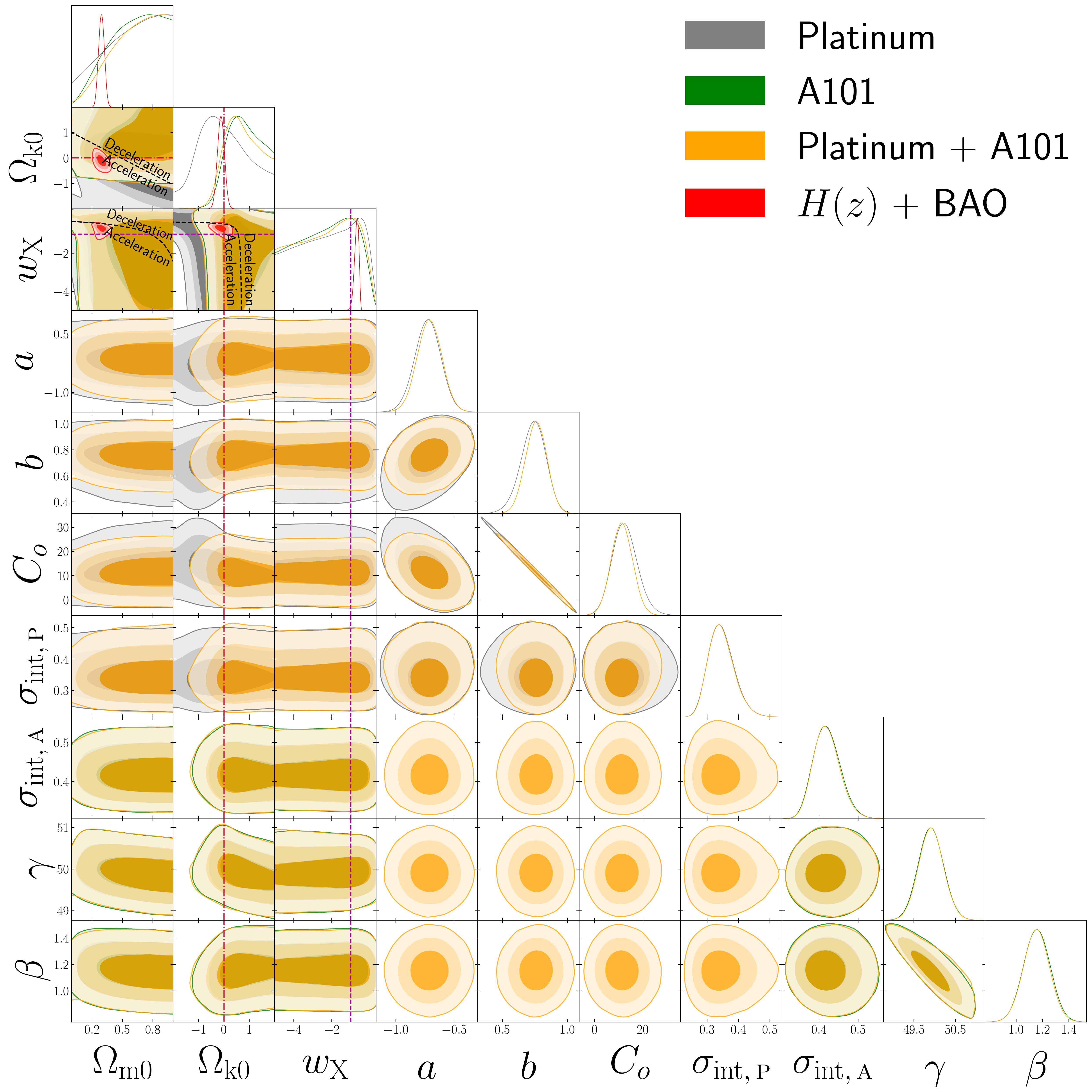}}\\
 \subfloat[]{%
    \includegraphics[width=0.5\textwidth,height=0.5\textwidth]{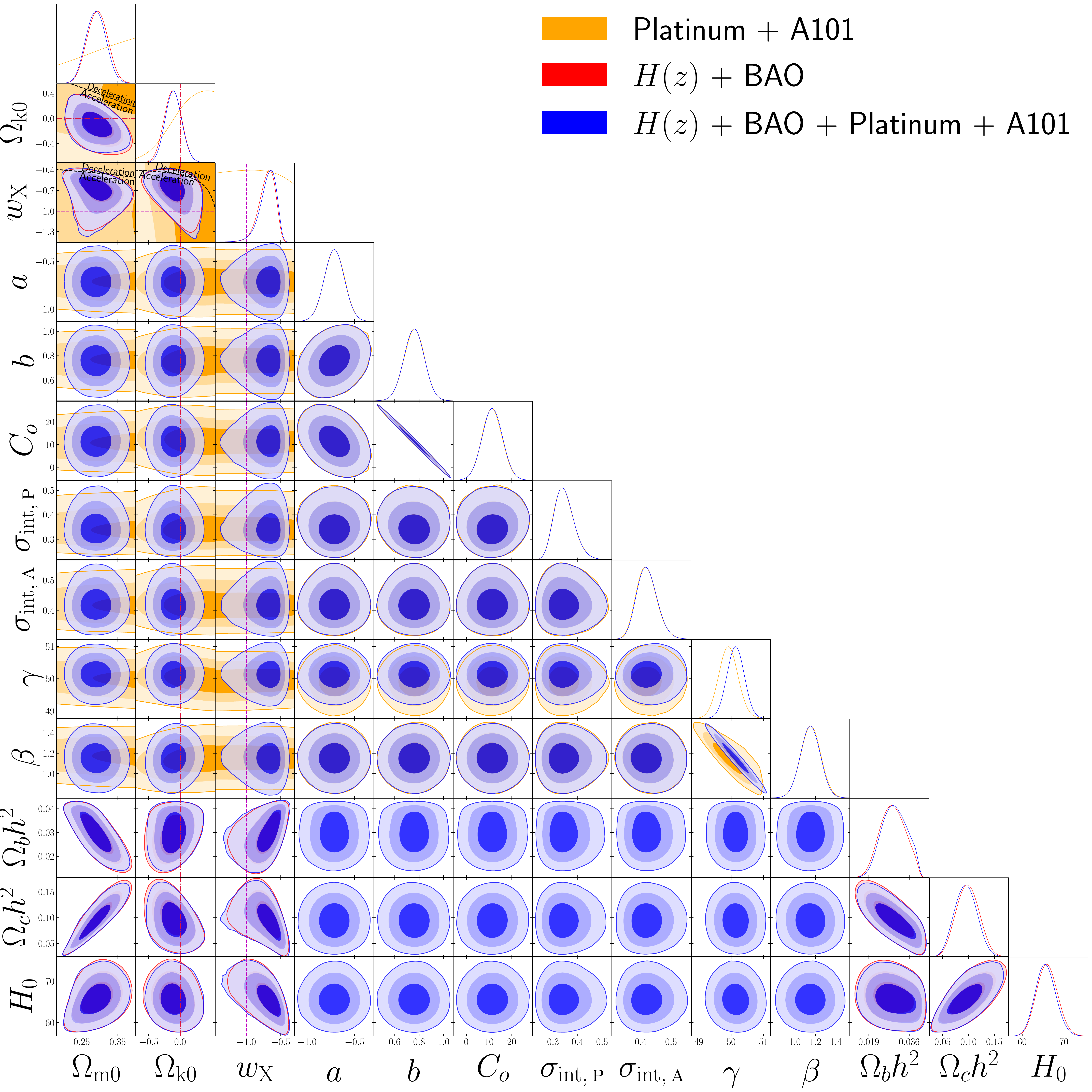}}
 \subfloat[Cosmological parameters zoom in]{%
    \includegraphics[width=0.5\textwidth,height=0.5\textwidth]{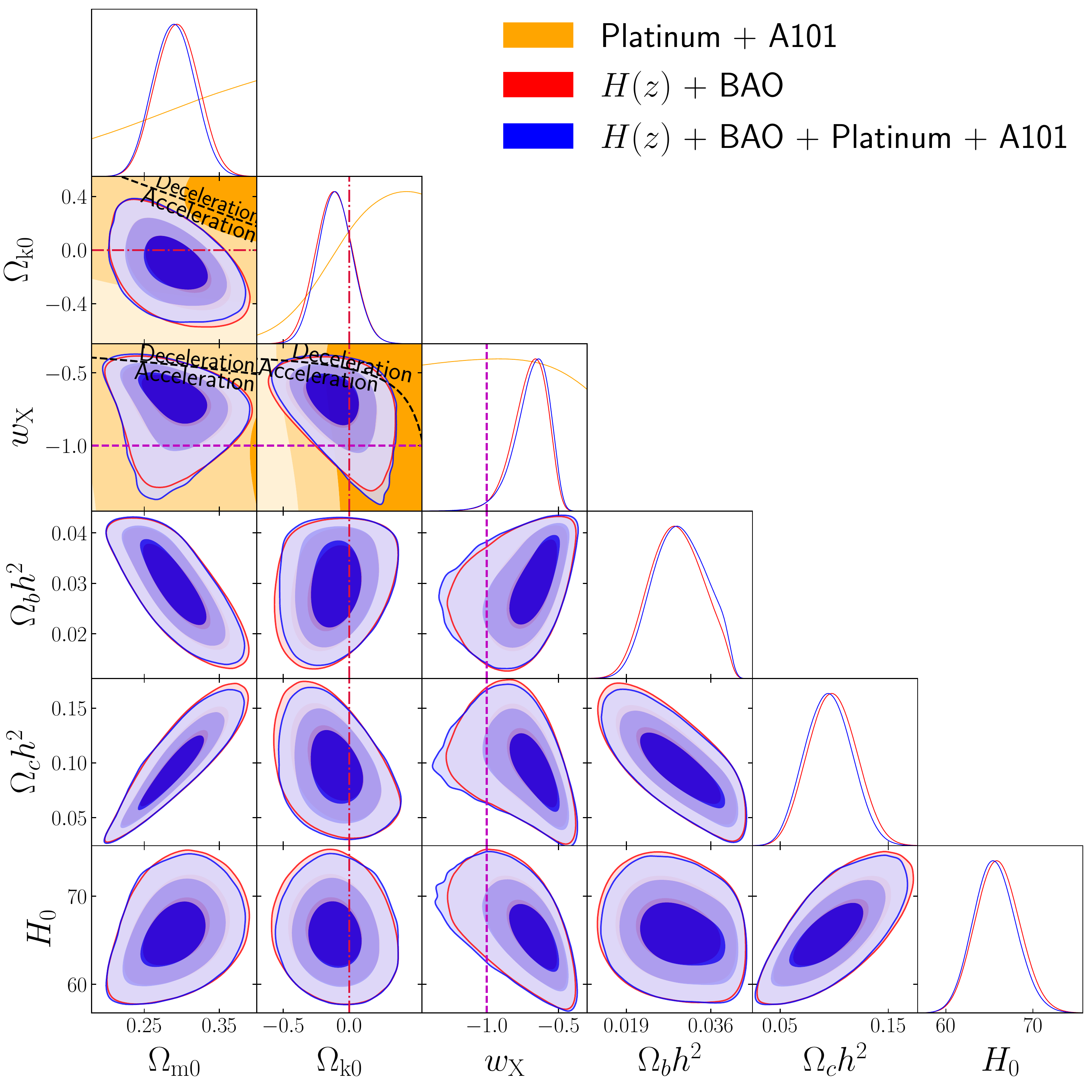}}\\
\caption{Same as Fig.\ \ref{fig3C8} but for non-flat XCDM. The zero-acceleration black dashed lines are computed for the third cosmological parameter set to the $H(z)$ + BAO data best-fitting values listed in Table \ref{tab:BFPC8}, and divide the parameter space into regions associated with currently-accelerating (either below left or below) and currently-decelerating (either above right or above) cosmological expansion. The crimson dash-dot lines represent flat hypersurfaces, with closed spatial hypersurfaces either below or to the left. The magenta dashed lines represent $w_{\rm X}=-1$, i.e.\ non-flat \lcdm.}
\label{fig4C8}
\end{figure*}

\begin{figure*}
\centering
\centering
 \subfloat[]{%
    \includegraphics[width=0.5\textwidth,height=0.5\textwidth]{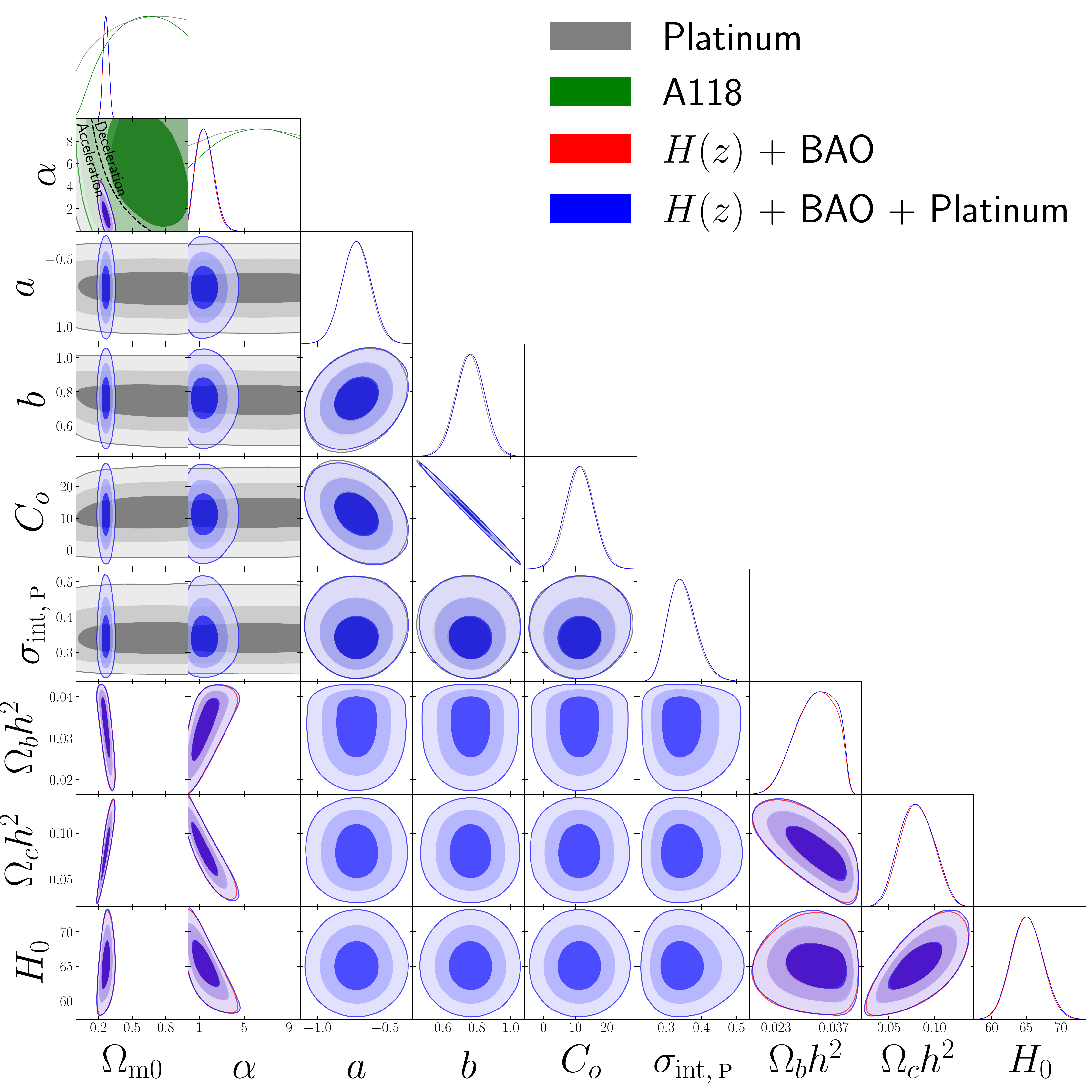}}
 \subfloat[]{%
    \includegraphics[width=0.5\textwidth,height=0.5\textwidth]{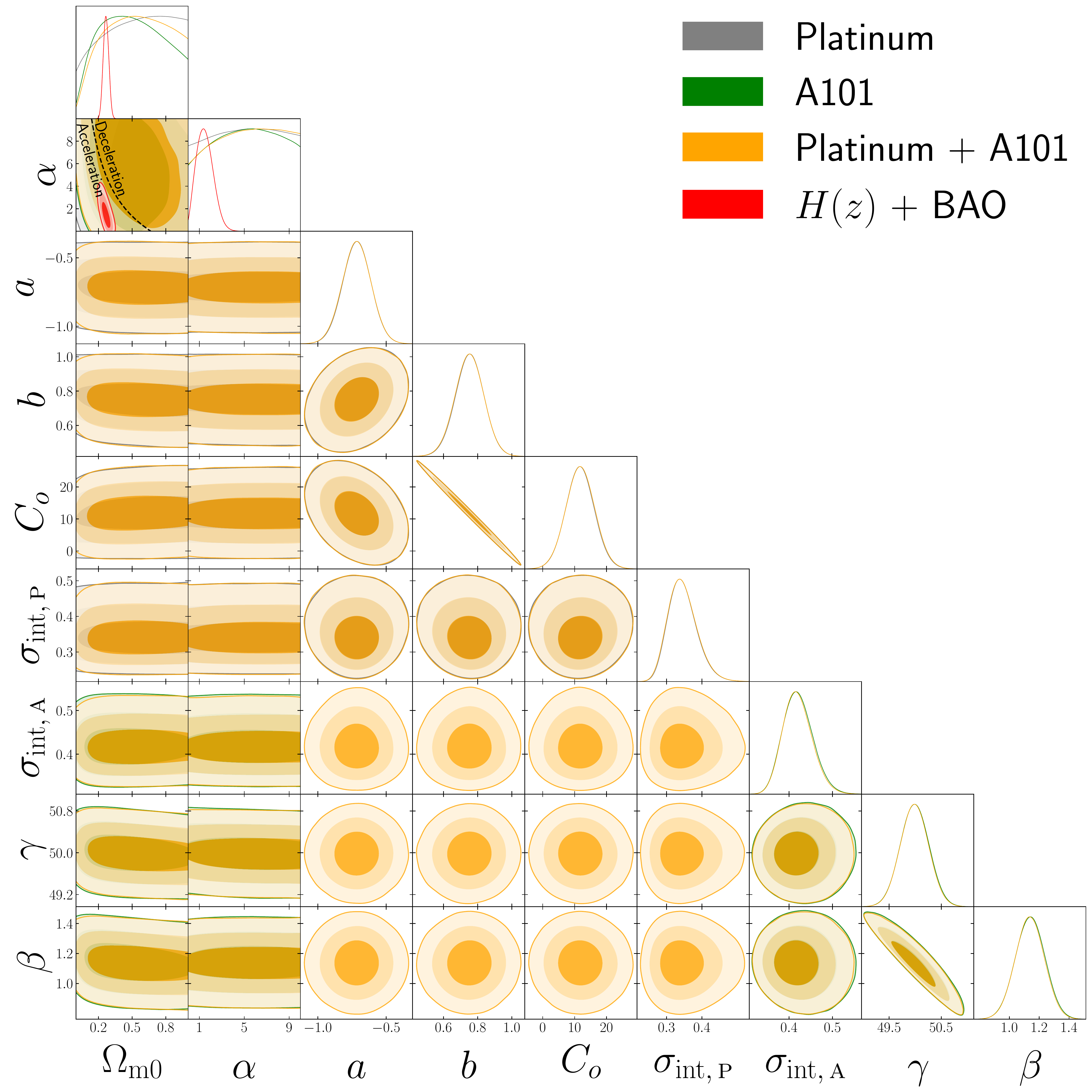}}\\
 \subfloat[]{%
    \includegraphics[width=0.5\textwidth,height=0.5\textwidth]{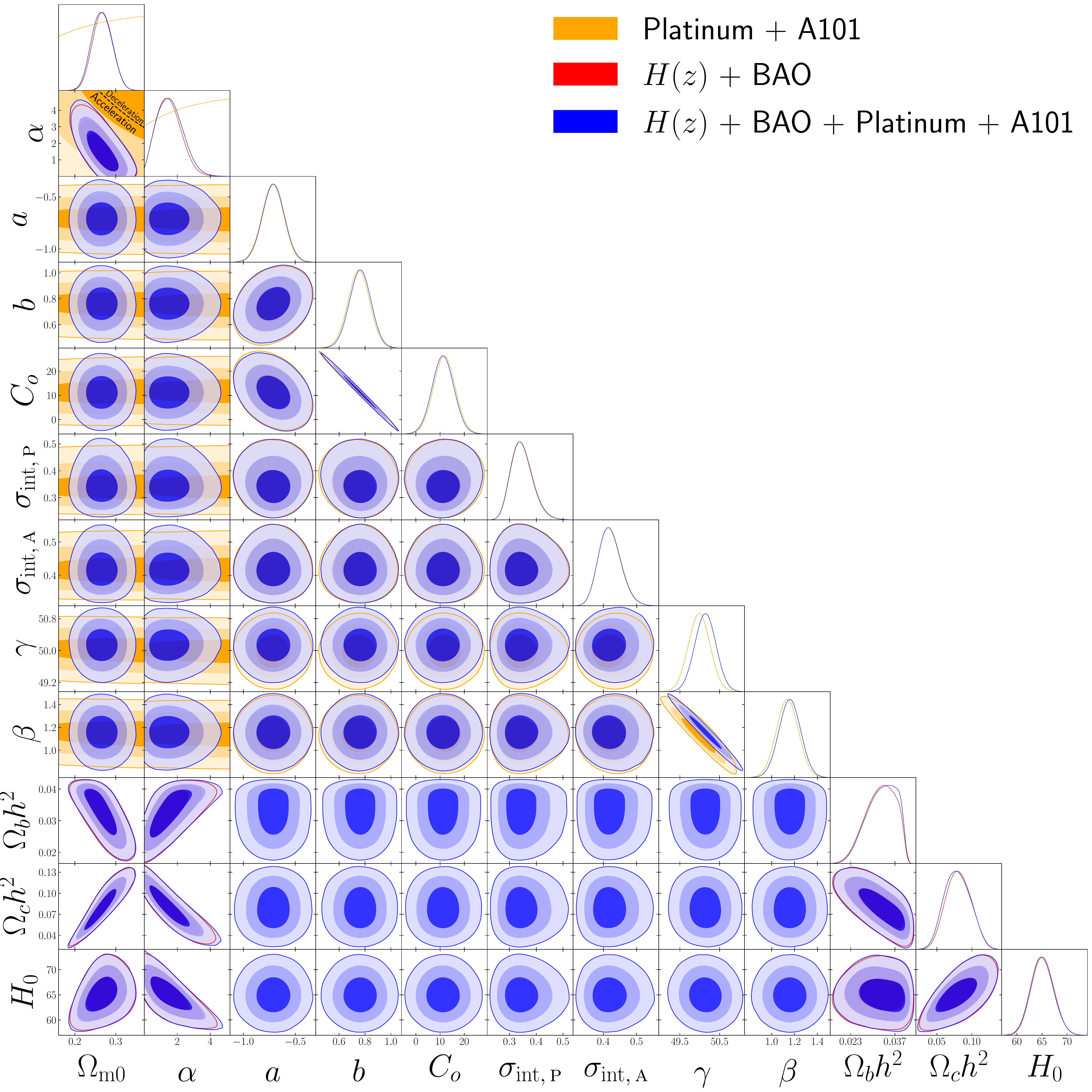}}
 \subfloat[Cosmological parameters zoom in]{%
    \includegraphics[width=0.5\textwidth,height=0.5\textwidth]{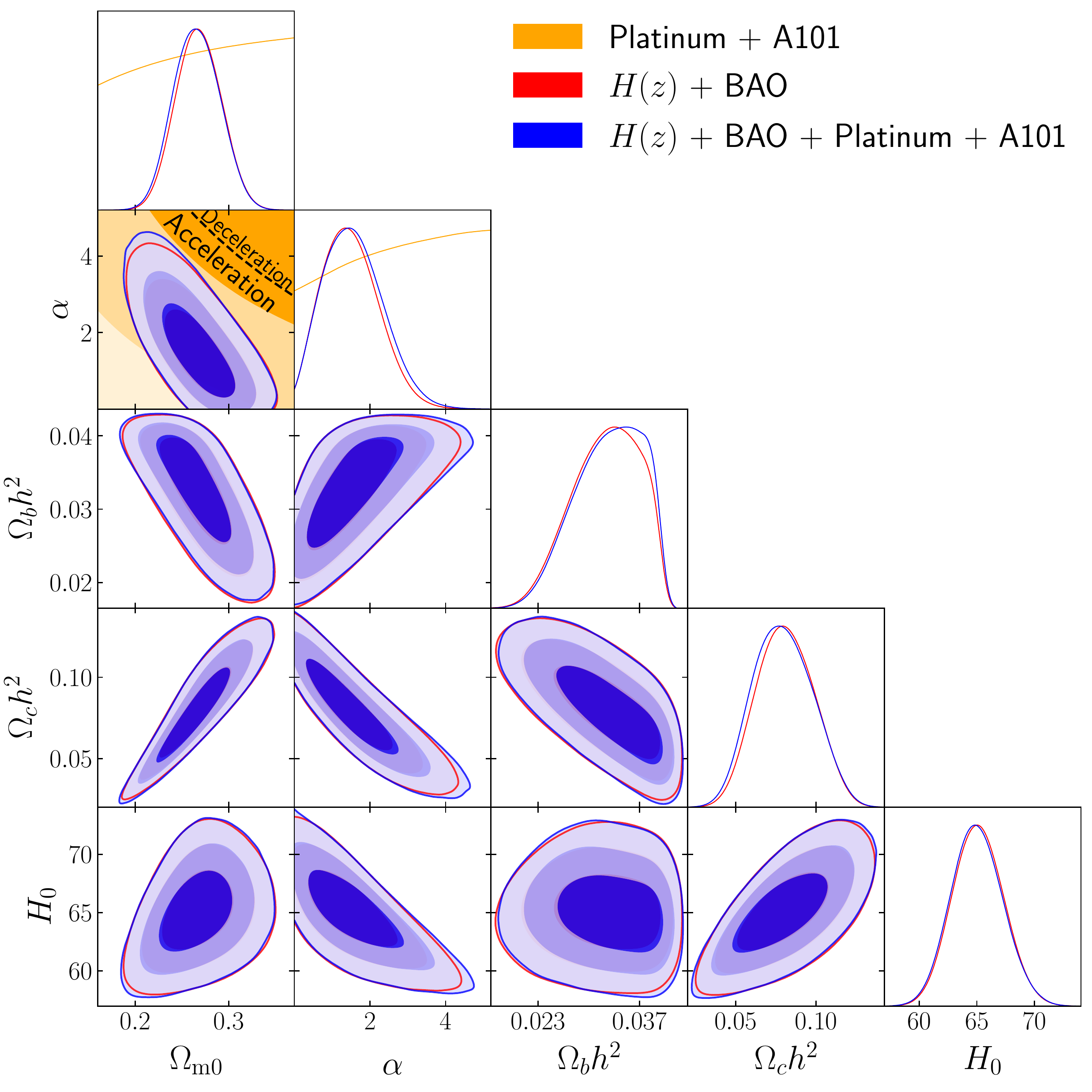}}\\
\caption{One-dimensional likelihood distributions and 1$\sigma$, 2$\sigma$, and 3$\sigma$ two-dimensional likelihood confidence contours for flat \pcdm\ from various combinations of data. The zero-acceleration black dashed lines divide the parameter space into regions associated with currently-accelerating (below left) and currently-decelerating (above right) cosmological expansion. The $\alpha = 0$ axes correspond to flat \lcdm.}
\label{fig5C8}
\end{figure*}

\begin{figure*}
\centering
 \subfloat[]{%
    \includegraphics[width=0.5\textwidth,height=0.5\textwidth]{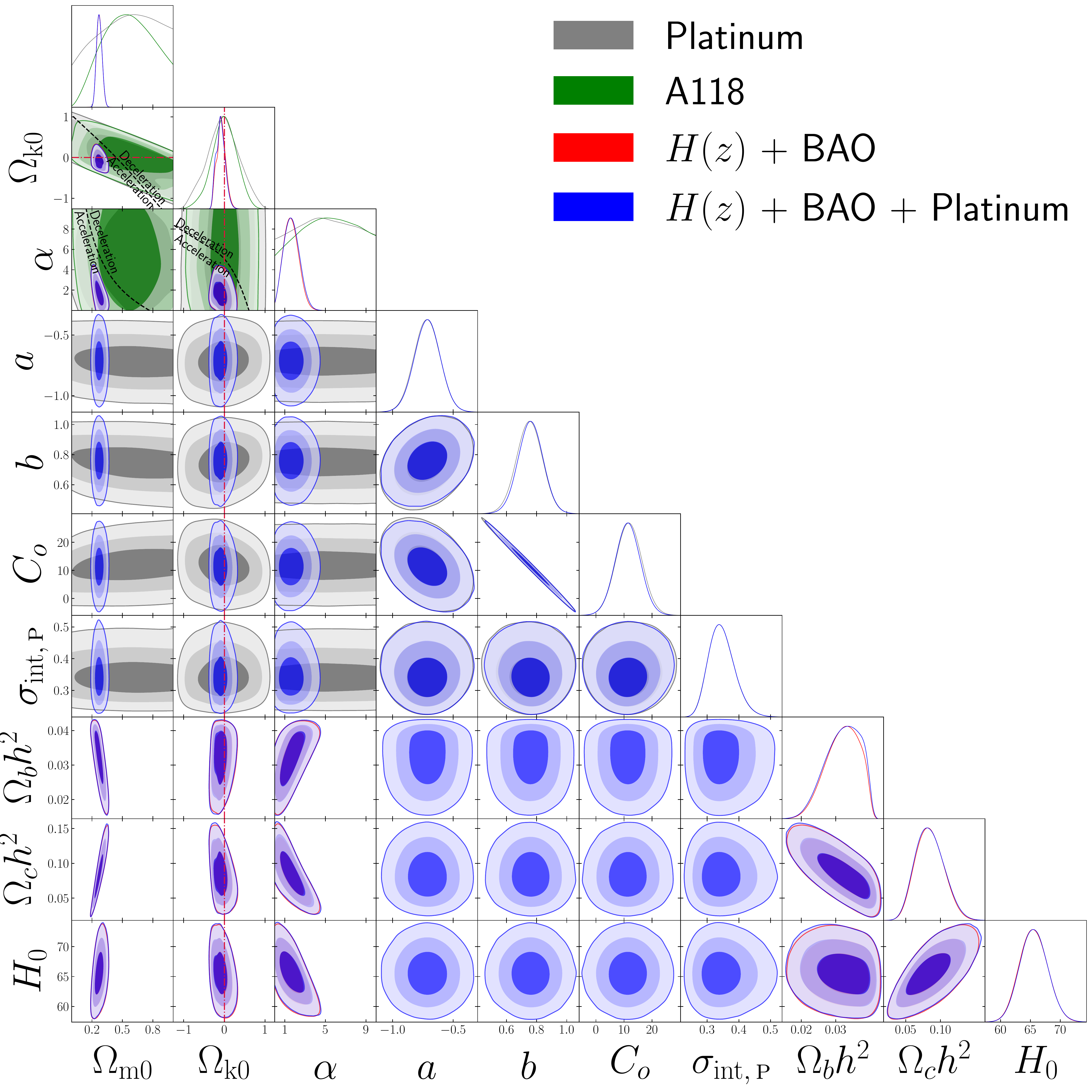}}
 \subfloat[]{%
    \includegraphics[width=0.5\textwidth,height=0.5\textwidth]{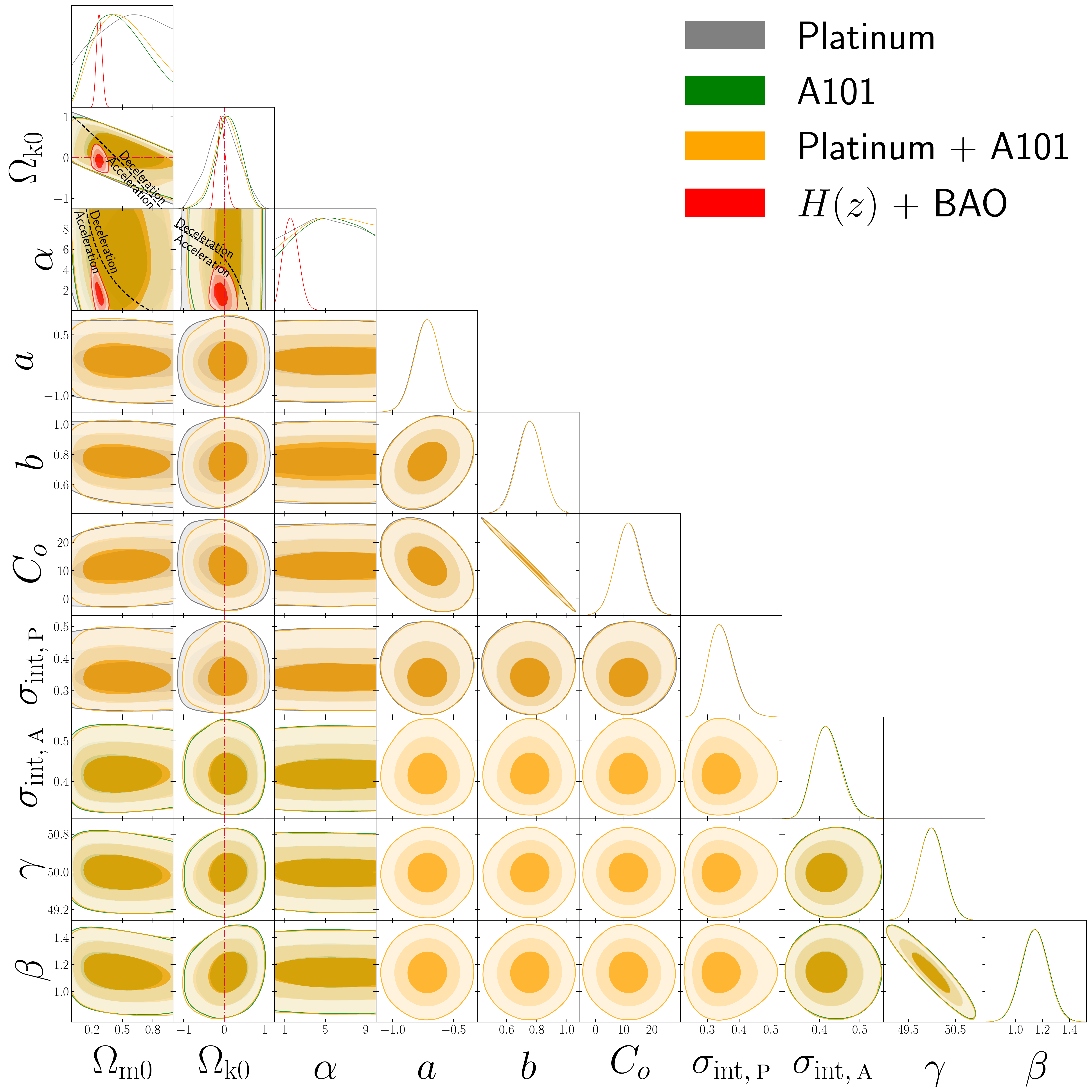}}\\
 \subfloat[]{%
    \includegraphics[width=0.5\textwidth,height=0.5\textwidth]{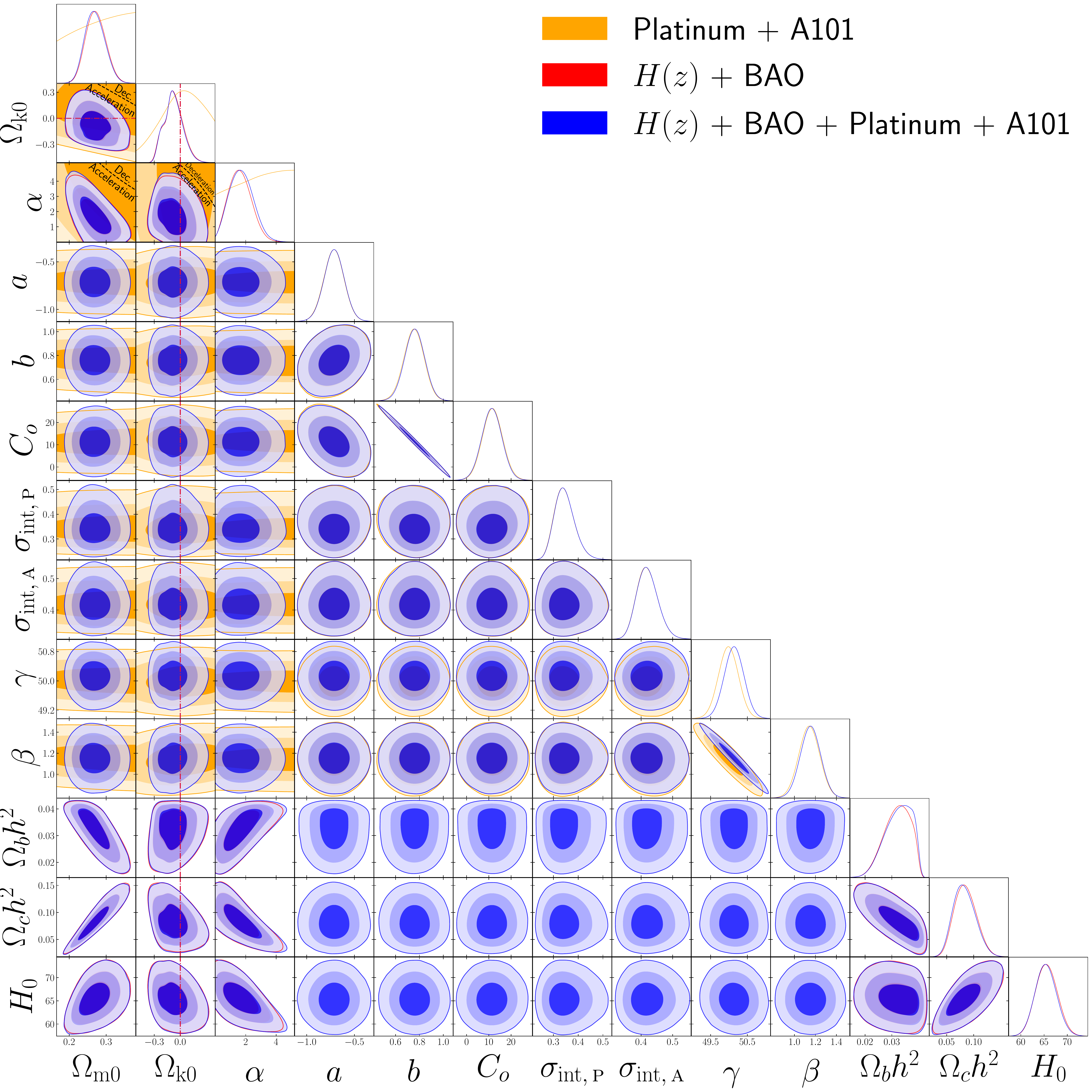}}
 \subfloat[Cosmological parameters zoom in]{%
    \includegraphics[width=0.5\textwidth,height=0.5\textwidth]{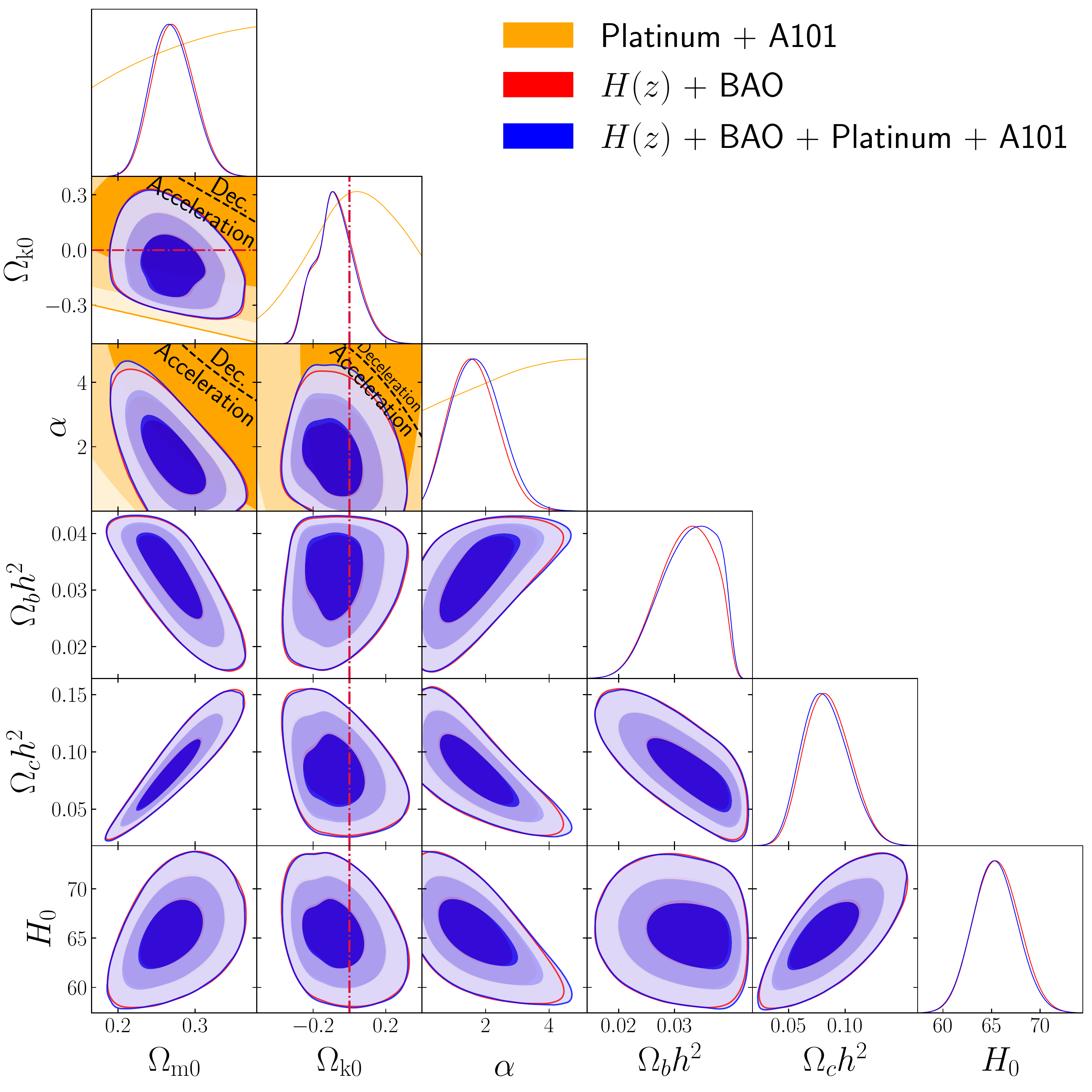}}\\
\caption{Same as Fig.\ \ref{fig5C8} but for non-flat \pcdm. The zero-acceleration black dashed lines are computed for the third cosmological parameter set to the $H(z)$ + BAO data best-fitting values listed in Table \ref{tab:BFPC8}, and divide the parameter space into regions associated with currently-accelerating (below left) and currently-decelerating (above right) cosmological expansion. The crimson dash-dot lines represent flat hypersurfaces, with closed spatial hypersurfaces either below or to the left. The $\alpha = 0$ axes correspond to non-flat \lcdm.}
\label{fig6C8}
\end{figure*}

\begin{figure*}
\centering
 \subfloat[Flat \lcdm]{%
    \includegraphics[width=0.5\textwidth,height=0.35\textwidth]{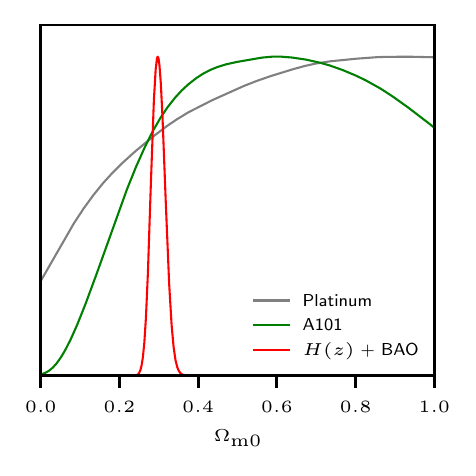}}
 \subfloat[Non-flat \lcdm]{%
    \includegraphics[width=0.5\textwidth,height=0.34\textwidth]{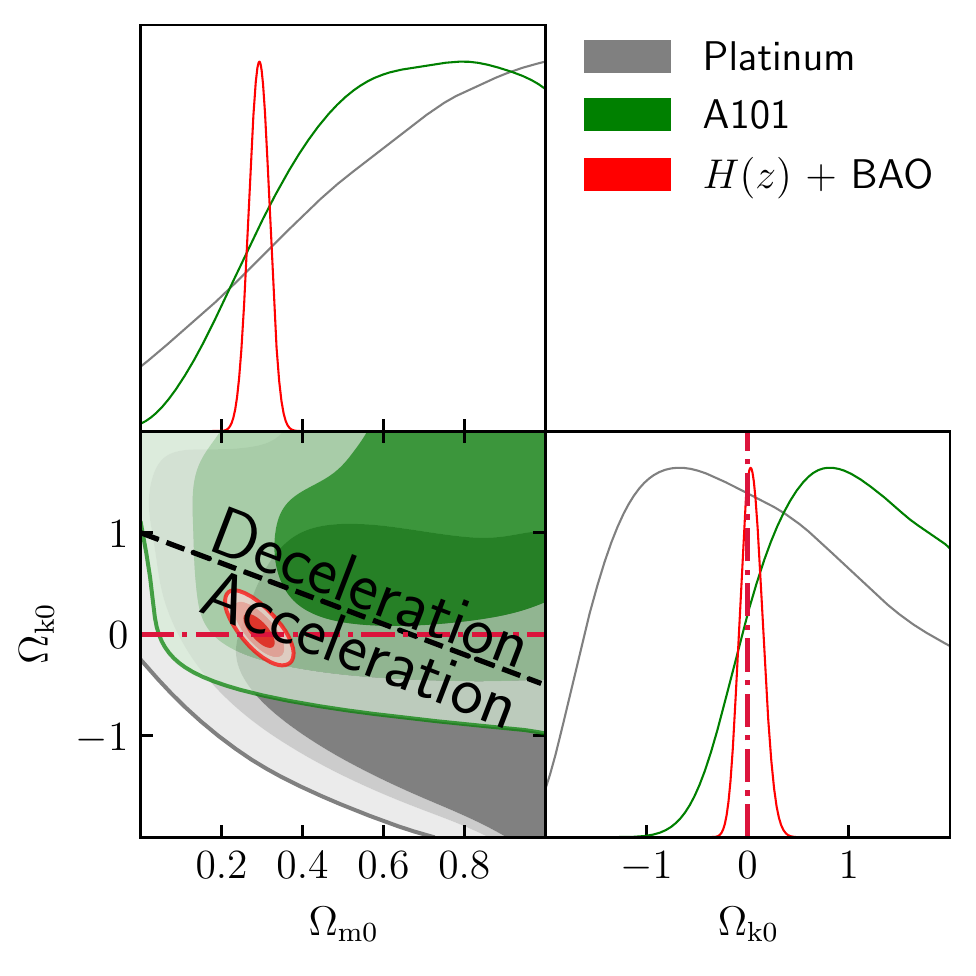}}\\
 \subfloat[Flat XCDM]{%
    \includegraphics[width=0.5\textwidth,height=0.34\textwidth]{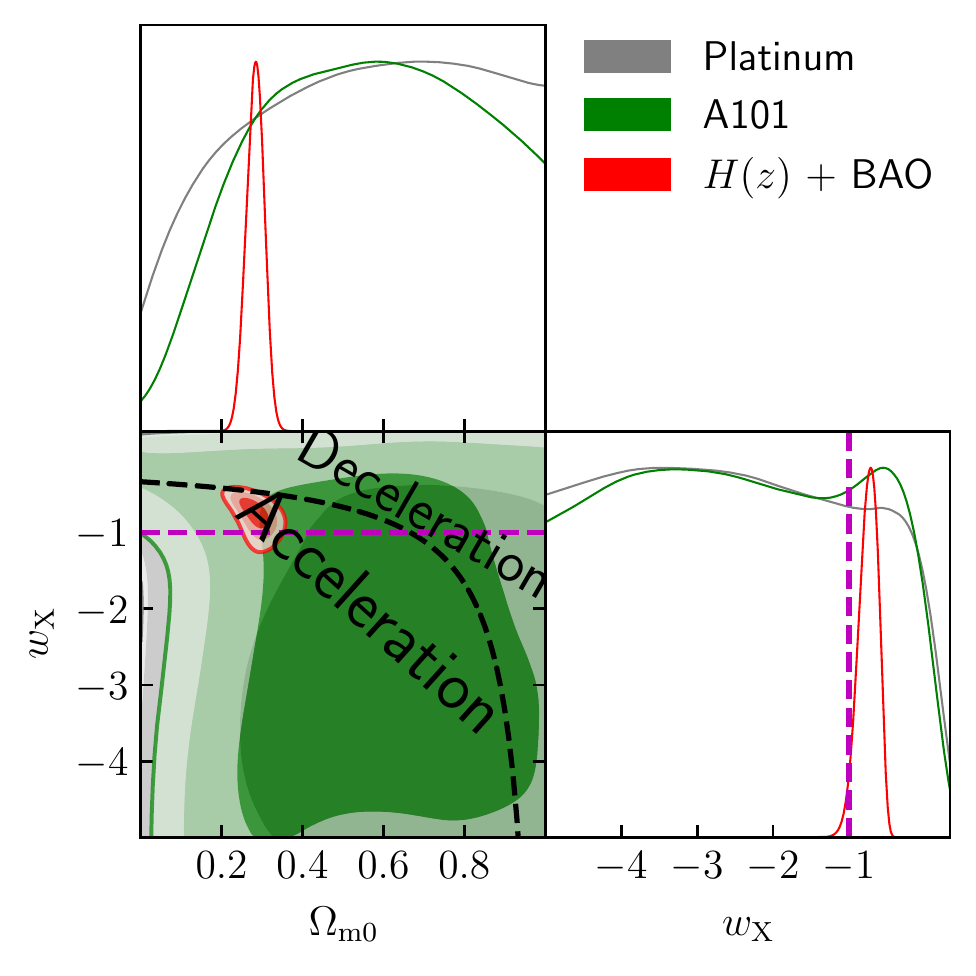}}
 \subfloat[Non-flat XCDM]{%
    \includegraphics[width=0.5\textwidth,height=0.34\textwidth]{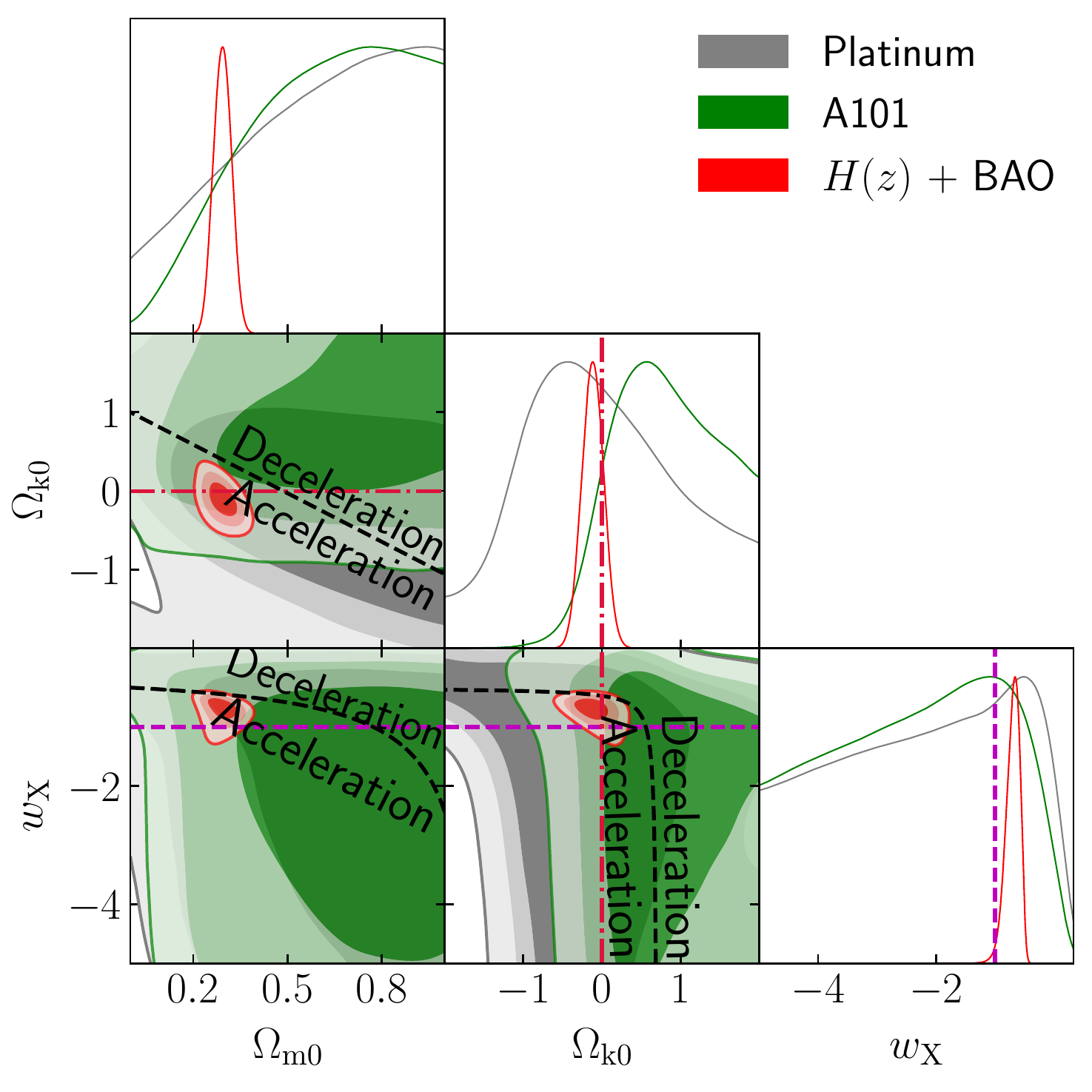}}\\
 \subfloat[Flat \pcdm]{%
    \includegraphics[width=0.5\textwidth,height=0.34\textwidth]{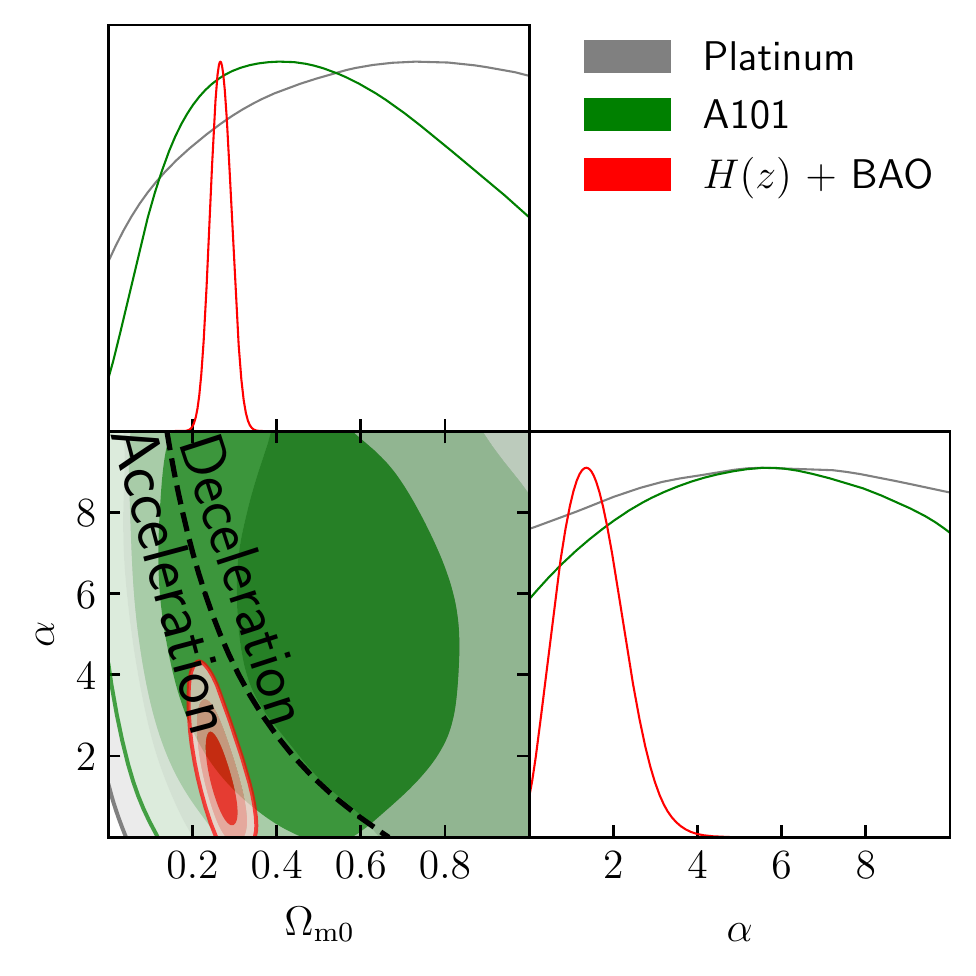}}
 \subfloat[Non-flat \pcdm]{%
    \includegraphics[width=0.5\textwidth,height=0.34\textwidth]{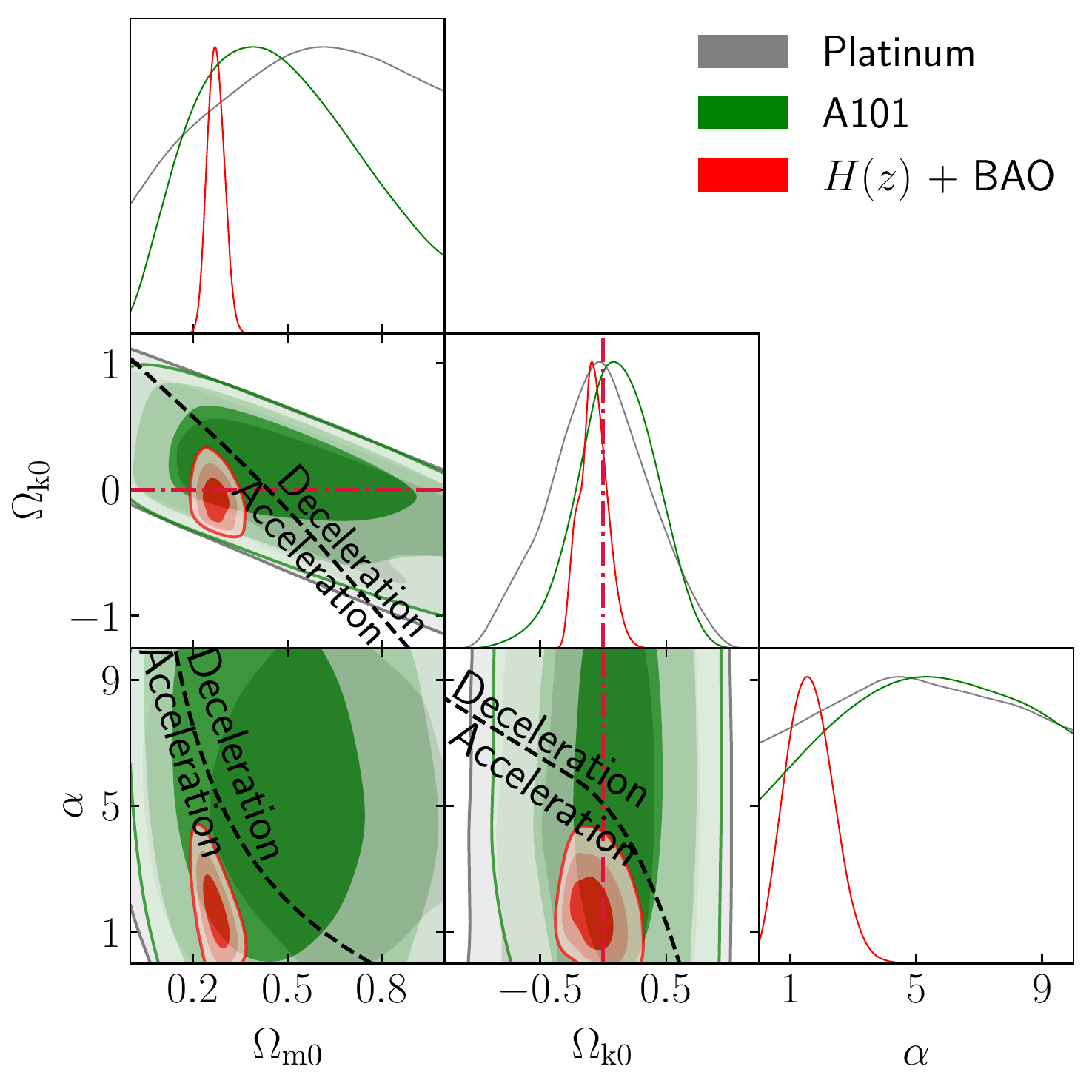}}\\
\caption{One-dimensional likelihood distributions and 1$\sigma$, 2$\sigma$, and 3$\sigma$ two-dimensional likelihood confidence contours for some cosmological parameters of the six cosmological models from Platinum data (gray), A101 data (green), and $H(z)$ + BAO data (red), which more clearly show the cosmological parameter overlaps between Platinum/A101 data and $H(z)$ + BAO data.}
\label{fig7C8}
\end{figure*}

We show the posterior one-dimensional (1D) probability distributions and two-dimensional (2D) confidence regions of cosmological-model and GRB-correlation parameters for the six cosmological models in Figs. \ref{fig1C8}--\ref{fig7C8}, in gray (Platinum), green (A118 and A101), orange (Platinum + A101), red [$H(z)$ + BAO], and blue [$H(z)$ + BAO + Platinum, in short HzBP, and $H(z)$ + BAO + Platinum + A101, in short HzBPA101]. The unmarginalized best-fitting parameter values, as well as the values of maximum likelihood $\mathcal{L}_{\rm max}$, $AIC$, $BIC$, $DIC$, $\Delta AIC$, $\Delta BIC$, and $\Delta DIC$, for all models and data combinations, are listed in Table \ref{tab:BFPC8}. We list the marginalized posterior mean parameter values and uncertainties ($\pm 1\sigma$ error bars and 1 or 2$\sigma$ limits), for all models and data combinations, in Table \ref{tab:1d_BFPC8}.

\begin{sidewaystable*}
\centering
\resizebox{\columnwidth}{!}{%
\begin{threeparttable}
\caption{Unmarginalized best-fitting parameter values for all models from various combinations of data.}\label{tab:BFPC8}
\begin{tabular}{lccccccccccccccccccccc}
\toprule
Model & Data set & $\Omega_{b}h^2$\,\tnote{a} & $\Omega_{c}h^2$ & $\Omega_{\mathrm{m0}}$ & $\Omega_{\mathrm{k0}}$ & $w_{\mathrm{X}}$/$\alpha$\tnote{b} & $H_0$\tnote{c} & $\sigma_{\mathrm{int,\,\textsc{p}}}$ & $a$ & $b$ & $C_{o}$ & $\sigma_{\mathrm{int,\,\textsc{a}}}$ & $\gamma$ & $\beta$ & $-2\ln\mathcal{L}_{\mathrm{max}}$ & $AIC$ & $BIC$ & $DIC$ & $\Delta AIC$ & $\Delta BIC$ & $\Delta DIC$ \\
\midrule
 & Platinum & -- & 0.4560 & 0.982 & -- & -- & -- & 0.322 & $-0.716$ & 0.743 & 12.19 & -- & -- & -- & 32.99 & 42.99 & 52.55 & 42.80 & 0.00 & 0.00 & 0.00\\
 & A118 & -- & 0.4088 & 0.886 & -- & -- & -- & -- & -- & -- & -- & 0.402 & 50.02 & 1.099 & 128.72 & 136.72 & 147.81 & 136.05 & 0.00 & 0.00 & 0.00\\
 & A101 & -- & 0.2547 & 0.571 & -- & -- & -- & -- & -- & -- & -- & 0.409 & 50.04 & 1.135 & 112.81 & 120.81 & 131.27 & 120.16 & 0.00 & 0.00 & 0.00\\
Flat & Plat. + A101 & -- & 0.3505 & 0.767 & -- & -- & -- & 0.324 & $-0.694$ & 0.753 & 11.60 & 0.413 & 49.98 & 1.125 & 146.08 & 162.08 & 186.22 & 162.07 & 0.00 & 0.00 & 0.00\\
\lcdm & $H(z)$ + BAO & 0.0240 & 0.1177 & 0.298 & -- & -- & 69.11 & -- & -- & -- & -- & -- & -- & -- & 23.66 & 29.66 & 34.87 & 30.23 & 0.00 & 0.00 & 0.00\\
 & HzBP\tnote{d} & 0.0243 & 0.1169 & 0.298 & -- & -- & 69.04 & 0.325 & $-0.715$ & 0.757 & 11.53 & -- & -- & -- & 57.33 & 73.33 & 88.99 & 72.58 & 0.00 & 0.00 & 0.00\\
 & HzBPA101\tnote{e} & 0.0236 & 0.1155 & 0.296 & -- & -- & 68.71 & 0.324 & $-0.707$ & 0.760 & 11.35 & 0.412 & 50.16 & 1.153 & 170.63 & 190.63 & 223.26 & 191.94 & 0.00 & 0.00 & 0.00\\
\midrule
 & Platinum & -- & 0.1178 & 0.292 & $-0.992$ & -- & -- & 0.292 & $-0.700$ & 0.829 & 7.59 & -- & -- & -- & 24.51 & 36.51 & 47.98 & 51.00 & $-6.48$ & $-4.57$ & 8.20\\
 & A118 & -- & 0.4625 & 0.995 & 1.094 & -- & -- & -- & -- & -- & -- & 0.401 & 49.90 & 1.117 & 127.96 & 137.96 & 151.82 & 136.72 & 1.24 & 4.01 & 0.67\\
 & A101 & -- & 0.4606 & 0.991 & 1.993 & -- & -- & -- & -- & -- & -- & 0.405 & 49.78 & 1.151 & 111.08 & 121.08 & 134.16 & 120.04 & 0.27 & 2.89 & $-0.12$\\
Non-flat & Plat. + A101 & -- & 0.4300 & 0.929 & 1.857 & -- & -- & 0.326 & $-0.712$ & 0.746 & 11.99 & 0.400 & 49.75 & 1.167 & 144.65 & 162.65 & 189.80 & 161.75 & 0.57 & 3.58 & $-0.32$\\
\lcdm  & $H(z)$ + BAO & 0.0247 & 0.1141 & 0.295 & 0.023 & -- & 68.78 & -- & -- & -- & -- & -- & -- & -- & 23.59 & 31.59 & 38.54 & 32.21 & 1.93 & 3.67 & 1.99\\
 & HzBP\tnote{d} & 0.0245 & 0.1149 & 0.294 & 0.010 & -- & 68.99 & 0.328 & $-0.719$ & 0.752 & 11.81 & -- & -- & -- & 57.31 & 73.31 & 93.48 & 74.65 & $-0.02$ & 4.49 & 2.07\\
 & HzBPA101\tnote{e} & 0.0237 & 0.1115 & 0.295 & 0.032 & -- & 67.91 & 0.328 & $-0.726$ & 0.750 & 11.91 & 0.410 & 50.13 & 1.164 & 170.70 & 192.70 & 228.59 & 193.50 & 2.07 & 5.33 & 1.56\\
\midrule
 & Platinum & -- & 0.0809 & 0.216 & -- & 0.137 & -- & 0.323 & $-0.724$ & 0.735 & 12.59 & -- & -- & -- & 32.81 & 44.81 & 56.28 & 42.83 & 1.82 & 3.73 & 0.04\\
 & A118 & -- & $-0.0198$ & 0.011 & -- & $-0.139$ & -- & -- & -- & -- & -- & 0.402 & 50.04 & 1.112 & 128.42 & 138.42 & 152.28 & 136.65 & 1.70 & 4.47 & 0.61\\
 & A101 & -- & $-0.0221$ & 0.006 & -- & $-0.251$ & -- & -- & -- & -- & -- & 0.406 & 50.01 & 1.148 & 111.99 & 121.99 & 135.06 & 121.23 & 1.18 & 3.79 & 1.08\\
Flat & Plat. + A101 & -- & 0.1570 & 0.372 & -- & $-0.193$ & -- & 0.317 & $-0.712$ & 0.753 & 11.65 & 0.405 & 50.02 & 1.115 & 145.74 & 163.74 & 190.90 & 162.40 & 1.66 & 4.68 & 0.33\\
XCDM & $H(z)$ + BAO & 0.0309 & 0.0870 & 0.280 & -- & $-0.694$ & 65.11 & -- & -- & -- & -- & -- & -- & -- & 19.65 & 27.65 & 34.60 & 28.11 & $-2.01$ & $-0.27$ & $-2.11$\\
 & HzBP\tnote{d} & 0.0302 & 0.0899 & 0.284 & -- & $-0.711$ & 65.22 & 0.320 & $-0.709$ & 0.758 & 11.46 & -- & -- & -- & 53.26 & 69.26 & 89.44 & 70.25 & $-4.07$ & 0.45 & $-2.33$\\
 & HzBPA101\tnote{e} & 0.0301 & 0.0891 & 0.282 & -- & $-0.709$ & 69.13 & 0.325 & $-0.724$ & 0.756 & 11.63 & 0.408 & 50.17 & 1.150 & 166.30 & 188.30 & 224.19 & 188.98 & $-2.33$ & 0.93 & $-2.96$\\
\midrule
 & Platinum & -- & 0.0918 & 0.239 & $-0.760$ & $-1.123$ & -- & 0.286 & $-0.761$ & 0.788 & 10.04 & -- & -- & -- & 24.58 & 38.58 & 51.97 & 51.01 & $-4.41$ & $-0.58$ & 8.21\\
 & A118 & -- & 0.4644 & 0.999 & 0.998 & $-1.137$ & -- & -- & -- & -- & -- & 0.398 & 49.91 & 1.113 & 127.97 & 139.97 & 156.60 & 137.16 & 3.25 & 8.79 & 1.11\\
 & A101 & -- & 0.4444 & 0.958 & 1.849 & $-1.222$ & -- & -- & -- & -- & -- & 0.410 & 49.73 & 1.164 & 111.14 & 123.14 & 138.83 & 120.59 & 2.33 & 7.56 & 0.44\\
Non-flat & Plat. + A101 & -- & 0.4120 & 0.892 & 1.526 & $-1.208$ & -- & 0.335 & $-0.707$ & 0.758 & 11.34 & 0.405 & 49.77 & 1.163 & 144.69 & 164.69 & 194.86 & 162.75 & 2.61 & 8.64 & 0.67\\
XCDM & $H(z)$ + BAO & 0.0295 & 0.0962 & 0.294 & $-0.159$ & $-0.648$ & 65.62 & -- & -- & -- & -- & -- & -- & -- & 18.31 & 28.31 & 37.00 & 28.96 & $-1.35$ & 2.13 & $-1.26$\\
 & HzBP\tnote{d} & 0.0301 & 0.0933 & 0.290 & $-0.164$ & $-0.640$ & 65.39 & 0.324 & $-0.709$ & 0.746 & 12.04 & -- & -- & -- & 51.90 & 69.90 & 92.60 & 70.74 & $-3.43$ & 3.61 & $-1.85$\\
 & HzBPA101\tnote{e} & 0.0299 & 0.0883 & 0.284 & $-0.112$ & $-0.640$ & 64.65 & 0.319 & $-0.714$ & 0.767 & 11.01 & 0.414 & 50.19 & 1.137 & 165.30 & 189.30 & 228.45 & 189.85 & $-1.33$ & 5.19 & $-2.09$\\
\midrule
 & Platinum & -- & 0.4615 & 0.993 & -- & 4.896 & -- & 0.322 & $-0.723$ & 0.744 & 12.14 & -- & -- & -- & 32.99 & 44.99 & 56.46 & 42.37 & 2.00 & 3.91 & $-0.43$\\
 & A118 & -- & 0.2119 & 0.484 & -- & 9.617 & -- & -- & -- & -- & -- & 0.400 & 50.04 & 1.105 & 128.56 & 138.56 & 152.41 & 135.74 & 1.84 & 4.60 & $-0.31$\\
 & A101 & -- & 0.0764 & 0.207 & -- & 9.922 & -- & -- & -- & -- & -- & 0.409 & 49.99 & 1.152 & 112.26 & 122.26 & 135.33 & 119.96 & 1.45 & 4.06 & $-0.20$\\
Flat & Plat. + A101 & -- & 0.0936 & 0.242 & -- & 7.330 & -- & 0.326 & $-0.723$ & 0.755 & 11.62 & 0.410 & 50.04 & 1.135 & 145.75 & 163.75 & 190.90 & 161.62 & 1.67 & 4.68 & $-0.46$\\
\pcdm  & $H(z)$ + BAO & 0.0332 & 0.0789 & 0.265 & -- & 1.455 & 65.24 & -- & -- & -- & -- & -- & -- & -- & 19.49 & 27.49 & 34.44 & 26.96 & $-2.17$ & $-0.43$ & $-3.27$\\
 & HzBP\tnote{d} & 0.0343 & 0.0746 & 0.259 & -- & 1.633 & 64.98 & 0.331 & $-0.714$ & 0.765 & 11.10 & -- & -- & -- & 53.12 & 69.12 & 89.29 & 69.20 & $-4.21$ & 0.30 & $-3.39$\\
 & HzBPA101\tnote{e} & 0.0374 & 0.0663 & 0.246 & -- & 1.968 & 65.20 & 0.327 & $-0.703$ & 0.759 & 11.38 & 0.412 & 50.10 & 1.172 & 166.28 & 188.28 & 224.17 & 187.99 & $-2.35$ & 0.91 & $-3.95$\\
\midrule
 & Platinum & -- & 0.4366 & 0.942 & $-0.905$ & 0.061 & -- & 0.325 & $-0.735$ & 0.725 & 13.16 & -- & -- & -- & 32.53 & 46.53 & 59.92 & 43.06 & 3.54 & 7.37 & 0.26\\
 & A118 & -- & 0.3331 & 0.731 & 0.234 & 5.269 & -- & -- & -- & -- & -- & 0.402 & 50.02 & 1.111 & 128.42 & 140.42 & 157.04 & 136.67 & 3.70 & 9.23 & 0.63\\
 & A101 & -- & 0.2367 & 0.534 & 0.460 & 8.680 & -- & -- & -- & -- & -- & 0.406 & 49.99 & 1.151 & 111.89 & 123.89 & 139.58 & 120.48 & 3.08 & 8.31 & 0.33\\
Non-flat & Plat. + A101 & -- & 0.3063 & 0.677 & 0.317 & 9.746 & -- & 0.319 & $-0.728$ & 0.742 & 12.27 & 0.412 & 49.96 & 1.140 & 145.40 & 165.40 & 195.57 & 162.44 & 3.32 & 9.35 & 0.36\\
\pcdm  & $H(z)$ + BAO & 0.0344 & 0.0786 & 0.263 & $-0.149$ & 2.014 & 65.71 & -- & -- & -- & -- & -- & -- & -- & 18.15 & 28.15 & 36.84 & 27.39 & $-1.51$ & 1.97 & $-2.84$\\
 & HzBP\tnote{d} & 0.0326 & 0.0851 & 0.271 & $-0.176$ & 1.826 & 66.10 & 0.322 & $-0.727$ & 0.752 & 11.78 & -- & -- & -- & 51.79 & 69.79 & 92.49 & 69.44 & $-3.54$ & 3.50 & $-3.14$\\
 & HzBPA101\tnote{e} & 0.0353 & 0.0730 & 0.257 & $-0.135$ & 2.199 & 65.09 & 0.321 & $-0.729$ & 0.758 & 11.52 & 0.403 & 50.16 & 1.144 & 165.08 & 189.08 & 228.23 & 188.65 & $-1.55$ & 4.97 & $-3.29$\\
\bottomrule
\end{tabular}
\begin{tablenotes}[flushleft]
\item [a] In the four GRB-only cases $\Omega_{b}h^2$ is set to 0.0245.
\item [b] \wx\ corresponds to flat/non-flat XCDM and $\alpha$ corresponds to flat/non-flat \pcdm.
\item [c] \hunit. In the four GRB-only cases $H_0$ is set to 70 \hunit.
\item [d] $H(z)$ + BAO + Platinum.
\item [e] $H(z)$ + BAO + Platinum + A101.
\end{tablenotes}
\end{threeparttable}%
}
\end{sidewaystable*}

\begin{sidewaystable*}
\centering
\resizebox*{\columnwidth}{0.74\columnwidth}{%
\begin{threeparttable}
\caption{One-dimensional marginalized posterior mean values and uncertainties ($\pm 1\sigma$ error bars or $2\sigma$ limits) of the parameters for all models from various combinations of data.}\label{tab:1d_BFPC8}
\begin{tabular}{lcccccccccccccc}
\toprule
Model & Data set & $\Omega_{b}h^2$\,\tnote{a} & $\Omega_{c}h^2$ & $\Omega_{\mathrm{m0}}$ & $\Omega_{\mathrm{k0}}$ & $w_{\mathrm{X}}$/$\alpha$\tnote{b} & $H_0$\tnote{c} & $\sigma_{\mathrm{int,\,\textsc{p}}}$ & $a$ & $b$ & $C_{o}$ & $\sigma_{\mathrm{int,\,\textsc{a}}}$ & $\gamma$ & $\beta$ \\
\midrule
 & Platinum & -- & -- & $>0.411$\tnote{d} & -- & -- & -- & $0.347^{+0.033}_{-0.046}$ & $-0.714\pm0.104$ & $0.756\pm0.083$ & $11.52\pm4.45$ & -- & -- & -- \\
 & A118 & -- & -- & $>0.256$ & -- & -- & -- & -- & -- & -- & -- & $0.412^{+0.027}_{-0.033}$ & $50.09\pm0.25$ & $1.109\pm0.089$ \\
 & A101 & -- & -- & $0.584^{+0.298}_{-0.240}$ & -- & -- & -- & -- & -- & -- & -- & $0.422^{+0.030}_{-0.037}$ & $50.03\pm0.28$ & $1.140\pm0.098$ \\
Flat & Plat. + A101 & -- & -- & $0.614^{+0.380}_{-0.130}$ & -- & -- & -- & $0.347^{+0.033}_{-0.046}$ & $-0.717\pm0.102$ & $0.751\pm0.083$ & $11.80\pm4.45$ & $0.421^{+0.029}_{-0.036}$ & $50.03\pm0.27$ & $1.138\pm0.095$ \\
\lcdm & $H(z)$ + BAO & $0.0243\pm0.0029$ & $0.1184^{+0.0077}_{-0.0084}$ & $0.299^{+0.016}_{-0.018}$ & -- & -- & $69.27\pm1.85$ & -- & -- & -- & -- & -- & -- & -- \\
 & HzBP\tnote{e} & $0.0242\pm0.0029$ & $0.1184^{+0.0076}_{-0.0084}$ & $0.299^{+0.016}_{-0.018}$ & -- & -- & $69.24\pm1.81$ & $0.347^{+0.033}_{-0.045}$ & $-0.711\pm0.104$ & $0.762\pm0.081$ & $11.23\pm4.35$ & -- & -- & -- \\
 & HzBPA101\tnote{f} & $0.0242^{+0.0027}_{-0.0030}$ & $0.1187^{+0.0075}_{-0.0083}$ & $0.300^{+0.016}_{-0.018}$ & -- & -- & $69.20\pm1.79$ & $0.347^{+0.032}_{-0.045}$ & $-0.711\pm0.104$ & $0.762\pm0.080$ & $11.23\pm4.32$ & $0.421^{+0.029}_{-0.036}$ & $50.13\pm0.26$ & $1.159\pm0.094$ \\
\midrule
 & Platinum & -- & -- & $>0.491$\tnote{d} & $-0.023^{+0.869}_{-1.401}$ & -- & -- & $0.347^{+0.034}_{-0.046}$ & $-0.727\pm0.109$ & $0.738^{+0.102}_{-0.088}$ & $12.49^{+4.71}_{-5.46}$ & -- & -- & -- \\
 & A118 & -- & -- & $>0.295$ & $0.701^{+0.636}_{-0.844}$ & -- & -- & -- & -- & -- & -- & $0.411^{+0.027}_{-0.033}$ & $50.00\pm0.26$ & $1.123\pm0.088$ \\
 & A101 & -- & -- & $>0.212$ & $0.886^{+0.819}_{-0.614}$ & -- & -- & -- & -- & -- & -- & $0.419^{+0.030}_{-0.037}$ & $49.93\pm0.27$ & $1.156\pm0.096$ \\
Non-flat & Plat. + A101 & -- & -- & $>0.235$ & $0.840^{+0.758}_{-0.720}$ & -- & -- & $0.346^{+0.031}_{-0.044}$ & $-0.711\pm0.100$ & $0.759\pm0.080$ & $11.33^{+4.28}_{-4.27}$ & $0.418^{+0.029}_{-0.036}$ & $49.93\pm0.27$ & $1.154\pm0.095$ \\
\lcdm & $H(z)$ + BAO & $0.0255^{+0.0040}_{-0.0048}$ & $0.1127\pm0.0195$ & $0.293\pm0.025$ & $0.039^{+0.100}_{-0.114}$ & -- & $68.76\pm2.36$ & -- & -- & -- & -- & -- & -- & -- \\
 & HzBP\tnote{e} & $0.0253^{+0.0039}_{-0.0048}$ & $0.1135^{+0.0192}_{-0.0193}$ & $0.294\pm0.024$ & $0.034^{+0.098}_{-0.113}$ & -- & $68.80\pm2.32$ & $0.347^{+0.033}_{-0.046}$ & $-0.710\pm0.104$ & $0.763\pm0.080$ & $11.18\pm4.30$ & -- & -- & -- \\
 & HzBPA101\tnote{f} & $0.0255^{+0.0038}_{-0.0047}$ & $0.1121^{+0.0181}_{-0.0182}$ & $0.293\pm0.023$ & $0.043^{+0.094}_{-0.107}$ & -- & $68.62\pm2.24$ & $0.347^{+0.033}_{-0.045}$ & $-0.710\pm0.102$ & $0.764\pm0.079$ & $11.15\pm4.27$ & $0.421^{+0.029}_{-0.036}$ & $50.13\pm0.26$ & $1.161\pm0.093$ \\
\midrule
 & Platinum & -- & -- & $>0.386$\tnote{d} & -- & $<-0.078$ & -- & $0.347^{+0.033}_{-0.046}$ & $-0.717\pm0.103$ & $0.749\pm0.083$ & $11.90\pm4.45$ & -- & -- & -- \\
 & A118 & -- & -- & $0.599^{+0.360}_{-0.166}$ & -- & $-2.412^{+1.790}_{-1.730}$ & -- & -- & -- & -- & -- & $0.412^{+0.028}_{-0.034}$ & $50.15^{+0.27}_{-0.30}$ & $1.106\pm0.090$ \\
 & A101 & -- & -- & $0.558^{+0.285}_{-0.280}$ & -- & $-2.423^{+1.765}_{-1.716}$ & -- & -- & -- & -- & -- & $0.422^{+0.030}_{-0.037}$ & $50.10^{+0.29}_{-0.33}$ & $1.134\pm0.097$ \\
Flat & Plat. + A101 & -- & -- & $0.576^{+0.382}_{-0.182}$ & -- & $<-0.089$ & -- & $0.346^{+0.033}_{-0.046}$ & $-0.718\pm0.105$ & $0.749\pm0.084$ & $11.94\pm4.54$ & $0.421^{+0.029}_{-0.036}$ & $50.08^{+0.28}_{-0.32}$ & $1.133\pm0.094$ \\
XCDM & $H(z)$ + BAO & $0.0299^{+0.0046}_{-0.0052}$ & $0.0926^{+0.0192}_{-0.0170}$ & $0.283^{+0.023}_{-0.021}$ & -- & $-0.750^{+0.149}_{-0.104}$ & $65.83^{+2.34}_{-2.63}$ & -- & -- & -- & -- & -- & -- & -- \\
 & HzBP\tnote{e} & $0.0301^{+0.0046}_{-0.0053}$ & $0.0917^{+0.0196}_{-0.0171}$ & $0.283^{+0.023}_{-0.021}$ & -- & $-0.743^{+0.153}_{-0.101}$ & $65.73^{+2.31}_{-2.63}$ & $0.347^{+0.033}_{-0.046}$ & $-0.711\pm0.104$ & $0.763\pm0.081$ & $11.24\pm4.32$ & -- & -- & -- \\
 & HzBPA101\tnote{f} & $0.0303^{+0.0047}_{-0.0053}$ & $0.0904^{+0.0196}_{-0.0173}$ & $0.282^{+0.023}_{-0.021}$ & -- & $-0.731^{+0.150}_{-0.096}$ & $65.54^{+2.26}_{-2.58}$ & $0.347^{+0.032}_{-0.046}$ & $-0.711\pm0.102$ & $0.762\pm0.080$ & $11.21\pm4.38$ & $0.420^{+0.029}_{-0.036}$ & $50.14\pm0.26$ & $1.159\pm0.093$ \\
\midrule
 & Platinum & -- & -- & $>0.457$\tnote{d} & $0.011^{+0.821}_{-1.158}$ & $-2.212^{+2.279}_{-1.046}$ & -- & $0.346^{+0.034}_{-0.046}$ & $-0.727\pm0.109$ & $0.736^{+0.102}_{-0.087}$ & $12.62^{+4.65}_{-5.51}$ & -- & -- & -- \\
 & A118 & -- & -- & $>0.257$ & $0.569^{+0.474}_{-0.790}$ & $-2.289^{+1.998}_{-1.087}$ & -- & -- & -- & -- & -- & $0.412^{+0.027}_{-0.033}$ & $50.01\pm0.27$ & $1.120\pm0.088$ \\
 & A101 & -- & -- & $>0.193$ & $0.824^{+0.644}_{-0.749}$ & $-2.338^{+2.038}_{-1.083}$ & -- & -- & -- & -- & -- & $0.421^{+0.030}_{-0.037}$ & $49.91\pm0.30$ & $1.157\pm0.098$ \\
Non-flat & Plat. + A101 & -- & -- & $>0.207$ & $0.743^{+0.618}_{-0.800}$ & $-2.305^{+2.139}_{-1.049}$ & -- & $0.346^{+0.033}_{-0.046}$ & $-0.711\pm0.103$ & $0.759\pm0.081$ & $11.33^{+4.38}_{-4.34}$ & $0.420^{+0.029}_{-0.036}$ & $49.91\pm0.29$ & $1.154\pm0.096$ \\
XCDM & $H(z)$ + BAO & $0.0291^{+0.0055}_{-0.0053}$ & $0.0986^{+0.0218}_{-0.0220}$ & $0.294\pm0.028$ & $-0.116^{+0.135}_{-0.136}$ & $-0.702^{+0.139}_{-0.083}$ & $65.98^{+2.36}_{-2.58}$ & -- & -- & -- & -- & -- & -- & -- \\
 & HzBP\tnote{e} & $0.0292^{+0.0051}_{-0.0057}$ & $0.0983^{+0.0215}_{-0.0218}$ & $0.294\pm0.028$ & $-0.128^{+0.135}_{-0.134}$ & $-0.695^{+0.141}_{-0.080}$ & $65.93^{+2.27}_{-2.53}$ & $0.347^{+0.033}_{-0.046}$ & $-0.713\pm0.103$ & $0.758\pm0.082$ & $11.46\pm4.44$ & -- & -- & -- \\
 & HzBPA101\tnote{f} & $0.0298^{+0.0053}_{-0.0055}$ & $0.0948^{+0.0212}_{-0.0214}$ & $0.290\pm0.027$ & $-0.106\pm0.128$ & $-0.690^{+0.144}_{-0.076}$ & $65.65^{+2.24}_{-2.53}$ & $0.346^{+0.033}_{-0.045}$ & $-0.713\pm0.103$ & $0.758\pm0.081$ & $11.44\pm4.38$ & $0.421^{+0.029}_{-0.036}$ & $50.14\pm0.26$ & $1.151\pm0.094$ \\
\midrule
 & Platinum & -- & -- & $>0.379$\tnote{d} & -- & -- & -- & $0.346^{+0.032}_{-0.044}$ & $-0.715\pm0.102$ & $0.753\pm0.081$ & $11.68\pm4.33$ & -- & -- & -- \\
 & A118 & -- & -- & $0.576^{+0.355}_{-0.202}$ & -- & -- & -- & -- & -- & -- & -- & $0.411^{+0.027}_{-0.033}$ & $50.05\pm0.25$ & $1.110\pm0.089$ \\
 & A101 & -- & -- & $0.506^{+0.239}_{-0.348}$ & -- & -- & -- & -- & -- & -- & -- & $0.421^{+0.030}_{-0.037}$ & $49.99\pm0.27$ & $1.140\pm0.097$ \\
Flat & Plat. + A101 & -- & -- & $0.541^{+0.301}_{-0.297}$ & -- & -- & -- & $0.346^{+0.033}_{-0.045}$ & $-0.716\pm0.103$ & $0.751\pm0.082$ & $11.77\pm4.39$ & $0.420^{+0.029}_{-0.036}$ & $49.99\pm0.26$ & $1.137\pm0.095$ \\
\pcdm & $H(z)$ + BAO & $0.0324^{+0.0060}_{-0.0034}$ & $0.0809^{+0.0180}_{-0.0178}$ & $0.268\pm0.024$ & -- & $1.492^{+0.623}_{-0.854}$ & $65.12\pm2.18$ & -- & -- & -- & -- & -- & -- & -- \\
 & HzBP\tnote{e} & $0.0325^{+0.0071}_{-0.0030}$ & $0.0805^{+0.0177}_{-0.0195}$ & $0.267\pm0.024$ & -- & $1.509^{+0.639}_{-0.904}$ & $65.15^{+2.13}_{-2.34}$ & $0.347^{+0.033}_{-0.045}$ & $-0.711\pm0.104$ & $0.762\pm0.081$ & $11.24\pm4.37$ & -- & -- & -- \\
 & HzBPA101\tnote{f} & $0.0328^{+0.0069}_{-0.0023}$ & $0.0794^{+0.0174}_{-0.0198}$ & $0.266^{+0.024}_{-0.025}$ & -- & $1.568^{+0.652}_{-0.911}$ & $65.03^{+2.10}_{-2.30}$ & $0.347^{+0.033}_{-0.046}$ & $-0.711\pm0.103$ & $0.762\pm0.082$ & $11.23^{+4.39}_{-4.38}$ & $0.420^{+0.029}_{-0.036}$ & $50.14\pm0.26$ & $1.158\pm0.093$ \\
\midrule
 & Platinum & -- & -- & $0.537^{+0.427}_{-0.187}$ & $-0.026^{+0.391}_{-0.384}$ & -- & -- & $0.346^{+0.033}_{-0.046}$ & $-0.717\pm0.105$ & $0.751\pm0.084$ & $11.78^{+4.51}_{-4.52}$ & -- & -- & -- \\
 & A118 & -- & -- & $0.559^{+0.258}_{-0.249}$ & $-0.002^{+0.300}_{-0.292}$ & $5.185^{+3.757}_{-2.564}$ & -- & -- & -- & -- & -- & $0.412^{+0.027}_{-0.033}$ & $50.05\pm0.25$ & $1.110\pm0.090$ \\
 & A101 & -- & -- & $0.477^{+0.199}_{-0.310}$ & $0.107^{+0.306}_{-0.294}$ & $5.205^{+3.974}_{-2.507}$ & -- & -- & -- & -- & -- & $0.421^{+0.030}_{-0.037}$ & $49.99\pm0.27$ & $1.146\pm0.098$ \\
Non-flat & Plat. + A101 & -- & -- & $0.514^{+0.225}_{-0.300}$ & $0.061^{+0.308}_{-0.297}$ & -- & -- & $0.346^{+0.033}_{-0.045}$ & $-0.715\pm0.103$ & $0.754\pm0.083$ & $11.64\pm4.42$ & $0.420^{+0.030}_{-0.036}$ & $49.98\pm0.26$ & $1.142\pm0.097$ \\
\pcdm & $H(z)$ + BAO & $0.0320^{+0.0060}_{-0.0037}$ & $0.0847^{+0.0179}_{-0.0218}$ & $0.272^{+0.025}_{-0.028}$ & $-0.076^{+0.106}_{-0.114}$ & $1.623^{+0.670}_{-0.821}$ & $65.53^{+2.30}_{-2.29}$ & -- & -- & -- & -- & -- & -- & -- \\
 & HzBP\tnote{e} & $0.0321^{+0.0069}_{-0.0031}$ & $0.0847^{+0.0179}_{-0.0226}$ & $0.272^{+0.024}_{-0.029}$ & $-0.090^{+0.092}_{-0.130}$ & $1.674^{+0.697}_{-0.850}$ & $65.57\pm2.27$ & $0.347^{+0.033}_{-0.046}$ & $-0.713\pm0.104$ & $0.760\pm0.081$ & $11.37\pm4.35$ & -- & -- & -- \\
 & HzBPA101\tnote{f} & $0.0324^{+0.0073}_{-0.0024}$ & $0.0829^{+0.0175}_{-0.0223}$ & $0.270^{+0.024}_{-0.029}$ & $-0.079^{+0.104}_{-0.112}$ & $1.714^{+0.713}_{-0.856}$ & $65.39\pm2.24$ & $0.346^{+0.032}_{-0.045}$ & $-0.712\pm0.103$ & $0.760\pm0.082$ & $11.36\pm4.40$ & $0.420^{+0.029}_{-0.036}$ & $50.14\pm0.26$ & $1.153\pm0.094$ \\
\bottomrule
\end{tabular}
\begin{tablenotes}[flushleft]
\item [a] In the four GRB-only cases $\Omega_{b}h^2$ is set to 0.0245.
\item [b] \wx\ corresponds to flat/non-flat XCDM and $\alpha$ corresponds to flat/non-flat \pcdm.
\item [c] \hunit. In the four GRB-only cases $H_0$ is set to 70 \hunit.
\item [d] This is the 1$\sigma$ limit. The 2$\sigma$ limit is set by the prior and not shown here.
\item [e] $H(z)$ + BAO + Platinum.
\item [f] $H(z)$ + BAO + Platinum + A101.
\end{tablenotes}
\end{threeparttable}%
}
\end{sidewaystable*}

\subsection{Constraints from Platinum, A118, A101, and Platinum + A101 data}
\label{PA}

As in \cite{Khadkaetal_2021b} and \cite{CaoKhadkaRatra2022}, in the four GRB-only cases here we set $H_0=70$ \hunit\ and $\Omega_{b}=0.05$.

The constraints on the Platinum GRB correlation parameters in the six different cosmological models are mutually consistent, so the three-dimensional Dainotti correlation Platinum data set is standardizable. The constraints on the intrinsic scatter parameter $\sigma_{\rm int,\,\textsc{p}}$ are almost identical, ranging from $0.346^{+0.032}_{-0.044}$ (flat \pcdm) to $0.347^{+0.034}_{-0.046}$ (non-flat \lcdm). The constraints on the slope $a$ range from a low of $-0.727\pm0.109$ (non-flat \lcdm\ and non-flat XCDM) to a high of $-0.714\pm0.104$ (flat \lcdm), the constraints on the slope $b$ range from a low of $0.736^{+0.102}_{-0.087}$ (non-flat XCDM) to a high of $0.756\pm0.083$ (flat \lcdm), and the constraints on the intercept $C_{o}$ range from a low of $11.52\pm4.45$ (flat \lcdm) to a high of $12.62^{+4.65}_{-5.51}$ (non-flat XCDM), with central values of each pair being $0.09\sigma$, $0.15\sigma$, and $0.16\sigma$ away from each other, respectively. We note that a compilation of (the two-parameter) Dainotti-correlation GRBs, the 31 MD-LGRBs of \cite{Wangetal_2021}, have a somewhat smaller intrinsic dispersion, $\sigma_{\rm int} \sim 0.303-0.306$, Table 5 of \cite{CaoKhadkaRatra2022}, than the $\sigma_{\rm int,\,\textsc{p}} \sim 0.346-0.347$ of the 50 Platinum GRBs.\footnote{There are 12 common GRBs between the Platinum and MD-LGRB data sets (060605, 060906, 061222A, 070306, 080310, 081008, 120404A, 160121A, 160227A, 180329B, 190106A, and 190114A). There are also common GRBs between the Platinum and the (two-parameter) Dainotti-correlated GW-LGRB compilation of \cite{Huetal2021} [091029, 120118B, 131105A, 170202A, 170705A, and 151027A (same name but different redshifts)]. Because of the significant number of overlapping GRBs in these cases, we believe it would not be that useful to perform joint analyses of the Platinum and truncated (to remove the overlapping GRBs) MD-LGRB or GW-LGRB data sets.} A possible reason for the smaller scatter of the 2-parameter MD-LGRB sample is due to how the sample was chosen. In principle one can retain fewer GRBs that lie exactly on the plane, or are much closer to the plane, thus reducing the scatter. However, further investigation is needed in order to draw a definite conclusion. Probably as a consequence of the smaller $\sigma_{\rm int}$, the MD-LGRB constraints on \om\ and \ok\ are more restrictive than the constraints from Platinum data.

Compared with the analyses in \cite{CaoKhadkaRatra2022} where we neglect massive neutrinos (setting $\onh=0$), here we include a non-zero \onhs. The constraints on the A118 GRB correlation parameters here are almost identical to those listed in Table 8 of \cite{CaoKhadkaRatra2022} and are cosmological-model-independent implying that the A118 GRBs are standardizable. For the 118 GRBs A118 data $\sigma_{\rm int,\,\textsc{a}} \sim 0.411-0.412$, larger than the $\sigma_{\rm int,\,\textsc{p}} \sim 0.346-0.347$ of the 50 Platinum GRBs. The constraints on the A101 GRB correlation parameters are also cosmological-model-independent, so A101 GRBs are also standardizable. The constraints on the A101 intrinsic scatter parameter $\sigma_{\rm int,\,\textsc{a}}$ range from a low of $0.419^{+0.030}_{-0.037}$ (non-flat \lcdm) to a high of $0.422^{+0.030}_{-0.037}$ (flat \lcdm\ and flat XCDM), which are slightly ($0.2\sigma$ at most) higher than those from A118 data; the constraints on the slope $\beta$ range from a low of $1.140\pm0.098$ (flat \lcdm) to a high of $1.157\pm0.098$ (non-flat XCDM), which are slightly ($0.28\sigma$ at most) higher than those from A118 data; and the constraints on the intercept $\gamma$ range from a low of $49.91\pm0.30$ (non-flat XCDM) to a high of $50.10^{+0.29}_{-0.33}$ (flat XCDM), which are slightly ($0.25\sigma$ at most) lower than those from A118 data; with central values of each pair being $0.06\sigma$, $0.12\sigma$, and $0.43\sigma$ away from each other, respectively.

Below we discuss cosmological parameter constraints in more detail. From Fig.\ \ref{fig7C8} we see that the overlap of the constraints on the cosmological parameters indicate that Platinum data and A101 data cosmological constraints are mutually consistent and so these two data sets can be jointly analyzed. The Platinum + A101 data combination is also standardizable with cosmological-model-independent constraints on GRB correlation parameters that are consistent (well within $1\sigma$) with those from both Platinum data and A101 data individually.

We find that in the flat \lcdm\ model and the flat XCDM parametrization all GRB data more favor currently accelerating cosmological expansion. In the non-flat \lcdm\ model, all but the Platinum GRBs more favor currently decelerating cosmological expansion. In the non-flat XCDM parametrization, in the $\Ok-\Om$ parameter subspace, all but the Platinum GRBs more favor currently decelerating cosmological expansion, while in the $\wX-\Om$ ($\wX-\Ok$) parameter subspace, all (all but the A101) GRBs more favor currently accelerating cosmological expansion. In the flat and non-flat \pcdm\ models, currently decelerating cosmological expansion is slightly more favored.

Cosmological constraints from Platinum data are less restrictive than those from A118 and A101 data, which contain a little more than twice as many GRBs than the Platinum compilation. Comparing the Platinum + A101 constraints with the Platinum constraints and with the A101 constraints, we see that A101 data, with double the number of GRBs compared to Platinum, play a more dominant role in the combination.

In the flat and non-flat \lcdm\ models, the Platinum, A118, A101, and Platinum + A101 $2\sigma$ constraints on \om\ are mutually consistent and also consistent with those of $H(z)$ + BAO, with the $2\sigma$ lower limits in the flat (non-flat) \lcdm\ model being None (None), $>0.256$ ($>0.295$), $>0.191$ ($>0.212$), and $>0.216$ ($>0.235$), respectively. In the flat \lcdm\ model, the A101 constraint on \om\ is more consistent (than the Platinum and Platinum + A101 constraints) with that from $H(z)$ + BAO data, differing by only $1.18\sigma$. In both models Platinum + A101 data favor a higher $1\sigma$ lower limit of \om\ than do Platinum data or A101 data. In the non-flat \lcdm\ model, the constraints on \ok\ from the four GRB data sets are mutually consistent within $1\sigma$, with all but Platinum data slightly favoring open spatial hypersurfaces. Platinum and A118 data are consistent with flat hypersurfaces within $1\sigma$, while A101 and Platinum + A101 data are $1.44\sigma$ and $1.17\sigma$, respectively, away from flat. The posterior mean value of \ok\ from A101 data is $1.36\sigma$ away from that from $H(z)$ + BAO data, while those from the other three GRB data sets are consistent with that from $H(z)$ + BAO data within $1\sigma$.

In the flat and non-flat XCDM parametrizations, the Platinum, A118, A101, and Platinum + A101 $2\sigma$ constraints on \om\ are mutually consistent and also consistent with those of $H(z)$ + BAO, with the $2\sigma$ lower limits in the flat (non-flat) XCDM parametrization being None (None), $>0.181$ ($>0.257$), $>0.148$ ($>0.193$), and $>0.151$ ($>0.207$), respectively. In flat XCDM, the A101 constraint on \om\ is consistent with that from $H(z)$ + BAO within $1\sigma$. Platinum + A101 data also favor a higher $1\sigma$ lower limit of \om\ than do Platinum or A101 data in both XCDM parametrizations. The constraints on \wx\ are weak, thus affected by the \wx\ prior, and consistent with each other. They mildly favor phantom dark energy, but $\Lambda$ is less than $1\sigma$ away, except for the flat XCDM Platinum and Platinum + A101 cases (where $\Lambda$ is still within $2\sigma$), as is the case for the \wx\ constraints from $H(z)$ + BAO data.\footnote{These are computed 2$\sigma$ constraints, not necessarily twice the 1$\sigma$ ones, and are not shown in the table.} In non-flat XCDM, the constraints on \ok\ from the four GRB data sets are mutually consistent within $1\sigma$, and they slightly favor open hypersurfaces. Platinum, A118, and Platinum + A101 data are consistent with flat hypersurfaces within $1\sigma$, while A101 data are $1.10\sigma$ away from flat. The posterior mean value of \ok\ from A101 data is $1.24\sigma$ away from that from $H(z)$ + BAO data, while those from the other three GRB data sets are consistent with that from $H(z)$ + BAO data within $1\sigma$.

In the flat and non-flat \pcdm\ models, the Platinum, A118, A101, and Platinum + A101 constraints on \om\ are mutually consistent within $1\sigma$ and the A101 and Platinum + A101 (Platinum and A118) constraints are within 1$\sigma$ (2$\sigma$) of those from $H(z)$ + BAO data. Only A118 and A101 data provide (very weak) constraints on $\alpha$ with $\Lambda$ being more than $1\sigma$ away but within $2\sigma$. In non-flat \pcdm, the constraints on \ok\ from the four GRB data sets are mutually consistent and consistent with that from $H(z)$ + BAO data and flat hypersurfaces are within $1\sigma$.

From the $\Delta AIC$ and $\Delta BIC$ values listed in Table \ref{tab:BFPC8}, in the Platinum case, non-flat \lcdm\ is the most favored model, while the evidence against other models are positive, strong, and very strong (non-flat \pcdm). In the A118, A101, and Platinum + A101 cases, the flat \lcdm\ model is the most favored model and, except for non-flat XCDM and non-flat \pcdm\ (with strong $BIC$ evidence against them), the evidence against other models are either weak or positive. However, based on $\Delta DIC$, in theses cases, flat \pcdm\ is the most favored model; in the A118, A101, and Platinum + A101 cases, the evidence against other models are weak; whereas in the Platinum case, the evidence against other models are either weak or strong (non-flat \lcdm\ and non-flat XCDM).

\subsection{Constraints from $H(z)$ + BAO data in combination with Platinum (HzBP) and Platinum + A101 (HzBPA101) data}
\label{HzBPA}

Since the $H(z)$ + BAO, Platinum, and Platinum + A101 cosmological constraints are mutually consistent, we jointly analyze combinations of these data to determine more restrictive constraints on cosmological and GRB correlation parameters. We find that the new constraints on GRB correlation parameters are more restrictive but change less than $1\sigma$ compared to those from the GRB-only analyses. The correlation parameter constraints remain cosmological-model-independent, indicating again that these GRBs are standardizable. In what follows we discuss how the $H(z)$ + BAO data cosmological parameter constraints are altered when these GRB data are jointly analyzed with $H(z)$ + BAO data.

Compared to the $H(z)$ + BAO constraints, the HzBP constraints on \om\ are almost unchanged in all models, but there are small changes in other cosmological parameter constraints. The constraints on $H_0$ are slightly tightened, with posterior means being $0.01-0.03\sigma$ smaller or larger from model to model; the constraints on \wx\ are slightly shifted away from $\Lambda$, with posterior means being $\sim0.04\sigma$ larger; the constraints on $\alpha$ are slightly shifted away from $\Lambda$, with posterior means being $0.02\sigma$ ($0.05\sigma$) larger in flat (non-flat) \pcdm; and the constraints on \ok\ are slightly shifted away from flat, with posterior means being $0.06\sigma$ ($0.10\sigma$) smaller in non-flat XCDM (\pcdm).

Compared to $H(z)$ + BAO, the HzBPA101 data provide mildly different constraints on \om\ in all models, but the effects on other cosmological parameter constraints are more noticeable. The constraints on $H_0$ are slightly tightened, with posterior means being $0.03-0.10\sigma$ smaller from model to model; the constraints on \wx\ are slightly shifted away from $\Lambda$, with posterior means being $0.11\sigma$ ($0.08\sigma$) larger in flat (non-flat) XCDM; the constraints on $\alpha$ are slightly shifted away from $\Lambda$, with posterior means being $0.07\sigma$ ($0.08\sigma$) larger in flat (non-flat) \pcdm; and the constraints on \ok\ are slightly shifted towards (away) from flat, with posterior means being $0.05\sigma$ ($0.02\sigma$) larger (smaller) in non-flat XCDM (\pcdm), respectively.

From the $\Delta AIC$, $\Delta BIC$, and $\Delta DIC$ values listed in Table \ref{tab:BFPC8}, in both the HzBP and the HzBPA101 cases, following the patterns of the $H(z)$ + BAO case, flat \pcdm\ is the most favored model but the evidence against other models are only either weak or positive. As expected, the better-established $H(z)$ + BAO data play the dominant role in these combined analyses and therefore currently accelerating cosmological expansion is favored.

\section{Conclusion}
\label{makereference8.5}

We have analyzed the 50 Platinum GRBs, that obey the three-parameter fundamental plane (or Dainotti) relation, using six different cosmological models. By simultaneously constraining cosmological model and GRB correlation parameters our approach circumvents the circularity problem. We find that the Platinum GRB correlation parameters are cosmological-model-independent so the Platinum sample is standardizable through the three-parameter Dainotti correlation and can be used to constrain cosmological parameters. Since the cosmological constraints from Platinum data are consistent with those from $H(z)$ + BAO data, we have combined Platinum and $H(z)$ + BAO data to perform a joint (HzBP) analysis and find mild changes of the cosmological parameter constraints relative to those from $H(z)$ + BAO data, with the central values in the two cases agreeing to two significant figures.

We have also reanalyzed the A118 GRBs that obey the Amati ($E_{\rm p}-E_{\rm iso}$) correlation \citep{Khadkaetal_2021b,CaoKhadkaRatra2022}, because of different, improved modeling of neutrino physics here, and, by excluding the 17 overlapping (with Platinum) GRBs from the larger A118 data set to form the truncated A101 data set, we have also performed a joint analysis of Platinum and A101 data. The twice as large A101 data set plays a dominant role in the joint Platinum + A101 analysis, and the cosmological constraints from Platinum + A101 data are closer to those from A101 data. The results show that the Platinum + A101 GRBs are also standardizable and their cosmological constraints are also more consistent (than the A118 ones) with those from $H(z)$ + BAO data. The joint analyses of $H(z)$ + BAO and Platinum + A101 (HzBPA101) data show that Platinum + A101 data do have more impact on the joint constraints than do Platinum data, however, $H(z)$ + BAO data still play the dominant role.

Current compilations of GRBs provide some improvements on the cosmological constraints determined using $H(z)$ + BAO data. Importantly, GRBs are the only currently reliable probes of the $z \sim 3-8$ part of cosmological redshift space and so are well worth improving upon. With the upcoming SVOM mission \citep{Cordier2019} in 2023, and possibly Theseus \citep{Amatietal2021} in 2037 if approved in the new ESA call, there soon will be larger GRB data sets over a wider range of redshifts that will allow for improved GRB cosmological constraints.


\cleardoublepage


\chapter{Using lower-redshift, non-CMB, data to constrain the Hubble constant and other cosmological parameters}
\label{makereference9}

This chapter is based on \cite{CaoRatra2022}.

\section{Introduction} 
\label{makereference9.1}

The expansion of the Universe is currently accelerating. This is well-supported by many observations but the underlying theory remains obscure. If general relativity is valid on cosmological scales, a dark energy that has negative pressure is thought to be responsible for the accelerated cosmological expansion. In the well-known spatially-flat \lcdm\ model \citep{peeb84}, dark energy is a cosmological constant $\Lambda$ and contributes $\sim70\%$ of the current cosmological energy budget \citep[see, e.g.][]{Farooq_Ranjeet_Crandall_Ratra_2017, scolnic_et_al_2018, planck2018b, eBOSS_2020}. However, potential observational discrepancies \citep[see, e.g.][]{DiValentinoetal2021a,PerivolaropoulosSkara2021,Abdallaetal2022} motivate consideration of other cosmological models besides flat \lcdm. In our analyses here we also allow for non-zero spatial curvature\footnote{The \textit{Planck} TT,TE,EE+lowE+lensing cosmic microwave background (CMB) anisotropy data favor positive spatial curvature over flatness \citep{planck2018b}.} as well as dark energy dynamics.

Many observations have been used to compare the goodness of fit of cosmological models and determine cosmological parameter constraints. These include CMB anisotropy data \citep[see, e.g.][]{planck2018b} that largely probe the high-redshift, $z\sim1100$, Universe, as well as lower-$z$ cosmological measurements that we make use of here, such as reverberation-measured $\mathrm{H}\beta$ quasar (QSO) and Mg\,\textsc{ii} QSO observations that reach to $z \sim 1.9$ \citep[see, e.g.][]{Czernyetal2021, Zajaceketal2021, Yuetal2021, Khadkaetal_2021a, Khadkaetal2022a}\footnote{Current $\mathrm{H}\beta$ QSO data probe to $z \sim 0.9$ and the resulting cosmological parameter constraints from these data are in $\sim 2\sigma$ tension with those from better-established cosmological probes \citep{Khadkaetal2022a} so we do not use these data in our analyses here.}, Hubble parameter [$H(z)$] data that reach to $z\sim2$ \citep[see, e.g.][]{moresco_et_al_2016, Farooq_Ranjeet_Crandall_Ratra_2017, Ryanetal2019,  CaoRyanRatra2022}, type Ia supernova (SN Ia) observations that reach to $z\sim 2.3$ \citep[see, e.g.][]{scolnic_et_al_2018, DES_2019d}, baryon acoustic oscillation (BAO) measurements that reach to $z\sim 2.3$ \citep[see, e.g.][]{eBOSS_2020, CaoRyanRatra2022}, \hii\ starburst galaxy apparent magnitude data that reach to $z \sim 2.4$ \citep[see, e.g.][]{Mania_2012, Chavez_2014, GonzalezMoranetal2021, CaoRyanRatra2022, Johnsonetal2022, Mehrabietal2022}, QSO angular size (QSO-AS) measurements that reach to $z \sim 2.7$ \citep[see, e.g.][]{Cao_et_al2017b, Ryanetal2019, CaoRyanRatra2020, CaoRyanRatra2022, Zhengetal2021, Lian_etal_2021}, QSO flux observations that reach to $z \sim 7.5$ \citep{RisalitiLusso2015, RisalitiLusso2019, KhadkaRatra2020a, KhadkaRatra2020b, KhadkaRatra2021, KhadkaRatra2022, Lussoetal2020, Yangetal2020, ZhaoXia2021, Lietal2021, Lian_etal_2021, Luongoetal2021, Rezaeietal2022, DainottiBardiacchi2022}\footnote{We do not use these data in this paper since the latest \cite{Lussoetal2020} QSO flux compilation assumes a UV--X-ray correlation model that is invalid above $z \sim 1.5-1.7$ \citep{KhadkaRatra2021, KhadkaRatra2022}.}, and gamma-ray burst (GRB) data that reach to $z \sim 8.2$ \citep[see, e.g.][]{Wang_2016, Wangetal_2021, Dainottietal2016, Dainottietal2017, Dainottietal2020, Dirirsa2019, Amati2019, KhadkaRatra2020c, Huetal2021, Daietal_2021, Demianskietal_2021, Khadkaetal_2021b, LuongoMuccino2021, CaoKhadkaRatra2022, CaoDainottiRatra2022, CaoDainottiRatra2022b, Liuetal2022, DainottiNielson2022}\footnote{Only a subset containing 118 Amati-correlated GRBs are suitable for cosmological purposes \citep{KhadkaRatra2020c, Caoetal_2021, Khadkaetal_2021b}, and these are the Amati-correlated GRBs we use in our analyses here.}.

In this paper, we use most of the aforementioned non-CMB data sets to jointly constrain cosmological parameters. In \cite{CaoRyanRatra2021}, by using $H(z)$ + BAO + SN data (SN refers to Pantheon and DES-3yr SN Ia data, discussed in Sec.\ \ref{makereference9.2} below), we estimated summary values of the current non-relativistic matter density parameter $\Om=0.294\pm0.020$ and the Hubble constant $H_0=68.8\pm1.8$ \hunit. In \cite{CaoRyanRatra2022}, by using $H(z)$ + BAO + SN + QSO-AS + \hiig\ data, summary values of $\Om=0.293\pm0.021$ and $H_0=69.7\pm1.2$ \hunit\ were obtained. Compared to our earlier analysis, the addition of QSO-AS and \hiig\ data results in similar constraints on \om\ with a slightly larger $1\sigma$ uncertainty and more restrictive ($1\sigma$ uncertainty reduced by 50\%) $H_0$ constraints, with a higher central value of $H_0$ ($0.42\sigma$ higher). 

In our analysis here we improve on our earlier work by more correctly accounting for the neutrinos. We also use updated BAO and $H(z)$ data and now also include \mq\ and A118 GRB data.\footnote{We also examined constraints from mutually consistent Platinum + A101 GRB data used in \cite{CaoDainottiRatra2022} and jointly analyzed them with QSO-AS, \hiig, and \mq\ data. Cosmological constraints from the joint QSO-AS + \hiig\ + \mq\ + Platinum + A101 data are similar to those from the QSO-AS + \hiig\ + \mq\ + A118 data, so we decided to perform further analyses with the latter that constrain fewer non-cosmological parameters.} Here the joint analyses of $H(z)$ + BAO + SN + QSO-AS + \hiig\ + \mq\ + A118 data provide model-independent values of $\Om=0.295\pm0.017$ and $H_0=69.7\pm1.2$ \hunit. Our $H_0$ measurement is in better agreement with the median statistics $H_0$ estimate of \cite{chenratmed} than with the local expansion rate $H_0$ estimate of \cite{Riess_2021}. Flat \lcdm\ is favored the most, but mild dark energy dynamics or a little spatial curvature energy density is not ruled out. Although here we use updated $H(z)$ and BAO data, and add \mq\ and A118 data, the constraint on $H_0$ is identical to that of \cite{CaoRyanRatra2022}\footnote{\mq\ and A118 data do not have the power to constrain $H_0$ and updated $H(z)$ and BAO data we use here provide similar constraints to those from older BAO and $H(z)$ data.}, whereas the new constraint on \om\ is more restrictive ($1\sigma$ uncertainty reduced by $\sim24$\%) and $\sim0.15\sigma$ higher.

This paper is organized as follows. In Chapter \ref{sec:models} we described the cosmological models/parametrizations used in our analyses. In Sec.\ \ref{makereference9.2} we describe the data sets used in our analyses, with the methods we use summarized in Sec.\ \ref{makereference9.3}. We discuss our cosmological parameter constraints results in Sec.\ \ref{makereference9.4} and summarize our conclusions in Sec.\ \ref{makereference9.5}.

\section{Data}
\label{makereference9.2}

In this paper we use updated $H(z)$ and BAO data, as well as other data sets, to constrain cosmological parameters. These are summarized next.

\begin{table}
\centering
\begin{threeparttable}
\caption{Updated $H(z)$ data.}\label{tab:hzC9}
\setlength{\tabcolsep}{5pt}
\begin{tabular}{lcc}
\toprule
$z$ & $H(z)$\tnote{a} & Reference\\
\midrule
0.07 & $69.0\pm19.6$ & \cite{73}\\
0.09 & $69.0\pm12.0$ & \cite{69}\\
0.12 & $68.6\pm26.2$ & \cite{73}\\
0.17 & $83.0\pm8.0$ & \cite{69}\\
0.179 & $75.0\pm4.0$ & \cite{70}\\
0.199 & $75.0\pm5.0$ & \cite{70}\\
0.2 & $72.9\pm29.6$ & \cite{73}\\
0.27 & $77.0\pm14.0$ & \cite{69}\\
0.28 & $88.8\pm36.6$ & \cite{73}\\
0.352 & $83.0\pm14.0$ & \cite{70}\\
0.3802 & $83.0\pm13.5$ &  \cite{moresco_et_al_2016}\\
0.4 & $95.0\pm17.0$ & \cite{69}\\
0.4004 & $77.0\pm10.2$ &  \cite{moresco_et_al_2016}\\
0.4247 & $87.1\pm11.2$ &  \cite{moresco_et_al_2016}\\
0.4497 & $92.8\pm12.9$ &  \cite{moresco_et_al_2016}\\
0.47 & $89.0\pm50.0$ & \cite{15}\\
0.4783 & $80.9\pm9.0$ &  \cite{moresco_et_al_2016}\\
0.48 & $97.0\pm62.0$ & \cite{71}\\
0.593 & $104.0\pm13.0$ & \cite{70}\\
0.68 & $92.0\pm8.0$ & \cite{70}\\
0.75 & $98.8\pm33.6$ & \cite{Borghi_etal_2022}\\
0.781 & $105.0\pm12.0$ & \cite{70}\\
0.875 & $125.0\pm17.0$ & \cite{70}\\
0.88 & $90.0\pm40.0$ & \cite{71}\\
0.9 & $117.0\pm23.0$ & \cite{69}\\
1.037 & $154.0\pm20.0$ & \cite{70}\\
1.3 & $168.0\pm17.0$ & \cite{69}\\
1.363 & $160.0\pm33.6$ & \cite{72}\\
1.43 & $177.0\pm18.0$ & \cite{69}\\
1.53 & $140.0\pm14.0$ & \cite{69}\\
1.75 & $202.0\pm40.0$ & \cite{69}\\
1.965 & $186.5\pm50.4$ & \cite{72}\\
\bottomrule
\end{tabular}
\begin{tablenotes}[flushleft]
\item[a] \hunit.
\end{tablenotes}
\end{threeparttable}
\end{table}

\begin{table}
\centering
\begin{threeparttable}
\caption{Updated BAO data.}\label{tab:baoC9}
\setlength{\tabcolsep}{3.5pt}
\begin{tabular}{lccc}
\toprule
$z$ & Measurement\tnote{a} & Value & Reference\\
\midrule
$0.122$ & $D_V\left(r_{s,{\rm fid}}/r_s\right)$ & $539\pm17$ & \cite{Carter_2018}\\
$0.38$ & $D_M/r_s$ & 10.23406 & \cite{eBOSSG_2020}\tnote{b}\\
$0.38$ & $D_H/r_s$ & 24.98058 & \cite{eBOSSG_2020}\tnote{b}\\
$0.51$ & $D_M/r_s$ & 13.36595 & \cite{eBOSSG_2020}\tnote{b}\\
$0.51$ & $D_H/r_s$ & 22.31656 & \cite{eBOSSG_2020}\tnote{b}\\
$0.698$ & $D_M/r_s$ & 17.85823691865007 & \tnote{c}\\
$0.698$ & $D_H/r_s$ & 19.32575373059217 & \tnote{c}\\
$0.81$ & $D_A/r_s$ & $10.75\pm0.43$ & \cite{DES_2019b}\\
$1.48$ & $D_M/r_s$ & 30.6876 & \tnote{d}\\
$1.48$ & $D_H/r_s$ & 13.2609 & \tnote{d}\\
$2.334$ & $D_M/r_s$ & 37.5 & \tnote{e}\\
$2.334$ & $D_H/r_s$ & 8.99 & \tnote{e}\\
\bottomrule
\end{tabular}
\begin{tablenotes}[flushleft]
\item[a] $D_V$, $r_s$, $r_{s, {\rm fid}}$, $D_M$, $D_H$, and $D_A$ have units of Mpc.
\item[b] The four measurements from \cite{eBOSSG_2020} are correlated; see equation \eqref{CovM2C9} for their correlation matrix.
\item[c] The two measurements from \cite{eBOSSG_2020} and \cite{eBOSSL_2021} are correlated; see equation \eqref{CovM3C9} for their correlation matrix.
\item[d] The two measurements from \cite{eBOSSQ_2020} and \cite{eBOSSQ_2021} are correlated; see equation \eqref{CovM4C9} for their correlation matrix.
\item[e] The two measurements from \cite{duMas2020} are correlated; see equation \eqref{CovM1C9} for their correlation matrix.
\end{tablenotes}
\end{threeparttable}
\end{table}

\begin{itemize}

\item[]{$\textbf{ \emph{H(z)}}$ \bf data}. There are 32 $H(z)$ measurements listed in Table \ref{tab:hzC9}, spanning the redshift range $0.07 \leq z \leq 1.965$. Compared with what is given in table 1 of \cite{Ryan_1}, the updated $H(z)$ data here have one additional data point from \cite{Borghi_etal_2022}.

\item[]{\bf BAO data}. There are 12 BAO measurements listed in Table \ref{tab:baoC9}, spanning the redshift range $0.122 \leq z \leq 2.334$. The covariance matrices for given BAO data are summarized below.

The covariance matrix $\textbf{C}$ for BAO data from \cite{duMas2020} is
\be
\label{CovM1C9}
    \begin{bmatrix}
    1.3225 & -0.1009 \\
    -0.1009 & 0.0380
    \end{bmatrix},
\ee
for BAO data from \cite{eBOSSG_2020} $\textbf{C}$ is
\be
\label{CovM2C9}
    \begin{bmatrix}
    0.02860520 & -0.04939281 & 0.01489688 & -0.01387079\\
    -0.04939281 & 0.5307187 & -0.02423513 & 0.1767087\\
    0.01489688 & -0.02423513 & 0.04147534 & -0.04873962\\
    -0.01387079 & 0.1767087 & -0.04873962 & 0.3268589
    \end{bmatrix},
\ee
for BAO data from \cite{eBOSSG_2020} and \cite{eBOSSL_2021} $\textbf{C}$ is
\be
\label{CovM3C9}
    \begin{bmatrix}
    0.1076634008565565 & -0.05831820341302727\\
    -0.05831820341302727 & 0.2838176386340292 
    \end{bmatrix},
\ee
and for BAO data from \cite{eBOSSQ_2020} and \cite{eBOSSQ_2021} $\textbf{C}$ is
\be
\label{CovM4C9}
    \begin{bmatrix}
    0.63731604 & 0.1706891\\
    0.1706891 & 0.30468415
    \end{bmatrix}.
\ee

\item[]{\bf SN Ia data}. As in \cite{CaoRyanRatra2022}, we use SN Ia data that consist of 1048 Pantheon \citep{scolnic_et_al_2018} and 20 binned DES 3yr \citep{DES_2019d} SNe Ia, spanning the redshift ranges $0.01 < z < 2.3$ and $0.015 \leq z \leq 0.7026$, respectively.

\item[]{\bf QSO angular size (QSO-AS) data}. There are 120 QSO-AS measurements listed in table 1 of \cite{Cao_et_al2017b}, spanning the redshift range $0.462 \leq z \leq 2.73$. The measured quantities are $z$ and the angular size $\theta(z)$ with the characteristic linear size of QSOs in the sample, $l_{\rm m}$, as a free parameter to be constrained. The angular size $\theta(z)=l_{\rm m}/D_A(z)$, where $D_A(z)$ is the angular diameter distance. A detailed description of the use of these data can be found in \cite{CaoRyanRatra2022}.

\item[]{\bf \hiig\ data}. There are 181 \hiig\ measurements listed in table A3 of \cite{GonzalezMoranetal2021}, with 107 low-$z$ data from \cite{Chavez_2014} recalibrated by \cite{GonzalezMoran2019}, spanning the redshift range $0.0088 \leq z \leq 0.16417$, and 74 high-$z$ data spanning the redshift range $0.63427 \leq z \leq 2.545$. The measured quantities are $z$, \hiig\ flux $F(\mathrm{H}\beta)$, and velocity dispersion $\sigma$.

\item[]{\bf \mq\ sample}. The \mq\ sample consists of 78 QSOs listed in table A1 of \cite{Khadkaetal_2021a}, spanning the redshift range $0.0033 \leq z \leq 1.89$. \mq\ data obey the radius-luminosity ($R-L$) relation and the measured quantities are the time delay $\tau$ and QSO flux $F_{3000}$ measured at 3000 \(\text{\r{A}}\).

\item[]{\bf A118 sample}. The A118 sample includes 118 long GRBs listed in table 7 of \cite{Khadkaetal_2021b}, spanning the redshift range $0.3399 \leq z \leq 8.2$. A118 data obey the Amati (or $E_{\rm p}-E_{\rm iso}$) correlation and the measured quantities are $z$, rest-frame spectral peak energy $E_{\rm p}$, and measured bolometric fluence $S_{\rm bolo}$, computed in the standard rest-frame energy band $1-10^4$ keV.\footnote{As noted in \citet{Liuetal2022}, the $E_{\rm p}$ value for GRB081121 reported in table 5 of \cite{Dirirsa2019}, and used in our analysis here, is incorrect. One should instead use the correct value provided in table 4 of \cite{Wang_2016}, $E_{\rm p}=871\pm123$ keV. However, since this data point has negligible effect on the cosmological-model and GRB-correlation parameter constraints and the conclusions remain unchanged after correcting it, we do not revise our Amati-correlated GRB results here and in \cite{Caoetal_2021,CaoKhadkaRatra2022,CaoDainottiRatra2022}. In future analyses we will use the correct \cite{Wang_2016} value.}

\item[]{\bf Platinum + A101 sample}. The Platinum sample includes 50 long GRBs listed in table A1 of \cite{CaoDainottiRatra2022}, spanning the redshift range $0.553 \leq z \leq 5.0$. The A101 sample includes 101 long GRBs with common GRBs between the Platinum and the A118 samples excluded, spanning the redshift range $0.3399 \leq z \leq 8.2$. The Platinum GRBs obey the three-dimensional Dainotti correlation and the measured quantities are $z$, characteristic time scale $T^{*}_{X}$, the measured $\gamma$-ray energy flux $F_{X}$ at $T^{*}_{X}$, the prompt peak flux $F_{\rm peak}$ over a 1 s interval, and the X-ray spectral index of the plateau phase $\beta^{\prime}$. 

\end{itemize}

\section{Data Analysis Methodology}
\label{makereference9.3}

In this paper we determine constraints on the cosmological model parameters, and non-cosmological parameters related to different data sets, by maximizing the likelihood function, $\mathcal{L}$. These analyses are performed by using the Markov chain Monte Carlo (MCMC) code \textsc{MontePython} \citep{Audrenetal2013}, with the physics coded in the \textsc{class} code. In Table \ref{tab:priorsC9}, we list the flat prior ranges of the constrained free parameters.

The detailed descriptions for the likelihood functions of $H(z)$, BAO, \hiig, QSO-AS, and SN Ia data can be found in \cite{CaoRyanRatra2020, CaoRyanRatra2021, Caoetal_2021}, whereas those of Platinum, A118/A101, and \mq\ data can be found in \cite{CaoDainottiRatra2022} and \cite{Khadkaetal_2021a}. One can also find the definitions of the Akaike Information Criterion (AIC) and the Bayesian Information Criterion (BIC) as well as the deviance information criterion (DIC) in \cite{CaoDainottiRatra2022}.\footnote{Unlike AIC and BIC, DIC estimates the effective number of free parameters.} We compute $\Delta \mathrm{AIC}$, $\Delta \mathrm{BIC}$, and $\Delta \mathrm{DIC}$ differences for the other five cosmological models relative to the flat \lcdm\ reference model values. Negative (positive) values of $\Delta \mathrm{AIC}$, $\Delta \mathrm{BIC}$, or $\Delta \mathrm{DIC}$ indicate that the model under investigation fits the data compilation better (worse) than does the reference model. Relative to the model with minimum AIC(BIC/DIC), $\Delta \mathrm{AIC(BIC/DIC)} \in (0, 2]$ is defined to be weak evidence against the model under investigation, $\Delta \mathrm{AIC(BIC/DIC)} \in (2, 6]$ is positive evidence against the model under investigation, $\Delta \mathrm{AIC(BIC/DIC)} \in (6, 10] $ is strong evidence against the model under investigation, and $\Delta \mathrm{AIC(BIC/DIC)}>10$ is very strong evidence against the model under investigation.

As in \cite{CaoDainottiRatra2022}, we assume one massive and two massless neutrino species, with the effective number of relativistic neutrino species $N_{\rm eff} = 3.046$ and the total neutrino mass $\sum m_{\nu}=0.06$ eV. Therefore, here the current value of the non-relativistic neutrino physical energy density parameter, $\onh=\sum m_{\nu}/(93.14\ \rm eV)$, is not a free parameter, and along with the current values of the observationally-constrained baryonic (\obhs) and cold dark matter (\ochs) physical energy density parameters, \om\ is derived as $\Om = (\onh + \obh + \och)/{h^2}$, where $h$ is the Hubble constant in units of 100 \hunit.

\begin{table}
\centering
\begin{threeparttable}
\caption{Flat priors of the constrained parameters.}
\label{tab:priorsC9}
\setlength{\tabcolsep}{3.5pt}
\begin{tabular}{lcc}
\toprule
Parameter & & Prior\\
\midrule
 & Cosmological Parameters & \\
\midrule
$H_0$\,\tnote{a} &  & [None, None]\\
\obhs\,\tnote{b} &  & [0, 1]\\
\ochs\,\tnote{c} &  & [0, 1]\\
\ok &  & [-2, 2]\\
$\alpha$ &  & [0, 10]\\
\wx &  & [-5, 0.33]\\
\midrule
 & Non-Cosmological Parameters & \\
\midrule
$k$ &  & [0, 5]\\
$b_{\mathrm{\textsc{m}}}$ &  & [0, 10]\\
$\sigma_{\rm int}$ &  & [0, 5]\\
$a$ &  & [-5, 5]\\
$b_{\mathrm{\textsc{p}}}$ &  & [-5, 5]\\
$C_{o}$ &  & [-50, 50]\\
$\beta$ &  & [0, 5]\\
$\gamma$ &  & [0, 300]\\
$l_{\rm m}$ &  & [None, None]\\
\bottomrule
\end{tabular}
\begin{tablenotes}[flushleft]
\item [a] \hunit. In the \mq\ + A118 case, $H_0$ is set to be 70 \hunit, while in other cases, the prior range is irrelevant (unbounded).
\item [b] In the \mq\ + A118 case, \obhs\ is set to be 0.0245, i.e. $\Omega_{b}=0.05$.
\item [c] In the \mq\ + A118 case, $\Om\in[0,1]$ is ensured.
\end{tablenotes}
\end{threeparttable}%
\end{table}

\section{Results}
\label{makereference9.4}

The posterior one-dimensional probability distributions and two-dimensional confidence regions of the cosmological and non-cosmological parameters are shown in Figs.\ \ref{fig1C9}--\ref{fig6C9}, in red (QSO-AS + \hiig\ and $H(z)$ + BAO + SN), green (QSO-AS + \hiig\ + \mq\ + A118), orange (\mq\ + A118 and QSO-AS + \hiig\ + \mq\ + Platinum + A101, QHMPA101), and blue ($H(z)$ + BAO + SN + QSO-AS + \hiig\ + \mq\ + A118, HzBSNQHMA). The unmarginalized best-fitting parameter values, as well as the corresponding $-2\ln\mathcal{L}_{\rm max}$, AIC, BIC, DIC, $\Delta \mathrm{AIC}$, $\Delta \mathrm{BIC}$, and $\Delta \mathrm{DIC}$ values, for all models and data combinations, are listed in Table \ref{tab:BFPC9}, whereas the marginalized posterior mean parameter values and uncertainties ($\pm 1\sigma$ error bars or $2\sigma$ limits), for all models and data combinations, are listed in Table \ref{tab:1d_BFPC9}.\footnote{We use \textsc{python} package \textsc{getdist} \citep{Lewis_2019} to analyze the samples and generate the plots.}

In the non-flat \lcdm\ and flat and non-flat \pcdm\ models, \mq\ + A118 data mildly favor currently decelerating cosmological expansion, which is most likely caused by the choice of fixed $\Omega_b$ and $H_0$ values. All other data combinations more favor currently accelerating cosmological expansion.

\subsection{Constraints from $H(z)$, BAO, and SN Ia data}
 \label{subsec:HzBSN}

The updated $H(z)$ + BAO results derived here are quite similar to the $H(z)$ + BAO results given in \cite{CaoDainottiRatra2022}, so we do not discuss them in detail. $H(z)$ + BAO + SN is a more important data combination, so here we discuss these constraints in more detail. While the computation of the $H(z)$ + BAO + SN results reported in \cite{CaoRyanRatra2022} neglected the late-time contribution of non-relativistic neutrinos, in this paper, where we account for the contributions of one massive and two massless neutrino species, we find very similar constraints. 

The constraints from $H(z)$ + BAO + SN data on \om\ range from a low of $0.287\pm0.017$ (flat \pcdm) to a high of $0.304^{+0.014}_{-0.015}$ (flat \lcdm), with a difference of $0.75\sigma$. 

The $H_0$ constraints range from a low of $68.29\pm1.78$ \hunit\ (flat \pcdm) to a high of $69.04\pm1.77$ \hunit\ (flat \lcdm), with a difference of $0.30\sigma$, which are $0.09\sigma$ (flat \pcdm) and $0.31\sigma$ (flat \lcdm) higher than the median statistics estimate of $H_0=68\pm2.8$ \hunit\ \citep{chenratmed}, and $2.23\sigma$ (flat \pcdm) and $1.89\sigma$ (flat \lcdm) lower than the local Hubble constant measurement of $H_0 = 73.2 \pm 1.3$ \hunit\ \citep{Riess_2021}.

The constraints on \ok\ are $0.040\pm0.070$, $-0.001\pm0.098$, and $-0.038^{+0.071}_{-0.085}$ for non-flat \lcdm, XCDM, and \pcdm, respectively. Although non-flat hypersurfaces are mildly favored, flat hypersurfaces are well within 1$\sigma$.

There is a slight preference for dark energy dynamics. For flat (non-flat) XCDM, $w_{\rm X}=-0.941\pm0.064$ ($w_{\rm X}=-0.948^{+0.098}_{-0.068}$), with central values being $0.92\sigma$ ($0.76\sigma$) higher than $w_{\rm X}=-1$; and for flat (non-flat) \pcdm, $\alpha=0.324^{+0.122}_{-0.264}$ ($\alpha=0.382^{+0.151}_{-0.299}$), with central values being $1.23\sigma$ ($1.28\sigma$) away from $\alpha=0$.

\subsection{Constraints from QSO-AS, \hiig, \mq, A118, and Platinum + A101 data}
\label{subsec:QHMA}

Given our improved treatment of neutrinos in this paper, compared to our earlier analyses, we have reanalyzed data we had previously studied. 

As shown in \cite{CaoRyanRatra2022}, QSO-AS data alone do not deal well with $H_0$, so an unbounded prior range for $H_0$ makes it hard for the computation to converge and results in an unreasonably high $H_0$ value and so an unreasonably low $l_{\rm m}$ value. However, we expect constraints on the other cosmological parameters consistent with those given in \cite{CaoRyanRatra2022}. Constraints from \hiig\ data are consistent with what are given in \cite{CaoRyanRatra2022}. Constraints from \mq\ data are consistent with those described in \cite{Khadkaetal_2021a} while those from A118 and Platinum + A101 data are consistent with those in \cite{CaoDainottiRatra2022}. We find that cosmological parameter constraints from those four data sets are mutually consistent so they can be used to do joint analyses. As expected, cosmological parameter constraints from the joint QSO-AS + \hiig\ data and \mq\ + A118 data are indeed mutually consistent, as seen in Tables \ref{tab:BFPC9} and \ref{tab:1d_BFPC9}. We do not discuss these results in detail since there are no significant changes compared to those derived in our earlier analyses. We consider the joint analyses results of QSO-AS + \hiig\ + \mq\ + A118 data to be more useful and discuss these in more detail next.\footnote{We note that cosmological parameter constraints from Platinum, A101, and Platinum + A101 data are also consistent with those from QSO-AS, \hiig, and \mq\ data, so we also investigate the joint QSO-AS + \hiig\ + \mq\ + Platinum + A101 (QHPMAP101) data combination. As seen in Tables \ref{tab:BFPC9} and \ref{tab:1d_BFPC9}, we find no significant differences between the QHPMAP101 cosmological constraints and those from the QSO-AS + \hiig\ + \mq\ + A118 data combination that contains fewer non-cosmological parameters.}

The constraints from QSO-AS + \hiig\ + \mq\ + A118 data on \om\ range from a low of $0.175^{+0.075}_{-0.081}$ (flat \pcdm) to a high of $0.314^{+0.051}_{-0.044}$ (flat XCDM), with a difference of $1.60\sigma$. Following the pattern of \hiig\ data, the \om\ difference is relatively large.

The $H_0$ constraints range from a low of $70.38\pm1.84$ \hunit\ (non-flat \pcdm) to a high of $73.14^{+2.14}_{-2.48}$ \hunit\ (flat XCDM), with a difference of $0.89\sigma$, which are $0.71\sigma$ (non-flat \pcdm) and $1.37\sigma$ (flat XCDM) higher than the median statistics estimate of $H_0=68\pm2.8$ \hunit\ \citep{chenratmed}, and $1.25\sigma$ (non-flat \pcdm) and $0.02\sigma$ (flat XCDM) lower than the local Hubble constant measurement of $H_0 = 73.2 \pm 1.3$ \hunit\ \citep{Riess_2021}.

The constraints on \ok\ are $-0.139^{+0.116}_{-0.228}$, $0.054^{+0.227}_{-0.238}$, and $0.044^{+0.104}_{-0.256}$ for non-flat \lcdm, XCDM, and \pcdm, respectively. As opposed to $H(z)$ + BAO + SN results, non-flat \lcdm\ mildly favors closed hypersurfaces, whereas non-flat XCDM and non-flat \pcdm\ mildly favor open hypersurfaces. However, flat hypersurfaces are well within 1$\sigma$.

There are mild preferences for dark energy dynamics. For flat (non-flat) XCDM, $w_{\rm X}=-1.836^{+0.804}_{-0.419}$ ($w_{\rm X}=-2.042^{+1.295}_{-0.451}$), with central values being $1.04\sigma$ ($0.80\sigma$) lower than $w_{\rm X}=-1$; and for flat (non-flat) \pcdm, $\alpha<6.756$ ($\alpha<7.239$), with $\alpha=0$ being within $1\sigma$.

\subsection{Constraints from $H(z)$ + BAO + SN + QSO-AS + \hiig\ + \mq\ + A118 (HzBSNQHMA) data}
\label{subsec:HzBSNQHMA}

\begin{figure*}
\centering
 \subfloat[]{%
    \includegraphics[width=0.5\textwidth,height=0.5\textwidth]{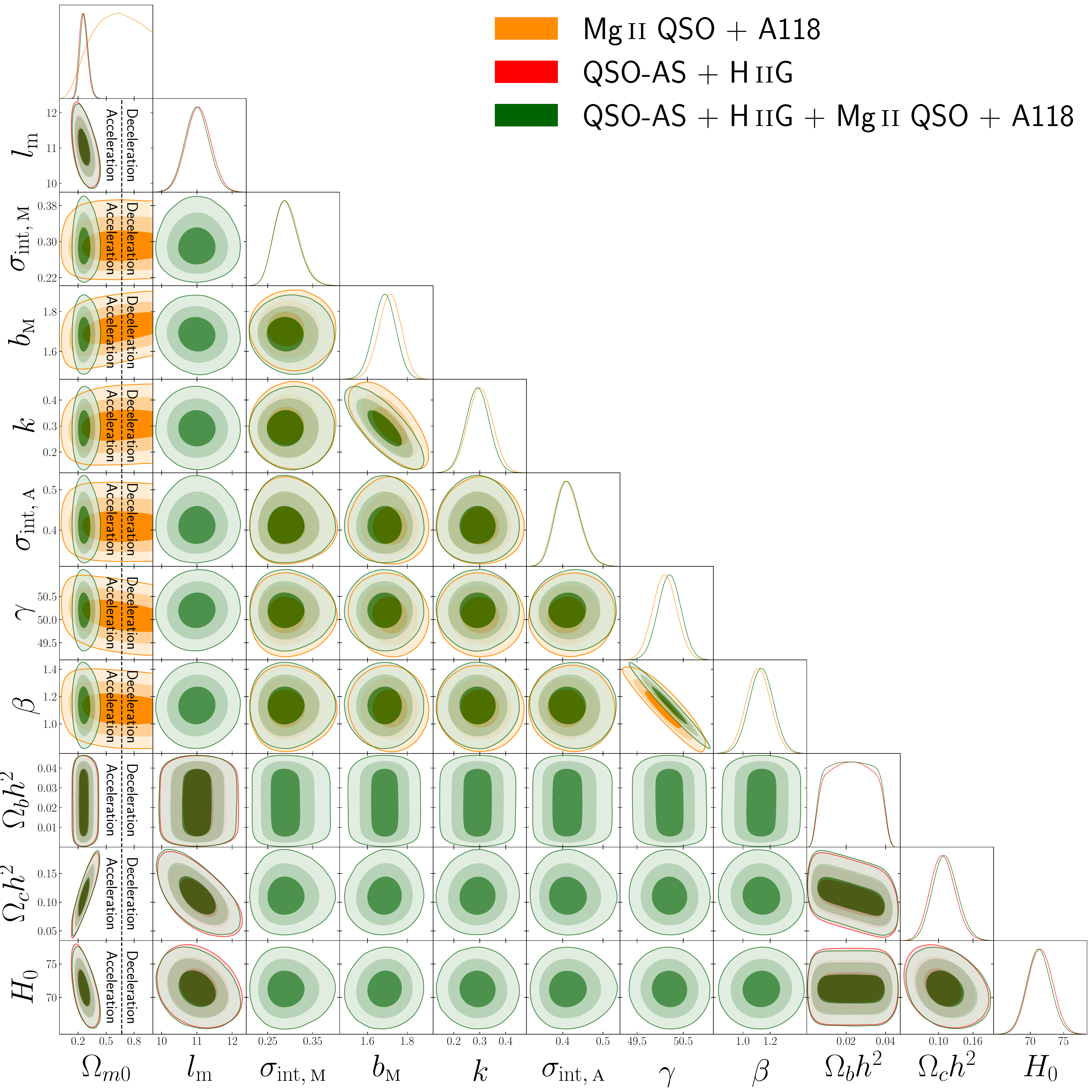}}
 \subfloat[]{%
    \includegraphics[width=0.5\textwidth,height=0.5\textwidth]{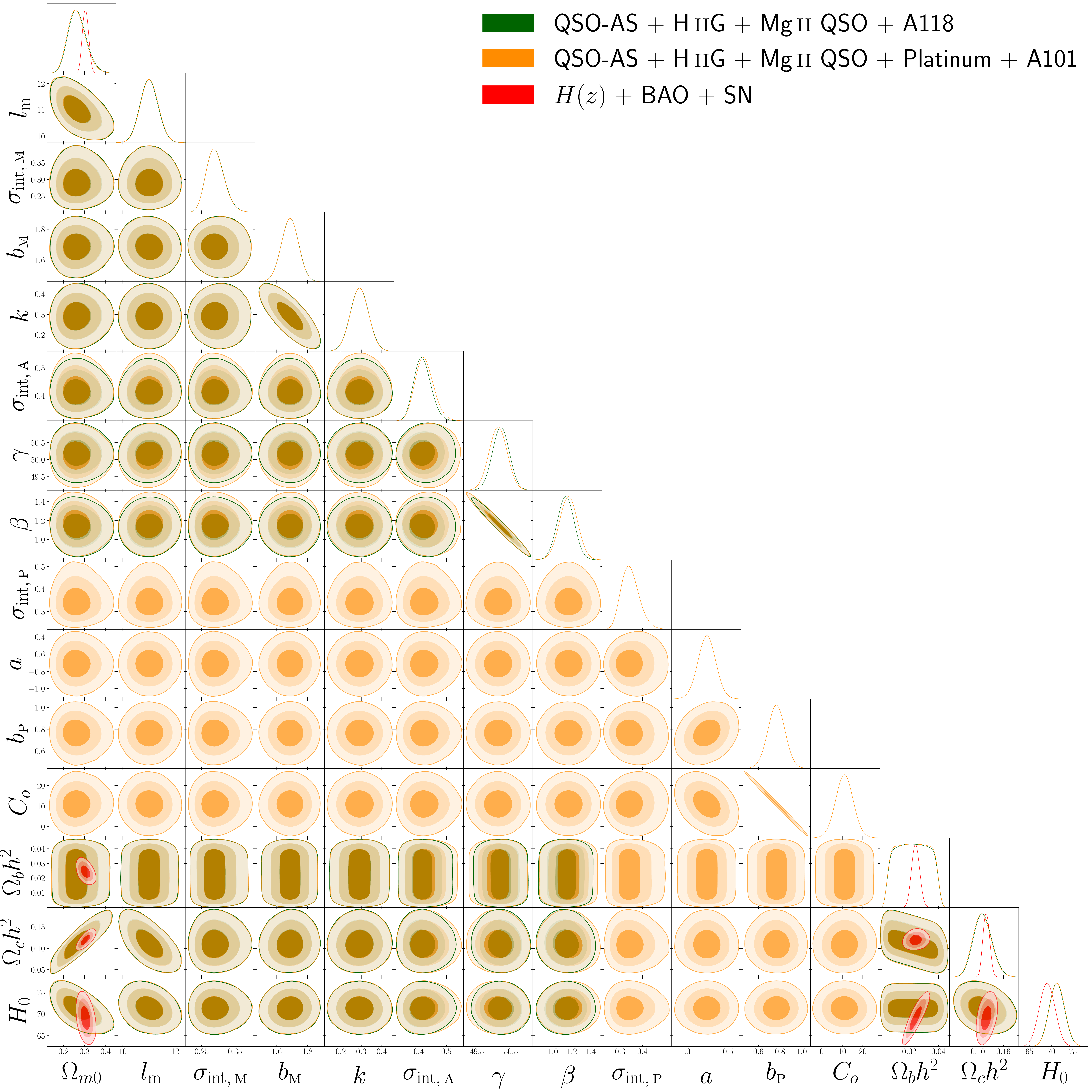}}\\
 \subfloat[]{%
    \includegraphics[width=0.5\textwidth,height=0.5\textwidth]{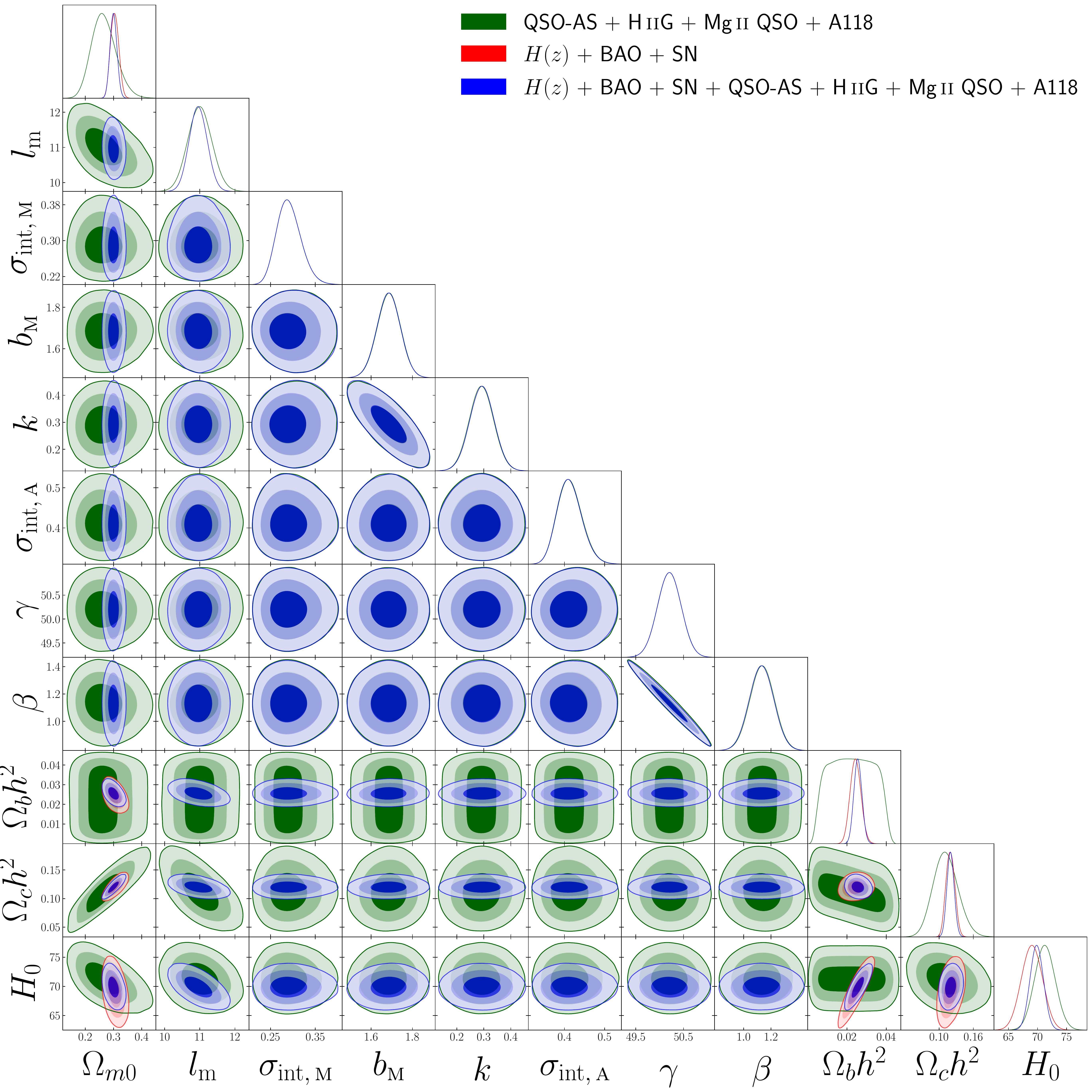}}
 \subfloat[]{%
    \includegraphics[width=0.5\textwidth,height=0.5\textwidth]{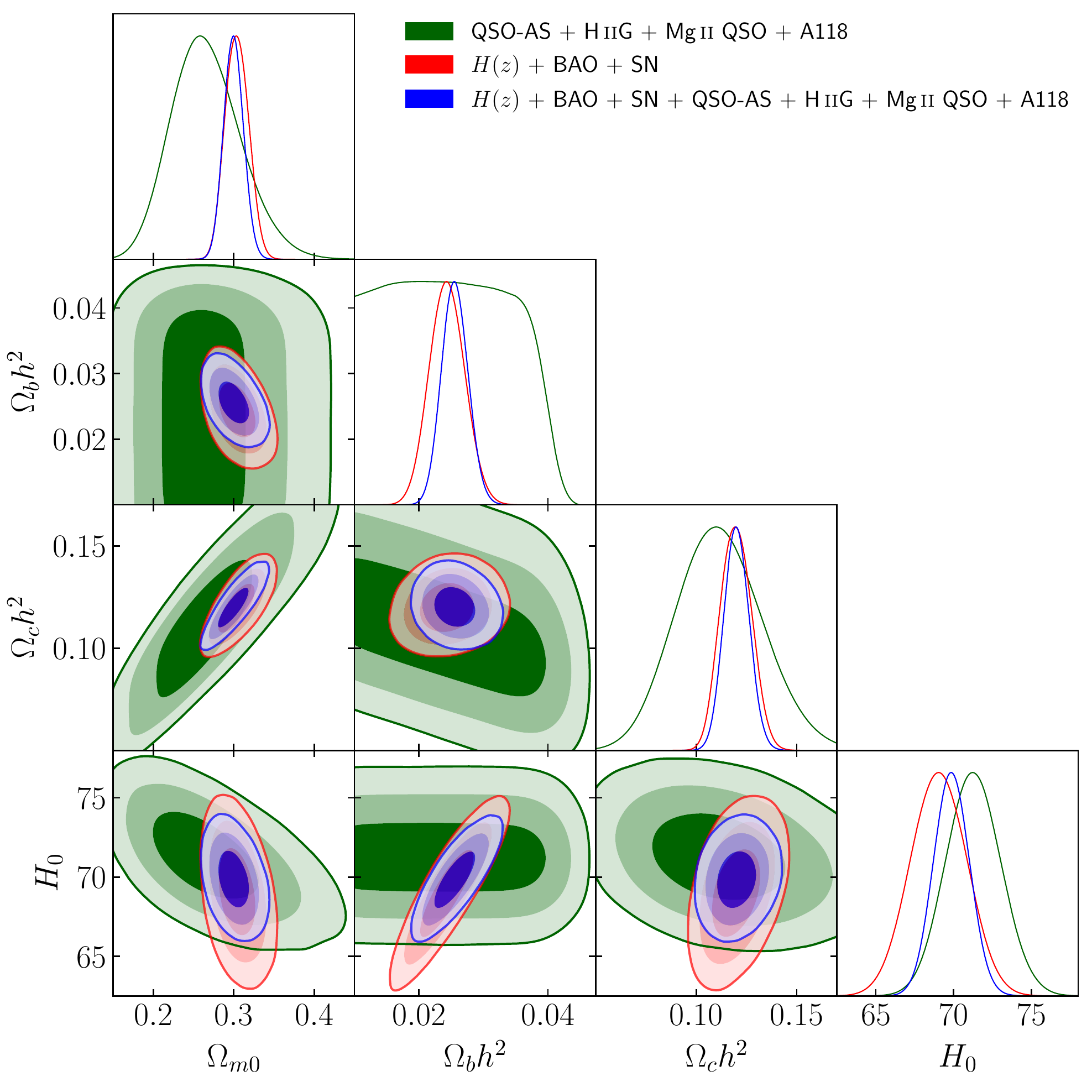}}\\
\caption{One-dimensional likelihood distributions and 1$\sigma$, 2$\sigma$, and 3$\sigma$ two-dimensional likelihood confidence contours for flat \lcdm\ from various combinations of data. The zero-acceleration black dashed lines in some (a) panels divide the parameter space into regions associated with currently-accelerating (left) and currently-decelerating (right) cosmological expansion.}
\label{fig1C9}
\end{figure*}

\begin{figure*}
\centering
 \subfloat[]{%
    \includegraphics[width=0.5\textwidth,height=0.5\textwidth]{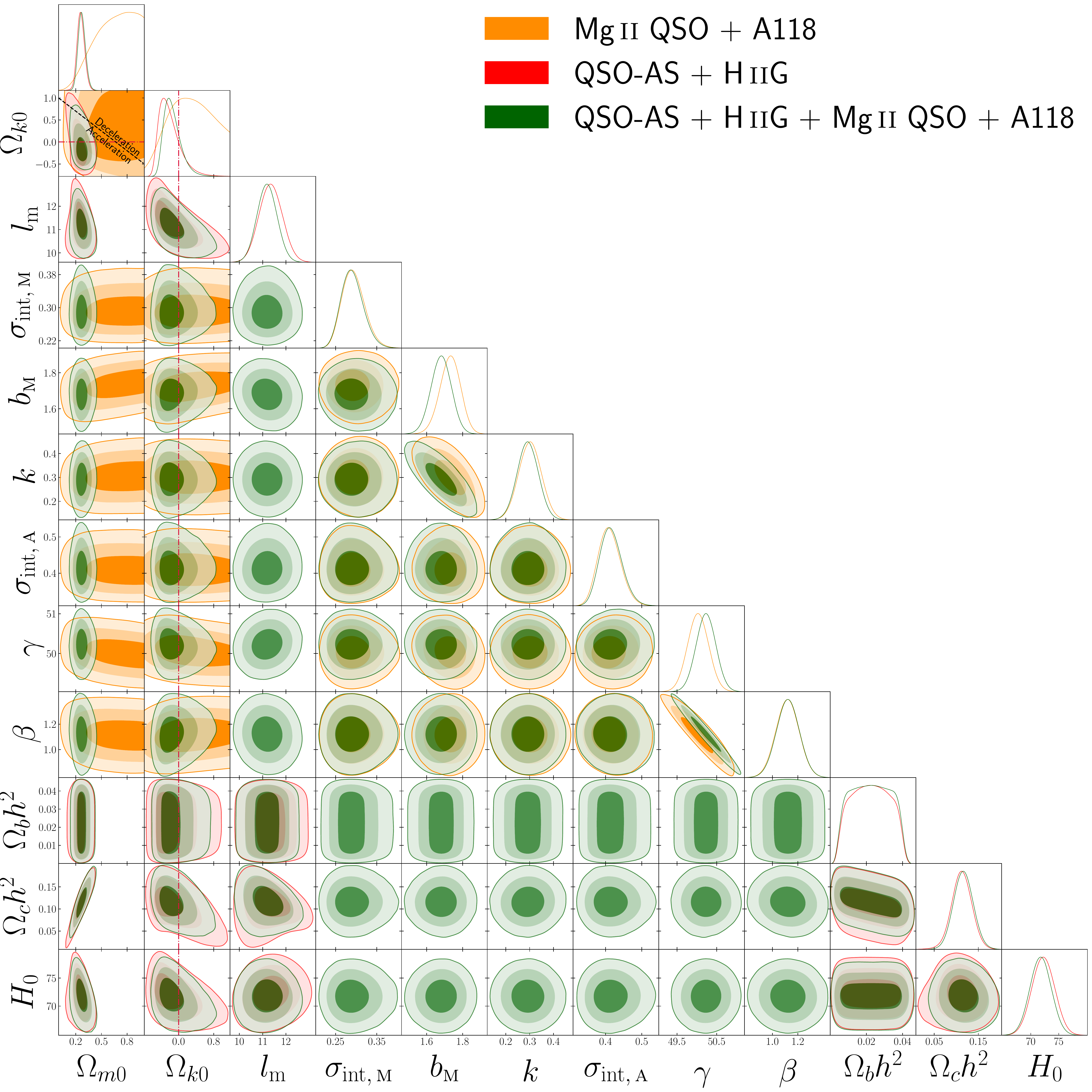}}
 \subfloat[]{%
    \includegraphics[width=0.5\textwidth,height=0.5\textwidth]{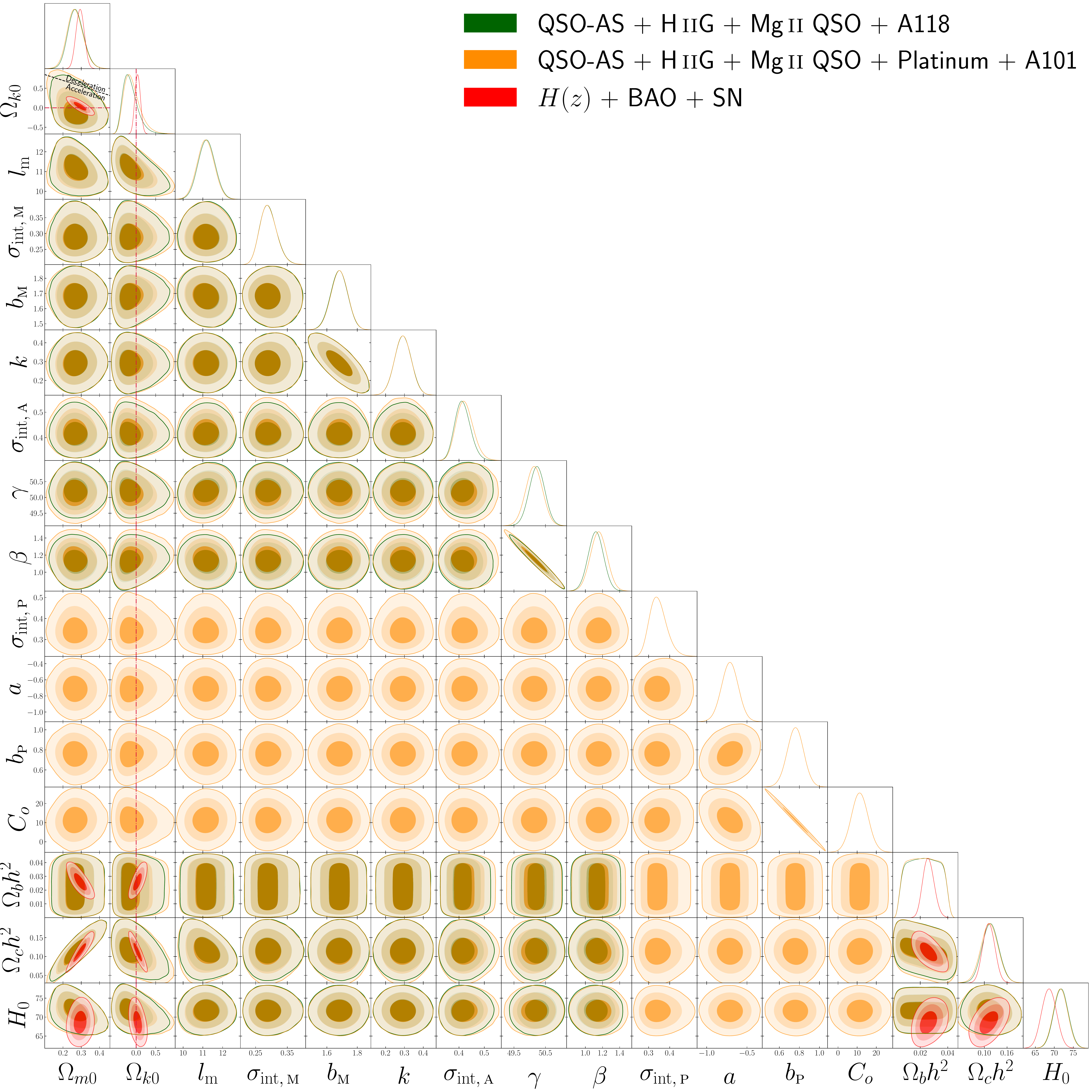}}\\
 \subfloat[]{%
    \includegraphics[width=0.5\textwidth,height=0.5\textwidth]{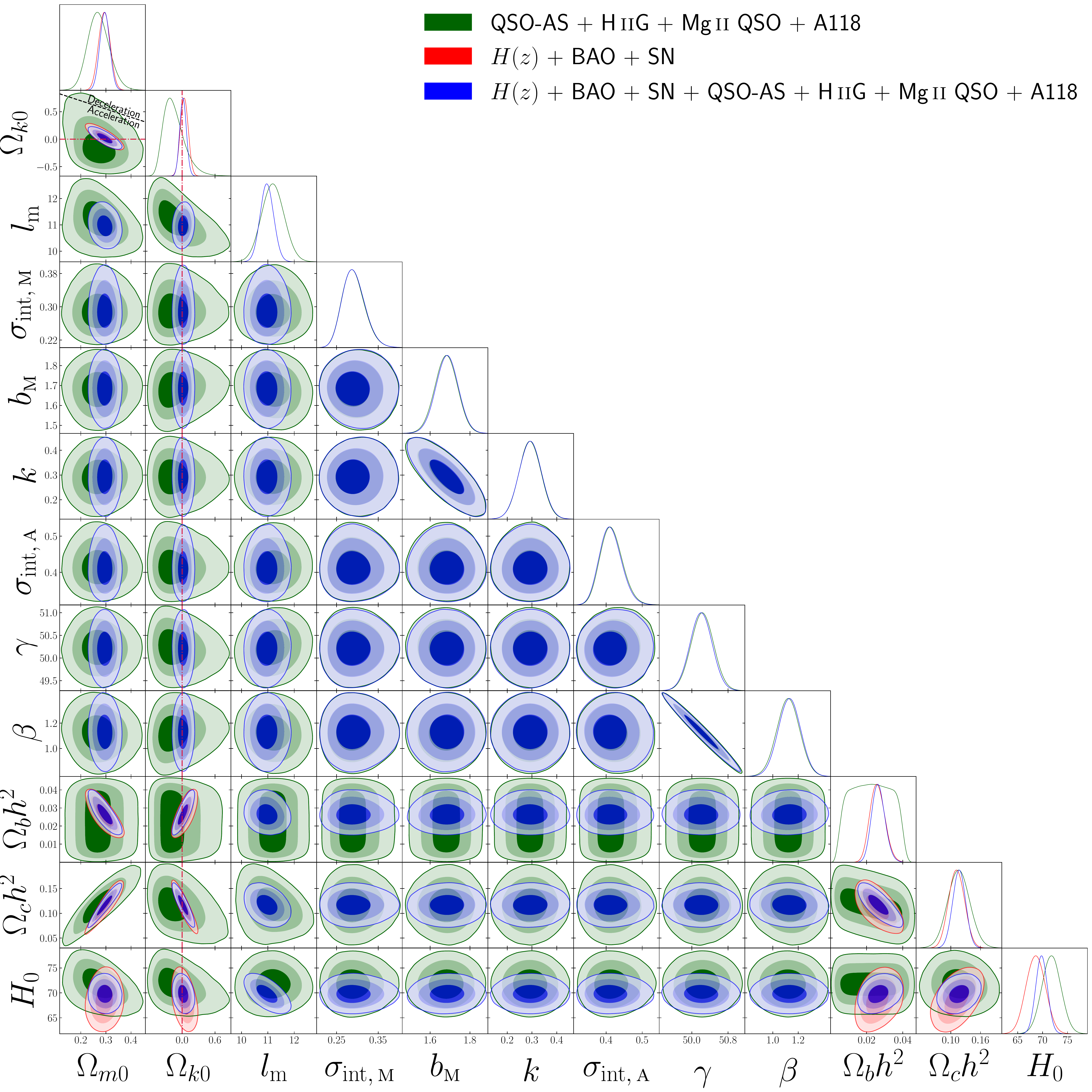}}
 \subfloat[]{%
    \includegraphics[width=0.5\textwidth,height=0.5\textwidth]{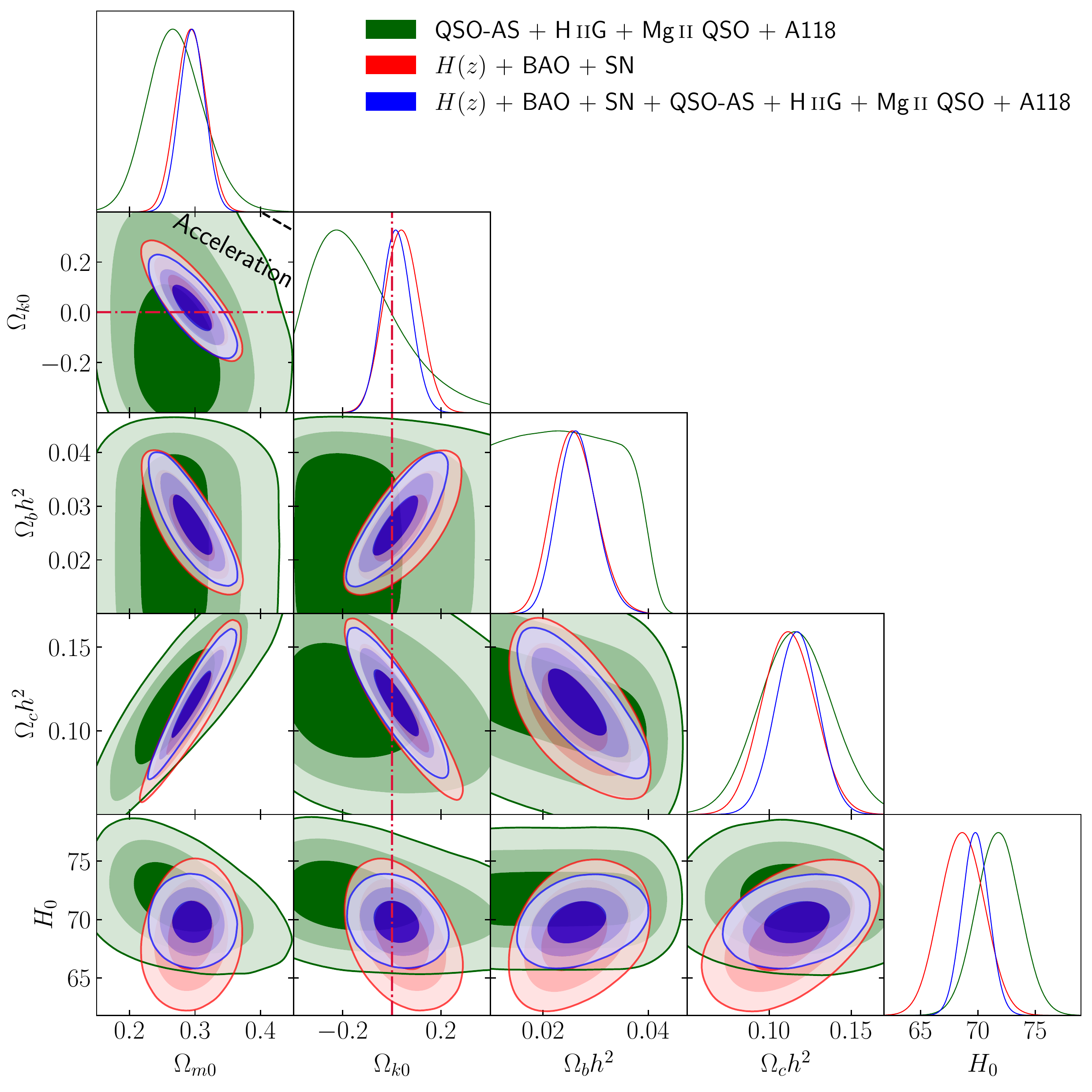}}\\
\caption{Same as Fig.\ \ref{fig1C9} but for non-flat \lcdm. The zero-acceleration black dashed lines divide the parameter space into regions associated with currently-accelerating (below left) and currently-decelerating (above right) cosmological expansion.}
\label{fig2C9}
\end{figure*}

\begin{figure*}
\centering
 \subfloat[]{%
    \includegraphics[width=0.5\textwidth,height=0.5\textwidth]{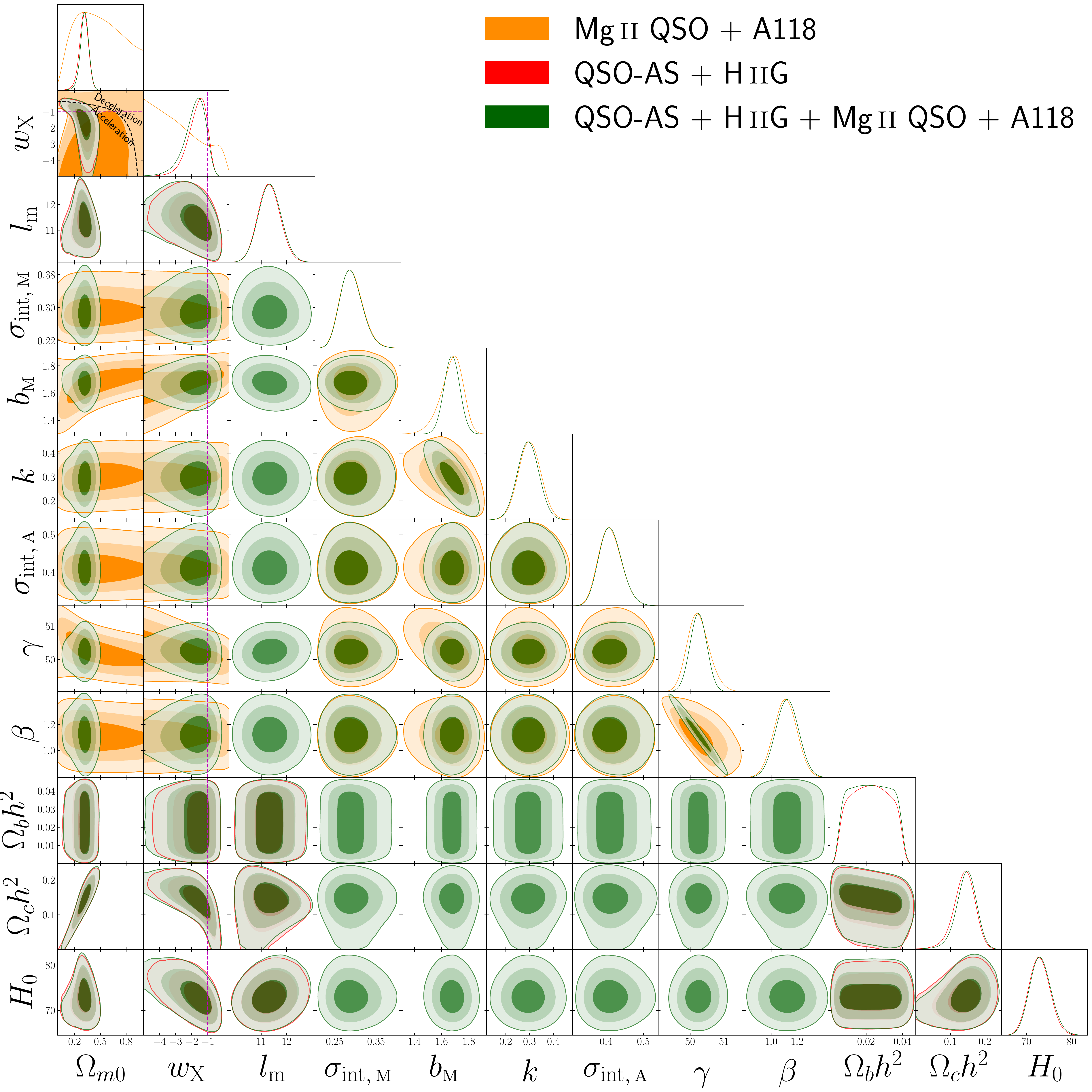}}
 \subfloat[]{%
    \includegraphics[width=0.5\textwidth,height=0.5\textwidth]{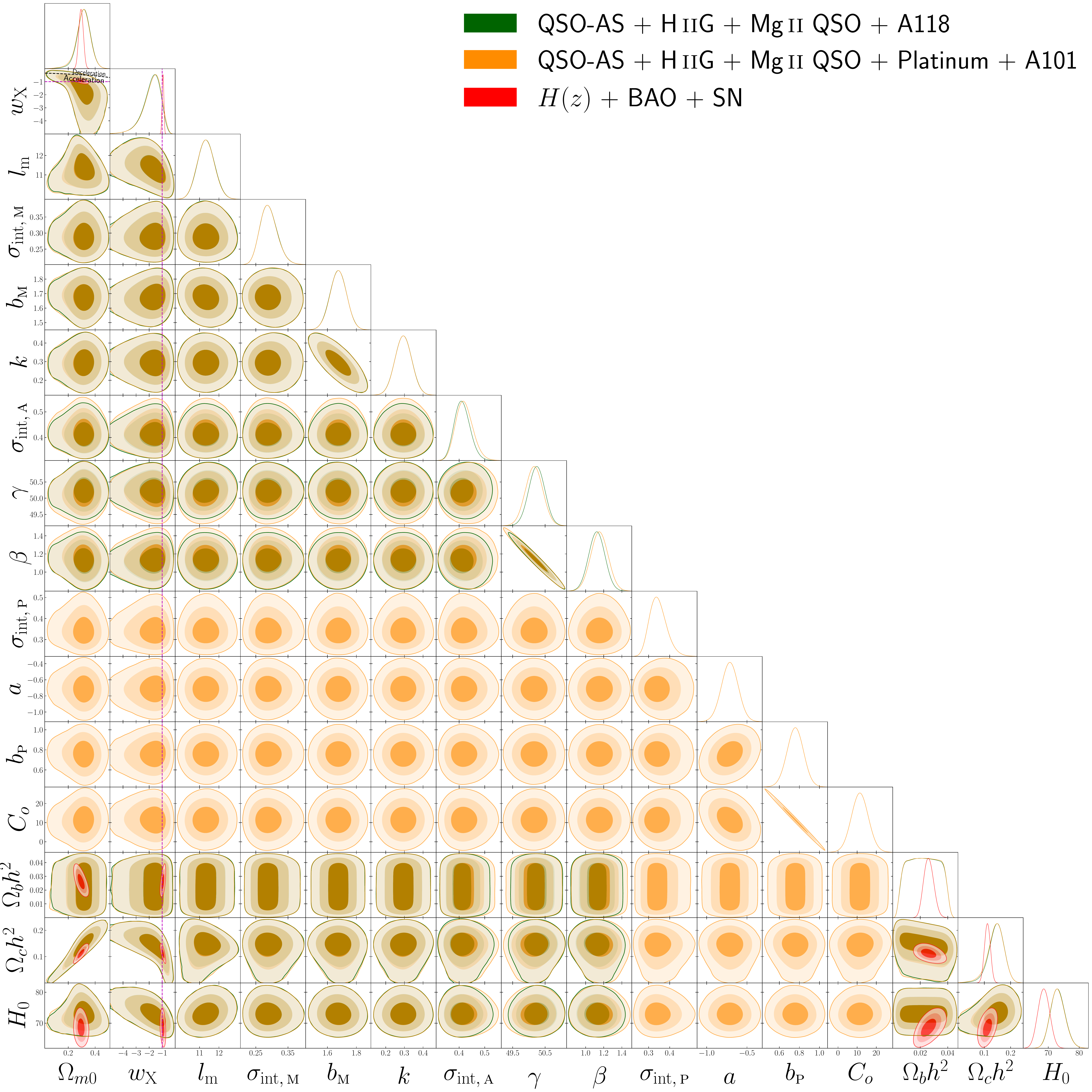}}\\
 \subfloat[]{%
    \includegraphics[width=0.5\textwidth,height=0.5\textwidth]{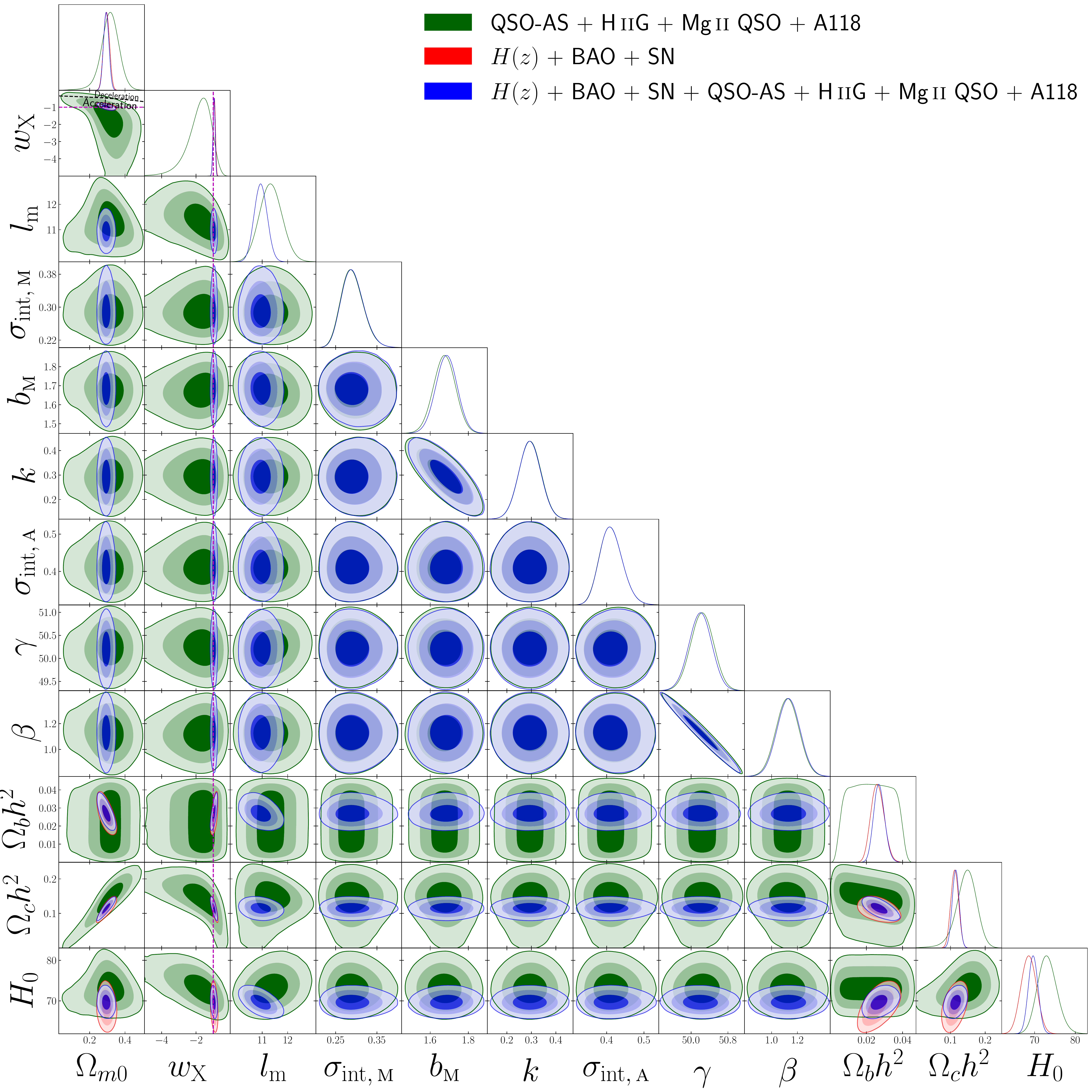}}
 \subfloat[]{%
    \includegraphics[width=0.5\textwidth,height=0.5\textwidth]{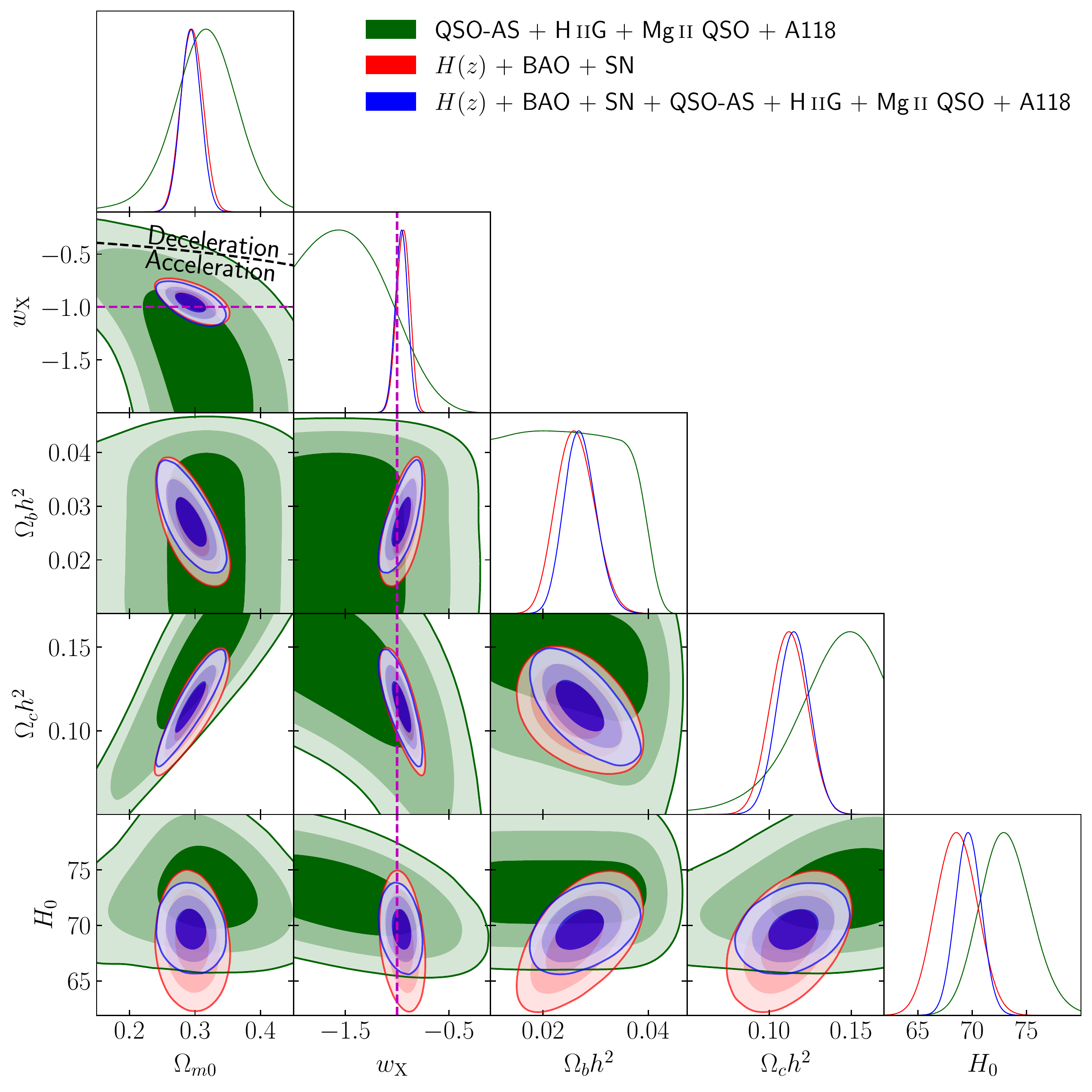}}\\
\caption{One-dimensional likelihood distributions and 1$\sigma$, 2$\sigma$, and 3$\sigma$ two-dimensional likelihood confidence contours for flat XCDM from various combinations of data. The zero-acceleration black dashed lines divide the parameter space into regions associated with currently-accelerating (either below left or below) and currently-decelerating (either above right or above) cosmological expansion. The magenta dashed lines represent $w_{\rm X}=-1$, i.e.\ flat \lcdm.}
\label{fig3C9}
\end{figure*}

\begin{figure*}
\centering
 \subfloat[]{%
    \includegraphics[width=0.5\textwidth,height=0.5\textwidth]{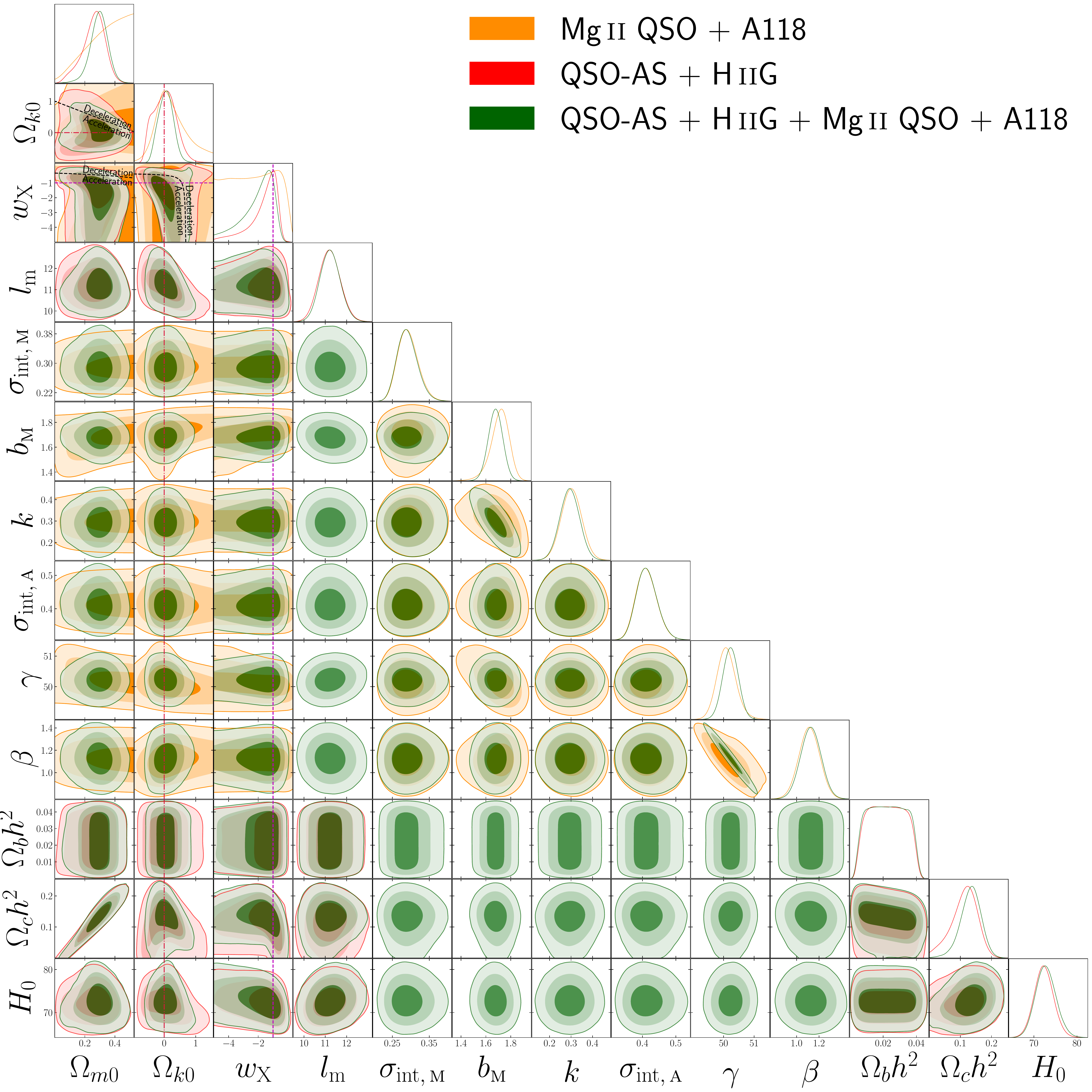}}
 \subfloat[]{%
    \includegraphics[width=0.5\textwidth,height=0.5\textwidth]{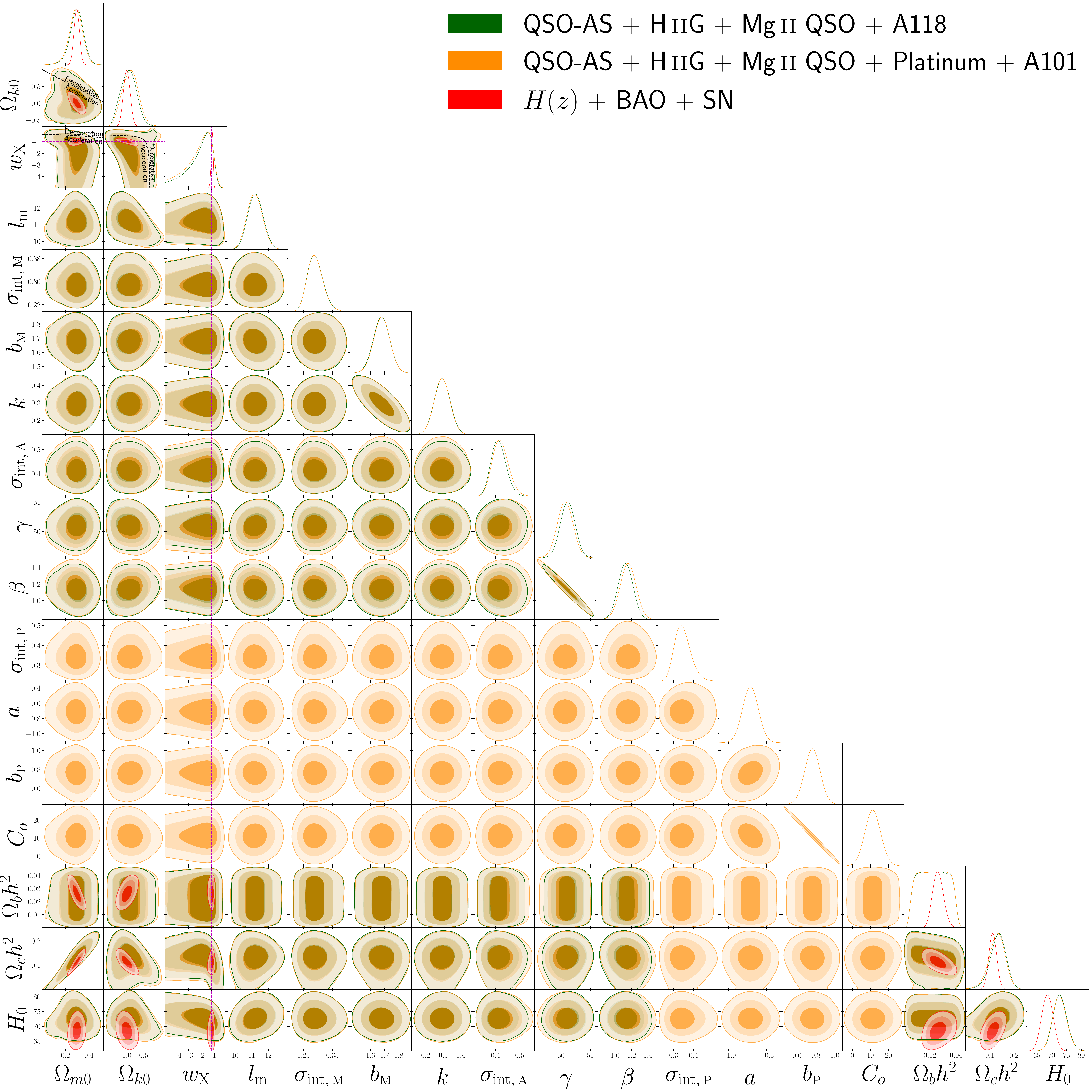}}\\
 \subfloat[]{%
    \includegraphics[width=0.5\textwidth,height=0.5\textwidth]{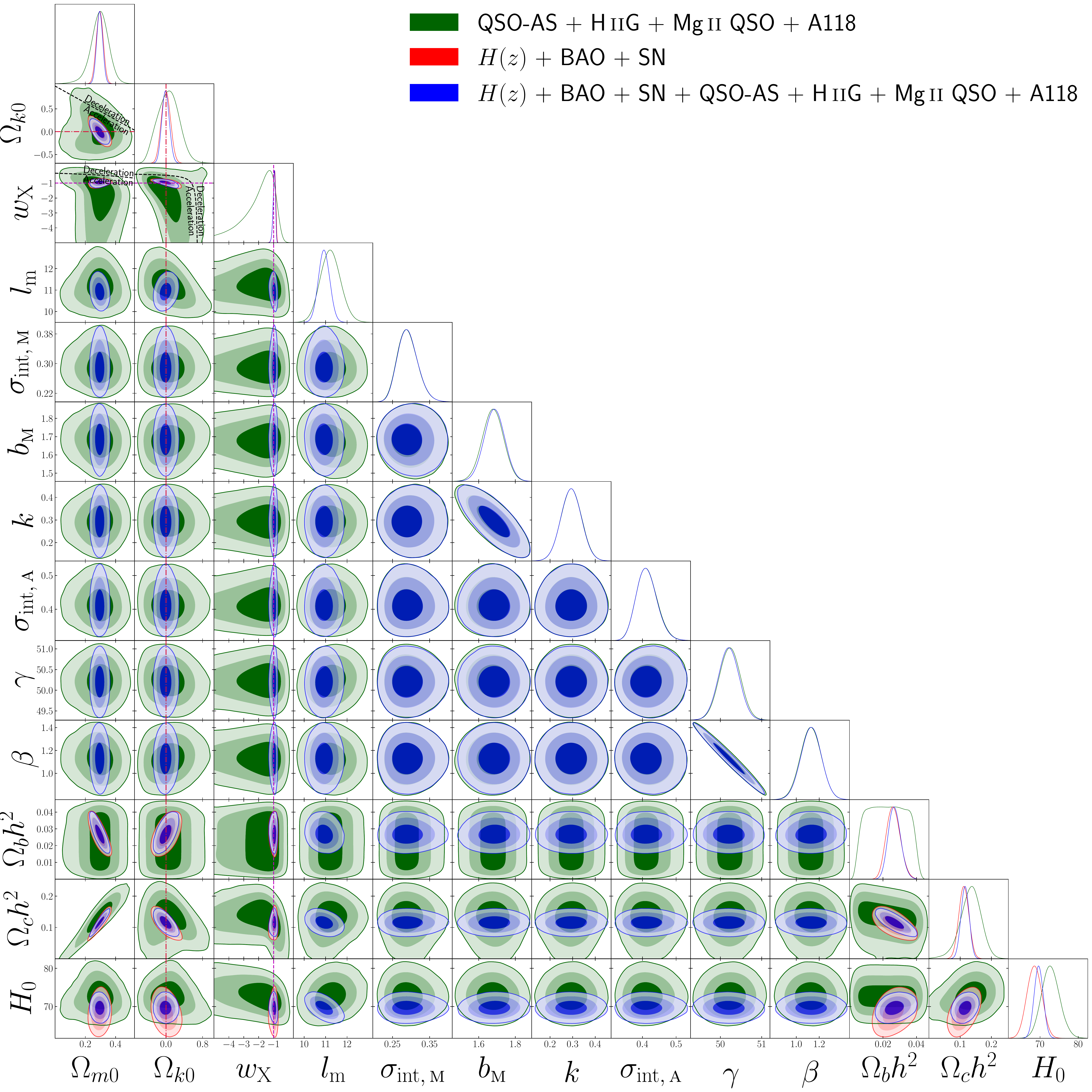}}
 \subfloat[]{%
    \includegraphics[width=0.5\textwidth,height=0.5\textwidth]{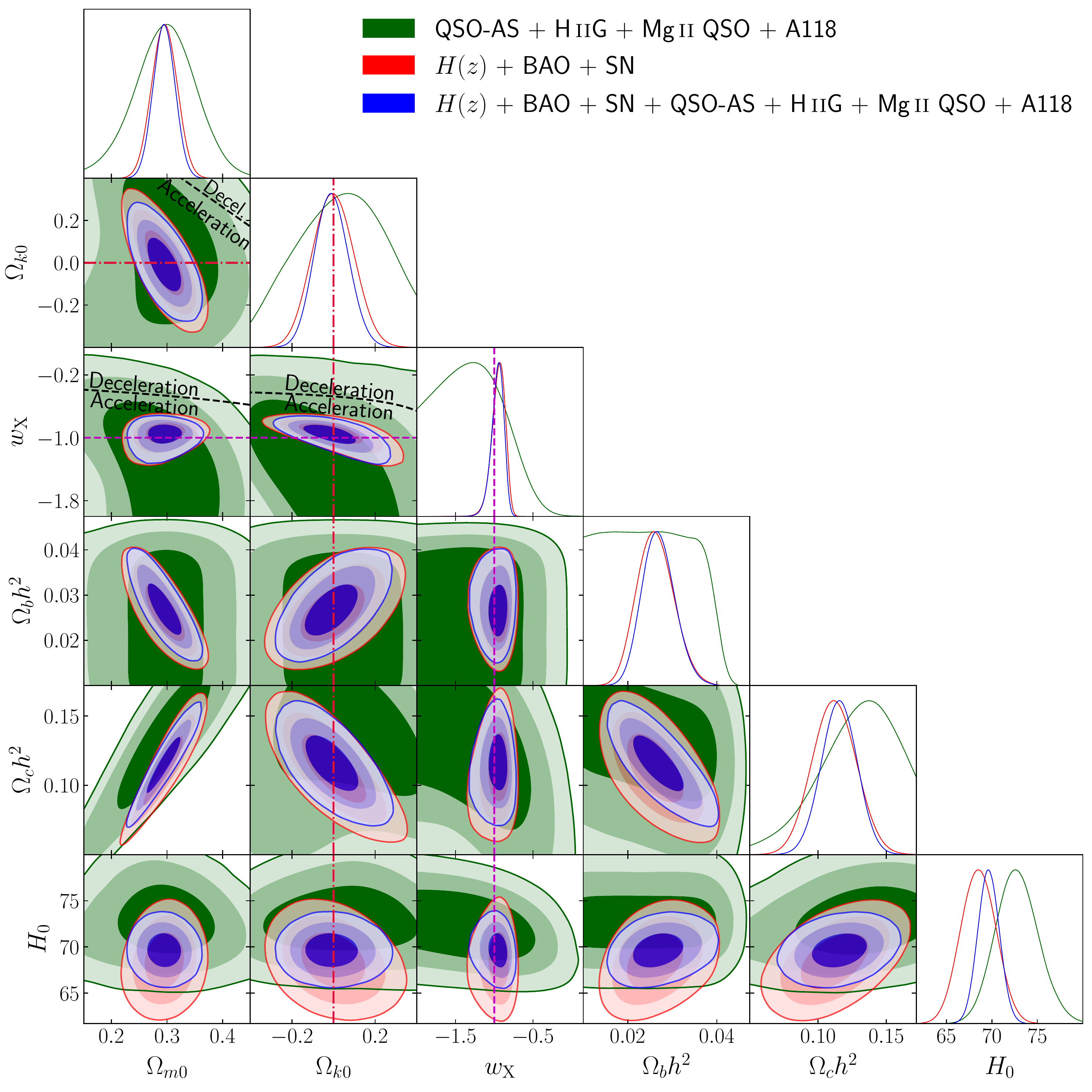}}\\
\caption{Same as Fig.\ \ref{fig3C9} but for non-flat XCDM. The zero-acceleration black dashed lines are computed for the third cosmological parameter set to the $H(z)$ + BAO data best-fitting values listed in Table \ref{tab:BFPC9}, and divide the parameter space into regions associated with currently-accelerating (either below left or below) and currently-decelerating (either above right or above) cosmological expansion. The crimson dash-dot lines represent flat hypersurfaces, with closed spatial hypersurfaces either below or to the left. The magenta dashed lines represent $w_{\rm X}=-1$, i.e.\ non-flat \lcdm.}
\label{fig4C9}
\end{figure*}

\begin{figure*}
\centering
\centering
 \subfloat[]{%
    \includegraphics[width=0.5\textwidth,height=0.5\textwidth]{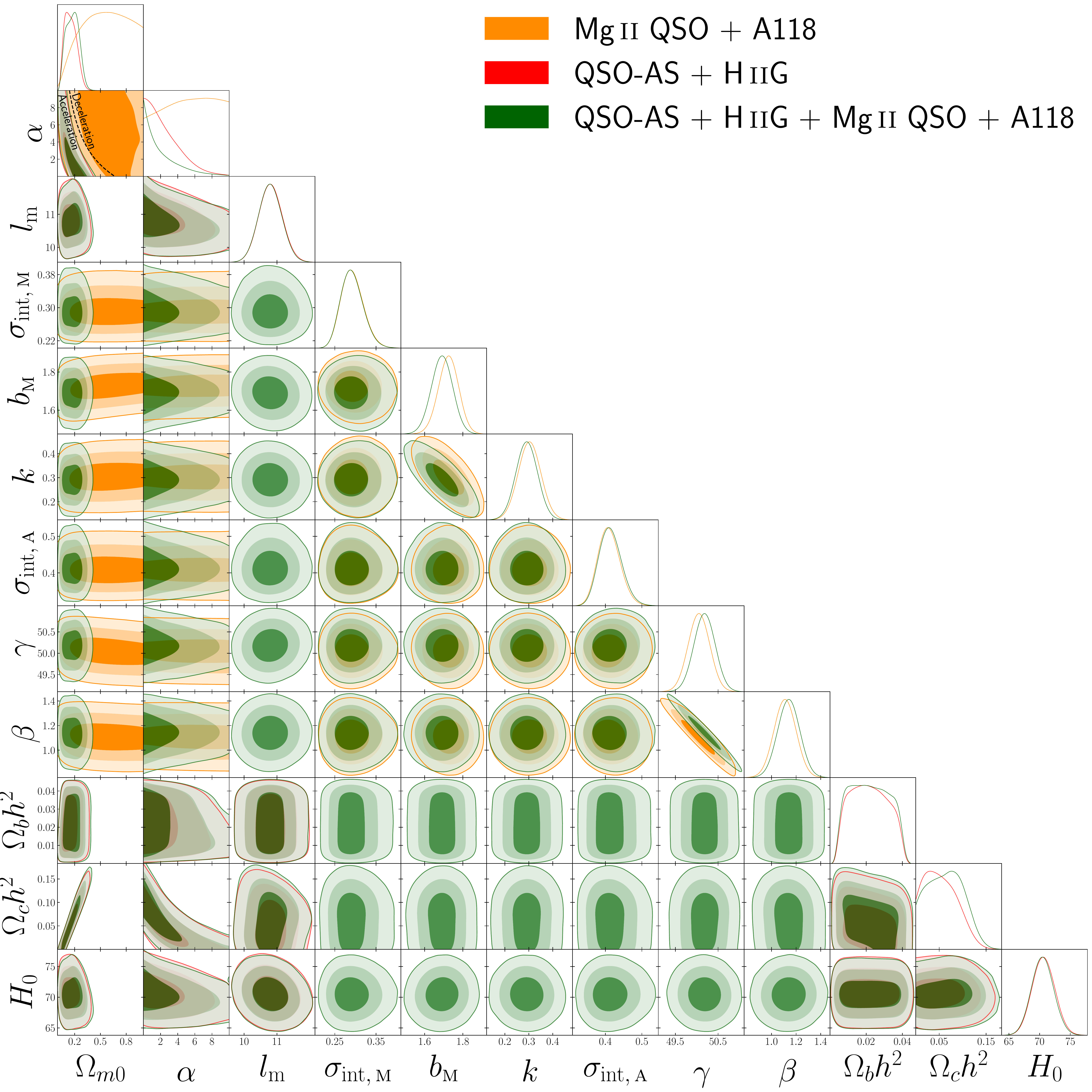}}
 \subfloat[]{%
    \includegraphics[width=0.5\textwidth,height=0.5\textwidth]{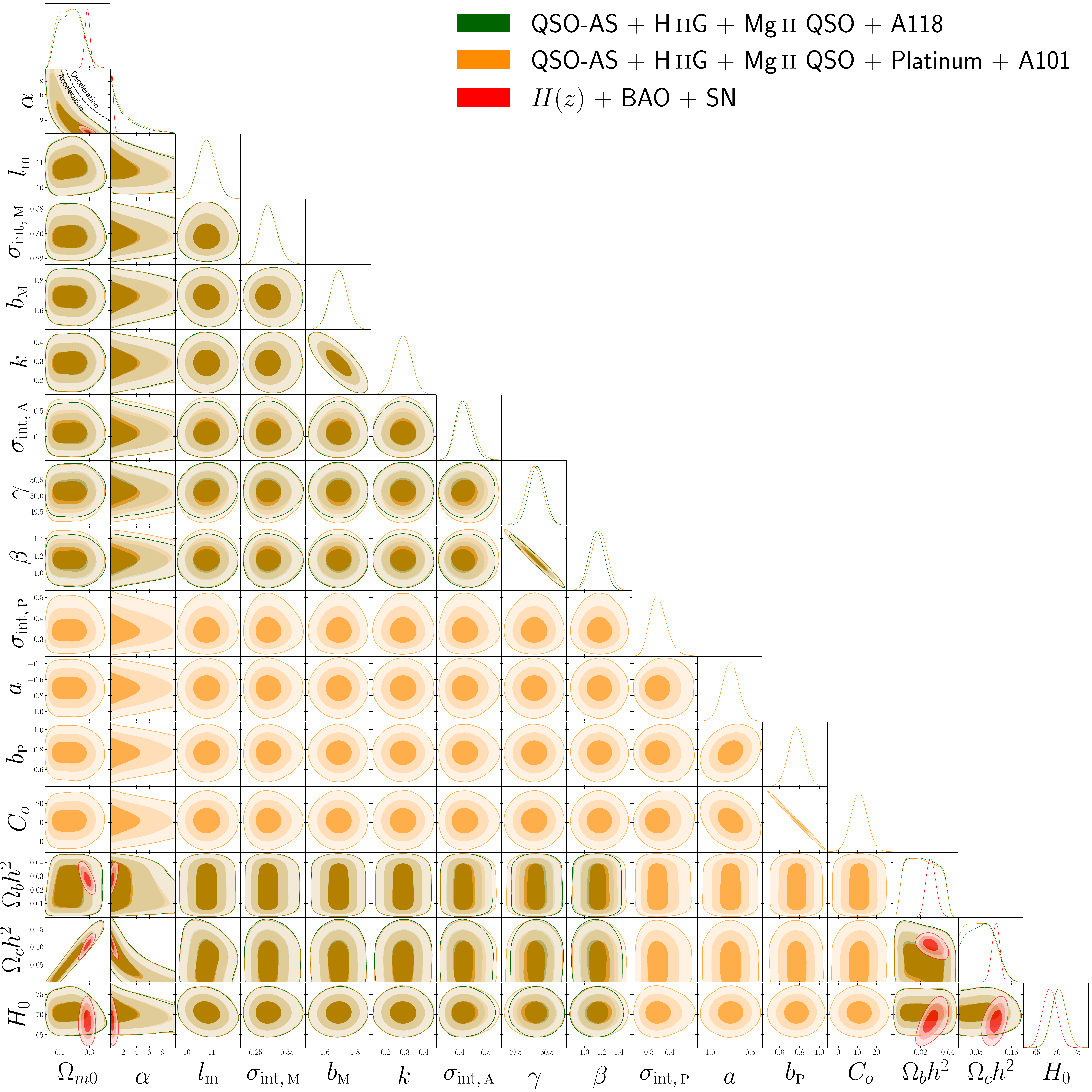}}\\
 \subfloat[]{%
    \includegraphics[width=0.5\textwidth,height=0.5\textwidth]{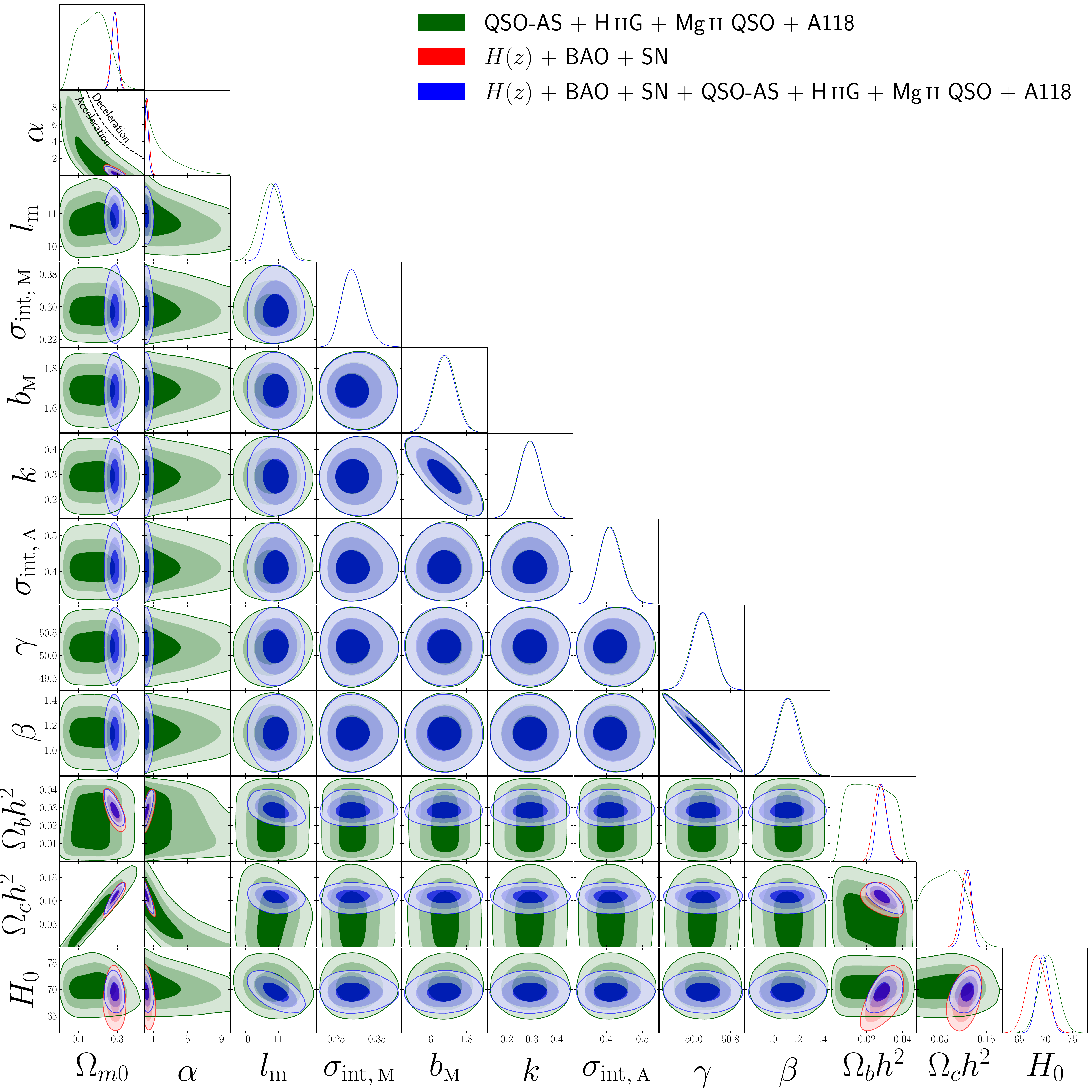}}
 \subfloat[]{%
    \includegraphics[width=0.5\textwidth,height=0.5\textwidth]{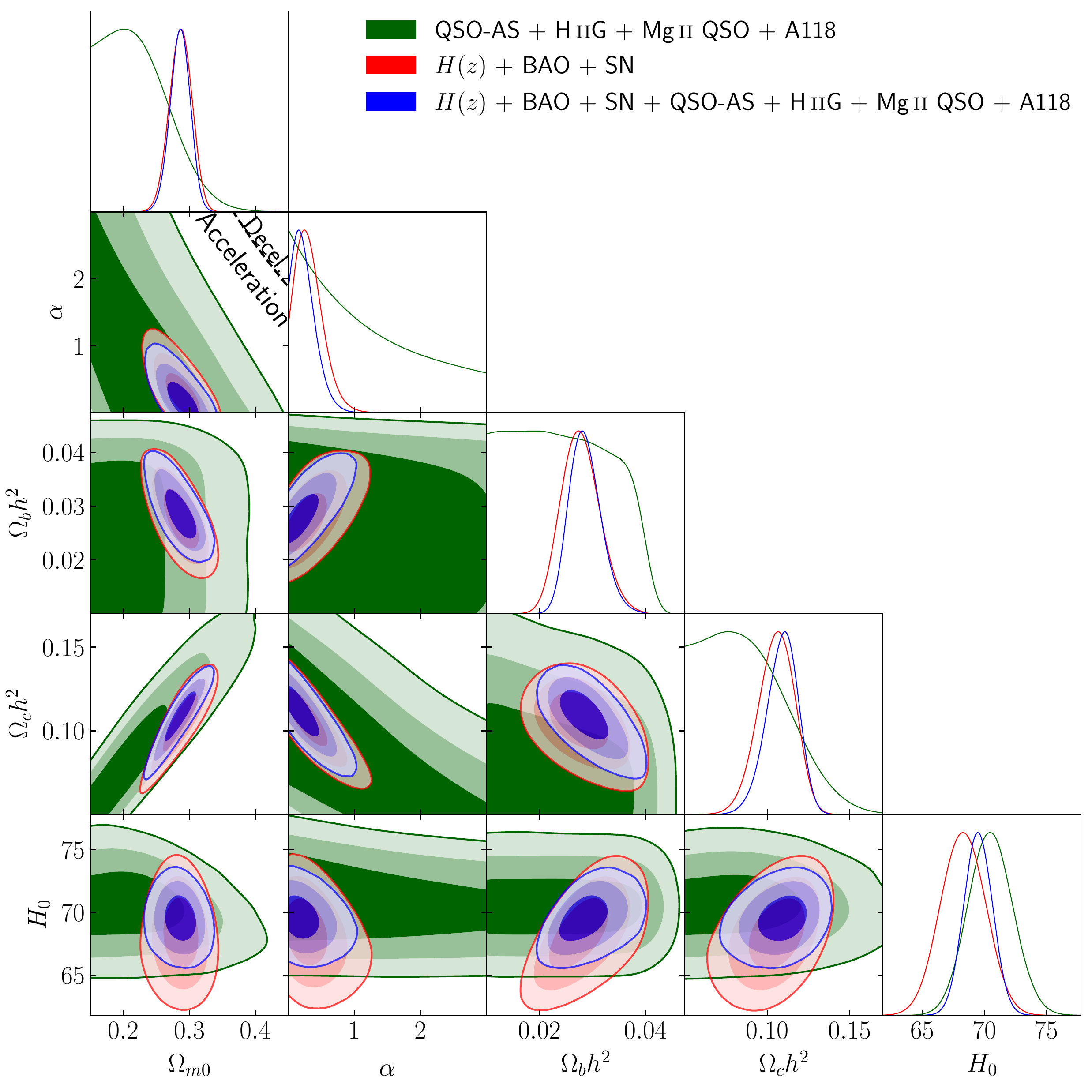}}\\
\caption{One-dimensional likelihood distributions and 1$\sigma$, 2$\sigma$, and 3$\sigma$ two-dimensional likelihood confidence contours for flat \pcdm\ from various combinations of data. The zero-acceleration black dashed lines divide the parameter space into regions associated with currently-accelerating (below left) and currently-decelerating (above right) cosmological expansion. The $\alpha = 0$ axes correspond to flat \lcdm.}
\label{fig5C9}
\end{figure*}

\begin{figure*}
\centering
 \subfloat[]{%
    \includegraphics[width=0.5\textwidth,height=0.5\textwidth]{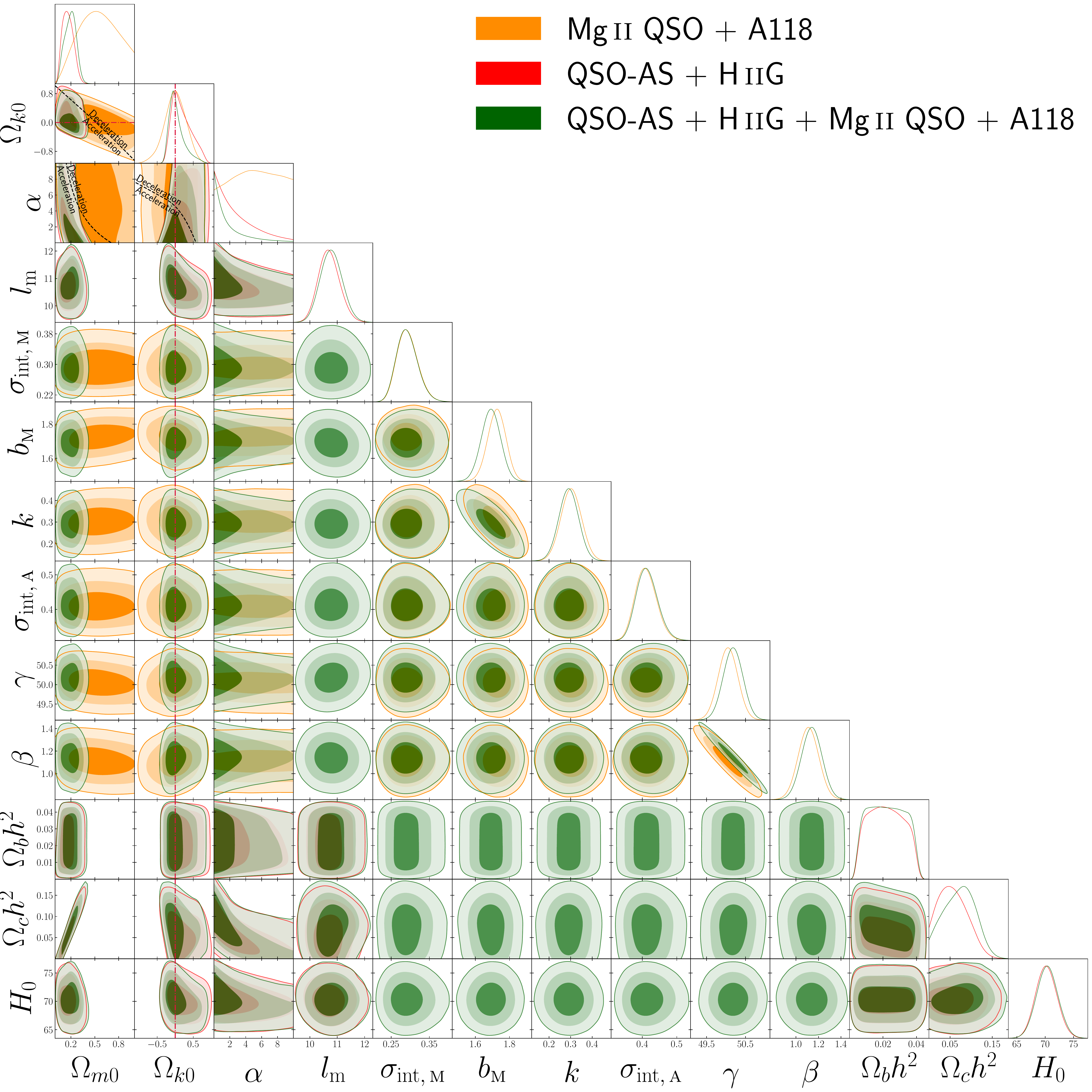}}
 \subfloat[]{%
    \includegraphics[width=0.5\textwidth,height=0.5\textwidth]{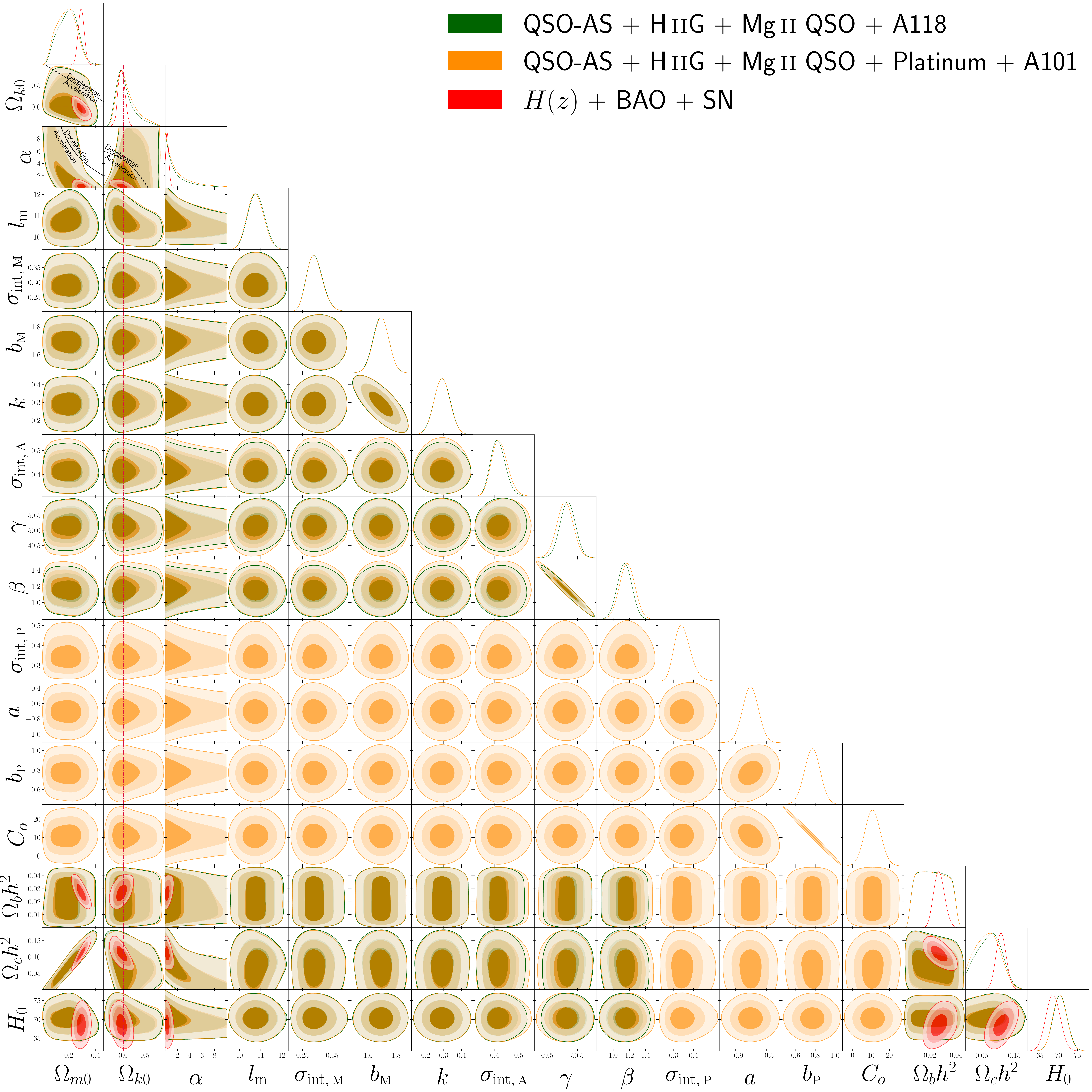}}\\
 \subfloat[]{%
    \includegraphics[width=0.5\textwidth,height=0.5\textwidth]{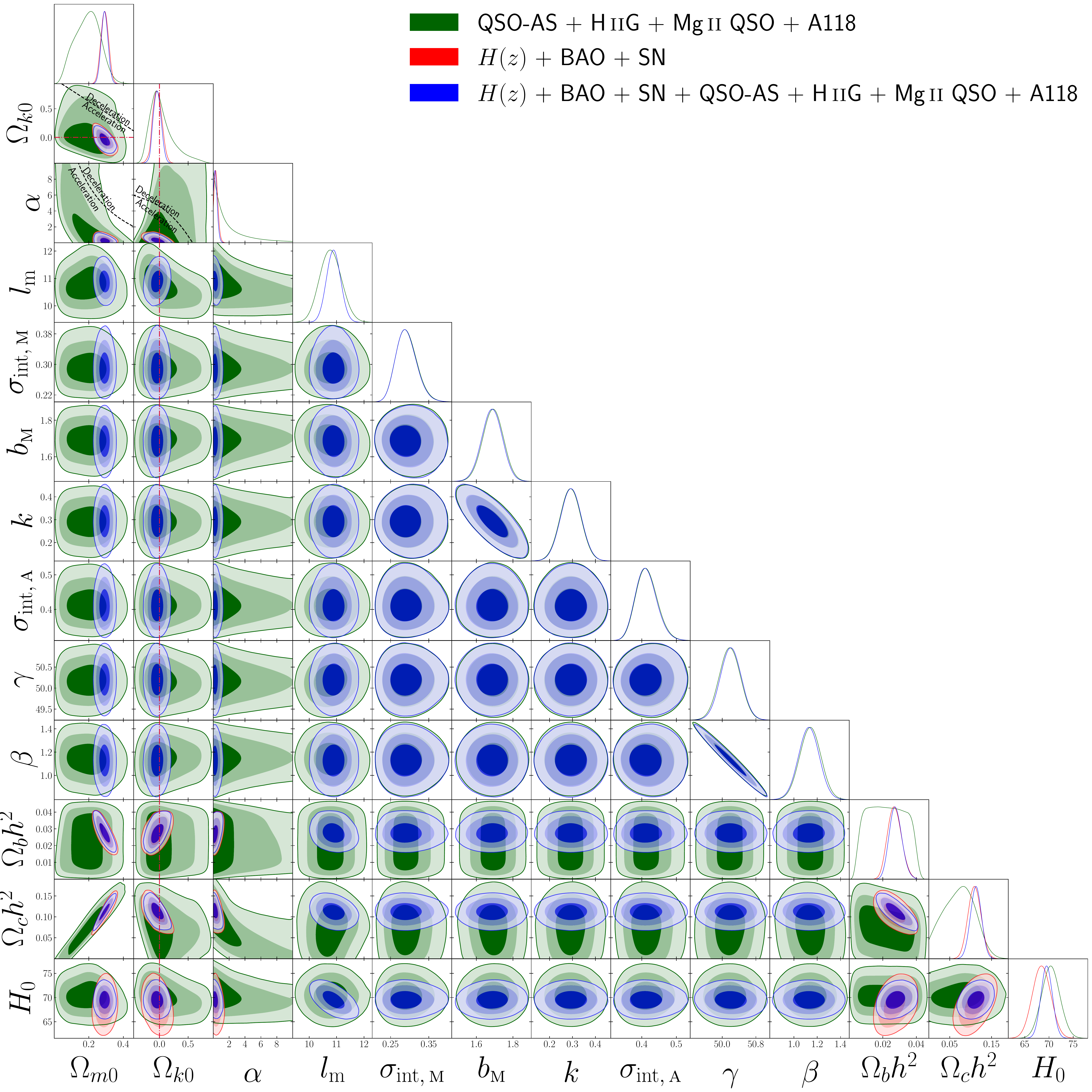}}
 \subfloat[]{%
    \includegraphics[width=0.5\textwidth,height=0.5\textwidth]{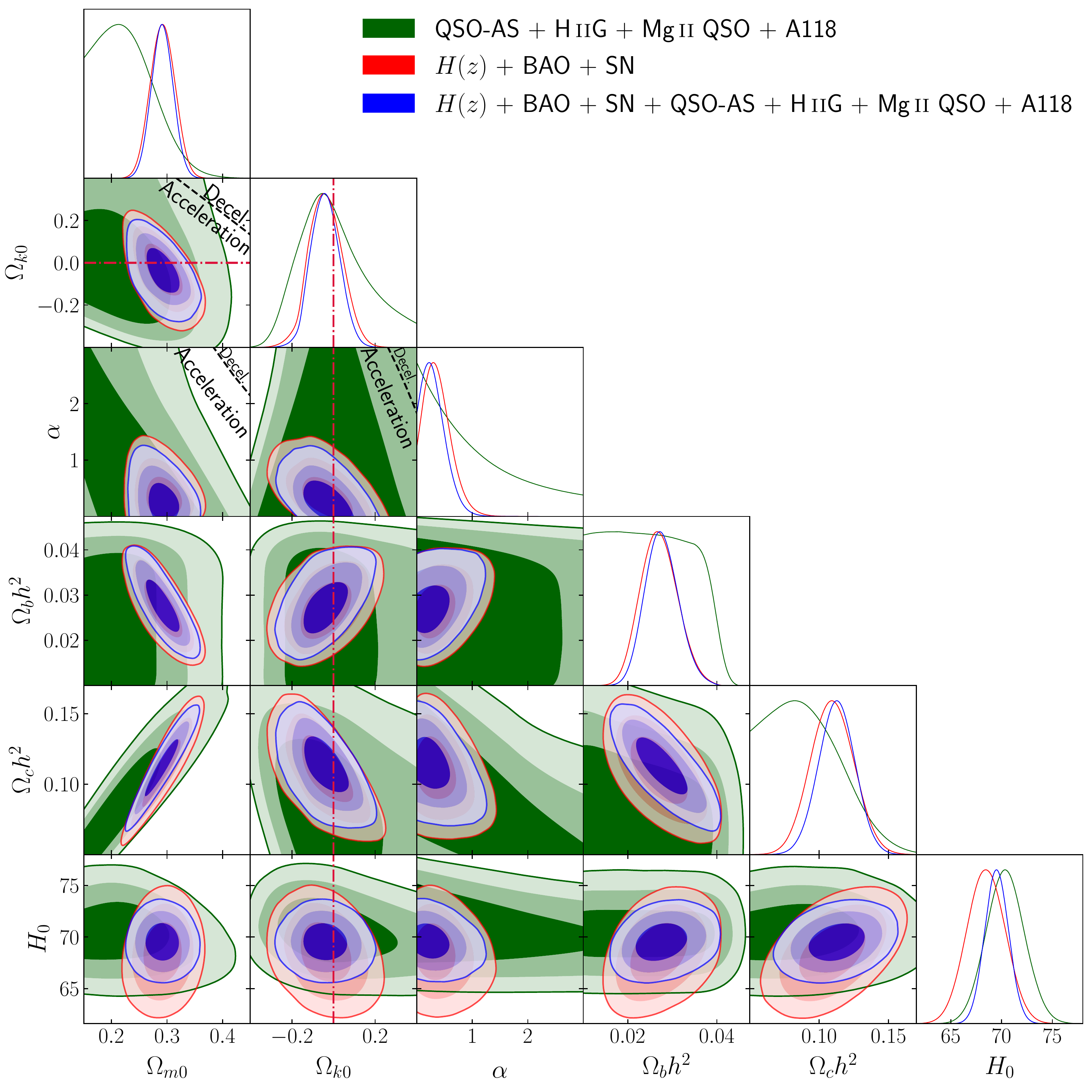}}\\
\caption{Same as Fig.\ \ref{fig5C9} but for non-flat \pcdm. The zero-acceleration black dashed lines are computed for the third cosmological parameter set to the $H(z)$ + BAO data best-fitting values listed in Table \ref{tab:BFPC9}, and divide the parameter space into regions associated with currently-accelerating (below left) and currently-decelerating (above right) cosmological expansion. The crimson dash-dot lines represent flat hypersurfaces, with closed spatial hypersurfaces either below or to the left. The $\alpha = 0$ axes correspond to non-flat \lcdm.}
\label{fig6C9}
\end{figure*}

Cosmological parameter constraints from $H(z)$ + BAO + SN data are consistent with those from QSO-AS + \hiig\ + \mq\ + A118 data. From model to model, there are differences, ranging from $-0.41\sigma$ (flat XCDM) to $1.45\sigma$ (flat \pcdm), between \om\ constraints from $H(z)$ + BAO + SN data and those from QSO-AS + \hiig\ + \mq\ + A118 data; 
and there are differences, ranging from $0.73\sigma$ (non-flat \pcdm) to $1.48\sigma$ (flat XCDM), between $H_0$ constraints from QSO-AS + \hiig\ + \mq\ + A118 data and those from $H(z)$ + BAO + SN data. 
For the XCDM parametrizations, $H(z)$ + BAO + SN data slightly prefer non-phantom dark energy dynamics, whereas QSO-AS + \hiig\ + \mq\ + A118 data prefer phantom dark energy dynamics, however, their differences are within $1\sigma$. As can be seen in the (d) panels of Figs.\ \ref{fig1C9}--\ref{fig6C9}, the two-dimensional posterior cosmological constraints from $H(z)$ + BAO + SN data and QSO-AS + \hiig\ + \mq\ + A118 data are significantly more mutually consistent than the less than $1.5\sigma$ differences between the maximum and minimum one-dimensional posterior mean values discussed above. Consequently we can combine these data in a joint HzBSNQHMA data analysis; we discuss the results from this analysis next.

The constraints from HzBSNQHMA data on \om\ range from a low of $0.286\pm0.015$ (flat \pcdm) to a high of $0.300\pm0.012$ (flat \lcdm), with a difference of $0.73\sigma$.

The $H_0$ constraints range from a low of $69.50\pm1.14$ \hunit\ (flat \pcdm) to a high of $69.87\pm1.13$ \hunit\ (flat \lcdm), with a difference of $0.23\sigma$, which are $0.50\sigma$ (flat \pcdm) and $0.62\sigma$ (flat \lcdm) higher than the median statistics estimate of $H_0=68\pm2.8$ \hunit\ \citep{chenratmed}, and $2.14\sigma$ (flat \pcdm) and $1.93\sigma$ (flat \lcdm) lower than the local Hubble constant measurement of $H_0 = 73.2 \pm 1.3$ \hunit\ \citep{Riess_2021}.\footnote{Other local determinations of $H_0$ result in somewhat lower central values with somewhat larger error bars \citep{rigault_etal_2015,zhangetal2017,Dhawan,FernandezArenas,Breuvaletal_2020, Efstathiou_2020, Khetan_et_al_2021,rameez_sarkar_2021, Freedman2021}. Our $H_0$ determinations here are consistent with earlier median statistics estimates \citep{gott_etal_2001, Calabreseetal2012} and with other recent $H_0$ determinations \citep{chen_etal_2017,DES_2018,Gomez-ValentAmendola2018, planck2018b,dominguez_etal_2019,Cuceu_2019,zeng_yan_2019,schoneberg_etal_2019, Blum_et_al_2020, Lyu_et_al_2020, Philcox_et_al_2020, Birrer_et_al_2020, Denzel_et_al_2020,Pogosianetal_2020,Kimetal_2020,Harvey_2020,Boruahetal_2021, Zhang_Huang_2021,lin_ishak_2021,Wuetal2022}.}

The constraints on \ok\ are $0.018\pm0.059$, $-0.009^{+0.077}_{-0.083}$, and $-0.040^{+0.064}_{-0.072}$ for non-flat \lcdm, XCDM, and \pcdm, respectively. Following the same pattern as $H(z)$ + BAO + SN data results, flat hypersurfaces are also well within 1$\sigma$.

There is a slight preference for dark energy dynamics. For flat (non-flat) XCDM, $w_{\rm X}=-0.959\pm0.059$ ($w_{\rm X}=-0.959^{+0.090}_{-0.063}$), with central values being $0.69\sigma$ ($0.65\sigma$) higher than $w_{\rm X}=-1$; and for flat (non-flat) \pcdm, $\alpha=0.249^{+0.069}_{-0.239}$ ($\alpha=0.316^{+0.101}_{-0.292}$), with central values being $1.04\sigma$ ($1.08\sigma$) away from $\alpha=0$.

\subsection{Model Comparison}
 \label{subsec:comp}

From the AIC, BIC, and DIC values listed in Table \ref{tab:BFPC9}, we find the following results:
\begin{itemize}
    \item[1)]{\bf AIC} $H(z)$ + BAO data favor flat \pcdm\ the most, \mq\ + A118 data favor flat XCDM the most, QSO-AS + \hiig\ data favor non-flat XCDM the most, and the other data combinations favor flat \lcdm\ the most. However the evidence against the rest of the models/parametrizations is either only weak or positive.
    
    \item[2)] {\bf BIC} All data combinations favor flat \lcdm\ the most. $H(z)$ + BAO data only provide weak or positive evidence against other models/parametrizations.
    
    Both \mq\ + A118 and QSO-AS + \hiig\ data provide strong (very strong) evidence against non-flat XCDM (non-flat \pcdm) and positive evidence against the others. 
    
    QSO-AS + \hiig\ + \mq\ + A118 data provide strong (very strong) evidence against flat \pcdm\ (non-flat XCDM and non-flat \pcdm) and positive evidence against non-flat \lcdm\ and flat XCDM. 
    
    $H(z)$ + BAO + SN data provide strong (very strong) evidence against non-flat \lcdm\ (non-flat XCDM and non-flat \pcdm) and positive evidence against flat XCDM and flat \pcdm. 
    
    $H(z)$ + BAO + SN + QSO-AS + \hiig\ + \mq\ + A118 data provide very strong evidence against non-flat XCDM and non-flat \pcdm, and strong evidence against the others. 
    
    QSO-AS + \hiig\ + \mq\ + Platinum + A101 data provide strong (very strong) evidence against flat \pcdm\ (non-flat XCDM and non-flat \pcdm) and positive evidence against non-flat \lcdm\ and flat XCDM.

    \item[3)] {\bf DIC} $H(z)$ + BAO, \mq\ + A118, and $H(z)$ + BAO + SN data favor flat \pcdm\ the most, and the other data combinations favor flat \lcdm\ the most. There is strong evidence against non-flat XCDM from QSO-AS + \hiig\ data, strong evidence against non-flat \pcdm\ from QSO-AS + \hiig, QSO-AS + \hiig\ + \mq\ + A118, and QSO-AS + \hiig\ + \mq\ + Platinum + A101 data, and weak or positive evidence against the others from the remaining data sets.
\end{itemize}

Perhaps the most reliable summary conclusion is that, based on DIC, the $H(z)$ + BAO + SN + QSO-AS + \hiig\ + \mq\ + A118 data combination does not provide strong evidence against any of the cosmological models/parametrizations.

\section{Conclusion}
\label{makereference9.5}

In this paper we use many of the most up-to-date available non-CMB data sets to determine cosmological constraints. We analyze 32 $H(z)$, 12 BAO, 1048 Pantheon SN Ia, 20 binned DES-3yr SN Ia, 120 QSO-AS, 181 \hiig, 78 \mq, 118 (101) A118 (A101) GRB, and 50 Platinum GRB measurements and find that the cosmological constraints from each data set are mutually consistent. We find very small differences between cosmological constraints determined from QSO-AS + \hiig\ + \mq\ + A118 data and those from QSO-AS + \hiig\ + \mq\ + Platinum + A101 data, so report only the cosmological constraints from joint $H(z)$ + BAO + SN + QSO-AS + \hiig\ + \mq\ + A118 (HzBSNQHMA) data.

The HzBSNQHMA data provide a fairly restrictive summary value\footnote{As in \cite{CaoRyanRatra2021, CaoRyanRatra2022}, the summary central value is computed from the mean of the two central-most of the six mean values and the summary uncertainty is computed from the quadrature sum of the systematic uncertainty, defined to be half of the difference between the two central-most mean values, and the statistical uncertainty, defined to be the average of the error bars of the two central-most results.} of $\Om=0.295\pm0.017$ that agrees well with many other recent measurements and a fairly restrictive summary value of $H_0=69.7\pm1.2$ \hunit\ that is in better agreement with the result of \cite{chenratmed} than with the result of \cite{Riess_2021}.\footnote{Our model-independent $H_0$ error bar is slightly smaller than that of \cite{Riess_2021} and is a factor of 2.2 larger than that of the flat \lcdm\ model \citep{planck2018b} TT,TE,EE+lowE+lensing CMB anisotropy $H_0$ error bar while our $\Om$ error bar is a factor of 2.3 larger than the corresponding \textit{Planck} flat \lcdm\ one.} Our $H_0$ measurement here lies in the middle of the flat \lcdm\ model result of \cite{planck2018b} and the local expansion rate result of \cite{Riess_2021}, slightly closer to the former. Based on DIC, the HzBSNQHMA data compilation prefers flat \lcdm\ the most, but does not rule out mild dark energy dynamics or a little spatial curvature energy density (evidence against them is either weak or positive). 

We hope that in the near future the quality and amount of the types of lower-redshift, non-CMB, data we have used here will improve enough to result in cosmological parameter error bars comparable to those from \textit{Planck} CMB anisotropy data.

\begin{sidewaystable*}
\centering
\resizebox*{\columnwidth}{0.74\columnwidth}{%
\begin{threeparttable}
\caption{Unmarginalized best-fitting parameter values for all models from various combinations of data.}\label{tab:BFPC9}
\begin{tabular}{lccccccccccccccccccccccccc}
\toprule
Model & Data set & $\Omega_{b}h^2$ & $\Omega_{c}h^2$ & $\Omega_{m0}$ & $\Omega_{k0}$ & $w_{\mathrm{X}}$/$\alpha$\tnote{a} & $H_0$\tnote{b} & $l_{\mathrm{m}}$\tnote{c} & $\sigma_{\mathrm{int,\,\textsc{m}}}$ & $b_{\mathrm{\textsc{m}}}$ & $k$ & $\sigma_{\mathrm{int,\,\textsc{a}}}$ & $\gamma$ & $\beta$ & $\sigma_{\mathrm{int,\,\textsc{p}}}$ & $a$ & $b_{\mathrm{\textsc{p}}}$ & $C_{o}$ & $-2\ln\mathcal{L}_{\mathrm{max}}$ & AIC & BIC & DIC & $\Delta \mathrm{AIC}$ & $\Delta \mathrm{BIC}$ & $\Delta \mathrm{DIC}$ \\
\midrule
 & $H(z)$ + BAO & 0.0244 & 0.1181 & 0.301 & -- & -- & 68.98 & -- & -- & -- & -- & -- & -- & -- & -- & -- & -- & -- & 25.64 & 31.64 & 36.99 & 32.32 & 0.00 & 0.00 & 0.00\\
 & \mq\ + A118 & -- & 0.2526 & 0.567 & -- & -- & -- & -- & 0.286 & 1.711 & 0.300 & 0.398 & 50.12 & 1.105 & -- & -- & -- & -- & 159.64 & 173.64 & 196.59 & 173.37 & 0.00 & 0.00 & 0.00\\
Flat & QSO-AS + \hiig & 0.0332 & 0.0947 & 0.251 & -- & -- & 71.53 & 11.06 & -- & -- & -- & -- & -- & -- & -- & -- & -- & -- & 786.45 & 794.45 & 809.28 & 792.69 & 0.00 & 0.00 & 0.00\\
\lcdm & QSO-AS + \hiig\ + \mq\ + A118 & 0.0183 & 0.1127 & 0.258 & -- & -- & 71.50 & 11.04 & 0.282 & 1.682 & 0.294 & 0.404 & 50.23 & 1.124 & -- & -- & -- & -- & 947.34 & 967.34 & 1009.43 & 966.23 & 0.00 & 0.00 & 0.00\\
 & $H(z)$ + BAO + SN & 0.0242 & 0.1191 & 0.304 & -- & -- & 68.86 & -- & -- & -- & -- & -- & -- & -- & -- & -- & -- & -- & 1082.39 & 1088.39 & 1103.44 & 1089.92 & 0.00 & 0.00 & 0.00\\
 & HzNBSNQHMA\tnote{e} & 0.0258 & 0.1207 & 0.300 & -- & -- & 70.06 & 10.93 & 0.286 & 1.671 & 0.296 & 0.406 & 50.26 & 1.110 & -- & -- & -- & -- & 2031.30 & 2051.30 & 2105.13 & 2051.86 & 0.00 & 0.00 & 0.00\\
 & QHMPA101\tnote{e} & 0.0197 & 0.1144 & 0.264 & -- & -- & 71.41 & 11.02 & 0.283 & 1.673 & 0.315 & 0.410 & 50.11 & 1.169 & 0.321 & $-0.746$ & 0.773 & 10.81 & 964.62 & 992.62 & 1052.44 & 990.65 & 0.00 & 0.00 & 0.00\\
\midrule
 & $H(z)$ + BAO & 0.0260 & 0.1098 & 0.292 & 0.048 & -- & 68.35 & -- & -- & -- & -- & -- & -- & -- & -- & -- & -- & -- & 25.30 & 33.30 & 40.43 & 33.87 & 1.66 & 3.44 & 1.54\\
 & \mq\ + A118 & -- & 0.2692 & 0.601 & 0.252 & -- & -- & -- & 0.281 & 1.713 & 0.308 & 0.401 & 50.07 & 1.117 & -- & -- & -- & -- & 159.65 & 175.65 & 201.87 & 174.46 & 2.01 & 5.28 & 1.09\\
Non-flat & QSO-AS + \hiig & 0.0228 & 0.1145 & 0.260 & $-0.360$ & -- & 72.91 & 11.60 & -- & -- & -- & -- & -- & -- & -- & -- & -- & -- & 784.18 & 794.18 & 812.71 & 793.24 & $-0.27$ & 3.43 & 0.55\\
\lcdm & QSO-AS + \hiig\ + \mq\ + A118 & 0.0261 & 0.1186 & 0.278 & $-0.254$ & -- & 72.25 & 11.29 & 0.282 & 1.678 & 0.299 & 0.410 & 50.27 & 1.106 & -- & -- & -- & -- & 945.97 & 967.97 & 1014.27 & 966.91 & 0.63 & 4.84 & 0.69\\
 & $H(z)$ + BAO + SN & 0.0255 & 0.1121 & 0.295 & 0.035 & -- & 68.53 & -- & -- & -- & -- & -- & -- & -- & -- & -- & -- & -- & 1082.11 & 1090.11 & 1110.16 & 1091.17 & 1.72 & 6.72 & 1.24\\
 & HzBSNQHMA\tnote{d} & 0.0261 & 0.1182 & 0.297 & 0.008 & -- & 69.90 & 10.94 & 0.281 & 1.674 & 0.301 & 0.400 & 50.20 & 1.126 & -- & -- & -- & -- & 2031.26 & 2053.26 & 2112.48 & 2053.69 & 1.96 & 7.35 & 1.84\\
 & QHMPA101\tnote{e} & 0.0246 & 0.1161 & 0.276 & $-0.221$ & -- & 71.56 & 11.37 & 0.287 & 1.682 & 0.304 & 0.411 & 50.20 & 1.132 & 0.329 & $-0.674$ & 0.790 & 9.64 & 963.83 & 993.83 & 1057.92 & 992.04 & 1.21 & 5.48 & 1.39\\
\midrule
 & $H(z)$ + BAO & 0.0296 & 0.0951 & 0.290 & -- & $-0.754$ & 65.79 & -- & -- & -- & -- & -- & -- & -- & -- & -- & -- & -- & 22.39 & 30.39 & 37.52 & 30.63 & $-1.25$ & 0.53 & $-1.69$\\
 & \mq\ + A118 & -- & 0.0448 & 0.143 & -- & $-4.972$ & -- & -- & 0.279 & 1.554 & 0.292 & 0.407 & 50.65 & 1.126 & -- & -- & -- & -- & 156.84 & 172.84 & 199.07 & 174.36 & $-0.80$ & 2.48 & 0.99\\
Flat & QSO-AS + \hiig & 0.0174 & 0.1308 & 0.285 & -- & $-1.280$ & 72.32 & 11.23 & -- & -- & -- & -- & -- & -- & -- & -- & -- & -- & 786.05 & 796.05 & 814.58 & 795.03 & 1.60 & 5.30 & 2.34\\
XCDM & QSO-AS + \hiig\ + \mq\ + A118 & 0.0069 & 0.1558 & 0.307 & -- & $-1.505$ & 72.94 & 11.32 & 0.282 & 1.679 & 0.290 & 0.401 & 50.24 & 1.122 & -- & -- & -- & -- & 946.23 & 968.23 & 1014.52 & 967.53 & 0.89 & 5.09 & 1.31\\
 & $H(z)$ + BAO + SN & 0.0258 & 0.1115 & 0.295 & -- & $-0.940$ & 68.37 & -- & -- & -- & -- & -- & -- & -- & -- & -- & -- & -- & 1081.34 & 1089.34 & 1109.40 & 1090.43 & 0.95 & 5.96 & 0.50\\
 & HzBSNQHMA\tnote{d} & 0.0266 & 0.1162 & 0.296 & -- & $-0.975$ & 69.62 & 10.94 & 0.280 & 1.696 & 0.278 & 0.406 & 50.24 & 1.118 & -- & -- & -- & -- & 2030.88 & 2052.88 & 2112.10 & 2053.30 & 1.58 & 6.97 & 1.44\\
 & QHMPA101\tnote{e} & 0.0176 & 0.1450 & 0.311 & -- & $-1.573$ & 72.46 & 11.41 & 0.276 & 1.686 & 0.294 & 0.417 & 50.12 & 1.168 & 0.326 & $-0.682$ & 0.756 & 11.45 & 963.77 & 993.77 & 1057.86 & 992.26 & 1.15 & 5.42 & 1.61\\
\midrule
 & $H(z)$ + BAO & 0.0289 & 0.0985 & 0.296 & $-0.053$ & $-0.730$ & 65.76 & -- & -- & -- & -- & -- & -- & -- & -- & -- & -- & -- & 22.13 & 32.13 & 41.05 & 32.51 & 0.49 & 4.06 & 0.19\\
 & \mq\ + A118 & -- & 0.0169 & 0.086 & $-0.027$ & $-5.000$ & -- & -- & 0.279 & 1.473 & 0.298 & 0.396 & 50.91 & 1.118 & -- & -- & -- & -- & 156.52 & 174.52 & 204.03 & 177.50 & 0.88 & 7.44 & 4.13\\
Non-flat & QSO-AS + \hiig & 0.0300 & 0.0031 & 0.065 & $-0.560$ & $-0.651$ & 71.87 & 11.45 & -- & -- & -- & -- & -- & -- & -- & -- & -- & -- & 781.18 & 793.18 & 815.43 & 799.59 & $-1.27$ & 6.15 & 6.90\\
XCDM & QSO-AS + \hiig\ + \mq\ + A118 & 0.0365 & 0.1224 & 0.307 & $-0.087$ & $-1.278$ & 72.14 & 11.33 & 0.292 & 1.677 & 0.287 & 0.398 & 50.23 & 1.123 & -- & -- & -- & -- & 946.08 & 970.08 & 1020.59 & 969.33 & 2.74 & 11.16 & 3.11\\
 & $H(z)$ + BAO + SN & 0.0255 & 0.1155 & 0.300 & $-0.030$ & $-0.922$ & 68.68 & -- & -- & -- & -- & -- & -- & -- & -- & -- & -- & -- & 1081.28 & 1091.28 & 1116.35 & 1092.47 & 2.89 & 12.91 & 2.55\\
 & HzBSNQHMA\tnote{d} & 0.0260 & 0.1158 & 0.295 & $-0.016$ & $-0.947$ & 69.53 & 10.94 & 0.288 & 1.697 & 0.277 & 0.409 & 50.15 & 1.151 & -- & -- & -- & -- & 2030.78 & 2054.78 & 2119.38 & 2055.42 & 3.48 & 14.25 & 3.56\\
 & QHMPA101\tnote{e} & 0.0354 & 0.1154 & 0.292 & $-0.048$ & $-1.221$ & 72.03 & 11.31 & 0.272 & 1.680 & 0.294 & 0.313 & 50.06 & 1.178 & 0.313 & $-0.736$ & 0.769 & 10.98 & 963.91 & 995.91 & 1064.28 & 993.76 & 3.29 & 11.84 & 3.11\\
\midrule
 & $H(z)$ + BAO & 0.0310 & 0.0900 & 0.280 & -- & 1.010 & 65.89 & -- & -- & -- & -- & -- & -- & -- & -- & -- & -- & -- & 22.31 & 30.31 & 37.45 & 29.90 & $-1.33$ & 0.46 & $-2.42$\\
 & \mq\ + A118 & -- & 0.2760 & 0.615 & -- & 0.353 & -- & -- & 0.279 & 1.714 & 0.308 & 0.403 & 50.09 & 1.105 & -- & -- & -- & -- & 159.69 & 175.69 & 201.91 & 172.93 & 2.05 & 5.32 & $-0.44$\\
Flat & QSO-AS + \hiig & 0.0198 & 0.1066 & 0.249 & -- & 0.000 & 71.42 & 11.10 & -- & -- & -- & -- & -- & -- & -- & -- & -- & -- & 786.46 & 796.46 & 815.00 & 796.31 & 2.01 & 5.72 & 3.62\\
\pcdm & QSO-AS + \hiig\ + \mq\ + A118 & 0.0098 & 0.1196 & 0.253 & -- & 0.020 & 71.63 & 10.98 & 0.283 & 1.683 & 0.289 & 0.416 & 50.18 & 1.145 & -- & -- & -- & -- & 947.60 & 969.60 & 1015.89 & 970.30 & 2.26 & 6.46 & 4.07\\
 & $H(z)$ + BAO + SN & 0.0263 & 0.1097 & 0.292 & -- & 0.203 & 68.39 & -- & -- & -- & -- & -- & -- & -- & -- & -- & -- & -- & 1081.22 & 1089.22 & 1109.28 & 1089.91 & 0.83 & 5.84 & $-0.01$\\
 & HzBSNQHMA\tnote{d} & 0.0274 & 0.1116 & 0.289 & -- & 0.150 & 69.51 & 10.97 & 0.280 & 1.685 & 0.292 & 0.409 & 50.23 & 1.121 & -- & -- & -- & -- & 2030.52 & 2052.52 & 2111.74 & 2053.18 & 1.22 & 6.61 & 1.33\\
 & QHMPA101\tnote{e} & 0.0388 & 0.0884 & 0.251 & -- & 0.105 & 71.31 & 11.13 & 0.287 & 1.679 & 0.293 & 0.402 & 50.11 & 1.168 & 0.317 & $-0.664$ & 0.775 & 10.43 & 964.97 & 994.97 & 1059.07 & 994.25 & 2.35 & 6.63 & 3.60\\
\midrule
 & $H(z)$ + BAO & 0.0306 & 0.0920 & 0.284 & $-0.058$ & 1.200 & 65.91 & -- & -- & -- & -- & -- & -- & -- & -- & -- & -- & -- & 22.05 & 32.05 & 40.97 & 31.30 & 0.41 & 3.98 & $-1.02$\\
 & \mq\ + A118 & -- & 0.2233 & 0.507 & 0.101 & 6.181 & -- & -- & 0.280 & 1.723 & 0.303 & 0.401 & 50.04 & 1.115 & -- & -- & -- & -- & 159.63 & 177.63 & 207.13 & 173.84 & 3.99 & 10.54 & 0.47\\
Non-flat & QSO-AS + \hiig & 0.0338 & 0.0979 & 0.251 & $-0.250$ & 0.000 & 72.53 & 11.47 & -- & -- & -- & -- & -- & -- & -- & -- & -- & -- & 784.61 & 796.61 & 818.85 & 801.32 & 2.16 & 9.57 & 8.63\\
\pcdm & QSO-AS + \hiig\ + \mq\ + A118 & 0.0073 & 0.1290 & 0.265 & $-0.228$ & 0.115 & 71.91 & 11.29 & 0.288 & 1.673 & 0.300 & 0.412 & 50.23 & 1.120 & -- & -- & -- & -- & 946.09 & 970.09 & 1020.59 & 972.93 & 2.75 & 11.16 & 6.70\\
 & $H(z)$ + BAO + SN & 0.0261 & 0.1119 & 0.295 & $-0.023$ & 0.253 & 68.56 & -- & -- & -- & -- & -- & -- & -- & -- & -- & -- & -- & 1081.12 & 1091.12 & 1116.19 & 1091.27 & 2.73 & 12.75 & 1.35\\
 & HzBSNQHMA\tnote{d} & 0.0251 & 0.1207 & 0.300 & $-0.056$ & 0.195 & 69.84 & 10.91 & 0.284 & 1.699 & 0.279 & 0.411 & 50.19 & 1.132 & -- & -- & -- & -- & 2030.76 & 2054.76 & 2119.36 & 2055.13 & 3.46 & 14.23 & 3.28\\
 & QHMPA101\tnote{e} & 0.0370 & 0.0977 & 0.260 & $-0.217$ & 0.066 & 72.15 & 11.23 & 0.274 & 1.689 & 0.277 & 0.408 & 50.08 & 1.180 & 0.335 & $-0.703$ & 0.777 & 10.42 & 963.77 & 995.77 & 1064.14 & 997.14 & 3.15 & 11.70 & 6.49\\
\bottomrule
\end{tabular}
\begin{tablenotes}[flushleft]
\item [a] \wx\ corresponds to flat/non-flat XCDM and $\alpha$ corresponds to flat/non-flat \pcdm.
\item [b] \hunit. In these GRB cases, $\Omega_b$ and $H_0$ are set to be 0.05 and 70 \hunit, respectively.
\item [c] pc.
\item [d] $H(z)$ + BAO + SN + QSO-AS + \hiig\ + \mq\ + A118.
\item [e] QSO-AS + \hiig\ + \mq\ + Platinum + A101.
\end{tablenotes}
\end{threeparttable}%
}
\end{sidewaystable*}

\begin{sidewaystable*}
\centering
\resizebox*{\columnwidth}{0.74\columnwidth}{%
\begin{threeparttable}
\caption{One-dimensional marginalized posterior mean values and uncertainties ($\pm 1\sigma$ error bars or $2\sigma$ limits) of the parameters for all models from various combinations of data.}\label{tab:1d_BFPC9}
\begin{tabular}{lcccccccccccccccccc}
\toprule
Model & Data set & $\Omega_{b}h^2$ & $\Omega_{c}h^2$ & $\Omega_{m0}$ & $\Omega_{k0}$ & $w_{\mathrm{X}}$/$\alpha$\tnote{a} & $H_0$\tnote{b} & $l_{\mathrm{m}}$\tnote{c} & $\sigma_{\mathrm{int,\,\textsc{m}}}$ & $b_{\mathrm{\textsc{m}}}$ & $k$ & $\sigma_{\mathrm{int,\,\textsc{a}}}$ & $\gamma$ & $\beta$ & $\sigma_{\mathrm{int,\,\textsc{p}}}$ & $a$ & $b_{\mathrm{\textsc{p}}}$ & $C_{o}$ \\
\midrule
 & $H(z)$ + BAO & $0.0247\pm0.0030$ & $0.1186^{+0.0076}_{-0.0083}$ & $0.301^{+0.016}_{-0.018}$ & -- & -- & $69.14\pm1.85$ & -- & -- & -- & -- & -- & -- & -- & -- & -- & -- & -- \\
 & \mq\ + A118 & -- & -- & $0.609^{+0.294}_{-0.210}$ & -- & -- & -- & -- & $0.293^{+0.023}_{-0.030}$ & $1.712\pm0.056$ & $0.301\pm0.047$ & $0.411^{+0.027}_{-0.033}$ & $50.10\pm0.25$ & $1.112\pm0.088$ & -- & -- & -- & -- \\
Flat & QSO-AS + \hiig & $0.0225\pm0.0113$ & $0.1076^{+0.0197}_{-0.0224}$ & $0.257^{+0.037}_{-0.047}$ & -- & -- & $71.52\pm1.79$ & $11.04\pm0.34$ & -- & -- & -- & -- & -- & -- & -- & -- & -- & -- \\
\lcdm & QSO-AS + \hiig\ + \mq\ + A118 & $0.0224\pm0.0116$ & $0.1118^{+0.0197}_{-0.0223}$ & $0.266^{+0.037}_{-0.047}$ & -- & -- & $71.28\pm1.73$ & $11.00\pm0.33$ & $0.292^{+0.023}_{-0.029}$ & $1.686\pm0.055$ & $0.291\pm0.044$ & $0.414^{+0.027}_{-0.033}$ & $50.20\pm0.24$ & $1.135\pm0.087$ & -- & -- & -- & -- \\
 & $H(z)$ + BAO + SN & $0.0244\pm0.0027$ & $0.1199\pm0.0076$ & $0.304^{+0.014}_{-0.015}$ & -- & -- & $69.04\pm1.77$ & -- & -- & -- & -- & -- & -- & -- & -- & -- & -- & -- \\
 & HzBSNQHMA\tnote{d} & $0.0256\pm0.0020$ & $0.1201\pm0.0061$ & $0.300\pm0.012$ & -- & -- & $69.87\pm1.13$ & $10.96\pm0.25$ & $0.292^{+0.023}_{-0.029}$ & $1.684\pm0.055$ & $0.293\pm0.044$ & $0.413^{+0.026}_{-0.033}$ & $50.20\pm0.24$ & $1.131\pm0.087$ & -- & -- & -- & -- \\
 & QHMPA101\tnote{e} & $0.0225^{+0.0117}_{-0.0118}$ & $0.1105^{+0.0197}_{-0.0223}$ & $0.264^{+0.036}_{-0.047}$ & -- & -- & $71.34^{+1.73}_{-1.74}$ & $11.02\pm0.34$ & $0.291^{+0.023}_{-0.029}$ & $1.686\pm0.055$ & $0.291\pm0.044$ & $0.421^{+0.029}_{-0.036}$ & $50.12\pm0.26$ & $1.164\pm0.094$ & $0.346^{+0.032}_{-0.045}$ & $-0.710\pm0.101$ & $0.765\pm0.078$ & $11.06\pm4.19$ \\
\midrule
 & $H(z)$ + BAO & $0.0266^{+0.0039}_{-0.0045}$ & $0.1088\pm0.0166$ & $0.291\pm0.023$ & $0.059^{+0.081}_{-0.091}$ & -- & $68.37\pm2.10$ & -- & -- & -- & -- & -- & -- & -- & -- & -- & -- & -- \\
 & \mq\ + A118 & -- & -- & $>0.274$ & $0.521^{+0.512}_{-0.872}$ & -- & -- & -- & $0.294^{+0.023}_{-0.030}$ & $1.730^{+0.058}_{-0.053}$ & $0.302\pm0.047$ & $0.411^{+0.027}_{-0.033}$ & $50.03\pm0.26$ & $1.118\pm0.089$ & -- & -- & -- & -- \\
Non-flat & QSO-AS + \hiig & $0.0224\pm0.0111$ & $0.1122^{+0.0223}_{-0.0218}$ & $0.260^{+0.039}_{-0.045}$ & $-0.196^{+0.112}_{-0.295}$ & -- & $72.25\pm1.99$ & $11.35\pm0.49$ & -- & -- & -- & -- & -- & -- & -- & -- & -- & -- \\
\lcdm & QSO-AS + \hiig\ + \mq\ + A118 & $0.0225\pm0.0116$ & $0.1162\pm0.0218$ & $0.271^{+0.038}_{-0.045}$ & $-0.139^{+0.116}_{-0.228}$ & -- & $71.77\pm1.87$ & $11.19\pm0.42$ & $0.291^{+0.022}_{-0.029}$ & $1.680\pm0.055$ & $0.292\pm0.044$ & $0.415^{+0.027}_{-0.033}$ & $50.23\pm0.24$ & $1.122\pm0.088$ & -- & -- & -- & -- \\
 & $H(z)$ + BAO + SN & $0.0260^{+0.0037}_{-0.0043}$ & $0.1119\pm0.0157$ & $0.294\pm0.022$ & $0.040\pm0.070$ & -- & $68.62\pm1.90$ & -- & -- & -- & -- & -- & -- & -- & -- & -- & -- & -- \\
 & HzBSNQHMA\tnote{d} & $0.0265^{+0.0032}_{-0.0038}$ & $0.1168\pm0.0127$ & $0.295\pm0.019$ & $0.018\pm0.059$ & -- & $69.79\pm1.14$ & $10.96\pm0.25$ & $0.292^{+0.023}_{-0.029}$ & $1.685\pm0.055$ & $0.293\pm0.044$ & $0.413^{+0.027}_{-0.033}$ & $50.20\pm0.24$ & $1.131\pm0.086$ & -- & -- & -- & -- \\
 & QHMPA101\tnote{e} & $0.0224\pm0.0117$ & $0.1138^{+0.0226}_{-0.0224}$ & $0.267^{+0.039}_{-0.046}$ & $-0.096^{+0.127}_{-0.251}$ & -- & $71.65^{+1.86}_{-1.85}$ & $11.15\pm0.42$ & $0.291^{+0.023}_{-0.029}$ & $1.682\pm0.055$ & $0.292\pm0.044$ & $0.424^{+0.029}_{-0.036}$ & $50.14\pm0.26$ & $1.154\pm0.095$ & $0.346^{+0.032}_{-0.045}$ & $-0.712\pm0.102$ & $0.760\pm0.080$ & $11.37\pm4.34$ \\
\midrule
 & $H(z)$ + BAO & $0.0295^{+0.0042}_{-0.0050}$ & $0.0969^{+0.0178}_{-0.0152}$ & $0.289\pm0.020$ & -- & $-0.784^{+0.140}_{-0.107}$ & $66.22^{+2.31}_{-2.54}$ & -- & -- & -- & -- & -- & -- & -- & -- & -- & -- & -- \\
 & \mq\ + A118 & -- & -- & $0.493^{+0.209}_{-0.353}$ & -- & $<-0.218$ & -- & -- & $0.291^{+0.023}_{-0.030}$ & $1.670^{+0.095}_{-0.062}$ & $0.301\pm0.046$ & $0.412^{+0.027}_{-0.033}$ & $50.25^{+0.28}_{-0.36}$ & $1.110\pm0.088$ \\
Flat & QSO-AS + \hiig & $0.0224\pm0.0112$ & $0.1391^{+0.0333}_{-0.0256}$ & $0.305^{+0.056}_{-0.047}$ & -- & $-1.683^{+0.712}_{-0.387}$ & $72.92^{+2.15}_{-2.40}$ & $11.31\pm0.43$ & -- & -- & -- & -- & -- & -- & -- & -- & -- & -- \\
XCDM & QSO-AS + \hiig\ + \mq\ + A118 & $0.0224\pm0.0117$ & $0.1449^{+0.0314}_{-0.0246}$ & $0.314^{+0.051}_{-0.044}$ & -- & $-1.836^{+0.804}_{-0.419}$ & $73.14^{+2.14}_{-2.48}$ & $11.34\pm0.43$ & $0.291^{+0.023}_{-0.029}$ & $1.675\pm0.056$ & $0.294\pm0.044$ & $0.413^{+0.027}_{-0.033}$ & $50.24\pm0.24$ & $1.124\pm0.087$ & -- & -- & -- & -- \\
 & $H(z)$ + BAO + SN & $0.0262^{+0.0033}_{-0.0037}$ & $0.1120\pm0.0110$ & $0.295\pm0.016$ & -- & $-0.941\pm0.064$ & $68.55\pm1.85$ & -- & -- & -- & -- & -- & -- & -- & -- & -- & -- & -- \\
 & HzBSNQHMA\tnote{d} & $0.0271^{+0.0027}_{-0.0031}$ & $0.1147^{+0.0098}_{-0.0097}$ & $0.294\pm0.015$ & -- & $-0.959\pm0.059$ & $69.66\pm1.16$ & $10.94\pm0.25$ & $0.292^{+0.023}_{-0.029}$ & $1.685\pm0.055$ & $0.293\pm0.044$ & $0.413^{+0.027}_{-0.033}$ & $50.20\pm0.24$ & $1.131\pm0.087$ & -- & -- & -- & -- \\
 & QHMPA101\tnote{e} & $0.0224\pm0.0118$ & $0.1429^{+0.0333}_{-0.0248}$ & $0.310^{+0.054}_{-0.044}$ & -- & $-1.801^{+0.814}_{-0.421}$ & $73.09^{+2.13}_{-2.48}$ & $11.33\pm0.43$ & $0.290^{+0.022}_{-0.029}$ & $1.675\pm0.056$ & $0.294\pm0.044$ & $0.421^{+0.029}_{-0.036}$ & $50.16\pm0.26$ & $1.152\pm0.094$ & $0.346^{+0.032}_{-0.045}$ & $-0.713\pm0.102$ & $0.758\pm0.081$ & $11.45\pm4.35$ \\
\midrule
 & $H(z)$ + BAO & $0.0294^{+0.0047}_{-0.0050}$ & $0.0980^{+0.0186}_{-0.0187}$ & $0.292\pm0.025$ & $-0.027\pm0.109$ & $-0.770^{+0.149}_{-0.098}$ & $66.13^{+2.35}_{-2.36}$ & -- & -- & -- & -- & -- & -- & -- & -- & -- & -- & -- \\
 & \mq\ + A118 & -- & -- & $>0.170$ & $0.309^{+0.347}_{-0.650}$ & $<-0.132$ & -- & -- & $0.293^{+0.023}_{-0.030}$ & $1.711^{+0.081}_{-0.058}$ & $0.302\pm0.047$ & $0.413^{+0.027}_{-0.033}$ & $50.10^{+0.28}_{-0.32}$ & $1.114\pm0.091$ & -- & -- & -- & -- \\
Non-flat & QSO-AS + \hiig & $0.0224\pm0.0114$ & $0.1122^{+0.0473}_{-0.0326}$ & $0.258^{+0.086}_{-0.057}$ & $0.018^{+0.345}_{-0.383}$ & $-1.670^{+1.063}_{-0.245}$ & $72.34\pm2.16$ & $11.20\pm0.49$ & -- & -- & -- & -- & -- & -- & -- & -- & -- & -- \\
XCDM & QSO-AS + \hiig\ + \mq\ + A118 & $0.0225\pm0.0118$ & $0.1327^{+0.0348}_{-0.0286}$ & $0.294^{+0.059}_{-0.051}$ & $0.054^{+0.227}_{-0.238}$ & $-2.042^{+1.295}_{-0.451}$ & $72.79^{+2.18}_{-2.42}$ & $11.23^{+0.44}_{-0.49}$ & $0.291^{+0.023}_{-0.030}$ & $1.678\pm0.057$ & $0.294\pm0.045$ & $0.414^{+0.027}_{-0.033}$ & $50.23\pm0.25$ & $1.127\pm0.089$ & -- & -- & -- & -- \\
 & $H(z)$ + BAO + SN & $0.0262^{+0.0037}_{-0.0043}$ & $0.1119\pm0.0157$ & $0.295\pm0.022$ & $-0.001\pm0.098$ & $-0.948^{+0.098}_{-0.068}$ & $68.53\pm1.90$ & -- & -- & -- & -- & -- & -- & -- & -- & -- & -- & -- \\
 & HzBSNQHMA\tnote{d} & $0.0269^{+0.0033}_{-0.0039}$ & $0.1155^{+0.0128}_{-0.0127}$ & $0.295\pm0.019$ & $-0.009^{+0.077}_{-0.083}$ & $-0.959^{+0.090}_{-0.063}$ & $69.65\pm1.16$ & $10.93\pm0.26$ & $0.292^{+0.023}_{-0.029}$ & $1.685\pm0.055$ & $0.293\pm0.044$ & $0.413^{+0.027}_{-0.033}$ & $50.20\pm0.24$ & $1.130\pm0.087$ & -- & -- & -- & -- \\
 & QHMPA101\tnote{e} & $0.0225\pm0.0118$ & $0.1283^{+0.0354}_{-0.0290}$ & $0.286^{+0.060}_{-0.052}$ & $0.104^{+0.232}_{-0.244}$ & $-2.131^{+1.385}_{-0.524}$ & $72.70^{+2.19}_{-2.43}$ & $11.19\pm0.46$ & $0.291^{+0.023}_{-0.029}$ & $1.679\pm0.056$ & $0.293\pm0.044$ & $0.421^{+0.029}_{-0.036}$ & $50.14\pm0.26$ & $1.159\pm0.095$ & $0.346^{+0.032}_{-0.045}$ & $-0.711\pm0.102$ & $0.761\pm0.080$ & $11.28\pm4.32$ \\
\midrule
 & $H(z)$ + BAO & $0.0320^{+0.0054}_{-0.0041}$ & $0.0855^{+0.0175}_{-0.0174}$ & $0.275\pm0.023$ & -- & $1.267^{+0.536}_{-0.807}$ & $65.47^{+2.22}_{-2.21}$ & -- & -- & -- & -- & -- & -- & -- & -- & -- & -- & -- \\
 & \mq\ + A118 & -- & -- & $0.561^{+0.332}_{-0.241}$ & -- & -- & -- & -- & $0.293^{+0.023}_{-0.030}$ & $1.725\pm0.052$ & $0.303\pm0.046$ & $0.411^{+0.026}_{-0.032}$ & $50.05\pm0.25$ & $1.110\pm0.087$ & -- & -- & -- & -- \\
Flat & QSO-AS + \hiig & $0.0217^{+0.0081}_{-0.0138}$ & $0.0543^{+0.0225}_{-0.0471}$ & $0.154^{+0.053}_{-0.086}$ & -- & $<6.506$ & $70.64\pm1.80$ & $10.81\pm0.34$ & -- & -- & -- & -- & -- & -- & -- & -- & -- & -- \\
\pcdm & QSO-AS + \hiig\ + \mq\ + A118 & $0.0219^{+0.0076}_{-0.0165}$ & $0.0644^{+0.0384}_{-0.0410}$ & $0.175^{+0.075}_{-0.081}$ & -- & $<6.756$ & $70.50\pm1.76$ & $10.80\pm0.34$ & $0.292^{+0.023}_{-0.029}$ & $1.691\pm0.055$ & $0.291\pm0.044$ & $0.414^{+0.027}_{-0.033}$ & $50.18\pm0.24$ & $1.140\pm0.088$ & -- & -- & -- & -- \\
 & $H(z)$ + BAO + SN & $0.0278^{+0.0032}_{-0.0039}$ & $0.1054^{+0.0117}_{-0.0100}$ & $0.287\pm0.017$ & -- & $0.324^{+0.122}_{-0.264}$ & $68.29\pm1.78$ & -- & -- & -- & -- & -- & -- & -- & -- & -- & -- & -- \\
 & HzBSNQHMA\tnote{d} & $0.0286^{+0.0025}_{-0.0033}$ & $0.1089^{+0.0103}_{-0.0083}$ & $0.286\pm0.015$ & -- & $0.249^{+0.069}_{-0.239}$ & $69.50\pm1.14$ & $10.92\pm0.25$ & $0.292^{+0.023}_{-0.029}$ & $1.686\pm0.054$ & $0.293\pm0.044$ & $0.413^{+0.027}_{-0.033}$ & $50.20\pm0.24$ & $1.132\pm0.086$ & -- & -- & -- & -- \\
 & QHMPA101\tnote{e} & $0.0217^{+0.0098}_{-0.0164}$ & $0.0607^{+0.0296}_{-0.0493}$ & $0.167^{+0.067}_{-0.088}$ & -- & $<7.149$ & $70.53\pm1.76$ & $10.80\pm0.34$ & $0.292^{+0.023}_{-0.029}$ & $1.691\pm0.054$ & $0.291\pm0.044$ & $0.421^{+0.029}_{-0.036}$ & $50.10\pm0.26$ & $1.172\pm0.093$ & $0.347^{+0.032}_{-0.045}$ & $-0.707\pm0.102$ & $0.768\pm0.078$ & $10.89\pm4.21$ \\
\midrule
 & $H(z)$ + BAO & $0.0320^{+0.0057}_{-0.0038}$ & $0.0865^{+0.0172}_{-0.0198}$ & $0.277^{+0.023}_{-0.026}$ & $-0.034^{+0.087}_{-0.098}$ & $1.360^{+0.584}_{-0.819}$ & $65.53\pm2.19$ & -- & -- & -- & -- & -- & -- & -- & -- & -- & -- & -- \\
 & \mq\ + A118 & -- & -- & $0.549^{+0.241}_{-0.268}$ & $-0.001^{+0.296}_{-0.286}$ & -- & -- & -- & $0.293^{+0.023}_{-0.030}$ & $1.725\pm0.052$ & $0.303\pm0.047$ & $0.412^{+0.027}_{-0.033}$ & $50.05\pm0.25$ & $1.111\pm0.089$ & -- & -- & -- & -- \\
Non-flat & QSO-AS + \hiig & $0.0219^{+0.0093}_{-0.0130}$ & $0.0576^{+0.0268}_{-0.0424}$ & $0.163^{+0.058}_{-0.081}$ & $0.181^{+0.180}_{-0.339}$ & $<7.875$ & $70.21\pm1.83$ & $10.70^{+0.36}_{-0.41}$ & -- & -- & -- & -- & -- & -- & -- & -- & -- & -- \\
\pcdm & QSO-AS + \hiig\ + \mq\ + A118 & $0.0222^{+0.0098}_{-0.0135}$ & $0.0725^{+0.0398}_{-0.0368}$ & $0.193^{+0.077}_{-0.071}$ & $0.044^{+0.104}_{-0.256}$ & $<7.239$ & $70.38\pm1.84$ & $10.79^{+0.38}_{-0.41}$ & $0.292^{+0.024}_{-0.030}$ & $1.691\pm0.055$ & $0.292\pm0.045$ & $0.415^{+0.028}_{-0.033}$ & $50.18\pm0.25$ & $1.140\pm0.089$ & -- & -- & -- & -- \\
 & $H(z)$ + BAO + SN & $0.0271^{+0.0038}_{-0.0043}$ & $0.1095\pm0.0152$ & $0.292\pm0.022$ & $-0.038^{+0.071}_{-0.085}$ & $0.382^{+0.151}_{-0.299}$ & $68.48\pm1.85$ & -- & -- & -- & -- & -- & -- & -- & -- & -- & -- & -- \\
 & HzBSNQHMA\tnote{d} & $0.0277^{+0.0034}_{-0.0040}$ & $0.1126\pm0.0128$ & $0.291\pm0.019$ & $-0.040^{+0.064}_{-0.072}$ & $0.316^{+0.101}_{-0.292}$ & $69.52\pm1.15$ & $10.89\pm0.25$ & $0.292^{+0.023}_{-0.030}$ & $1.685\pm0.055$ & $0.294\pm0.044$ & $0.413^{+0.027}_{-0.033}$ & $50.20\pm0.24$ & $1.128\pm0.087$ & -- & -- & -- & -- \\
 & QHMPA101\tnote{e} & $0.0221^{+0.0094}_{-0.0138}$ & $0.0677^{+0.0368}_{-0.0389}$ & $0.183^{+0.072}_{-0.076}$ & $0.084^{+0.119}_{-0.275}$ & $<7.625$ & $70.33\pm1.83$ & $10.76^{+0.37}_{-0.41}$ & $0.292^{+0.023}_{-0.030}$ & $1.692\pm0.054$ & $0.291\pm0.044$ & $0.421^{+0.029}_{-0.036}$ & $50.09\pm0.26$ & $1.172\pm0.094$ & $0.347^{+0.033}_{-0.046}$ & $-0.707\pm0.102$ & $0.769\pm0.078$ & $10.87\pm4.22$ \\
\bottomrule
\end{tabular}
\begin{tablenotes}[flushleft]
\item [a] \wx\ corresponds to flat/non-flat XCDM and $\alpha$ corresponds to flat/non-flat \pcdm.
\item [b] \hunit. In \mq, A118, and \mq\ + A118 cases, $\Omega_b$ and $H_0$ are set to be 0.05 and 70 \hunit, respectively.
\item [c] pc.
\item [d] $H(z)$ + BAO + SN + QSO-AS + \hiig\ + \mq\ + A118.
\item [e] QSO-AS + \hiig\ + \mq\ + Platinum + A101.
\end{tablenotes}
\end{threeparttable}%
}
\end{sidewaystable*}


\cleardoublepage


\chapter{Gamma-ray burst data strongly favour the three-parameter fundamental plane (Dainotti) correlation over the two-parameter one}
\label{makereference10}

This chapter is based on \cite{CaoDainottiRatra2022b}.

\section{Introduction} 
\label{makereference10.1}

In the standard general-relativistic spatially-flat \lcdm\ model \citep{peeb84}, dark energy is a time-independent cosmological constant $\Lambda$ that sources $\sim70\%$ of the current cosmological energy budget and the observed currently accelerating cosmological expansion. The predictions of this model are consistent with most current cosmological observations, such as Hubble parameter [$H(z)$], type Ia supernova (SNIa) apparent magnitude, cosmic microwave background (CMB) anisotropy, and baryon acoustic oscillation (BAO) measurements \citep[see, e.g.][]{Farooq_Ranjeet_Crandall_Ratra_2017, scolnic_et_al_2018, planck2018b, eBOSS_2020}. The measurements, however, are not yet decisive enough \citep[see, e.g.][]{Dainottietal2021a,Dainottietal2022,DiValentinoetal2021a,PerivolaropoulosSkara2021,Abdallaetal2022} to disallow other cosmological models. Here we also consider dynamical dark energy models as well as models with non-zero spatial curvature.

In addition to the better-established cosmological probes mentioned above, developing cosmological probes can play a significant role in paring down the theoretical options. Developing cosmological probes that are under active debate now include \hii\ starburst galaxy measurements that reach to $z \sim 2.4$ \citep{Mania_2012, Chavez_2014, GonzalezMoranetal2021, CaoRyanRatra2020, CaoRyanRatra2021, Johnsonetal2022, Mehrabietal2022}, quasar (QSO) angular size observations that reach to $z \sim 2.7$ \citep{Cao_et_al2017a, Ryanetal2019, Zhengetal2021, Lian_etal_2021, CaoRyanRatra2022}, reverberation-mapped QSO observations that reach to redshift $z \sim 3.4$ \citep{Czernyetal2021, Zajaceketal2021, Yuetal2021, Khadkaetal_2021a, Khadkaetal2022a, Khadkaetal2022b, Cao:2022pdv}, QSO flux measurements that reach to $z \sim 7.5$ \citep{RisalitiLusso2015, RisalitiLusso2019, KhadkaRatra2020a, KhadkaRatra2020b, KhadkaRatra2021, KhadkaRatra2022, Lussoetal2020, ZhaoXia2021, Rezaeietal2022, Luongoetal2021, Leizerovichetal2021, Colgainetal2022, DainottiBardiacchi2022},\footnote{The latest \cite{Lussoetal2020} QSO flux compilation assumes a model for the QSO UV--X-ray correlation that is not valid above a much lower redshift, $z \sim 1.5-1.7$ (i.e., above these redshifts the assumed QSO UV--X-ray luminosities correlation relation is different in different cosmological models), meaning that these QSOs can be used to determine only much lower-$z$ cosmological constraints \citep{KhadkaRatra2021, KhadkaRatra2022}.} and --- the main subject of our paper --- gamma-ray burst (GRB) observations that reach to $z \sim 8.2$ \citep{CardoneCapozzielloDainotti2009, Cardoneetal2010, samushia_ratra_2010, Dainottietal2013a, Postnikovetal2014, Wangetal2015, Wang_2016, DainottiaDelVecchio2017, Dainottietal2018, DainottiAmati2018, Wangetal_2021, Dirirsa2019, Amati2019, KhadkaRatra2020c, Huetal2021, Demianskietal_2021, Khadkaetal_2021b, Luongoetal2021, LuongoMuccino2021, Caoetal_2021, Liuetal2022, DainottiNielson2022, DainottiSarracino2022}. In this paper the highest redshift GRB we use is at $z = 9.4$ \citep{Cucchiaraetal2011}, but GRBs might be detectable to $z = 20$ \citep{Lamb2000}, because they are in principle free from dust extinction. The highest low-$z$ better-established BAO and SNIa observations reach to $z\sim 2.3$, while the high-$z$ better-established CMB anisotropy measurements largely probe $z \sim 1100$, mostly leaving the exploration of the intermediate part of redshift space to the developing probes. 

In this paper we study long GRBs, those GRBs with burst duration longer than 2 s. The measured quantities for the GRBs are the redshift $z$, the characteristic time scale $T^{*}_{X}$ which marks the end of the plateau emission, the measured X-ray energy flux $F_{X}$ at $T^{*}_{X}$, the measured $\gamma$-ray energy flux $F_{\rm peak}$ in the peak of the prompt emission over a 1 s interval, and the X-ray photon indices of the plateau phase $\alpha_{\rm plateau}$ and of the prompt emission $\alpha_{\rm prompt}$.  We make use of the 50 Platinum GRBs, spanning the redshift range $0.553 \leq z \leq 5.0$, introduced in \citet{Dainottietal2020}, that we previously studied \citep{CaoDainottiRatra2022}, and a new LGRB95 sample consisting of 95 long GRBs, spanning $0.297 \leq z \leq 9.4$ and also taken from \citet{Dainottietal2020}, as well as the combined LGRB145 data set of 145 GRBs, to test whether they are better described by the three-dimensional (3D) fundamental plane (Dainotti) correlation between the peak prompt luminosity, the luminosity at the end of the plateau emission, and its rest frame duration \citep{Srinivasaragavan2020ApJ...903...18S,Dainottietal2016,Dainottietal2017,Dainottietal2020,Dainottietal2021,Dainotti2021ApJS..255...13D} or by the two-dimensional (2D) Dainotti correlation between the luminosity at the end of the plateau emission and its rest frame duration \citep{Dainottietal2008,Dainottietal2010,Dainottietal2011,Dainottietal2013a,Dainotti15a,Dainottietal2017},\footnote{The 2D and 3D Dainotti correlation relations are discussed in Sec.\ \ref{makereference10.3}} and to constrain cosmological-model and GRB-correlation parameters. The Platinum sample is a compilation of the higher-quality (lower intrinsic dispersion) GRBs considered in \citet{Dainottietal2020}, and are tabulated in Table \ref{tab:P50aB}.\footnote{This is a correction of table A1 of \cite{CaoDainottiRatra2022} that incorrectly accounted for the GRB $K$-corrections. In Sec.\ \ref{makereference10.3} we discuss our improved method of accounting for the spectral evolution of GRBs and both the prompt and afterglow photon indices. These corrections are not appreciable and do not affect any of the qualitative conclusions of \cite{CaoDainottiRatra2022}.} The remaining 95 long GRBs considered in \citet{Dainottietal2020} constitute the LGRB95 sample listed in Table \ref{tab:LGRB95}. LGRB95 data have a $\sim50$\% larger intrinsic scatter parameter, which contains the unknown systematic errors, than the Platinum data.

Based on information criteria, we discover that Platinum, LGRB95, and LGRB145 data strongly prefer the 3D Dainotti correlation over the 2D one. Although LGRB95 data have $\sim50$\% larger intrinsic scatter parameter values than Platinum data, they provide consistent but slightly tighter cosmological-model and GRB-correlation parameter constraints than do Platinum data, perhaps solely due to the larger number of data points, 95 versus 50. LGRB145 data provide tighter constraints on GRB-correlation parameters than those from the individual GRB data sets. 

Our paper is organized as follows. We use the cosmological models described in Chapter \ref{sec:models} and describe the data sets we use in Sec.\ \ref{makereference10.2}. We outline our analyses methods in Sec.\ \ref{makereference10.3} and present results in Sec.\ \ref{makereference10.4}. Our summary and conclusions are in Sec.\ \ref{makereference10.5}.

\section{Data}
\label{makereference10.2}

In this paper we analyze three different GRB data sets to study two-parameter or two-dimensional (2D) Dainotti correlation and the three-parameter or 3D fundamental-plane (Dainotti) correlation. These contain only long GRBs, with burst duration longer than 2 s, and are taken from the compilation of \citet{Dainottietal2020}. For these data sets, the measured quantities for a GRB are the redshift $z$, the characteristic time scale $T^{*}_{X}$ which marks the end of the plateau emission, the measured X-ray energy flux $F_{X}$ at $T^{*}_{X}$, the prompt peak $\gamma$-ray energy flux $F_{\rm peak}$ over a 1 s interval, and the X-ray photon indices of the plateau phase $\alpha_{\rm plateau}$ and of the prompt emission $\alpha_{\rm prompt}$. The data sets we use here are summarized next.

\begin{itemize}

\item[]{\it Platinum sample}. This includes 50 long GRBs that have a plateau phase with angle $< 41^\circ$, that do not flare during the plateau phase, and that have a plateau phase that lasts longer than 500 s. The first criterion follows from the evidence that those with angle $> 41^\circ$ are outliers of the Gaussian distribution; the second criterion eliminates flaring-contaminated cases; and, the third criterion eliminates cases where prompt emission might mask the plateau \citep{Willingaleetal2007, Willingaleetal2010}. The Platinum GRBs are listed in Table \ref{tab:P50aB} of Appendix \ref{AppendixB}, which is a correction of table A1 of \cite{CaoDainottiRatra2022}. This sample spans the redshift range $0.553 \leq z \leq 5.0$.

\item[]{\it LGRB95 sample}. This sample includes the remaining 95 long GRBs from the compilation of \citet{Dainottietal2020}. As discussed below, this GRB data set has a larger intrinsic scatter parameter $\sigma_{\rm int}$ than the Platinum GRBs. These GRBs are listed in Table \ref{tab:LGRB95} of Appendix \ref{AppendixB}. This sample spans the redshift range $0.297 \leq z \leq 9.4$. 

\item[]{\it LGRB145 sample}. This sample is a combination of the Platinum sample and the LGRB95 sample and spans the redshift range $0.297 \leq z \leq 9.4$. 

\end{itemize}

\section{Data Analysis Methodology}
\label{makereference10.3}

The 3D fundamental plane (or 3D Dainotti) correlation \citep{Dainottietal2016, Dainottietal2017, Dainottietal2020, Dainottietal2021} is
\begin{equation}
    \label{eq:3DC10}
    \log L_{X} = C_{o}  + a\log T^{*}_{X} + b\log L_{\rm peak},
\end{equation}
where the X-ray source rest-frame luminosity
\be
\label{eq:LxC10}
    L_{X}=4\pi D_L^2F_{X}K_{\rm plateau},
\ee
with the power-law (PL) plateau $K$-correction
\be
\label{eq:kplC10}
K_{\rm plateau}=(1+z)^{\alpha_{\rm plateau}-2},
\ee
the peak prompt luminosity 
\be
\label{eq:LpeakC10}
    L_{\rm peak}=4\pi D_L^2F_{\rm peak}K_{\rm prompt},
\ee
with the prompt $K$-correction
\be
\label{eq:kprC10}
    K_{\mathrm{prompt}} = \frac{\int^{150/(1+z)}_{15/(1+z)} Ef(E)dE}{\int^{150}_{15} Ef(E)dE},
\ee
where $E$ is the photon energy and the differential photon spectrum \citep{Sakamotoetal2011}
\be
    f(E) = 
    \begin{cases}
    K^{\mathrm{PL}}_{50}\big(\frac{E}{50\ \mathrm{keV}}\big)^{\alpha^{\mathrm{PL}}} & \text{if}\ \mathrm{PL}, \\
    \vspace{1mm}
    K^{\mathrm{CPL}}_{50}\big(\frac{E}{50\ \mathrm{keV}}\big)^{\alpha^{\mathrm{CPL}}}\exp \Big[\frac{-E\left(2+\alpha^{\mathrm{CPL}}\right)}{E_{\mathrm{peak}}}\Big] & \text{if}\ \mathrm{CPL}.
    \end{cases}
\ee
Here $K^{\mathrm{PL}}_{50}$ and $K^{\mathrm{CPL}}_{50}$ are the normalization at 50 keV in units of photons $\rm cm^{-2}\ s^{-1}\ keV^{-1}$ in the PL and cutoff power-law (CPL) models, respectively, $\alpha^{\mathrm{PL}}$ and $\alpha^{\mathrm{CPL}}$ are the PL and CPL photon indices, respectively, and $E_{\mathrm{peak}}$ is the peak energy in the $\nu F_{\nu}$ spectrum in units of keV, where $\nu$ is the photon frequency proportional to $E$ and $F_{\nu}$ is the photon energy flux per unit frequency. When $\Delta\chi^2\equiv\chi^2_{\mathrm{PL}}-\chi^2_{\mathrm{CPL}}>6$, the CPL model is used to compute the prompt $K$-correction, otherwise the PL model is used. In the preceding equations, $\{C_{o},a,b\}$ are the GRB correlation parameters to be constrained, $T^{*}_{X}$ (s) is the time at the end of the plateau emission, $F_{X}$ and $F_{\rm peak}$ are the measured X-ray and $\gamma$-ray energy flux (erg cm$^{-2}$ s$^{-1}$) at $T^{*}_{X}$ and in the peak of the prompt emission over a 1 s interval, respectively. The luminosity distance is defined in equation \eqref{eq:DLC8}.

Here we have improved upon the analysis of \citet{CaoDainottiRatra2022}, that assumed the GRB $K$-corrections are the same throughout the burst duration, by also considering the prompt emission photon index. We consider the sliced photon index in the spectrum starting from the time of the beginning of plateau emission to the time of the end of plateau emission. We use the photon counting (PC) mode for the majority of cases and the window timing (WT) mode for only a few cases where we do not have the PC mode. This procedure differs from previous analyses in which photon indices were computed using an average of both WT and PC modes.

\begin{table}
\centering
\begin{threeparttable}
\caption{Flat priors of the constrained parameters.}
\label{tab:priorsC10}
\setlength{\tabcolsep}{3.5pt}
\begin{tabular}{lcc}
\toprule
Parameter & & Prior\\
\midrule
 & Cosmological Parameters & \\
\midrule
$\Omega_{c}$ &  & [-0.051315, 0.948685]\\
\ok &  & [-2, 2]\\
$\alpha$ &  & [0, 10]\\
\wx &  & [-5, 0.33]\\
\midrule
 & GRB Correlation Parameters & \\
\midrule
$a$ &  & [-2, -0.001]\\
$b$\tnote{a} &  & [0, 2]\\
$C_{o}$ &  & [-10, 60]\\
$\sigma_{\rm int}$ &  & [0, 5]\\
\bottomrule
\end{tabular}
\begin{tablenotes}[flushleft]
\item [a] In two-dimensional GRB analyses, $b=0$.
\end{tablenotes}
\end{threeparttable}%
\end{table}

The 3D fundamental plane (or 3D Dainotti) correlation reduces to the 2D Dainotti correlation when $b = 0$. Note that the 3D fundamental plane Dainotti relation is a combination of this 2D Dainotti $L_X-T^*_X$ correlation and another 2D Dainotti correlation between the peak prompt luminosity and the luminosity at the end of the plateau emission \citep{DainottiOstrowskiWillingale2011,Dainottietal2015}.

The natural log of the likelihood function \citep{D'Agostini_2005} is
\be
\label{eq:LH_GRB}
    \ln\mathcal{L}_{\rm GRB}= -\frac{1}{2}\Bigg[\chi^2_{\rm GRB}+\sum^{N}_{i=1}\ln\left(2\pi\sigma^2_{\mathrm{tot},i}\right)\Bigg],
\ee
where, in the 3D fundamental plane relation case,
\be
\label{eq:chi2_GRB}
    \chi^2_{\rm GRB} = \sum^{N}_{i=1}\bigg[\frac{(\log L_{X,i} - C_{o}  - a\log T^{*}_{X,i} - b\log L_{\mathrm{peak},i})^2}{\sigma^2_{\mathrm{tot},i}}\bigg],
\ee
with
\be
\label{eq:sigma}
\sigma^2_{\mathrm{tot},i}=\sigma_{\rm int}^2+\sigma_{{\log L_{X,i}}}^2+a^2\sigma_{{\log T^{*}_{X,i}}}^2+b^2\sigma_{{\log L_{\mathrm{peak},i}}}^2.
\ee
$N$ is the number of data points and $\sigma_{\rm int}$ is the intrinsic scatter parameter that contains the unknown systematic uncertainty. Note that $\sigma_{{\log L_{X}}}=\sigma_{\log\mathrm{FK}_{\mathrm{plateau}}}$ and $\sigma_{{\log L_{\mathrm{peak}}}}=\sigma_{\log\mathrm{FK}_{\mathrm{prompt}}}$, where $\log\mathrm{FK}_{\mathrm{plateau}}\equiv\log F_{X}+\log K_{\rm plateau}$ and $\log\mathrm{FK}_{\mathrm{prompt}}\equiv\log F_{\rm peak}+\log K_{\rm prompt}$. In the 2D Dainotti correlation case we fix $b = 0$ in equations \eqref{eq:chi2_GRB} and \eqref{eq:sigma}. 

We avoid the circularity problem by simultaneously constraining cosmological-model and GRB-correlation parameters, and if the GRB correlation parameters are independent of the cosmological models used in the analysis then the GRBs are standardizable \citep{KhadkaRatra2020c}. The simultaneous fitting technique also allows for the determination of GRB-only cosmological constraints, unlike the cosmological constraints determined from GRBs that have been calibrated using other data (which are then correlated with data used in the calibration process), that can be directly compared to (or combined with) constraints determined from other data.

We list the flat priors of the free cosmological and GRB correlation parameters in Table \ref{tab:priorsC10}. Since these GRB data sets cannot constrain $\Omega_b$ and $H_0$, we set $\Omega_b=0.05$ and $H_0=70$ \hunit\ in our analyses. By maximizing the likelihood functions, we obtain the unmarginalized best-fitting values and posterior distributions of all free cosmological-model and GRB-correlation parameters. We use the Markov chain Monte Carlo (MCMC) code \textsc{MontePython} \citep{Audrenetal2013,Brinckmann2019} that interacts with the \textsc{class} code cosmological model physics. We use the \textsc{python} package \textsc{getdist} \citep{Lewis_2019} to perform our analyses.

The definitions of the Akaike Information Criterion (AIC), the Bayesian Information Criterion (BIC), and the deviance information criterion (DIC) can be found in our previous papers \citep[see, e.g.][]{CaoKhadkaRatra2022,CaoDainottiRatra2022}. $\Delta \mathrm{AIC}$, $\Delta \mathrm{BIC}$, and $\Delta \mathrm{DIC}$ are the differences between the AIC, BIC, and DIC values of the other five cosmological models and those of the flat \lcdm\ reference model, while $\Delta \mathrm{AIC}^{\prime}$, $\Delta \mathrm{BIC}^{\prime}$, and $\Delta \mathrm{DIC}^{\prime}$ are the differences between values of 2D and 3D Dainotti correlations in the same cosmological model. Negative (positive) values of these $\Delta \mathrm{IC}$s indicate that the model under investigation fits the data better (worse) than does the reference model. Relative to the model with the minimum IC, $\Delta \mathrm{IC} \in (0, 2]$ is defined to be weak evidence against the model under investigation, $\Delta \mathrm{IC} \in (2, 6]$ is positive evidence against the model under investigation, $\Delta \mathrm{IC} \in (6, 10] $ is strong evidence against the model under investigation, and $\Delta \mathrm{IC}>10$ is very strong evidence against the model under investigation.

As in \cite{CaoRatra2022}, we assume one massive and two massless neutrino species, with the non-relativistic neutrino physical energy density parameter $\onh=\sum m_{\nu}/(93.14\ \rm eV)=0.06\ \rm eV/(93.14\ \rm eV)$, where $h$ is the Hubble constant in units of 100 \hunit. The non-relativistic matter density parameter $\Om = (\onh + \obh + \och)/{h^2}$, where the current value of the baryonic matter energy density parameter is set to $\Omega_b=0.05$\footnote{Since GRB data are unable to constrain $\Omega_b$.} and the current value of the cold dark matter energy density parameter ($\Omega_c$) is constrained as a free cosmological parameter. 

\begin{figure*}
\centering
 \subfloat[]{%
    \includegraphics[width=0.5\textwidth,height=0.5\textwidth]{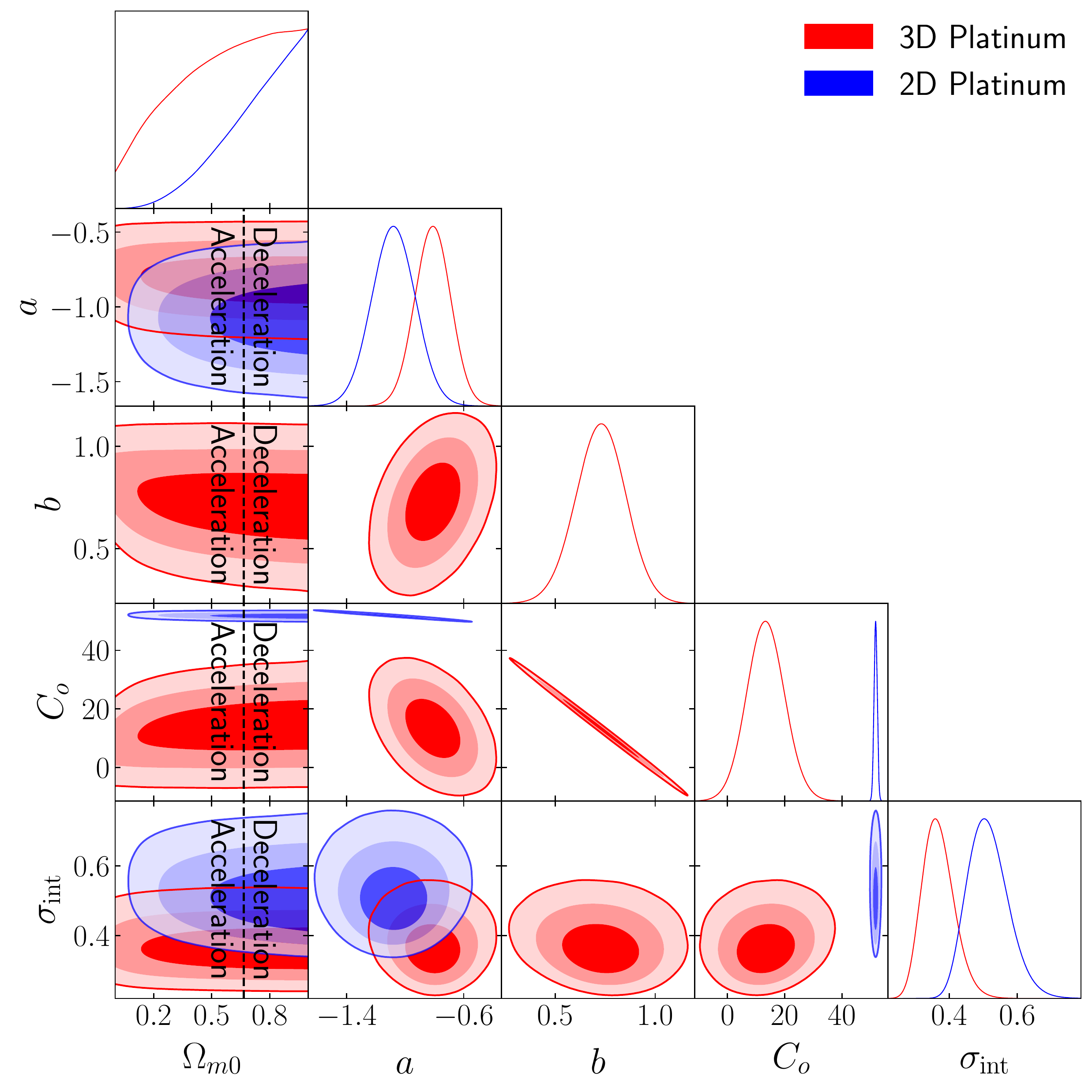}}
 \subfloat[]{%
    \includegraphics[width=0.5\textwidth,height=0.5\textwidth]{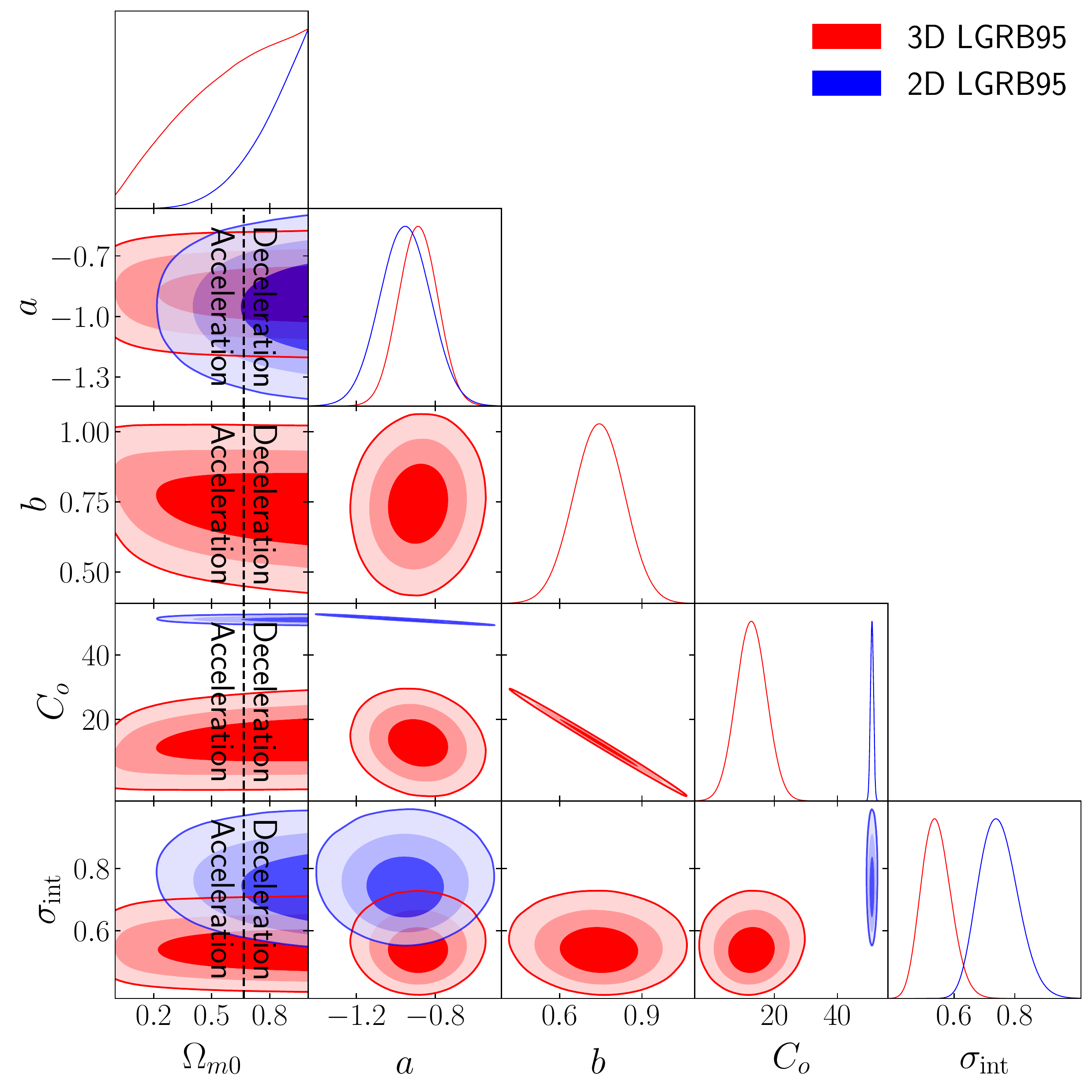}}\\
 \subfloat[]{%
    \includegraphics[width=0.5\textwidth,height=0.5\textwidth]{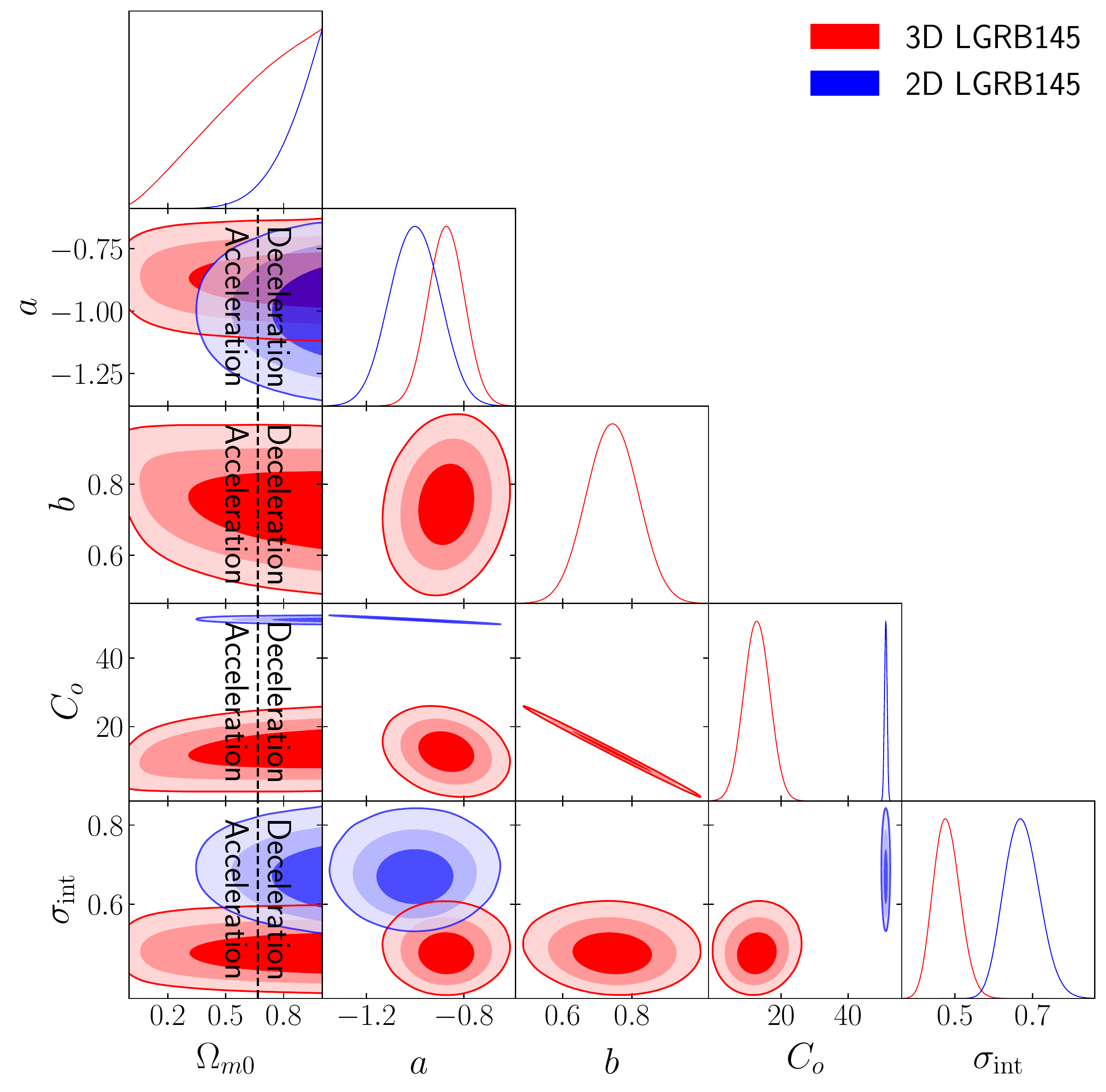}}
 \subfloat[]{%
    \includegraphics[width=0.5\textwidth,height=0.5\textwidth]{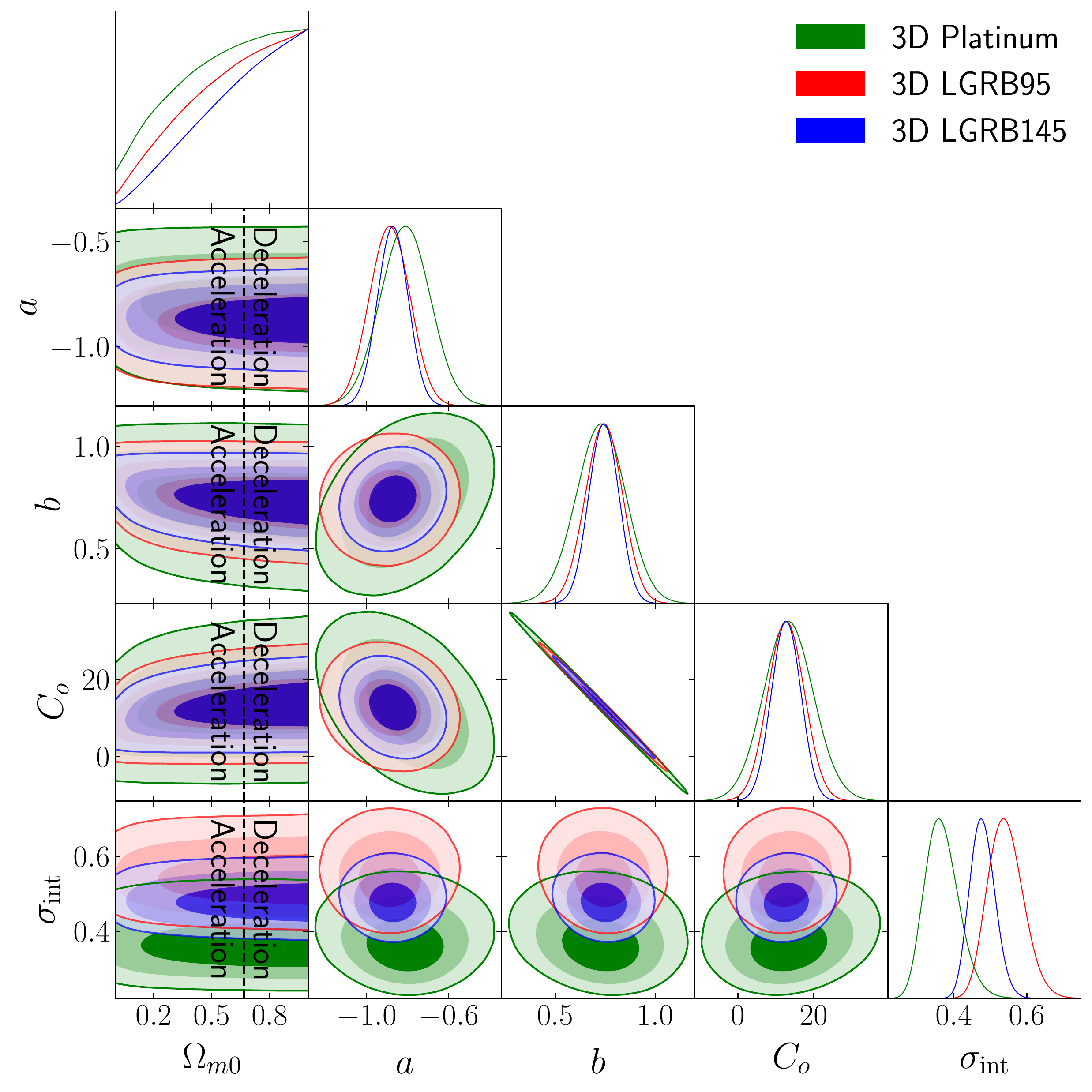}}\\
\caption{One-dimensional likelihood distributions and 1$\sigma$, 2$\sigma$, and 3$\sigma$ two-dimensional likelihood confidence contours for flat \lcdm\ from various combinations of data. The zero-acceleration black dashed lines in panels (a) and (b) divide the parameter space into regions associated with currently-accelerating (left) and currently-decelerating (right) cosmological expansion.}
\label{fig1C10}
\end{figure*}

\begin{figure*}
\centering
 \subfloat[]{%
    \includegraphics[width=0.5\textwidth,height=0.5\textwidth]{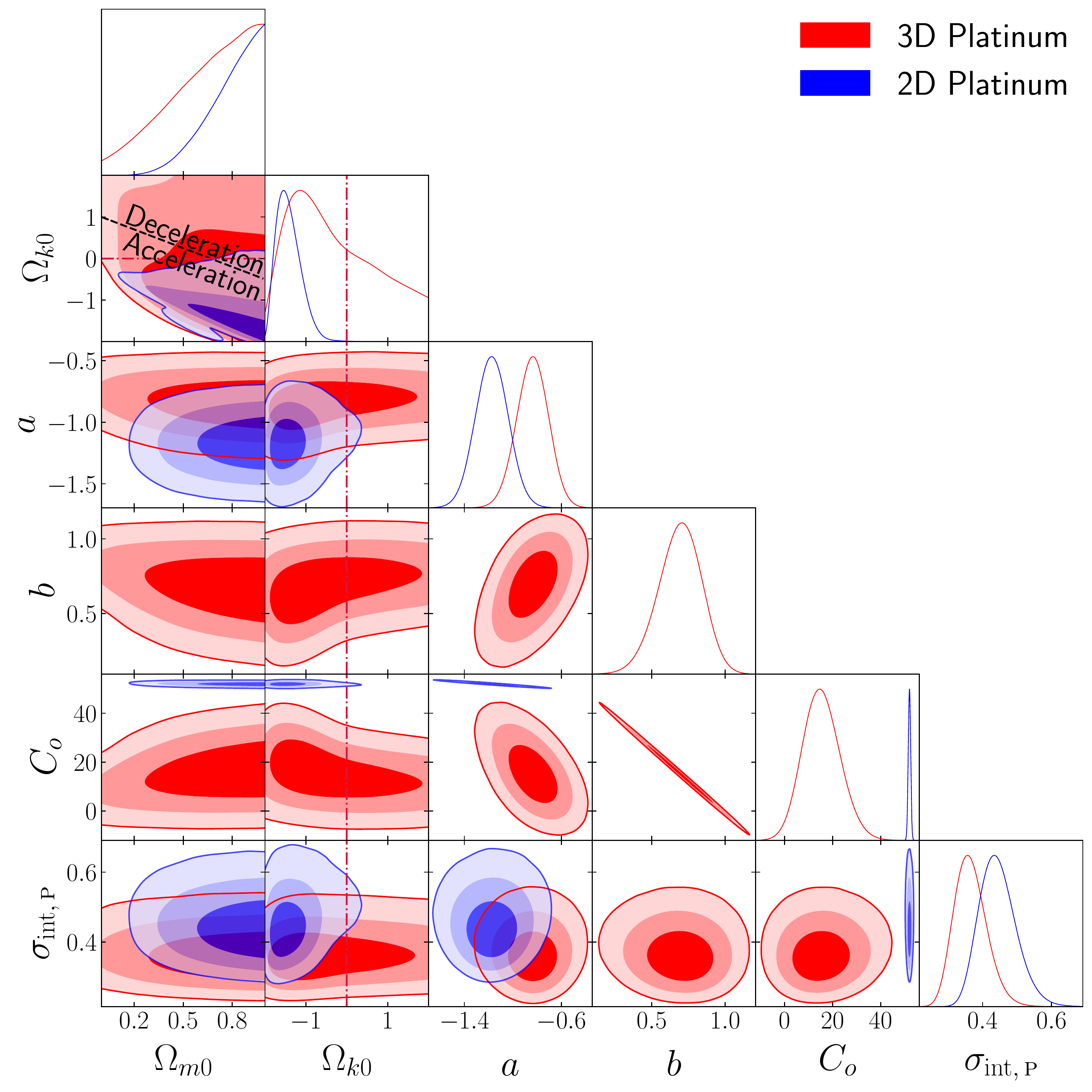}}
 \subfloat[]{%
    \includegraphics[width=0.5\textwidth,height=0.5\textwidth]{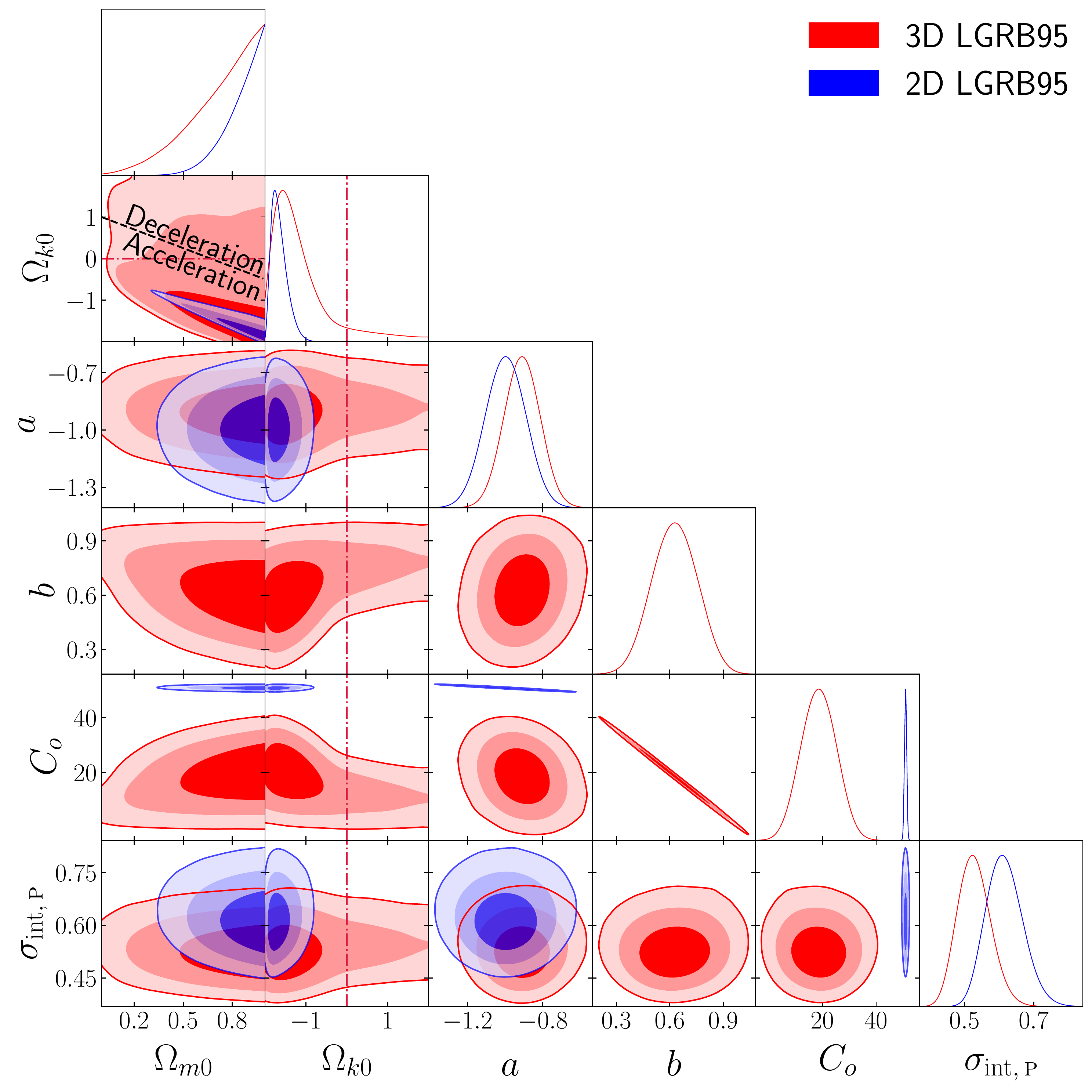}}\\
 \subfloat[]{%
    \includegraphics[width=0.5\textwidth,height=0.5\textwidth]{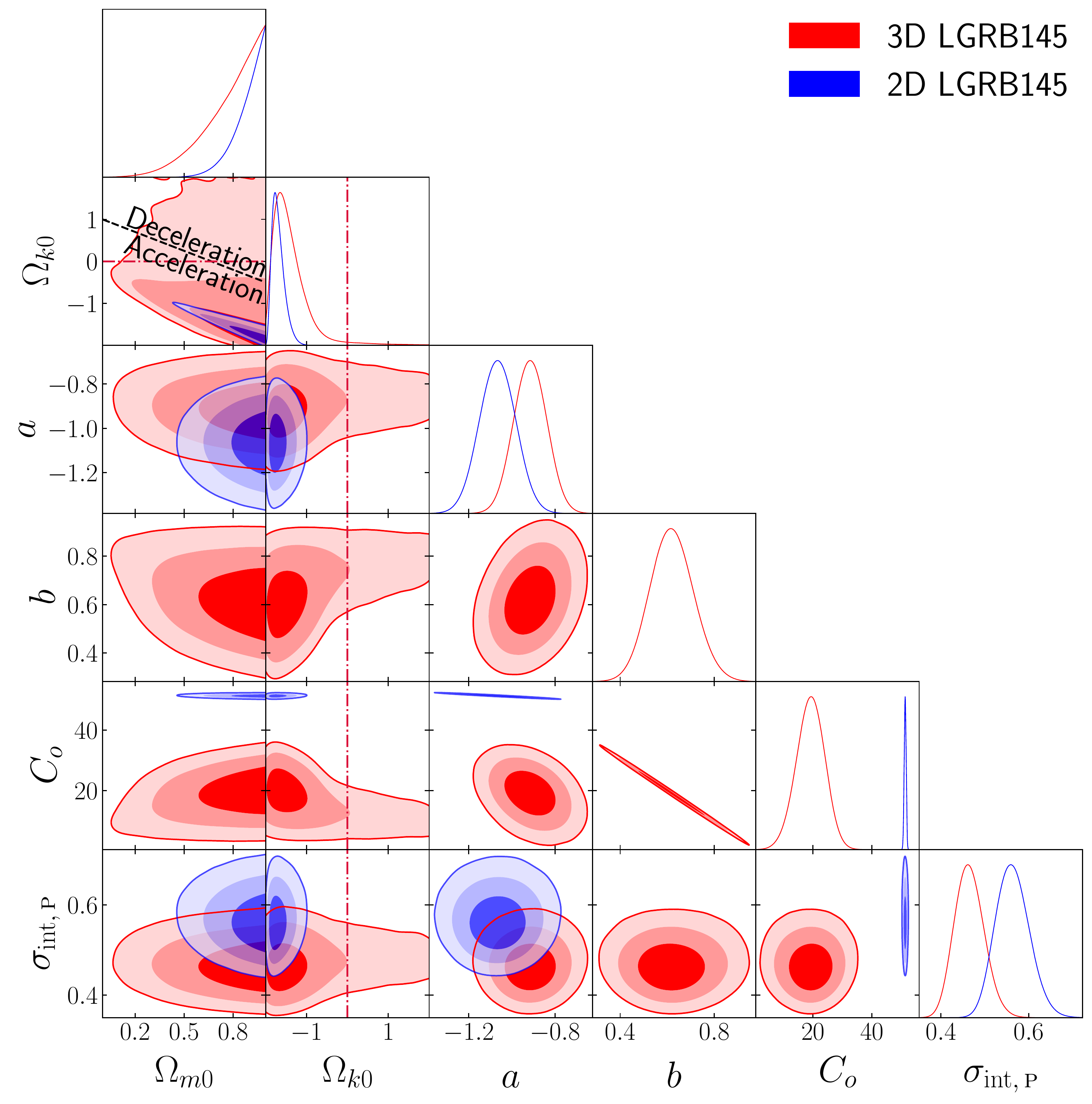}}
 \subfloat[]{%
    \includegraphics[width=0.5\textwidth,height=0.5\textwidth]{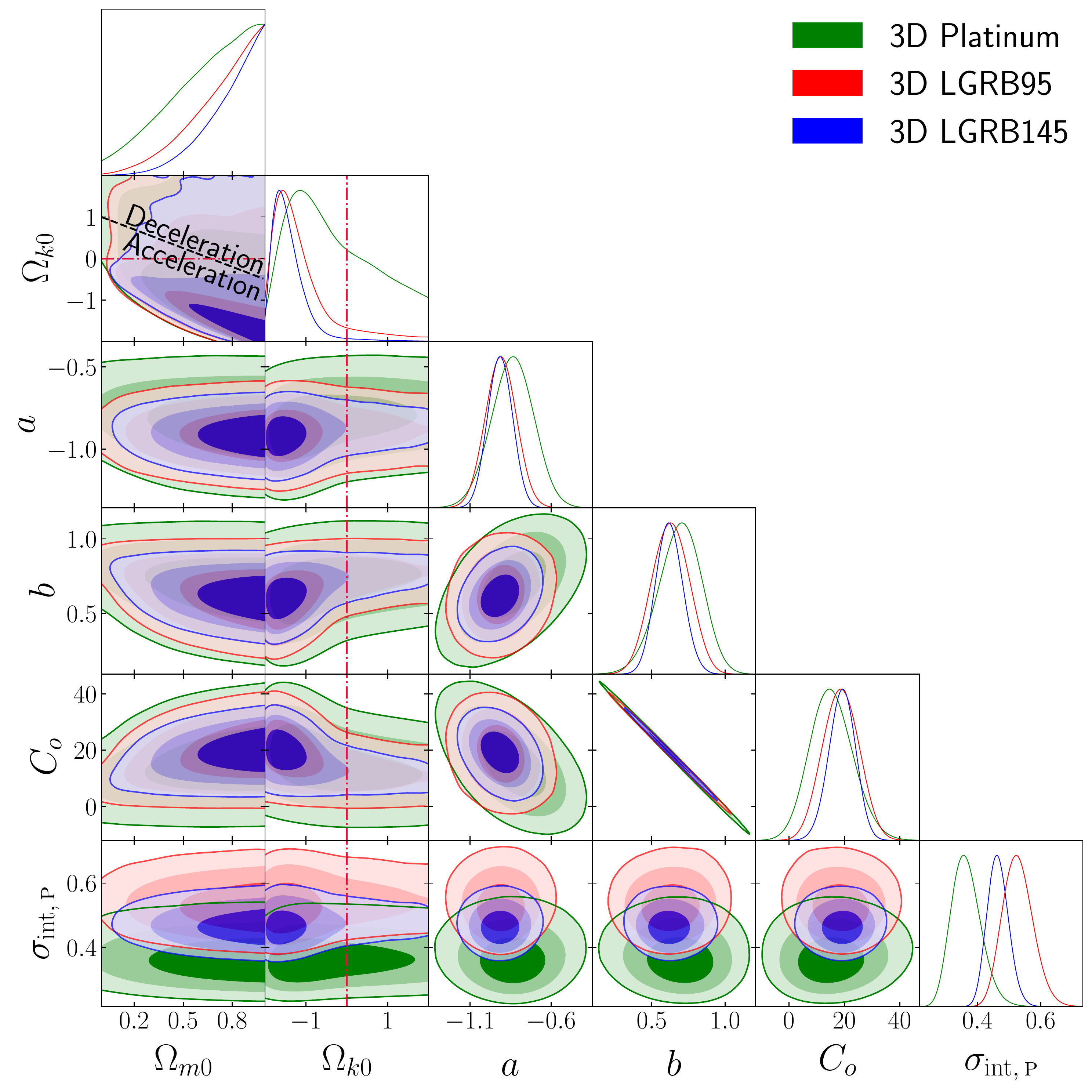}}\\
\caption{Same as Fig.\ \ref{fig1C10} but for non-flat \lcdm. The zero-acceleration black dashed lines divide the parameter space into regions associated with currently-accelerating (below left) and currently-decelerating (above right) cosmological expansion.}
\label{fig2C10}
\end{figure*}

\begin{figure*}
\centering
 \subfloat[]{%
    \includegraphics[width=0.5\textwidth,height=0.5\textwidth]{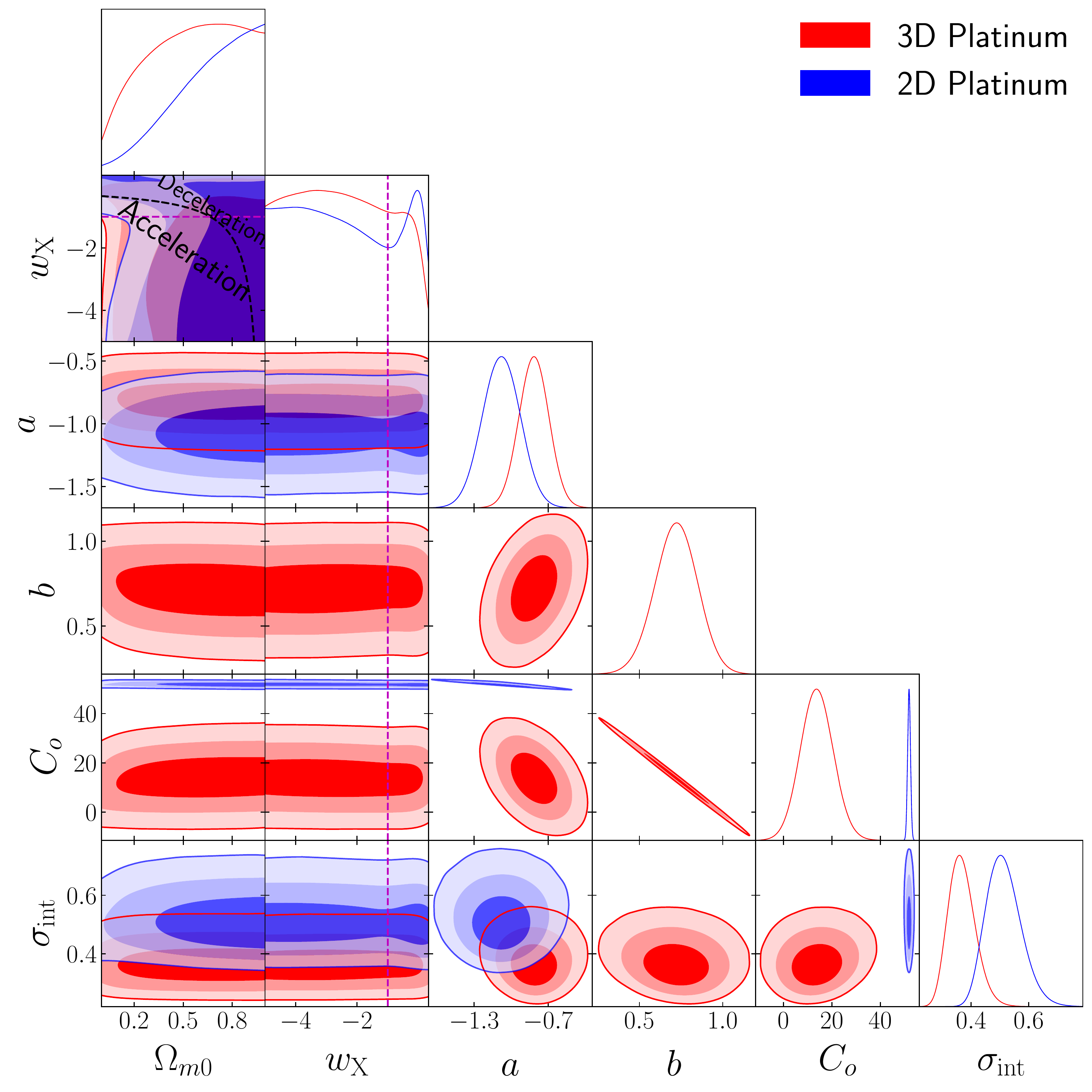}}
 \subfloat[]{%
    \includegraphics[width=0.5\textwidth,height=0.5\textwidth]{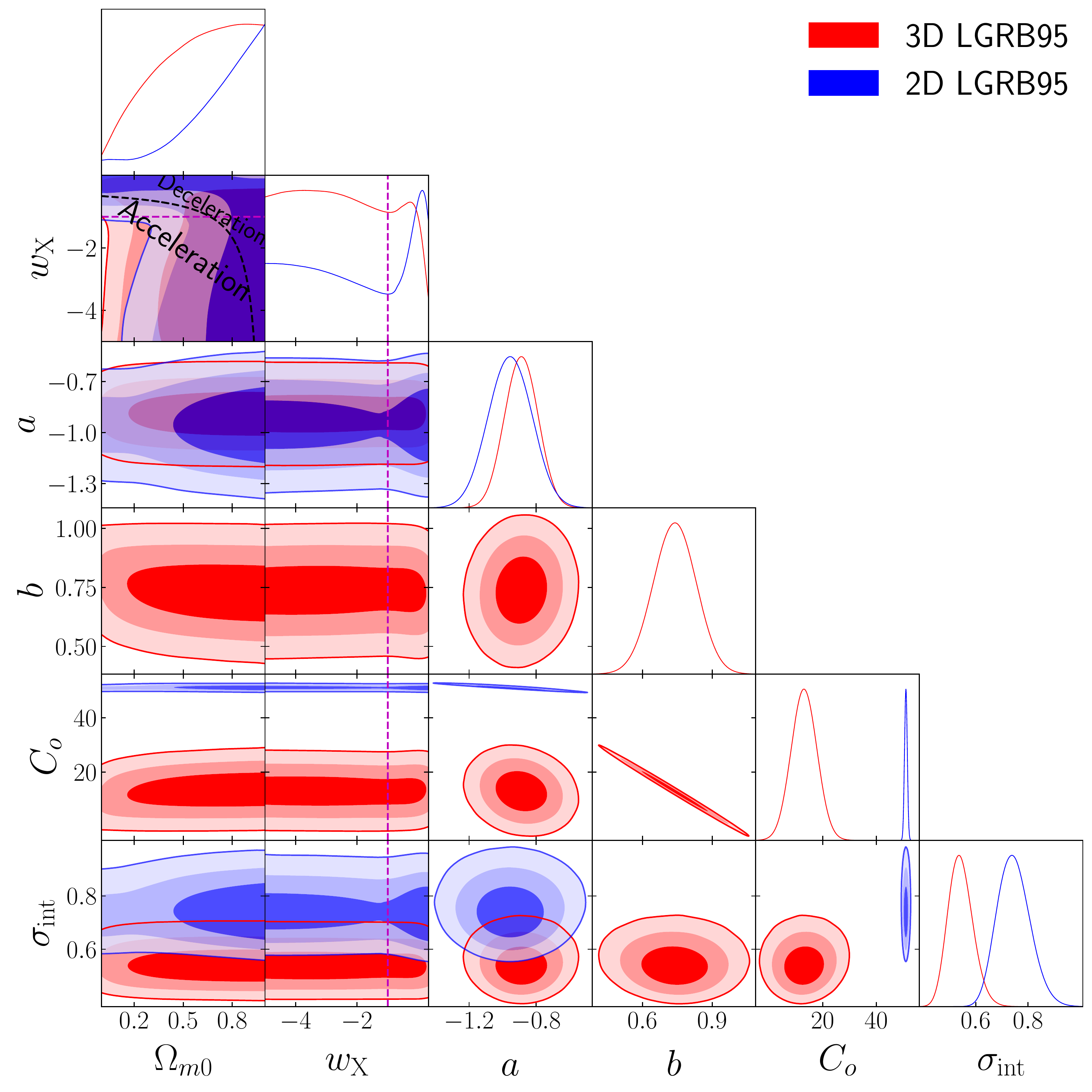}}\\
 \subfloat[]{%
    \includegraphics[width=0.5\textwidth,height=0.5\textwidth]{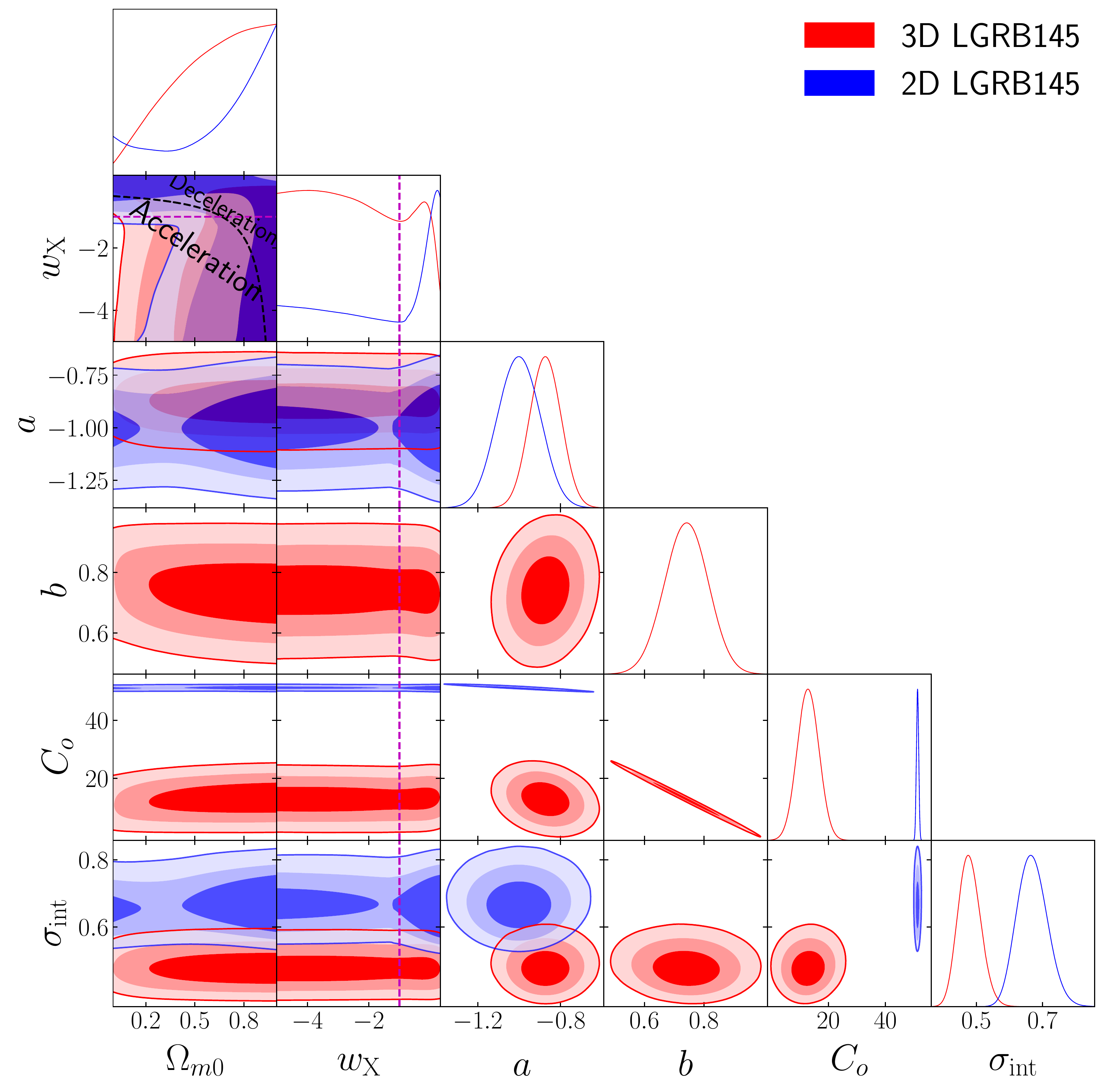}}
 \subfloat[Cosmological parameters zoom in]{%
    \includegraphics[width=0.5\textwidth,height=0.5\textwidth]{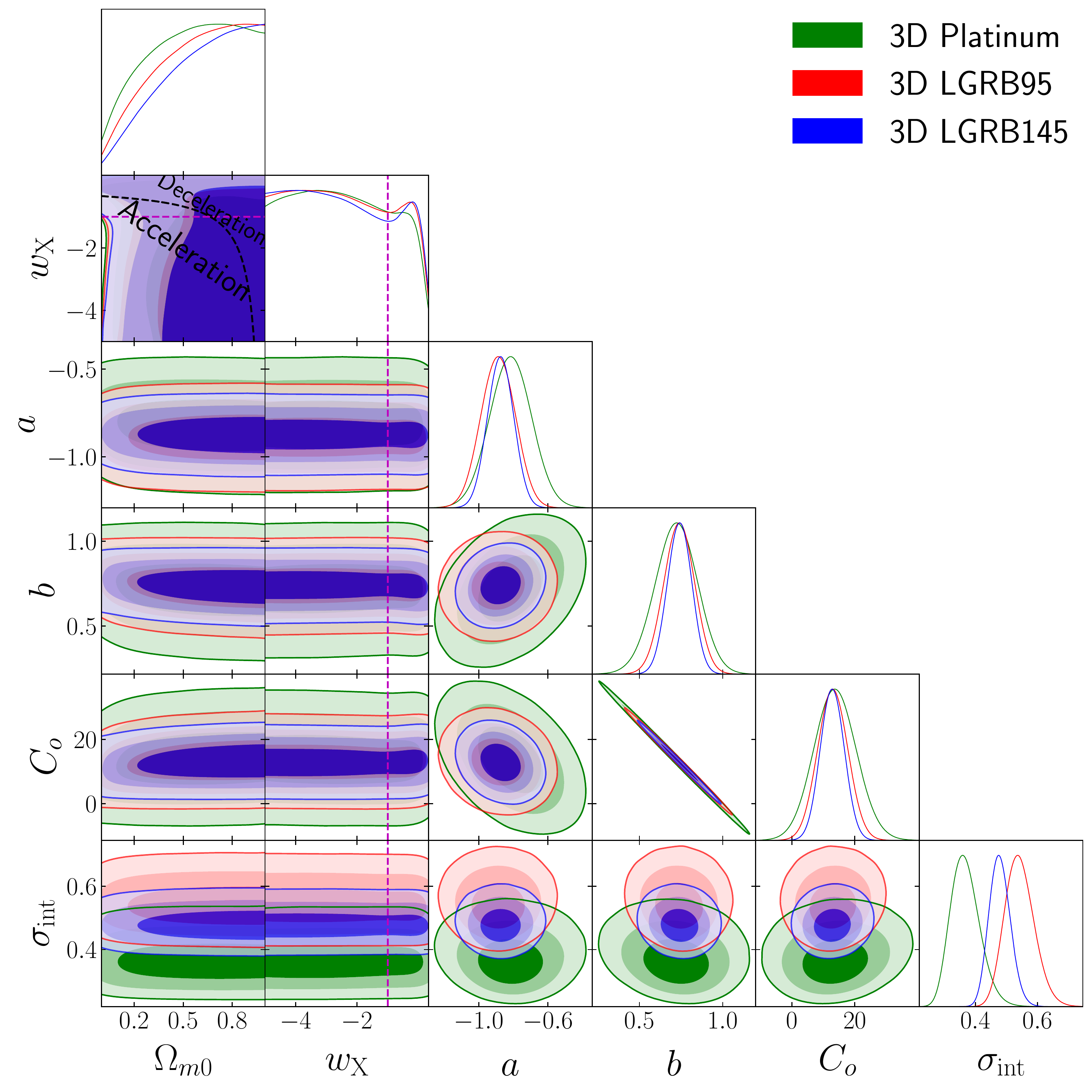}}\\
\caption{One-dimensional likelihood distributions and 1$\sigma$, 2$\sigma$, and 3$\sigma$ two-dimensional likelihood confidence contours for flat XCDM from various combinations of data. The zero-acceleration black dashed lines divide the parameter space into regions associated with currently-accelerating (either below left or below) and currently-decelerating (either above right or above) cosmological expansion. The magenta dashed lines represent $w_{\rm X}=-1$, i.e.\ flat \lcdm.}
\label{fig3C10}
\end{figure*}

\begin{figure*}
\centering
 \subfloat[]{%
    \includegraphics[width=0.5\textwidth,height=0.5\textwidth]{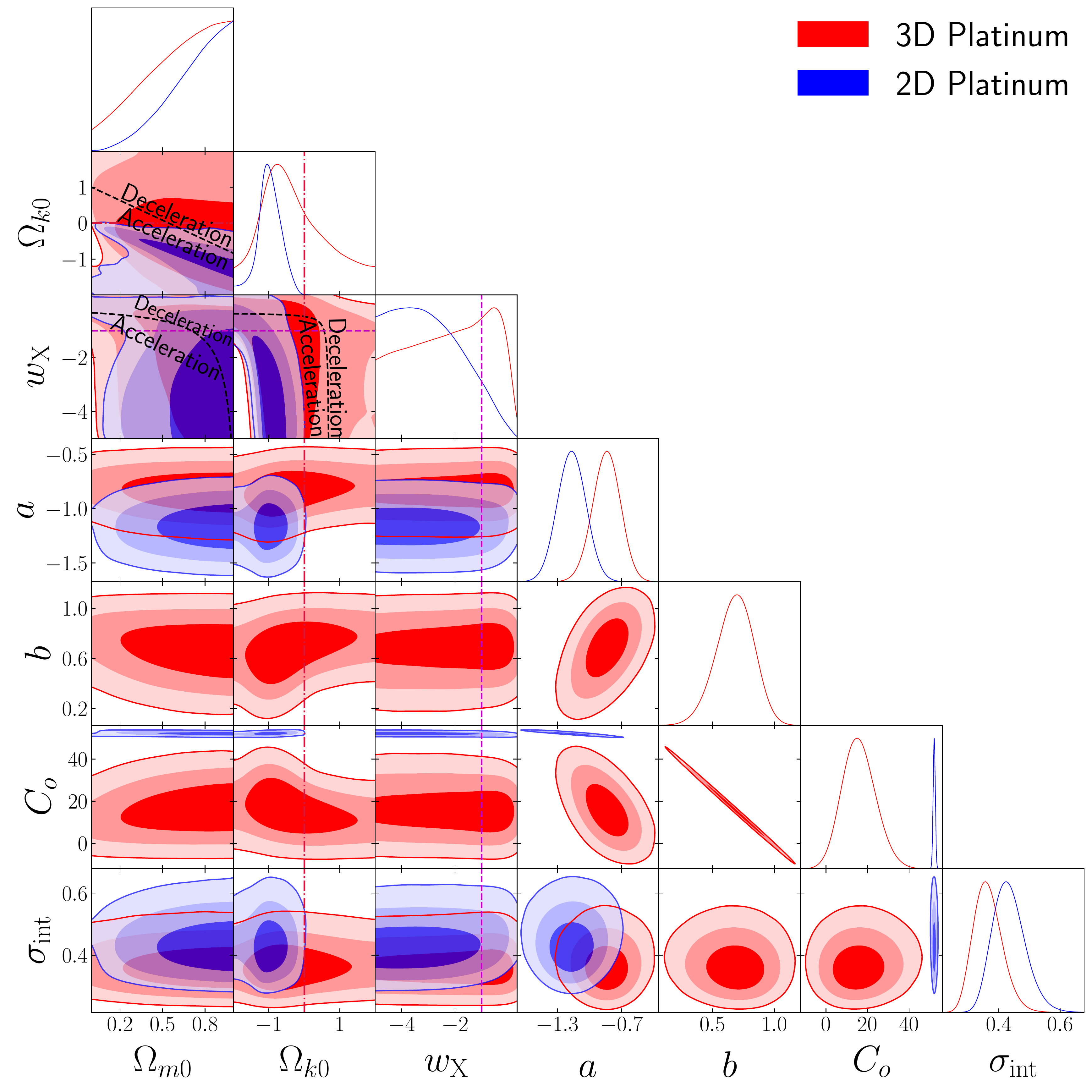}}
 \subfloat[]{%
    \includegraphics[width=0.5\textwidth,height=0.5\textwidth]{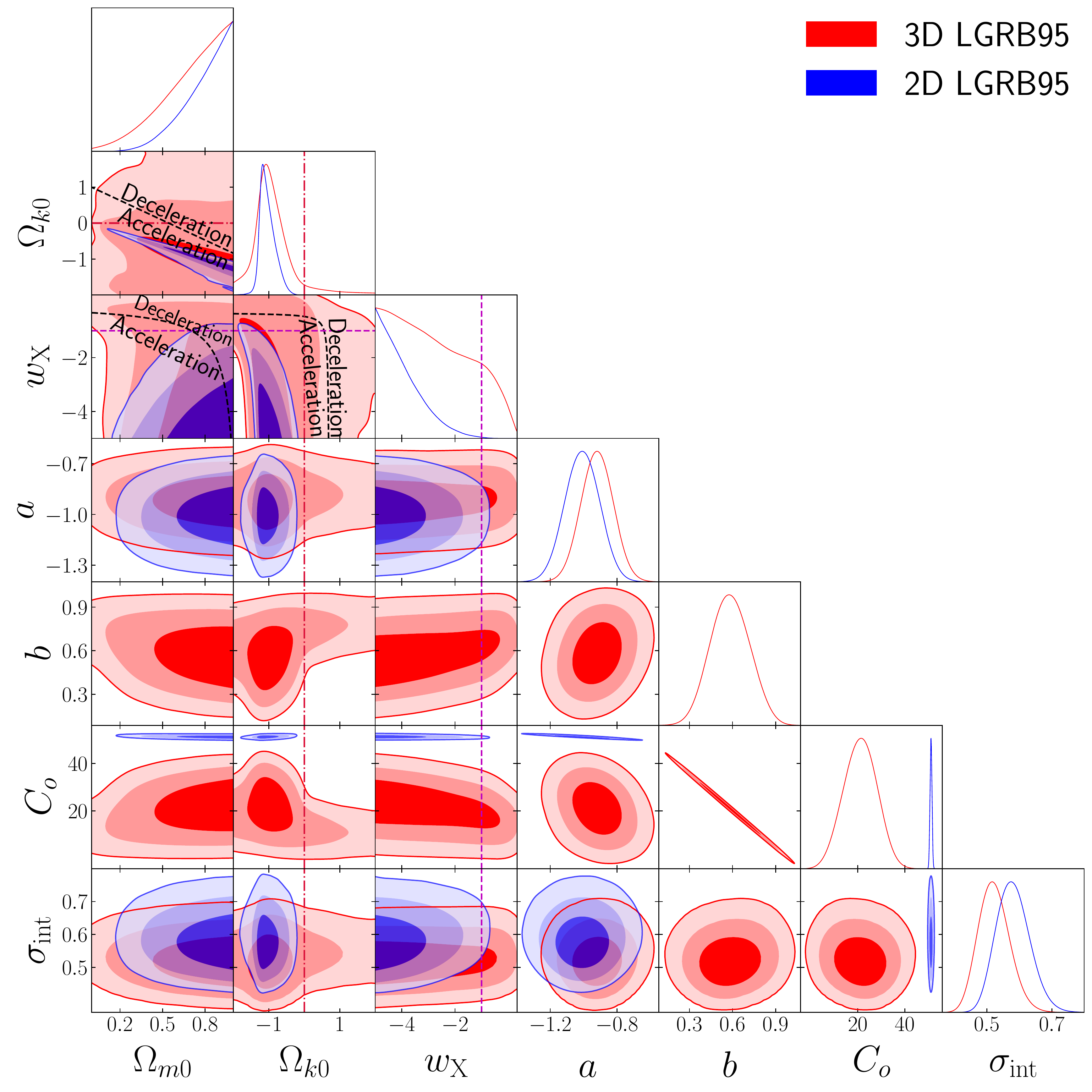}}\\
 \subfloat[]{%
    \includegraphics[width=0.5\textwidth,height=0.5\textwidth]{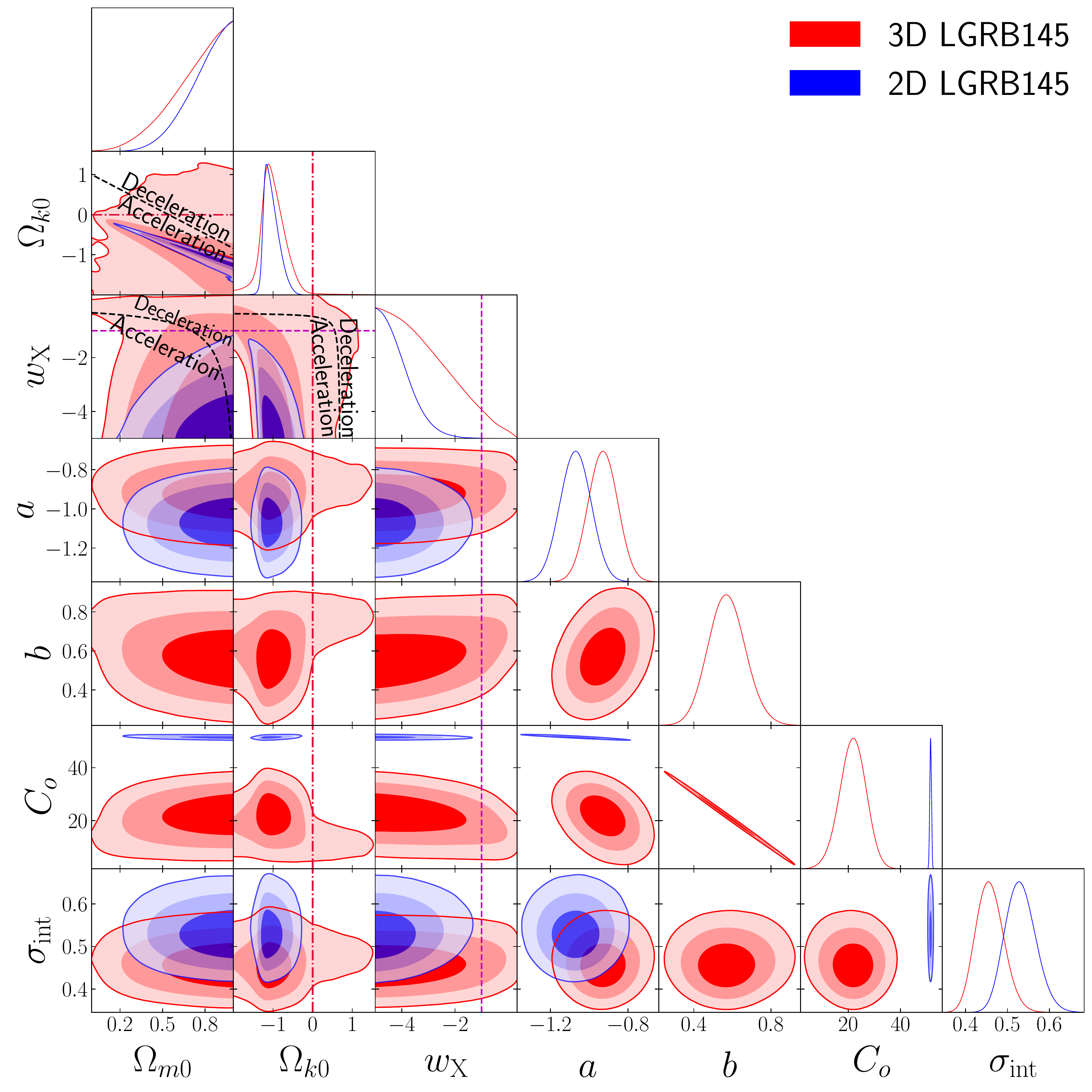}}
 \subfloat[Cosmological parameters zoom in]{%
    \includegraphics[width=0.5\textwidth,height=0.5\textwidth]{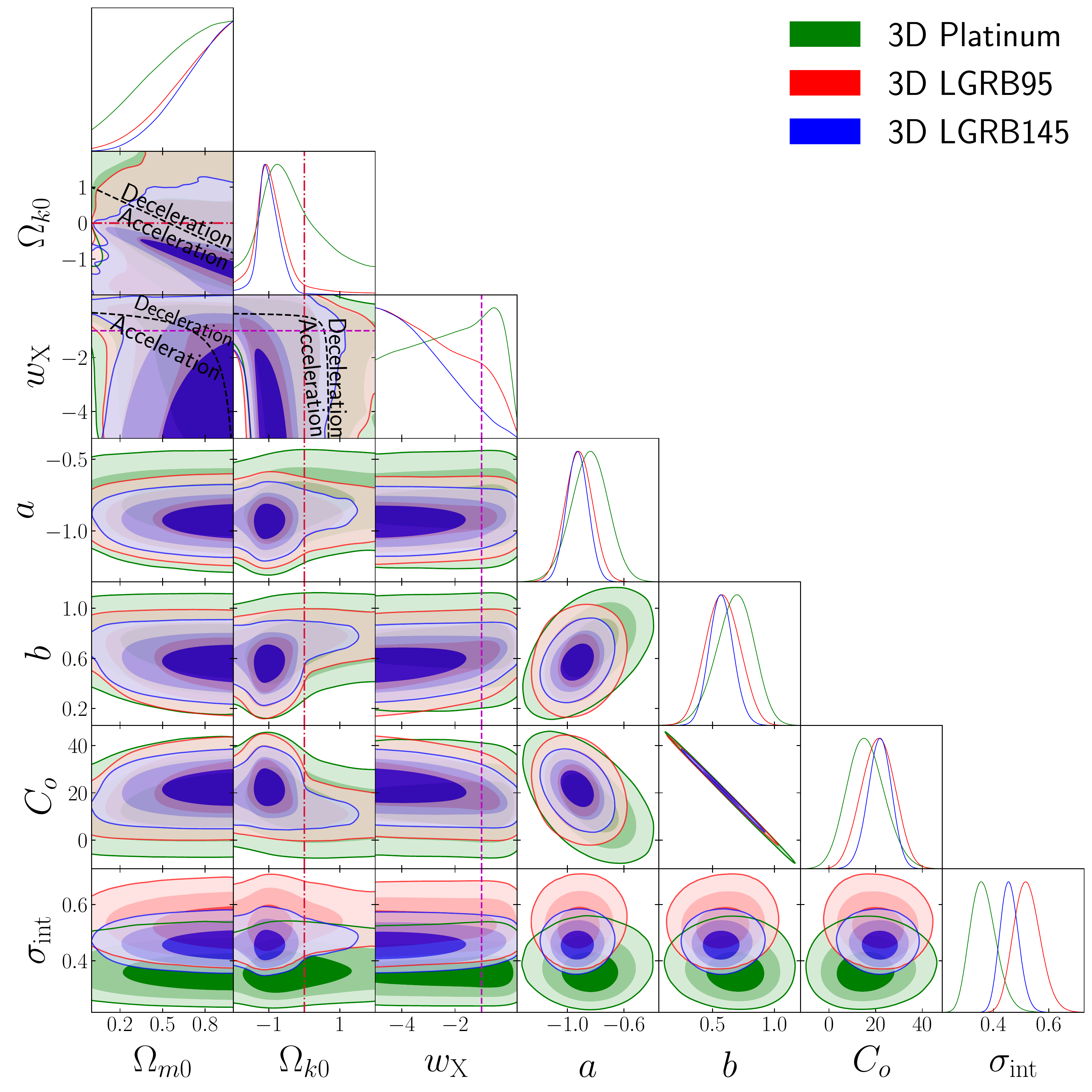}}\\
\caption{Same as Fig.\ \ref{fig3C10} but for non-flat XCDM. The zero-acceleration black dashed lines are computed for the third cosmological parameter set to the $H(z)$ + BAO data best-fitting values listed in table 4 of \protect\cite{CaoRatra2022}, and divide the parameter space into regions associated with currently-accelerating (either below left or below) and currently-decelerating (either above right or above) cosmological expansion. The crimson dash-dot lines represent flat hypersurfaces, with closed spatial hypersurfaces either below or to the left. The magenta dashed lines represent $w_{\rm X}=-1$, i.e.\ non-flat \lcdm.}
\label{fig4C10}
\end{figure*}

\begin{figure*}
\centering
\centering
 \subfloat[]{%
    \includegraphics[width=0.5\textwidth,height=0.5\textwidth]{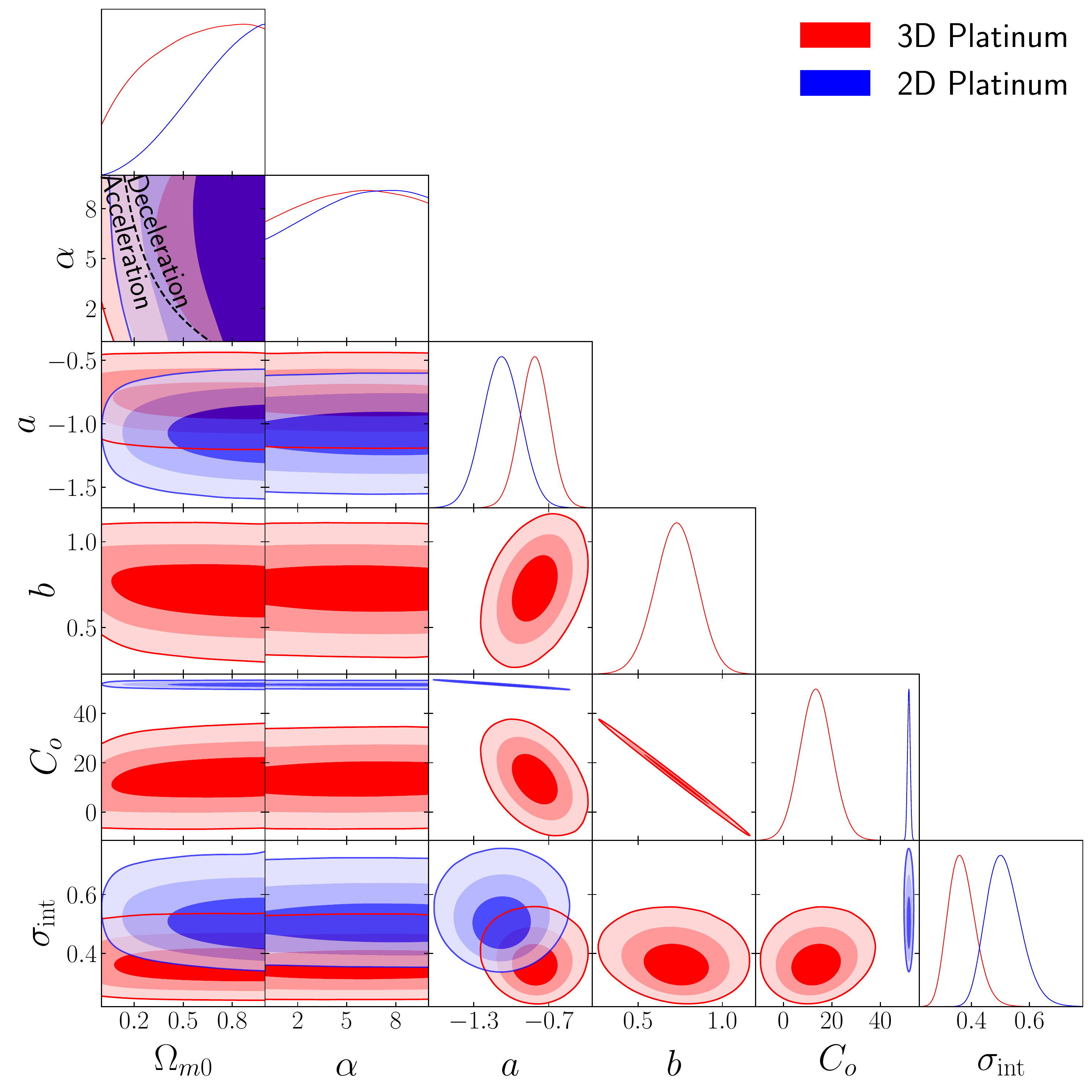}}
 \subfloat[]{%
    \includegraphics[width=0.5\textwidth,height=0.5\textwidth]{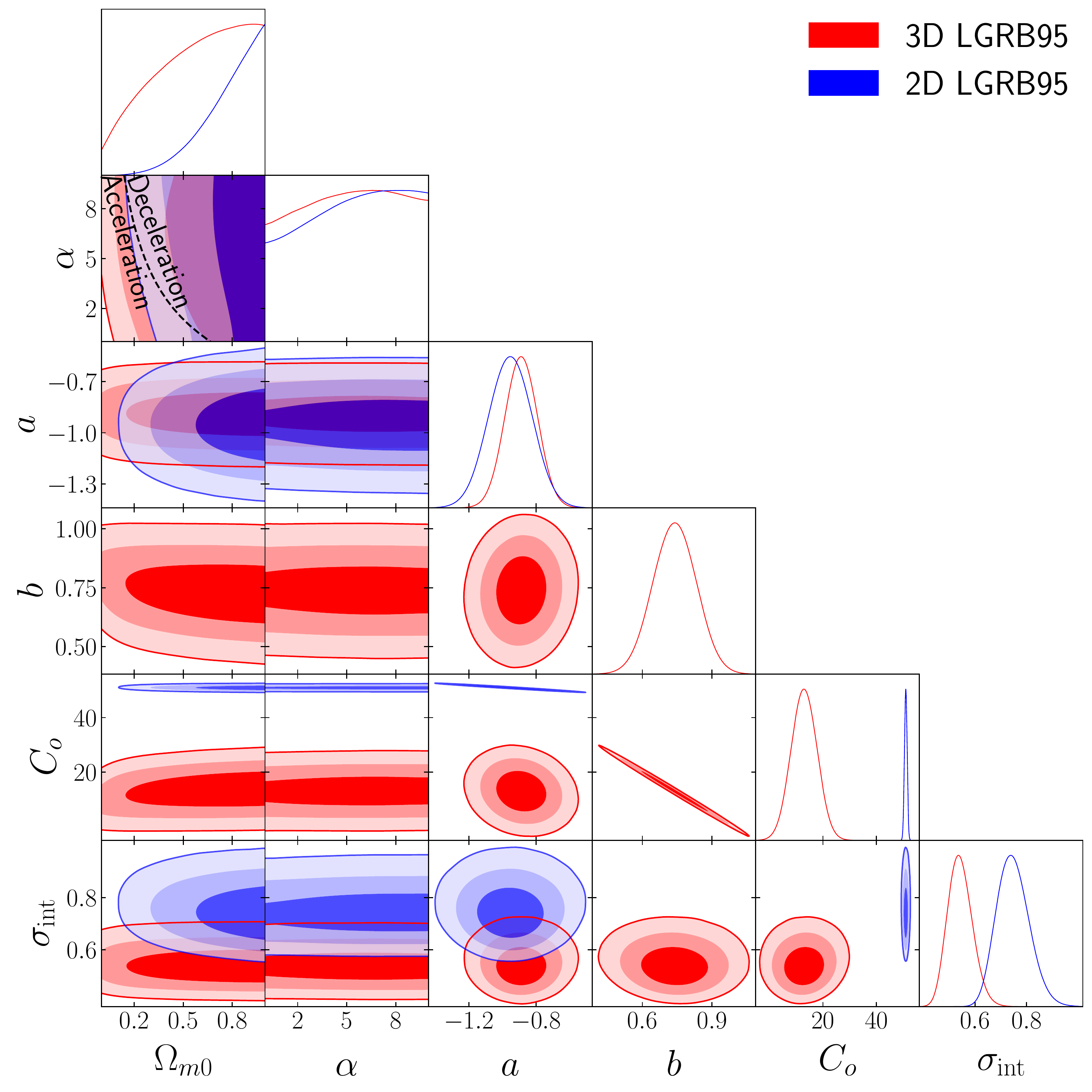}}\\
 \subfloat[]{%
    \includegraphics[width=0.5\textwidth,height=0.5\textwidth]{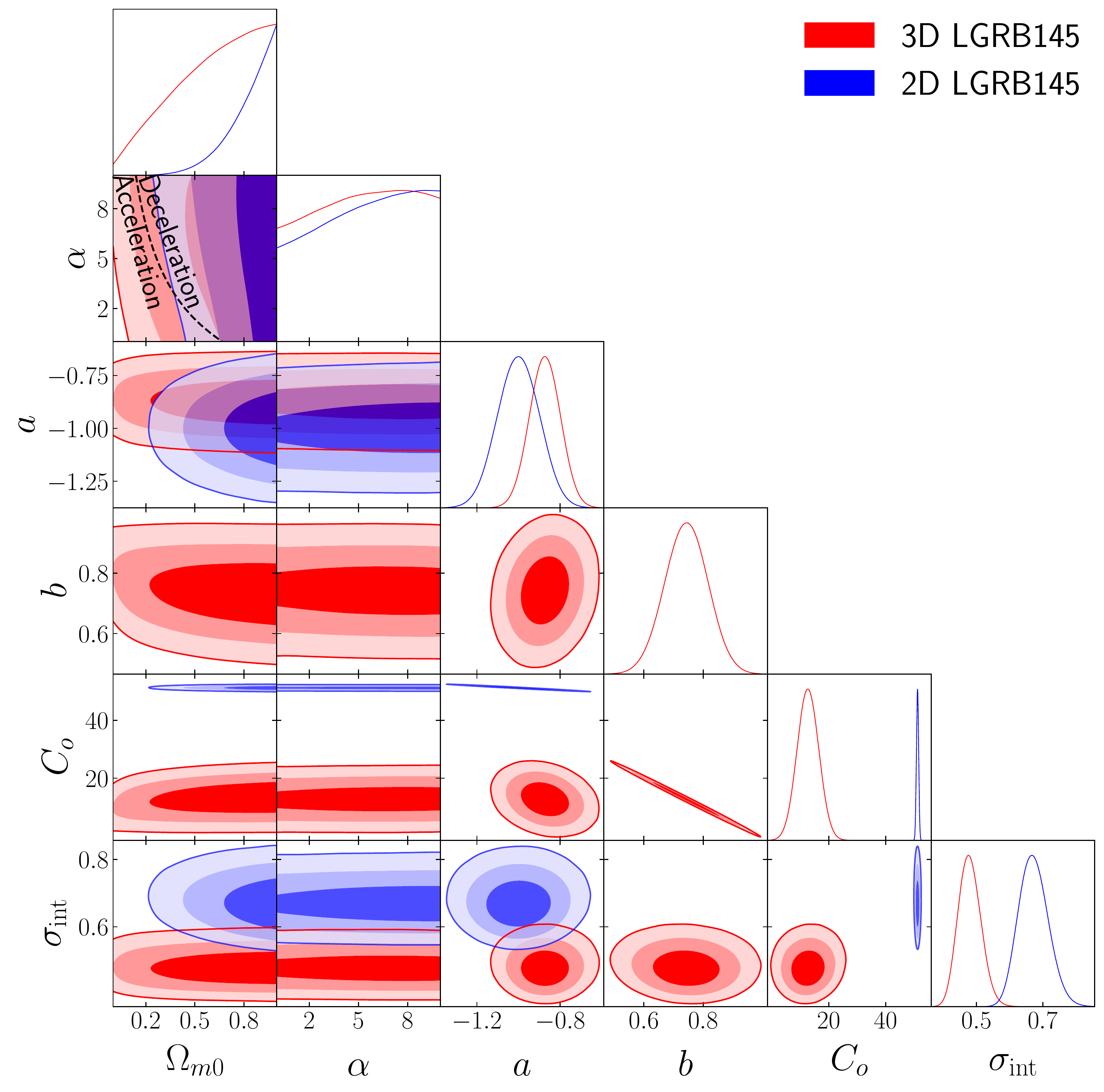}}
 \subfloat[Cosmological parameters zoom in]{%
    \includegraphics[width=0.5\textwidth,height=0.5\textwidth]{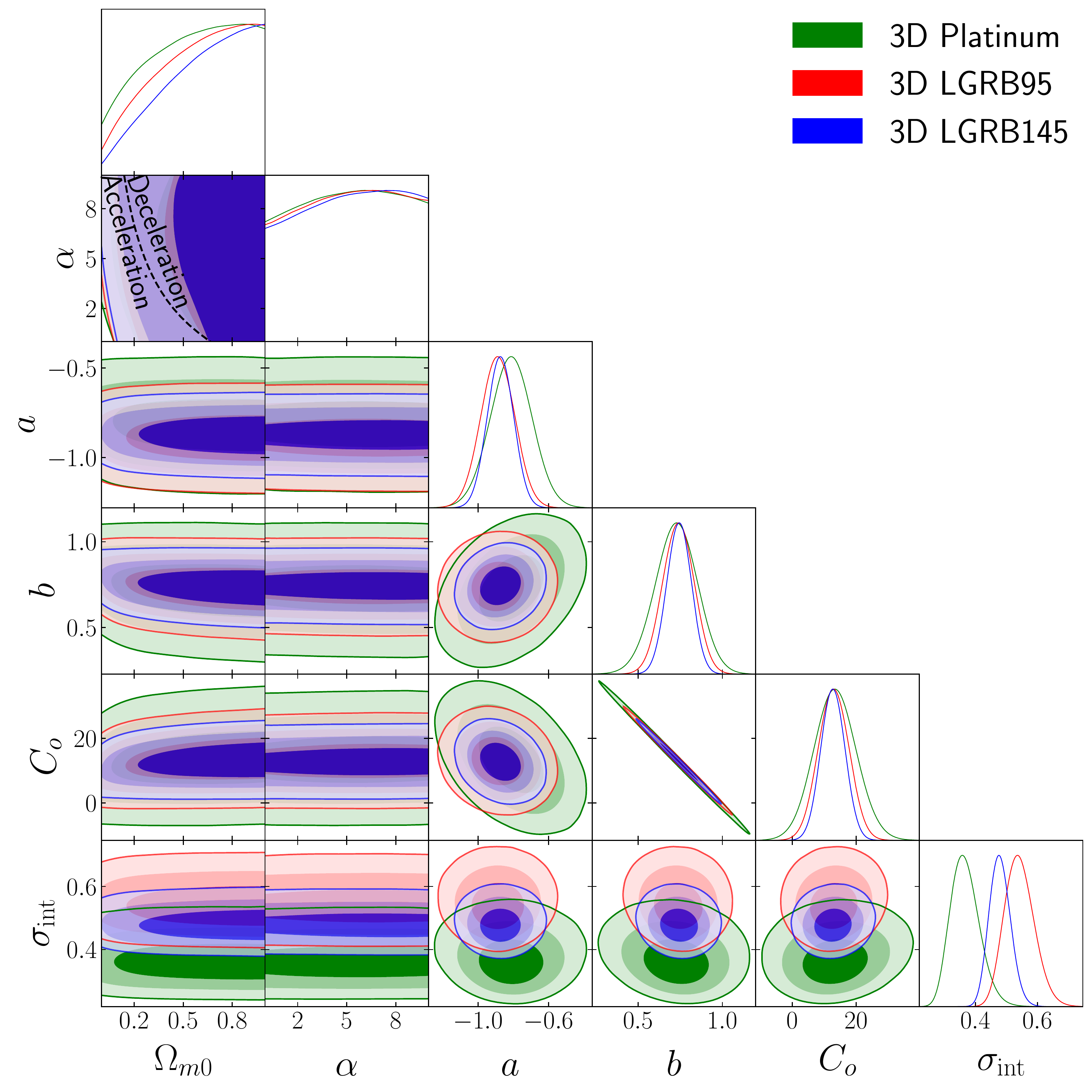}}\\
\caption{One-dimensional likelihood distributions and 1$\sigma$, 2$\sigma$, and 3$\sigma$ two-dimensional likelihood confidence contours for flat \pcdm\ from various combinations of data. The zero-acceleration black dashed lines divide the parameter space into regions associated with currently-accelerating (below left) and currently-decelerating (above right) cosmological expansion. The $\alpha = 0$ axes correspond to flat \lcdm.}
\label{fig5C10}
\end{figure*}

\begin{figure*}
\centering
 \subfloat[]{%
    \includegraphics[width=0.5\textwidth,height=0.5\textwidth]{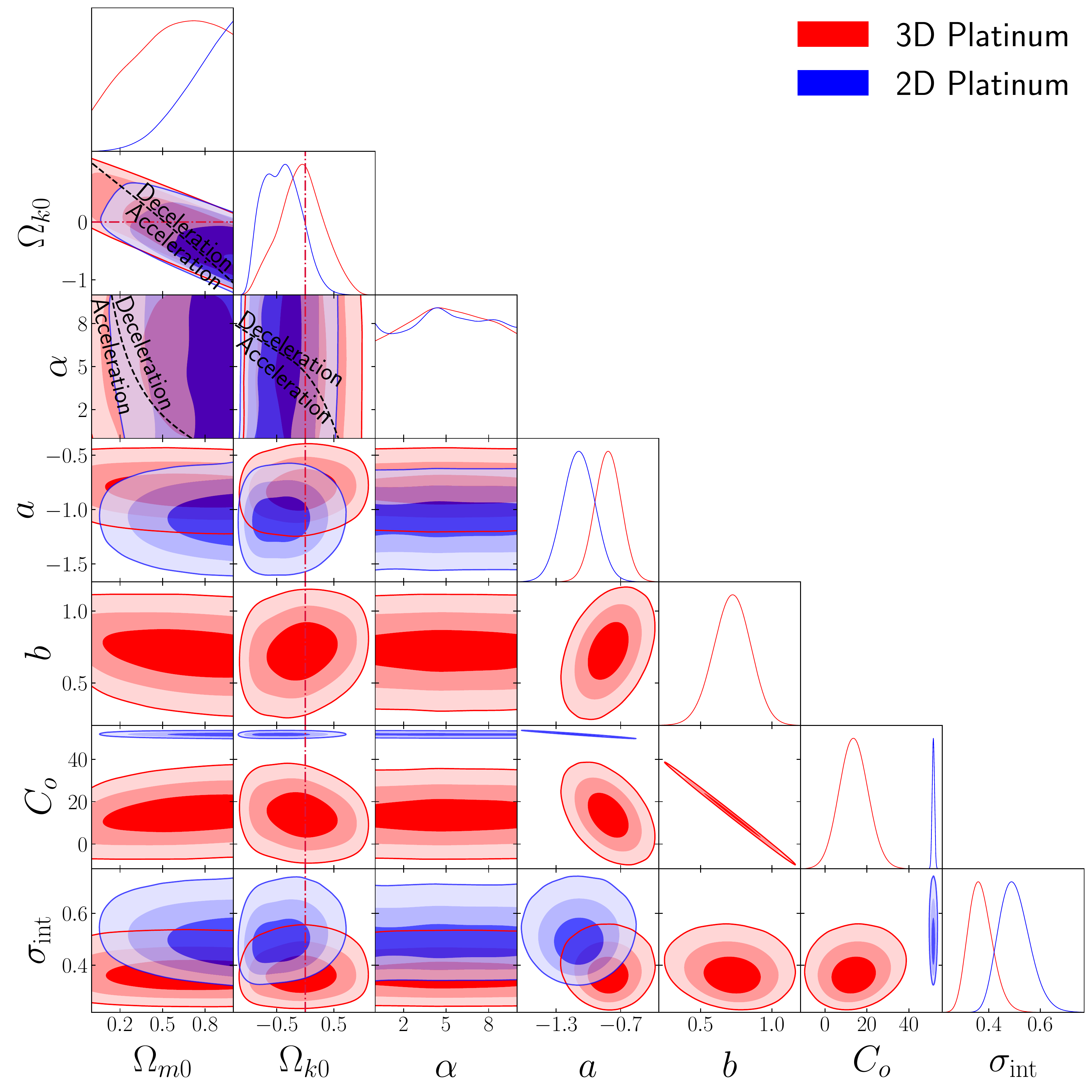}}
 \subfloat[]{%
    \includegraphics[width=0.5\textwidth,height=0.5\textwidth]{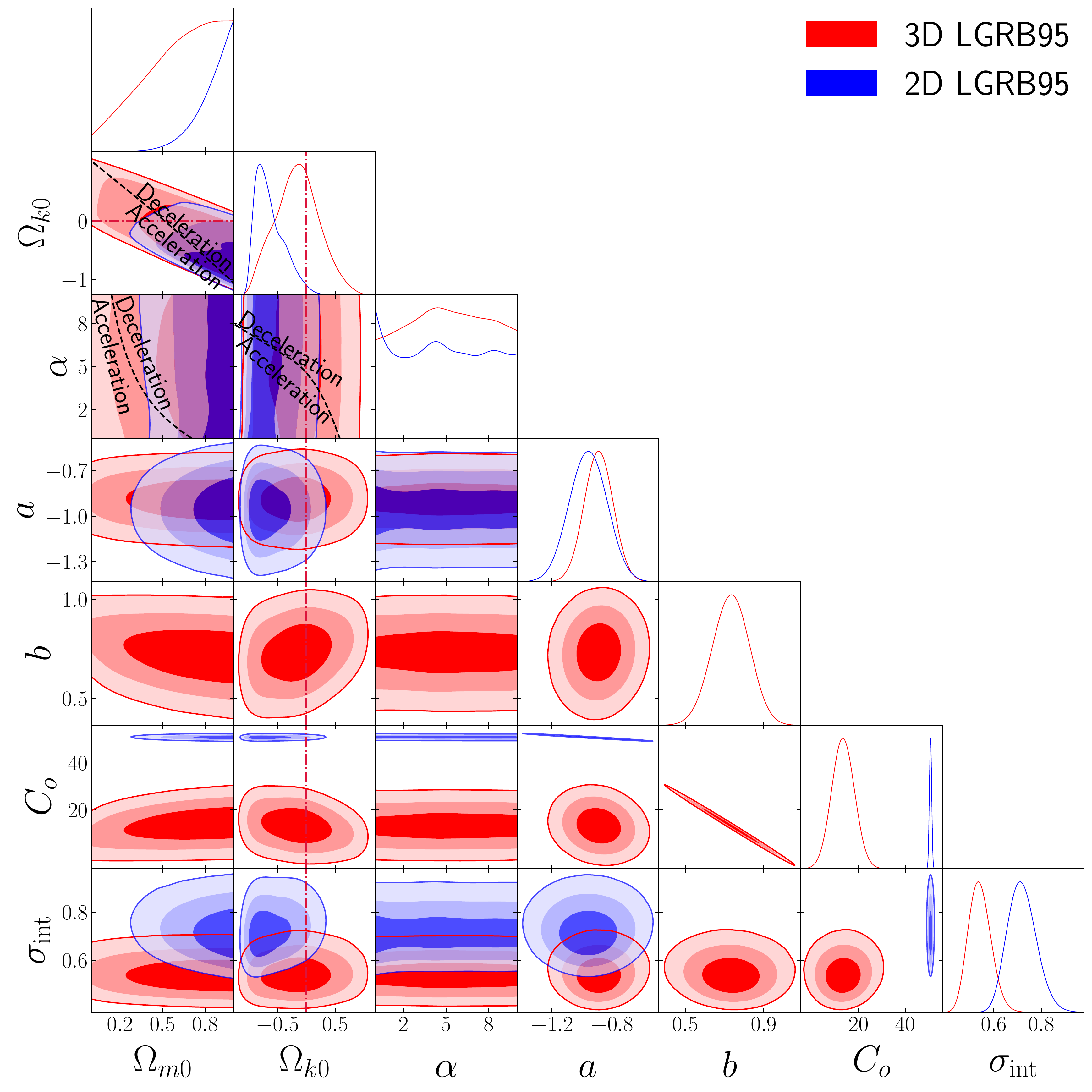}}\\
 \subfloat[]{%
    \includegraphics[width=0.5\textwidth,height=0.5\textwidth]{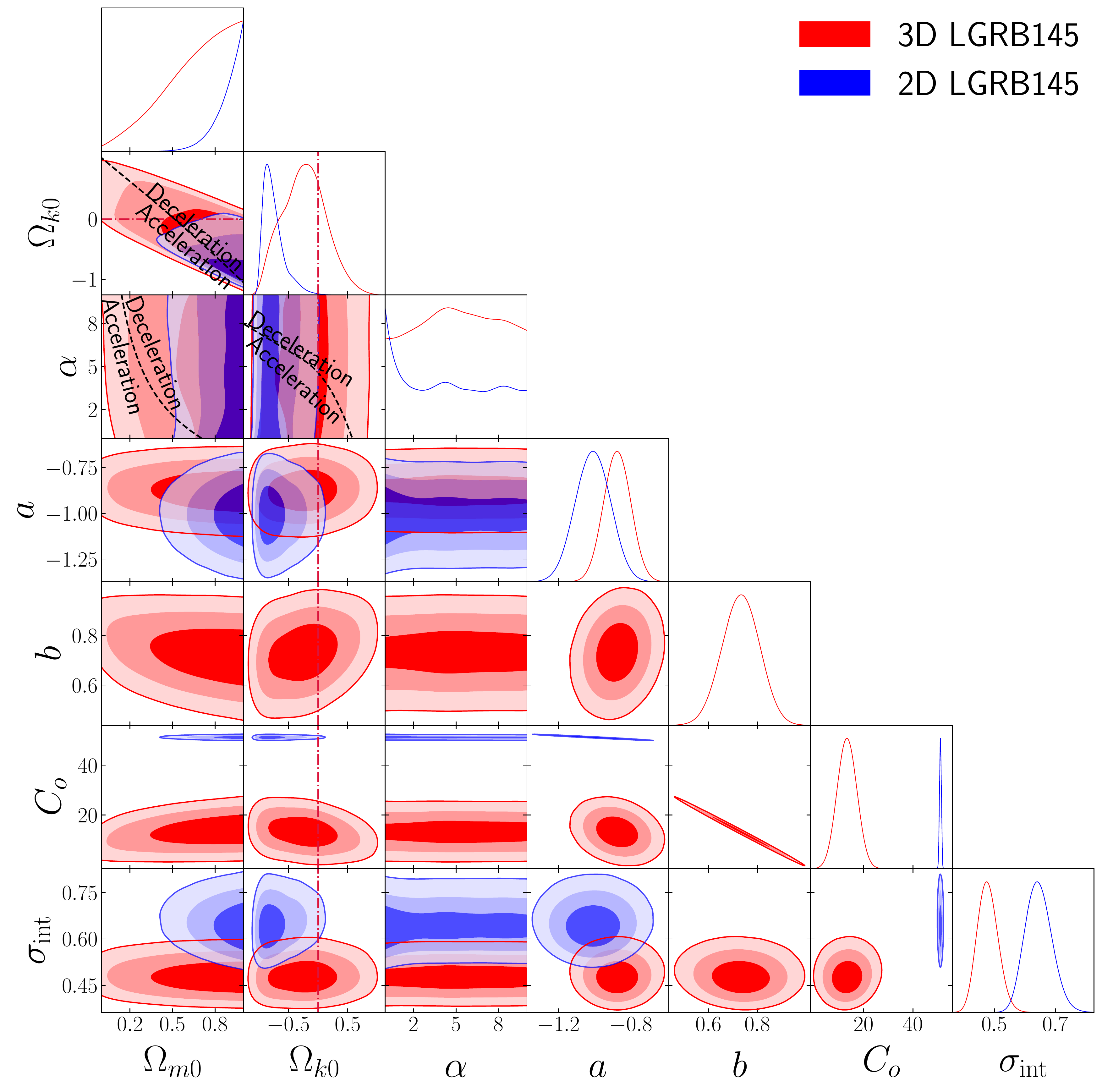}}
 \subfloat[Cosmological parameters zoom in]{%
    \includegraphics[width=0.5\textwidth,height=0.5\textwidth]{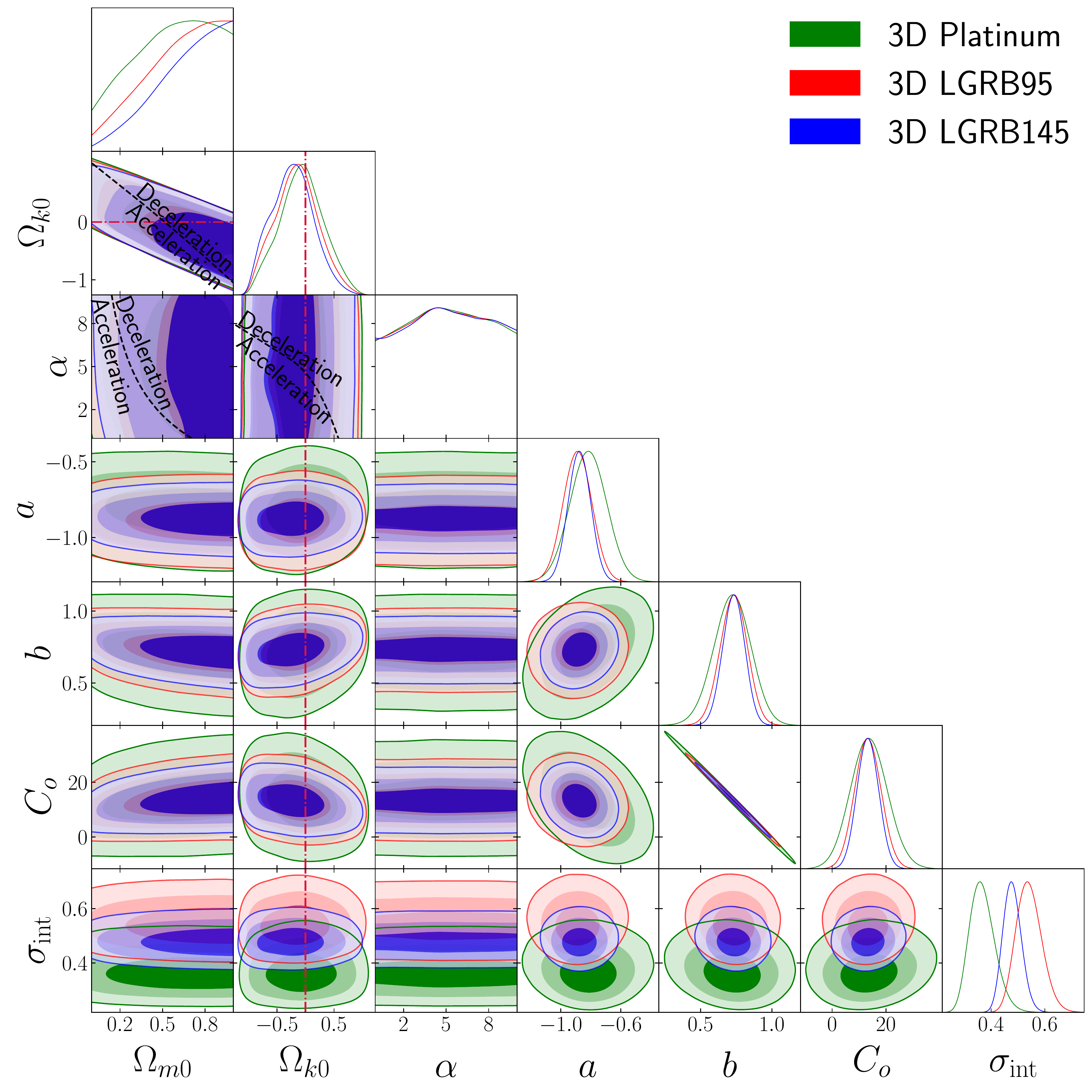}}\\
\caption{Same as Fig.\ \ref{fig5C10} but for non-flat \pcdm. The zero-acceleration black dashed lines are computed for the third cosmological parameter set to the $H(z)$ + BAO data best-fitting values listed in table 4 of \protect\cite{CaoRatra2022}, and divide the parameter space into regions associated with currently-accelerating (below left) and currently-decelerating (above right) cosmological expansion. The crimson dash-dot lines represent flat hypersurfaces, with closed spatial hypersurfaces either below or to the left. The $\alpha = 0$ axes correspond to non-flat \lcdm.}
\label{fig6C10}
\end{figure*}

\begin{figure}
\centering
    \includegraphics[width=0.45\textwidth,height=0.45\textwidth]{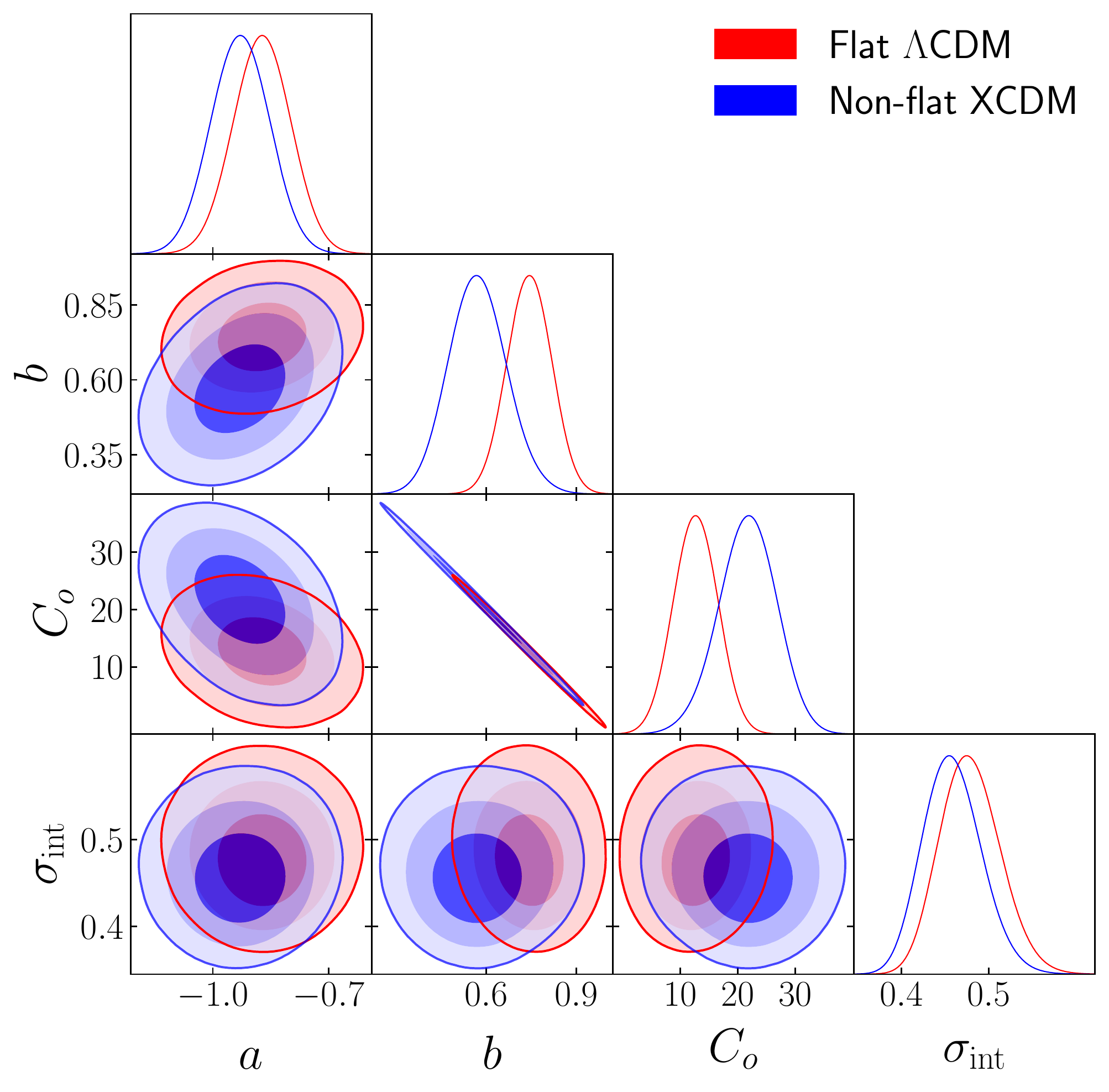}\\
\caption{Comparisons between 3D Dainotti correlation parameters of flat \lcdm\ and those of non-flat XCDM from LGRB145 data, where their two-dimensional likelihood confidence contours are mutually consistent within 1$\sigma$.}
\label{fig7C10}
\end{figure}

\begin{figure*}
\centering
 \subfloat[2D LGRB95]{%
    \includegraphics[width=0.45\textwidth,height=0.45\textwidth]{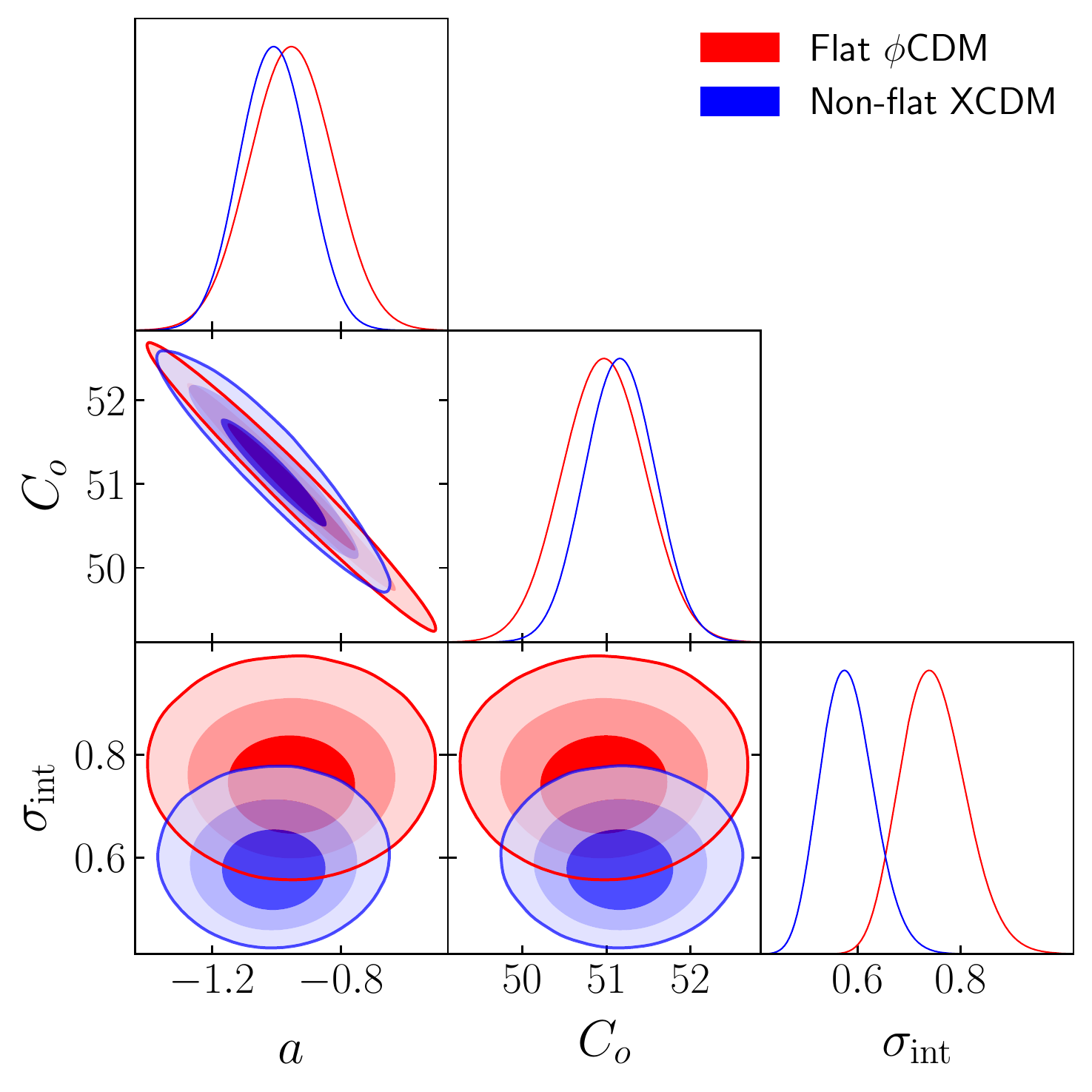}}
 \subfloat[2D LGRB145]{%
    \includegraphics[width=0.45\textwidth,height=0.45\textwidth]{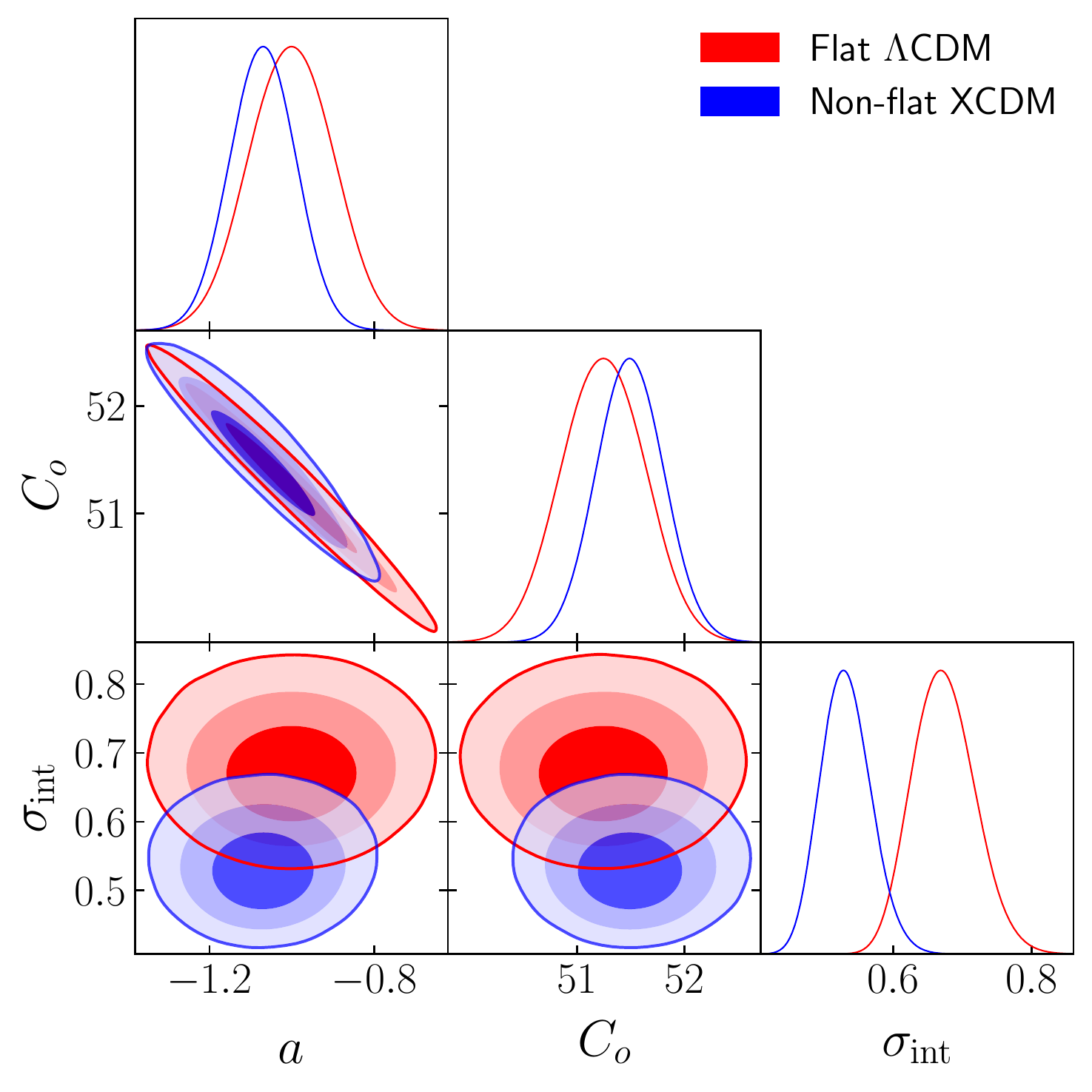}}\\
\caption{Comparisons between 2D Dainotti correlation parameters of different cosmological models.}
\label{fig8C10}
\end{figure*}

\section{Results}
\label{makereference10.4}

The posterior one-dimensional probability distributions and two-dimensional confidence regions of cosmological-model and GRB-correlation parameters for the six cosmological models are shown in Figs.\ \ref{fig1C10}--\ref{fig6C10}, where in panels (a)--(c) of each figure results of the 3D and 2D Dainotti correlation analyses are shown in red and blue, respectively; while in panel (d) of each figure 3D Dainotti correlation results are shown in green (Platinum), red (LGRB95), and blue (LGRB145). The unmarginalized best-fitting parameter values, as well as the values of maximum likelihood $\mathcal{L}_{\rm max}$, AIC, BIC, DIC, $\Delta \mathrm{AIC}^{\prime}$, $\Delta \mathrm{BIC}^{\prime}$, $\Delta \mathrm{DIC}^{\prime}$, $\Delta \mathrm{AIC}$, $\Delta \mathrm{BIC}$, and $\Delta \mathrm{DIC}$, for all models and data sets, are listed in Table \ref{tab:BFPC10}. We list the marginalized posterior mean parameter values and uncertainties ($\pm 1\sigma$ error bars and 1 or 2$\sigma$ limits), for all models and data sets, in Table \ref{tab:1d_BFPC10}. 

In \cite{CaoDainottiRatra2022} we showed that the 3D Dainotti correlation parameters in all six cosmological models, determined from the Platinum data set, were mutually consistent, confirming that the 3D Dainotti correlation Platinum GRBs are standardizable.\footnote{The corrected results in Tables \ref{tab:BFPC10} and \ref{tab:1d_BFPC10} show that the conclusions of \cite{CaoDainottiRatra2022} do not change.} When we compare the 3D Dainotti correlation parameters results in Table \ref{tab:1d_BFPC10} for the six different cosmological models, we see that the 3D Dainotti correlation LGRB95 GRBs are standardizable.\footnote{Note that although the highest (non-flat XCDM) and lowest (flat \lcdm) values of 1D marginalized $b$ and $C_{o}$ constraints from LGRB145 data differ by $1.44\sigma$ and $1.45\sigma$, respectively, as shown in Fig.\ \ref{fig7C10} their 1$\sigma$ 2D marginalized contours overlap so the 3D Dainotti correlation LGRB145 GRBs are also standardizable.} Similarly, the determined 2D Dainotti correlation parameters, $a$ and $C_o$, are independent (within the errors) of the cosmological model used in the analysis, for the Platinum, LGRB95, and LGRB145 GRBs. However, unlike the 3D Dainotti correlation cases and the 2D Dainotti Platinum case, from the cosmological models which result in the largest and smallest $\sigma_{\rm int}$ values, in the 2D Dainotti correlation LGRB95 and LGRB145 cases, although 1D marginalized $\sigma_{\rm int}$ constraints are in $>2\sigma$ tension, as shown in panels (a) and (b) of Fig.\ \ref{fig8C10}, their 2D marginalized contours are within $2\sigma$. These results indicate that the 2D Dainotti correlation LGRB95 and LGRB145 data need more careful study.

For the 3D Platinum data set, the constraints on the intrinsic scatter parameter $\sigma_{\rm int}$ range from a low of $0.365^{+0.038}_{-0.052}$ (non-flat XCDM) to a high of $0.369^{+0.038}_{-0.052}$ (flat \lcdm), with a difference of $0.06\sigma$. The constraints on the slope $a$ range from a low of $-0.842\pm0.127$ (non-flat XCDM) to a high of $-0.812\pm0.119$ (flat \lcdm), with a difference of $0.17\sigma$. The constraints on the slope $b$ range from a low of $0.680^{+0.160}_{-0.136}$ (non-flat XCDM) to a high of $0.728\pm0.122$ (flat \lcdm), with a difference of $0.24\sigma$. The constraints on the intercept $C_{o}$ range from a low of $13.38\pm6.45$ (flat \lcdm) to a high of $15.93^{+7.22}_{-8.52}$ (non-flat XCDM), with a difference of $0.24\sigma$.

For the 3D LGRB95 data set, the constraints on the intrinsic scatter parameter $\sigma_{\rm int}$ range from a low of $0.524^{+0.042}_{-0.051}$ (non-flat XCDM) to a high of $0.543^{+0.042}_{-0.052}$ (flat \lcdm), with a difference of $0.28\sigma$, which are larger ($>2\sigma$) than those from Platinum data. The constraints on the slope $a$ range from a low of $-0.918\pm0.095$ (non-flat XCDM) to a high of $-0.887\pm0.097$ (flat \lcdm), with a difference of $0.23\sigma$, which are lower ($\sim0.5\sigma$) than those from Platinum data. The constraints on the slope $b$ range from a low of $0.588^{+0.140}_{-0.138}$ (non-flat XCDM) to a high of $0.743\pm0.091$ (flat \lcdm), with a difference of $0.93\sigma$, which are either lower ($\sim-0.3/0.5\sigma$) or higher ($\sim0.1\sigma$) than those from Platinum data. The constraints on the intercept $C_{o}$ range from a low of $12.73\pm4.69$ (flat \lcdm) to a high of $20.75^{+7.14}_{-7.22}$ (non-flat XCDM), with a difference of $0.93\sigma$, which are either higher ($\sim0.3/0.8\sigma$) or lower ($\sim-0.1\sigma$) than those from Platinum data. Although the constraints of $\sigma_{\rm int}$ from LGRB95 data are $>2\sigma$ larger than those from Platinum data, which means that LGRB95 data do not fit the 3D Dainotti correlation as well as Platinum data do, the LGRB95 and Platinum 3D Dainotti correlation parameters ($a$, $b$, and $C_o$) are mutually consistent within $1\sigma$, so they both obey same 3D Dainotti correlation and so can be jointly analyzed.

For the joint 3D LGRB145 data set, the constraints on the intrinsic scatter parameter $\sigma_{\rm int}$ range from a low of $0.458^{+0.030}_{-0.035}$ (non-flat XCDM) to a high of $0.480^{+0.030}_{-0.036}$ (flat \pcdm), with a difference of $0.47\sigma$. The constraints on the slope $a$ range from a low of $-0.929\pm0.074$ (non-flat XCDM) to a high of $-0.872\pm0.073$ (flat \lcdm), with a difference of $0.55\sigma$. The constraints on the slope $b$ range from a low of $0.572\pm0.094$ (non-flat XCDM) to a high of $0.743\pm0.073$ (flat \lcdm), with a difference of $1.44\sigma$. The constraints on the intercept $C_{o}$ range from a low of $12.71\pm3.78$ (flat \lcdm) to a high of $21.68^{+4.93}_{-4.91}$ (non-flat XCDM), with a difference of $1.45\sigma$. Although the constraints of $b$ and $C_{o}$ from LGRB145 data are $\sim1.4\sigma$ away, their 2D contours overlap within $1\sigma$.

These results show that for both GRB samples the correlation slope $a$ remains consistent within 1$\sigma$ with the value of the correlation slope corrected for selection biases, $a^{\prime}=(-0.75 \pm 0.11,\ -0.69 \pm 0.07)$ for the Gold and Long GRBs, respectively \citep{Dainottietal2017}, highlighting that the physics of the correlation, i.e. that the energy reservoir remains constant, is consistently maintained, independent of the sample and cosmology used (here we do account for the  selection biases correction as in \citealp{DainottiNielson2022}). Also, the positive correlation between $L_{\rm peak}$ and $L_X$ is maintained at the 1$\sigma$ level when compared with the intrinsic correlation corrected for selection biases which yield $b=(0.7\pm 0.07,\ 0.64\pm 0.11)$ for the Long and Gold GRBs, respectively \citep{Dainottietal2017}. Again, regardless of the sample and cosmological model used, the underlying physics of the correlation is preserved confirming the reliability of our results.

In comparison with the cosmological parameter constraints from Platinum data, LGRB95 data provide slightly tighter constraints. For \om\ constraints, LGRB95 data provide higher 1 or 2$\sigma$ lower limits than most of those from Platinum data with only 1$\sigma$ lower limits. For \ok\ constraints, LGRB95 data provide more restrictive and lower posterior mean values than those from Platinum data. Closed hypersurfaces are favoured but except for non-flat \pcdm, flatness is more than $1\sigma$ away. LGRB95 data provide lower 2$\sigma$ upper limits of \wx\ than those provided by Platinum data, while they do not constrain $\alpha$. It is worth noting that LGRB95 data provide slightly more restrictive constraints on both the cosmological-model and the 3D Dainotti correlation parameters, than do Platinum data, likely a consequence of the larger number of data points, 95 versus 50, more than compensating for the larger $\sigma_{\rm int}$ value, $\sim 0.52-0.54$ versus $\sim 0.37$. 

Although the cosmological parameter constraints from joint LGRB145 data are more restrictive than those from individual data sets, due to it containing more data points, the resulting constraints are not yet comparable with those from SNIa data. However, LGRB145 data favour higher values of \om\ and lower values of \ok\ (closed geometry) than do Platinum data and LGRB95 data. LGRB145 data provide 2$\sigma$ lower limits of \om, ranging from $>0.157$ (flat \pcdm) to $>0.444$ (non-flat \lcdm). In the non-flat \lcdm\ model, LGRB145 data provide tighter and lower \ok\ values of $-1.396^{+0.125}_{-0.487}$ (1$\sigma$) and $<-0.520$ (2$\sigma$), away from flatness to $>2\sigma$. In the non-flat XCDM parametrization, LGRB145 data provide tighter and lower \ok\ values of $-0.991^{+0.240}_{-0.307}$ (1$\sigma$) and $-0.991^{+0.682}_{-0.673}$ (2$\sigma$), away from flatness to $>2\sigma$. In the non-flat \pcdm\ model, LGRB145 data provide tighter and lower \ok\ value of $-0.226^{+0.357}_{-0.388}$ (1$\sigma$), with flatness within $1\sigma$ ($0.63\sigma$).

Based on AIC and BIC, non-flat XCDM is favoured the most by both LGRB95 and LGRB145 data, with the evidence against non-flat XCDM and non-flat \lcdm\ being either weak or positive, and with the evidence against the remaining models being either strong or very strong. However, based on DIC, non-flat \lcdm\ and non-flat XCDM are the most favoured model by LGRB95 and LGRB145 data, with positive evidence against the remaining models, and with positive evidence against non-flat \lcdm\ and either strong or very strong evidence against the remaining models, respectively.

From the AIC, BIC, and DIC results we find that the 3D Dainotti correlation is very strongly favoured over the 2D Dainotti correlation by all three of the GRB data sets. Therefore, although some of the cosmological parameter constraints are more restrictive in the 2D Dainotti correlation cases, possibly because there is one less free parameter to constrain in the 2D correlation cases, we do not discuss them in detail. Leaving aside the 2D Dainotti correlation LGRB145 data set, we briefly discuss the results from the 2D correlation Platinum and LGRB95 data. Overall, these GRB data used with the 2D Dainotti correlation prefer higher values of \om\ and lower values of \ok\ and \wx\ (non-flat XCDM), whereas they do not provide restrictive constraints on \wx\ in flat XCDM (except for Platinum data) and $\alpha$ in \pcdm\ models. However, the constraints on the Platinum and LGRB95 2D Dainotti correlation parameters $a$ and $C_{o}$ are cosmological model-independent, with the 2D $a$ values being more negative and less restrictive and the 2D $C_o$ values being larger and more restrictive than those from the corresponding 3D Dainotti correlation data sets.

\section{Summary and Conclusion}
\label{makereference10.5}

In addition to 50 Platinum GRBs, we use LGRB95 data that contains 95 long GRBs, as well as the joint 145 GRB data compilation, to study whether the 2D or 3D Dainotti correlation is more favoured by data, as well as to constrain cosmological-model and GRB-correlation parameters, in six flat and non-flat dark energy cosmological models. Based on AIC, BIC, and DIC results, we find that the 3D Dainotti correlation is much more strongly favoured than the 2D one by the GRB data sets we study. 

We also find that LGRB95 data obey the 3D Dainotti correlation and are standardizable. Platinum and LGRB95 data provide mutually consistent constraints on both cosmological-model and GRB-correlation parameters, and also provide cosmological-model independent 3D Dainotti correlation parameter constraints. Therefore, we can combine Platinum with LGRB95 data to form the LGRB145 data set and use it for similar analyses. We find that while LGRB95 data have $\sim42-49$\% larger values of intrinsic scatter parameter $\sigma_{\rm int}\sim0.524-0.543$ than $\sigma_{\rm int}\sim0.365-0.369$ of Platinum data, they provide somewhat tighter constraints on cosmological-model and GRB-correlation parameters, perhaps mostly due to the larger number of data points, 95 versus 50. We recommend that when compiling GRB data for the purpose of constraining cosmological parameters, given the quality of current GRB data, attention be placed on also expanding the sample size, in addition to attempting to reduce the value of $\sigma_{\rm int}$ of the compilation.\footnote{This recommendation also holds for QSO and related data.}

LGRB95 data favour higher values of \om\ and lower values of \ok\ than do Platinum data, whereas the joint LGRB145 data favour even higher and lower values of \om\ and \ok\ than both Platinum and LGRB95 data, respectively. All these GRB data do not provide restrictive constraints on \wx\ and $\alpha$. LGRB145 data also provide tighter constraints on GRB-correlation parameters and the intrinsic scatter parameter.

Given the current paucity of GRB data it is therefore necessary to increase the sample size to allow the 3D correlation to have cosmological constraints comparable to those from SNIa data. A detailed study on simulating GRB constraints based on the Platinum sample to determine the number of Platinum-quality GRBs needed to reach constraints similar to those from a number of recent SNIa data sets is presented in \citet{DainottiNielson2022}. To achieve similar GRB constraints one needs to wait for more GRB data from future missions, and one can use machine learning techniques and lightcurves reconstruction on these larger data sets that can enable smaller scatter (47.5\%) on the 2D and 3D correlation parameters. 

We note that a major restriction on the use of more current GRBs as cosmological tools is the lack of redshift for many GRBs. Only 26\% of the total number of GRBs observed by Swift have reliable redshifts. Work on the inference of redshifts is underway \citep{Dainottietal2019} and does not require waiting for a new mission. Once reliable redshifts are determined for the GRBs with X-ray plateaus, we anticipate having a Platinum-quality sample twice as large as the current one as well as also anticipate doubling the size of the LGRB95-quality sample.

Current GRB data alone cannot provide very restrictive cosmological constraints comparable to those from better-established probes such as CMB, BAO, $H(z)$, or SNIa measurements, but one can do joint analyses of GRB data with these data to get more restrictive cosmological parameter constraints \citep{Xuetal2021a, CaoKhadkaRatra2022, CaoDainottiRatra2022, CaoRatra2022}.

We also look forward to a larger, better-quality, compilation of GRB data from the \textit{SVOM} mission scheduled to be launched in 2023 \citep{Atteiaetal2022}, and possibly the \textit{THESEUS} mission \citep{Amatietal2021} in 2037. In conjunction with machine learning techniques, these new data should provide significantly more restrictive GRB cosmological parameter constraints that could be comparable with those from SNIa data.

\begin{sidewaystable*}
\centering
\resizebox*{\columnwidth}{0.74\columnwidth}{%
\begin{threeparttable}
\caption{Unmarginalized best-fitting parameter values for all models from different data sets.\tnote{a}}\label{tab:BFPC10}
\begin{tabular}{lccccccccccccccccccc}
\toprule
Model & Data set & $\Omega_{c}h^2$ & $\Omega_{\mathrm{m0}}$ & $\Omega_{\mathrm{k0}}$ & $w_{\mathrm{X}}$/$\alpha$\tnote{b} & $\sigma_{\mathrm{int}}$ & $a$ & $b$ & $C_{o}$ & $-2\ln\mathcal{L}_{\mathrm{max}}$ & AIC & BIC & DIC & $\Delta \mathrm{AIC}^{\prime}$ & $\Delta \mathrm{BIC}^{\prime}$ & $\Delta \mathrm{DIC}^{\prime}$ & $\Delta \mathrm{AIC}$ & $\Delta \mathrm{BIC}$ & $\Delta \mathrm{DIC}$ \\
\midrule
 & 3D Platinum & 0.4555 & 0.981 & -- & -- & 0.341 & $-0.817$ & 0.711 & 14.23 & 41.72 & 51.72 & 61.28 & 51.88 & -- & -- & -- & -- & -- & --\\
 & 3D LGRB95 & 0.4644 & 0.999 & -- & -- & 0.522 & $-0.891$ & 0.729 & 13.44 & 156.64 & 166.64 & 179.41 & 167.20 & -- & -- & -- & -- & -- & --\\
Flat & 3D LGRB145 & 0.4635 & 0.997 & -- & -- & 0.468 & $-0.882$ & 0.733 & 13.27 & 209.18 & 219.18 & 234.06 & 219.88 & -- & -- & -- & -- & -- & --\\
\cmidrule{2-20}
\lcdm & 2D Platinum & 0.4648 & 1.000 & -- & -- & 0.478 & $-1.079$ & -- & 51.66 & 69.58 & 77.58 & 85.23 & 79.28 & 25.86 & 23.95 & 27.40 & -- & -- & --\\
 & 2D LGRB95 & 0.4642 & 0.999 & -- & -- & 0.720 & $-0.957$ & -- & 50.94 & 210.61 & 218.61 & 228.83 & 220.95 & 51.97 & 49.42 & 53.75 & -- & -- & --\\
 & 2D LGRB145 & 0.4647 & 1.000 & -- & -- & 0.658 & $-1.000$ & -- & 51.20 & 294.75 & 302.75 & 314.66 & 305.16 & 83.58 & 80.60 & 85.28 & -- & -- & --\\
\midrule
 & 3D Platinum & 0.0307 & 0.114 & $-0.495$ & -- & 0.307 & $-0.858$ & 0.667 & 16.65 & 33.89 & 45.89 & 57.36 & 57.84 & -- & -- & -- & $-5.83$ & $-3.92$ & 5.96\\
& 3D LGRB95 & 0.4641 & 0.999 & $-1.915$ & -- & 0.491 & $-0.923$ & 0.541 & 23.03 & 147.73 & 157.73 & 175.06 & 164.80 & -- & -- & -- & $-6.90$ & $-4.35$ & $-2.39$\\
Non-flat & 3D LGRB145 & 0.4606 & 0.991 & $-1.889$ & -- & 0.447 & $-0.940$ & 0.578 & 21.28 & 197.47 & 209.47 & 227.34 & 213.18 & -- & -- & -- & $-9.70$ & $-6.72$ & $-6.70$\\
\cmidrule{2-20}
\lcdm & 2D Platinum & 0.4644 & 0.999 & $-1.819$ & -- & 0.408 & $-1.189$ & -- & 51.88 & 54.84 & 64.84 & 74.40 & 67.11 & 18.95 & 17.04 & 9.27 & $-12.75$ & $-10.83$ & $-12.17$\\
 & 2D LGRB95 & 0.4648 & 1.000 & $-1.898$ & -- & 0.590 & $-1.000$ & -- & 50.85 & 174.93 & 184.93 & 197.70 & 187.31 & 25.20 & 22.65 & 22.51 & $-33.68$ & $-31.12$ & $-33.64$\\
 & 2D LGRB145 & 0.4639 & 0.998 & $-1.880$ & -- & 0.547 & $-1.070$ & -- & 51.22 & 246.10 & 256.10 & 270.98 & 258.28 & 45.62 & 43.65 & 45.09 & $-46.66$ & $-43.68$ & $-46.89$\\
\midrule
 & 3D Platinum & 0.0041 & 0.060 & -- & 0.131 & 0.348 & $-0.826$ & 0.701 & 14.72 & 41.37 & 53.37 & 64.84 & 51.95 & -- & -- & -- & 1.65 & 3.56 & 0.07\\
 & 3D LGRB95 & $-0.0188$ & 0.013 & -- & 0.121 & 0.523 & $-0.884$ & 0.697 & 15.02 & 155.96 & 167.96 & 183.28 & 167.34 & -- & -- & -- & 1.32 & 3.88 & 0.14\\
Flat & 3D LGRB145 & $-0.0246$ & 0.001 & -- & 0.135 & 0.459 & $-0.864$ & 0.718 & 13.94 & 208.12 & 220.12 & 237.98 & 220.31 & -- & -- & -- & 0.95 & 3.92 & 0.43\\
\cmidrule{2-20}
XCDM & 2D Platinum & $-0.0242$ & 0.002 & -- & 0.140 & 0.468 & $-1.086$ & -- & 51.59 & 67.15 & 77.15 & 86.71 & 81.30 & 23.78 & 21.87 & 29.35 & $-0.43$ & 1.48 & 2.02\\
 & 2D LGRB95 & $-0.0132$ & 0.024 & -- & 0.141 & 0.699 & $-0.974$ & -- & 50.92 & 205.84 & 215.84 & 228.61 & 224.19 & 47.89 & 45.33 & 56.85 & $-2.77$ & $-0.21$ & 3.24\\
 & 2D LGRB145 & $-0.0149$ & 0.021 & -- & 0.141 & 0.636 & $-1.017$ & -- & 51.17 & 287.71 & 297.71 & 312.59 & 308.85 & 77.59 & 74.61 & 88.54 & $-5.04$ & $-2.07$ & 3.69\\
\midrule
 & 3D Platinum & 0.1032 & 0.262 & $-1.446$ & $-0.696$ & 0.310 & $-0.803$ & 0.706 & 14.34 & 35.58 & 49.58 & 62.96 & 57.71 & -- & -- & -- & $-2.14$ & $-1.68$ & 3.83\\
 & 3D LGRB95 & 0.0721 & 0.199 & $-1.518$ & $-0.605$ & 0.471 & $-0.943$ & 0.616 & 19.24 & 140.12 & 154.12 & 171.99 & 168.13 & -- & -- & -- & $-12.52$ & $-7.41$ & 0.93\\
Non-flat & 3D LGRB145 & 0.4611 & 0.992 & $-1.228$ & $-4.971$ & 0.441 & $-0.941$ & 0.517 & 24.48 & 192.09 & 206.09 & 226.93 & 210.37 & -- & -- & -- & $-13.09$ & $-7.13$ & $-9.51$\\
\cmidrule{2-20}
XCDM & 2D Platinum & 0.4568 & 0.983 & $-1.182$ & $-4.983$ & 0.389 & $-1.182$ & -- & 52.04 & 51.73 & 63.73 & 75.20 & 65.25 & 14.16 & 12.24 & 9.53 & $-13.85$ & $-10.02$ & $-14.03$\\
 & 2D LGRB95 & 0.4609 & 0.992 & $-1.232$ & $-4.912$ & 0.553 & $-1.009$ & -- & 51.05 & 163.56 & 175.56 & 190.88 & 178.23 & 21.44 & 18.89 & 10.10 & $-43.05$ & $-37.94$ & $-42.71$\\
 & 2D LGRB145 & 0.4633 & 0.997 & $-1.232$ & $-4.998$ & 0.511 & $-1.078$ & -- & 51.41 & 229.92 & 241.92 & 259.78 & 243.77 & 35.83 & 32.85 & 33.40 & $-60.83$ & $-54.88$ & $-61.40$\\
\midrule
 & 3D Platinum & 0.4564 & 0.983 & -- & 8.469 & 0.342 & $-0.828$ & 0.698 & 14.90 & 41.72 & 53.72 & 65.19 & 51.56 & -- & -- & -- & 2.00 & 3.91 & $-0.32$\\
 & 3D LGRB95 & 0.4614 & 0.993 & -- & 4.276 & 0.524 & $-0.894$ & 0.724 & 13.70 & 156.64 & 168.64 & 183.97 & 166.90 & -- & -- & -- & 2.01 & 4.56 & $-0.30$\\
Flat & 3D LGRB145 & 0.4598 & 0.990 & -- & 8.582 & 0.466 & $-0.871$ & 0.732 & 13.28 & 209.19 & 221.19 & 239.05 & 219.62 & -- & -- & -- & 2.01 & 4.99 & $-0.26$\\
\cmidrule{2-20}
 \pcdm & 2D Platinum & 0.4647 & 1.000 & -- & 6.040 & 0.477 & $-1.079$ & -- & 51.65 & 69.58 & 79.58 & 89.14 & 79.15 & 25.86 & 23.95 & 27.59 & 2.00 & 3.91 & $-0.13$\\
 & 2D LGRB95 & 0.4648 & 1.000 & -- & 4.187 & 0.723 & $-0.949$ & -- & 50.91 & 210.61 & 220.61 & 233.38 & 220.64 & 51.96 & 49.41 & 53.74 & 2.00 & 4.55 & $-0.31$\\
 & 2D LGRB145 & 0.4647 & 1.000 & -- & 9.417 & 0.654 & $-1.010$ & -- & 51.23 & 294.76 & 304.76 & 319.65 & 304.90 & 83.58 & 80.60 & 85.28 & 2.01 & 4.98 & $-0.26$\\
\midrule
 & 3D Platinum & 0.4359 & 0.941 & $-0.939$ & 0.024 & 0.333 & $-0.818$ & 0.670 & 16.30 & 40.75 & 54.75 & 68.13 & 52.57 & -- & -- & -- & 3.03 & 6.85 & 0.69\\
 & 3D LGRB95 & 0.4566 & 0.983 & $-0.969$ & 0.381 & 0.515 & $-0.896$ & 0.670 & 16.47 & 154.66 & 168.66 & 186.54 & 168.06 & -- & -- & -- & 2.03 & 7.14 & 0.86\\
Non-flat & 3D LGRB145 & 0.4634 & 0.997 & $-0.970$ & 0.352 & 0.461 & $-0.877$ & 0.694 & 15.20 & 206.38 & 220.38 & 241.22 & 221.01 & -- & -- & -- & 1.20 & 7.16 & 1.13\\
\cmidrule{2-20}
\pcdm & 2D Platinum & 0.4634 & 0.997 & $-0.985$ & 0.000 & 0.449 & $-1.093$ & -- & 51.67 & 63.73 & 75.73 & 87.21 & 80.15 & 20.98 & 19.07 & 27.58 & $-1.85$ & 1.98 & 0.87\\
 & 2D LGRB95 & 0.4647 & 1.000 & $-0.991$ & 0.003 & 0.679 & $-0.981$ & -- & 51.04 & 198.63 & 210.63 & 225.96 & 218.39 & 41.97 & 39.42 & 50.33 & $-7.98$ & $-2.87$ & $-2.56$\\
 & 2D LGRB145 & 0.4628 & 0.996 & $-0.994$ & 0.000 & 0.611 & $-1.004$ & -- & 51.20 & 277.29 & 289.29 & 307.15 & 299.79 & 68.91 & 65.93 & 78.78 & $-13.47$ & $-7.51$ & $-5.37$\\
\bottomrule
\end{tabular}
\begin{tablenotes}[flushleft]
\item [a] $\Omega_{b}$ is set to 0.05 and $H_0$ is set to 70 \hunit.
\item [b] \wx\ corresponds to flat/non-flat XCDM and $\alpha$ corresponds to flat/non-flat \pcdm.
\end{tablenotes}
\end{threeparttable}%
}
\end{sidewaystable*}

\begin{sidewaystable*}
\centering
\resizebox*{\columnwidth}{0.74\columnwidth}{%
\begin{threeparttable}
\caption{One-dimensional marginalized posterior mean values and uncertainties ($\pm 1\sigma$ error bars or $2\sigma$ limits) of the parameters for all models from different data sets.\tnote{a}}\label{tab:1d_BFPC10}
\begin{tabular}{lcccccccc}
\toprule
Model & Data set & $\Omega_{\mathrm{m0}}$ & $\Omega_{\mathrm{k0}}$ & $w_{\mathrm{X}}$/$\alpha$\tnote{b} & $\sigma_{\mathrm{int}}$ & $a$ & $b$ & $C_{o}$ \\
\midrule
 & 3D Platinum & $>0.441$\tnote{c} & -- & -- & $0.369^{+0.038}_{-0.052}$ & $-0.812\pm0.119$ & $0.728\pm0.122$ & $13.38\pm6.45$\\
 & 3D LGRB95 & $>0.155$ & -- & -- & $0.543^{+0.042}_{-0.052}$ & $-0.887\pm0.097$ & $0.743\pm0.091$ & $12.73\pm4.69$\\
Flat & 3D LGRB145 & $>0.213$ & -- & -- & $0.479^{+0.031}_{-0.036}$ & $-0.872\pm0.073$ & $0.743\pm0.073$ & $12.71\pm3.78$\\\cmidrule{2-9}
\lcdm & 2D Platinum & $>0.372$ & -- & -- & $0.515^{+0.049}_{-0.066}$ & $-1.080\pm0.146$ & -- & $51.75\pm0.55$\\
 & 2D LGRB95 & $>0.526$ & -- & -- & $0.747^{+0.055}_{-0.069}$ & $-0.953\pm0.125$ & -- & $50.98\pm0.48$\\
 & 2D LGRB145 & $>0.628$ & -- & -- & $0.673^{+0.040}_{-0.048}$ & $-1.001\pm0.100$ & -- & $51.24\pm0.38$\\
\midrule
 & 3D Platinum & $>0.170$ & $-0.304^{+0.630}_{-1.452}$ & -- & $0.366^{+0.038}_{-0.052}$ & $-0.839\pm0.128$ & $0.690^{+0.153}_{-0.133}$ & $15.36^{+7.02}_{-8.12}$\\
 & 3D LGRB95 & $>0.321$ & $-1.111^{+0.112}_{-0.790}$ & -- & $0.528^{+0.042}_{-0.051}$ & $-0.911\pm0.095$ & $0.628\pm0.125$ & $18.63^{+6.42}_{-6.45}$\\
Non-flat & 3D LGRB145 & $>0.444$ & $-1.396^{+0.125}_{-0.487}$ & -- & $0.465^{+0.030}_{-0.035}$ & $-0.915\pm0.075$ & $0.620^{+0.087}_{-0.094}$ & $19.12\pm4.69$\\\cmidrule{2-9}
\lcdm & 2D Platinum & $>0.475$ & $-1.415^{+0.196}_{-0.367}$ & -- & $0.445^{+0.043}_{-0.059}$ & $-1.173\pm0.132$ & -- & $51.93\pm0.49$\\
 & 2D LGRB95 & $>0.622$ & $-1.648^{+0.104}_{-0.240}$ & -- & $0.616^{+0.046}_{-0.057}$ & $-0.997\pm0.106$ & -- & $50.91\pm0.41$\\
 & 2D LGRB145 & $>0.700$ & $-1.693^{+0.084}_{-0.186}$ & -- & $0.563^{+0.034}_{-0.041}$ & $-1.066\pm0.083$ & -- & $51.26\pm0.31$\\
\midrule
 & 3D Platinum & $>0.407$\tnote{c} & -- & $-2.470^{+2.578}_{-2.342}$ & $0.368^{+0.038}_{-0.052}$ & $-0.816\pm0.119$ & $0.720\pm0.124$ & $13.78\pm6.58$\\
 & 3D LGRB95 & $>0.128$ & -- & $<-0.023$ & $0.543^{+0.041}_{-0.051}$ & $-0.888\pm0.097$ & $0.738\pm0.091$ & $13.01\pm4.69$\\
Flat & 3D LGRB145 & $>0.159$ & -- & $<-0.002$ & $0.479^{+0.030}_{-0.036}$ & $-0.874\pm0.074$ & $0.741\pm0.071$ & $12.87^{+3.72}_{-3.71}$\\\cmidrule{2-9}
XCDM & 2D Platinum & $>0.208$ & -- & $<0.065$ & $0.514^{+0.049}_{-0.067}$ & $-1.081\pm0.150$ & -- & $51.79\pm0.57$\\
 & 2D LGRB95 & $>0.172$ & -- & -- & $0.746^{+0.054}_{-0.067}$ & $-0.955\pm0.127$ & -- & $51.02\pm0.50$\\
 & 2D LGRB145 & $>0.587$\tnote{c} & -- & -- & $0.669^{+0.041}_{-0.048}$ & $-1.002\pm0.100$ & -- & $51.25\pm0.39$\\
\midrule
 & 3D Platinum & $>0.499$\tnote{c} & $-0.251^{+0.697}_{-1.117}$ & $-2.210^{+2.295}_{-0.985}$ & $0.365^{+0.038}_{-0.052}$ & $-0.842\pm0.127$ & $0.680^{+0.160}_{-0.136}$ & $15.93^{+7.22}_{-8.52}$\\
 & 3D LGRB95 & $>0.267$ & $-0.861^{+0.310}_{-0.511}$ & $<-0.327$ & $0.524^{+0.042}_{-0.051}$ & $-0.918\pm0.095$ & $0.588^{+0.140}_{-0.138}$ & $20.75^{+7.14}_{-7.22}$\\
Non-flat & 3D LGRB145 & $>0.363$ & $-0.991^{+0.240}_{-0.307}$ & $<-0.964$ & $0.458^{+0.030}_{-0.035}$ & $-0.929\pm0.074$ & $0.572\pm0.094$ & $21.68^{+4.93}_{-4.91}$\\\cmidrule{2-9}
XCDM & 2D Platinum & $>0.327$ & $-0.987^{+0.322}_{-0.282}$ & $<-0.746$ & $0.434^{+0.043}_{-0.059}$ & $-1.166\pm0.130$ & -- & $52.08\pm0.50$\\
 & 2D LGRB95 & $>0.477$ & $-1.035^{+0.135}_{-0.239}$ & $<-2.386$ & $0.581^{+0.045}_{-0.055}$ & $-1.008\pm0.102$ & -- & $51.16\pm0.40$\\
 & 2D LGRB145 & $>0.513$ & $-1.029^{+0.105}_{-0.226}$ & $<-2.960$ & $0.531^{+0.032}_{-0.039}$ & $-1.070\pm0.078$ & -- & $51.49\pm0.31$\\
\midrule
 & 3D Platinum & $>0.406$\tnote{c} & -- & -- & $0.368^{+0.038}_{-0.052}$ & $-0.813\pm0.116$ & $0.726\pm0.123$ & $13.45\pm6.48$\\
 & 3D LGRB95 & $>0.453$\tnote{c} & -- & -- & $0.543^{+0.041}_{-0.051}$ & $-0.888\pm0.094$ & $0.739\pm0.091$ & $12.95\pm4.72$\\
Flat & 3D LGRB145 & $>0.157$ & -- & -- & $0.480^{+0.030}_{-0.036}$ & $-0.873\pm0.073$ & $0.743\pm0.071$ & $12.73\pm3.71$\\\cmidrule{2-9}
\pcdm & 2D Platinum & $>0.287$ & -- & -- & $0.513^{+0.048}_{-0.066}$ & $-1.077\pm0.149$ & -- & $51.71\pm0.55$\\
 & 2D LGRB95 & $>0.390$ & -- & -- & $0.748^{+0.054}_{-0.068}$ & $-0.953\pm0.126$ & -- & $50.97\pm0.48$\\
 & 2D LGRB145 & $>0.548$ & -- & -- & $0.672^{+0.040}_{-0.047}$ & $-1.000\pm0.099$ & -- & $51.23\pm0.37$\\
\midrule
 & 3D Platinum & $>0.425$\tnote{c} & $-0.065^{+0.384}_{-0.382}$ & -- & $0.368^{+0.038}_{-0.052}$ & $-0.817\pm0.120$ & $0.721\pm0.127$ & $13.71\pm6.72$\\
 & 3D LGRB95 & $>0.150$ & $-0.153^{+0.371}_{-0.392}$ & -- & $0.541^{+0.041}_{-0.051}$ & $-0.889\pm0.094$ & $0.732\pm0.093$ & $13.29^{+4.79}_{-4.78}$\\
Non-flat & 3D LGRB145 & $>0.215$ & $-0.226^{+0.357}_{-0.388}$ & -- & $0.479^{+0.030}_{-0.036}$ & $-0.876\pm0.074$ & $0.732\pm0.075$ & $13.28\pm3.91$\\\cmidrule{2-9}
\pcdm & 2D Platinum & $>0.405$ & $-0.402^{+0.256}_{-0.404}$ & -- & $0.501^{+0.049}_{-0.067}$ & $-1.089\pm0.146$ & -- & $51.69\pm0.55$\\
 & 2D LGRB95 & $>0.594$ & $-0.628^{+0.104}_{-0.324}$ & -- & $0.719^{+0.053}_{-0.065}$ & $-0.958\pm0.123$ & -- & $50.90\pm0.47$\\
 & 2D LGRB145 & $>0.688$ & $-0.749^{+0.062}_{-0.210}$ & -- & $0.645^{+0.039}_{-0.046}$ & $-1.010\pm0.095$ & -- & $51.17\pm0.36$\\
\bottomrule
\end{tabular}
\begin{tablenotes}[flushleft]
\item [a] $\Omega_{b}$ is set to 0.05 and $H_0$ is set to 70 \hunit.
\item [b] \wx\ corresponds to flat/non-flat XCDM and $\alpha$ corresponds to flat/non-flat \pcdm.
\item [c] This is the 1$\sigma$ limit. The 2$\sigma$ limit is set by the prior and not shown here.
\end{tablenotes}
\end{threeparttable}%
}
\end{sidewaystable*}


\cleardoublepage


\chapter{Standardizing reverberation-measured \civ\ time-lag quasars, and using them with standardized \mii\ quasars to constrain cosmological parameters}
\label{makereference11}

This chapter is based on \cite{Cao:2022pdv}.

\section{Introduction}
\label{makereference11.1}

The well-established observed currently accelerated expansion of the Universe has motivated many theoretical cosmological models. In the standard general-relativistic spatially-flat \lcdm\ cosmological model \citep{peeb84} dark energy in the form of a time-independent cosmological constant $\Lambda$ powers the currently accelerated cosmological expansion and contributes $\sim70\%$ of the current cosmological energy budget, with non-relativistic cold dark matter (CDM) and baryonic matter contributing $\sim25\%$ and $\sim5\%$, respectively. Although the flat \lcdm\ model makes predictions consistent with most observations \citep[see, e.g.][]{scolnic_et_al_2018, Yuetal2018, planck2018b, eBOSS_2020}, it has some potential observational discrepancies \citep{DiValentinoetal2021a, PerivolaropoulosSkara2021, Abdallaetal2022} that motivate us to also study dynamical dark energy models as well as models with non-zero spatial curvature.

There are better-established cosmological probes, such as baryon acoustic oscillation (BAO), type Ia supernova apparent magnitude, and Hubble parameter [$H(z)$] measurements that, when jointly analyzed, provide fairly restrictive cosmological parameter constraints \citep[see, e.g.][]{CaoRatra2022}. Cosmological probes that are now under active development can tighten these constraints. Amongst these developing probes are \hii\ starburst galaxy apparent magnitude data that reach to redshift $z \sim 2.4$ \citep{Mania_2012, Chavez_2014, GonzalezMoranetal2021, CaoRyanRatra2020, CaoRyanRatra2021, Johnsonetal2022, Mehrabietal2022}, quasar (QSO) angular size measurements that reach to $z \sim 2.7$ \citep{Cao_et_al2017a, Ryanetal2019, Zhengetal2021, Lian_etal_2021, CaoRyanRatra2022}, QSO flux observations that reach to $z \sim 7.5$ \citep{RisalitiLusso2015, RisalitiLusso2019, KhadkaRatra2020a, KhadkaRatra2020b, KhadkaRatra2021, KhadkaRatra2022, Lussoetal2020, ZhaoXia2021, Rezaeietal2022, Luongoetal2021, Leizerovichetal2021, Colgainetal2022, DainottiBardiacchi2022},\footnote{Note however that the \cite{Lussoetal2020} QSO flux compilation assumes a model for the QSO UV--X-ray correlation that is invalid above redshifts $z \sim 1.5-1.7$ so this is the limit to which the \cite{Lussoetal2020} QSOs can be used to determine cosmological constraints \citep{KhadkaRatra2021, KhadkaRatra2022}.} gamma-ray burst data that reach to $z \sim 8.2$ \citep{Wang_2016, Wangetal_2021, Dirirsa2019, Demianskietal_2021, KhadkaRatra2020c, Huetal2021, Khadkaetal_2021b, LuongoMuccino2021, Caoetal_2021, CaoDainottiRatra2022b, CaoKhadkaRatra2022, CaoDainottiRatra2022, DainottiNielson2022, Liuetal2022}, and --- the main subject of our paper --- reverberation-mapped (RM) QSO data that reach to $z \sim 3.4$ \citep[][and this paper]{Czernyetal2021, Zajaceketal2021, Khadkaetal_2021a, Khadkaetal2022a, Khadkaetal2022a}. 

\citet{Khadkaetal_2021a, Khadkaetal2022a, Khadkaetal2022a} derived cosmological constraints from RM H$\beta$ and \mq\ data. In this paper, we show that RM \cq\ data, that extend to higher $z$, are standardizable and derive the first cosmological constraints from \civ\ data. 

With its ionization potential energy of $\sim 64.5$ eV, \civ\ ($\lambda$1549) belongs to the high-ionization line (HIL) component of the QSO broad-line region \citep[BLR; ][]{1988MNRAS.232..539C,2005MNRAS.356.1029B,Karasetal2021}, which can partially form an outflow that is manifested by the blueshifted centroid of the line \citep[see e.g.][]{2021MNRAS.503.3145B} as well as by the frequent blueward line-emission asymmetry \citep{2005MNRAS.356.1029B}. It is therefore not yet established whether all of the \civ\ material is virialized and if the standard reverberation mapping (hereafter RM\footnote{Depending on the context, we use the abbreviation RM for both reverberation mapping as a method and for reverberation-mapped quasars interchangeably.}) of HILs can lead to the reliable measurements of the SMBH masses, as was previously done using the low-ionization lines (LILs) for several hundreds of objects (mainly using the Balmer line H$\beta$ and the resonance \mii\ line).

However, the BLR radius-luminosity ($R-L$) relationship for the \civ\ line, in the flat $\Lambda$CDM model, appears to now be well established with a significant positive correlation and a relatively small dispersion \citep{2007ApJ...659..997K,2018ApJ...865...56L,2019ApJ...887...38G,2021ApJ...915..129K}, which allows for the possibility of using this relation for constraining cosmological parameters as we previously did for H$\beta$ and \mii\ lines \citep{Mehrabietal2021, 2019FrASS...6...75P, Zajaceketal2021,Czernyetal2021,Khadkaetal_2021a,Khadkaetal2022a,Khadkaetal2022a}.

In this paper, we use 38 high-quality \cq\ measurements, which span eight orders of magnitude in luminosity $\sim 10^{40-48}$ erg s$^{-1}$ and the redshift range $0.001064 \leq z \leq 3.368$, to constrain, for the first time, cosmological-model and $R-L$ relation parameters, in six general relativistic dark energy cosmological models, using a more correct technique to account for the asymmetric errors of the time-lag measurements. We find that the \civ\ $R-L$ relation parameters are independent of cosmological model, so \cq\ data are standardizable through the \civ\ $R-L$ relation. \mq\ data were also found to be standardizable through the \mii\ $R-L$ relation \citep{Khadkaetal_2021a, Khadkaetal2022a}. We find that cosmological constraints from \civ\ and \mii\ QSO data are mutually consistent and are also consistent with those from $H(z)$ + BAO data. Although the cosmological constraints from the joint analysis of \civ\ + \mii\ QSO data are weak, jointly analyzing \civ\ + \mii\ data with $H(z)$ + BAO data results in a mild ($<0.1\sigma$) tightening of the $H(z)$ + BAO cosmological constraints.

This paper is organized as follows. We describe the cosmological models/parametrizations we study in Chapter \ref{sec:models}. In Sections \ref{makereference11.2} and \ref{makereference11.3}, we outline the data sets and the analysis methods we use, respectively. Our constrained cosmological parameter and $R-L$ relation parameter results are presented and discussed in Sections \ref{makereference11.4} and \ref{makereference11.5}. We summarize our conclusions in Section \ref{makereference11.6}.

\section{Data}
\label{makereference11.2}

In this work, we use \cq, \mq, and $H(z)$ + BAO data, as well as combinations of these data sets, to constrain cosmological-model and QSO $R-L$ relation parameters. These data sets are summarized next, with the emphasis on the \cq\ sample.

\begin{itemize}
\item[]{\it \cq\ data.} Here we describe the sample of high-quality 38 \civ\ RM QSOs. The first time-lag measurements of the broad \civ\ line were inferred by \citet{2005ApJ...632..799P, 2006ApJ...641..638P} for 4 sources. The source NGC4151 was monitored by \citet{2006ApJ...647..901M}. The \civ\ time-lag for the intensively monitored NGC5548 was determined by \citet{2015ApJ...806..128D}. \citet{2018ApJ...865...56L} performed RM for 17 high-luminosity QSOs for over 10 years, out of which 8 QSOs were reported to have statistically significant ($>1\sigma$) \civ\ time-lag measurements. \citet{2019MNRAS.487.3650H} report 2 \civ\ detections for quasars at $z=1.905$ and $z=2.593$ from the photometric Dark Energy Survey Supernova Program (DES-SN) and the spectroscopic Australian Dark Energy Survey (Oz-DES). Within the Sloan Digital Sky Survey Reverberation Mapping project (SDSS-RM), \citet{2019ApJ...887...38G} determined \civ\ time-lag measurements for 48 QSOs with an average false-positive rate of $10\%$. Of these, 16 QSOs pass the highest-quality criteria, in the redshift range of $1.4<z<2.8$ and the monochromatic luminosity range $44.5<\log{[L_{1350}\,({\rm erg\,s^{-1}})]}<45.6$. \citet{2019ApJ...883L..14S} showed that adding more photometric data points and spectroscopic data ($9+5$ years of photometric and spectroscopic measurements, respectively, in comparison to $4+4$ years of spectroscopic and photometric monitoring performed by \citeauthor{2019ApJ...887...38G}, \citeyear{2019ApJ...887...38G}) results in the significant detection of 3 more \civ\ time-lag measurements. Using 20 years of photometric and spectrophotometric data, \citet{2021ApJ...915..129K} report significant \civ\ time-delays for 3 QSOs at redshifts $z=2.172$, $2.646$, and $3.192$.\\

\begin{figure}
    \centering
    \includegraphics[width=\columnwidth]{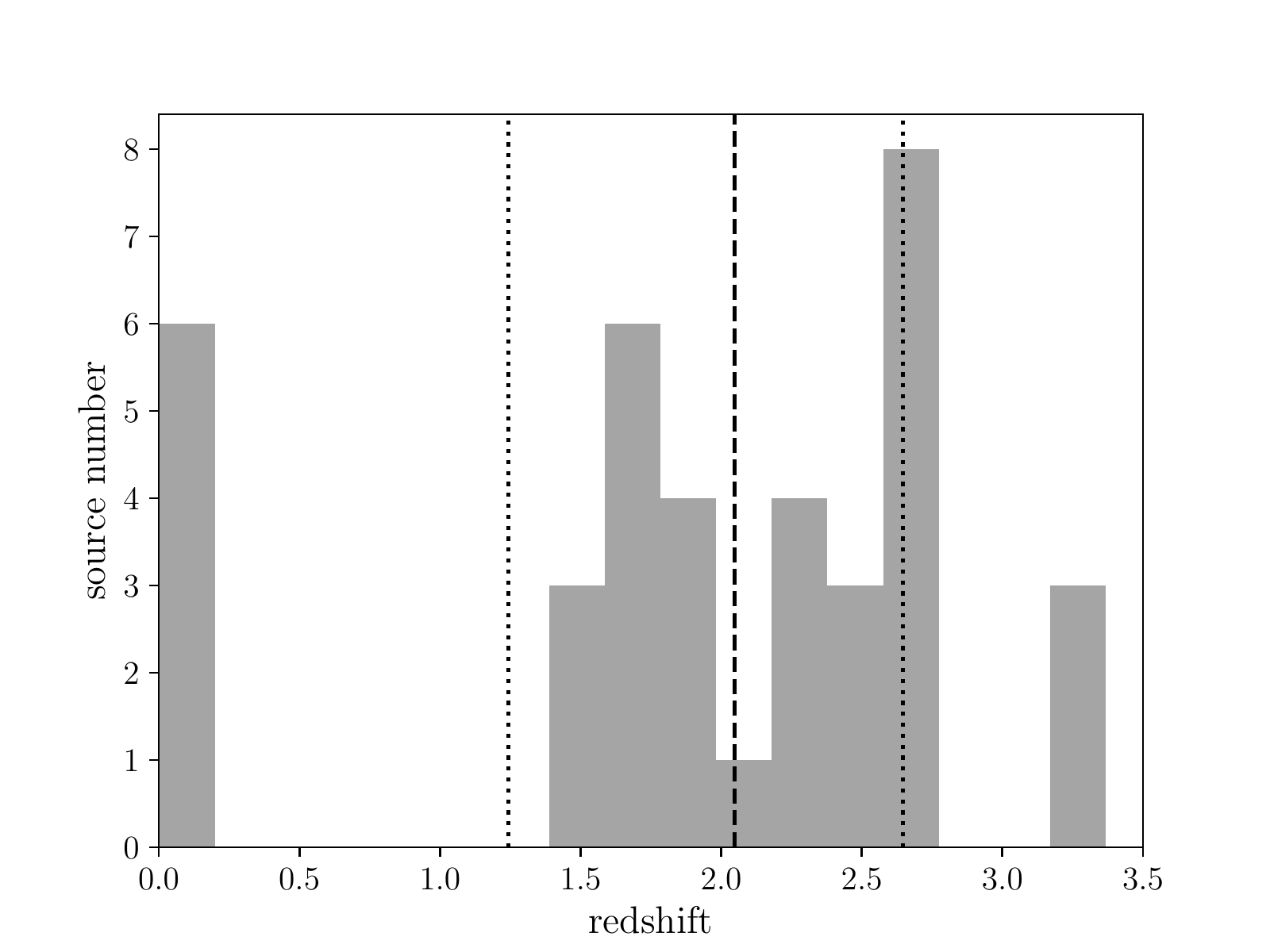}
    \caption{The redshift distribution for the ``golden'' sample of 38 \civ\ QSOs from \citet{2021ApJ...915..129K}. The dashed line stands for the redshift median at $2.048$, while the dotted lines represent 16\% and 84\% percentiles at $1.244$ and $2.647$, respectively. The histogram bin size is $\Delta z\simeq 0.2$.}
    \label{fig_redshift}
\end{figure}

\citet{2021ApJ...915..129K} compiled (with a few corrections) available \civ\ RM sources, finding 38 QSOs with significantly measured \civ\ time-delays, including their 3 QSOs. The time-delays were determined by using either the interpolated cross-correlation function (ICCF) or the $z$-transformed dicrete correlation function (zDCF) or a combination of both, which ensures a relative homogeneity of the sample in terms of the time-delay methodology in comparison with, e.g., the H$\beta$ sample where several different methods are applied and combined, see e.g. \citet{2019AN....340..577Z} or \citet{2020A&A...642A..59R} for overviews. We refer to this sample of 38 \civ\ QSOs as ``golden'' and it covers the redshift range $0.001064 \leq z \leq 3.368$, with the median redshift of $2.048$ and 16-\% and 84-\% percentile redshifts of $1.244$ and $2.647$, respectively. The redshift distribution is shown in Fig.~\ref{fig_redshift}. The sources belonging to the golden sample are listed in Table~\ref{tab:civdata}, including their redshift, flux density at 1350\,\AA\,, monochromatic luminosity at 1350\,\AA\, (computed assuming a flat $\Lambda$CDM model with $H_0=70\,{\rm km\,s^{-1}\,Mpc^{-1}}$, $\Om=0.3$, and $\Omega_{\Lambda}=0.7$), and the rest-frame \civ\ time-lag $\tau$ (typically with asymmetrical error bars). 

The correlation between the rest-frame \civ\ time-delay and the UV monochromatic luminosity at $1350\,$\AA\, is significant, with the Pearson correlation coefficient $r=0.898$ ($p=2.082 \times 10^{-14}$) and the Spearman rank-order correlation coefficient $s=0.799$ ($p=1.751 \times 10^{-9}$). Given the large correlation coefficient, we fit the golden dataset with the power-law relation $\log{\tau}=\beta_{\rm C}+\gamma_{\rm C} \log{(L_{1350}/10^{44}\,{\rm erg\,s^{-1}})} $, where $\log\equiv\log_{10}$, and find the best-fit intercept $\beta_{\rm C}=1.04 \pm 0.07$ and the best-fit slope $\gamma_{\rm C}=0.42 \pm 0.03$, for which the individual time-delay errors were neglected in the Levenberg-Marquardt algorithm. This results in an intrinsic scatter of $\sigma=0.32$ dex and $\chi^2=31.8$ (36 degrees of freedom). When we consider individual symmetrized time-delay errors (see the discussion in the paragraph below eq.\ (\ref{eq:sigma_mq})), we obtain $\gamma_{\rm C}=0.56\pm 0.04$ and $\beta_{\rm C}=0.98\pm 0.07$ with $\sigma=0.41$ dex and $\chi^2=16.6$. See Fig.~\ref{fig_civ_radius_luminosity} for the best-fitting relations, which were inferred using the \texttt{curve\_fit} function from the \texttt{scipy} library. To add information about the accretion state of each QSO from our sample in the $R-L$ relation, we estimate the Eddington ratio $\lambda_{\rm Edd}=L_{\rm bol}/L_{\rm Edd}$, where $L_{\rm bol}=\kappa_{\rm bol}L_{1350}$ is the bolometric luminosity calculated using the luminosity-dependent bolometric factor $\kappa_{\rm bol}$ according to \citet{2019MNRAS.488.5185N}, and $L_{\rm Edd}$ is the Eddington luminosity \citep{2014A&A...565A..17Z,2017FoPh...47..553E,2020ApJ...903..140Z}. To obtain $L_{\rm Edd}$, the supermassive black hole (SMBH) mass was calculated using the virial relation $M_{\bullet}=f_{\rm vir} c\tau \text{FWHM}^2/G$, where the virial factor $f_{\rm vir}$ was estimated using the fitted formula that inversely scales with the full width at half maximum, FWHM, see \citet{2018NatAs...2...63M}. The FWHM and $\tau$ values were adopted from the compilation of \citet{2021ApJ...915..129K}. In Fig.~\ref{fig_civ_radius_luminosity} each source is coloured by the corresponding value of $\log{\lambda_{\rm Edd}}$.

\begin{figure}
    \centering
    \includegraphics[width=\columnwidth]{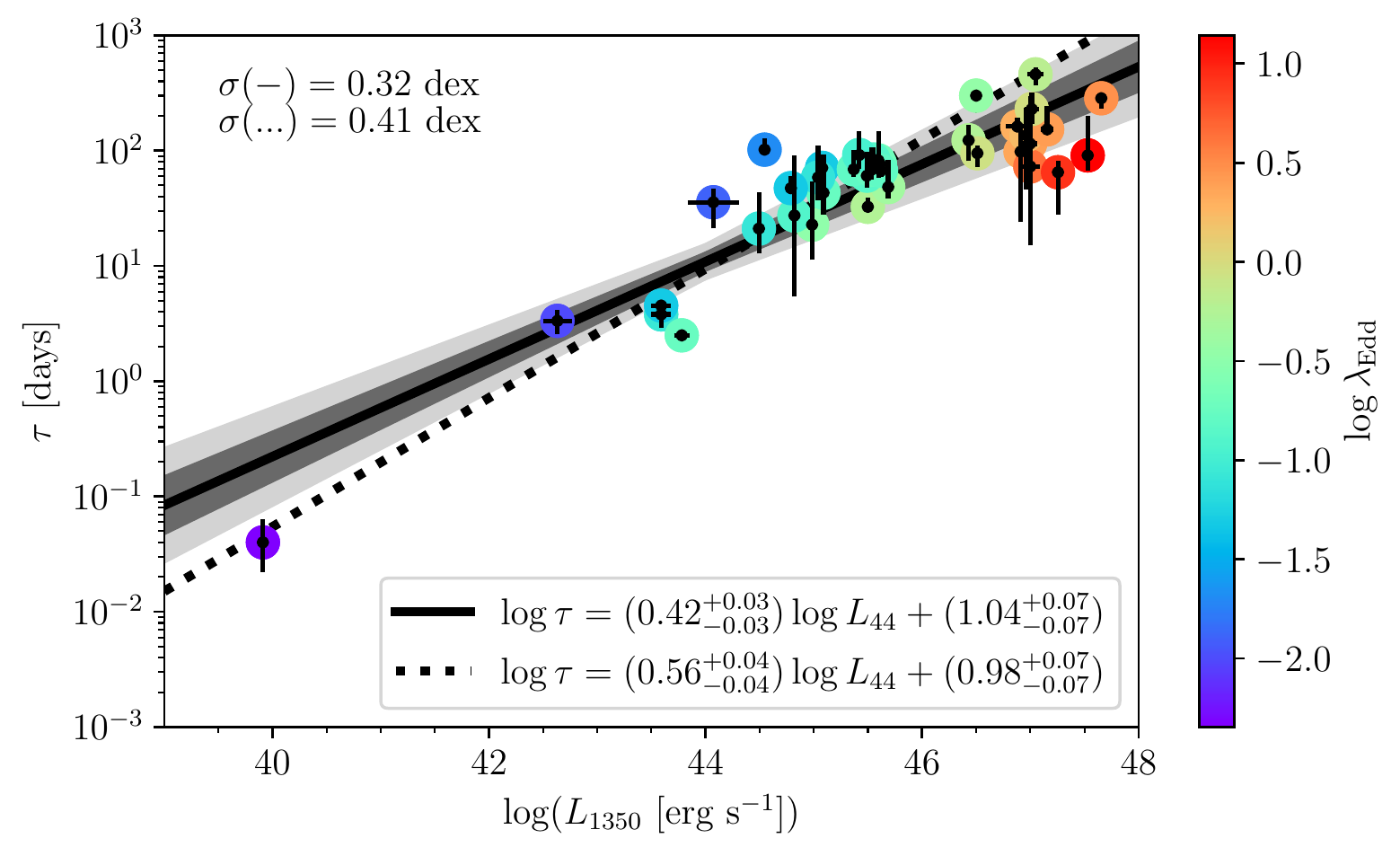}
    \caption{Radius-luminosity relationship for the ``golden'' sample of 38 \civ\ QSOs \citep[see also][]{2021ApJ...915..129K}. The luminosity at 1350\,\AA\, is based on the flat $\Lambda$CDM model (with $H_0=70\,{\rm km\,s^{-1}\,Mpc^{-1}}$, $\Om=0.3$, and $\Omega_{\Lambda}=0.7$). The points are colour-coded with respect to the calculated Eddington ratio $\lambda_{\rm Edd}=L_{\rm bol}/L_{\rm Edd}$ of the source. When we neglect individual time-delay errors, the best-fit relation (black solid line) determined by the Levenberg-Marquardt algorithm has a slope of $\gamma_{\rm C}=0.42\pm 0.03$ and a scatter of $\sigma=0.32$ dex. The dark and light gray areas around the best-fit solid line are one and two sigma confidence regions. Considering individual time-delay errors yields the best-fit relation (dotted line) with a larger slope of $\gamma_{\rm C}=0.56\pm 0.04$ and the scatter of data around this relation is $\sigma=0.41$ dex.}
    \label{fig_civ_radius_luminosity}
\end{figure}

 Apart from the ``golden'' \civ\ AGN sample reported by \citet{2021ApJ...915..129K}, there are additional lower-quality \civ\ time-lag detections reported in literature. \citet{2019ApJ...887...38G} report additional 32 \civ\ detections with a lower quality rating of 1, 2, and 3.\footnote{\citet{2019ApJ...887...38G}  assign the quality flag of 1 to the lowest-quality time-lag detections, while 5 is the highest quality rating. There are different factors considered in the rating scheme, such as the clear correlation between the continuum and \civ\ emission-line light curves, overall consistency between JAVELIN, CREAM, and ICCF time-lag detections, and the presence of multiple significant time-lag peaks.} As a follow-up of the Oz-DES RM project, \citet{2022MNRAS.509.4008P} designed a quality-cut methodology and out of 6 test sources, 2 at redshifts $1.93$ and $2.74$ pass all the quality criteria. These additional 34 sources will be considered in a future study.

\item[]{\it \mq\ data.} As listed in table A1 of \cite{Khadkaetal_2021a}, the \mq\ sample consists of 78 QSOs and spans the redshift range $0.0033 \leq z \leq 1.89$. Detailed descriptions of \mq\ data can be found in \cite{Khadkaetal_2021a} and \cite{2020ApJ...903...86M}, where it was shown that they obey the $R-L$ relation, with measured quantities being time-delay $\tau^{\prime}$ and QSO flux $F_{3000}$ measured at 3000 \(\text{\r{A}}\); see also \cite{2022A&A...667A..42P} for an updated \mq\ $R-L$ relation in the fixed flat $\Lambda$CDM cosmology. We note that the \mq\ sample is relatively homogeneous since 57 significant time-delay detections (hence 73\% of \mq\ sources) were determined by \citet{2020ApJ...901...55H}, who applied a consistent time-delay method to all the sources based on JAVELIN \citep{2011ApJ...735...80Z}, which was compared with the CREAM results that were generally consistent \citep{2016MNRAS.456.1960S}. However, the time-delay uncertainties for this sample are not completely homogeneous and a consistent treatment of the continuum and the line-emission light-curve correlation and the time-delay inference is needed to homogenize the sample of the best \mq\ time delays and their uncertainties.

\item[]{$H(z)\ +\ BAO\ data$.} There are 32 $H(z)$ and 12 BAO measurements listed in Tables 1 and 2 of \cite{CaoRatra2022}, spanning the redshift ranges $0.07 \leq z \leq 1.965$ and $0.122 \leq z \leq 2.334$, respectively.

\end{itemize}

\section{Data Analysis Methodology}
\label{makereference11.3}

We utilize the $R-L$ relation parametrization according to \citet{Bentzetal2013}, where we replace the monochromatic luminosity and rest-frame time-delay taking into account the \civ-region emission properties
\begin{equation}
    \label{eq:civ}
    \log{\frac{\tau}{\rm days}}=\beta_{\rm\textsc{c}}+\gamma_{\rm\textsc{c}} \log{\frac{L_{1350}}{10^{44}\,{\rm erg\ s^{-1}}}},
\end{equation}
where $\tau$, $\beta_{\rm\textsc{c}}$, and $\gamma_{\rm\textsc{c}}$ are the \civ\ time-lag, the intercept parameter, and the slope parameter, respectively, and the monochromatic luminosity at 1350\,\AA\
\be
\label{eq:L1350}
    L_{1350}=4\pi D_L^2F_{1350},
\ee
with measured quasar flux $F_{1350}$ at 1350\,\AA\ in units of $\rm erg\ s^{-1}\ cm^{-2}$. The luminosity distance in equation \eqref{eq:DLC8} is a function of redshift $z$ and the cosmological parameters.

The \mq\ $R-L$ relation is
\begin{equation}
    \label{eq:mq}
    \log{\frac{\tau^{\prime}}{\rm days}}=\beta_{\rm\textsc{m}}+\gamma_{\rm\textsc{m}} \log{\frac{L_{3000}}{10^{44}\,{\rm erg\ s^{-1}}}},
\end{equation}
where $\tau^{\prime}$, $\beta_{\rm\textsc{m}}$, and $\gamma_{\rm\textsc{m}}$ are the \mii\ time-lag, the intercept parameter, and the slope parameter, respectively, and the monochromatic luminosity at 3000\,\AA
\be
\label{eq:L3000}
    L_{3000}=4\pi D_L^2F_{3000},
\ee
with measured quasar flux $F_{3000}$ at 3000\,\AA\ in units of $\rm erg\ s^{-1}\ cm^{-2}$.

The natural log of the \civ\ likelihood function \citep{D'Agostini_2005} is
\be
\label{eq:LH_civ}
    \ln\mathcal{L}_{\rm C\,\textsc{iv}}= -\frac{1}{2}\Bigg[\chi^2_{\rm C\,\textsc{iv}}+\sum^{N}_{i=1}\ln\left(2\pi\sigma^2_{\mathrm{tot,\textsc{c}},i}\right)\Bigg],
\ee
where
\be
\label{eq:chi2_civ}
    \chi^2_{\rm C\,\textsc{iv}} = \sum^{N}_{i=1}\bigg[\frac{(\log \tau_{\mathrm{obs},i} - \beta_{\rm\textsc{c}}  -\gamma_{\rm\textsc{c}}\log L_{1350,i})^2}{\sigma^2_{\mathrm{tot,\textsc{c}},i}}\bigg]
\ee
with total uncertainty
\be
\label{eq:sigma_civ}
\sigma^2_{\mathrm{tot,\textsc{c}},i}=\sigma_{\rm int,\,\textsc{c}}^2+\sigma_{{\log \tau_{\mathrm{obs},i}}}^2+\gamma_{\rm\textsc{c}}^2\sigma_{{\log F_{1350,i}}}^2,
\ee
where $\sigma_{\rm int,\,\textsc{c}}$ is the \cq\ intrinsic scatter parameter which also contains the unknown systematic uncertainty, and $N$ is the number of data points.

The natural log of the \mii\ likelihood function is
\be
\label{eq:LH_mq}
    \ln\mathcal{L}_{\rm Mg\,\textsc{ii}}= -\frac{1}{2}\Bigg[\chi^2_{\rm Mg\,\textsc{ii}}+\sum^{N}_{i=1}\ln\left(2\pi\sigma^2_{\mathrm{tot,\textsc{m}},i}\right)\Bigg],
\ee
where
\be
\label{eq:chi2_mq}
    \chi^2_{\rm Mg\,\textsc{ii}} = \sum^{N}_{i=1}\bigg[\frac{(\log \tau^{\prime}_{\mathrm{obs},i} - \beta_{\rm\textsc{m}}  -\gamma_{\rm\textsc{m}}\log L_{3000,i})^2}{\sigma^2_{\mathrm{tot,\textsc{m}},i}}\bigg]
\ee
with total uncertainty
\be
\label{eq:sigma_mq}
\sigma^2_{\mathrm{tot,\textsc{m}},i}=\sigma_{\rm int,\,\textsc{m}}^2+\sigma_{{\log \tau^{\prime}_{\mathrm{obs},i}}}^2+\gamma_{\rm\textsc{m}}^2\sigma_{{\log F_{3000,i}}}^2,
\ee
where $\sigma_{\rm int,\,\textsc{m}}$ is the \mq\ intrinsic scatter parameter which also contains the unknown systematic uncertainty.

The $\tau$ error bars are typically asymmetric. As used for \mq{s} in \cite{Khadkaetal_2021a} and \cite{CaoRatra2022}, in what follows, $\tau$ symmetrized errors mean that we are using symmetrized $\sigma_{\tau}$'s defined as $\sigma_{\tau}=[2\sigma_{\tau,+}\sigma_{\tau,-}/(\sigma_{\tau,+}+\sigma_{\tau,-})+\sqrt{\sigma_{\tau,+}\sigma_{\tau,-}}]/2$, where $\sigma_{\tau,+}$ and $\sigma_{\tau,-}$ are the upper and lower errors of $\tau$, respectively. On the other hand, $\tau$ asymmetric errors mean that we directly use asymmetric $\sigma_{\tau,+}$ and $\sigma_{\tau,-}$ as follows: when the theoretical prediction for $\log{\tau}$ is larger (smaller) than the observed value, $\sigma_{\tau}=\sigma_{\tau,+}$ ($\sigma_{\tau}=\sigma_{\tau,-}$).

The detailed descriptions for the likelihood functions of $H(z)$ and BAO data can be found in \cite{CaoRyanRatra2020}. 

\begin{table}
\centering
\setlength{\tabcolsep}{3.5pt}
\begin{threeparttable}
\caption{Flat priors of the constrained parameters.}
\label{tab:priorsC11}
\begin{tabular}{lcc}
\toprule
Parameter & & Prior\\
\midrule
 & Cosmological-Model Parameters & \\
\midrule
$H_0$\,\tnote{a} &  & [None, None]\\
\obhs\,\tnote{b} &  & [0, 1]\\
\ochs\,\tnote{c} &  & [0, 1]\\
\ok &  & [-2, 2]\\
$\alpha$ &  & [0, 10]\\
\wx &  & [-5, 0.33]\\
\midrule
 & $R-L$ Relation Parameters & \\
\midrule
$\gamma$ &  & [0, 5]\\
$\beta$ &  & [0, 10]\\
$\sigma_{\rm int}$ &  & [0, 5]\\
\bottomrule
\end{tabular}
\begin{tablenotes}[flushleft]
\item [a] \hunit. In the \cq\ and \mq\ alone cases, $H_0$ is set to be 70 \hunit, while in other cases, the prior range is irrelevant (unbounded).
\item [b] In the \cq\ and \mq\ alone cases, \obhs\ is set to be 0.0245, i.e. $\Omega_{b}=0.05$.
\item [c] In the \cq\ and \mq\ alone cases, $\Om\in[0,1]$ is ensured.
\end{tablenotes}
\end{threeparttable}%
\end{table}

We list the flat priors of the free cosmological-model and $R-L$ relation parameters in Table \ref{tab:priorsC11}. By maximizing the likelihood functions, we obtain the unmarginalized best-fitting values and posterior distributions of all free cosmological-model and $R-L$ relation parameters. The Markov chain Monte Carlo (MCMC) code \textsc{MontePython} \citep{Audrenetal2013,Brinckmann2019}, the \textsc{class} code, and the \textsc{python} package \textsc{getdist} \citep{Lewis_2019} are used to perform our analyses.

One can find the definitions of the Akaike Information Criterion (AIC), the Bayesian Information Criterion (BIC), and the Deviance Information Criterion (DIC) in our previous paper \citep[see, e.g.][]{CaoDainottiRatra2022}. $\Delta \mathrm{AIC}$, $\Delta \mathrm{BIC}$, and $\Delta \mathrm{DIC}$ are computed as the differences between the AIC, BIC, and DIC values of the other five cosmological dark energy models and those of the flat \lcdm\ reference model. Positive (negative) values of these $\Delta \mathrm{IC}$s show that the model under investigation fits the data worse (better) than does the flat \lcdm\ reference model. In comparison with the model with the minimum IC, $\Delta \mathrm{IC} \in (0, 2]$ indicates weak evidence against the model under investigation, $\Delta \mathrm{IC} \in (2, 6]$ indicates positive evidence against the model under investigation, $\Delta \mathrm{IC} \in (6, 10] $ indicates strong evidence against the model under investigation, and $\Delta \mathrm{IC}>10$ indicates very strong evidence against the model under investigation.

We assume one massive and two massless neutrino species with the non-relativistic neutrino physical energy density parameter $\onh=\sum m_{\nu}/(93.14\ \rm eV)=0.06\ \rm eV/(93.14\ \rm eV)$, where $h$ is the Hubble constant $H_0$ in units of 100 \hunit. The non-relativistic matter density parameter $\Om = (\onh + \obh + \och)/{h^2}$, where \obhs\ and \ochs\ are the current values of the observationally-constrained baryonic and cold dark matter physical energy density parameters, respectively.\footnote{In the \cq\ and \mq\ alone cases, the current value of the baryonic matter energy density parameter and the Hubble constant are set to $\Omega_b=0.05$ and $H_0=70$ \hunit, respectively, because these data alone are unable to constrain $\Omega_b$ and $H_0$.} Including neutrino species is more accurate even though it only has a mild effect on the constraints from \mii\ and \civ\ data.

\section{Results}
\label{makereference11.4}

\begin{figure*}
\centering
 \subfloat[]{%
    \includegraphics[width=0.33\textwidth,height=0.4\textwidth]{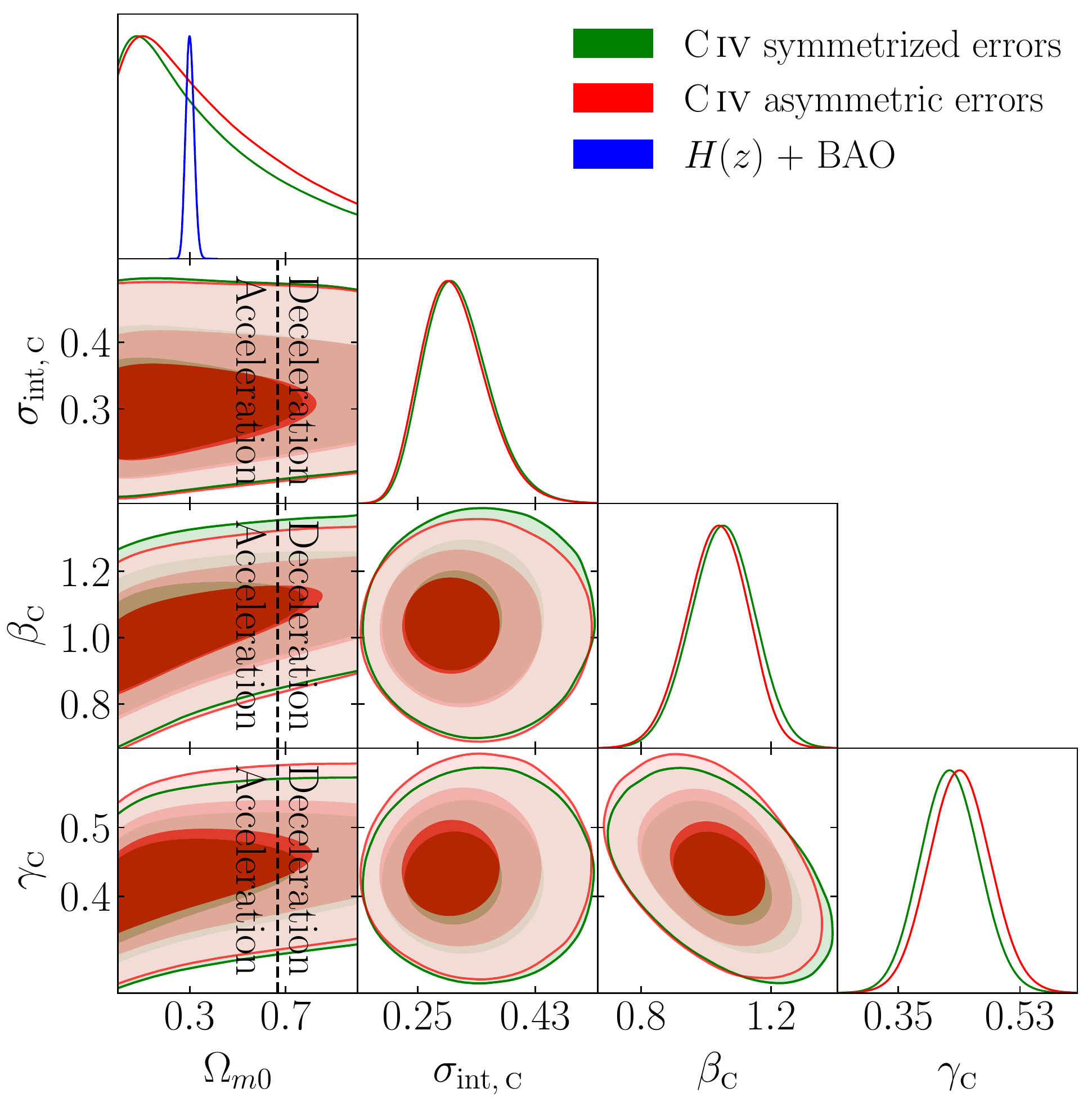}}
 \subfloat[]{%
    \includegraphics[width=0.33\textwidth,height=0.4\textwidth]{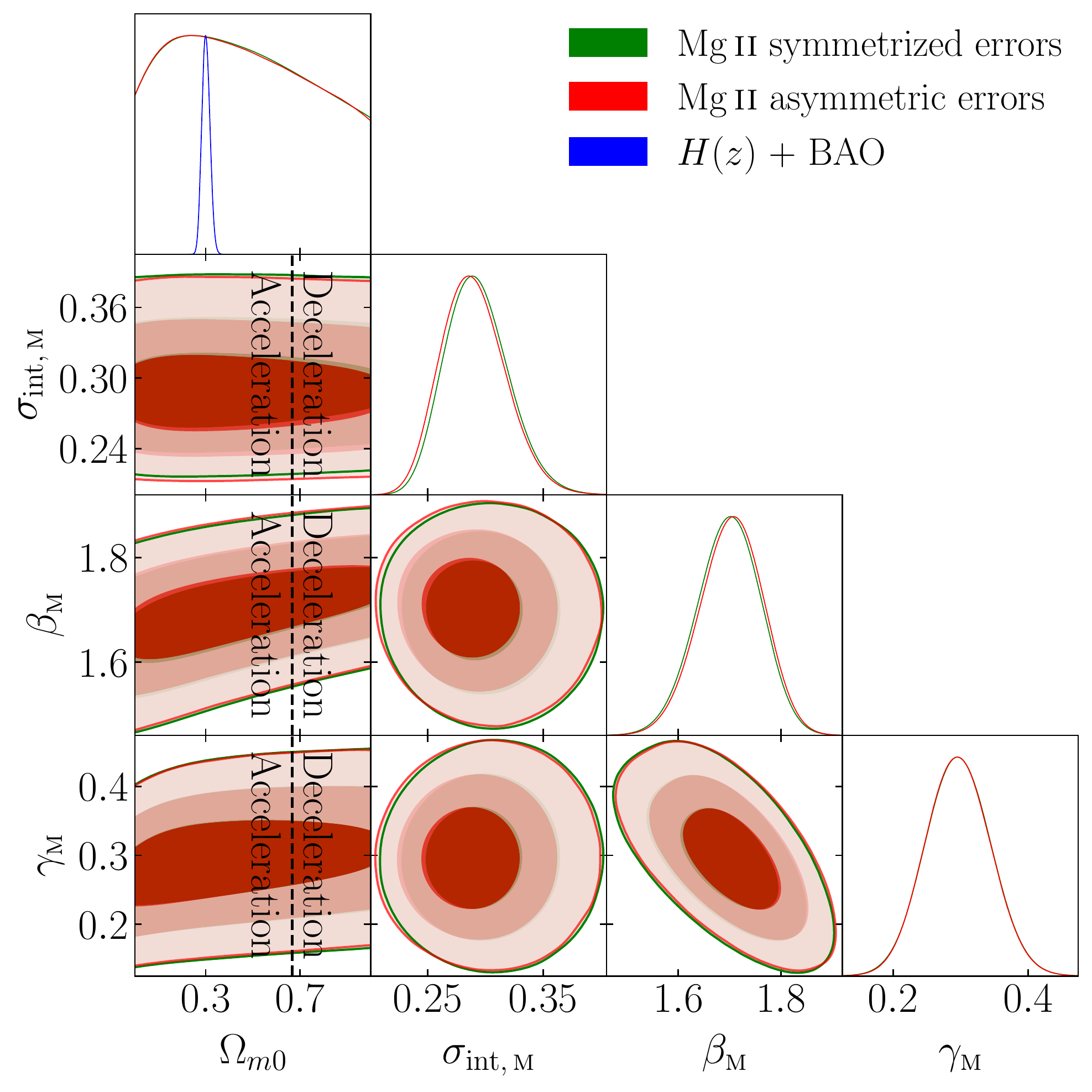}}
 \subfloat[]{%
    \includegraphics[width=0.33\textwidth,height=0.4\textwidth]{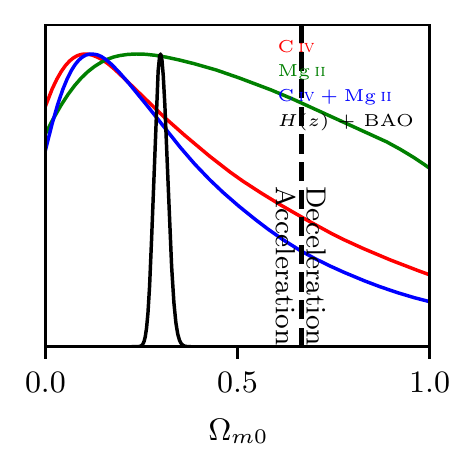}}\\
 \subfloat[]{%
    \includegraphics[width=0.5\textwidth,height=0.55\textwidth]{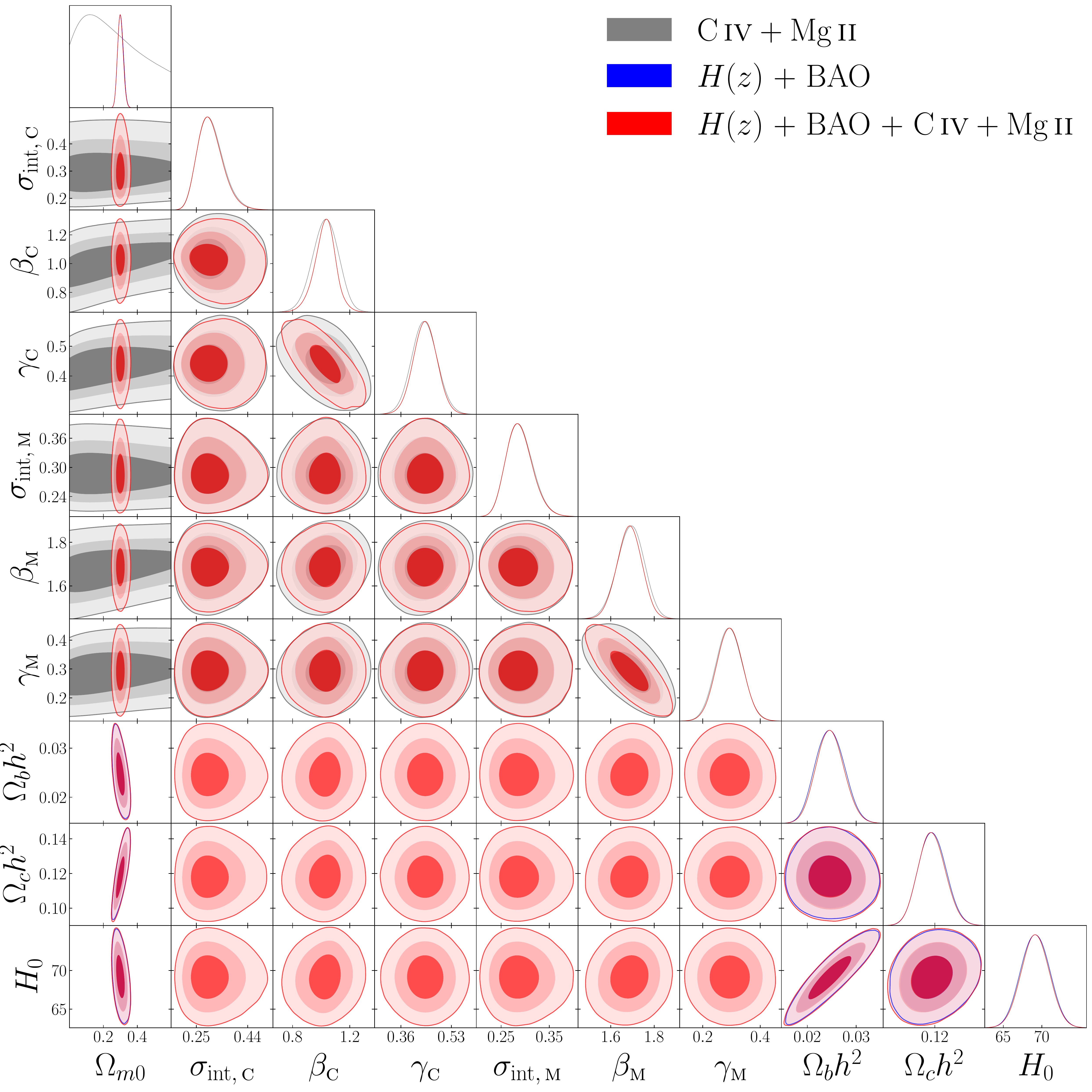}}
 \subfloat[]{%
    \includegraphics[width=0.5\textwidth,height=0.55\textwidth]{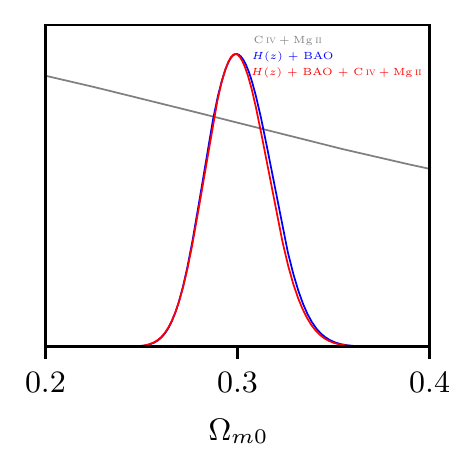}}\\
\caption{One-dimensional likelihood distributions and 1$\sigma$, 2$\sigma$, and 3$\sigma$ two-dimensional likelihood confidence contours for flat \lcdm\ from various combinations of data. The zero-acceleration black dashed lines in panels (a) and (b) divide the parameter space into regions associated with currently-accelerating (left) and currently-decelerating (right) cosmological expansion.}
\label{fig1C11}
\end{figure*}

\begin{figure*}
\centering
 \subfloat[]{%
    \includegraphics[width=0.33\textwidth,height=0.4\textwidth]{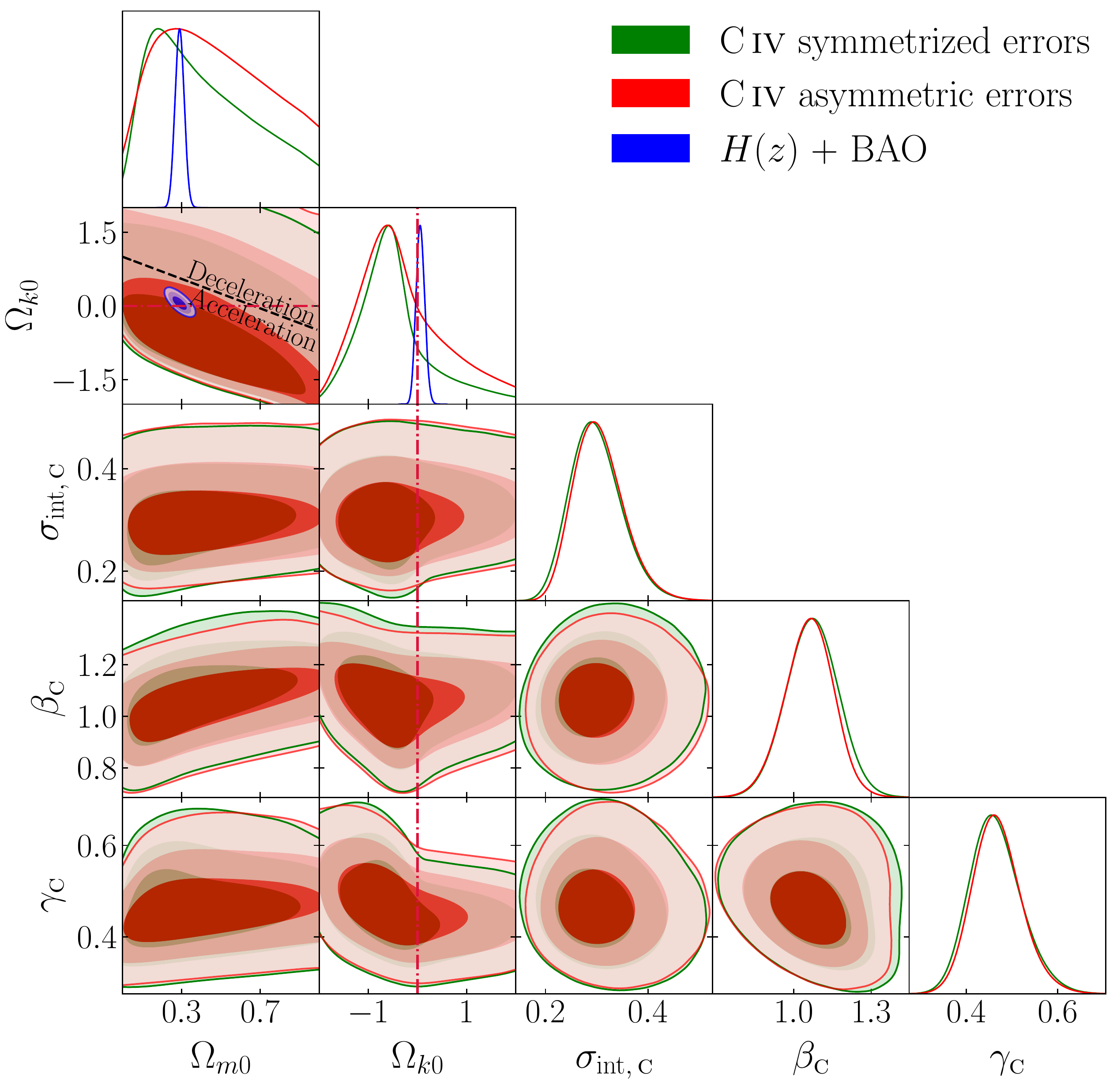}}
 \subfloat[]{%
    \includegraphics[width=0.33\textwidth,height=0.4\textwidth]{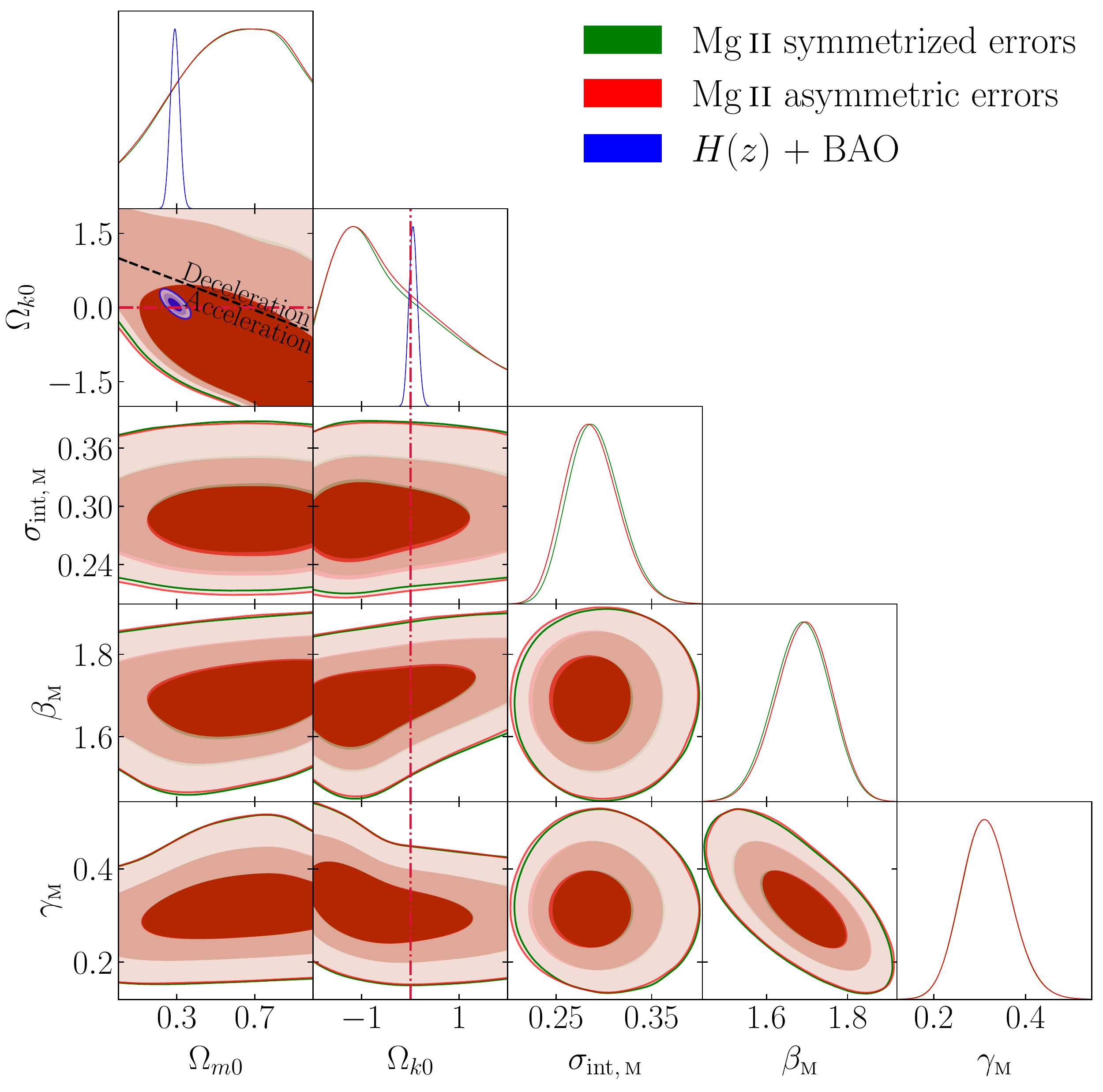}}
 \subfloat[]{%
    \includegraphics[width=0.33\textwidth,height=0.4\textwidth]{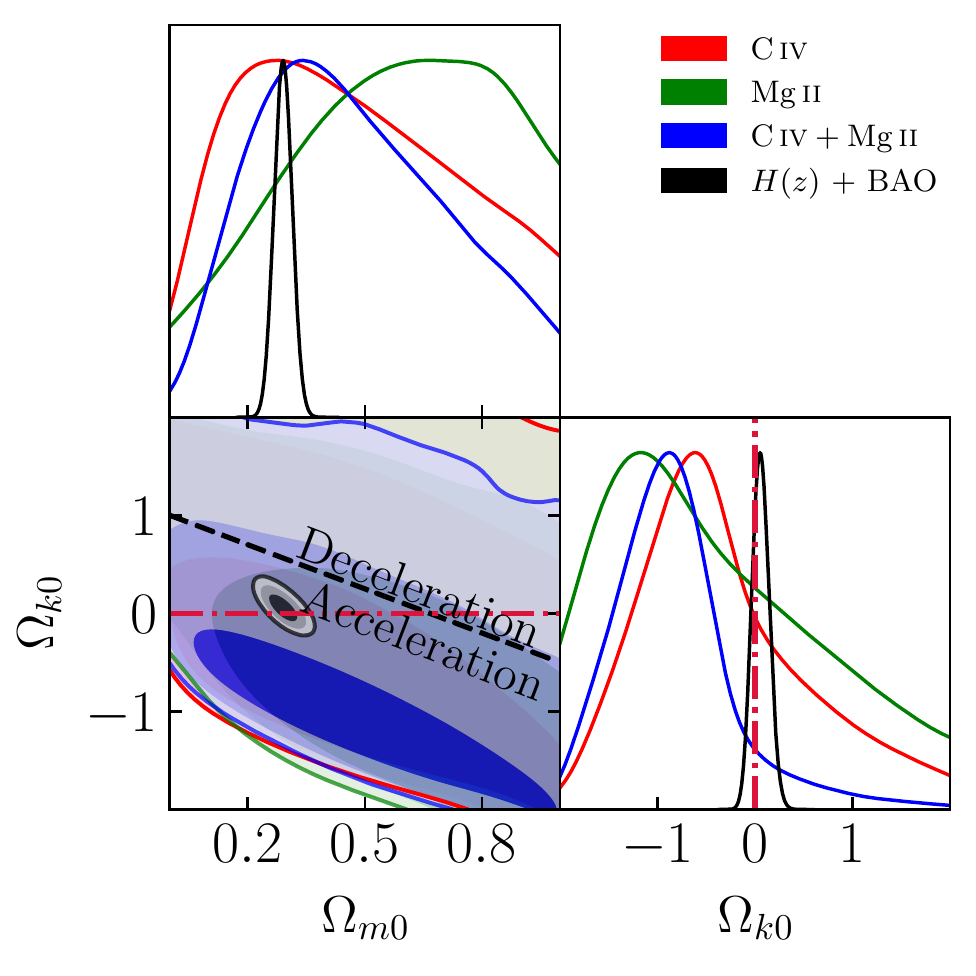}}\\
 \subfloat[]{%
    \includegraphics[width=0.5\textwidth,height=0.55\textwidth]{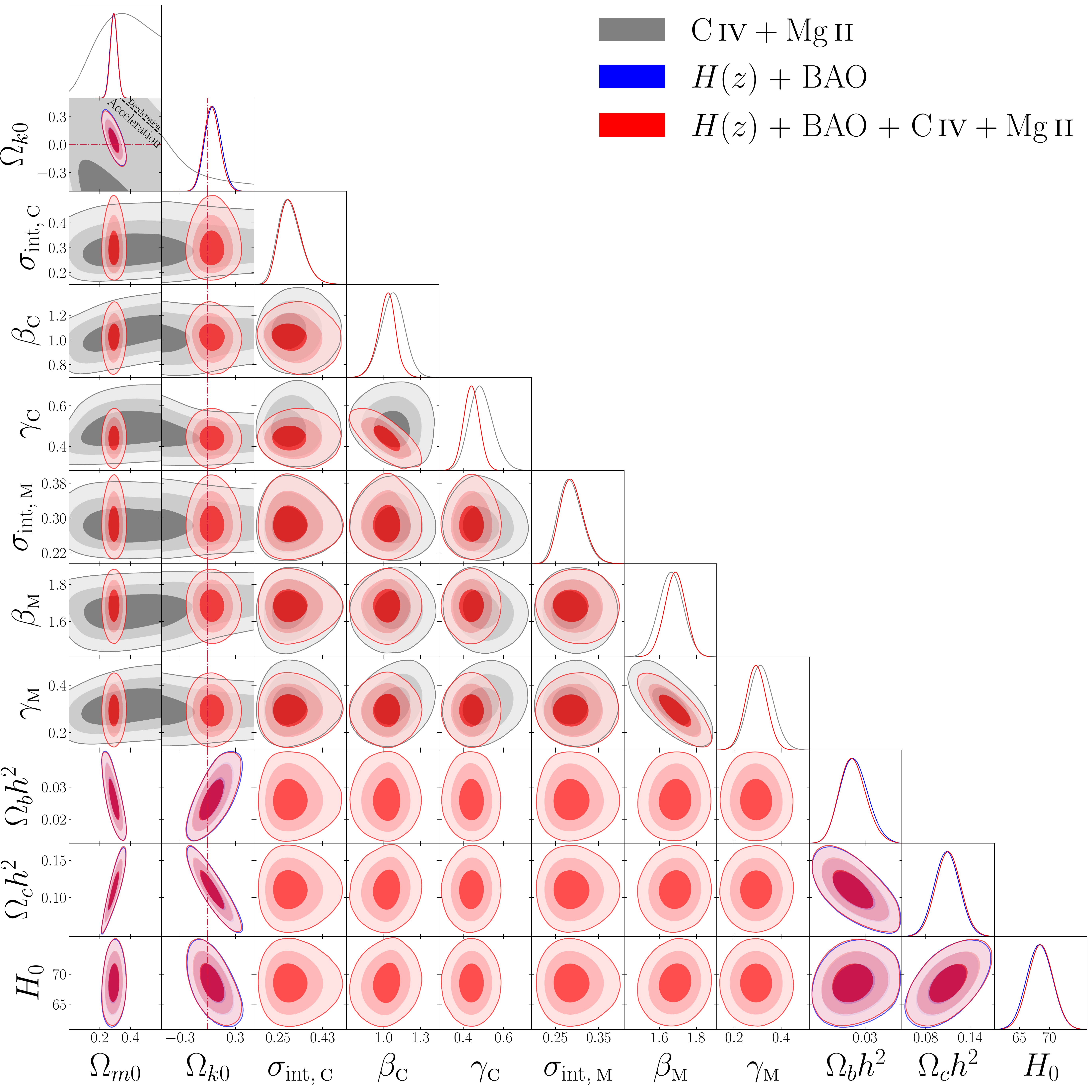}}
 \subfloat[]{%
    \includegraphics[width=0.5\textwidth,height=0.55\textwidth]{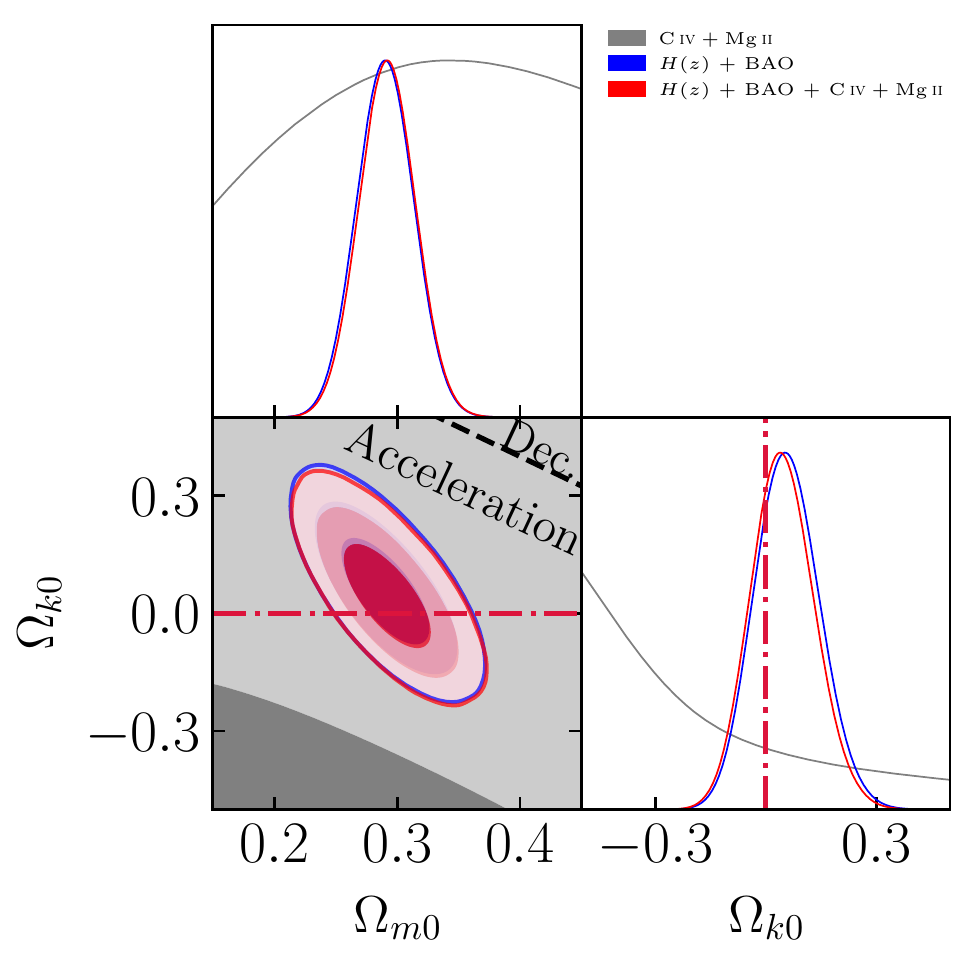}}\\
\caption{Same as Fig.\ \ref{fig1C11} but for non-flat \lcdm. The zero-acceleration black dashed lines divide the parameter space into regions associated with currently-accelerating (below left) and currently-decelerating (above right) cosmological expansion.}
\label{fig2C11}
\end{figure*}

\begin{figure*}
\centering
 \subfloat[]{%
    \includegraphics[width=0.33\textwidth,height=0.4\textwidth]{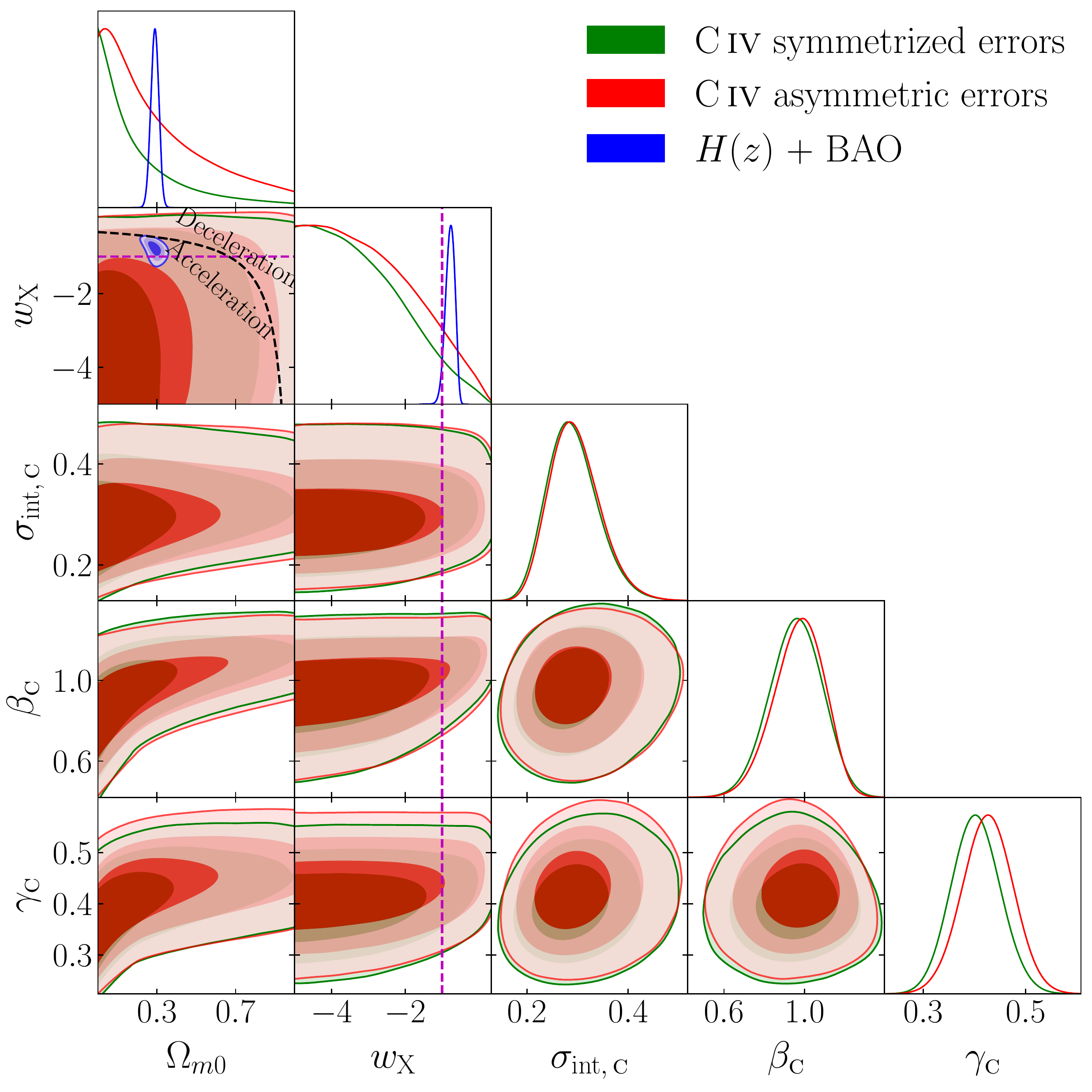}}
 \subfloat[]{%
    \includegraphics[width=0.33\textwidth,height=0.4\textwidth]{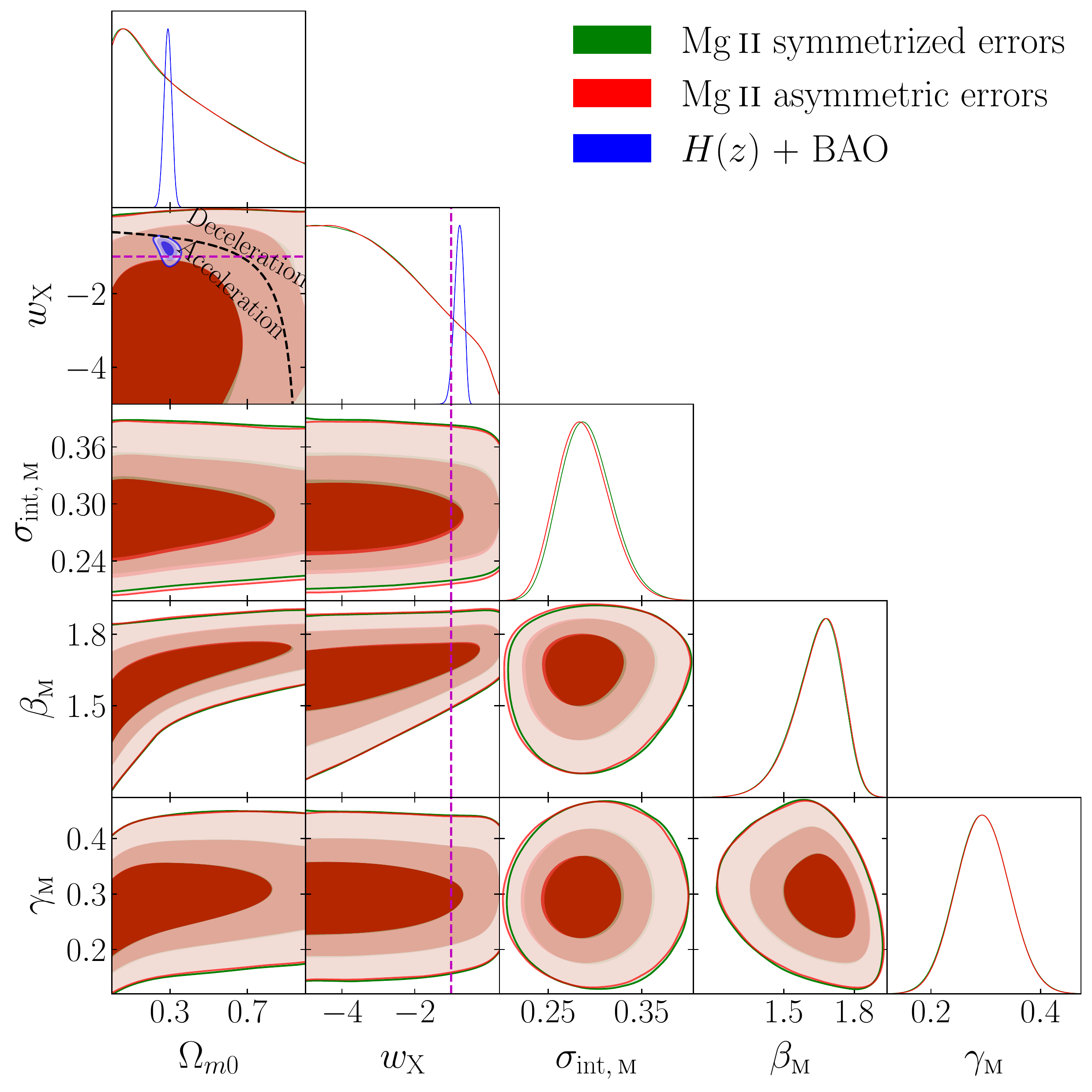}}
 \subfloat[]{%
    \includegraphics[width=0.33\textwidth,height=0.4\textwidth]{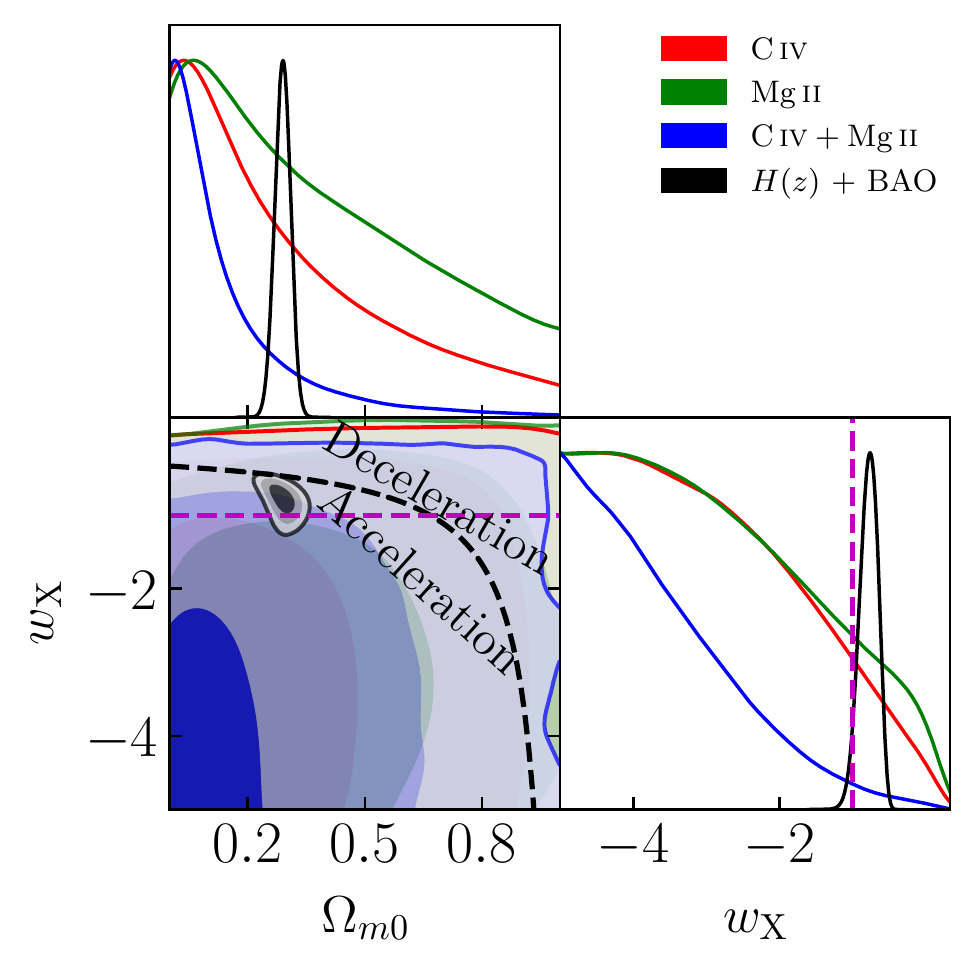}}\\
 \subfloat[]{%
    \includegraphics[width=0.5\textwidth,height=0.55\textwidth]{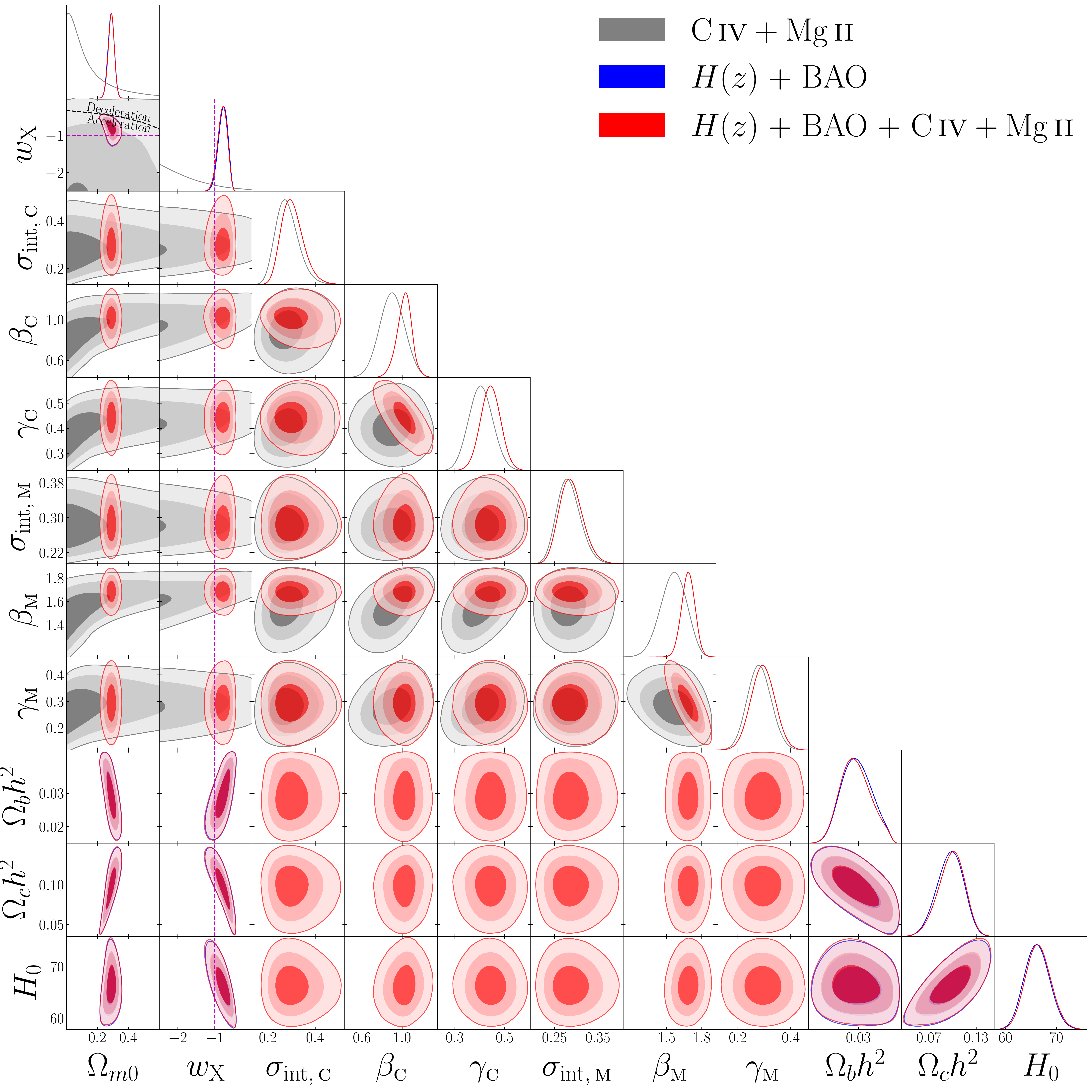}}
 \subfloat[]{%
    \includegraphics[width=0.5\textwidth,height=0.55\textwidth]{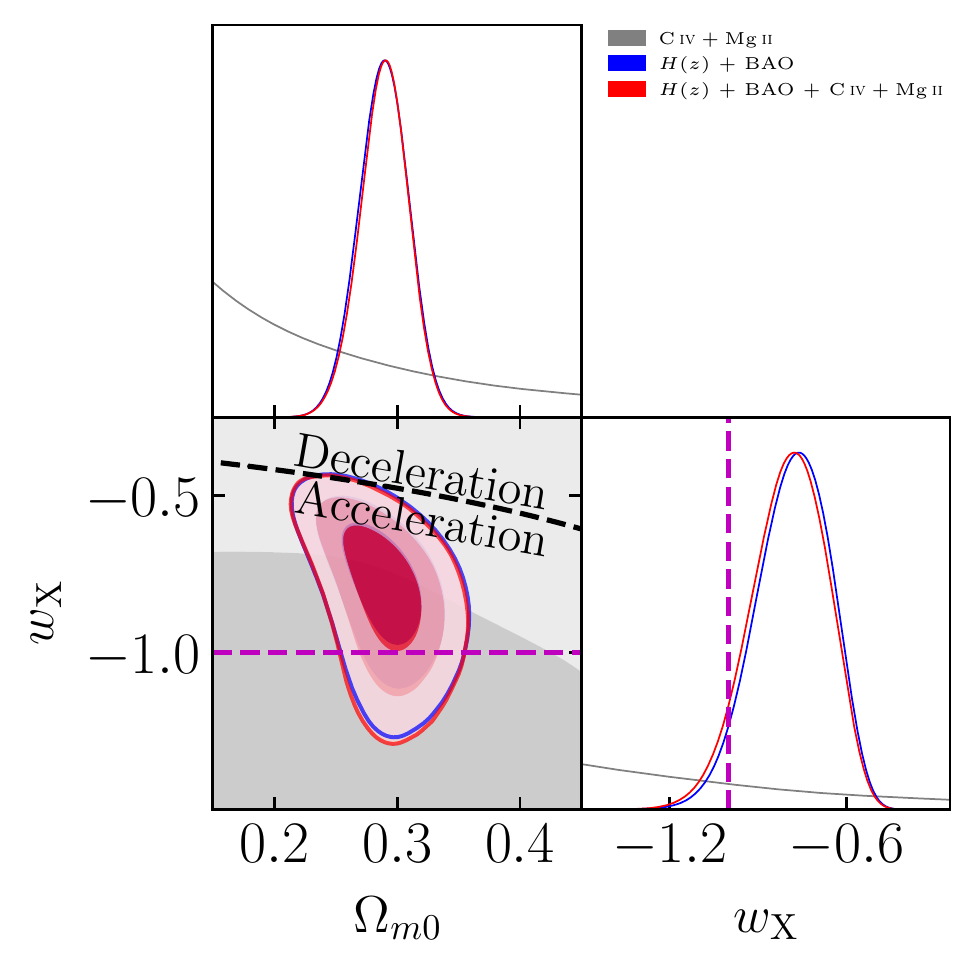}}\\
\caption{One-dimensional likelihood distributions and 1$\sigma$, 2$\sigma$, and 3$\sigma$ two-dimensional likelihood confidence contours for flat XCDM from various combinations of data. The zero-acceleration black dashed lines divide the parameter space into regions associated with currently-accelerating (either below left or below) and currently-decelerating (either above right or above) cosmological expansion. The magenta dashed lines represent $w_{\rm X}=-1$, i.e.\ flat \lcdm.}
\label{fig3C11}
\end{figure*}

\begin{figure*}
\centering
 \subfloat[]{%
    \includegraphics[width=0.33\textwidth,height=0.4\textwidth]{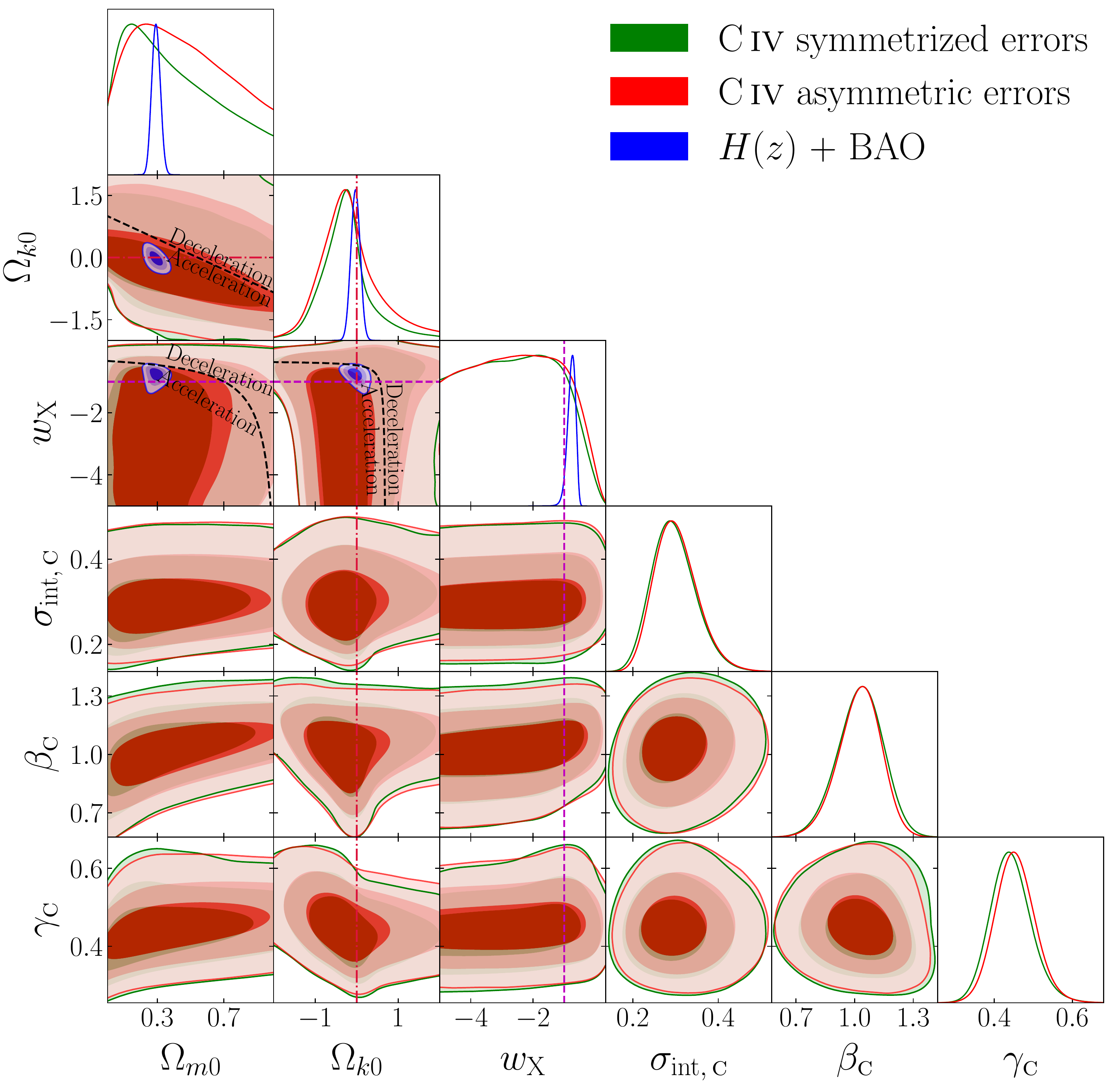}}
 \subfloat[]{%
    \includegraphics[width=0.33\textwidth,height=0.4\textwidth]{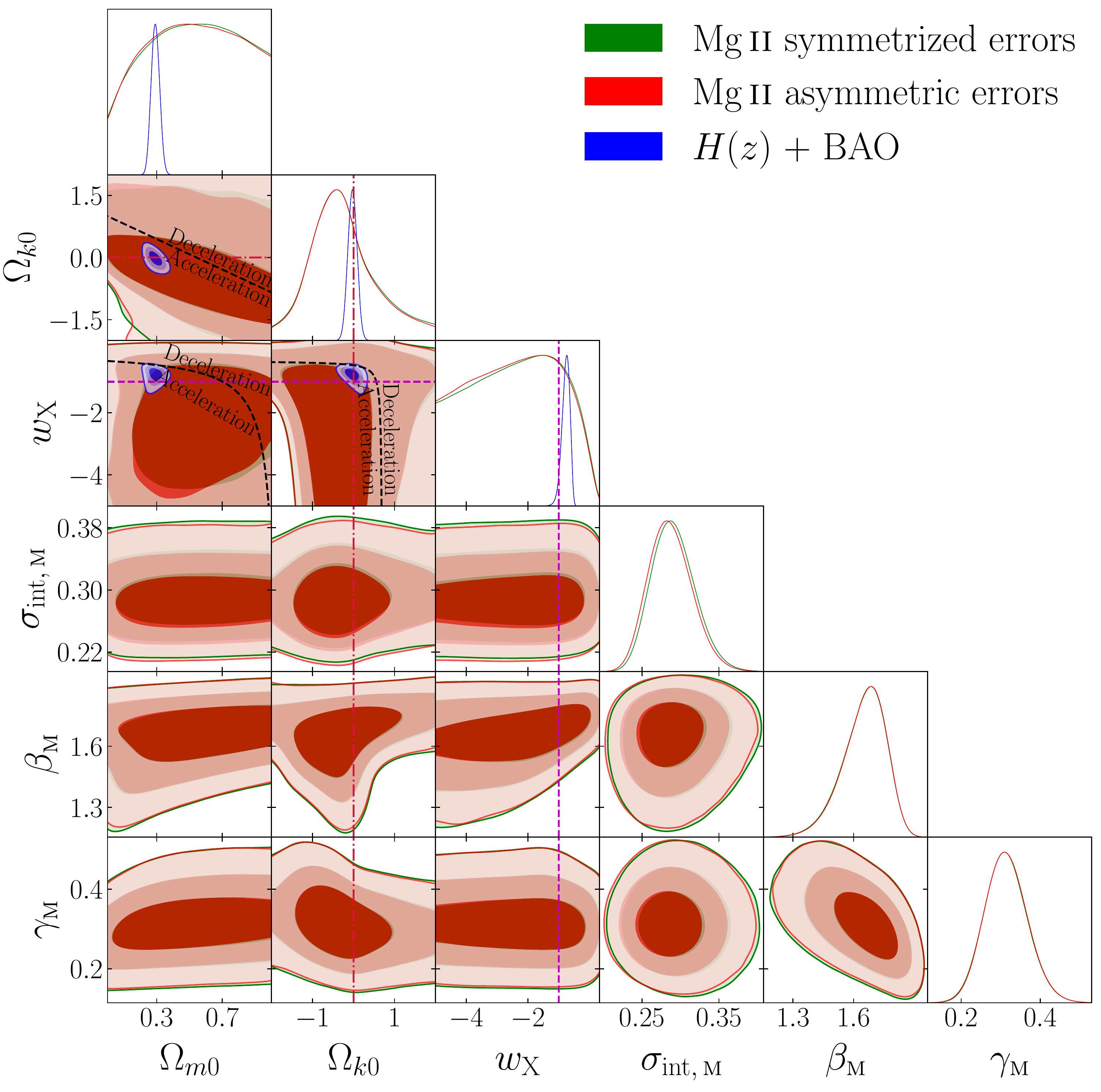}}
 \subfloat[]{%
    \includegraphics[width=0.33\textwidth,height=0.4\textwidth]{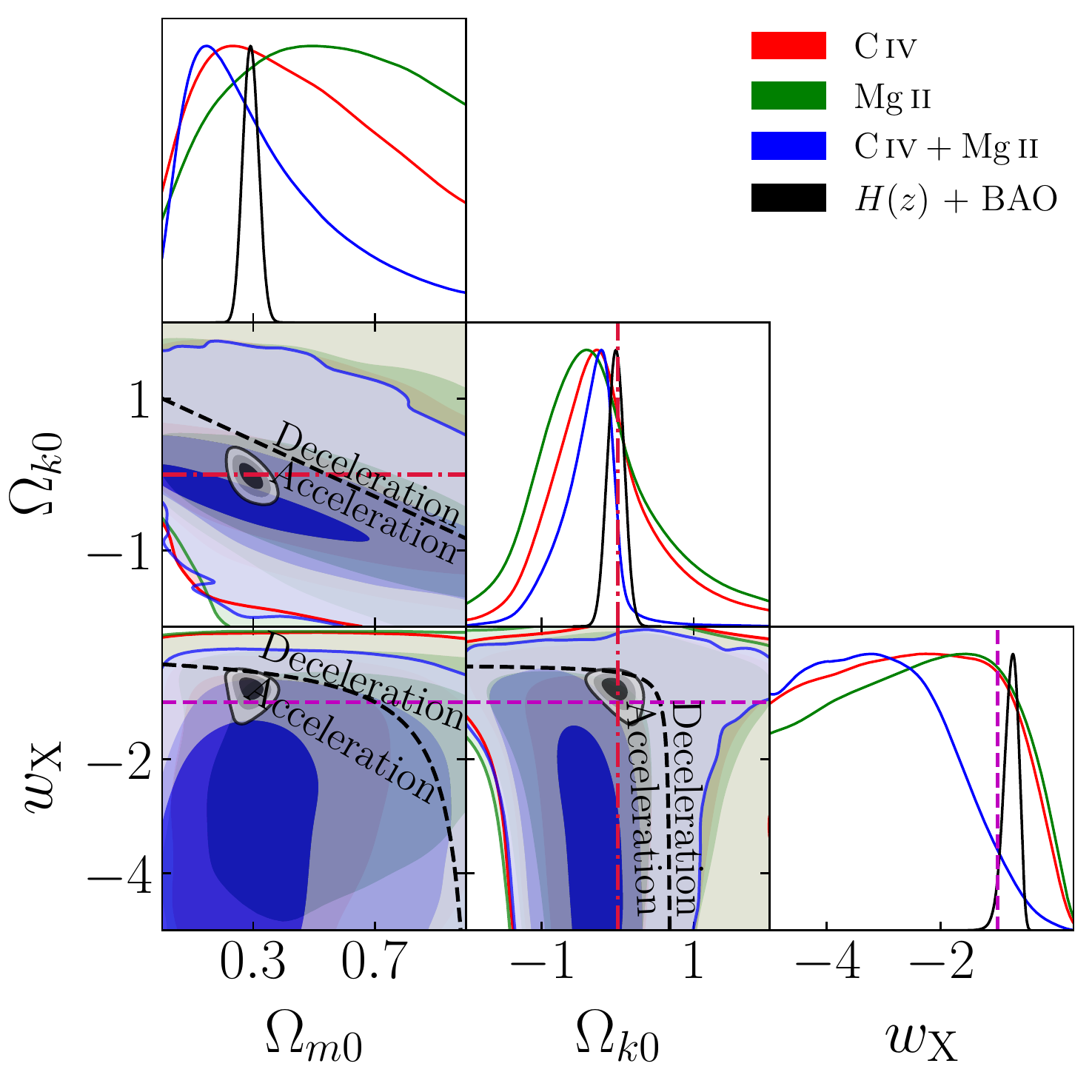}}\\
 \subfloat[]{%
    \includegraphics[width=0.5\textwidth,height=0.55\textwidth]{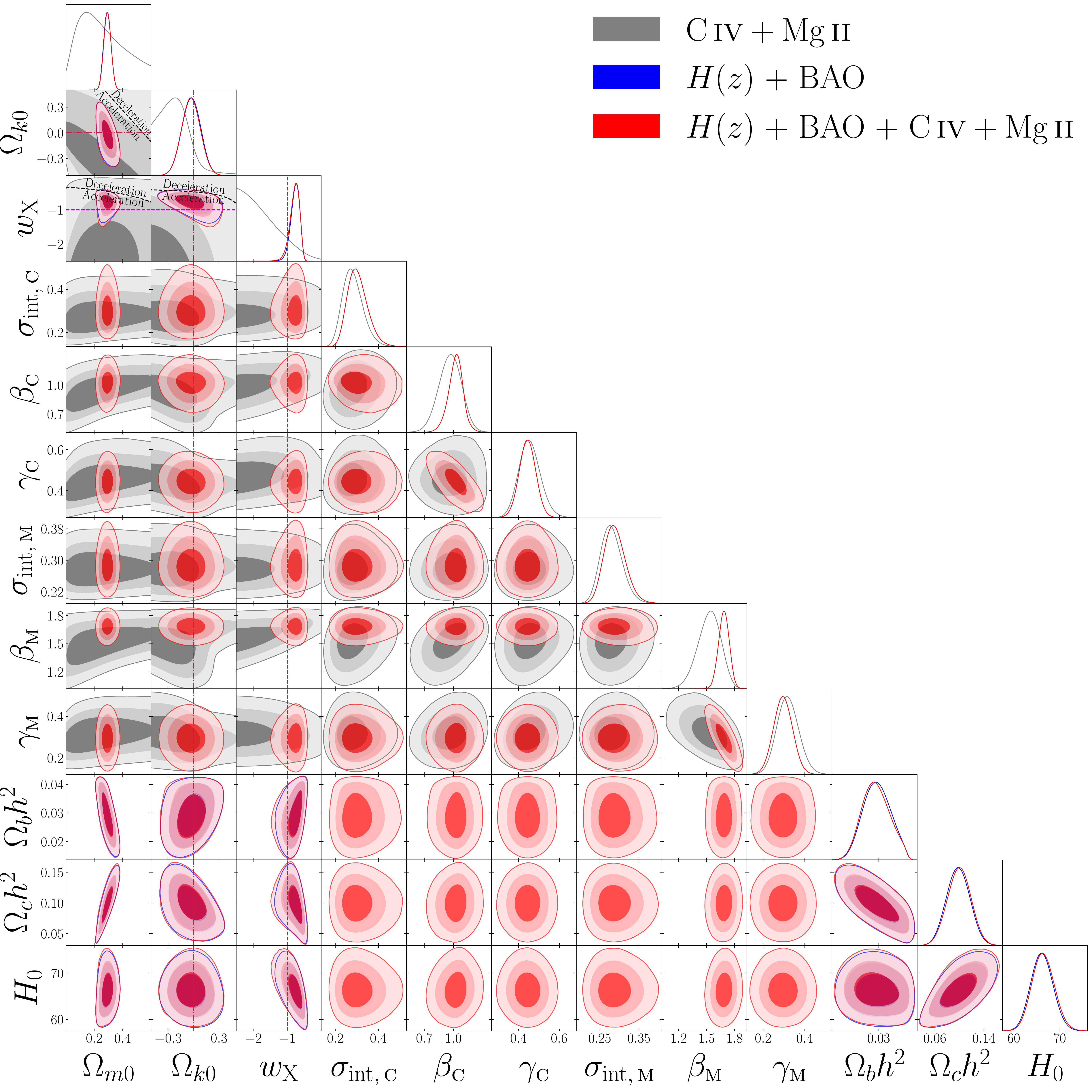}}
 \subfloat[]{%
    \includegraphics[width=0.5\textwidth,height=0.55\textwidth]{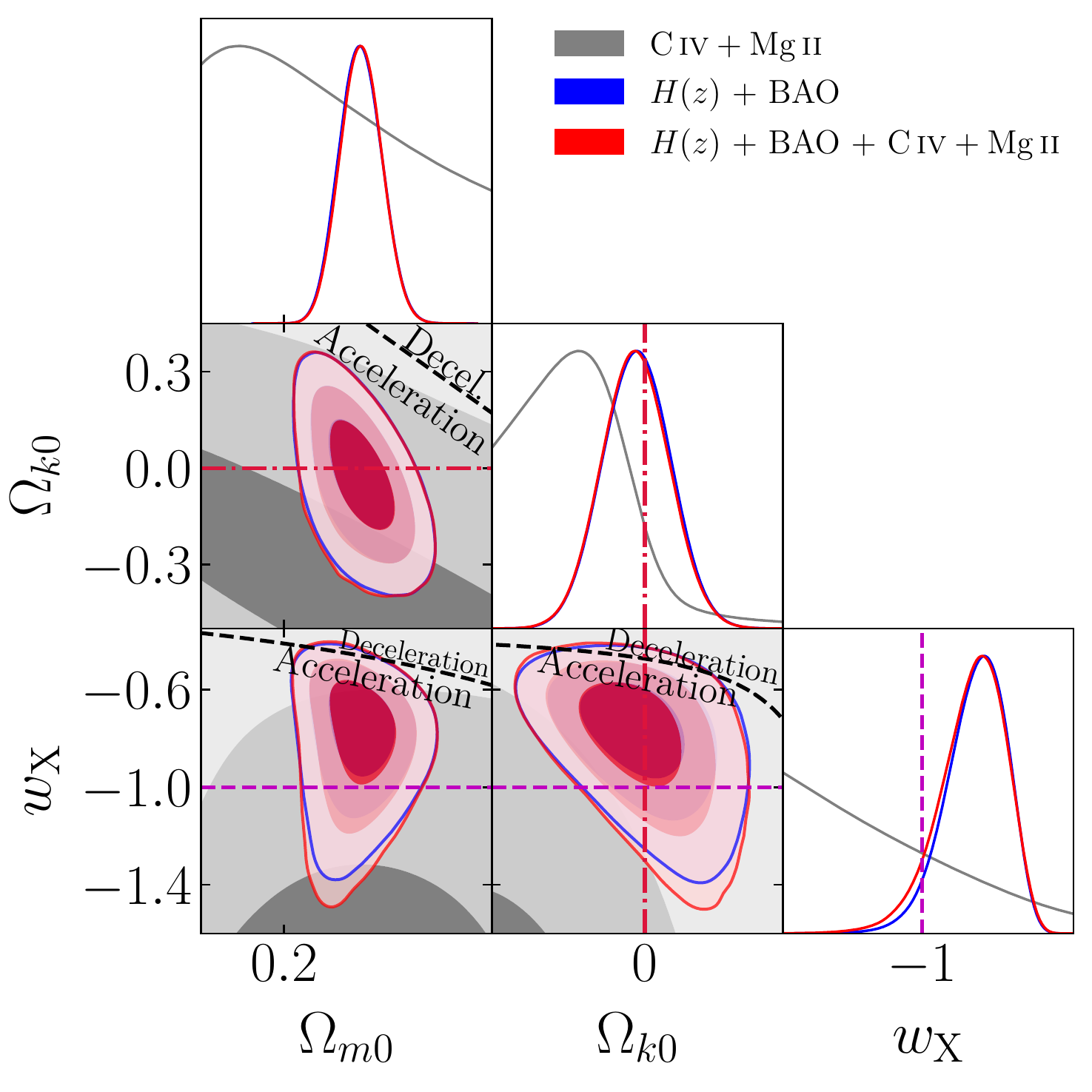}}\\
\caption{Same as Fig.\ \ref{fig3C11} but for non-flat XCDM. The zero-acceleration black dashed lines are computed for the third cosmological parameter set to the $H(z)$ + BAO data best-fitting values listed in Table \ref{tab:BFPC11}, and divide the parameter space into regions associated with currently-accelerating (either below left or below) and currently-decelerating (either above right or above) cosmological expansion. The crimson dash-dot lines represent flat hypersurfaces, with closed spatial hypersurfaces either below or to the left. The magenta dashed lines represent $w_{\rm X}=-1$, i.e.\ non-flat \lcdm.}
\label{fig4C11}
\end{figure*}

\begin{figure*}
\centering
\centering
 \subfloat[]{%
    \includegraphics[width=0.33\textwidth,height=0.4\textwidth]{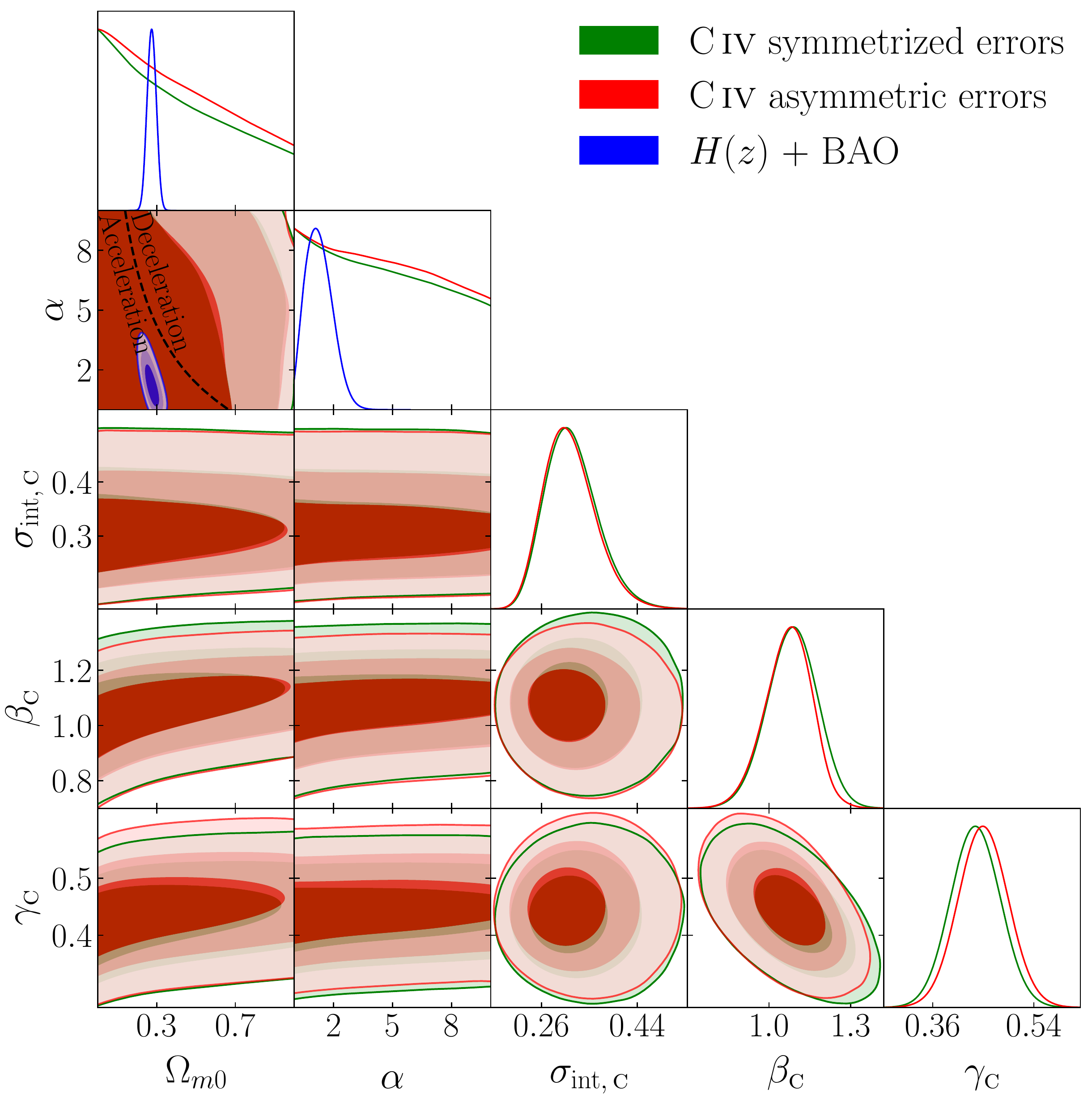}}
 \subfloat[]{%
    \includegraphics[width=0.33\textwidth,height=0.4\textwidth]{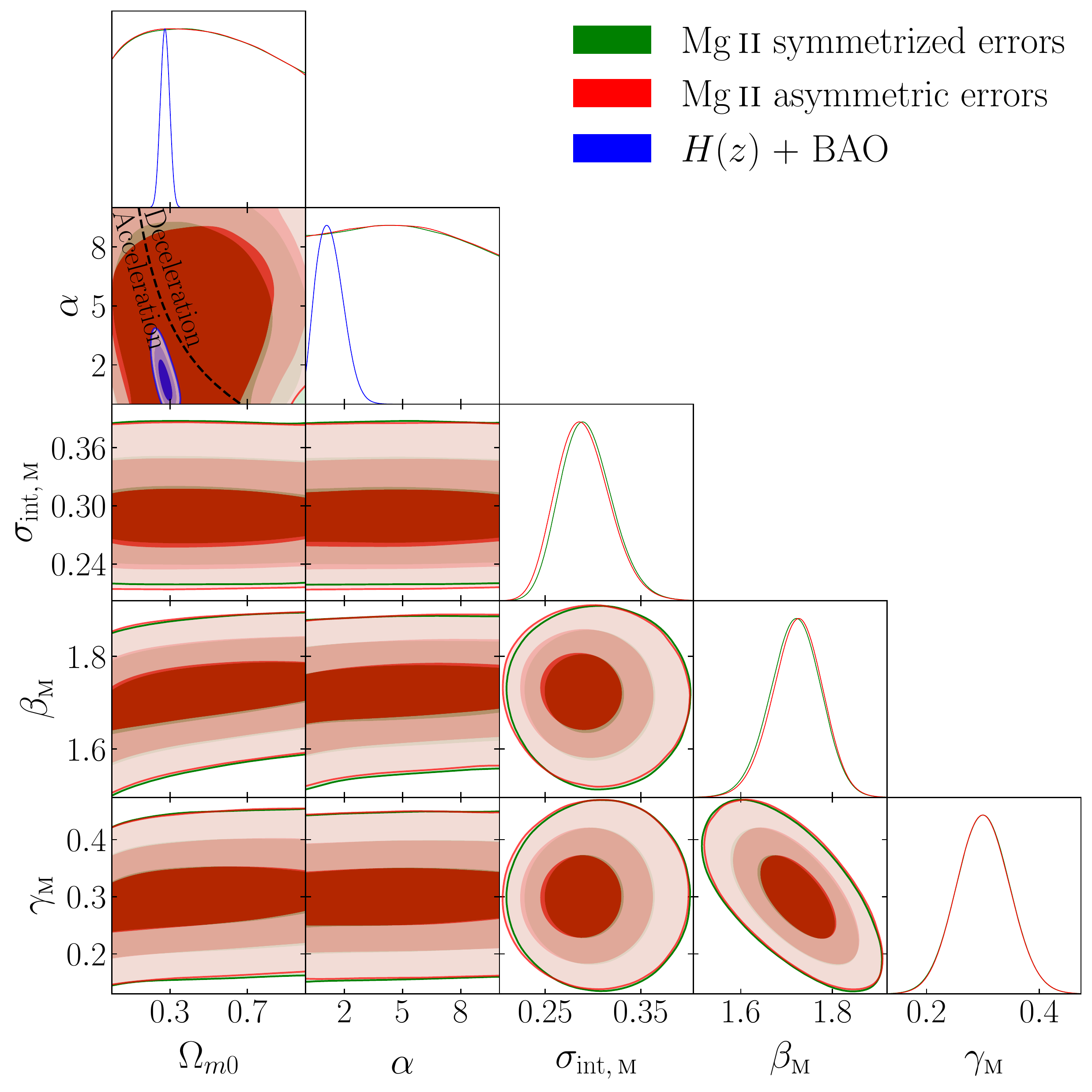}}
 \subfloat[]{%
    \includegraphics[width=0.33\textwidth,height=0.4\textwidth]{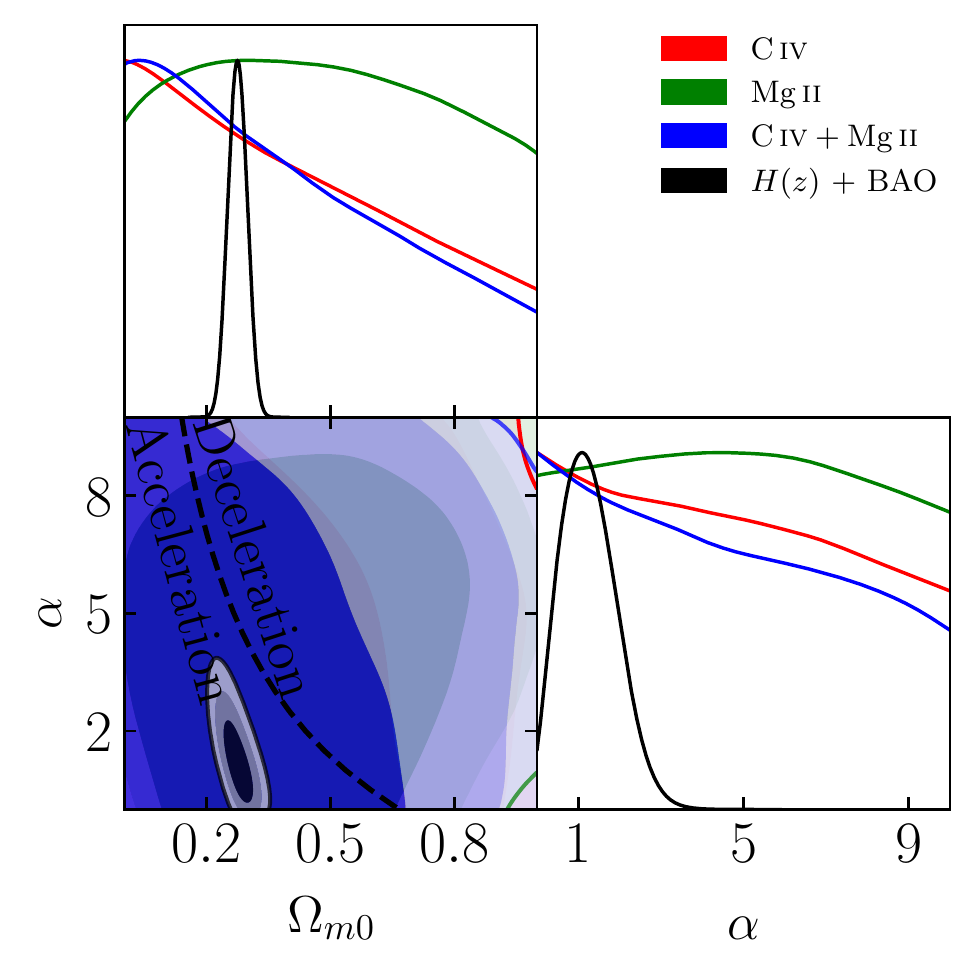}}\\
 \subfloat[]{%
    \includegraphics[width=0.5\textwidth,height=0.55\textwidth]{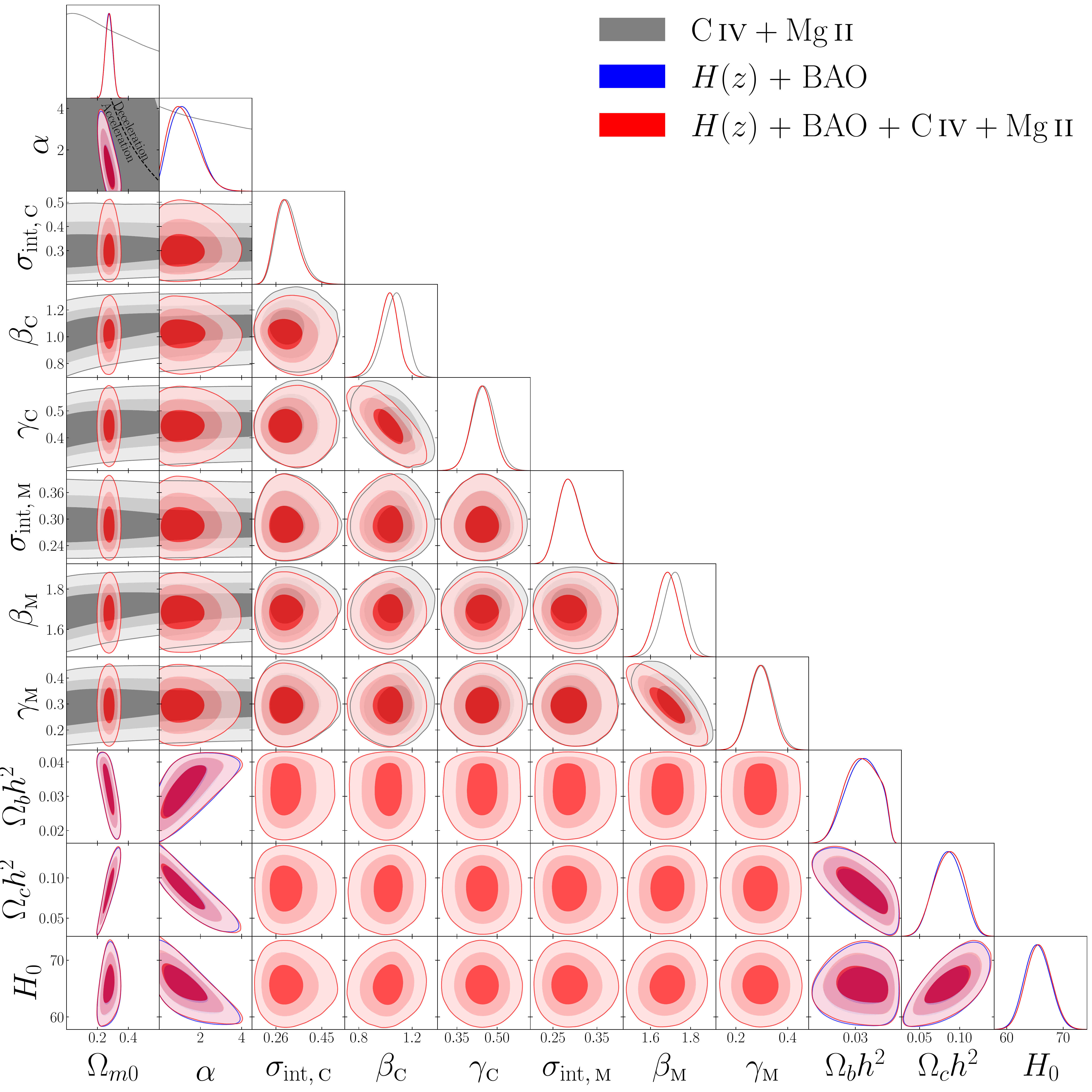}}
 \subfloat[]{%
    \includegraphics[width=0.5\textwidth,height=0.55\textwidth]{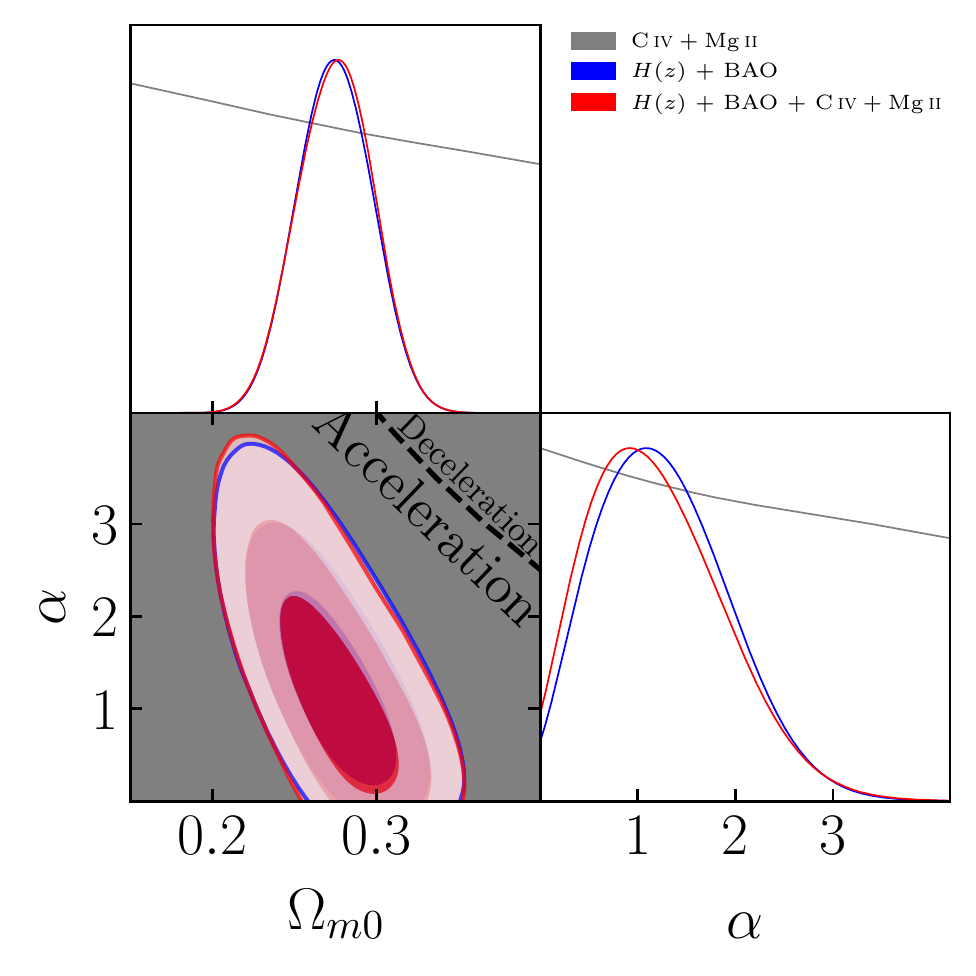}}\\
\caption{One-dimensional likelihood distributions and 1$\sigma$, 2$\sigma$, and 3$\sigma$ two-dimensional likelihood confidence contours for flat \pcdm\ from various combinations of data. The zero-acceleration black dashed lines divide the parameter space into regions associated with currently-accelerating (below left) and currently-decelerating (above right) cosmological expansion. The $\alpha = 0$ axes correspond to flat \lcdm.}
\label{fig5C11}
\end{figure*}

\begin{figure*}
\centering
 \subfloat[]{%
    \includegraphics[width=0.33\textwidth,height=0.4\textwidth]{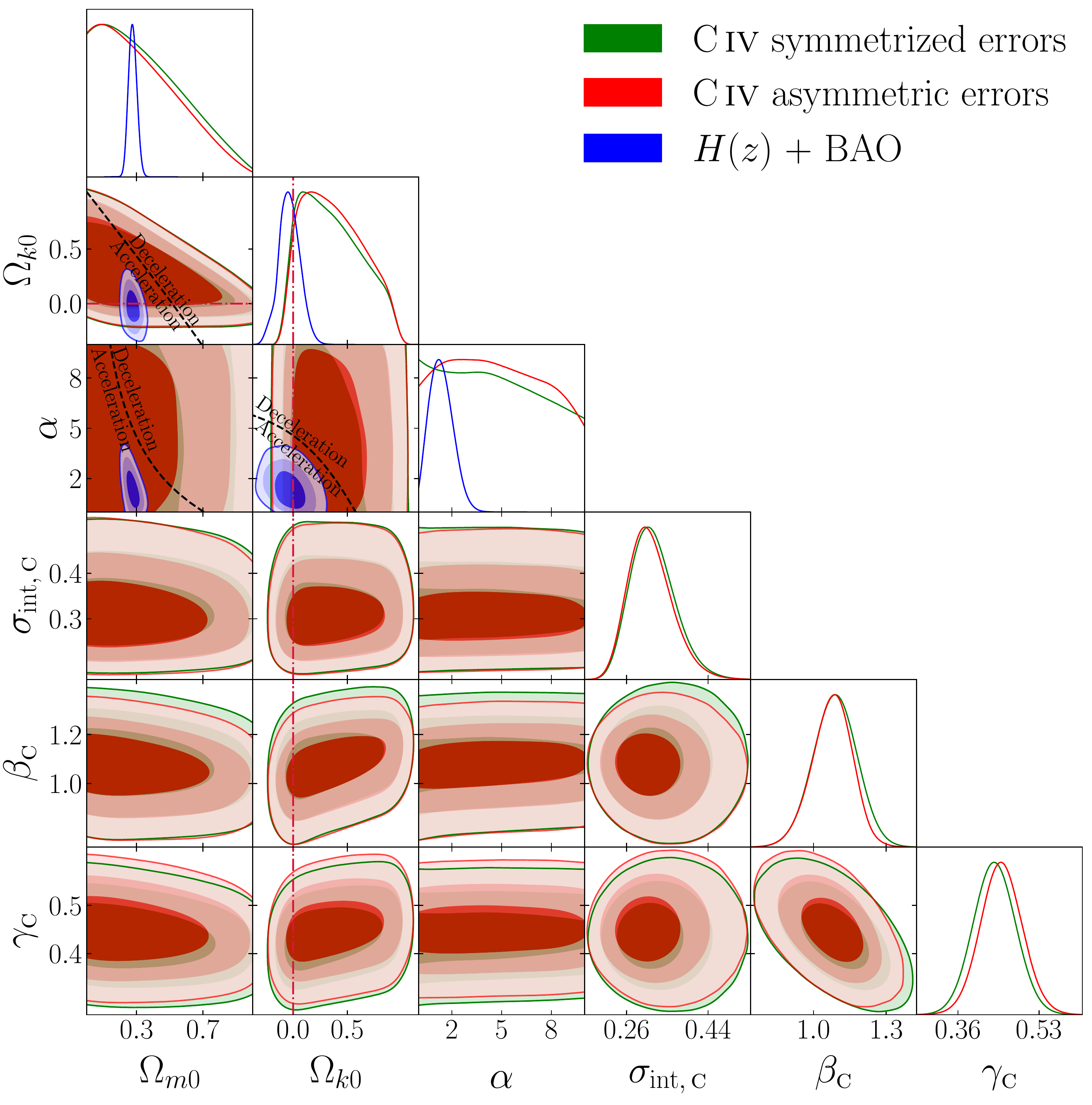}}
 \subfloat[]{%
    \includegraphics[width=0.33\textwidth,height=0.4\textwidth]{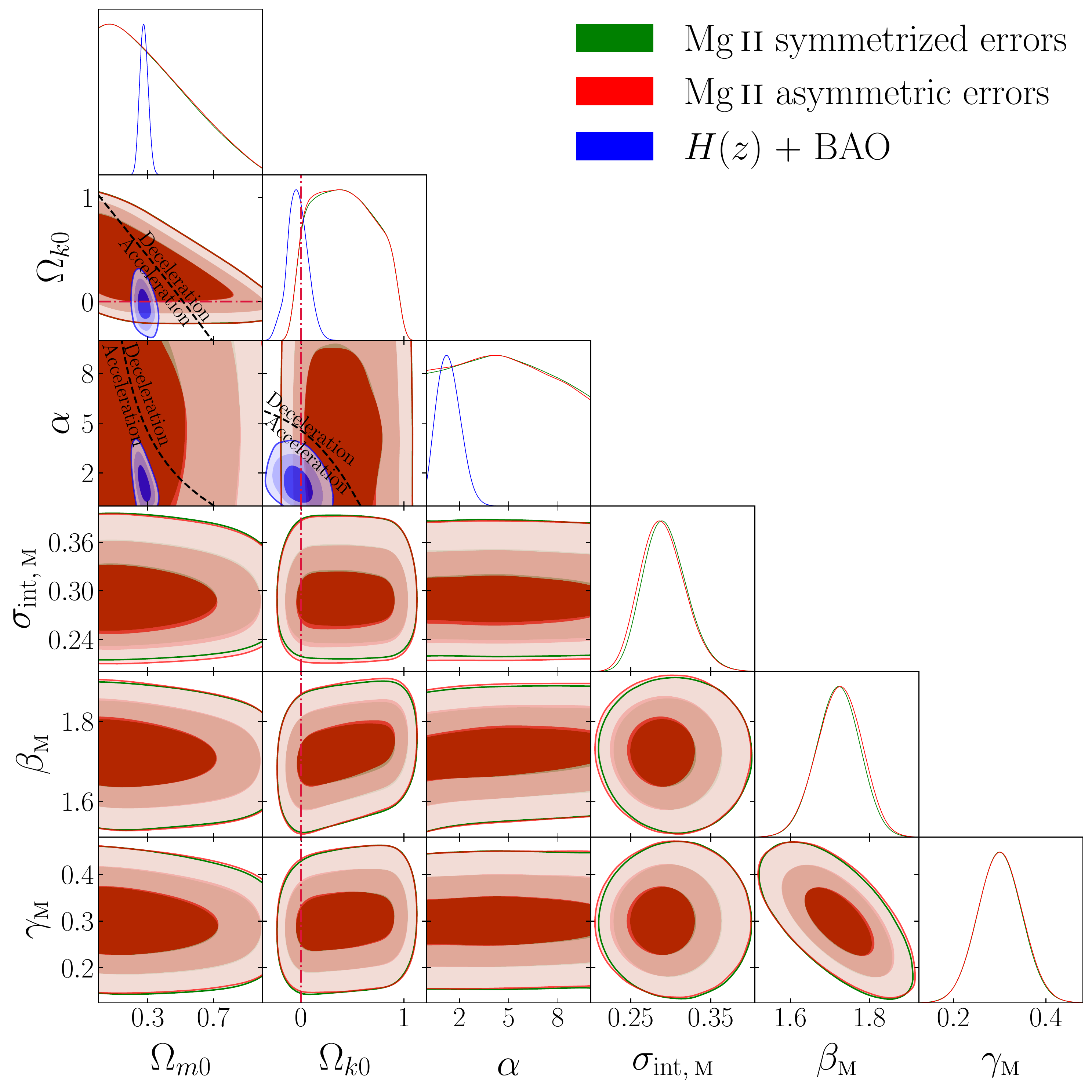}}
 \subfloat[]{%
    \includegraphics[width=0.33\textwidth,height=0.4\textwidth]{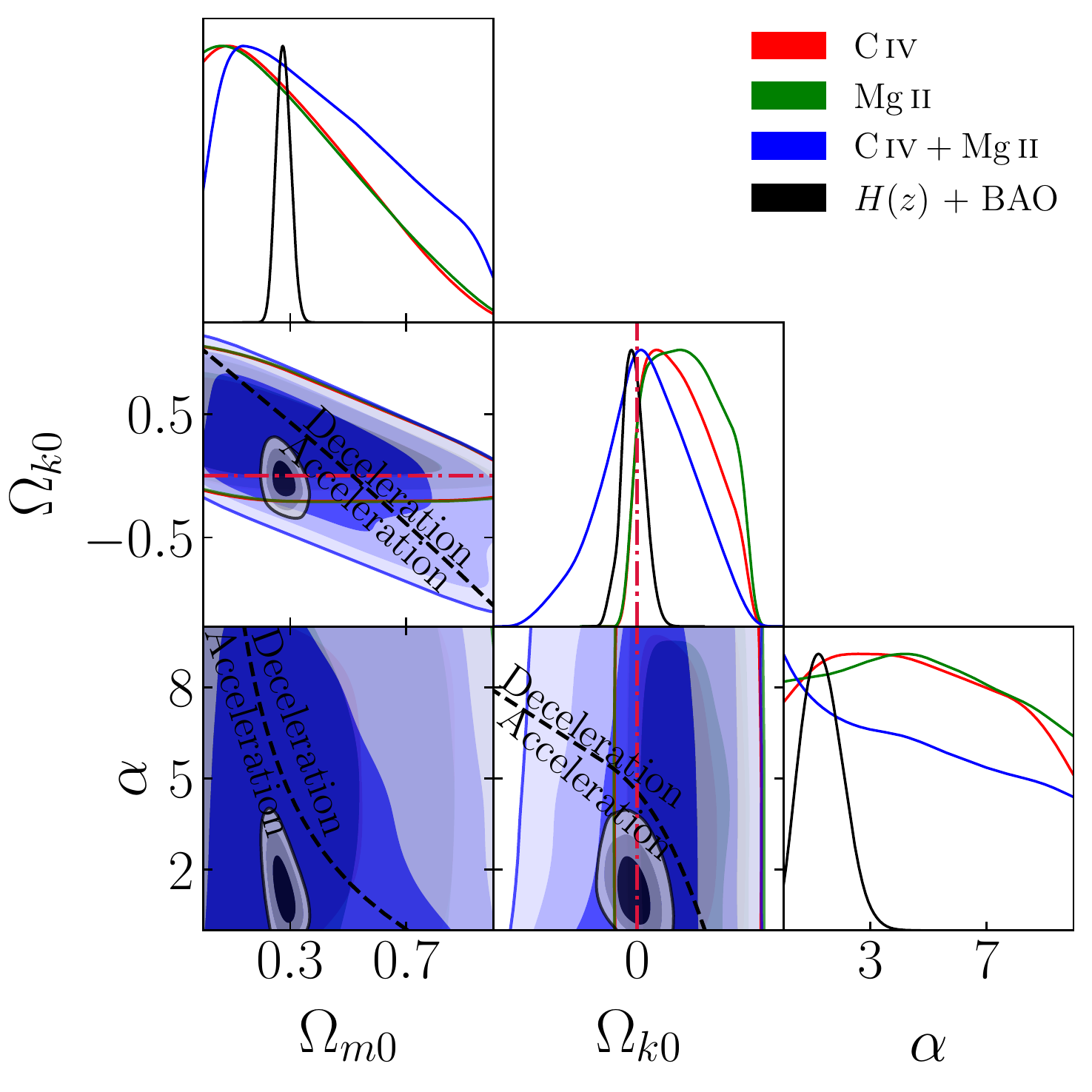}}\\
 \subfloat[]{%
    \includegraphics[width=0.5\textwidth,height=0.55\textwidth]{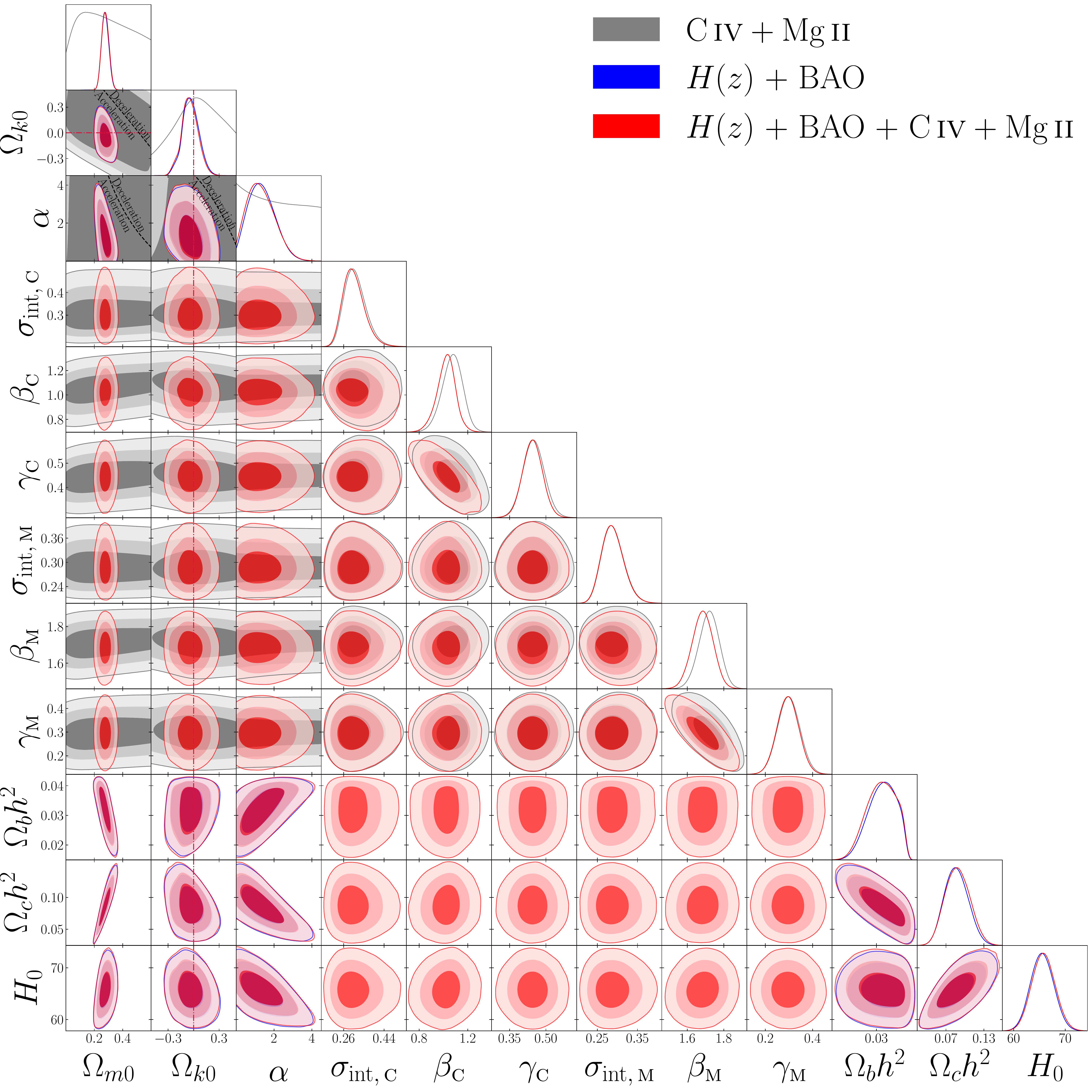}}
 \subfloat[]{%
    \includegraphics[width=0.5\textwidth,height=0.55\textwidth]{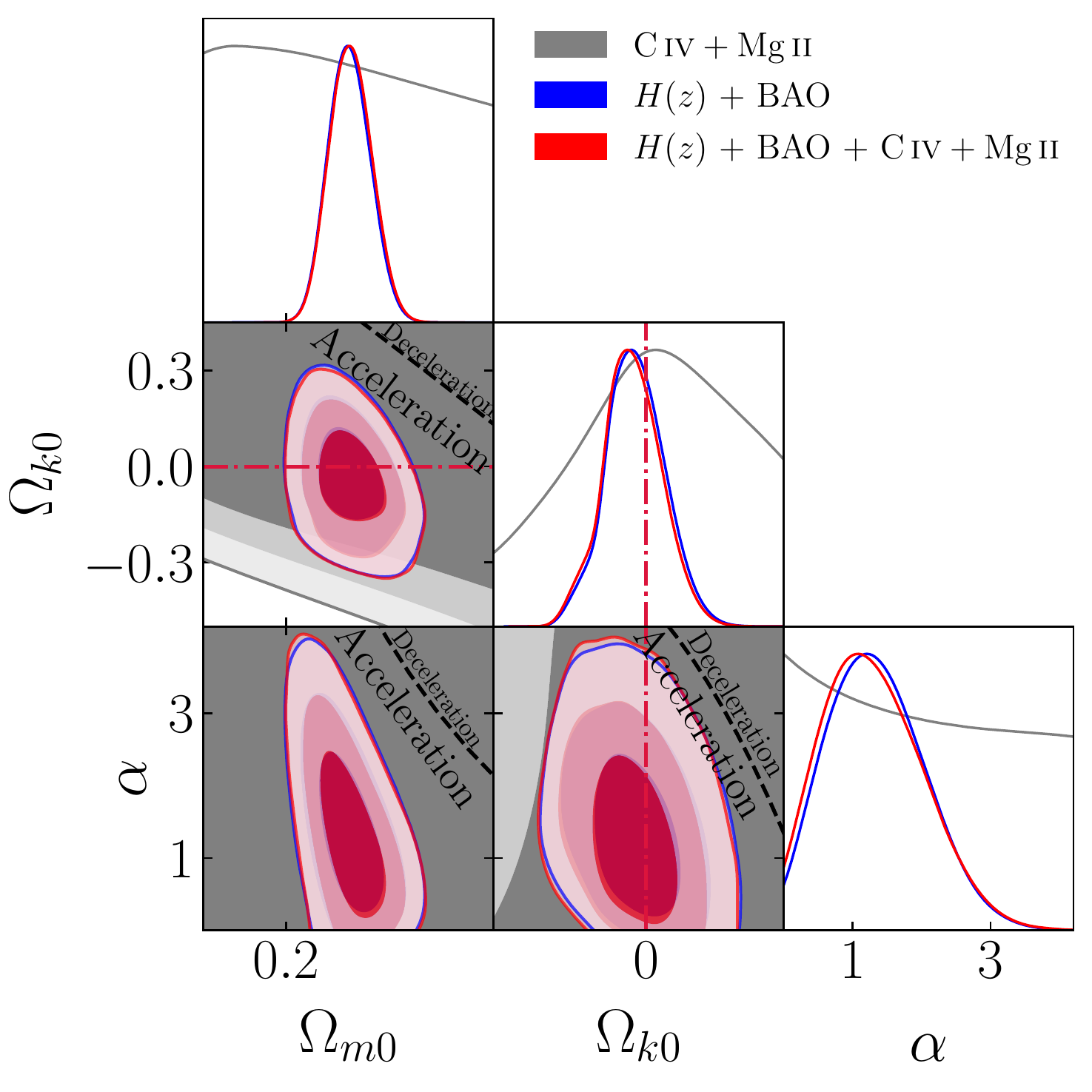}}\\
\caption{Same as Fig.\ \ref{fig5C11} but for non-flat \pcdm. The zero-acceleration black dashed lines are computed for the third cosmological parameter set to the $H(z)$ + BAO data best-fitting values listed in Table \ref{tab:BFPC11}, and divide the parameter space into regions associated with currently-accelerating (below left) and currently-decelerating (above right) cosmological expansion. The crimson dash-dot lines represent flat hypersurfaces, with closed spatial hypersurfaces either below or to the left. The $\alpha = 0$ axes correspond to non-flat \lcdm.}
\label{fig6C11}
\end{figure*}

The posterior one-dimensional probability distributions and two-dimensional confidence regions of cosmological-model and $R-L$ relation parameters for the six cosmological models are shown in Figs.\ \ref{fig1C11}--\ref{fig6C11}, where in panel (a) of each figure results of the \civ\ data analyses with symmetrized errors and asymmetric errors are shown in green and red, respectively; in panel (b) of each figure results of the \mii\ data analyses with symmetrized errors and asymmetric errors are shown in green and red, respectively; in panel (c) of each figure the results of the \civ, \mii, and joint \civ\ + \mii\ data analyses with asymmetric errors, and the $H(z)$ + BAO data analysis are shown in red, green, blue, and black, respectively; and in panels (d) and (e) of each figure the results of the joint (asymmetric errors) \civ\ + \mii, $H(z)$ + BAO, and $H(z)$ + BAO + \civ\ + \mii\ data analyses are shown in grey, blue, and red, respectively. The unmarginalized best-fitting parameter values, as well as the values of maximum likelihood $\mathcal{L}_{\rm max}$, AIC, BIC, DIC, $\Delta \mathrm{AIC}$, $\Delta \mathrm{BIC}$, and $\Delta \mathrm{DIC}$, for all models and data sets, are listed in Table \ref{tab:BFPC11}. The marginalized posterior mean parameter values and uncertainties ($\pm 1\sigma$ error bars and 1 or 2$\sigma$ limits), for all models and data sets, are listed in Table \ref{tab:1d_BFPC11}. 

In all six cosmological models, all data combinations more favour currently accelerating cosmological expansion. This is also the case with \mii\ QSO data \citep{Khadkaetal_2021a, Khadkaetal2022a}, but differs from what happens with H$\beta$ QSO data, which more favour currently decelerated cosmological expansion \citep{Khadkaetal2022a}.

\subsection{Constraints from \civ, \mii, and \civ\ + \mii\ QSO data}
 \label{subsec:CMQ}

As shown in panels (a) and (b) of Figs.\ \ref{fig1C11}--\ref{fig6C11} and Table \ref{tab:1d_BFPC11}, we find that results from \cq\ and \mq\ symmetrized errors data analyses are only mildly different (less than 1$\sigma$) from those of \cq\ and \mq\ asymmetric errors data analyses. Since the analyses with asymmetric errors are the more correct ones, we summarize the asymmetric errors \civ\ results in what follows. The symmetric errors \mii\ results are discussed in \cite{Khadkaetal_2021a}. 

The \om\ constraints from \civ\ data range from a low of $<0.840$ (2$\sigma$, flat XCDM) to a high of $0.467^{+0.199}_{-0.378}$ (1$\sigma$, non-flat \lcdm). 

The \ok\ constraints from \civ\ data are $-0.330^{+0.534}_{-1.060}$, $-0.168^{+0.451}_{-0.789}$, and $0.096^{+0.359}_{-0.337}$ for non-flat \lcdm, XCDM, and \pcdm, respectively. Although \civ\ data favour closed hypersurfaces in non-flat \lcdm\ and non-flat XCDM, and favour open hypersurfaces in non-flat \pcdm, flat hypersurfaces are well within 1$\sigma$. \civ\ data only provide very weak constraints of \wx\ and $\alpha$. 

From panels (c) of Figs.\ \ref{fig1C11}--\ref{fig6C11}, we see that the cosmological-model parameter constraints from \cq\ data and from \mq\ data are mutually consistent, as are the cosmological constraints from \cq\ and $H(z)$ + BAO data and from \mq\ and $H(z)$ + BAO data.\footnote{\citet{Khadkaetal_2021a, Khadkaetal2022a} had earlier shown that the symmetric errors \mii\ cosmological constraints are mutually consistent with those from $H(z)$ + BAO data. This differs from the H$\beta$ QSOs cosmological constraints, which are $\sim 2\sigma$ inconsistent with those from $H(z)$ + BAO data \citep{Khadkaetal2022a}.} It is therefore reasonable to perform joint analyses of \cq\ and \mq\ data. As shown in panels (d) and (e) of Figs.\ \ref{fig1C11}--\ref{fig6C11}, the cosmological-model parameter constraints from \civ\ + \mq\ data and from $H(z)$ + BAO data are mutually consistent at $\lesssim 1.5\sigma$,\footnote{I.e., in the two-dimensional contour plots all or most of the 1$\sigma$ $H(z)$ + BAO data contour always lie inside the 2$\sigma$ \civ\ + \mq\ data contour.} so these data sets can be jointly analyzed. 

The constraints on the \civ\ $R-L$ relation parameters in the six different cosmological models are mutually consistent, so the $R-L$ relation \civ\ data set is standardizable. The constraints on the intrinsic scatter parameter of the \civ\ $R-L$ relation, $\sigma_{\rm int,\,\textsc{c}}$, range from a low of $0.296^{+0.041}_{-0.056}$ (flat XCDM) to a high of $0.314^{+0.038}_{-0.055}$ (flat \pcdm), with a difference of $0.26\sigma$, which are $\sim0.1-0.4\sigma$ larger than those of \mii\ ($\sigma_{\rm int,\,\textsc{m}}$). The constraints on the slope parameter of the \civ\ $R-L$ relation, $\gamma_{\rm\textsc{c}}$, range from a low of $0.426\pm0.049$ (flat XCDM) to a high of $0.468^{+0.046}_{-0.056}$ (non-flat \lcdm), with a difference of $0.56\sigma$. The constraints on the intercept parameter of the \civ\ $R-L$ relation, $\beta_{\rm\textsc{c}}$, range from a low of $0.970^{+0.136}_{-0.108}$ (flat XCDM) to a high of $1.076^{+0.087}_{-0.074}$ (non-flat \pcdm), with a difference of $0.68\sigma$.

A summary value of the measured \civ\ $R-L$ relation slope, $\gamma_{\rm\textsc{c}} \simeq 0.45 \pm 0.04$, indicates that it is about $1\sigma$ lower than and consistent with the prediction of simple photoionization theory \citep[$\gamma = 0.5$,][]{Karasetal2021, Panda2022}, on the other hand a summary value of the measured \mii\ $R-L$ relation slope, $\gamma_{\rm\textsc{m}} \simeq 0.3 \pm 0.05$ \citep[also see][]{Khadkaetal_2021a, Khadkaetal2022a}, indicates that it is about $4\sigma$ lower than $\gamma = 0.5$.

The \om\ constraints from (asymmetric errors) \civ\ + \mii\ data range from a low of $<0.563$ (2$\sigma$, flat XCDM) to a high of $<0.537$ (1$\sigma$, flat \pcdm).

The \ok\ constraints from \civ\ + \mii\ data are $-0.818^{+0.391}_{-0.637}$ ($<0.474$, 2$\sigma$), $-0.410^{+0.368}_{-0.222}$ ($-0.410^{+0.698}_{-0.846}$, 2$\sigma$), and $0.088^{+0.384}_{-0.364}$  ($0.088^{+0.732}_{-0.722}$, 2$\sigma$) for non-flat \lcdm, XCDM, and \pcdm, respectively. \civ\ + \mii\ data favour closed hypersurfaces in non-flat \lcdm\ and non-flat XCDM, being $>1\sigma$ (but $<2\sigma$) and $\sim1.1\sigma$ away from flat hypersurfaces, respectively, and favour open hypersurfaces in non-flat \pcdm, with flat hypersurfaces being within 1$\sigma$. \civ\ + \mii\ data still provide weak constraints on \wx\ and $\alpha$, but \wx\ constraints are more than 2$\sigma$ away from $\wX=-1$ (\lcdm).

\civ\ + \mii\ data provide consistent (within $1\sigma$) constraints on the intrinsic scatter, slope, and intercept parameters of both \civ\ and \mii\ $R-L$ relations, which confirms that \civ\ and \mii\ QSOs are standardizable through different $R-L$ relations.

\subsection{Constraints from $H(z)$ + BAO + \civ\ + \mii\ data}
\label{subsec:HzBCM}

The mutually consistent cosmological-model parameter constraints allow us to jointly analyze $H(z)$ + BAO and \civ\ + \mii\ data. In what follows we summarize the cosmological-model parameter constraints from $H(z)$ + BAO + \civ\ + \mii\ data and contrast them with those from $H(z)$ + BAO data.

The $H(z)$ + BAO + \civ\ + \mii\ data provide \om\ constraints ranging from a low of $0.275\pm0.023$ (flat \pcdm) to a high of $0.301^{+0.015}_{-0.017}$ (flat \lcdm), with a difference of $0.91\sigma$, which only slightly differ from those determined using only $H(z)$ + BAO data.

The $H_0$ constraints from $H(z)$ + BAO + \civ\ + \mii\ data range from a low of $65.68^{+2.20}_{-2.19}$ \hunit\ (flat \pcdm) to a high of $69.15\pm1.77$ \hunit\ (flat \lcdm), with a difference of $1.23\sigma$, which are $0.65\sigma$ (flat \pcdm) lower than and $0.35\sigma$ (flat \lcdm) higher than the median statistics estimate of $H_0=68\pm2.8$ \hunit\ \citep{chenratmed}, and $2.94\sigma$ (flat \pcdm) and $1.84\sigma$ (flat \lcdm) lower than the local Hubble constant measurement of $H_0 = 73.2 \pm 1.3$ \hunit\ \citep{Riess_2021}. The $H(z)$ + BAO + \civ\ + \mii\ data provide $H_0$ constraints that are slightly higher ($\sim0.1\sigma$ at most) and mostly more restrictive ($\sim4\%$ at most) than those from $H(z)$ + BAO data.

The \ok\ constraints from $H(z)$ + BAO + \civ\ + \mii\ data are $0.047^{+0.079}_{-0.089}$, $-0.031\pm0.108$, and $-0.044^{+0.090}_{-0.094}$ for non-flat \lcdm, XCDM, and \pcdm, respectively, which are slightly lower ($\sim0.1\sigma$ at most) than those from $H(z)$ + BAO data. Similar to the $H(z)$ + BAO data results, non-flat \lcdm\ mildly favours open hypersurfaces, whereas non-flat XCDM and non-flat \pcdm\ mildly favour closed hypersurfaces. However, flat hypersurfaces are well within 1$\sigma$.

Dark energy dynamics is favoured. For flat (non-flat) XCDM, $w_{\rm X}=-0.799^{+0.143}_{-0.111}$ ($w_{\rm X}=-0.787^{+0.165}_{-0.102}$), with central values being $1.81\sigma$ ($<2\sigma$) higher than $w_{\rm X}=-1$; and for flat (non-flat) \pcdm, $\alpha=1.202^{+0.490}_{-0.862}$ ($\alpha=1.320^{+0.572}_{-0.869}$), with central values being $1.39\sigma$ ($1.52\sigma$) away from $\alpha=0$. The addition of \civ\ + \mii\ data to $H(z)$ + BAO data bring \wx\ and $\alpha$ values lower, and closer to \lcdm\ model values.

As expected, the constraints on the \civ\ and \mii\ $R-L$ relation parameters are consistent with those from the individual data sets and the \civ\ + \mii\ data.

\subsection{Model Comparison}
 \label{subsec:compC11}

From the AIC, BIC, and DIC values listed in Table \ref{tab:BFPC11}, we find the following results (from the more correct \civ\ and \mii\ asymmetric errors analyses):
\begin{itemize}
    \item[1)]{\bf AIC.} $H(z)$ + BAO and $H(z)$ + BAO + \civ\ + \mii\ data favour flat \pcdm\ the most, and the evidence against the rest of the models/parametrizations is either only weak or positive. 
    
    \civ, \mii, and \civ\ + \mii\ data favour non-flat XCDM the most, however, in the \civ\ case, the evidence against non-flat \lcdm\ and flat XCDM is only weak, the evidence against flat \lcdm\ is positive, and the evidence against flat and non-flat \pcdm\ is strong; in the \mii\ case, the evidence against non-flat \lcdm\ and flat XCDM is positive, the evidence against flat \lcdm\ and flat \pcdm\ is strong, and the evidence against non-flat \pcdm\ is very strong; and in the \civ\ + \mii\ case, the evidence against flat XCDM is positive, the evidence against non-flat \lcdm\ is strong, and other models are very strongly disfavoured.
    
    \item[2)] {\bf BIC.} $H(z)$ + BAO and $H(z)$ + BAO + \civ\ + \mii\ data favour flat \lcdm\ the most, and in the former case, the evidence against the rest of the models/parametrizations is either only weak or positive, while in the latter case, the evidence against the rest of the models/parametrizations is either positive or strong (non-flat XCDM and non-flat \pcdm). 
    
    \civ\ data favour non-flat \lcdm\ the most, and the evidence against flat and non-flat XCDM is only weak, the evidence against flat \lcdm\ is positive, and the evidence against flat and non-flat \pcdm\ is strong.
    
    \mii\ and \civ\ + \mii\ data favour non-flat XCDM the most, however, in the \mii\ case, the evidence against non-flat \lcdm, flat XCDM, and non-flat XCDM is only weak, the evidence against flat \pcdm\ is strong, and the evidence against non-flat \pcdm\ is very strong; and in the \civ\ + \mii\ case, the evidence against non-flat \lcdm\ and flat XCDM is positive, the evidence against flat \lcdm\ is strong, and non-flat XCDM and non-flat \pcdm\ are very strongly disfavoured.

    \item[3)] {\bf DIC.} $H(z)$ + BAO and $H(z)$ + BAO + \civ\ + \mii\ data favour flat \pcdm\ the most, and the evidence against the rest of the models/parametrizations is either only weak or positive. 
    
    \civ\ and \mii\ data favour flat \lcdm\ the most, however, in the former case, the evidence against the rest of the models/parametrizations is either only weak or positive, whereas in the latter case, the evidence against flat and non-flat \pcdm\ is only weak, the evidence against non-flat \lcdm\ and flat XCDM is positive, and the evidence against non-flat XCDM is strong.
    
    \civ\ + \mii\ data favour non-flat XCDM the most, and the evidence against the rest of the models/parametrizations is either only weak or positive. 
\end{itemize}

Based on the more reliable DIC \citep{eacc21461d7e4419a9dee07b7fa8f657,90}, except for the \mii\ data set, these data sets do not provide strong evidence against any of the cosmological models/parametrizations.

\section{Discussion}
\label{makereference11.5}

We have shown that QSOs with measured \civ\ time-delays can be standardized and so can be used as cosmological probes. \mii\ QSOs are also standardizable and so can be jointly analyzed with \civ\ QSOs to constrain cosmological model parameters. This is not true for current H$\beta$ QSOs \citep{Khadkaetal2022a} and more work is needed to clarify the H$\beta$ QSO situation.

However, we find a 2.3$\sigma$ difference in the measured slopes of the \civ\ and \mii\ $R-L$ relations, in the flat $\Lambda$CDM model asymmetric error bars results of Table \ref{tab:1d_BFPC11}, or from 
the summary values of the measured \civ\ and \mii\ $R-L$ relation slopes, $\gamma_{\rm\textsc{c}} \simeq 0.45 \pm 0.04$ and $\gamma_{\rm\textsc{m}} \simeq 0.3 \pm 0.05$. And while the \civ\ slope is only about $1\sigma$ lower than the $\gamma = 0.5$ slope predicted by simple photoionization theory \citep{Bentzetal2013, Karasetal2021, Panda2022} or a dust-based model of the BLR \citep{czerny2011,2021ApJ...920...30N,2022ApJ...931...39M,2022A&A...663A..77N}, the \mii\ slope is about $4\sigma$ lower than $\gamma = 0.5$, which is more statistically significant. 

In this section we examine potential (selection effect produced) differences between the \civ\ and \mii\ compilations and conclude that the ones we study here are not very significant. We begin by computing the Eddington ratio $\lambda_{\rm Edd}$ for the 78 \mii\ sources and the 38 \civ\ sources. We use the definition $\lambda_{\rm Edd}=L_{\rm bol}/L_{\rm Edd}$ where the bolometric luminosity is estimated as a multiple of the corresponding monochromatic luminosity ($L_{1350}$ for the \civ\ sample and $L_{3000}$ for the \mii\ sample; for the calculations in this section we adopt the fixed flat $\Lambda$CDM model with $\Om=0.3$, $\Omega_{\Lambda}=0.7$, and $H_0=70\,{\rm km\,s^{-1}\,Mpc^{-1}}$), taking into account the monochromatic luminosity-dependent bolometric correction factors according to \citet{2019MNRAS.488.5185N}. Subsequently, we evaluate the correlations between $\lambda_{\rm Edd}$ and the monochromatic luminosity and the redshift. 

\begin{figure}
    \centering
    \includegraphics[width=\columnwidth]{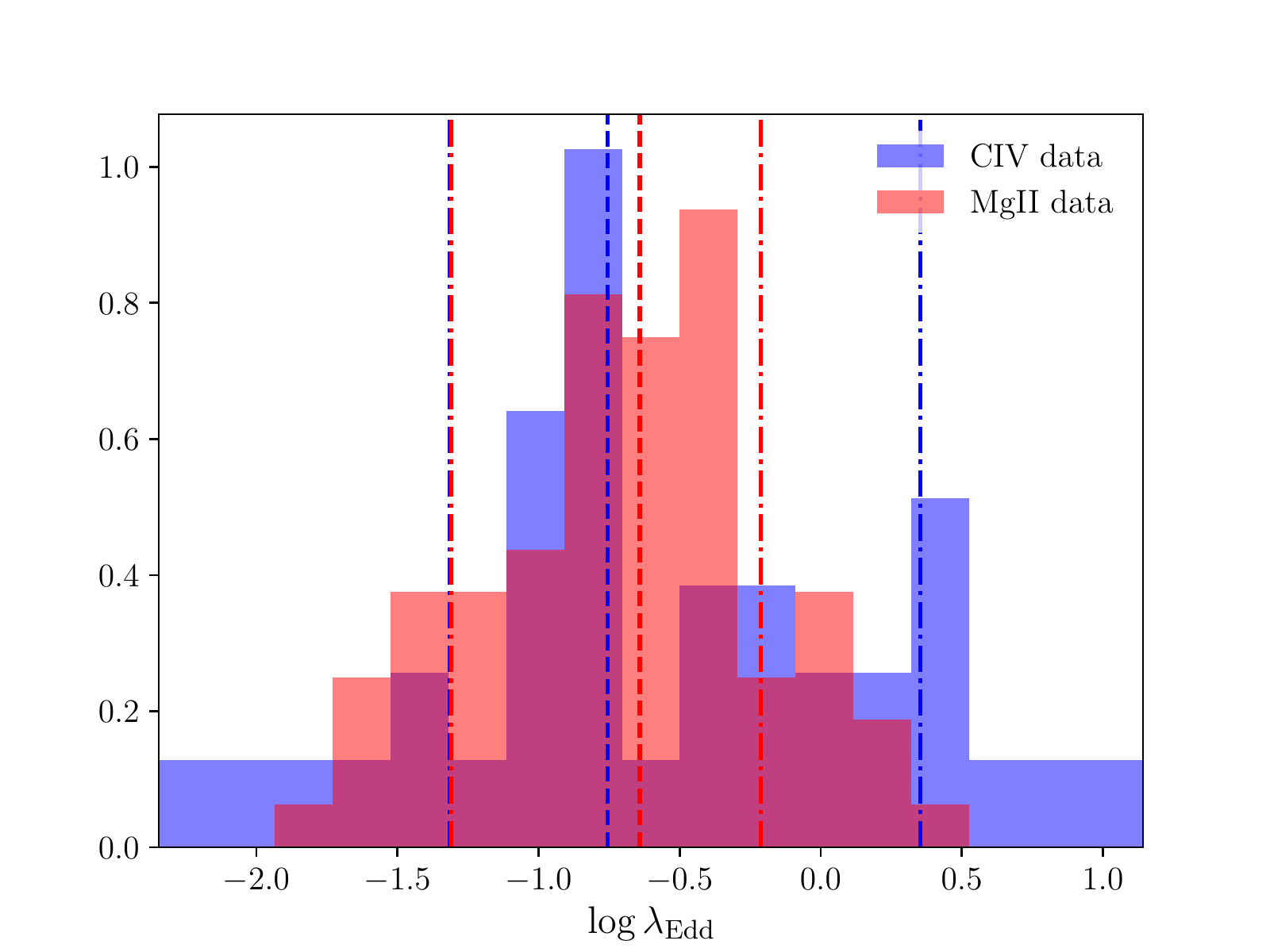}
    \caption{Normalized distributions of the Eddington ratio ($\log{\lambda_{\rm Edd}}$) for the \civ\ sample (blue histogram) and for the \mii\ sample (pink histogram). Vertical dashed lines stand for median values, while the dot-dashed lines represent 16- and 84-\% percentiles. The bin size is $\Delta(\log{\lambda_{\rm Edd}})\sim 0.205$.}
    \label{fig_lambda_dist}
\end{figure}

The Eddington-ratio normalized distributions for the \civ\ and \mii\ samples are shown in Fig.~\ref{fig_lambda_dist}, represented by blue and pink histograms, respectively. The vertical dashed lines stand for the distribution medians, while vertical dot-dashed lines mark 16- and 84-\% percentiles. For the \mii\ sample, the $\lambda_{\rm Edd}$ distribution median is $-0.64$, while 16- and 84-\% percentiles are $-1.31$ and $-0.21$, respectively. For the \civ\ sample, the median of the $\lambda_{\rm Edd}$ distribution is $-0.75$, while 16- and 84-\% percentiles are $-1.32$ and $0.35$, respectively.  The median $\lambda_{\rm Edd}$ of the \mii\ sources is larger than the median $\lambda_{\rm Edd}$ of the \civ\ sources. However, the \civ\ distribution is skewed significantly towards higher $\lambda_{\rm Edd}$ values, with a total range of $(-2.35, 1.14)$, in comparison with the range of $(-1.93,0.50)$ for the \mii\ sample. Also note that the Eddington ratios of the super-Eddington sources generally have large error bars, being consistent with the Eddington limit as well.

\begin{figure*}
    \centering
    \includegraphics[width=0.48\textwidth]{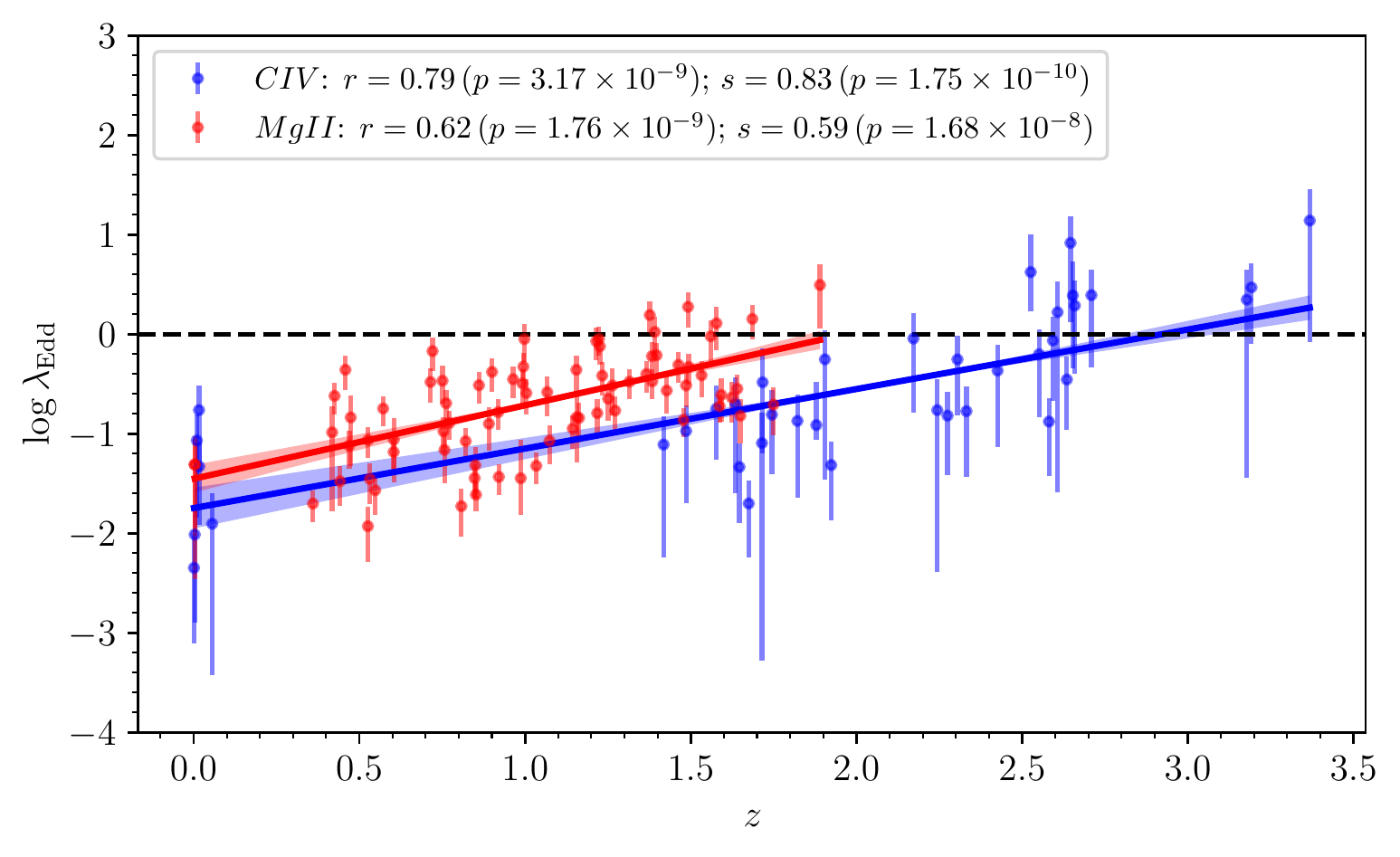}
    \includegraphics[width=0.48\textwidth]{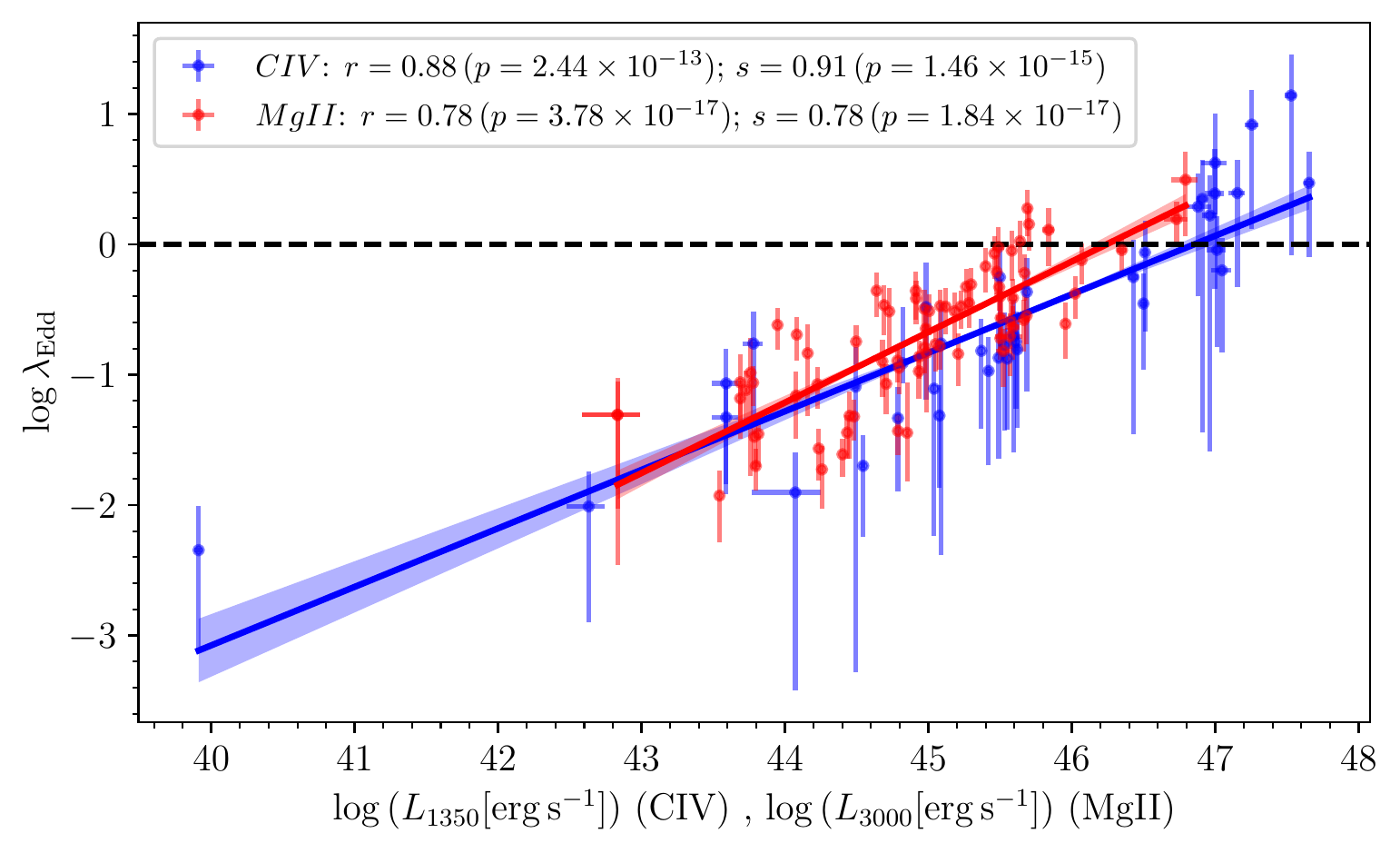}
    \caption{The Eddington ration $\lambda_{\rm Edd}=L_{\rm bol}/L_{\rm Edd}$ as a function of the source redshift (left panel) and as a function of the monochromatic luminosities $L_{1350}$ or $L_{3000}$ (right panel) for \civ\ (blue points) and \mii\ (red points) sources. In both panels, we include the Pearson ($r$) and the Spearman rank-order correlation coefficients ($s$), which indicate significant positive correlations between $\lambda_{\rm Edd}$ and $z$ as well as $\lambda_{\rm Edd}$ and $L_{1350}$ (or $L_{3000}$). The horizontal dashed line stands for the Eddington limit ($\log{\lambda_{\rm Edd}}=0$). The solid blue and red lines stand for the best-fitting linear relations for the \civ\ and \mii\ datasets, respectively, see Eqs.~\eqref{eq_lambda_z} and Eqs.~\eqref{eq_lambda_lum} for the best-fitting slopes and intercepts including 1$\sigma$ uncertainties.}
    \label{fig_lambda_correlations}
\end{figure*}

The $\lambda_{\rm Edd}$-$z$ and $\lambda_{\rm Edd}$-$L_{1350}$ or $\lambda_{\rm Edd}$-$L_{3000}$  correlations are positive, see Fig.~\ref{fig_lambda_correlations}, left and right panels, respectively. The correlation significance is evaluated using Pearson and Spearman rank-order correlation coefficients, see the legend in Fig.~\ref{fig_lambda_correlations}. For both \civ\ and \mii\ samples, the $\lambda_{\rm Edd}$-$z$ correlation is positive and significant, being slightly stronger and more significant for \civ\ sources. The correlation between $\lambda_{\rm Edd}$ and the corresponding monochromatic luminosity is also positive and significant for both samples. Here we stress that the correlation between the Eddington ratio $\lambda_{\rm Edd}$ and the corresponding monochromatic luminosities is enhanced intrinsically due to the definition of $\lambda_{\rm Edd}=L_{\rm bol}/L_{\rm Edd}$, where the bolometric luminosity is proportional to the monochromatic luminosity. However, the relative comparison of the correlation slopes provides hints about the similarities/differences of the two samples. 

The slopes and the intercepts of the best-fitting linear relations between $\lambda_{\rm Edd}$ and $z$ are
\begin{align}
    \log{\lambda_{\rm Edd}}(\mathrm{\civ})&=(0.60 \pm 0.10)z-(1.75\pm 0.20)\,,\notag\\
    \log{\lambda_{\rm Edd}}(\mathrm{\mii})&=(0.74 \pm 0.12)z-(1.45\pm 0.14)\,,\label{eq_lambda_z}
\end{align}
while for the best-fitting linear relations between $\lambda_{\rm Edd}$ and the monochromatic luminosity we obtain
\begin{align}
    \log{\lambda_{\rm Edd}}(\mathrm{\civ})&=(0.45 \pm 0.04)\log{L_{1350}}-(21.03\pm 1.97)\,,\notag\\
    \log{\lambda_{\rm Edd}}(\mathrm{\mii})&=(0.54 \pm 0.05)\log{L_{3000}}-(25.01\pm 2.25)\,.\label{eq_lambda_lum}
\end{align}
The best-fitting relations are depicted in Fig.~\ref{fig_lambda_correlations} including 1$\sigma$ uncertainties of the best-fitting parameters. Compared to the \civ\ case, the \mii\ Eddington ratios for the sample of 78 sources appear to increase more steeply with both redshift and monochromatic luminosity, though the \mii\ and \civ\ slopes are consistent within 1.0$\sigma$ (for the $\lambda_{\rm Edd}$ vs. $z$ correlations) and 1.4$\sigma$ (for the $\lambda_{\rm Edd}$ vs. monochromatic luminosity correlations), respectively. The differences in slopes may merely be due to the limited number of sources in each sample, and hence a selection effect.

\begin{figure}
    \centering
    \includegraphics[width=\columnwidth]{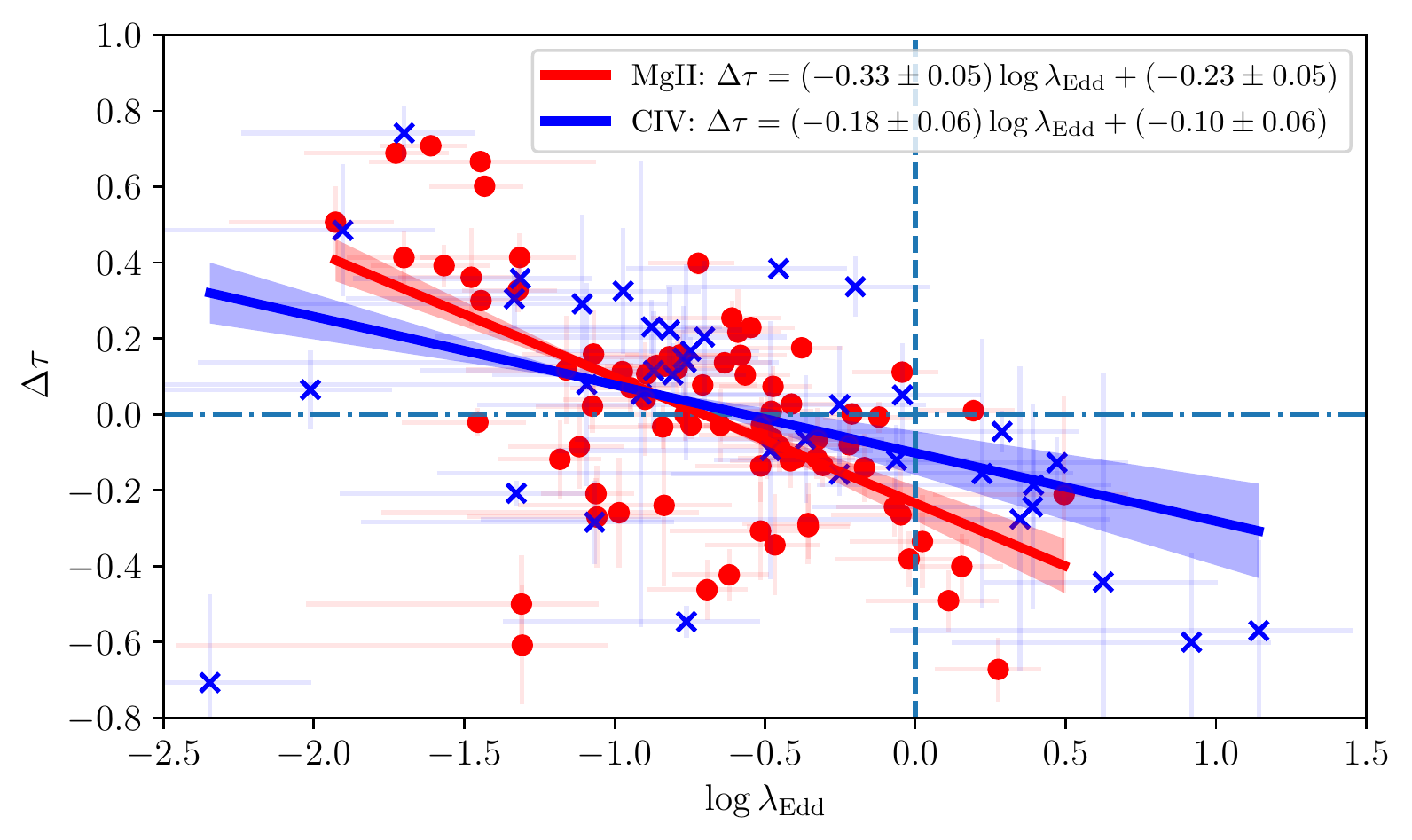}
    \caption{(Anti)correlation between $\Delta \tau \equiv \log{(\tau/\tau_{\rm RL})}$ and the Eddington ratio $\lambda_{\rm Edd}$ for both \civ\ sources (blue points) and \mii\ sources (red points). The correlation is more signifcant for the \mii\ sample with a Spearman rank-order correlation coefficient of $-0.53$  $(p=4.76\times 10^{-7})$ in comparison with $-0.50$ ($p=0.0015$) for the \civ\ sample. The slope of the anticorrelation is steeper for \mii\ sources, see Eq.~\eqref{eq_delta_tau} for the comparison.}
    \label{fig_delta_tau}
\end{figure}

However, when we study how the time-delay difference between the measured value and the value predicted from the best-fit $R-L$ relation (in the fixed flat $\Lambda$CDM model) --- $\Delta \tau \equiv \log{(\tau/\tau_{\rm RL})}$ \citep{Mehrabietal2021,2020ApJ...896..146Z} --- correlates with the Eddington ratio, the (anti)correlation is more significant and steeper for \mii\ sources. This is shown in Fig.~\ref{fig_delta_tau} for both \civ\ (blue points) and \mii\ sources (red points). As a caveat, we note that unless the SMBH mass is inferred independently of the RM, $\Delta \tau$ and $\log{\lambda_{\rm Edd}}$ are intrinsically correlated as $\Delta \tau \propto -\log{\tau}$, which implies anticorrelation.\footnote{It is generally thought advisable to use independent accretion-rate proxies, such as the relative \Feii\ strength \rfe\, the variability factor $F_{\rm var}$ \citep{2020ApJ...903...86M}, or the shape of the ionizing continuum \citep{Panda_etal_2019,2020ApJ...899...73F, Ferland_etal_2020}, rather than the Eddington ratio, due to the interdependency on the rest-frame time delay and the monochromatic luminosity that appear in the $R-L$ correlation. However, here we focus on a relative comparison between the \mii\ and \civ\ samples instead of the absolute values of the correlation slopes and normalizations.} Therefore, we can only assess the different behaviour between $\Delta \tau$ and $\lambda_{\rm Edd}$ based on the relative comparison of the correlation coefficients and the correlation slopes for the two samples. The anticorrelation between $\Delta \tau$ and $\log{\lambda_{\rm Edd}}$ is more significant for the \mii\ sample with a Spearman rank-order correlation coefficient of $s=-0.53$ $(p=4.76\times 10^{-7})$, while for the \civ\ sample we obtain $s=-0.50$ ($p=0.0015$). When we fit a linear function to the anticorrelations we get
\begin{align}
    \Delta \tau (\mathrm{\civ})&=(-0.18 \pm 0.06)\log{\lambda_{\rm Edd}}+(-0.10\pm 0.06)\,,\notag\\
    \Delta \tau (\mathrm{\mii})&=(-0.33 \pm 0.05)\log{\lambda_{\rm Edd}}+(-0.23\pm 0.05)\,.\label{eq_delta_tau}
\end{align}
Hence, the anticorrelation is significantly steeper for the \mii\ sample (by nearly a factor of two) in comparison with the current \civ\ sample. A combination of steeper $\Delta \tau$ -- $\log{\lambda_{\rm Edd}}$ anticorrelation with a slightly steeper correlation between $\log{\lambda_{\rm Edd}}$ and the monochromatic luminosity for the \mii\ sources can qualitatively account for the flatter \mii\ $R-L$ relation with respect to the \civ\ one. However, with a future increase in the number of \mii\ and \civ\ RM sources, both towards lower and higher redshifts, the difference in the correlation slopes is expected to become progressively weaker, since the general trends appear to be consistent between the \mii\ and the \civ\ samples and the differences can be traced to a few outlying sources. 

There may also be a potential problem in the way $\lambda_{\rm Edd}$ was measured in the \civ\ QSOs. Such measurements are uncertain, particularly for \civ. The {\civ}-emitting material is likely not completely virialized as revealed by its blueshift and blueward asymmetry with respect to low-ionization lines, which hints at an outflow approaching the observer. Therefore, the SMBH mass determination may be biased \citep{2005MNRAS.356.1029B}. The anticorrelation between the virial factor and the \civ\ FWHM is also the most uncertain among broad lines due to the smallest correlation coefficient as investigated by \citet{2018NatAs...2...63M}, which results in larger errorbars of \civ\ Eddington ratios in comparison with \mii\ ones in Fig.~\ref{fig_lambda_correlations}. Looking at how the \civ\ sample was selected might suggest that the histogram in Fig.~\ref{fig_lambda_dist} may not well represent the true \civ\ $\lambda_{\rm Edd}$ values. One of the criteria used when selecting the final \civ\ QSOs from the SDSS-RM sample is the variability of the light curve. Sources with no significant variability were excluded from the final sample to ensure a reliable time-lag estimate \citep{2019ApJ...887...38G}. Several papers claim that there is a negative correlation between the variability and the Eddington ratio \citep[e.g.][and references therein]{2022A&A...664A.117D}. This suggests that the criterion of the variability applied to the SDSS-RM sample excludes highly-radiating QSOs. Such QSOs flatten the $R-L$ relation and increase the scatter in the case of the H$\beta$ $R-L$ relation, which is not observed in the \civ\ $R-L$ relation. In addition, \citet{2021ApJ...915..129K} claim that their three sources do not show strong outflows, which are usually found in QSOs with high Eddington ratios \citep{2017MNRAS.465.2120C, 2018A&A...618A.179M}. We performed a visual inspection of the rest of the sources and most of them show spectral features associated with low Eddington ratios (symmetric profiles in \civ\ $\lambda$1549, moderate He\,\textsc{ii} $\lambda1640$ contribution, strong contribution of C\,\textsc{iii}] $\lambda1909$, and low contributions of \Feii\ and Fe\,\textsc{iii}). All these facts, in contradiction to Fig.~\ref{fig_lambda_dist}, suggest that the \civ\ sample mainly includes low Eddington sources. However, a detailed analysis of variability and spectral properties is needed to confirm the accretion state of sources in the \civ\ sample.

The \civ\ sample spans eight orders of luminosity ($L_{1350}\sim 10^{40-48}$ erg s$^{-1}$) and covers a large redshift range ($0.001<z<3.37$), and is standardizable and suitable for constraining cosmological models. The \mii\ sample has smaller luminosity and redshift ranges ($L_{3000}\sim 10^{43-47}$ erg s$^{-1}$, $0<z<1.8$), but is standardizable and also suitable for constraining cosmological parameters \citep{Khadkaetal_2021a, Khadkaetal2022a}, and can be jointly analyzed with the \civ\ sample. The same is not true for the H$\beta$ sample. Although the H$\beta$ sample spans almost five orders of luminosity ($L_{5100}=10^{41.5-46}$ erg s$^{-1}$), the redshift range is narrow ($0.002<z<0.9$), but most importantly the current H$\beta$ sample appears to be not standardizable \citep{Khadkaetal2022a}. Therefore, we cannot jointly use the \mii, \civ\ and H$\beta$ samples, which is unfortunate since they together contain more sources over large luminosity and redshift ranges, all of which are beneficial if such QSOs are to be useful for cosmological purposes.

With future missions and surveys, such as the Black Hole Mapper by the SDSS-V  \citep{SDSS_BH_Mapper_2017arXiv171103234K}, long-term spectroscopic and photometric measurements of reverberation lags for the continuum and the ``major" emission lines (\civ, \mii, and H$\beta$/H$\alpha$) will be obtained for a large sample of quasars, adding wide-area, multi-epoch optical spectroscopy to the era of time-domain imaging. With the photometric Legacy Survey of Space and Time performed by the Vera C.\ Rubin observatory \citep{LSST_2019ApJ...873..111I}, many more RM objects will be obtained, e.g.\ the Rubin Deep Drilling Field should result in a few thousand measurements at redshifts $\sim 1 - 2$ \citep{Ksubm22, 2018arXiv181106542B}. This will lead to the decrease in the statistical error. The source monitoring will be performed photometrically in six broad optical bands, which will be used to probe the continuum accretion-disc response as well as the BLR response using the photometric RM \citep[see e.g., ][for the assessment of the photometric RM for BLR RM and the future application in cosmology]{2019FrASS...6...75P,2020mbhe.confE..10M}. Currently, we obtain a difference with respect to BAO+$H(z)$ data constraints by at most $0.1\sigma$ when 116 \civ\ + \mii\ sources are included, hence a QSO dataset larger by an order of magnitude can naturally influence the overall cosmological constraints more. It is more challenging to lower the dispersion in the individual time-delay measurements. Using a consistent methodology for all the RM QSOs to infer time-delays will be necessary to avoid systematic errors introduced by combining different subsets of RM sources into one large dataset. Specifically, surveys and RM projects typically select sources in a narrow luminosity range, which can result in a problem. For instance, if the time-delay method used for higher-luminosity sources is systematically susceptible to yield smaller lags in comparison with the method used for lower-luminosity sources, it can lead to a systematically smaller slope of the R-L relation compared to the case when the same method is applied to all the sources across several orders of magnitude in luminosity.

\section{Conclusions}
\label{makereference11.6}

In this paper, for the first time, we use 38 highest-quality \cq\ data to simultaneously constrain cosmological model parameters, in six cosmological models, and \civ\ $R-L$ relation parameters. We use a new technique we have developed to more correctly take into account the asymmetric time-lag $\tau$ error bars and applied it to both \civ\ and \mii\ QSO data \citep{Khadkaetal_2021a} for the first time. We find that similar to \mq\ data, \cq\ data are also standardizable through the $R-L$ relation since the \civ\ $R-L$ relation parameters are cosmological model-independent. Unlike the H$\beta$ QSOs cosmological constraints \citep{Khadkaetal2022a}, those from \civ\ and \mii\ QSO data are consistent with cosmological constraints from better established $H(z)$ + BAO data.

The mutually consistent cosmological parameter constraints from \civ, \mii, and $H(z)$ + BAO data allow us to perform joint analyses on \civ\ + \mii\ data as well as on $H(z)$ + BAO + \civ\ + \mii\ data. Although the joint \civ\ + \mii\ cosmological constraints are still weak, they do slightly ($\sim0.1\sigma$ at most) alter the $H(z)$ + BAO constraints when jointly analyzed.

The Rubin Observatory Legacy Survey of Space and Time \citep{LSST_2019ApJ...873..111I}, as well as the SDSS-V Black Hole Mapper \citep{SDSS_BH_Mapper_2017arXiv171103234K}, will find many more \civ\ and \mii\  QSOs. These new QSOs will result in significantly more restrictive \civ\ (and \mii) cosmological constraints than the first ones we have derived here.

\begin{sidewaystable*}
\centering
\resizebox*{\columnwidth}{0.75\columnwidth}{%
\begin{threeparttable}
\caption{Unmarginalized best-fitting parameter values for all models from various combinations of data.}\label{tab:BFPC11}
\begin{tabular}{lcccccccccccccccccccc}
\toprule
Model & Data set & $\Omega_{b}h^2$ & $\Omega_{c}h^2$ & $\Omega_{m0}$ & $\Omega_{k0}$ & $w_{\mathrm{X}}$/$\alpha$\tnote{a} & $H_0$\tnote{b} & $\sigma_{\mathrm{int,\,\textsc{c}}}$ & $\gamma_{\rm\textsc{c}}$ & $\beta_{\rm\textsc{c}}$ & $\sigma_{\mathrm{int,\,\textsc{m}}}$ & $\gamma_{\rm\textsc{m}}$ & $\beta_{\rm\textsc{m}}$ & $-2\ln\mathcal{L}_{\mathrm{max}}$ & AIC & BIC & DIC & $\Delta \mathrm{AIC}$ & $\Delta \mathrm{BIC}$ & $\Delta \mathrm{DIC}$ \\
\midrule
 & $H(z)$ + BAO & 0.0244 & 0.1181 & 0.301 & -- & -- & 68.98  & -- & -- & -- & -- & -- & -- & 25.64 & 31.64 & 36.99 & 32.32 & 0.00 & 0.00 & 0.00\\
 & \mq\ symm & -- & 0.0518 & 0.157 & -- & -- & -- & -- & -- & -- & 0.282 & 0.285 & 1.670 & 30.16 & 38.16 & 47.59 & 37.68 & 0.00 & 0.00 & 0.00\\
Flat & \mq\ asymm & -- & 0.0472 & 0.148 & -- & -- & -- & -- & -- & -- & 0.278 & 0.282 & 1.676 & 30.18 & 38.18 & 47.61 & 37.86 & 0.00 & 0.00 & 0.00\\%
\lcdm & \cq\ symm & -- & $-0.0007$ & 0.050 & -- & -- & -- & 0.275 & 0.403 & 0.989 & -- & -- & -- & 20.61 & 28.61 & 35.16 & 31.10 & 0.00 & 0.00 & 0.00\\
 & \cq\ asymm & -- & $-0.0026$ & 0.046 & -- & -- & -- & 0.265 & 0.411 & 0.980 & -- & -- & -- & 20.51 & 28.51 & 35.06 & 32.74 & 0.00 & 0.00 & 0.00\\
 & \cq\ asymm + \mq\ asymm & -- & 0.0082 & 0.068 & -- & -- & -- & 0.274 & 0.412 & 0.995 & 0.280 & 0.286 & 1.647 & 50.94 & 64.94 & 84.21 & 69.10 & 0.00 & 0.00 & 0.00\\
 & $H(z)$ + BAO + \cq\ asymm + \mq\ asymm & 0.0245 & 0.1148 & 0.295 & -- & -- & 68.86 & 0.284 & 0.421 & 1.079 & 0.279 & 0.288 & 1.688 & 78.31 & 96.31 & 123.98 & 98.13 & 0.00 & 0.00 & 0.00\\
\midrule
 & $H(z)$ + BAO & 0.0260 & 0.1098 & 0.292 & 0.048 & -- & 68.35 & -- & -- & -- & -- & -- & -- & 25.30 & 33.30 & 40.43 & 33.87 & 1.66 & 3.44 & 1.54\\
 & \mq\ symm & -- & 0.1703 & 0.399 & $-1.135$ & -- & -- & -- & -- & -- & 0.275 & 0.352 & 1.625 & 25.39 & 35.39 & 47.17 & 40.12 & $-2.77$ & $-0.42$ & 2.44\\
Non-flat & \mq\ asymm & -- & 0.1708 & 0.400 & $-1.134$ & -- & -- & -- & -- & -- & 0.270 & 0.349 & 1.635 & 25.62 & 35.62 & 47.40 & 40.25 & $-2.56$ & $-0.20$ & 2.38\\
\lcdm & \cq\ symm & -- & 0.0147 & 0.081 & $-0.358$ & -- & -- & 0.237 & 0.486 & 0.968 & -- & -- & -- & 9.38 & 19.38 & 27.57 & 36.94 & $-9.23$ & $-7.59$ & 5.84\\
 & \cq\ asymm & -- & 0.0378 & 0.128 & $-0.471$ & -- & -- & 0.242 & 0.498 & 0.939 & -- & -- & -- & 14.44 & 24.44 & 32.63 & 36.77 & $-4.07$ & $-2.43$ & 4.03\\
 & \cq\ asymm + \mq\ asymm & -- & 0.0791 & 0.213 & $-0.678$ & -- & -- & 0.269 & 0.512 & 1.018 & 0.279 & 0.293 & 1.642 & 42.92 & 58.92 & 80.95 & 68.83 & $-6.01$ & $-3.26$ & $-0.27$\\
 & $H(z)$ + BAO + \cq\ asymm + \mq\ asymm & 0.0260 & 0.1112 & 0.293 & 0.036 & -- & 68.64 & 0.278 & 0.430 & 1.045 & 0.278 & 0.293 & 1.684 & 78.09 & 98.09 & 128.84 & 99.73 & 1.79 & 4.86 & 1.60\\
\midrule
 & $H(z)$ + BAO & 0.0296 & 0.0951 & 0.290 & -- & $-0.754$ & 65.79 & -- & -- & -- & -- & -- & -- & 22.39 & 30.39 & 37.52 & 30.63 & $-1.25$ & 0.53 & $-1.69$\\
 & \mq\ symm & -- & $-0.0234$ & 0.003 & -- & $-4.949$ & -- & -- & -- & -- & 0.270 & 0.241 & 1.354 & 24.39 & 34.39 & 46.18 & 41.19 & $-3.77$ & $-1.41$ & 3.51\\
Flat & \mq\ asymm & -- & $-0.0226$ & 0.005 & -- & $-4.983$ & -- & -- & -- & -- & 0.267 & 0.249 & 1.361 & 24.35 & 34.35 & 46.13 & 41.14 & $-3.83$ & $-1.48$ & 3.28\\
XCDM & \cq\ symm & -- & $-0.0237$ & 0.003 & -- & $-4.990$ & -- & 0.234 & 0.321 & 0.727 & -- & -- & -- & 12.69 & 22.69 & 30.88 & 31.64 & $-5.92$ & $-4.29$ & 0.53\\
 & \cq\ asymm & -- & $-0.0203$ & 0.010 & -- & $-4.988$ & -- & 0.232 & 0.355 & 0.736 & -- & -- & -- & 15.58 & 25.58 & 33.77 & 33.67 & $-2.93$ & $-1.29$ & 0.93\\
 & \cq\ asymm + \mq\ asymm & -- & $-0.0212$ & 0.008 & -- & $-4.875$ & -- & 0.225 & 0.337 & 0.757 & 0.262 & 0.248 & 1.399 & 40.24 & 56.24 & 78.27 & 66.94 & $-8.70$ & $-5.95$ & $-2.16$\\
 & $H(z)$ + BAO + \cq\ asymm + \mq\ asymm & 0.0283 & 0.1007 & 0.297 & -- & $-0.792$ & 66.05 & 0.283 & 0.425 & 1.074 & 0.279 & 0.286 & 1.691 & 75.85 & 95.85 & 126.60 & 97.19 & $-0.46$ & 2.62 & $-0.94$\\
\midrule
 & $H(z)$ + BAO & 0.0289 & 0.0985 & 0.296 & $-0.053$ & $-0.730$ & 65.76 & -- & -- & -- & -- & -- & -- & 22.13 & 32.13 & 41.05 & 32.51 & 0.49 & 4.06 & 0.19\\
 & \mq\ symm & -- & $-0.0103$ & 0.030 & $-0.060$ & $-3.207$ & -- & -- & -- & -- & 0.257 & 0.284 & 1.382 & 18.29 & 30.29 & 44.43 & 46.23 & $-7.87$ & $-3.16$ & 8.54\\
Non-flat & \mq\ asymm & -- & $-0.0016$ & 0.048 & $-0.095$ & $-2.947$ & -- & -- & -- & -- & 0.260 & 0.303 & 1.395 & 19.66 & 31.66 & 45.80 & 45.06 & $-6.52$ & $-1.81$ & 7.19\\
XCDM & \cq\ symm & -- & $-0.0161$ & 0.019 & $-0.035$ & $-3.815$ & -- & 0.220 & 0.428 & 0.827 & -- & -- & -- & 6.26 & 18.26 & 28.09 & 38.74 & $-10.35$ & $-7.08$ & 7.64\\
 & \cq\ asymm & -- & $-0.0130$ & 0.025 & $-0.042$ & $-4.329$ & -- & 0.228 & 0.386 & 0.828 & -- & -- & -- & 12.08 & 24.08 & 33.90 & 37.86 & $-4.43$ & $-1.16$ & 5.12\\
 & \cq\ asymm + \mq\ asymm & -- & $-0.0149$ & 0.021 & $-0.034$ & $-5.000$ & -- & 0.226 & 0.374 & 0.808 & 0.270 & 0.285 & 1.262 & 32.43 & 50.43 & 75.21 & 68.46 & $-14.51$ & $-9.00$ & $-0.71$\\
 & $H(z)$ + BAO + \cq\ asymm + \mq\ asymm & 0.0281 & 0.1046 & 0.296 & $-0.065$ & $-0.779$ & 67.10 & 0.276 & 0.434 & 1.064 & 0.274 & 0.298 & 1.676 & 75.81 & 97.81 & 131.64 & 98.92 & 1.51 & 7.66 & 0.79\\
\midrule
 & $H(z)$ + BAO & 0.0310 & 0.0900 & 0.280 & -- & 1.010 & 65.89 & -- & -- & -- & -- & -- & -- & 22.31 & 30.31 & 37.45 & 29.90 & $-1.33$ & 0.46 & $-2.42$\\
 & \mq\ symm & -- & 0.0414 & 0.136 & -- & 0.018 & -- & -- & -- & -- & 0.282 & 0.282 & 1.668 & 30.17 & 40.17 & 51.96 & 38.18 & 2.01 & 4.37 & 0.50\\
Flat & \mq\ asymm & -- & 0.0532 & 0.160 & -- & 0.047 & -- & -- & -- & -- & 0.279 & 0.284 & 1.679 & 30.20 & 40.20 & 51.99 & 38.40 & 2.02 & 4.38 & 0.52\\
\pcdm & \cq\ symm & -- & 0.0003 & 0.052 & -- & 0.030 & -- & 0.274 & 0.399 & 0.996 & -- & -- & -- & 20.67 & 30.67 & 38.85 & 33.47 & 2.05 & 3.69 & 2.37\\
 & \cq\ asymm & -- & $-0.0059$ & 0.039 & -- & 0.061 & -- & 0.270 & 0.410 & 0.977 & -- & -- & -- & 20.59 & 30.59 & 38.78 & 34.56 & 2.08 & 3.72 & 1.82\\
 & \cq\ asymm + \mq\ asymm & -- & $-0.0078$ & 0.035 & -- & 0.014 & -- & 0.263 & 0.399 & 0.984 & 0.283 & 0.265 & 1.662 & 51.19 & 67.19 & 89.21 & 72.04 & 2.25 & 5.00 & 2.94\\
 & $H(z)$ + BAO + \cq\ asymm + \mq\ asymm & 0.0289 & 0.0985 & 0.288 & -- & 0.663 & 66.68 & 0.275 & 0.430 & 1.050 & 0.278 & 0.304 & 1.668 & 75.78 & 95.78 & 126.54 & 96.48 & $-0.52$ & 2.55 & $-1.65$\\
\midrule
 & $H(z)$ + BAO & 0.0306 & 0.0920 & 0.284 & $-0.058$ & 1.200 & 65.91 & -- & -- & -- & -- & -- & -- & 22.05 & 32.05 & 40.97 & 31.30 & 0.41 & 3.98 & $-1.02$\\
 & \mq\ symm & -- & 0.1167 & 0.290 & $-0.277$ & 0.000 & -- & -- & -- & -- & 0.283 & 0.290 & 1.678 & 29.81 & 41.81 & 55.95 & 38.70 & 3.65 & 8.36 & 1.02\\
Non-flat & \mq\ asymm & -- & 0.1354 & 0.328 & $-0.310$ & 0.042 & -- & -- & -- & -- & 0.275 & 0.292 & 1.675 & 29.91 & 41.91 & 56.05 & 38.65 & 3.73 & 8.45 & 0.78\\
\pcdm & \cq\ symm & -- & 0.0489 & 0.151 & $-0.146$ & 0.065 & -- & 0.272 & 0.417 & 1.010 & -- & -- & -- & 20.42 & 32.42 & 42.24 & 34.07 & 3.80 & 7.08 & 2.96\\
 & \cq\ asymm & -- & 0.0367 & 0.126 & $-0.125$ & 0.121 & -- & 0.268 & 0.423 & 1.006 & -- & -- & -- & 20.41 & 32.41 & 42.24 & 34.92 & 3.91 & 7.18 & 2.18\\
 & \cq\ asymm + \mq\ asymm & -- & 0.0611 & 0.176 & $-0.173$ & 0.115 & -- & 0.264 & 0.421 & 1.034 & 0.271 & 0.289 & 1.659 & 50.74 & 68.74 & 93.52 & 72.86 & 3.80 & 9.31 & 3.76\\
 & $H(z)$ + BAO + \cq\ asymm + \mq\ asymm & 0.0313 & 0.0882 & 0.278 & $-0.044$ & 1.256 & 65.76 & 0.281 & 0.455 & 1.035 & 0.276 & 0.292 & 1.687 & 75.73 & 97.73 & 131.55 & 97.61 & 1.42 & 7.57 & $-0.52$\\
\bottomrule
\end{tabular}
\begin{tablenotes}[flushleft]
\item [a] \wx\ corresponds to flat/non-flat XCDM and $\alpha$ corresponds to flat/non-flat \pcdm.
\item [b] \hunit. $\Omega_b$ and $H_0$ are set to be 0.05 and 70 \hunit, respectively.
\end{tablenotes}
\end{threeparttable}%
}
\end{sidewaystable*}

\begin{sidewaystable*}
\centering
\resizebox*{\columnwidth}{0.75\columnwidth}{%
\begin{threeparttable}
\caption{One-dimensional marginalized posterior mean values and uncertainties ($\pm 1\sigma$ error bars or $2\sigma$ limits) of the parameters for all models from various combinations of data.}\label{tab:1d_BFPC11}
\begin{tabular}{lccccccccccccc}
\toprule
Model & Data set & $\Omega_{b}h^2$ & $\Omega_{c}h^2$ & $\Omega_{m0}$ & $\Omega_{k0}$ & $w_{\mathrm{X}}$/$\alpha$\tnote{a} & $H_0$\tnote{b} & $\sigma_{\mathrm{int,\,\textsc{c}}}$ & $\gamma_{\rm\textsc{c}}$ & $\beta_{\rm\textsc{c}}$ & $\sigma_{\mathrm{int,\,\textsc{m}}}$ & $\gamma_{\rm\textsc{m}}$ & $\beta_{\rm\textsc{m}}$ \\
\midrule
 & $H(z)$ + BAO & $0.0247\pm0.0030$ & $0.1186^{+0.0076}_{-0.0083}$ & $0.301^{+0.016}_{-0.018}$ & -- & -- & $69.14\pm1.85$ & -- & -- & -- & -- & -- & -- \\
 & \mq\ symm & -- & -- & $0.470^{+0.199}_{-0.426}$ & -- & -- & -- & -- & -- & -- & $0.293^{+0.023}_{-0.030}$ & $0.297\pm0.047$ & $1.698^{+0.063}_{-0.058}$ \\
Flat & \mq\ asymm & -- & -- & $0.470^{+0.196}_{-0.430}$ & -- & -- & -- & -- & -- & -- & $0.290^{+0.024}_{-0.030}$ & $0.296\pm0.047$ & $1.703^{+0.064}_{-0.057}$ \\
\lcdm & \cq\ symm & -- & -- & $<0.471$\tnote{c} & -- & -- & -- & $0.311^{+0.039}_{-0.056}$ & $0.427\pm0.043$ & $1.051\pm0.096$ & -- & -- & -- \\
 & \cq\ asymm & -- & -- & $<0.503$\tnote{c} & -- & -- & -- & $0.307^{+0.039}_{-0.056}$ & $0.441\pm0.044$ & $1.034^{+0.097}_{-0.087}$ & -- & -- & -- \\
 & \cq\ asymm + \mq\ asymm & -- & -- & $<0.444$\tnote{c} & -- & -- & -- & $0.305^{+0.037}_{-0.054}$ & $0.440\pm0.042$ & $1.030\pm0.089$ & $0.289^{+0.023}_{-0.030}$ & $0.292\pm0.045$ & $1.691\pm0.061$ \\
 & $H(z)$ + BAO + \cq\ asymm + \mq\ asymm & $0.0247\pm0.0028$ & $0.1183^{+0.0073}_{-0.0080}$ & $0.301^{+0.015}_{-0.017}$ & -- & -- & $69.15\pm1.77$ & $0.303^{+0.036}_{-0.053}$ & $0.442\pm0.039$ & $1.026^{+0.075}_{-0.065}$ & $0.288^{+0.023}_{-0.029}$ & $0.294\pm0.044$ & $1.686\pm0.056$ \\
\midrule
 & $H(z)$ + BAO & $0.0266^{+0.0039}_{-0.0045}$ & $0.1088\pm0.0166$ & $0.291\pm0.023$ & $0.059^{+0.081}_{-0.091}$ & -- & $68.37\pm2.10$ & -- & -- & -- & -- & -- & -- \\
 & \mq\ symm & -- & -- & $0.568^{+0.359}_{-0.196}$ & $-0.427^{+0.546}_{-1.453}$ & -- & -- & -- & -- & -- & $0.291^{+0.024}_{-0.030}$ & $0.315^{+0.049}_{-0.056}$ & $1.687\pm0.067$ \\
Non-flat & \mq\ asymm & -- & -- & $0.567^{+0.365}_{-0.192}$ & $-0.419^{+0.560}_{-1.449}$ & -- & -- & -- & -- & -- & $0.288^{+0.024}_{-0.031}$ & $0.316^{+0.048}_{-0.057}$ & $1.692^{+0.070}_{-0.064}$ \\
\lcdm & \cq\ symm & -- & -- & $0.417^{+0.153}_{-0.353}$ & $-0.478^{+0.346}_{-0.851}$ & -- & -- & $0.300^{+0.041}_{-0.055}$ & $0.465^{+0.047}_{-0.062}$ & $1.072\pm0.098$ & -- & -- & -- \\
 & \cq\ asymm & -- & -- & $0.467^{+0.199}_{-0.378}$ & $-0.330^{+0.534}_{-1.060}$ & -- & -- & $0.305^{+0.039}_{-0.055}$ & $0.468^{+0.046}_{-0.056}$ & $1.059^{+0.097}_{-0.086}$ & -- & -- & -- \\
 & \cq\ asymm + \mq\ asymm & -- & -- & $0.473^{+0.187}_{-0.311}$ & $-0.818^{+0.391}_{-0.637}$ & -- & -- & $0.299^{+0.036}_{-0.053}$ & $0.491^{+0.050}_{-0.064}$ & $1.073^{+0.093}_{-0.094}$ & $0.285^{+0.023}_{-0.030}$ & $0.314^{+0.048}_{-0.052}$ & $1.662\pm0.065$ \\
 & $H(z)$ + BAO + \cq\ asymm + \mq\ asymm & $0.0262^{+0.0037}_{-0.0044}$ & $0.1105\pm0.0165$ & $0.292\pm0.022$ & $0.047^{+0.079}_{-0.089}$ & -- & $68.55\pm2.05$ & $0.303^{+0.035}_{-0.052}$ & $0.441\pm0.039$ & $1.023^{+0.075}_{-0.064}$ & $0.288^{+0.023}_{-0.030}$ & $0.294\pm0.043$ & $1.685\pm0.055$ \\
\midrule
 & $H(z)$ + BAO & $0.0295^{+0.0042}_{-0.0050}$ & $0.0969^{+0.0178}_{-0.0152}$ & $0.289\pm0.020$ & -- & $-0.784^{+0.140}_{-0.107}$ & $66.22^{+2.31}_{-2.54}$ & -- & -- & -- & -- & -- & -- \\
 & \mq\ symm & -- & -- & $<0.515$\tnote{c} & -- & $<-0.371$ & -- & -- & -- & -- & $0.291^{+0.024}_{-0.030}$ & $0.294\pm0.048$ & $1.640^{+0.118}_{-0.073}$ \\
Flat & \mq\ asymm & -- & -- & $<0.512$\tnote{c} & -- & $<-0.370$ & -- & -- & -- & -- & $0.288^{+0.024}_{-0.030}$ & $0.295\pm0.047$ & $1.644^{+0.118}_{-0.072}$ \\
XCDM & \cq\ symm & -- & -- & $<0.737$ & -- & $<-0.926$ & -- & $0.292^{+0.041}_{-0.056}$ & $0.403\pm0.047$ & $0.957\pm0.128$ & -- & -- & -- \\
 & \cq\ asymm & -- & -- & $<0.840$ & -- & $<-0.636$ & -- & $0.296^{+0.041}_{-0.056}$ & $0.426\pm0.049$ & $0.970^{+0.136}_{-0.108}$ & -- & -- & -- \\
 & \cq\ asymm + \mq\ asymm & -- & -- & $<0.563$ & -- & $<-1.509$ & -- & $0.282^{+0.038}_{-0.054}$ & $0.405\pm0.047$ & $0.900^{+0.122}_{-0.121}$ & $0.283^{+0.023}_{-0.029}$ & $0.282^{+0.042}_{-0.046}$ & $1.557^{+0.117}_{-0.101}$ \\
 & $H(z)$ + BAO + \cq\ asymm + \mq\ asymm & $0.0292^{+0.0040}_{-0.0050}$ & $0.0983^{+0.0181}_{-0.0144}$ & $0.290^{+0.020}_{-0.018}$ & -- & $-0.799^{+0.143}_{-0.111}$ & $66.45^{+2.28}_{-2.53}$ & $0.305^{+0.036}_{-0.053}$ & $0.443\pm0.039$ & $1.021^{+0.077}_{-0.065}$ & $0.289^{+0.023}_{-0.030}$ & $0.295\pm0.044$ & $1.684\pm0.056$ \\
\midrule
 & $H(z)$ + BAO & $0.0294^{+0.0047}_{-0.0050}$ & $0.0980^{+0.0186}_{-0.0187}$ & $0.292\pm0.025$ & $-0.027\pm0.109$ & $-0.770^{+0.149}_{-0.098}$ & $66.13^{+2.35}_{-2.36}$ & -- & -- & -- & -- & -- & -- \\
 & \mq\ symm & -- & -- & $0.526^{+0.351}_{-0.268}$ & $-0.235^{+0.555}_{-0.936}$ & $-2.416^{+1.894}_{-1.260}$ & -- & -- & -- & -- & $0.291^{+0.024}_{-0.031}$ & $0.313^{+0.049}_{-0.055}$ & $1.649^{+0.124}_{-0.079}$ \\
 & \mq\ asymm & -- & -- & $0.523^{+0.341}_{-0.280}$ & $-0.245^{+0.544}_{-0.917}$ & $-2.448^{+1.834}_{-1.321}$ & -- & -- & -- & -- & $0.287^{+0.024}_{-0.030}$ & $0.314^{+0.049}_{-0.055}$ & $1.651^{+0.122}_{-0.079}$ \\
Non-flat & \cq\ symm & -- & -- & $0.399^{+0.140}_{-0.375}$ & $-0.247^{+0.360}_{-0.645}$ & $-2.636^{+1.523}_{-1.570}$ & -- & $0.297^{+0.042}_{-0.057}$ & $0.443^{+0.045}_{-0.055}$ & $1.028^{+0.121}_{-0.106}$ & -- & -- & -- \\
XCDM & \cq\ asymm & -- & -- & $0.452^{+0.186}_{-0.394}$ & $-0.168^{+0.451}_{-0.789}$ & $-2.573^{+1.568}_{-1.605}$ & -- & $0.302^{+0.040}_{-0.056}$ & $0.453^{+0.046}_{-0.051}$ & $1.025^{+0.115}_{-0.096}$ & -- & -- & -- \\
 & \cq\ asymm + \mq\ asymm & -- & -- & $0.338^{+0.101}_{-0.299}$ & $-0.410^{+0.368}_{-0.222}$ & $<-1.124$ & -- & $0.282^{+0.037}_{-0.053}$ & $0.456^{+0.047}_{-0.056}$ & $0.966^{+0.119}_{-0.110}$ & $0.280^{+0.023}_{-0.030}$ & $0.319^{+0.048}_{-0.054}$ & $1.526^{+0.132}_{-0.108}$ \\
 & $H(z)$ + BAO + \cq\ asymm + \mq\ asymm & $0.0290^{+0.0044}_{-0.0053}$ & $0.0997^{+0.0186}_{-0.0188}$ & $0.293\pm0.025$ & $-0.031\pm0.108$ & $-0.787^{+0.165}_{-0.102}$ & $66.41^{+2.26}_{-2.49}$ & $0.305^{+0.036}_{-0.053}$ & $0.444\pm0.040$ & $1.023^{+0.077}_{-0.065}$ & $0.289^{+0.023}_{-0.030}$ & $0.295\pm0.045$ & $1.685\pm0.056$ \\
\midrule
 & $H(z)$ + BAO & $0.0320^{+0.0054}_{-0.0041}$ & $0.0855^{+0.0175}_{-0.0174}$ & $0.275\pm0.023$ & -- & $1.267^{+0.536}_{-0.807}$ & $65.47^{+2.22}_{-2.21}$ & -- & -- & -- & -- & -- & -- \\
 & \mq\ symm & -- & -- & -- & -- & -- & -- & -- & -- & -- & $0.294^{+0.023}_{-0.029}$ & $0.301\pm0.046$ & $1.719\pm0.054$ \\
Flat & \mq\ asymm & -- & -- & -- & -- & -- & -- & -- & -- & -- & $0.290^{+0.024}_{-0.030}$ & $0.301\pm0.046$ & $1.724^{+0.055}_{-0.051}$ \\
\pcdm & \cq\ symm & -- & -- & $<0.553$\tnote{c} & -- & $<6.291$\tnote{c} & -- & $0.318^{+0.039}_{-0.055}$ & $0.436\pm0.043$ & $1.086\pm0.090$ & -- & -- & -- \\
 & \cq\ asymm & -- & -- & $<0.565$\tnote{c} & -- & -- & -- & $0.314^{+0.038}_{-0.055}$ & $0.450\pm0.044$ & $1.070^{+0.092}_{-0.075}$ & -- & -- & -- \\
 & \cq\ asymm + \mq\ asymm & -- & -- & $<0.537$\tnote{c} & -- & $<6.202$\tnote{c} & -- & $0.312^{+0.037}_{-0.054}$ & $0.449\pm0.043$ & $1.069^{+0.091}_{-0.074}$ & $0.289^{+0.023}_{-0.030}$ & $0.299\pm0.046$ & $1.717^{+0.059}_{-0.053}$ \\
 & $H(z)$ + BAO + \cq\ asymm + \mq\ asymm & $0.0318^{+0.0053}_{-0.0045}$ & $0.0866^{+0.0190}_{-0.0171}$ & $0.275\pm0.023$ & -- & $1.202^{+0.490}_{-0.862}$ & $65.68^{+2.20}_{-2.19}$ & $0.306^{+0.036}_{-0.053}$ & $0.444\pm0.040$ & $1.019^{+0.078}_{-0.066}$ & $0.289^{+0.023}_{-0.030}$ & $0.295\pm0.044$ & $1.683\pm0.056$ \\
\midrule
 & $H(z)$ + BAO & $0.0320^{+0.0057}_{-0.0038}$ & $0.0865^{+0.0172}_{-0.0198}$ & $0.277^{+0.023}_{-0.026}$ & $-0.034^{+0.087}_{-0.098}$ & $1.360^{+0.584}_{-0.819}$ & $65.53\pm2.19$ & -- & -- & -- & -- & -- & -- \\
 & \mq\ symm & -- & -- & $0.476^{+0.230}_{-0.396}$ & $0.042^{+0.390}_{-0.378}$ & -- & -- & -- & -- & -- & $0.294^{+0.023}_{-0.029}$ & $0.300\pm0.046$ & $1.721\pm0.054$ \\
Non-flat & \mq\ asymm & -- & -- & $0.473^{+0.216}_{-0.409}$ & $0.040^{+0.395}_{-0.383}$ & -- & -- & -- & -- & -- & $0.291^{+0.024}_{-0.030}$ & $0.301\pm0.047$ & $1.724\pm0.056$ \\
\pcdm & \cq\ symm & -- & -- & $<0.542$\tnote{c} & $0.104^{+0.381}_{-0.362}$ & -- & -- & $0.319^{+0.039}_{-0.055}$ & $0.437\pm0.043$ & $1.089\pm0.089$ & -- & -- & -- \\
 & \cq\ asymm & -- & -- & $0.427^{+0.153}_{-0.408}$ & $0.096^{+0.359}_{-0.337}$ & $4.747^{+2.065}_{-4.304}$ & -- & $0.313^{+0.037}_{-0.055}$ & $0.451\pm0.043$ & $1.076^{+0.087}_{-0.074}$ & -- & -- & -- \\
 & \cq\ asymm + \mq\ asymm & -- & -- & $<0.536$\tnote{c} & $0.088^{+0.384}_{-0.364}$ & $<6.162$\tnote{c} & -- & $0.312^{+0.037}_{-0.054}$ & $0.450\pm0.043$ & $1.072^{+0.088}_{-0.076}$ & $0.290^{+0.023}_{-0.030}$ & $0.299\pm0.046$ & $1.719\pm0.055$\\
 & $H(z)$ + BAO + \cq\ asymm + \mq\ asymm & $0.0317^{+0.0058}_{-0.0043}$ & $0.0884^{+0.0183}_{-0.0203}$ & $0.278^{+0.024}_{-0.026}$ & $-0.044^{+0.090}_{-0.094}$ & $1.320^{+0.572}_{-0.869}$ & $65.77\pm2.21$ & $0.306^{+0.036}_{-0.053}$ & $0.445\pm0.040$ & $1.021^{+0.079}_{-0.066}$ & $0.289^{+0.023}_{-0.030}$ & $0.296\pm0.045$ & $1.684\pm0.056$\\
\bottomrule
\end{tabular}
\begin{tablenotes}[flushleft]
\item [a] \wx\ corresponds to flat/non-flat XCDM and $\alpha$ corresponds to flat/non-flat \pcdm.
\item [b] \hunit. $\Omega_b$ and $H_0$ are set to be 0.05 and 70 \hunit, respectively.
\item [c] This is the 1$\sigma$ limit. The 2$\sigma$ limit is set by the prior and not shown here.
\end{tablenotes}
\end{threeparttable}%
}
\end{sidewaystable*}


\cleardoublepage


\chapter{Conclusion}
\label{makereference13}

In this study we have utilized various non-CMB data sets to constrain cosmological parameters in six different general relativistic dark energy models. By simultaneously constraining both the cosmological and correlation parameters, we discovered that certain correlations can standardize some GRBs and reverberation-measured QSOs, as their constraints are relatively model-independent.

More specifically, we found that the 118 long GRBs (A118) can be standardized through the Amati correlation; the 31 long and 5 short GRBs with the plateau phase dominated by magnetic dipole (MD) radiation, and 24 long GRBs with the plateau phase dominated by gravitational wave (GW) emission can be standardized through the two-parameter Dainotti correlation; the 50 Platinum and 95 long GRBs can be standardized through the three-parameter Dainotti correlation; the 78 reverberation-measured \mii\ QSOs can be standardized through the radius-luminosity ($R-L$) correlation; and the 38 reverberation-measured \civ\ QSOs can be standardized through another $R-L$ correlation.

Although the cosmological constraints obtained from these standardized probes are relatively weaker than those obtained from well-established $H(z)$ + BAO data, their constraints are still mutually consistent and, therefore, can be jointly analyzed. Among the GRB data sets, A118 provides slightly tighter constraints than the others, making it the preferred choice for joint analyses. However, we are still investigating other GRB data sets, and we expect that more and better quality GRBs will provide fresh insights on cosmological parameter values. 

QSO angular size, \hiig, and SN Ia data are complementary to the well-established $H(z)$ + BAO data. Although they provide weaker cosmological constraints than $H(z)$ + BAO data, they provide tighter constraints than the standardized GRBs and reverberation-measured QSOs. All these lower-redshift non-CMB data provide mutually consistent cosmological constraints and can be jointly analyzed.

Using advanced data analysis methods and more up-to-date measurements, we have obtained somewhat model-independent constraints on the Hubble constant ($H_0=69.7\pm1.2$ \hunit) and the non-relativistic matter density parameter ($\Om=0.3\pm0.02$). Nonetheless, the current cosmological parameter constraints from lower-redshift standardized non-CMB measurements are not comparable to those from \textit{Planck} CMB anisotropy data. We hope that in the future, the quality and quantity of non-CMB data will be significantly better, allowing us to measure cosmological parameter error bars comparable to those obtained from \textit{Planck} CMB anisotropy data.






\cleardoublepage
\phantomsection


\addcontentsline{toc}{chapter}{Bibliography}
\bibdata{references}
\bibliography{references}


\appendix

\cleardoublepage

\chapter{QSO-Flux}
\label{AppendixA}

QSOs obey a nonlinear relation between their luminosities in the X-ray and UV bands. Using a sample of 808 QSOs in the redshift range $0.061 \leq z \leq 6.280$, \cite{RisalitiLusso2015} confirmed that this relation can be written
\begin{equation}
\label{eq:LX-LUV}
  \log L_X=\beta+\gamma\log L_{UV},
\end{equation}
where $L_X$ and $L_{UV}$ are the X-ray and UV luminosities of the QSOs. To make contact with observations, equation \eqref{eq:LX-LUV} must be expressed in terms of the fluxes $F_X$ and $F_{UV}$ measured at fixed rest-frame wavelengths in the X-ray and UV bands, respectively. With this, equation (\ref{eq:LX-LUV}) becomes
\be
    \log F_X=\beta+(\gamma-1)\log 4\pi+\gamma\log F_{UV} +2(\gamma-1)\log D_{L}.
\ee
Here $D_L$ is the luminosity distance, which depends on the parameters of our cosmological models. We also treat the slope $\gamma$ and intercept $\beta$ as free parameters in our cosmological model fits.

For QSO-Flux data, the natural log of its likelihood function is
\begin{equation*}
    \ln\mathcal{L_{\rm QF}}=-\frac{1}{2}\sum^{N}_{i=1}\Bigg[\frac{\big[\log(F^{\rm{obs}}_X)_i-\log(F^{\rm{th}}_X)_i\big]^2}{s_i^2}+\ln(2\pi s_i^2)\Bigg], \label{eq:LH_QF} 
\end{equation*}
where $s^2_i = \sigma^2_i + \delta^2$. Here $\sigma_i$ is the uncertainty in $\log\left(F^{\rm obs}_X\right)_i$, and $\delta$ is the global intrinsic dispersion in the data (including the systematic uncertainties), which we treat as a free parameter in our cosmological model fits. We use the \cite{RisalitiLusso2019} compilation of 1598 QSO-Flux measurements in the range $0.036 \leq z\leq 5.1003$. The flat priors of cosmological parameters and the Amati relation parameters are in Sec. \ref{makereference4.3} and, as in \cite{KhadkaRatra2020b}, the flat priors of the parameters $\delta$, $\gamma$, and $\beta$ are non-zero over $0\leq\delta\leq e^{10}$, $-2\leq\gamma\leq2$, and $0\leq\beta\leq11$, respectively. 

\begin{figure*}
\centering
  \subfloat[]{%
    \includegraphics[width=3.25in,height=3.25in]{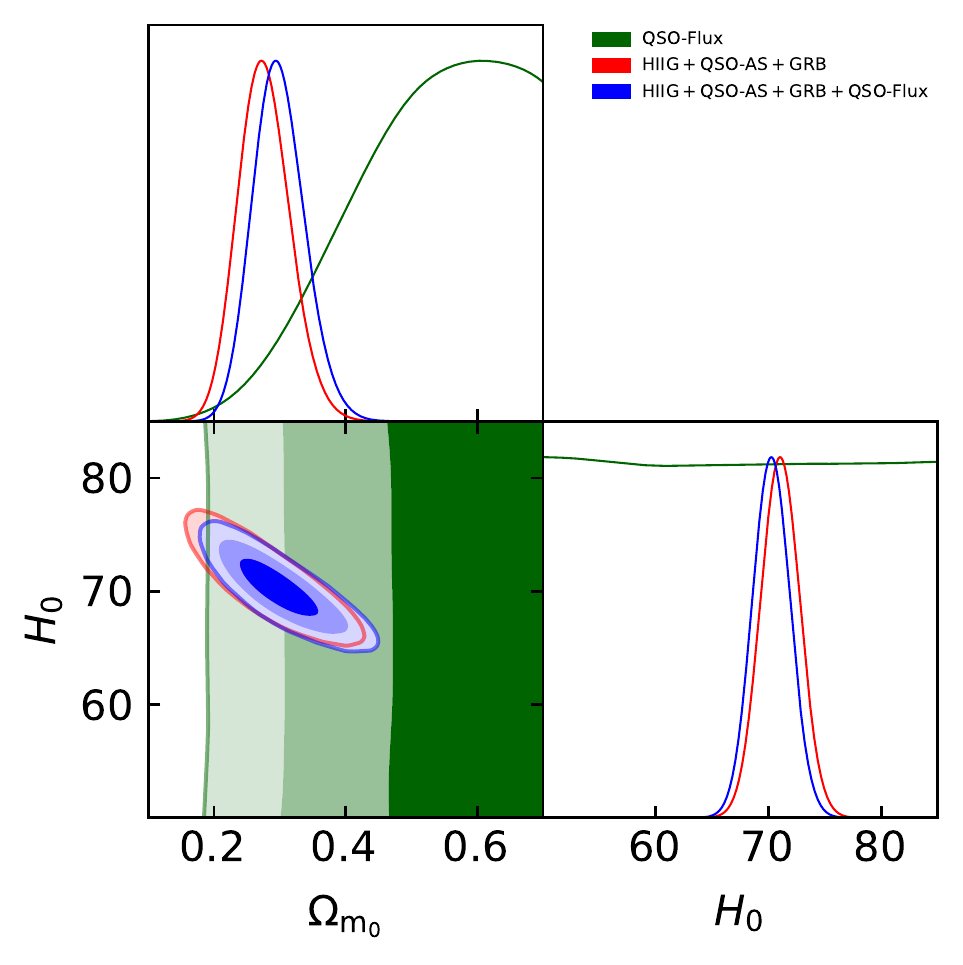}}
  \subfloat[]{%
    \includegraphics[width=3.25in,height=3.25in]{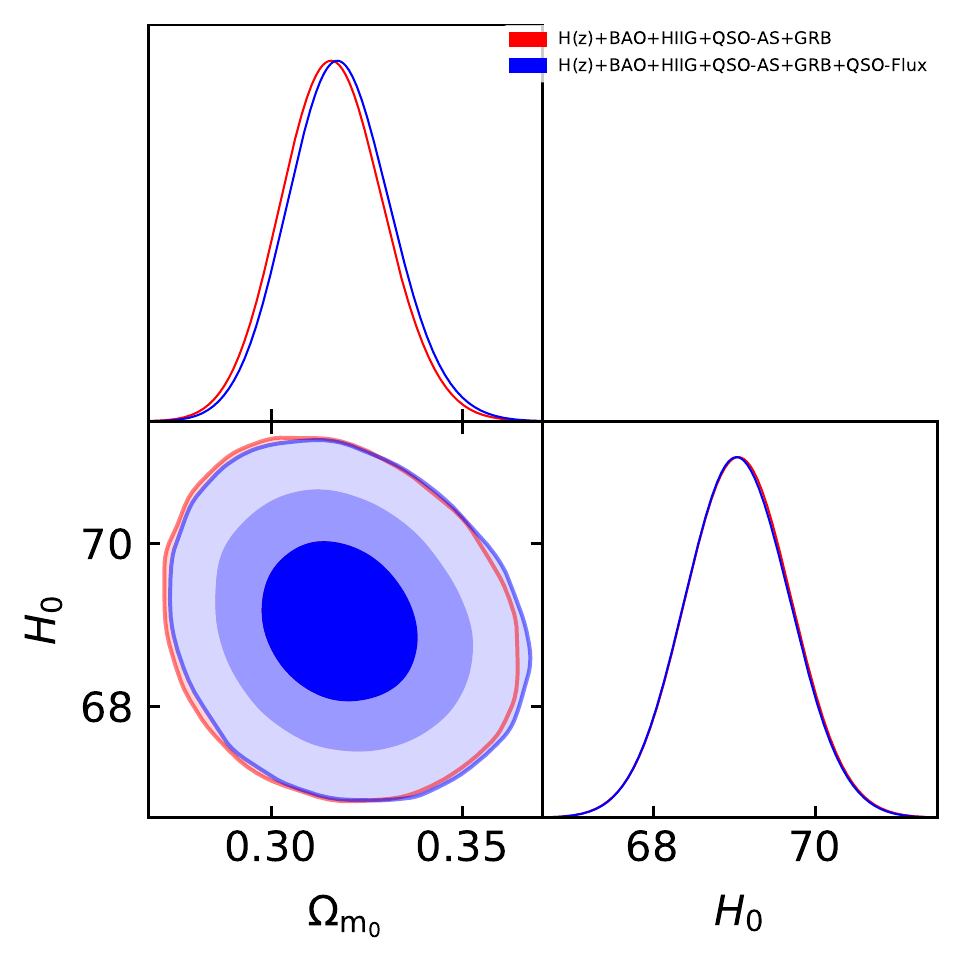}}\\
\caption{Same as Fig. \ref{fig1} (flat \lcdm) but for different combinations of data and showing only cosmological parameters.}
\label{fig01aA}
\end{figure*}

\begin{figure*}
\centering
  \subfloat[]{%
    \includegraphics[width=3.25in,height=3.25in]{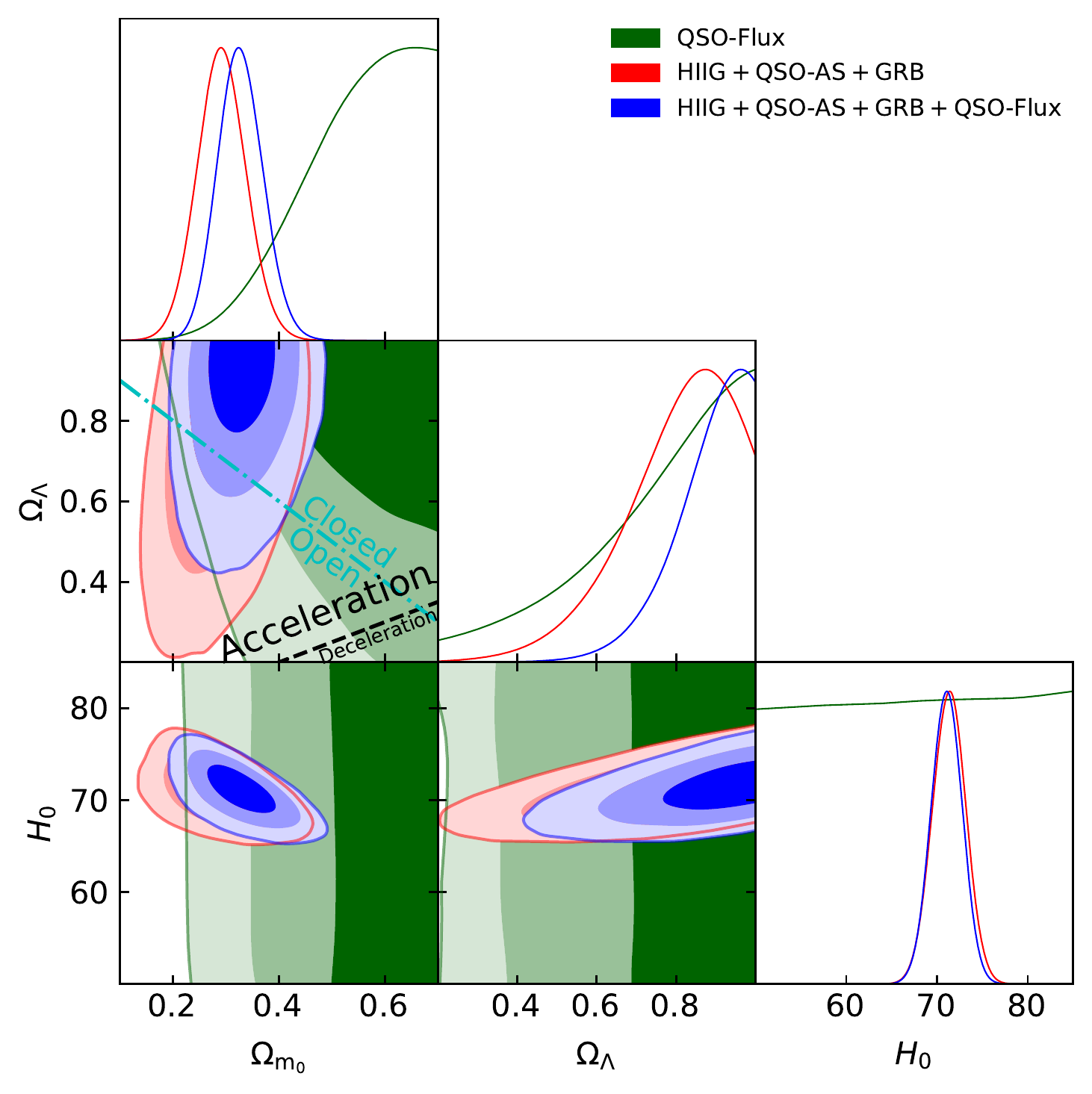}}
  \subfloat[]{%
    \includegraphics[width=3.25in,height=3.25in]{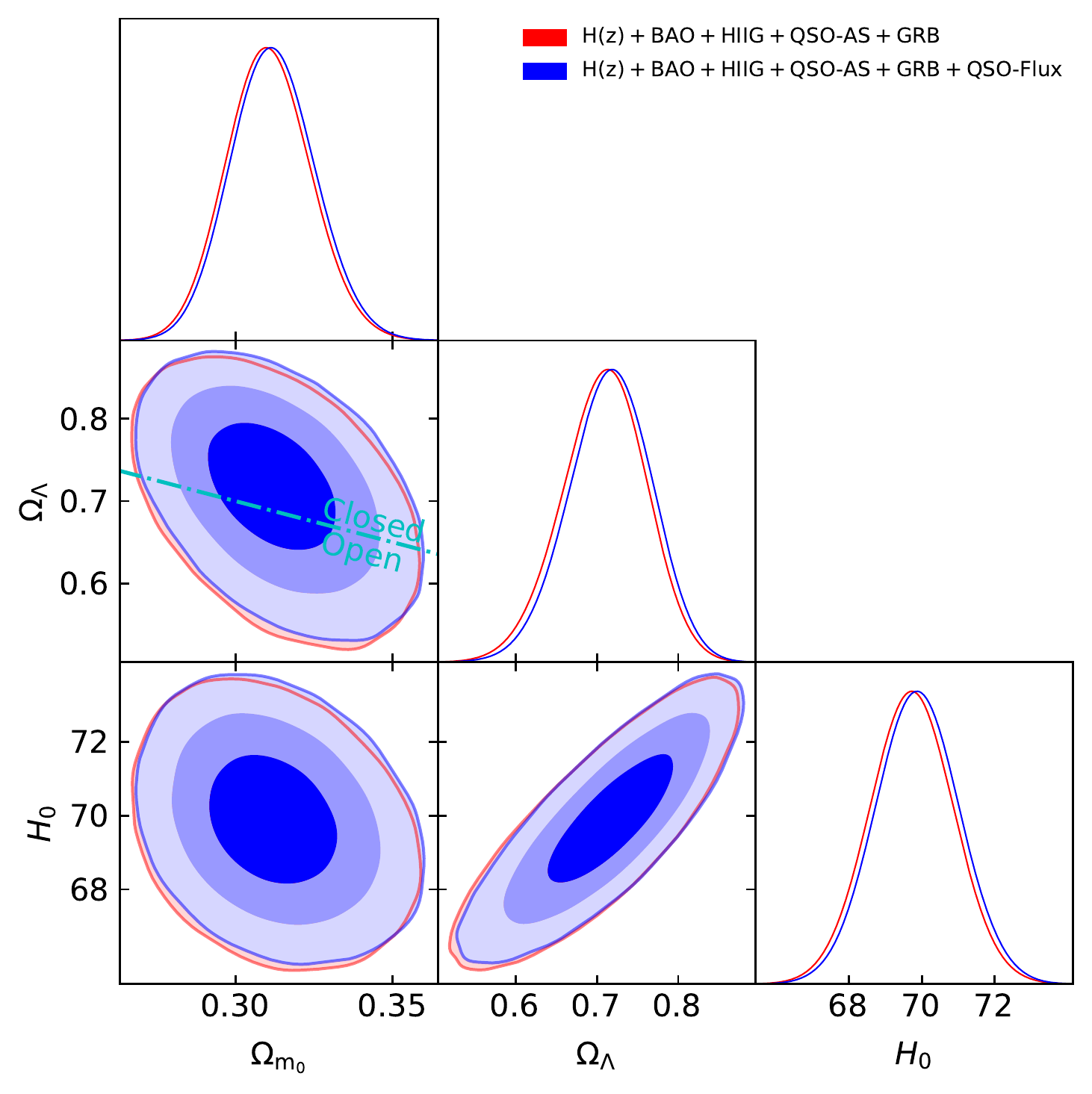}}\\
\caption{Same as Fig. \ref{fig2} (non-flat \lcdm) but for different combinations of data and showing only cosmological parameters.}
\label{fig02aA}
\end{figure*}

\begin{figure*}
\centering
  \subfloat[]{%
    \includegraphics[width=3.25in,height=3.25in]{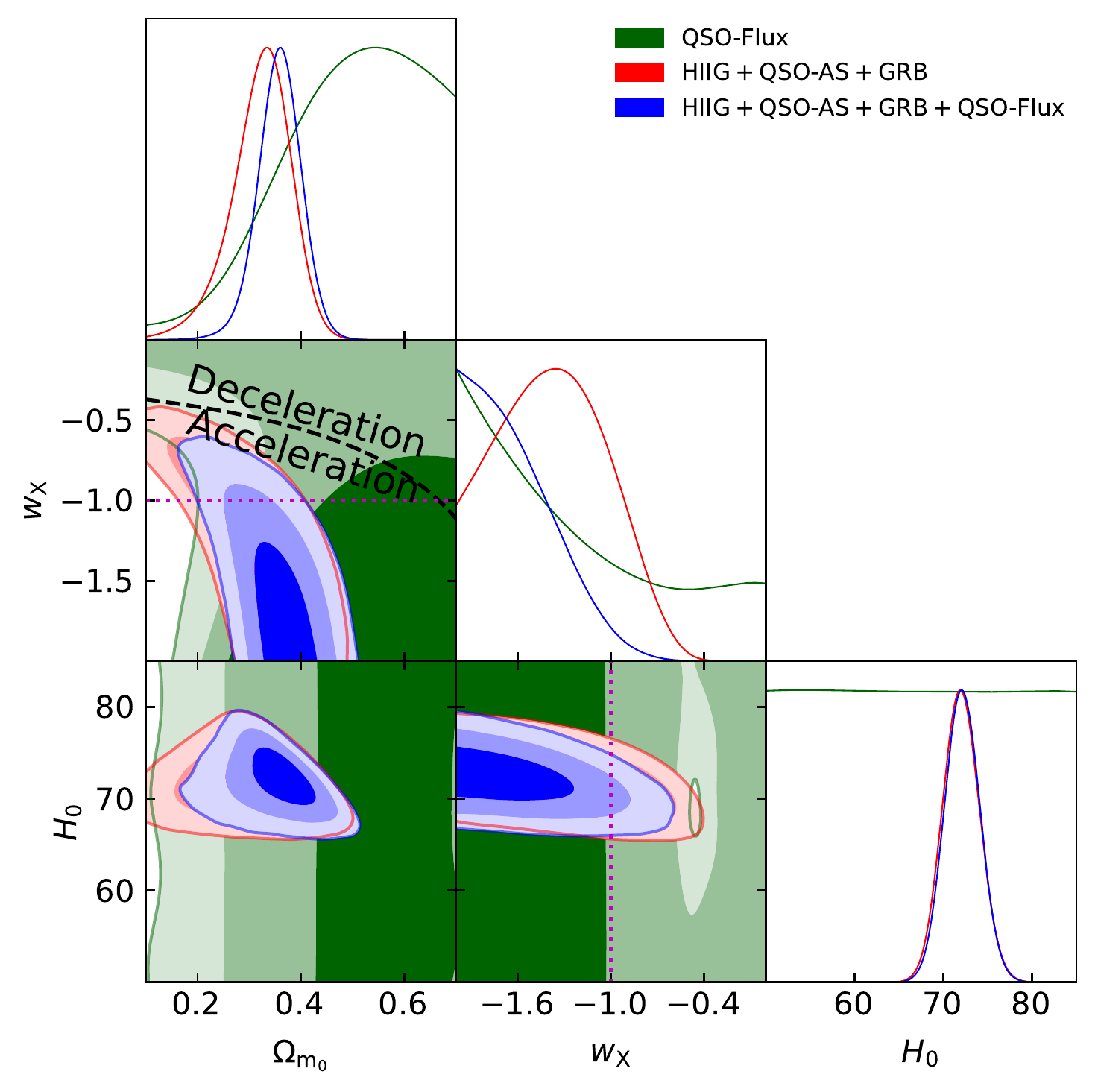}}
  \subfloat[]{%
    \includegraphics[width=3.25in,height=3.25in]{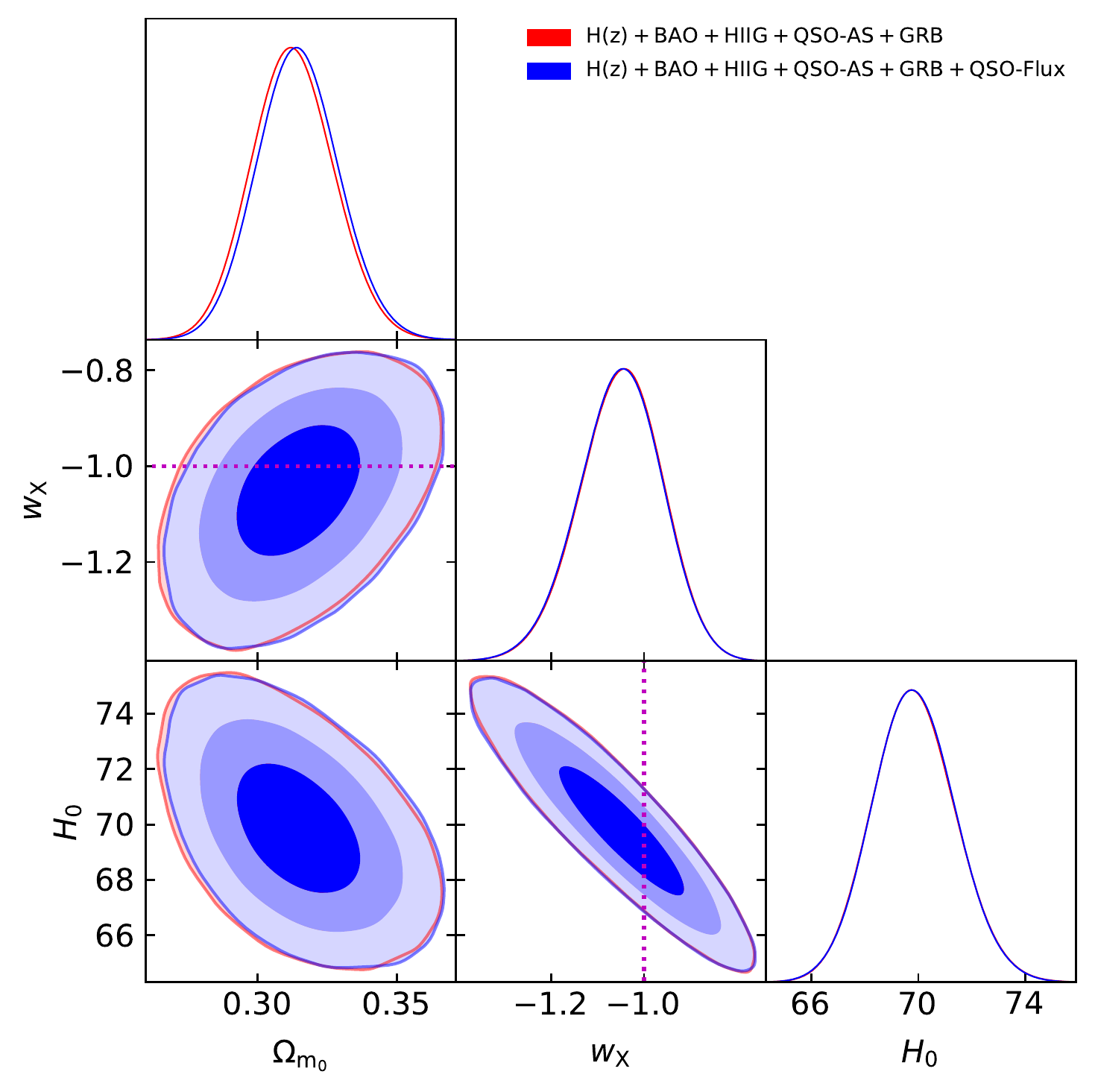}}\\
\caption{Same as Fig. \ref{fig3} (flat XCDM) but for different combinations of data and showing only cosmological parameters.}
\label{fig03aA}
\end{figure*}

\begin{figure*}
\centering
  \subfloat[]{%
    \includegraphics[width=3.25in,height=3.25in]{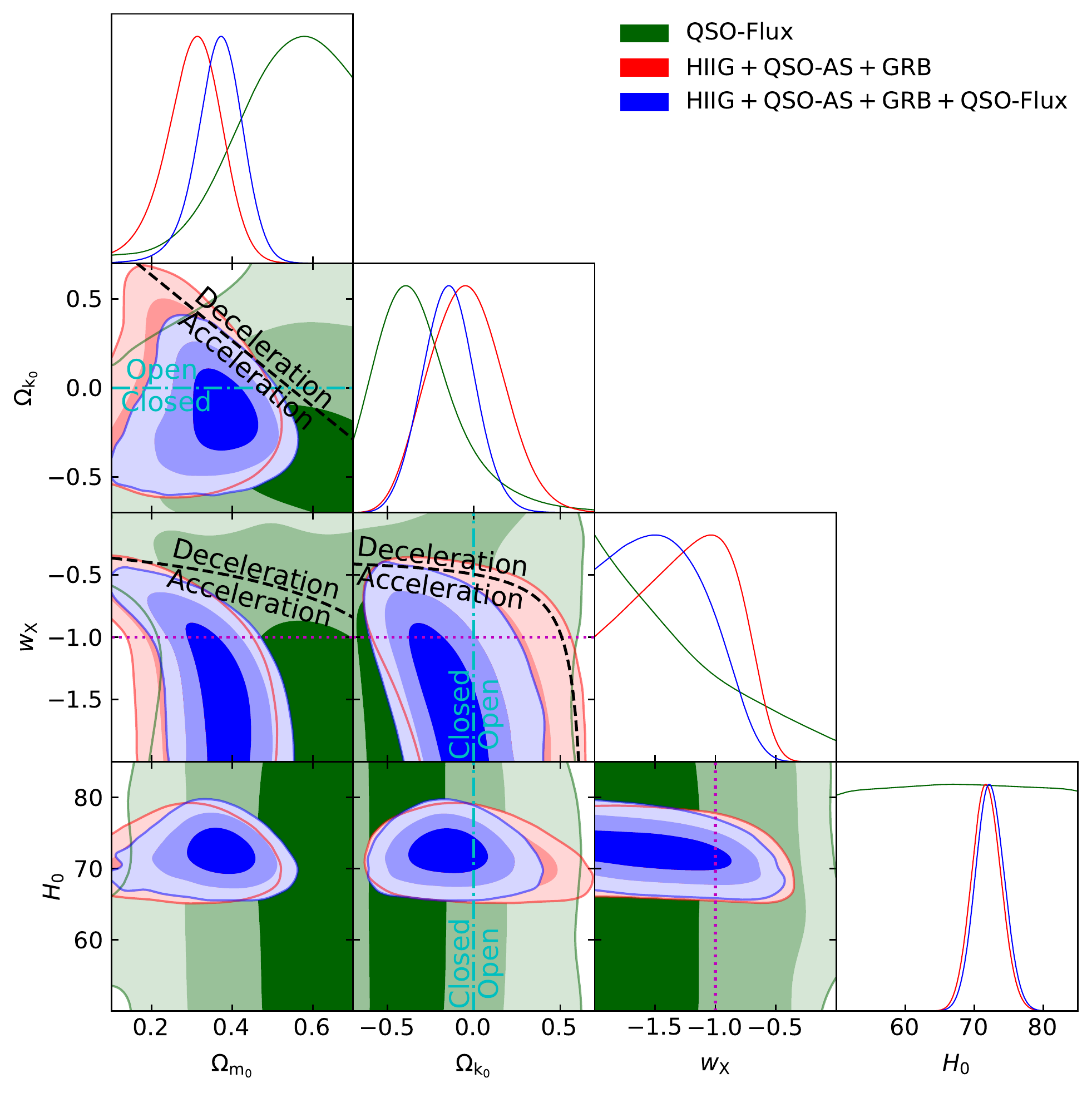}}
  \subfloat[]{%
    \includegraphics[width=3.25in,height=3.25in]{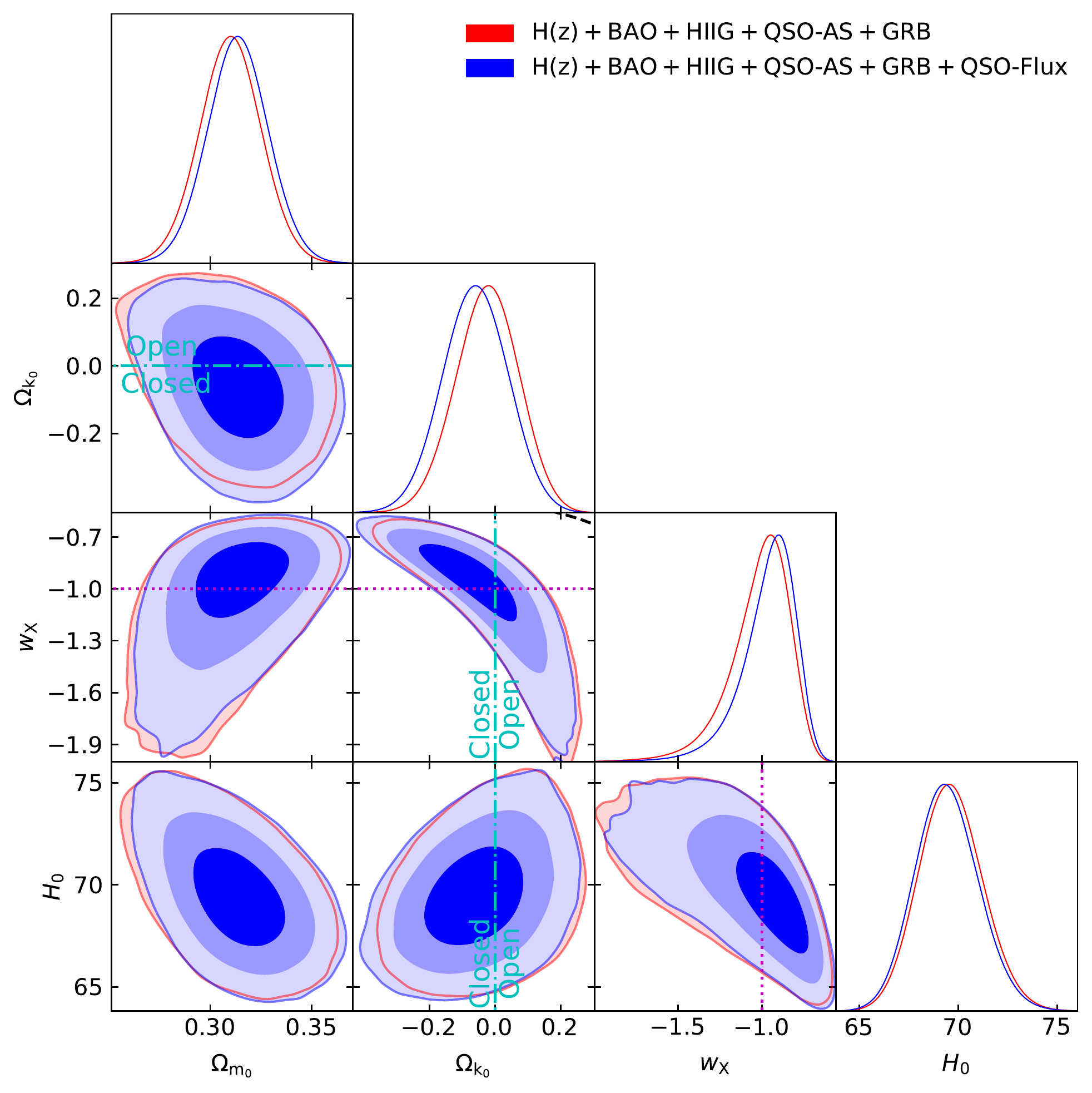}}\\
\caption{Same as Fig. \ref{fig4} (non-flat XCDM) but for different combinations of data and showing only cosmological parameters.}
\label{fig04aA}
\end{figure*}

As discussed in \cite{KhadkaRatra2020b} the QSO-Flux data alone favors large \om\ values for the physically-motivated flat and non-flat \lcdm\ and \pcdm\ models. \cite{RisalitiLusso2019} and \cite{KhadkaRatra2020b} note that this is largely a consequence of the $z\sim 2$--5 QSO data. While these large \om\ values differ from almost all other measurements of \om, the QSO-Flux data have larger error bars and their cosmological constraint contours are not in conflict with those from other data sets. For these reasons we have used the QSO-Flux data, but in this Section, and we have not computed QSO-Flux data results for the \pcdm\ cases (these being computationally demanding). We briefly summarize our constraints, listed in Tables \ref{tab:BFPaA} and \ref{tab:1d_BFPaA} and shown in Figs. \ref{fig01aA}--\ref{fig04aA}, below.

\section{QSO-Flux constraints}

Except for flat \lcdm, the constraints on \om\ in the QSO-Flux only case are 2$\sigma$ larger than those in the combined HzBHQASQFG case (see Sec. \ref{sec:A3}). QSO-Flux data cannot constrain $\alpha$, nor can they constrain $H_0$ (for the same reason that GRB data cannot constrain this parameter; see Section \ref{subsec:GRB}). QSO-Flux data set upper limits on $w_{\rm X}$ for flat and non-flat XCDM, with $w_{\rm X}=-1$ within the 1$\sigma$ range.

\section{\hiig, QSO-AS, QSO-Flux, and GRB (HQASQFG) constraints}

When adding QSO-Flux to HQASG data, the joint constraints favor larger \om\ and lower \ok. In non-flat $\Lambda$CDM closed geometry is favored at 3.24$\sigma$. The $H_0$ constraints are only mildly affected by the addition of the QSO-Flux data. The constraint on $w_{\rm X}$ changes from $-1.379^{+0.361}_{-0.375}$ in the HQASG case to $<-1.100$ (2$\sigma$ limit) in the HQASQFG case for flat XCDM, while for non-flat XCDM, the constraint on $w_{\rm X}$ in the HQASQFG case is 0.40$\sigma$ lower than that in the HQASG case and is 1.80$\sigma$ away from $w_{\rm X}=-1$.

\subsection{$H(z)$, BAO, \hiig, QSO-AS, QSO-Flux, and GRB (HzBHQASQFG) constraints}
\label{sec:A3}

When adding QSO-Flux to the HzBHQASG combination, the \om\ central values are only slightly larger because the $H(z)$ + BAO data dominate this compilation. The joint-constraint central $\Omega_{k0}$ values are lower, and consistent with flat geometry, while the constraints on $H_0$ from this combination are almost unaltered. The constraints on $w_{\rm X}$ are 0.02$\sigma$ lower and 0.23$\sigma$ higher for flat and non-flat XCDM, respectively, both being consistent with $w_{\rm X}=-1$ within 1$\sigma$.

\section{Model comparison}

From Table \ref{tab:BFPaA}, we see that the reduced $\chi^2$ of the QSO-Flux case for all models is near unity ($\sim1.01$) and that the reduced $\chi^2$ of cases that include QSO-Flux is brought down to $\sim1.24$--1.26 for all models. Based on the $BIC$ (see Table \ref{tab:BFPaA}), flat \lcdm\ is the most favored model, while based on the $AIC$, non-flat XCDM, flat XCDM, and flat \lcdm\ are the most favored models for the QSO-Flux, HQASQFG, and HzBHQASQFG combinations, respectively.\footnote{Note that based on the $\Delta \chi^2$ results of Table \ref{tab:BFPaA} flat \lcdm\ has the minimum $\chi^2$ in the QSO-Flux, HQASQFG, and HzBHQASQFG cases.} From $\Delta AIC$ and $\Delta BIC$, we find mostly weak or positive evidence against the models, and only in a few cases do we find strong evidence against our models. According to $\Delta BIC$, the evidence against non-flat XCDM is strong for the QSO-Flux data, and very strong for the HQASQFG and HzBHQASQFG data, and the evidence against non-flat \lcdm\ is strong for the HzBHQASQFG data. According to $\Delta AIC$, the evidence against flat XCDM is strong for the HzBHQASQFG data.

\begin{table*}
\centering
\resizebox*{1\columnwidth}{0.8\columnwidth}{%
\begin{threeparttable}
\caption{Unmarginalized best-fitting parameter values for all models from various combinations of data.}\label{tab:BFPaA}
\begin{tabular}{lccccccccccccccccccccc}
\toprule
 Model & Data set & $\Omega_{\mathrm{m_0}}$ & $\Omega_{\Lambda}$ & $\Omega_{\mathrm{k_0}}$ & $w_{\mathrm{X}}$ & $\alpha$ & $H_0$\tnote{c} & $\sigma_{\mathrm{ext}}$ & $a$ & $b$ & $\delta$ & $\gamma$ & $\beta$ & $\chi^2$ & $\nu$ & $-2\ln\mathcal{L}_{\mathrm{max}}$ & $AIC$ & $BIC$ & $\Delta\chi^2$ & $\Delta AIC$ & $\Delta BIC$ \\
\midrule
Flat \lcdm & QSO-Flux & 0.315 & 0.685 & -- & -- & -- & 68.69 & -- & -- & -- & -- & -- & -- & 1603.28 & 1593 & -50.13 & -40.13 & -13.24 & 0.00 & 1.62 & 0.00\\
 & HQASG\tnote{d} & 0.271 & 0.729 & -- & -- & -- & 71.13 & 0.407 & 50.18 & 1.138 & -- & -- & -- & 879.42 & 387 & 895.05 & 905.05 & 924.91 & 0.12 & 0.00 & 0.00\\
 & HQASQFG\tnote{e} & 0.305 & 0.695 & -- & -- & -- & 70.01 & 0.399 & 50.20 & 1.132 & 0.231 & 0.639 & 7.083 & 2480.01 & 1982 & 848.53 & 864.53 & 909.29 & 0.00 & 2.58 & 0.00\\
 & HzBHQASG\tnote{f} & 0.317 & 0.683 & -- & -- & -- & 69.06 & 0.404 & 50.19 & 1.134 & -- & -- & -- & 903.61 & 429 & 917.79 & 927.79 & 948.16 & 1.52 & 0.00 & 0.00\\
 & HzBHQASQFG\tnote{g} & 0.317 & 0.683 & -- & -- & -- & 69.06 & 0.399 & 50.23 & 1.119 & 0.232 & 0.637 & 7.144 & 2499.87 & 2024 & 870.31 & 886.31 & 931.25 & 0.00 & 0.00 & 0.00\\
\\
Non-flat \lcdm & QSO-Flux & 0.540 & 0.985 & $-0.525$ & -- & -- & 75.75 & -- & -- & -- & 0.230 & 0.611 & 7.888 & 1603.83 & 1592 & -53.25 & -41.25 & -8.99 & 0.55 & 0.50 & 4.25\\
 & HQASG\tnote{d} & 0.291 & 0.876 & $-0.167$ & -- & -- & 72.00 & 0.406 & 50.22 & 1.120 & -- & -- & -- & 879.30 & 386 & 894.02 & 906.02 & 929.85 & 0.00 & 0.97 & 4.94\\
 & HQASQFG\tnote{e} & 0.325 & 0.944 & $-0.269$ & -- & -- & 71.49 & 0.404 & 50.21 & 1.116 & 0.230 & 0.632 & 7.304 & 2486.97 & 1981 & 844.38 & 862.38 & 912.75 & 6.96 & 0.43 & 3.46\\
 & HzBHQASG\tnote{f} & 0.309 & 0.716 & $-0.025$ & -- & -- & 69.77 & 0.402 & 50.17 & 1.141 & -- & -- & -- & 904.47 & 428 & 917.17 & 929.17 & 953.61 & 2.38 & 1.38 & 5.45\\
 & HzBHQASQFG\tnote{g} & 0.309 & 0.709 & $-0.018$ & -- & -- & 69.59 & 0.412 & 50.21 & 1.128 & 0.231 & 0.637 & 7.151 & 2503.43 & 2023 & 869.71 & 887.71 & 938.26 & 3.56 & 1.40 & 7.01 \\
\\
Flat XCDM & QSO-Flux & 0.477 & -- & -- & $-1.988$ & -- & 60.86 & -- & -- & -- & 0.230 & 0.625 & 7.530 & 1604.18 & 1592 & -52.13 & -40.13 & -7.88 & 0.90 & 1.62 & 5.36\\
 & HQASG\tnote{d} & 0.320 & -- & -- & $-1.306$ & -- & 72.03 & 0.404 & 50.20 & 1.131 & -- & -- & -- & 880.47 & 386 & 894.27 & 906.27 & 930.10 & 1.17 & 1.22 & 5.19\\
 & HQASQFG\tnote{e} & 0.370 & -- & -- & $-1.980$ & -- & 73.66 & 0.399 & 50.18 & 1.129 & 0.231 & 0.632 & 7.301 & 2485.59 & 1981 & 843.95 & 861.95 & 912.31 & 5.58 & 0.00 & 3.02\\
 & HzBHQASG\tnote{f} & 0.313 & -- & -- & $-1.052$ & -- & 69.90 & 0.407 & 50.19 & 1.132 & -- & -- & -- & 902.09 & 428 & 917.55 & 929.55 & 953.99 & 0.00 & 1.76 & 5.83\\
 & HzBHQASQFG\tnote{g} & 0.313 & -- & -- & $-1.046$ & -- & 69.84 & 0.401 & 50.18 & 1.134 & 0.231 & 0.635 & 7.215 & 2506.25 & 2023 & 870.16 & 888.16 & 938.71 & 6.38 & 1.85 & 7.46\\
\\
Non-flat XCDM & QSO-Flux & 0.507 & -- & $-0.376$ & $-1.996$ & -- & 75.28 & -- & -- & -- & 0.229 & 0.611 & 7.934 & 1614.59 & 1591 & -55.75 & -41.75 & -4.12 & 11.31 & 0.00 & 9.12\\
 & HQASG\tnote{d} & 0.300 & -- & $-0.161$ & $-1.027$ & -- & 80.36 & 0.405 & 50.21 & 1.122 & -- & -- & -- & 879.48 & 385 & 894.01 & 908.01 & 935.81 & 0.18 & 2.96 & 10.90\\
 & HQASQFG\tnote{e} & 0.395 & -- & $-0.138$ & $-1.639$ & -- & 73.48 & 0.411 & 50.21 & 1.112 & 0.230 & 0.627 & 7.441 & 2486.77 & 1980 & 843.65 & 863.65 & 919.61 & 6.76 & 1.70 & 10.32\\
 & HzBHQASG\tnote{f} & 0.312 & -- & $-0.045$ & $-0.959$ & -- & 69.46 & 0.402 & 50.23 & 1.117 & -- & -- & -- & 904.17 & 427 & 917.07 & 931.07 & 959.58 & 2.08 & 3.28 & 11.42\\
 & HzBHQASQFG\tnote{g} & 0.316 & -- & $-0.089$ & $-0.891$ & -- & 69.05 & 0.410 & 50.23 & 1.111 & 0.230 & 0.633 & 7.247 & 2516.49 & 2022 & 869.25 & 889.25 & 945.41 & 16.62 & 2.94 & 14.16\\
\bottomrule
\end{tabular}
\begin{tablenotes}[flushleft]
\item [c] \hunit.
\item [d] \hiig\ + QSO-AS + GRB.
\item [e] \hiig\ + QSO-AS + GRB + QSO-Flux.
\item [f] $H(z)$ + BAO + \hiig\ + QSO-AS + GRB.
\item [g] $H(z)$ + BAO + \hiig\ + QSO-AS + GRB + QSO-Flux.
\end{tablenotes}
\end{threeparttable}%
}
\end{table*}

\begin{table*}
\centering
\resizebox*{1\columnwidth}{0.8\columnwidth}{%
\begin{threeparttable}
\caption{One-dimensional marginalized best-fitting parameter values and uncertainties ($\pm 1\sigma$ error bars or $2\sigma$ limits) for all models from various combinations of data.}\label{tab:1d_BFPaA}
\begin{tabular}{lccccccccccccc}
\toprule
 Model & Data set & $\Omega_{\mathrm{m_0}}$ & $\Omega_{\Lambda}$ & $\Omega_{\mathrm{k_0}}$ & $w_{\mathrm{X}}$ & $\alpha$ & $H_0$\tnote{c} & $\sigma_{\mathrm{ext}}$ & $a$ & $b$ & $\delta$ & $\gamma$ & $\beta$ \\
\midrule
Flat \lcdm & QSO-Flux & $>0.313$ & -- & -- & -- & -- & -- & -- & -- & -- & $0.231\pm0.004$ & $0.626\pm0.011$ & $7.469\pm0.321$ \\
 & HQASG\tnote{d} & $0.277^{+0.034}_{-0.041}$ & -- & -- & -- & -- & $71.03\pm1.67$ & $0.413^{+0.026}_{-0.032}$ & $50.19\pm0.24$ & $1.138\pm0.085$ & -- & -- & --\\
 & HQASQFG\tnote{e} & $0.299^{+0.036}_{-0.043}$ & -- & -- & -- & -- & $70.25^{+1.60}_{-1.61}$ & $0.412^{+0.027}_{-0.032}$ & $50.18\pm0.24$ & $1.136\pm0.085$ & $0.231^{+0.005}_{-0.004}$ & $0.639^{+0.009}_{-0.010}$ & $7.091^{+0.281}_{-0.279}$\\
 & HzBHQASG\tnote{f} & $0.316\pm0.013$ & -- & -- & -- & -- & $69.05^{+0.62}_{-0.63}$ & $0.412^{+0.026}_{-0.032}$ & $50.19\pm0.23$ & $1.133\pm0.085$ & -- & -- & --\\
 & HzBHQASQFG\tnote{g} & $0.318\pm0.013$ & -- & -- & -- & -- & $69.03\pm0.62$ & $0.412^{+0.026}_{-0.032}$ & $50.19\pm0.23$ & $1.133\pm0.084$ & $0.231\pm0.004$ & $0.637\pm0.009$ & $7.146\pm0.268$\\
\\
Non-flat \lcdm & QSO-Flux & $>0.353$ & $>0.357$ & $-0.303^{+0.131}_{-0.252}$ & -- & -- & -- & -- & -- & -- & $0.231^{+0.004}_{-0.005}$ & $0.618\pm0.012$ & $7.709\pm0.366$ \\
 & HQASG\tnote{d} & $0.292\pm0.044$ & $0.801^{+0.191}_{-0.055}$ & $-0.093^{+0.092}_{-0.190}$ & -- & -- & $71.33^{+1.75}_{-1.77}$ & $0.413^{+0.026}_{-0.032}$ & $50.19\pm0.24$ & $1.130\pm0.086$ & -- & -- & --\\
 & HQASQFG\tnote{e} & $0.327^{+0.039}_{-0.043}$ & $>0.691$ & $-0.204^{+0.063}_{-0.125}$ & -- & -- & $71.07\pm1.64$ & $0.413^{+0.027}_{-0.032}$ & $50.20\pm0.24$ & $1.120\pm0.086$ & $0.231\pm0.004$ & $0.632\pm0.010$ & $7.291^{+0.306}_{-0.305}$\\
 & HzBHQASG\tnote{f} & $0.311^{+0.012}_{-0.014}$ & $0.708^{+0.053}_{-0.046}$ & $-0.019^{+0.043}_{-0.048}$ & -- & -- & $69.72\pm1.10$ & $0.412^{+0.026}_{-0.032}$ & $50.19\pm0.23$ & $1.132\pm0.085$ & -- & -- & --\\
 & HzBHQASQFG\tnote{g} & $0.312^{+0.012}_{-0.013}$ & $0.716^{+0.052}_{-0.046}$ & $-0.028\pm0.045$ & -- & -- & $69.88\pm1.10$ & $0.412^{+0.025}_{-0.032}$ & $50.19\pm0.23$ & $1.131\pm0.084$ & $0.231\pm0.004$ & $0.637\pm0.009$ & $7.144\pm0.270$\\
\\
Flat XCDM & QSO-Flux & $0.496^{+0.192}_{-0.069}$ & -- & -- & $<-1.042$\tnote{h} & -- & -- & -- & -- & -- & $0.231\pm0.004$ & $0.624\pm0.011$ & $7.508\pm0.326$ \\
 & HQASG\tnote{d} & $0.322^{+0.062}_{-0.044}$ & -- & -- & $-1.379^{+0.361}_{-0.375}$ & -- & $72.00^{+1.99}_{-1.98}$ & $0.412^{+0.026}_{-0.032}$ & $50.20\pm0.24$ & $1.130\pm0.085$ & -- & -- & --\\
 & HQASQFG\tnote{e} & $0.358^{+0.040}_{-0.038}$ & -- & -- & $<-1.100$ & -- & $72.14\pm1.91$ & $0.411^{+0.026}_{-0.031}$ & $50.20\pm0.23$ & $1.125\pm0.084$ & $0.231\pm0.004$ & $0.633^{+0.009}_{-0.010}$ & $7.268^{+0.287}_{-0.288}$\\
 & HzBHQASG\tnote{f} & $0.313^{+0.014}_{-0.015}$ & -- & -- & $-1.050^{+0.090}_{-0.081}$ & -- & $69.85^{+1.42}_{-1.55}$ & $0.412^{+0.026}_{-0.032}$ & $50.19\pm0.24$ & $1.134\pm0.085$ & -- & -- & -- \\
 & HzBHQASQFG\tnote{g} & $0.315^{+0.013}_{-0.015}$ & -- & -- & $-1.052^{+0.091}_{-0.081}$ & -- & $69.85\pm1.48$ & $0.413^{+0.026}_{-0.032}$ & $50.19\pm0.24$ & $1.133\pm0.086$ & $0.231\pm0.004$ & $0.637\pm0.009$ & $7.135^{+0.270}_{-0.271}$\\
\\
Non-flat XCDM & QSO-Flux & $0.515^{+0.184}_{-0.050}$ & -- & $-0.310^{+0.137}_{-0.289}$ & $<-0.294$ & -- & -- & -- & -- & -- & $0.231^{+0.004}_{-0.005}$ & $0.615\pm0.013$ & $7.817^{+0.398}_{-0.400}$ \\
 & HQASG\tnote{d} & $0.303^{+0.073}_{-0.058}$ & -- & $-0.044^{+0.193}_{-0.217}$ & $-1.273^{+0.501}_{-0.321}$ & -- & $71.77\pm2.02$ & $0.413^{+0.026}_{-0.031}$ & $50.20\pm0.24$ & $1.129\pm0.085$ & -- & -- & --\\
 & HQASQFG\tnote{e} & $0.367^{+0.059}_{-0.048}$ & -- & $-0.146^{+0.143}_{-0.147}$ & $-1.433^{+0.241}_{-0.493}$ & -- & $72.27^{+2.01}_{-1.99}$ & $0.413^{+0.026}_{-0.032}$ & $50.21\pm0.24$ & $1.116\pm0.085$ & $0.231\pm0.004$ & $0.629^{+0.011}_{-0.010}$ & $7.382\pm0.321$\\
 & HzBHQASG\tnote{f} & $0.310\pm0.014$ & -- & $-0.024^{+0.092}_{-0.093}$ & $-1.019^{+0.202}_{-0.099}$ & -- & $69.63^{+1.45}_{-1.62}$ & $0.412^{+0.026}_{-0.031}$ & $50.19\pm0.23$ & $1.132\pm0.085$ & -- & -- & -- \\
 & HzBHQASQFG\tnote{g} & $0.314^{+0.014}_{-0.015}$ & -- & $-0.060^{+0.096}_{-0.095}$ & $-0.968^{+0.184}_{-0.087}$ & -- & $69.43^{+1.43}_{-1.63}$ & $0.412^{+0.026}_{-0.032}$ & $50.19\pm0.24$ & $1.130\pm0.085$ & $0.231\pm0.004$ & $0.636^{+0.009}_{-0.010}$ & $7.182^{+0.278}_{-0.281}$\\
\bottomrule
\end{tabular}
\begin{tablenotes}[flushleft]
\item [c] \hunit.
\item [d] \hiig\ + QSO-AS + GRB.
\item [e] \hiig\ + QSO-AS + GRB + QSO-Flux.
\item [f] $H(z)$ + BAO + \hiig\ + QSO-AS + GRB.
\item [g] $H(z)$ + BAO + \hiig\ + QSO-AS + GRB + QSO-Flux.
\item [h] This is the 1$\sigma$ limit. The $2\sigma$ limit is set by the prior, and is not shown here.
\end{tablenotes}
\end{threeparttable}%
}
\end{table*}


\cleardoublepage

\chapter{Corrected GRB data}
\label{AppendixB}

These data are taken from the appendix of \cite{CaoDainottiRatra2022b}, where the Platinum data listed in Table \ref{tab:P50aB} are correct and should be used to do cosmological analysis instead of those reported in Table \ref{tab:P50C8} from \cite{CaoDainottiRatra2022}.

\begin{sidewaystable*}
\centering
\resizebox*{\columnwidth}{0.74\columnwidth}{%
\begin{threeparttable}
\caption{Updated 50 Platinum GRB samples with 1$\sigma$ errors, where $F_{X}$, $F_{\rm peak}$, $\mathrm{FK}_{\mathrm{plateau}}$                       ($\log\mathrm{FK}_{\mathrm{plateau}}\equiv\log F_{X}+\log K_{\rm plateau}$), and ${FK}_{\mathrm{prompt}}$ ($\log\mathrm{FK}_{\mathrm{prompt}}\equiv\log F_{\rm peak}+\log K_{\rm prompt}$) have units of $\mathrm{erg\ cm}^{-2}\ \mathrm{s}^{-1}$, and $T^{*}_{X}$ and $E_{\mathrm{peak}}$ have units of s and keV, respectively. The last Ref.\ column lists the sources of GRBs, with ``C'' and ``B'' representing the \protect \href{https://swift.gsfc.nasa.gov/results/batgrbcat/index_tables.html}{third \textit{Swift} GRB Catalog} \citep{Lienetal2016} and \href{https://www.swift.ac.uk/burst_analyser/}{\textit{Swift} BAT burst analyser} \citep{Evansetal2010}, respectively. Central values are estimated to three (or four) significant figures with corresponding non-zero errors matching their decimal places and $\Delta\chi^2=\chi^2_{\mathrm{PL}}-\chi^2_{\mathrm{CPL}}$.}
\label{tab:P50aB}
\begin{tabular}{lccccccccccccc}
\toprule
GRB & $z$ & $\log T^{*}_{X}$ & $\log F_{X}$ & $\log F_{\rm peak}$ & $\alpha_{\mathrm{plateau}}$ & $K_{\mathrm{plateau}}$ & $\log\mathrm{FK}_{\mathrm{plateau}}$ & $\alpha^{\mathrm{PL/CPL}}_{\mathrm{prompt}}$ & $E_{\mathrm{peak}}$ & $K^{\mathrm{PL/CPL}}_{\mathrm{prompt}}$ & $\log\mathrm{FK}_{\mathrm{prompt}}$ & $\Delta\chi^2$ & Ref.\\
\midrule
060418 & 1.49 & $3.12\pm0.03$ & $-9.79\pm0.03$ & $-6.3^{+0.01}_{-0.02}$ & $1.93^{+0.13}_{-0.12}$ & $0.938^{+0.118}_{-0.097}$ & $-9.8\pm0.1$ & $-1.5^{+0.06}_{-0.05}$ & -- & $0.634^{+0.029}_{-0.034}$ & $-6.5^{+0.03}_{-0.04}$ & $0.88$ & C\\
060605 & 3.8 & $4.04\pm0.02$ & $-11.26\pm0.03$ & $-7.33^{+0.06}_{-0.07}$ & $1.98^{+0.07}_{-0.06}$ & $0.969^{+0.113}_{-0.087}$ & $-11.27^{+0.08}_{-0.07}$ & $-0.85^{+0.27}_{-0.25}$ & -- & $0.165^{+0.079}_{-0.057}$ & $-8.11^{+0.23}_{-0.25}$ & $0.02$ & C\\
060708 & 1.92 & $3.51\pm0.04$ & $-10.87\pm0.04$ & $-6.81\pm0.02$ & $2.4^{+0.3}_{-0.24}$ & $1.54^{+0.58}_{-0.35}$ & $-10.68^{+0.18}_{-0.15}$ & $-1.33\pm0.07$ & -- & $0.488^{+0.038}_{-0.036}$ & $-7.12\pm0.05$ & $3.14$ & C\\
060714 & 2.71 & $3.73\pm0.03$ & $-10.86\pm0.02$ & $-7.01^{+0.02}_{-0.03}$ & $1.93\pm0.1$ & $0.912^{+0.128}_{-0.112}$ & $-10.9\pm0.08$ & $-1.48\pm0.1$ & -- & $0.506^{+0.071}_{-0.062}$ & $-7.31^{+0.08}_{-0.09}$ & $4.14$ & C\\
060814 & 0.84 & $4.22\pm0.02$ & $-10.91\pm0.02$ & $-6.22\pm0.01$ & $1.96^{+0.07}_{-0.06}$ & $0.976^{+0.042}_{-0.035}$ & $-10.92\pm0.04$ & $-1.3\pm0.04$ & -- & $0.653\pm0.016$ & $-6.4\pm0.02$ & $2.35$ & C\\
060906 & 3.685 & $4.33\pm0.04$ & $-11.88^{+0.05}_{-0.06}$ & $-6.91^{+0.03}_{-0.05}$ & $2.06^{+0.11}_{-0.1}$ & $1.1^{+0.2}_{-0.16}$ & $-11.84^{+0.12}_{-0.13}$ & $-2\pm0.15$ & -- & $1^{+0.26}_{-0.21}$ & $-6.91^{+0.13}_{-0.15}$ & $0.66$ & C\\
061121 & 1.314 & $3.79\pm0.01$ & $-10.03\pm0.01$ & $-5.707^{+0.005}_{-0.006}$ & $1.87^{+0.08}_{-0.07}$ & $0.897^{+0.062}_{-0.051}$ & $-10.08\pm0.04$ & $-1.05\pm0.02$ & -- & $0.451^{+0.007}_{-0.008}$ & $-6.05\pm0.01$ & $-0.02$ & C\\
061222A & 2.088 & $3.92\pm0.01$ & $-9.95^{+0.01}_{-0.02}$ & $-6.135\pm0.009$ & $1.8\pm0.07$ & $0.798^{+0.066}_{-0.06}$ & $-10.05^{+0.04}_{-0.05}$ & $-0.487^{+0.132}_{-0.126}$ & $226^{+103}_{-36}$ & $0.269^{+0.047}_{-0.056}$ & $-6.71^{+0.08}_{-0.11}$ & $11.27$ & C\\
070110 & 2.352 & $4.27\pm0.03$ & $-11.05\pm0.03$ & $-7.32^{+0.05}_{-0.07}$ & $2\pm0.05$ & $1\pm0.06$ & $-11.05\pm0.06$ & $-1.46^{+0.2}_{-0.21}$ & -- & $0.52^{+0.151}_{-0.111}$ & $-7.6^{+0.16}_{-0.17}$ & $0.34$ & C\\
070306 & 1.4959 & $4.87\pm0.02$ & $-11.27\pm0.02$ & $-6.52^{+0.02}_{-0.01}$ & $1.85\pm0.05$ & $0.872^{+0.041}_{-0.039}$ & $-11.33\pm0.04$ & $-1.54\pm0.05$ & -- & $0.657\pm0.03$ & $-6.7^{+0.04}_{-0.03}$ & $3.16$ & C\\
070508 & 0.82 & $3.03\pm0.01$ & $-9.16\pm0.01$ & $-5.65\pm0.01$ & $1.79\pm0.02$ & $0.882^{+0.01}_{-0.011}$ & $-9.22\pm0.02$ & $-0.679^{+0.096}_{-0.092}$ & $213^{+60}_{-26}$ & $0.571^{+0.036}_{-0.048}$ & $-5.89^{+0.04}_{-0.05}$ & $18.84$ & C\\
070521 & 0.553 & $3.55\pm0.03$ & $-10.01\pm0.03$ & $-7.06^{+0.05}_{-0.08}$ & $1.73^{+0.18}_{-0.16}$ & $0.888^{+0.073}_{-0.06}$ & $-10.06\pm0.06$ & $-0.528^{+0.17}_{-0.161}$ & $184^{+95}_{-29}$ & $0.662^{+0.05}_{-0.078}$ & $-7.24^{+0.08}_{-0.13}$ & $9.49$ & C\\
070529 & 2.4996 & $3.09\pm0.04$ & $-10.25^{+0.04}_{-0.03}$ & $-6.95^{+0.06}_{-0.09}$ & $1.71^{+0.13}_{-0.1}$ & $0.695^{+0.123}_{-0.081}$ & $-10.41^{+0.11}_{-0.08}$ & $-1.49\pm0.24$ & -- & $0.528^{+0.185}_{-0.137}$ & $-7.23^{+0.19}_{-0.22}$ & $1.87$ & C\\
080310 & 2.4266 & $4.33\pm0.02$ & $-11.4^{+0.03}_{-0.02}$ & $-7.07^{+0.03}_{-0.05}$ & $1.91\pm0.06$ & $0.895^{+0.069}_{-0.064}$ & $-11.45^{+0.06}_{-0.05}$ & $-1.91^{+0.13}_{-0.15}$ & -- & $0.895^{+0.182}_{-0.132}$ & $-7.12^{+0.11}_{-0.12}$ & $3.43$ & C\\
080430 & 0.767 & $4.22\pm0.02$ & $-11.17^{+0.01}_{-0.02}$ & $-6.74\pm0.02$ & $1.96\pm0.09$ & $0.977^{+0.052}_{-0.048}$ & $-11.18^{+0.03}_{-0.04}$ & $-1.75\pm0.07$ & -- & $0.867^{+0.036}_{-0.034}$ & $-6.8\pm0.04$ & $0.15$ & C\\
080721 & 2.6 & $2.879\pm0.005$ & $-8.713^{+0.004}_{-0.005}$ & $-5.72^{+0.03}_{-0.02}$ & $1.72\pm0.01$ & $0.699\pm0.009$ & $-8.87\pm0.01$ & $-0.99\pm0.09$ & -- & $0.274^{+0.034}_{-0.03}$ & $-6.28^{+0.08}_{-0.07}$ & $-0.01$ & C\\
081008 & 1.967 & $3.86\pm0.03$ & $-10.79^{+0.03}_{-0.04}$ & $-6.98^{+0.03}_{-0.04}$ & $1.83^{+0.1}_{-0.06}$ & $0.831^{+0.096}_{-0.052}$ & $-10.87^{+0.08}_{-0.07}$ & $-1.34^{+0.12}_{-0.13}$ & -- & $0.488^{+0.074}_{-0.06}$ & $-7.29^{+0.09}_{-0.1}$ & $0.68$ & C\\
081221 & 2.26 & $2.93\pm0.02$ & $-9.35^{+0.01}_{-0.02}$ & $-5.823\pm0.007$ & $2.01\pm0.04$ & $1.01^{+0.05}_{-0.04}$ & $-9.34^{+0.03}_{-0.04}$ & $-1.17^{+0.23}_{-0.22}$ & -- & $0.375^{+0.111}_{-0.089}$ & $-6.25^{+0.12}_{-0.13}$ & $-25.22$ & C\\
090418A & 1.608 & $3.53\pm0.02$ & $-10.05^{+0.02}_{-0.01}$ & $-6.8^{+0.04}_{-0.05}$ & $1.9^{+0.12}_{-0.11}$ & $0.909^{+0.11}_{-0.091}$ & $-10.09^{+0.07}_{-0.05}$ & $-1.29\pm0.17$ & -- & $0.506^{+0.09}_{-0.076}$ & $-7.1^{+0.11}_{-0.12}$ & $-0.03$ & C\\
091018 & 0.971 & $2.9\pm0.02$ & $-9.63^{+0.02}_{-0.01}$ & $-6.26^{+0.01}_{-0.02}$ & $1.73\pm0.07$ & $0.833^{+0.04}_{-0.039}$ & $-9.71^{+0.04}_{-0.03}$ & $-1.3^{+0.2}_{-0.18}$ & $35.2^{+2.5}_{-3.7}$ & $1.07^{+0.06}_{-0.04}$ & $-6.23\pm0.03$ & $30.75$ & B\\
091020 & 1.71 & $2.95\pm0.03$ & $-9.71\pm0.02$ & $-6.44\pm0.02$ & $1.95^{+0.13}_{-0.12}$ & $0.951^{+0.132}_{-0.107}$ & $-9.7\pm0.1$ & $-1.18^{+0.07}_{-0.08}$ & -- & $0.442^{+0.036}_{-0.03}$ & $-6.8\pm0.05$ & $-0.03$ & C\\
091029 & 2.752 & $4.3\pm0.02$ & $-11.35^{+0.02}_{-0.01}$ & $-6.96^{+0.03}_{-0.04}$ & $2.03\pm0.07$ & $1.04^{+0.1}_{-0.09}$ & $-11.33^{+0.06}_{-0.05}$ & $-0.926^{+0.427}_{-0.362}$ & $53.2^{+16.3}_{-6}$ & $0.64^{+0.136}_{-0.199}$ & $-7.15^{+0.11}_{-0.2}$ & $6.66$ & C\\
100219A & 4.7 & $4.72\pm0.06$ & $-12.1\pm0.2$ & $-7.5^{+0.08}_{-0.12}$ & $1.45^{+0.1}_{-0.09}$ & $0.384^{+0.073}_{-0.056}$ & $-12.52^{+0.28}_{-0.27}$ & $-1.43^{+0.39}_{-0.37}$ & -- & $0.371^{+0.335}_{-0.183}$ & $-7.93^{+0.36}_{-0.42}$ & $0.00$ & C\\
110213A & 1.46 & $3.84^{+0.02}_{-0.01}$ & $-9.9^{+0.03}_{-0.02}$ & $-7.15^{+0.1}_{-0.14}$ & $1.81^{+0.05}_{-0.04}$ & $0.843^{+0.039}_{-0.03}$ & $-9.97^{+0.05}_{-0.04}$ & $-2.82^{+0.37}_{-0.59}$ & -- & $2.09^{+1.47}_{-0.59}$ & $-6.83^{+0.33}_{-0.28}$ & $1.41$ & C\\
110818A & 3.36 & $3.85\pm0.04$ & $-11.28^{+0.03}_{-0.04}$ & $-6.85^{+0.04}_{-0.05}$ & $1.84\pm0.12$ & $0.79^{+0.153}_{-0.128}$ & $-11.38^{+0.11}_{-0.12}$ & $-1.11^{+0.2}_{-0.19}$ & -- & $0.27^{+0.087}_{-0.069}$ & $-7.42^{+0.16}_{-0.18}$ & $-0.04$ & C\\
111008A & 5 & $3.99\pm0.02$ & $-10.67^{+0.02}_{-0.01}$ & $-6.27\pm0.03$ & $1.85\pm0.05$ & $0.764^{+0.072}_{-0.065}$ & $-10.79^{+0.06}_{-0.05}$ & $-1.26\pm0.11$ & -- & $0.266^{+0.057}_{-0.048}$ & $-6.85\pm0.12$ & $0.43$ & C\\
120118B & 2.943 & $3.68^{+0.09}_{-0.07}$ & $-10.96^{+0.08}_{-0.06}$ & $-6.86\pm0.04$ & $1.99^{+0.17}_{-0.16}$ & $0.986^{+0.259}_{-0.194}$ & $-10.97^{+0.18}_{-0.15}$ & $-1.9^{+0.14}_{-0.15}$ & -- & $0.872^{+0.199}_{-0.153}$ & $-6.92^{+0.13}_{-0.12}$ & $0.52$ & C\\
120404A & 2.88 & $3.8^{+0.06}_{-0.03}$ & $-11.03^{+0.06}_{-0.08}$ & $-7.12^{+0.06}_{-0.07}$ & $1.68^{+0.08}_{-0.07}$ & $0.648^{+0.074}_{-0.059}$ & $-11.22^{+0.11}_{-0.12}$ & $-1.84^{+0.22}_{-0.24}$ & -- & $0.805^{+0.31}_{-0.208}$ & $-7.21\pm0.2$ & $0.00$ & C\\
120811C & 2.67 & $3.16\pm0.05$ & $-10.14^{+0.03}_{-0.04}$ & $-6.58^{+0.01}_{-0.02}$ & $1.62^{+0.09}_{-0.08}$ & $0.61^{+0.076}_{-0.06}$ & $-10.36\pm0.08$ & $-1.02^{+0.27}_{-0.24}$ & $60.1^{+11}_{-5.1}$ & $0.619^{+0.096}_{-0.126}$ & $-6.79^{+0.07}_{-0.12}$ & $11.43$ & C\\
120922A & 3.1 & $3.6^{+0.13}_{-0.03}$ & $-10.6\pm0.1$ & $-6.97^{+0.02}_{-0.04}$ & $2.14^{+0.13}_{-0.12}$ & $1.22^{+0.24}_{-0.19}$ & $-10.51^{+0.18}_{-0.17}$ & $-1.95^{+0.11}_{-0.12}$ & -- & $0.932^{+0.172}_{-0.134}$ & $-7^{+0.09}_{-0.11}$ & $0.02$ & C\\
121128A & 2.2 & $3.23\pm0.02$ & $-9.62\pm0.02$ & $-5.98\pm0.01$ & $1.78^{+0.06}_{-0.04}$ & $0.774^{+0.056}_{-0.035}$ & $-9.73^{+0.05}_{-0.04}$ & $-0.495^{+0.139}_{-0.134}$ & $108^{+12}_{-8}$ & $0.373^{+0.048}_{-0.05}$ & $-6.41^{+0.06}_{-0.07}$ & $42.35$ & C\\
131030A & 1.29 & $2.85\pm0.02$ & $-9.29^{+0.01}_{-0.02}$ & $-5.576^{+0.007}_{-0.008}$ & $1.68^{+0.03}_{-0.02}$ & $0.767^{+0.019}_{-0.013}$ & $-9.4^{+0.02}_{-0.03}$ & $-0.585^{+0.116}_{-0.111}$ & $229^{+93}_{-34}$ & $0.416^{+0.046}_{-0.06}$ & $-5.96^{+0.05}_{-0.08}$ & $12.64$ & C\\
131105A & 1.686 & $3.95\pm0.03$ & $-10.91\pm0.02$ & $-6.52^{+0.05}_{-0.04}$ & $1.92\pm0.1$ & $0.924^{+0.096}_{-0.087}$ & $-10.94\pm0.06$ & $-1.15^{+0.18}_{-0.16}$ & -- & $0.432^{+0.074}_{-0.071}$ & $-6.89\pm0.12$ & $1.15$ & B\\
140206A & 2.7 & $3.59\pm0.01$ & $-9.73\pm0.01$ & $-5.772\pm0.007$ & $1.44^{+0.14}_{-0.08}$ & $0.481^{+0.096}_{-0.048}$ & $-10.05^{+0.09}_{-0.06}$ & $-0.536^{+0.11}_{-0.106}$ & $116^{+11}_{-8}$ & $0.31^{+0.038}_{-0.04}$ & $-6.28^{+0.06}_{-0.07}$ & $57.56$ & C\\
140419A & 3.956 & $3.68\pm0.01$ & $-10.01^{+0.02}_{-0.01}$ & $-6.38\pm0.01$ & $1.66\pm0.06$ & $0.58^{+0.059}_{-0.053}$ & $-10.25^{+0.06}_{-0.05}$ & $-1.03\pm0.05$ & -- & $0.212\pm0.017$ & $-7.05\pm0.04$ & $2.37$ & C\\
140506A & 0.889 & $3.31\pm0.04$ & $-9.9^{+0.03}_{-0.04}$ & $-6.06^{+0.02}_{-0.03}$ & $1.64^{+0.13}_{-0.12}$ & $0.795^{+0.069}_{-0.058}$ & $-10\pm0.1$ & $-0.53^{+0.393}_{-0.341}$ & $100.4^{+53.6}_{-13.3}$ & $0.664^{+0.087}_{-0.163}$ & $-6.24^{+0.07}_{-0.15}$ & $6.69$ & C\\
140509A & 2.4 & $3.59^{+0.07}_{-0.04}$ & $-11.1^{+0.06}_{-0.05}$ & $-6.93^{+0.06}_{-0.07}$ & $1.81^{+0.14}_{-0.12}$ & $0.793^{+0.148}_{-0.109}$ & $-11.2^{+0.13}_{-0.11}$ & $-1.49^{+0.23}_{-0.24}$ & -- & $0.536^{+0.183}_{-0.132}$ & $-7.2\pm0.19$ & $0.46$ & C\\
140629A & 2.3 & $2.86\pm0.04$ & $-9.85^{+0.03}_{-0.02}$ & $-6.46^{+0.03}_{-0.02}$ & $1.84\pm0.1$ & $0.826^{+0.105}_{-0.093}$ & $-9.9\pm0.1$ & $-1.35^{+0.08}_{-0.09}$ & -- & $0.46^{+0.052}_{-0.042}$ & $-6.8^{+0.08}_{-0.06}$ & $3.01$ & C\\
150314A & 1.758 & $2.49\pm0.01$ & $-8.61\pm0.01$ & $-5.42\pm0.01$ & $1.73\pm0.02$ & $0.76^{+0.016}_{-0.015}$ & $-8.73\pm0.02$ & $-0.465^{+0.108}_{-0.105}$ & $249^{+88}_{-36}$ & $0.297^{+0.04}_{-0.047}$ & $-5.95^{+0.07}_{-0.08}$ & $15.03$ & C\\
150403A & 2.06 & $3.197\pm0.004$ & $-8.821\pm0.003$ & $-5.77\pm0.01$ & $1.675\pm0.01$ & $0.695\pm0.007$ & $-8.98\pm0.01$ & $-0.92^{+0.04}_{-0.03}$ & -- & $0.299^{+0.01}_{-0.013}$ & $-6.3\pm0.03$ & $1.68$ & C\\
150910A & 1.36 & $3.85\pm0.02$ & $-9.97^{+0.84}_{-0.02}$ & $-7.08^{+0.09}_{-0.15}$ & $1.76^{+0.12}_{-0.11}$ & $0.814^{+0.088}_{-0.074}$ & $-10.06^{+0.88}_{-0.06}$ & $-1.46^{+0.54}_{-0.6}$ & -- & $0.629^{+0.424}_{-0.233}$ & $-7.28^{+0.31}_{-0.35}$ & $-0.05$ & C\\
151027A & 0.81 & $4\pm0.01$ & $-9.94^{+0.02}_{-0.01}$ & $-6.24^{+0.03}_{-0.02}$ & $1.77\pm0.06$ & $0.872^{+0.032}_{-0.03}$ & $-10\pm0.03$ & $-1.24\pm0.08$ & -- & $0.637^{+0.031}_{-0.029}$ & $-6.44^{+0.05}_{-0.04}$ & $0.00$ & C\\
160121A & 1.96 & $3.82\pm0.07$ & $-11.17\pm0.04$ & $-7.11^{+0.04}_{-0.05}$ & $2.02^{+0.16}_{-0.13}$ & $1.02^{+0.2}_{-0.13}$ & $-11.16^{+0.12}_{-0.1}$ & $-1.82^{+0.15}_{-0.16}$ & -- & $0.823^{+0.156}_{-0.124}$ & $-7.2\pm0.12$ & $2.09$ & C\\
160227A & 2.38 & $4.48\pm0.02$ & $-11^{+0.02}_{-0.01}$ & $-7.23\pm0.05$ & $1.65^{+0.06}_{-0.05}$ & $0.653^{+0.049}_{-0.039}$ & $-11.19^{+0.05}_{-0.04}$ & $-1.05^{+0.18}_{-0.16}$ & -- & $0.314^{+0.068}_{-0.061}$ & $-7.73^{+0.13}_{-0.15}$ & $4.74$ & B\\
160327A & 4.99 & $3.76\pm0.04$ & $-11.22\pm0.04$ & $-6.9\pm0.03$ & $1.78^{+0.12}_{-0.09}$ & $0.674^{+0.162}_{-0.1}$ & $-11.39^{+0.13}_{-0.11}$ & $-1.57\pm0.1$ & -- & $0.463^{+0.091}_{-0.076}$ & $-7.23\pm0.11$ & $1.95$ & C\\
170202A & 3.645 & $3.78\pm0.04$ & $-10.65^{+0.02}_{-0.03}$ & $-6.41\pm0.02$ & $1.99\pm0.09$ & $0.985^{+0.146}_{-0.127}$ & $-10.66^{+0.08}_{-0.09}$ & $-0.491^{+0.279}_{-0.25}$ & $110.4^{+45.8}_{-15}$ & $0.235^{+0.084}_{-0.093}$ & $-7.04^{+0.15}_{-0.24}$ & $9.64$ & C\\
170705A & 2.01 & $3.64\pm0.04$ & $-10.19^{+0.03}_{-0.02}$ & $-5.95\pm0.01$ & $1.62\pm0.1$ & $0.658^{+0.077}_{-0.069}$ & $-10.37^{+0.08}_{-0.07}$ & $-0.917^{+0.138}_{-0.132}$ & $161^{+83}_{-24}$ & $0.43^{+0.065}_{-0.087}$ & $-6.32^{+0.07}_{-0.11}$ & $9.03$ & C\\
180329B & 1.998 & $4.02\pm0.03$ & $-11.15\pm0.03$ & $-7.04^{+0.08}_{-0.12}$ & $1.78^{+0.09}_{-0.08}$ & $0.785^{+0.082}_{-0.066}$ & $-11.26\pm0.07$ & $-1.91^{+0.32}_{-0.39}$ & -- & $0.906^{+0.484}_{-0.268}$ & $-7.08\pm0.27$ & $-0.01$ & C\\
190106A & 1.86 & $4.19\pm0.02$ & $-10.43^{+0.01}_{-0.02}$ & $-6.47^{+0.01}_{-0.02}$ & $1.88^{+0.09}_{-0.04}$ & $0.882^{+0.087}_{-0.037}$ & $-10.49^{+0.05}_{-0.04}$ & $-0.893^{+0.237}_{-0.216}$ & $74.8^{+17.1}_{-7.3}$ & $0.614^{+0.082}_{-0.115}$ & $-6.68^{+0.06}_{-0.11}$ & $12.68$ & C\\
190114A & 3.37 & $3.75\pm0.03$ & $-10.75\pm0.02$ & $-7.39^{+0.09}_{-0.12}$ & $1.85^{+0.08}_{-0.07}$ & $0.802^{+0.1}_{-0.079}$ & $-10.85^{+0.07}_{-0.06}$ & $-1.51^{+0.4}_{-0.41}$ & -- & $0.485^{+0.404}_{-0.216}$ & $-7.7^{+0.35}_{-0.38}$ & $-0.01$ & C\\
\bottomrule
\end{tabular}
\end{threeparttable}%
}
\end{sidewaystable*}

\begin{landscape}
{\setlength\tabcolsep{2.5pt}\scriptsize 
\begin{longtable}{lccccccccccccc}
\caption{Same as Table \ref{tab:P50aB}, but for 95 long GRB samples. Note that PL photon indices of GRBs 151215A and 170405A are taken from the time-averaged BAT spectral analysis and PL peak energy flux of the latter is also time-averaged, since their 1 s peak analysis results are either unreasonable or out of range.
} 
\label{tab:LGRB95} \\

\toprule

\multicolumn{1}{l}{GRB} &
\multicolumn{1}{c}{$z$} &
\multicolumn{1}{c}{$\log T^{*}_{X}$} &
\multicolumn{1}{c}{$\log F_{X}$} &
\multicolumn{1}{c}{$\log F_{\rm peak}$} &
\multicolumn{1}{c}{$\alpha_{\mathrm{plateau}}$} &
\multicolumn{1}{c}{$K_{\mathrm{plateau}}$} &
\multicolumn{1}{c}{$\log\mathrm{FK}_{\mathrm{plateau}}$} &
\multicolumn{1}{c}{$\alpha^{\mathrm{PL/CPL}}_{\mathrm{prompt}}$} &
\multicolumn{1}{c}{$E_{\mathrm{peak}}$} &
\multicolumn{1}{c}{$K^{\mathrm{PL/CPL}}_{\mathrm{prompt}}$} &
\multicolumn{1}{c}{$\log\mathrm{FK}_{\mathrm{prompt}}$} &
\multicolumn{1}{c}{$\Delta\chi^2$} & 
\multicolumn{1}{c}{Ref.}\\

\midrule

\endfirsthead

\caption{ -- \textit{Continued from previous page}} \\
\toprule

\multicolumn{1}{l}{GRB} &
\multicolumn{1}{c}{$z$} &
\multicolumn{1}{c}{$\log T^{*}_{X}$} &
\multicolumn{1}{c}{$\log F_{X}$} &
\multicolumn{1}{c}{$\log F_{\rm peak}$} &
\multicolumn{1}{c}{$\alpha_{\mathrm{plateau}}$} &
\multicolumn{1}{c}{$K_{\mathrm{plateau}}$} &
\multicolumn{1}{c}{$\log\mathrm{FK}_{\mathrm{plateau}}$} &
\multicolumn{1}{c}{$\alpha^{\mathrm{PL/CPL}}_{\mathrm{prompt}}$} &
\multicolumn{1}{c}{$E_{\mathrm{peak}}$} &
\multicolumn{1}{c}{$K^{\mathrm{PL/CPL}}_{\mathrm{prompt}}$} &
\multicolumn{1}{c}{$\log\mathrm{FK}_{\mathrm{prompt}}$} &
\multicolumn{1}{c}{$\Delta\chi^2$} & 
\multicolumn{1}{c}{Ref.}\\

\midrule

\endhead

\endfoot
\bottomrule
\endlastfoot

050315 & 1.949 & $4.86\pm0.03$ & $-11.38\pm0.02$ & $-6.95\pm0.03$ & $1.86\pm0.05$ & $0.859^{+0.048}_{-0.045}$ & $-11.45\pm0.04$ & $-2.08^{+0.11}_{-0.12}$ & -- & $1.09^{+0.15}_{-0.12}$ & $-6.91^{+0.09}_{-0.08}$ & $2.84$ & C\\
050318 & 1.44 & $4.13\pm0.03$ & $-11.15^{+0.03}_{-0.04}$ & $-6.67\pm0.02$ & $1.97\pm0.07$ & $0.974^{+0.062}_{-0.059}$ & $-11.16^{+0.06}_{-0.07}$ & $-0.806^{+0.281}_{-0.254}$ & $67.2^{+13.7}_{-6.1}$ & $0.708^{+0.075}_{-0.112}$ & $-6.82^{+0.06}_{-0.09}$ & $12.75$ & C\\
050401 & 2.9 & $3.12\pm0.02$ & $-9.65^{+0.01}_{-0.02}$ & $-5.96\pm0.02$ & $2.1^{+0.36}_{-0.3}$ & $1.15^{+0.72}_{-0.39}$ & $-9.6\pm0.2$ & $-0.067^{+0.422}_{-0.363}$ & $108^{+36.1}_{-12.9}$ & $0.224^{+0.089}_{-0.097}$ & $-6.61^{+0.16}_{-0.27}$ & $10.33$ & C\\
050505 & 4.27 & $4.27\pm0.02$ & $-11.05\pm0.02$ & $-6.78\pm0.04$ & $1.98\pm0.05$ & $0.967^{+0.084}_{-0.077}$ & $-11.06\pm0.06$ & $-1.08\pm0.15$ & -- & $0.217^{+0.061}_{-0.048}$ & $-7.44\pm0.15$ & $1.91$ & C\\
050730 & 3.97 & $4.05\pm0.01$ & $-10.1^{+0.02}_{-0.01}$ & $-7.34^{+0.07}_{-0.08}$ & $1.47\pm0.03$ & $0.427^{+0.022}_{-0.02}$ & $-10.47^{+0.04}_{-0.03}$ & $-1.32^{+0.28}_{-0.27}$ & -- & $0.336^{+0.182}_{-0.121}$ & $-7.81^{+0.26}_{-0.28}$ & $0.00$ & C\\
050803 & 3.5 & $4.22\pm0.03$ & $-10.86\pm0.03$ & $-7.09^{+0.03}_{-0.04}$ & $1.78\pm0.05$ & $0.718^{+0.056}_{-0.052}$ & $-11\pm0.06$ & $-1.23^{+0.12}_{-0.13}$ & -- & $0.314^{+0.068}_{-0.052}$ & $-7.59\pm0.12$ & $-0.02$ & C\\
050826 & 0.297 & $4.76\pm0.08$ & $-12.3^{+0.1}_{-0.2}$ & $-7.5^{+0.08}_{-0.12}$ & $2.1^{+0.3}_{-0.24}$ & $1.03^{+0.08}_{-0.07}$ & $-12.29^{+0.13}_{-0.23}$ & $-1.25^{+0.42}_{-0.41}$ & -- & $0.823^{+0.092}_{-0.085}$ & $-7.59^{+0.13}_{-0.17}$ & $-0.02$ & C\\
050904 & 6.29 & $4.68\pm0.06$ & $-11.5^{+0.2}_{-0.1}$ & $-7.27^{+0.07}_{-0.09}$ & $1.83\pm0.04$ & $0.713^{+0.059}_{-0.054}$ & $-11.65^{+0.23}_{-0.13}$ & $-1.24^{+0.27}_{-0.26}$ & -- & $0.221^{+0.149}_{-0.092}$ & $-7.93^{+0.29}_{-0.32}$ & $1.24$ & C\\
051016B & 0.9364 & $4.26\pm0.04$ & $-11.59\pm0.03$ & $-7.08^{+0.03}_{-0.04}$ & $1.72\pm0.1$ & $0.831^{+0.057}_{-0.053}$ & $-11.67\pm0.06$ & $-1.91\pm0.13$ & -- & $0.942^{+0.085}_{-0.077}$ & $-7.11^{+0.07}_{-0.08}$ & $4.54$ & C\\
051109A & 2.346 & $3.91\pm0.03$ & $-10.52\pm0.03$ & $-6.52^{+0.04}_{-0.05}$ & $2.01\pm0.09$ & $1.01^{+0.12}_{-0.1}$ & $-10.51\pm0.08$ & $-1.45\pm0.15$ & -- & $0.515^{+0.102}_{-0.086}$ & $-6.81^{+0.12}_{-0.13}$ & $1.72$ & C\\
060115 & 3.53 & $3.7\pm0.07$ & $-11.3^{+0.05}_{-0.04}$ & $-7.14^{+0.04}_{-0.05}$ & $2.01^{+0.22}_{-0.11}$ & $1.02^{+0.4}_{-0.16}$ & $-11.29^{+0.19}_{-0.11}$ & $-1.27\pm0.15$ & -- & $0.332^{+0.084}_{-0.067}$ & $-7.62^{+0.14}_{-0.15}$ & $-0.01$ & C\\
060223A & 4.41 & $2.79\pm0.14$ & $-11^{+0.2}_{-0.1}$ & $-7^{+0.03}_{-0.04}$ & $1.84^{+0.16}_{-0.1}$ & $0.763^{+0.237}_{-0.118}$ & $-11.12^{+0.32}_{-0.17}$ & $-1.58\pm0.13$ & -- & $0.492^{+0.121}_{-0.097}$ & $-7.31\pm0.13$ & $0.45$ & C\\
060604 & 2.1357 & $4.46\pm0.03$ & $-11.77^{+0.03}_{-0.02}$ & $-7.63^{+0.11}_{-0.15}$ & $2.03\pm0.09$ & $1.03^{+0.12}_{-0.1}$ & $-11.76^{+0.07}_{-0.06}$ & $-1.75^{+0.46}_{-0.56}$ & -- & $0.751^{+0.674}_{-0.307}$ & $-7.75^{+0.39}_{-0.38}$ & $-0.03$ & B\\
060607A & 3.082 & $4.37\pm0.02$ & $-11.02^{+0.04}_{-0.05}$ & $-6.89^{+0.02}_{-0.03}$ & $1.51\pm0.04$ & $0.502^{+0.029}_{-0.027}$ & $-11.32^{+0.06}_{-0.08}$ & $-1.08^{+0.1}_{-0.09}$ & -- & $0.274^{+0.037}_{-0.036}$ & $-7.45^{+0.08}_{-0.09}$ & $-0.02$ & C\\
060908 & 1.8836 & $2.82\pm0.04$ & $-9.63^{+0.03}_{-0.02}$ & $-6.57^{+0.03}_{-0.02}$ & $2.02^{+0.18}_{-0.17}$ & $1.02^{+0.22}_{-0.17}$ & $-9.6\pm0.1$ & $-0.274^{+0.388}_{-0.338}$ & $123^{+88}_{-19}$ & $0.347^{+0.106}_{-0.149}$ & $-7.03^{+0.15}_{-0.26}$ & $6.25$ & C\\
060927 & 5.6 & $3.15\pm0.07$ & $-10.78\pm0.05$ & $-6.61\pm0.02$ & $1.65^{+0.17}_{-0.08}$ & $0.517^{+0.195}_{-0.073}$ & $-11.07^{+0.19}_{-0.12}$ & $-0.286^{+0.273}_{-0.247}$ & $130^{+56}_{-18}$ & $0.103^{+0.052}_{-0.048}$ & $-7.6^{+0.2}_{-0.3}$ & $10.04$ & C\\
061021 & 0.3463 & $3.24\pm0.02$ & $-10.04\pm0.02$ & $-6.26^{+0.01}_{-0.02}$ & $1.49^{+0.16}_{-0.09}$ & $0.859^{+0.042}_{-0.022}$ & $-10.11^{+0.04}_{-0.03}$ & $-1.13^{+0.04}_{-0.05}$ & -- & $0.772^{+0.012}_{-0.009}$ & $-6.37^{+0.02}_{-0.03}$ & $-0.12$ & C\\
061110A & 0.758 & $2.25\pm0.02$ & $-8.33\pm0.13$ & $-7.55^{+0.09}_{-0.1}$ & $2.97\pm0.04$ & $1.73\pm0.04$ & $-8.1\pm0.1$ & $-2.04^{+0.31}_{-0.36}$ & -- & $1.02^{+0.23}_{-0.16}$ & $-7.54\pm0.18$ & $0.20$ & C\\
070208 & 1.165 & $3.44\pm0.09$ & $-10.9\pm0.1$ & $-7.47^{+0.1}_{-0.16}$ & $2\pm0.24$ & $1^{+0.2}_{-0.17}$ & $-10.9\pm0.18$ & $-1.09^{+0.61}_{-0.5}$ & -- & $0.495^{+0.234}_{-0.186}$ & $-7.78^{+0.27}_{-0.37}$ & $0.00$ & C\\
070721B & 3.626 & $3.95^{+0.08}_{-0.04}$ & $-10.5\pm0.2$ & $-6.8^{+0.03}_{-0.05}$ & $1.52\pm0.07$ & $0.479^{+0.055}_{-0.048}$ & $-10.82\pm0.25$ & $-0.83\pm0.17$ & -- & $0.167^{+0.049}_{-0.039}$ & $-7.58^{+0.14}_{-0.16}$ & $0.46$ & C\\
070802 & 2.45 & $3.92\pm0.14$ & $-11.7\pm0.1$ & $-7.64^{+0.11}_{-0.17}$ & $1.78^{+0.2}_{-0.15}$ & $0.762^{+0.214}_{-0.13}$ & $-11.82^{+0.21}_{-0.18}$ & $-2.48^{+0.45}_{-0.9}$ & -- & $1.81^{+3.71}_{-0.77}$ & $-7.38^{+0.59}_{-0.41}$ & $-0.74$ & C\\
071003 & 1.60435 & $4.78\pm0.06$ & $-11.7\pm0.1$ & $-6.2^{+0.02}_{-0.01}$ & $1.83^{+0.08}_{-0.04}$ & $0.85^{+0.067}_{-0.032}$ & $-11.77^{+0.13}_{-0.12}$ & $-0.84^{+0.06}_{-0.07}$ & -- & $0.329^{+0.023}_{-0.018}$ & $-6.68^{+0.05}_{-0.03}$ & $-0.02$ & C\\
071117 & 1.3331 & $3.9\pm0.07$ & $-11.24^{+0.07}_{-0.06}$ & $-5.98\pm0.01$ & $1.9\pm0.18$ & $0.919^{+0.151}_{-0.13}$ & $-11.28^{+0.14}_{-0.13}$ & $-0.314^{+0.148}_{-0.139}$ & $143^{+26}_{-13}$ & $0.424^{+0.046}_{-0.059}$ & $-6.35^{+0.06}_{-0.07}$ & $28.95$ & C\\
080319C & 1.95 & $3.19^{+0.03}_{-0.02}$ & $-8.92^{+0.04}_{-0.03}$ & $-6.33\pm0.02$ & $1.49^{+0.06}_{-0.05}$ & $0.576^{+0.039}_{-0.03}$ & $-9.2\pm0.1$ & $-1.07^{+0.07}_{-0.06}$ & -- & $0.366^{+0.024}_{-0.027}$ & $-6.77\pm0.05$ & $0.45$ & C\\
080411 & 1.03 & $4.21\pm0.03$ & $-10.11^{+0.03}_{-0.04}$ & $-5.523\pm0.005$ & $1.86^{+0.09}_{-0.08}$ & $0.906^{+0.059}_{-0.05}$ & $-10.15^{+0.06}_{-0.07}$ & $-1.13\pm0.09$ & $101.9^{+9.8}_{-6}$ & $0.696^{+0.038}_{-0.043}$ & $-5.68\pm0.03$ & $41.12$ & B\\
080607 & 3.036 & $3.46\pm0.04$ & $-10.05^{+0.06}_{-0.07}$ & $-5.63^{+0.02}_{-0.01}$ & $1.9\pm0.3$ & $0.87^{+0.452}_{-0.298}$ & $-10.11^{+0.24}_{-0.25}$ & $-0.8^{+0.04}_{-0.05}$ & -- & $0.187^{+0.014}_{-0.01}$ & $-6.36^{+0.05}_{-0.03}$ & $0.72$ & C\\
080710 & 0.845 & $4.09\pm0.03$ & $-11.23\pm0.04$ & $-7.22^{+0.06}_{-0.08}$ & $1.72^{+0.09}_{-0.06}$ & $0.842^{+0.048}_{-0.03}$ & $-11.3\pm0.06$ & $-1.84^{+0.21}_{-0.24}$ & -- & $0.907^{+0.143}_{-0.11}$ & $-7.26^{+0.12}_{-0.13}$ & $1.78$ & C\\
080810 & 3.35 & $3.27\pm0.02$ & $-9.93\pm0.02$ & $-6.77^{+0.02}_{-0.03}$ & $2.07^{+0.05}_{-0.04}$ & $1.11^{+0.08}_{-0.06}$ & $-9.89\pm0.05$ & $-1.24^{+0.11}_{-0.1}$ & -- & $0.327^{+0.052}_{-0.049}$ & $-7.26^{+0.08}_{-0.1}$ & $-0.02$ & C\\
080928 & 1.69 & $4.08\pm0.03$ & $-11.13\pm0.04$ & $-6.8\pm0.02$ & $2\pm0.07$ & $1\pm0.07$ & $-11.13\pm0.07$ & $-1.5^{+0.07}_{-0.08}$ & -- & $0.61^{+0.05}_{-0.041}$ & $-7.02\pm0.05$ & $0.99$ & C\\
081028A & 3.038 & $4.75^{+0.02}_{-0.01}$ & $-11.4^{+0.04}_{-0.03}$ & $-7.39^{+0.06}_{-0.08}$ & $1.95\pm0.04$ & $0.933^{+0.053}_{-0.051}$ & $-11.43^{+0.06}_{-0.05}$ & $-1.64^{+0.23}_{-0.24}$ & -- & $0.605^{+0.241}_{-0.166}$ & $-7.61^{+0.2}_{-0.22}$ & $0.58$ & C\\
081029 & 3.847 & $4.48\pm0.03$ & $-11.86\pm0.05$ & $-7.51^{+0.09}_{-0.15}$ & $1.88^{+0.06}_{-0.05}$ & $0.827^{+0.083}_{-0.062}$ & $-11.94\pm0.09$ & $-1.74^{+0.39}_{-0.43}$ & -- & $0.663^{+0.645}_{-0.305}$ & $-7.69^{+0.38}_{-0.42}$ & $0.62$ & C\\
081203A & 2.1 & $4.02\pm0.02$ & $-10.92^{+0.03}_{-0.04}$ & $-6.61\pm0.02$ & $1.97^{+0.08}_{-0.07}$ & $0.967^{+0.091}_{-0.074}$ & $-10.94\pm0.07$ & $-1.17^{+0.07}_{-0.08}$ & -- & $0.391^{+0.037}_{-0.03}$ & $-7.02^{+0.06}_{-0.05}$ & $1.82$ & C\\
090102 & 1.55 & $3.15\pm0.03$ & $-9.46^{+0.03}_{-0.02}$ & $-6.27\pm0.04$ & $1.65\pm0.09$ & $0.721^{+0.063}_{-0.059}$ & $-9.6\pm0.1$ & $-0.91\pm0.14$ & -- & $0.36^{+0.051}_{-0.044}$ & $-6.71\pm0.1$ & $1.13$ & C\\
090205 & 4.7 & $3.71^{+0.08}_{-0.06}$ & $-11.34\pm0.08$ & $-7.53^{+0.08}_{-0.09}$ & $2.04^{+0.15}_{-0.12}$ & $1.07^{+0.32}_{-0.2}$ & $-11.31^{+0.19}_{-0.17}$ & $-1.98^{+0.27}_{-0.29}$ & -- & $0.966^{+0.634}_{-0.362}$ & $-7.55^{+0.3}_{-0.29}$ & $2.21$ & C\\
090423 & 8.2 & $3.92\pm0.05$ & $-11.28\pm0.05$ & $-6.91^{+0.03}_{-0.02}$ & $1.73^{+0.13}_{-0.12}$ & $0.549^{+0.184}_{-0.128}$ & $-11.54\pm0.17$ & $-1.65\pm0.09$ & -- & $0.46^{+0.102}_{-0.083}$ & $-7.25^{+0.12}_{-0.11}$ & $1.56$ & C\\
090429B & 9.4 & $3.2^{+0.11}_{-0.08}$ & $-10.9\pm0.1$ & $-6.94^{+0.03}_{-0.04}$ & $1.86^{+0.18}_{-0.16}$ & $0.72^{+0.378}_{-0.225}$ & $-11.04^{+0.28}_{-0.26}$ & $-1.7\pm0.1$ & -- & $0.495^{+0.131}_{-0.103}$ & $-7.25^{+0.13}_{-0.14}$ & $5.39$ & C\\
090516A & 4.1 & $4.33\pm0.02$ & $-11.27^{+0.02}_{-0.03}$ & $-5.75\pm0.03$ & $1.94\pm0.05$ & $0.907^{+0.077}_{-0.071}$ & $-11.31^{+0.05}_{-0.07}$ & $-1.68^{+0.11}_{-0.12}$ & -- & $0.594^{+0.128}_{-0.098}$ & $-5.98\pm0.11$ & $2.91$ & B\\
090519 & 3.85 & $3.45\pm0.15$ & $-11.8^{+0.2}_{-0.1}$ & $-7.16^{+0.06}_{-0.08}$ & $1.62^{+0.2}_{-0.16}$ & $0.549^{+0.204}_{-0.123}$ & $-12.06^{+0.34}_{-0.21}$ & $-0.31^{+0.44}_{-0.34}$ & -- & $0.069^{+0.05}_{-0.034}$ & $-8.32^{+0.29}_{-0.38}$ & $-0.18$ & C\\
090529A & 2.625 & $4.47\pm0.13$ & $-12.2\pm0.1$ & $-7.31^{+0.09}_{-0.12}$ & $1.64^{+0.3}_{-0.15}$ & $0.629^{+0.297}_{-0.11}$ & $-12.4^{+0.27}_{-0.18}$ & $-1.67^{+0.33}_{-0.34}$ & -- & $0.654^{+0.359}_{-0.227}$ & $-7.49^{+0.28}_{-0.3}$ & $2.57$ & C\\
090812 & 2.452 & $3.53\pm0.08$ & $-10.17^{+0.07}_{-0.06}$ & $-6.45^{+0.01}_{-0.02}$ & $1.64^{+0.16}_{-0.15}$ & $0.64^{+0.141}_{-0.108}$ & $-10.36^{+0.16}_{-0.14}$ & $-0.86\pm0.06$ & -- & $0.244\pm0.018$ & $-7.06^{+0.04}_{-0.05}$ & $-0.03$ & C\\
091109A & 3.076 & $3.38\pm0.05$ & $-11.15^{+0.04}_{-0.03}$ & $-7^{+0.07}_{-0.1}$ & $1.8\pm0.24$ & $0.755^{+0.303}_{-0.216}$ & $-11.27^{+0.19}_{-0.18}$ & $-1.36^{+0.34}_{-0.32}$ & -- & $0.407^{+0.231}_{-0.155}$ & $-7.39^{+0.27}_{-0.31}$ & $-0.02$ & C\\
091208B & 1.063 & $3.31\pm0.03$ & $-10.18\pm0.02$ & $-5.95\pm0.02$ & $1.97^{+0.12}_{-0.11}$ & $0.979^{+0.088}_{-0.075}$ & $-10.19^{+0.06}_{-0.05}$ & $-1.49\pm0.07$ & -- & $0.691^{+0.036}_{-0.034}$ & $-6.11\pm0.04$ & $-0.05$ & C\\
100425A & 1.755 & $3.6\pm0.07$ & $-11.5\pm0.04$ & $-7.13^{+0.05}_{-0.06}$ & $2.5^{+0.43}_{-0.36}$ & $1.66^{+0.91}_{-0.51}$ & $-11.28^{+0.23}_{-0.2}$ & $-2.36^{+0.19}_{-0.21}$ & -- & $1.44^{+0.34}_{-0.25}$ & $-6.97\pm0.14$ & $4.85$ & C\\
100513A & 4.772 & $3.75\pm0.08$ & $-11.57\pm0.06$ & $-7.32^{+0.06}_{-0.08}$ & $2.3\pm0.24$ & $1.69^{+0.89}_{-0.58}$ & $-11.34\pm0.24$ & $-1.29^{+0.27}_{-0.25}$ & -- & $0.288^{+0.158}_{-0.109}$ & $-7.86^{+0.25}_{-0.29}$ & $0.00$ & C\\
100728A & 1.567 & $4.03\pm0.03$ & $-10.18^{+0.05}_{-0.04}$ & $-6.28\pm0.01$ & $1.93^{+0.12}_{-0.11}$ & $0.936^{+0.112}_{-0.092}$ & $-10.21^{+0.1}_{-0.09}$ & $-0.7\pm0.05$ & -- & $0.294\pm0.014$ & $-6.81\pm0.03$ & $-0.07$ & C\\
100901A & 1.408 & $4.91^{+0.02}_{-0.01}$ & $-11.34\pm0.02$ & $-7.28^{+0.07}_{-0.11}$ & $2.07\pm0.04$ & $1.06^{+0.04}_{-0.03}$ & $-11.31\pm0.04$ & $-1.84^{+0.35}_{-0.4}$ & -- & $0.869^{+0.366}_{-0.23}$ & $-7.34^{+0.22}_{-0.24}$ & $0.00$ & C\\
110106B & 0.618 & $3.87\pm0.04$ & $-10.81^{+0.02}_{-0.03}$ & $-6.81^{+0.03}_{-0.05}$ & $1.96\pm0.13$ & $0.981^{+0.063}_{-0.06}$ & $-10.82^{+0.05}_{-0.06}$ & $-1.54^{+0.13}_{-0.14}$ & -- & $0.801^{+0.056}_{-0.048}$ & $-6.91^{+0.06}_{-0.08}$ & $2.94$ & C\\
110422A & 1.77 & $3.5\pm0.02$ & $-9.68^{+0.01}_{-0.02}$ & $-5.56\pm0.01$ & $1.74\pm0.06$ & $0.767^{+0.049}_{-0.045}$ & $-9.8^{+0.04}_{-0.05}$ & $-0.97\pm0.03$ & -- & $0.35^{+0.011}_{-0.01}$ & $-6.02\pm0.02$ & $4.95$ & C\\
110503A & 1.61 & $2.54\pm0.02$ & $-9.08\pm0.02$ & $-5.55^{+0.01}_{-0.02}$ & $1.69\pm0.04$ & $0.743^{+0.029}_{-0.028}$ & $-9.21\pm0.04$ & $-0.163^{+0.21}_{-0.194}$ & $130^{+26}_{-13}$ & $0.362^{+0.058}_{-0.071}$ & $-5.99^{+0.07}_{-0.11}$ & $22.68$ & C\\
110715A & 0.82 & $2.61\pm0.02$ & $-9.08^{+0.02}_{-0.01}$ & $-5.372\pm0.005$ & $1.75\pm0.04$ & $0.861^{+0.021}_{-0.02}$ & $-9.15^{+0.03}_{-0.02}$ & $-0.985^{+0.081}_{-0.078}$ & $152^{+26}_{-14}$ & $0.682^{+0.032}_{-0.039}$ & $-5.54\pm0.03$ & $29.78$ & C\\
110731A & 2.83 & $5.09\pm0.06$ & $-12.09^{+0.06}_{-0.05}$ & $-6.05\pm0.01$ & $1.43^{+0.26}_{-0.15}$ & $0.465^{+0.194}_{-0.085}$ & $-12.42^{+0.21}_{-0.14}$ & $-0.774^{+0.136}_{-0.131}$ & $135^{+33}_{-15}$ & $0.327^{+0.057}_{-0.064}$ & $-6.54^{+0.08}_{-0.1}$ & $17.81$ & C\\
111123A & 3.1516 & $4.42\pm0.04$ & $-11.61^{+0.05}_{-0.04}$ & $-7.14^{+0.05}_{-0.04}$ & $2.28^{+0.15}_{-0.14}$ & $1.49^{+0.35}_{-0.27}$ & $-11.44^{+0.14}_{-0.13}$ & $-1.33^{+0.17}_{-0.16}$ & -- & $0.385^{+0.099}_{-0.083}$ & $-7.55\pm0.15$ & $-0.01$ & C\\
120327A & 2.813 & $3.49\pm0.03$ & $-10.15^{+0.02}_{-0.03}$ & $-6.48^{+0.01}_{-0.02}$ & $1.65^{+0.08}_{-0.05}$ & $0.626^{+0.071}_{-0.041}$ & $-10.35^{+0.07}_{-0.06}$ & $-1.25^{+0.05}_{-0.06}$ & -- & $0.366^{+0.031}_{-0.023}$ & $-6.92\pm0.05$ & $0.66$ & C\\
120802A & 3.796 & $3.54\pm0.13$ & $-11.17\pm0.05$ & $-6.67\pm0.02$ & $1.99^{+0.18}_{-0.15}$ & $0.984^{+0.321}_{-0.206}$ & $-11.18^{+0.17}_{-0.15}$ & $-1.66^{+0.08}_{-0.07}$ & -- & $0.587^{+0.068}_{-0.069}$ & $-6.9^{+0.07}_{-0.08}$ & $2.87$ & C\\
120804A & 1.3 & $2.57\pm0.11$ & $-9.76^{+0.1}_{-0.11}$ & $-6.06^{+0.02}_{-0.01}$ & $1.9\pm0.24$ & $0.92^{+0.204}_{-0.167}$ & $-9.8\pm0.2$ & $-1.34\pm0.05$ & -- & $0.577^{+0.025}_{-0.023}$ & $-6.3^{+0.04}_{-0.03}$ & $3.76$ & C\\
120907A & 0.97 & $3.21\pm0.04$ & $-10.38^{+0.03}_{-0.02}$ & $-6.64^{+0.03}_{-0.04}$ & $1.76\pm0.1$ & $0.85^{+0.059}_{-0.056}$ & $-10.45^{+0.06}_{-0.05}$ & $-1.39\pm0.13$ & -- & $0.661^{+0.061}_{-0.056}$ & $-6.82^{+0.07}_{-0.08}$ & $-0.04$ & C\\
121024A & 2.298 & $4.17\pm0.04$ & $-11.31\pm0.04$ & $-6.9^{+0.04}_{-0.06}$ & $1.87\pm0.1$ & $0.856^{+0.109}_{-0.096}$ & $-11.38\pm0.09$ & $-1.07^{+0.19}_{-0.18}$ & -- & $0.33^{+0.079}_{-0.067}$ & $-7.38^{+0.13}_{-0.16}$ & $-0.03$ & C\\
121229A & 2.707 & $4.71\pm0.24$ & $-12.4^{+0.2}_{-0.1}$ & $-7.46^{+0.12}_{-0.19}$ & $2^{+0.3}_{-0.18}$ & $1^{+0.48}_{-0.21}$ & $-12.4^{+0.37}_{-0.2}$ & $-1.35^{+0.96}_{-0.66}$ & -- & $0.427^{+0.586}_{-0.306}$ & $-7.83^{+0.5}_{-0.74}$ & $-0.01$ & C\\
130408A & 3.76 & $3.58\pm0.04$ & $-10.33\pm0.05$ & $-6.46^{+0.06}_{-0.09}$ & $1.75^{+0.08}_{-0.06}$ & $0.677^{+0.09}_{-0.06}$ & $-10.5^{+0.1}_{-0.09}$ & $-1.78^{+0.23}_{-0.24}$ & -- & $0.709^{+0.323}_{-0.213}$ & $-6.61^{+0.22}_{-0.25}$ & $1.84$ & C\\
130418A & 1.218 & $2.59\pm0.04$ & $-9.25^{+0.04}_{-0.03}$ & $-7.47^{+0.09}_{-0.13}$ & $1.17^{+0.03}_{-0.02}$ & $0.516^{+0.013}_{-0.008}$ & $-9.54^{+0.05}_{-0.04}$ & $-2.57^{+0.34}_{-0.45}$ & -- & $1.57^{+0.68}_{-0.37}$ & $-7.27\pm0.25$ & $4.70$ & C\\
130420A & 1.297 & $3.57\pm0.05$ & $-10.87^{+0.03}_{-0.04}$ & $-6.66^{+0.02}_{-0.03}$ & $2.2\pm0.09$ & $1.18^{+0.09}_{-0.08}$ & $-10.8^{+0.06}_{-0.07}$ & $-1.03^{+0.31}_{-0.28}$ & $58^{+13.4}_{-5.5}$ & $0.821^{+0.07}_{-0.121}$ & $-6.75^{+0.06}_{-0.1}$ & $8.72$ & C\\
130505A & 2.27 & $3.25\pm0.01$ & $-8.92\pm0.01$ & $-5.52\pm0.03$ & $1.71\pm0.02$ & $0.709^{+0.017}_{-0.016}$ & $-9.07\pm0.02$ & $-0.82\pm0.1$ & -- & $0.247^{+0.031}_{-0.028}$ & $-6.13\pm0.08$ & $1.31$ & B\\
130511A & 1.3033 & $2.66\pm0.05$ & $-10.52\pm0.05$ & $-7.01\pm0.05$ & $1.54^{+0.17}_{-0.16}$ & $0.681^{+0.104}_{-0.085}$ & $-10.69\pm0.11$ & $-1.44\pm0.17$ & -- & $0.627^{+0.095}_{-0.083}$ & $-7.21\pm0.11$ & $0.08$ & C\\
130514A & 3.6 & $3.69\pm0.05$ & $-10.72\pm0.05$ & $-6.65^{+0.02}_{-0.03}$ & $1.7^{+0.24}_{-0.18}$ & $0.633^{+0.28}_{-0.152}$ & $-10.92^{+0.21}_{-0.17}$ & $-1.5\pm0.09$ & -- & $0.466^{+0.069}_{-0.06}$ & $-6.98^{+0.08}_{-0.09}$ & $2.35$ & C\\
130604A & 1.06 & $2.07\pm0.06$ & $-8.32^{+0.11}_{-0.1}$ & $-7.2^{+0.06}_{-0.08}$ & $2.05^{+0.19}_{-0.17}$ & $1.04^{+0.15}_{-0.12}$ & $-8.3\pm0.2$ & $-1.32^{+0.26}_{-0.25}$ & -- & $0.612^{+0.121}_{-0.105}$ & $-7.41^{+0.14}_{-0.16}$ & $-0.01$ & C\\
130606A & 5.913 & $3.3\pm0.05$ & $-10.18\pm0.04$ & $-6.63\pm0.02$ & $1.43^{+0.12}_{-0.06}$ & $0.332^{+0.087}_{-0.036}$ & $-10.66^{+0.14}_{-0.09}$ & $-1.01^{+0.07}_{-0.08}$ & -- & $0.147^{+0.025}_{-0.018}$ & $-7.46^{+0.09}_{-0.08}$ & $0.14$ & C\\
131117A & 4.042 & $2.87\pm0.08$ & $-10.84\pm0.05$ & $-7.33\pm0.06$ & $1.73^{+0.18}_{-0.12}$ & $0.646^{+0.219}_{-0.114}$ & $-11.03^{+0.18}_{-0.13}$ & $-1.68^{+0.19}_{-0.21}$ & -- & $0.596^{+0.241}_{-0.158}$ & $-7.56^{+0.21}_{-0.19}$ & $4.82$ & B\\
140213A & 1.208 & $3.83\pm0.04$ & $-10.13^{+0.04}_{-0.03}$ & $-5.79\pm0.01$ & $1.77^{+0.08}_{-0.07}$ & $0.833^{+0.055}_{-0.045}$ & $-10.21^{+0.07}_{-0.05}$ & $-1.71\pm0.03$ & -- & $0.795\pm0.019$ & $-5.89\pm0.02$ & $4.64$ & C\\
140423A & 3.26 & $3.91\pm0.03$ & $-10.85\pm0.03$ & $-6.73^{+0.03}_{-0.04}$ & $2\pm0.08$ & $1^{+0.12}_{-0.11}$ & $-10.85\pm0.08$ & $-1.19^{+0.11}_{-0.12}$ & -- & $0.309^{+0.059}_{-0.045}$ & $-7.24\pm0.11$ & $0.23$ & C\\
140430A & 1.6 & $4.14\pm0.08$ & $-11.36\pm0.05$ & $-6.7\pm0.02$ & $2.02^{+0.17}_{-0.13}$ & $1.02^{+0.18}_{-0.12}$ & $-11.35^{+0.12}_{-0.1}$ & $-1.39^{+0.07}_{-0.08}$ & -- & $0.558^{+0.045}_{-0.036}$ & $-6.95\pm0.05$ & $4.11$ & C\\
140512A & 0.725 & $3.54\pm0.01$ & $-9.504^{+0.009}_{-0.01}$ & $-6.29\pm0.01$ & $1.73\pm0.1$ & $0.863^{+0.048}_{-0.046}$ & $-9.57\pm0.03$ & $-1.24^{+0.04}_{-0.05}$ & -- & $0.661^{+0.018}_{-0.014}$ & $-6.47\pm0.02$ & $2.12$ & C\\
140614A & 4.233 & $3.7\pm0.06$ & $-10.96\pm0.07$ & $-7.48^{+0.09}_{-0.14}$ & $1.8^{+0.36}_{-0.3}$ & $0.718^{+0.585}_{-0.281}$ & $-11.1^{+0.33}_{-0.28}$ & $-0.78^{+0.83}_{-0.57}$ & -- & $0.133^{+0.208}_{-0.099}$ & $-8.36^{+0.5}_{-0.74}$ & $-0.05$ & C\\
141220A & 1.3195 & $2.78\pm0.03$ & $-9.87^{+0.02}_{-0.03}$ & $-6.09\pm0.02$ & $2.1^{+0.15}_{-0.13}$ & $1.09^{+0.14}_{-0.11}$ & $-9.8\pm0.1$ & $-0.249^{+0.372}_{-0.326}$ & $129^{+81}_{-20}$ & $0.438^{+0.1}_{-0.15}$ & $-6.45^{+0.11}_{-0.2}$ & $6.81$ & C\\
150120B & 0.46 & $3.39\pm0.06$ & $-10.79\pm0.04$ & $-7.11^{+0.05}_{-0.07}$ & $1.94^{+0.14}_{-0.13}$ & $0.978^{+0.053}_{-0.047}$ & $-10.8\pm0.06$ & $-1.94^{+0.21}_{-0.23}$ & -- & $0.978^{+0.088}_{-0.075}$ & $-7.12^{+0.09}_{-0.1}$ & $0.07$ & C\\
151021A & 2.33 & $3.16\pm0.03$ & $-9.64\pm0.04$ & $-6.12^{+0.03}_{-0.02}$ & $2.4^{+0.36}_{-0.3}$ & $1.62^{+0.87}_{-0.49}$ & $-9.4\pm0.2$ & $-1.42^{+0.09}_{-0.1}$ & -- & $0.498^{+0.063}_{-0.051}$ & $-6.42^{+0.08}_{-0.07}$ & $-0.02$ & C\\
151112A & 4.1 & $4.43\pm0.04$ & $-11.62\pm0.03$ & $-6.88^{+0.03}_{-0.04}$ & $2.24^{+0.11}_{-0.1}$ & $1.48^{+0.29}_{-0.22}$ & $-11.45^{+0.11}_{-0.1}$ & $-1.67^{+0.12}_{-0.13}$ & -- & $0.584^{+0.138}_{-0.104}$ & $-7.11^{+0.12}_{-0.13}$ & $1.98$ & B\\
151215A & 2.59 & $3.06\pm0.07$ & $-10.73\pm0.06$ & $-6.9^{+0.04}_{-0.05}$ & $2.2^{+0.16}_{-0.15}$ & $1.29^{+0.29}_{-0.22}$ & $-10.62^{+0.15}_{-0.14}$ & $-1.99^{+0.19}_{-0.21}$ & -- & $0.987^{+0.304}_{-0.213}$ & $-6.91^{+0.16}_{-0.15}$ & $1.26$ & B\\
160117B & 0.87 & $4.1\pm0.1$ & $-11.75\pm0.07$ & $-6.98\pm0.03$ & $1.65^{+0.12}_{-0.1}$ & $0.803^{+0.063}_{-0.048}$ & $-11.85\pm0.1$ & $-2.03^{+0.12}_{-0.11}$ & -- & $1.02\pm0.07$ & $-6.97\pm0.06$ & $1.38$ & C\\
160203A & 3.52 & $3.66\pm0.05$ & $-11.37\pm0.05$ & $-7.1^{+0.09}_{-0.13}$ & $1.9^{+0.43}_{-0.24}$ & $0.86^{+0.785}_{-0.261}$ & $-11.44^{+0.33}_{-0.21}$ & $-2^{+0.34}_{-0.43}$ & -- & $1^{+0.91}_{-0.4}$ & $-7.1^{+0.37}_{-0.35}$ & $0.00$ & C\\
160804A & 0.736 & $4.38\pm0.06$ & $-11.52^{+0.03}_{-0.04}$ & $-6.7\pm0.03$ & $1.92^{+0.11}_{-0.1}$ & $0.957^{+0.06}_{-0.052}$ & $-11.54\pm0.06$ & $-1.66^{+0.11}_{-0.12}$ & -- & $0.829^{+0.057}_{-0.049}$ & $-6.78\pm0.06$ & $1.60$ & C\\
161108A & 1.159 & $4.69\pm0.11$ & $-11.8\pm0.1$ & $-7.44^{+0.06}_{-0.08}$ & $1.84^{+0.15}_{-0.14}$ & $0.884^{+0.108}_{-0.09}$ & $-11.85\pm0.15$ & $-2.07^{+0.22}_{-0.25}$ & -- & $1.06^{+0.22}_{-0.17}$ & $-7.42^{+0.14}_{-0.15}$ & $3.11$ & C\\
161117A & 1.549 & $4.08\pm0.02$ & $-10.75\pm0.02$ & $-6.34\pm0.01$ & $1.97\pm0.07$ & $0.972^{+0.066}_{-0.061}$ & $-10.76\pm0.05$ & $-1.12^{+0.18}_{-0.17}$ & $74.4^{+13.8}_{-6.3}$ & $0.76^{+0.069}_{-0.106}$ & $-6.46^{+0.05}_{-0.08}$ & $14.08$ & B\\
170405A & 3.51 & $3.63\pm0.02$ & $-10.31\pm0.03$ & $-7.81^{+0.01}_{-0.02}$ & $1.61^{+0.11}_{-0.06}$ & $0.556^{+0.1}_{-0.048}$ & $-10.57^{+0.1}_{-0.07}$ & $-1.59\pm0.07$ & -- & $0.539^{+0.06}_{-0.054}$ & $-8.08^{+0.06}_{-0.07}$ & $-0.03$ & B\\
170519A & 0.818 & $4.15\pm0.03$ & $-11.06^{+0.03}_{-0.04}$ & $-7.27^{+0.06}_{-0.07}$ & $1.95\pm0.08$ & $0.971^{+0.047}_{-0.046}$ & $-11.07^{+0.05}_{-0.06}$ & $-1.4^{+0.22}_{-0.21}$ & -- & $0.699^{+0.093}_{-0.086}$ & $-7.43^{+0.11}_{-0.13}$ & $0.00$ & C\\
170531B & 2.366 & $3.8\pm0.07$ & $-11.4\pm0.05$ & $-7.29^{+0.07}_{-0.08}$ & $2.01^{+0.21}_{-0.18}$ & $1.01^{+0.3}_{-0.2}$ & $-11.4^{+0.16}_{-0.14}$ & $-1.72^{+0.27}_{-0.29}$ & -- & $0.712^{+0.3}_{-0.199}$ & $-7.44\pm0.22$ & $0.00$ & C\\
170903A & 0.886 & $4.41\pm0.04$ & $-11.26^{+0.02}_{-0.03}$ & $-6.59^{+0.03}_{-0.04}$ & $2.16\pm0.12$ & $1.11\pm0.08$ & $-11.22^{+0.05}_{-0.06}$ & $-1.85^{+0.12}_{-0.13}$ & -- & $0.909^{+0.078}_{-0.066}$ & $-6.63\pm0.07$ & $0.00$ & C\\
171222A & 2.409 & $4.78\pm0.23$ & $-12.1\pm0.1$ & $-7.41^{+0.09}_{-0.13}$ & $2.3\pm0.24$ & $1.44^{+0.5}_{-0.36}$ & $-11.94\pm0.23$ & $-2.13^{+0.36}_{-0.45}$ & -- & $1.17^{+0.87}_{-0.42}$ & $-7.34^{+0.33}_{-0.32}$ & $0.04$ & C\\
180205A & 1.409 & $3.21\pm0.05$ & $-10.31\pm0.04$ & $-6.63^{+0.02}_{-0.03}$ & $2.04^{+0.13}_{-0.12}$ & $1.04^{+0.12}_{-0.11}$ & $-10.3\pm0.09$ & $-1.72^{+0.08}_{-0.09}$ & -- & $0.782^{+0.064}_{-0.053}$ & $-6.74^{+0.05}_{-0.06}$ & $4.10$ & C\\
180404A & 1 & $4.2\pm0.08$ & $-11.62\pm0.04$ & $-7.03\pm0.04$ & $1.7^{+0.3}_{-0.18}$ & $0.812^{+0.188}_{-0.095}$ & $-11.71^{+0.13}_{-0.09}$ & $-1.67^{+0.13}_{-0.14}$ & -- & $0.796^{+0.081}_{-0.069}$ & $-7.13\pm0.08$ & $-0.04$ & C\\
180620B & 1.1175 & $4.67\pm0.04$ & $-10.98^{+0.02}_{-0.03}$ & $-6.48^{+0.01}_{-0.02}$ & $1.93\pm0.07$ & $0.949^{+0.051}_{-0.049}$ & $-11^{+0.04}_{-0.05}$ & $-1.1\pm0.06$ & -- & $0.509^{+0.023}_{-0.022}$ & $-6.77^{+0.03}_{-0.04}$ & $1.68$ & C\\
180624A & 2.855 & $4.17\pm0.06$ & $-11.54^{+0.05}_{-0.06}$ & $-7.05^{+0.05}_{-0.06}$ & $1.96^{+0.11}_{-0.1}$ & $0.947^{+0.152}_{-0.119}$ & $-11.56^{+0.11}_{-0.12}$ & $-1.87^{+0.18}_{-0.2}$ & -- & $0.839^{+0.26}_{-0.181}$ & $-7.13\pm0.17$ & $0.00$ & C\\
180720B & 0.654 & $3.62\pm0.004$ & $-8.849\pm0.003$ & $-6.2^{+0.06}_{-0.08}$ & $1.7\pm0.01$ & $0.86\pm0.004$ & $-8.915^{+0.006}_{-0.005}$ & $-0.664^{+0.162}_{-0.152}$ & $215^{+195}_{-41}$ & $0.625^{+0.056}_{-0.09}$ & $-6.4^{+0.1}_{-0.15}$ & $6.20$ & C\\
181010A & 1.39 & $3.16\pm0.04$ & $-9.93\pm0.02$ & $-6.9^{+0.03}_{-0.05}$ & $2.04^{+0.13}_{-0.12}$ & $1.04^{+0.12}_{-0.11}$ & $-9.9\pm0.1$ & $-1.18\pm0.14$ & -- & $0.489^{+0.064}_{-0.056}$ & $-7.21^{+0.08}_{-0.1}$ & $0.40$ & C\\
181020A & 2.938 & $4.29\pm0.03$ & $-10.42^{+0.06}_{-0.11}$ & $-6.12\pm0.01$ & $1.97\pm0.05$ & $0.96^{+0.068}_{-0.064}$ & $-10.44^{+0.09}_{-0.14}$ & $-0.86^{+0.04}_{-0.05}$ & -- & $0.21^{+0.014}_{-0.012}$ & $-6.8^{+0.04}_{-0.03}$ & $1.09$ & C\\
181110A & 1.505 & $3.77\pm0.03$ & $-10.4\pm0.1$ & $-6.62\pm0.02$ & $1.73^{+0.07}_{-0.04}$ & $0.78^{+0.052}_{-0.028}$ & $-10.51^{+0.13}_{-0.12}$ & $-1.85^{+0.07}_{-0.08}$ & -- & $0.871^{+0.067}_{-0.054}$ & $-6.68\pm0.05$ & $1.75$ & C\\
\end{longtable}%
}
\end{landscape}

\cleardoublepage

\chapter{Golden sample of \civ\ QSO data}
\label{AppendixC}

\addtolength{\tabcolsep}{0pt}
\LTcapwidth=\linewidth
{\setlength\tabcolsep{3pt}\footnotesize
\begin{longtable}{lccccc}
\caption{Sample of QSOs with a high-quality detection of the \civ\ time-delay. The sample is based on 38 sources compiled by \citet{2021ApJ...915..129K}. In the table, we list from the left to the right column: object name, redshift, flux density at 1350\,\AA, monochromatic luminosity at 1350\,\AA\, for the flat $\Lambda$CDM model ($H_0=70\,{\rm km\,s^{-1}\,Mpc^{-1}}$, $\Om=0.3$, $\Omega_{\Lambda}=0.7$), the rest-frame \civ\ time-lag (in days) determined for all the sources using either the ICCF or the zDCF method (or their combination), and the original reference. }
\label{tab:civdata}\\
\toprule
Object &  $z$ &  $\log \left(F_{1350}/{\rm erg}\,{\rm s^{-1}}{\rm cm^{-2}}\right)$  &  $\log \left(L_{1350}/{\rm erg}\,{\rm s^{-1}}\right)$  &  $\tau$ (days) & Reference\\
\midrule
\endfirsthead
\endhead
\bottomrule
\endfoot
NGC 4395 &  0.001064 &  $-11.4848 \pm     0.0272$ &    $39.9112 \pm     0.0272$ &    $0.040^{+0.024}_{-0.018}$  & \citet{2005ApJ...632..799P,2006ApJ...641..638P} \\
NGC 3783 & 0.00973  &  $-9.7341 \pm 0.0918$ &   $43.5899 \pm 0.0918 $&     $3.80^{+1.0}_{-0.9}$  & \citet{2005ApJ...632..799P,2006ApJ...641..638P}\\
NGC 7469 &  0.01632 &    $-9.9973 \pm 0.0712$  &  $43.7803 \pm 0.0712$ &    $2.5^{+0.3}_{-0.2}$    & \citet{2005ApJ...632..799P,2006ApJ...641..638P}\\
3C 390.3 &  0.0561 &   $-10.8036 \pm 0.2386$ &   $44.0719 \pm 0.2386$&    $35.7^{+11.4}_{-14.6}$ & \citet{2005ApJ...632..799P,2006ApJ...641..638P}\\
NGC 4151 &  0.00332 &    $-9.7544 \pm 0.1329$  &  $42.6314 \pm 0.1329$ &     $3.34^{+0.82}_{-0.77}$   & \citet{2006ApJ...647..901M}\\
NGC 5548  & 0.01676 &  $-10.2111 \pm 0.0894$  &  $43.5899 \pm  0.0894$ &     $4.53^{+0.35}_{-0.34}$ &  \citet{2015ApJ...806..128D}\\
CTS 286 &  2.551 &  $-11.6705 \pm 0.0719$  &  $47.0477 \pm     0.0719$&   $459^{+71}_{-92}$  & \citet{2018ApJ...865...56L} \\
CTS 406 &  3.178 &   $-12.0382 \pm 0.0402$ & $46.9101 \pm 0.0402$ &   $98^{+55}_{-74}$   & \citet{2018ApJ...865...56L}  \\
CTS 564 &  2.653 &   $-11.7615 \pm 0.0664$ & $46.9978 \pm 0.0664$ & $115^{+184}_{-29}$  & \citet{2018ApJ...865...56L} \\
CTS 650 &  2.659 &  $-11.8815 \pm 0.1068$  &  $46.8802 \pm 0.1068$ &  $162^{+33}_{-10}$   & \citet{2018ApJ...865...56L}  \\
CTS 953 &  2.526 &   $-11.7082 \pm 0.0868$ &   $46.9996 \pm 0.0868$ &   $73^{+115}_{-58}$  & \citet{2018ApJ...865...56L}  \\
CTS 1061 &  3.368 &   $-11.4788 \pm 0.0405$ &   $47.5299 \pm 0.0405$ &    $91^{+111}_{-24}$     & \citet{2018ApJ...865...56L}  \\
J 214355 &  2.607 &   $-11.7786 \pm    0.0485$ &   $46.9624 \pm    0.0485$ &  $136^{+100}_{-90}$    &  \citet{2018ApJ...865...56L} \\
J 221516 &  2.709 &   $-11.6263 \pm    0.0569$ &   $47.1550 \pm    0.0569$&   $153^{+91}_{-12}$   &  \citet{2018ApJ...865...56L} \\
DES J0228-04 & 1.905 &   $-11.9791 \pm    0.0405$ &   $46.4298 \pm    0.0405$ &   $123^{+43}_{-42}$   &  \citet{2019MNRAS.487.3650H} \\
DES J0033-42 &  2.593 &   $-12.2248 \pm  0.0201$ &   $46.5105  \pm   0.0201$ &    $95^{+16}_{-23}$    &  \citet{2019MNRAS.487.3650H}  \\
RMID 032 &  1.715 &   $-13.8040 \pm    0.0210$ &   $44.4928 \pm    0.0210$  &  $21.1^{+22.7}_{-8.3}$   & \citet{2019ApJ...887...38G}  \\
RMID 052 &  2.305 &   $-13.1121 \pm    0.0021$  &  $45.4990  \pm   0.0021$ &   $32.6^{+6.9}_{-2.1}$   & \citet{2019ApJ...887...38G}  \\
RMID 181 &  1.675 &   $-13.7265 \pm    0.0149$ &    $44.5451 \pm    0.0149$ &  $102.1^{+26.8}_{-10.0}$  & \citet{2019ApJ...887...38G}  \\
RMID 249 &  1.717 &   $-13.3140  \pm     0.0099$ &    $44.9841 \pm    0.0099$ &    $22.8^{+31.3}_{-11.5}$    & \citet{2019ApJ...887...38G}  \\
RMID 256 &  2.244 &   $-13.4939  \pm     0.0030$ &    $45.0888 \pm     0.0030$ &    $43.1^{+49.0}_{-15.1}$   & \citet{2019ApJ...887...38G}  \\
RMID 275 &  1.577 & $-12.5961  \pm     0.0010 $ &  $45.6110 \pm     0.0010$ &   $76.7^{+10.0}_{-3.9}$   & \citet{2019ApJ...887...38G}  \\
RMID 298 &  1.635  &   $-12.6497  \pm     0.0010$ &   $45.5960 \pm     0.0010$ &    $82.3^{+64.5}_{-24.5}$  & \citet{2019ApJ...887...38G}  \\
RMID 312 &  1.924  &   $-13.3424  \pm     0.0040$ &   $45.0770 \pm     0.0040$ &    $70.9^{+9.6}_{-3.3}$   & \citet{2019ApJ...887...38G}  \\
RMID 332 &  2.581 &   $-13.1795  \pm     0.0020$ &    $45.5510 \pm     0.0020$ &    $83.8^{+23.3}_{-6.5}$ & \citet{2019ApJ...887...38G}  \\
RMID 387 &  2.426 &   $-12.9782   \pm    0.0010$ &    $45.6870  \pm    0.0010$ &    $48.4^{+34.7}_{-10.1}$    & \citet{2019ApJ...887...38G}   \\
RMID 401 &  1.822 &   $-12.8714  \pm     0.0030$ &    $45.4900 \pm     0.0030$ &    $60.6^{+36.7}_{-13.0}$    & \citet{2019ApJ...887...38G}   \\
RMID 418 &  1.418 &   $-13.0533  \pm     0.0030$ &    $45.0398 \pm     0.0030$ &    $58.6^{+51.6}_{-21.3}$   & \citet{2019ApJ...887...38G}  \\
RMID 470 &  1.879 &   $-13.5732   \pm    0.0060$ &    $44.8210 \pm     0.0060$ &    $27.4^{+63.5}_{-22.0}$   & \citet{2019ApJ...887...38G}    \\
RMID 527 &  1.647 &  $-13.4655  \pm     0.0030$ &   $44.7880 \pm     0.0030$  &  $47.3^{+13.3}_{-5.0}$    & \citet{2019ApJ...887...38G}  \\
RMID 549 & 2.275 &   $-13.2283  \pm     0.0020$  &  $45.3690 \pm     0.0020$ &   $68.9^{+31.6}_{-9.6}$     & \citet{2019ApJ...887...38G}  \\
RMID 734 &  2.332 &   $-13.0935  \pm     0.0010$ & $45.5299 \pm     0.0010$ &   $68.0^{+38.2}_{-11.5}$  & \citet{2019ApJ...887...38G}  \\
RMID 363 &  2.635 &   $-12.2525   \pm    0.0206$ & $46.4997 \pm     0.0206$ &  $300.4^{+17.1}_{-4.7}$   & \citet{2019ApJ...883L..14S}  \\
RMID 372 &  1.745 &   $-12.6952  \pm     0.0198$ &    $45.6201 \pm     0.0198$ &    $67.0^{+20.4}_{-7.4}$   & \citet{2019ApJ...883L..14S}  \\
RMID 651 &  1.486 &  $-12.7234   \pm    0.0198$ &   $45.4200  \pm    0.0198$ &   $91.7^{+56.3}_{-22.7}$    & \citet{2019ApJ...883L..14S}  \\
S5 0836+71 &  2.172 &  $-11.5354   \pm    0.0680$ &    $47.0128 \pm     0.0680$ &   $230^{+91}_{-59}$   & \citet{2021ApJ...915..129K}  \\
SBS 1116+603  &  2.646 &   $-11.5013   \pm    0.0485$ &   $47.2553 \pm     0.0485$ &    $65^{+17}_{-37}$  &  \citet{2021ApJ...915..129K}   \\
SBS 1425+606  & 3.192 &   $-11.2978  \pm 0.0356$  &    $47.6551  \pm    0.0356$ &   $285^{+30}_{-53}$    & \citet{2021ApJ...915..129K}    \\
\end{longtable}}

\end{document}